\documentclass[12pt,twoside]{report}
\usepackage{comment}
\usepackage{fancyheadings}
\usepackage{defsign,defref}
\usepackage{amsmath,amssymb}
\usepackage{graphicx}
\usepackage{axodraw}
\usepackage{cite}
\newcommand{\clearemptydoublepage}{\newpage{\pagestyle{empty}\cleardoublepage}}
\renewcommand{\chaptermark}[1]%
  {\markboth{\chaptername\ \thechapter\quad #1}{}}

\begin{document}
\pagenumbering{roman}
%----------------------------------------------------------------------------
%\begin{titlepage}
\thispagestyle{empty}
\begin{center}
  \vspace*{2.5cm}
  {\Large
    Doctor Thesis
  }
  \\[5ex]
  {\bf\LARGE 
    Top quark pair production near the 
    \\[0.5ex]
    threshold in $e^+e^-$ collisions
  }
  \\[5ex]
  {\Large
    Takaaki {\sc Nagano}
  }
  \\[4ex]
  {\large\em
    Department of Physics, Tohoku University, Sendai 980-8578, Japan
  }
  \\[12cm]
  {\large \centering
    2000
%    1999
  }
\end{center}
%\end{titlepage}
%----------------------------------------------------------------------------
\newpage
\thispagestyle{empty}
%\begin{abstract}
\vspace*{\fill}
{
  \begin{center}
    \bf Abstract
  \end{center}
  The mass $m_t$ of the top quark $t$ is now measured to be $174\pm 5\GeV$.  
  Improvement of the present error $\Delta m_t = 5\GeV$ is important not 
  only for its own sake.  
%  If the precision $\Delta m_t = 5\GeV$ is improved, 
%  the precision of other theoretical predictions are also improved.  
  For example with $\Delta m_t \sim 0.2\GeV$, the Higgs mass $m_H$ is 
  predicted from electroweak precision measurements 
  with an error of $\Delta m_H/m_H \sim 20\%$, which is now $\sim 60\%$.  
  There is a consensus that $m_t$ is best measured at future 
  $e^+e^-$ Linear Colliders (LC) 
  by using the line shape of the total cross section 
  $\sigma_{\rm tot}(\sqrt{s}\sim 2\mt; e^+e^- \to t\tB)$ 
  near the $t\tB$ threshold.  
  However, it was reported recently that the Next-to-Next-to-Leading-Order 
  (NNLO) correction to 
  $\sigma_{\rm tot}$ is as large as the NLO correction to it.  
  This implies that 
  the convergence of the series is not good, 
%  the higher order corrections to $\sigma_{\rm tot}$ are large, 
  and this theoretical uncertainty affects the precision of the 
  determination of $m_t$ at LC.  
  In the thesis, 
  we improve the convergence of the perturbative series of 
  $\sigma_{\rm tot}$ by choosing appropriate expansion parameters for 
  both $m_t$ and the strong gauge coupling $\alpha_s$.  
  As for the mass parameter, the pole mass was used in the previous studies.  
  However, because of the infrared (IR) structure of QCD, 
  the pole mass of a quark is defined only with an ambiguity of 
  $\Order{\Lambda_{\rm QCD}}\sim 1\GeV$.  
  Instead, we use another mass parameter called potential-subtracted mass, 
  which is less affected by the IR region of QCD.  
  As for the gauge coupling, we resum a certain class of higher order 
  corrections by using renormalization group.  
  With these two prescriptions, the perturbative convergence of 
  $\sigma_{\rm tot}$ is much improved.  Correspondingly, the resulting 
  ambiguity for the determination of $m_t$ is 
  estimated to be $\sim 0.1\GeV$, which is not larger than the expected 
  statistical error at future LC.  
  We also study the top-quark momentum distribution 
  ${\rm d}\sigma/{\rm d}p_t$.  
  The NNLO correction for the line-shape of ${\rm d}\sigma/{\rm d}p_t$ turns 
  out to be fairly small compared to that for $\sigma_{\rm tot}$.  

  Aside from the contribution of Standard Model (SM), we also study 
  that of new physics.  
  Since Cabbibo-Kobayashi-Maskawa matrix elements for the top quark are 
  almost diagonal ($|V_{tb}|\simeq 1$), CP-violating effect in top-quark 
  sector is highly suppressed in SM.  
  In the thesis, 
  we study the effect of all types of anomalous CP violating top-quark 
  Electric Dipole Moment (EDM) interactions 
  near the threshold for the first time.  
  If they are measured to be non-zero, it would immediately 
  imply new physics.  
  We show that one can disentangle $t\tB g$-, $t\tB\gamma$- and $t\tB Z$-EDMs 
  by the dependence of CP-odd observables on the \scCM-energy and 
  on the degree of initial state polarization.  
  The sensitivity to chromo-EDM at LC turns out to be 
  better than that at LC in the open top region ($\sqrt{s} \gg 2\mt$), but 
  to be worse than that at LHC by a factor $\Order{1/10}$.  
%  However, the precise measurement can only be possible at LC.  
  The sensitivities to electroweak-EDMs near the threshold are similar 
  to those in the open top region at LC.  
}
\vspace*{\fill}
%\end{abstract}
%----------------------------------------------------------------------------
\newpage
\setcounter{page}{1}
\clearemptydoublepage
\tableofcontents
%----------------------------------------------------------------------------
%
\clearemptydoublepage
%
%
%--------------------------------------------------------------------------
\chapter{Introduction}
\label{ch:intro}
\pagenumbering{arabic}
%-------------------------------------------------------------------------
\section{Top quark}
As more experimental data is accumulated, 
it has become clearer that the Standard Model (SM) 
describes the physics around and below the ElectroWeak (EW) scale 
very well.  
One of the great triumph of the SM is that the top quark $t$ is discovered 
at Tevatron~\cite{A94} within the range predicted from 
EW precision measurements.  
Currently~\cite{C98}
\begin{align*}
  &  \mt = 174.3 \pm 5.1\GeV
  && \mbox{from direct measurement,}  \\
  &  \mt = 170 \pm 7 (+14)\GeV
  && \mbox{from indirect SM EW fit.}
\end{align*}
For the indirect fit, the central value and the first uncertainty are for 
$m_H = m_Z$, and the second is the shift from changing $m_H$ to $300\GeV$.  
Here $m_H$ is the Higgs mass.  

Despite of its success, few people consider the SM as the ultimate theory 
because of its ``un-naturalness''.  
There are several issues and possible solutions to them.  
First, the gauge symmetry of SM is a direct product of three distinct groups, 
and gauge anomaly cancellation is highly non-trivial.  
These ``un-natural'' aspects are explained if physics at higher energies is 
governed by a Grand Unified Theory (GUT).  
In fact, the three gauge couplings unify around $10^{16}\GeV$ 
in the minimal supersymmetric SM, 
if the soft SUSY breaking scale is around $1\TeV$.  
Weak-scale Super Symmetry (SUSY) itself is motivated to stabilize 
the EW scale against radiative corrections, which is also an 
``un-natural'' aspect of the SM.  

Another one is flavor repetition, which seems redundant, and 
its breaking by masses and mixings.  
In this respect, top quark is highly exceptional, since 
the mass of the next heaviest fermion $b$ is 
$m_b = 4.1\mbox{--}4.4\GeV$~\cite{C98}, which is 40 times smaller than $m_t$.  
Thus one can expect that top quark may play an important role for the 
physics around TeV scale, especially for flavor and Higgs physics.  
In fact, among the ingredients of the SM, 
the mechanism that breaks the electroweak symmetry is the only 
one that is not verified experimentally.  
The situation is similar also for top quark.  
Except its mass $m_t$, 
almost no property of the top quark is known experimentally.  
Thus there are still possibilities of its ``anomalous'' property.  
For example, it might not be merely a ``heavy quark'', 
which means that the generation is not repeating in fact; 
it might be a composite of subquarks, or it might condensate to 
trigger EW symmetry breaking~\cite{MTY89}, 
which means that there is a certain strong attractive force between $t\tB$.  
Or there might be a substantial mixing with a ``fourth generation quark'', 
which can be vector-like.  
On the other hand, even if top quark itself were merely a ``heavy quark'', 
it is expected to be sensitive to the mechanism of $\SU(2)_L\times \U(1)_Y$ 
breaking, because of its large mass and the large mass difference between 
$t$ and $b$ quarks.  
%For example in the decay process $t\to b W^+$, $W$ is dominated by its 
%longitudinal mode, which is a remnant of Goldstone mode, whose interaction 
%grows linearly with its momentum.  

Some properties of top quark will be measured at LHC, which operates before 
$e^+e^-$ linear colliders does.  
With its high \scCM\ energy and high luminosity, 
it is expected that certain clue to new physics will be discovered at LHC.  
However precise measurement at hadron colliders may not be possible, 
since the \scCM\ energy of each elementary process is not fixed.  
Thus it is lepton colliders that can measure the properties of top quark etc.\ 
precisely, after the discovery at Tevatron, or hadron colliders in general.  
%-------------------------------------------------------------------------
\subsection{Top quark pair production in $e^+e^-$ collisions near threshold}
Studies on top quark at $e^+e^-$ colliders can be classified into two 
groups depending on the \scCM\ energy $\sqrt{s}$; 
one is at the threshold region ($\sqrt{s}\sim 2\mt$), and the other is 
at the open top region ($\sqrt{s} \gg 2\mt$).  
One prominent feature of the threshold region, which makes the distinction 
above, is multiple gluon rescattering between $t$ and $\tB$, which is 
effectively not suppressed by powers of $\alpha_s$.  
%Thus QCD corrections are enhanced at the threshold region.  
In fact, top quark is an ideal probe of the short distance behavior of 
QCD.  
This is because of its large mass $\mt$ and its large decay width $\Gamma_t$, 
which are related~\cite{BDKKZ86,JK89}%
\footnote{
  More detailed expression is shown in \SecRef{sec:Gamma_t}.  
}
as follows: 
\begin{align*}
  \Gamma_t = \frac{G_F\mt^3}{8\pi\sqrt{2}} \simeq 1.5\GeV
  \sepperiod
\end{align*}
This expression holds at tree level in the limit $m_W/m_t \to 0$, and 
is for the SM, where almost every $t$ decays into $bW^+$.  
Here $|V_{tb}| \simeq 1$ is used, which is indeed the case 
when the unitarity of $V_{\rm CKM}$ among the first three generations is 
assumed.  
When the distance between $t$ and $\tB$ is shorter than the 
typical non-perturbative scale $\sim 1/\Lambda_{\rm QCD} \sim 1/(1\GeV)$, 
gluon exchanges between them can be approximated by the Coulomb potential%
\footnote{
  This expression is for a color-singlet $q\qB$.  
  For a color-octet state, it is 
  $V = (C_A/2-C_F)\alpha_s/r$, where $C_A=3$.  
}
$V = -C_F \alpha_s/r$, where $C_F=4/3$ is the quadratic Casimir for 
the fundamental representation of $\SU(3)$.  
Due to the large mass, the Bohr radius%
\footnote{
  It is defined in \SecRef{sec:G_for_pure_Coulomb}.  
  See also \FigRef{fig:alphaV_20-75-q}.  
}
$r_{\rmB}$ of the $t\tB$ pair 
is much smaller than the typical non-perturbative scale: 
$r_\rmB = 1/(C_F\alpha_s\mt/2) \simeq 1/(20\GeV) \ll 1/\Lambda_{\rm QCD}$.  
This means that the QCD potential between $t$ and $\tB$ is 
surely well within the perturbative region.  
And due to the large decay width, a $t\tB$ pair decays before 
it hadronizes~\cite{BDKKZ86}; 
that is, the fluctuation of top quark momentum%
\footnote{
  The peak momentum of the top quark is 
  $\sim |\sqrt{m_t(\sqrt{s}-2m_t+1\GeV+i\Gamma_t)}|$.  
  See \FigRef{fig:FT_scanE_p_peak}.  
  The fluctuation given here is for near the threshold.  
} 
$\sqrt{m_t\Gamma_t} \simeq 15\GeV \gg \Lambda_{\rm QCD}$ 
due to the off-shell effect acts as an IR cut-off.  
Thus a $t\tB$ pair produced in $e^+e^-$ collisions can be treated 
perturbatively from the production to the decay, 
even near the threshold, where the QCD effect is enhanced.  
Thus the threshold region is one of the best places for the 
study of perturbative QCD.  

There are many studies on $t\tB$ pair production near the threshold in 
$e^+e^-$ collisions~\cite{F94,FK87,FK91,SP91,FY91,JKT92,SFHMN93,MS93,S93,MY94,FMS94,MK94,J94,HJKT95,FKMC95,M96,HJKP97,PeSu98,HT98,MY98,Y99,NOS99,BSS99,HT99,KP99}, 
including ours.  
Some of them are reviewed briefly in the next two sections, and 
in the former halves of Chapters~\ref{ch:total_cr}~and~\ref{ch:EDMs} 
at some length.  
Concisely speaking, we (i) reduce the theoretical uncertainty for the 
determination of the top-quark mass using the process $e^+e^- \to t\tB$ 
near the threshold, and (ii) study the sensitivity of $e^+e^-$ colliders 
with $\sqrt{s}\simeq 2\mt$ to all the anomalous Electric Dipole Moments 
in $t$-$\tB$-gauge boson interactions for the first time.  
%-------------------------------------------------------------------------
\section{Overview of $t\tB$ production cross section}
%-------------------------------------------------------------------------
\subsection{LO and NLO calculations of $\sigma_{\rm tot}$ and ${\rm d}\sigma/{\rm d}p$}
In $e^+e^-$ collisions with $\sqrt{s}\simeq 2m_q$ where $q=c,b$, 
there are sharp resonances in the 
total production cross section $\sigma_{\rm tot}(s)$ 
due to the $q\qB$ bound states such as $J/\Psi$ or $\Upsilon$.  
The locations of those bound state poles are used to probe the QCD 
potential at $r \sim 1/(\alpha_s m_q)$, which is near the boundary of the 
perturbative description of QCD~\cite{QCDpot_cb}.  
On the other hand, 
due to the large decay width of top quark $\Gamma_t$, 
(would-be) $t\tB$ bound-states merge into 
a single slight rise of $\sigma_{\rm tot}(s)$ near the $1S$ resonance.  
Thus in order to calculate $\sigma_{\rm tot}(s)$ just below the threshold 
($\sqrt{s} \lesssim 2\mt$) reliably%
\footnote{
  The production cross section grows rapidly near the threshold, 
  because the leading contribution is $S$-wave.  
  The mass of top quark is determined by measuring this rise of 
  $\sigma_{\rm tot}$.  
  Thus the precise calculation of $\sigma_{\rm tot}$ near the threshold is 
  required for this reason.  
  Note also that the ``binding energy'' for the $1S$ state is 
  $\sim p_{\rm B}^2/m_t \sim 2\GeV$, where $p_{\rm B}=1/r_{\rm B}$ is 
  Bohr momentum.  See \SecRef{sec:G_for_pure_Coulomb}.  
}, 
one needs to sum over a host of resonances.  
This summation is conveniently done using 
the Green function for a $t\tB$ pair~\cite{FK87}: 
\begin{align*}
  G(\rBI,\rBI') &\equiv \braket{\rBI}{\frac{1}{H-(E+i\Gamma_t)}}{\rBI'}  \\
  &= \sum_{n} \frac{\psi_n(\rBI)\,\psi_n^*(\rBI')}{E_n-(E+i\Gamma_t)}
  + \int\!\!\frac{\diffn{k}{3}}{(2\pi)^3}
    \frac{\psi_{\vec{k}}(\rBI)\,\psi_{\vec{k}}^*(\rBI')}
         {E_{\vec{k}}-(E+i\Gamma_t)}  \sepcomma
\end{align*}
where $H$ is the Non-Relativistic (NR) Hamiltonian for $t\tB$, 
$E \equiv \sqrt{s}-2\mt$ is the NR energy, and 
$\psi_n$ and $\psi_{\vec{k}}$ are energy eigenfunctions for 
$E_n (<0)$ and $E_{\vec{k}}\equiv \vec{k}^2/\mt (>0)$, respectively.  
Since top quarks are non-relativistic near the threshold, it is 
convenient to use a non-relativistic approximation at the leading order.  
There is another advantage for the NR Green-function treatment.  
Near the threshold, it is known that the naive perturbative expansion 
in $\alpha_s$ breaks down (``threshold singularity'').  
%This phenomenon, called ``threshold singularity'', 
%originates from the momentum configurations of Feynman diagrams that all the 
%intermediate particles are simultaneously nearly on-shell, 
%which is possible near the threshold.  
In fact, $n$-th power of $\alpha_s$ is accompanied by 
$1/\beta_t^{~m}$ ($m\leq n$), where $\beta_t$ is the velocity of 
$t$ and $\tB$ at their \scCM\ frame.  
The situation is similar to certain perturbation series where 
powers of $\alpha_s$ are accompanied by powers of $\log(q/\mu)$; 
a stable perturbative expansion is obtained by summing up (potentially 
large) leading log's.  In other words, one should identify 
$\sum_n (\alpha_s(\mu)\,\ln(q/\mu))^n$ to be the leading order.  
Likewise, one should identify $\sum_n (\alpha_s/\beta_t)^n$ to be 
the leading order near the threshold.  
On the other hand, this summation is realized%
\footnote{
  The explicit expression is given in \EqRef{eq:R_LO_noGamma}.  
  Note that $\sqrt{E/\mt} = \beta_t + \Order{\beta_t^3}$.  
}
automatically in the Green function for Coulomb potential between 
$t$ and $\tB$.  
Thus the Green function method provides an efficient way 
(i) to sum up infinite number of broad resonances, and 
(ii) to sum up powers of $\alpha_s/\beta_t$.  
After these summations, 
the cross section is finite%
\footnote{
  See \FigRef{fig:Rfree}.  
}
even at the threshold, which corresponds%
\footnote{
  Due to finite decay width $\Gamma_t$, velocity $\beta_t$ 
  of top quark is not determined by the energy.  
  Its typical fluctuation is 
  $\beta_t \sim \sqrt{\Gamma_t/m_t} \simeq 0.1 \sim \alpha_s$.  
  See just below \EqRef{eq:R_free_LO_withGamma}.  
  This is consistent with our order counting.  
  The symbol $c$ is speed of light.  
}
to $\beta_t = 0$.  
Order counting becomes easier by noting 
$\alpha_s = \gs^2/(4\pi\hbar c) = \Order{1/c}$ and 
$\beta_t = v_t/c = \Order{1/c}$, which means $\alpha_s/\beta_t = \Order{1}$.  
This is the Leading Order (LO).  
Likewise the Next-to-Leading Order (NLO) is $\Order{1/c}$, and so on.  

The total cross section at the leading order was calculated in~\cite{SP91} 
with the QCD potential $V(r)$ that is Renormalization Group (RG) 
improved in the coordinate space.  
It triggered several studies both by experimentalists and 
by theorists~\cite{JKT92,SFHMN93,MS93}.  
A part of NLO corrections was taken into account in their analysis.  
%They elucidated large theoretical ambiguity due to the higher order 
%corrections.  
After a while, complete NLO corrections to $t\tB$ production and 
their decay cross section were calculated~\cite{S93,MY94}.  
They included the Final-State Interactions (FSIs), by which we mean 
gluon exchange between $t$ and $\bB$, $\tB$ and $b$, and $b$ and $\bB$.  
It was also shown that although these FSIs modify the momentum 
distribution of $t$, their corrections to the total cross section 
cancel to NLO.  

Top--anti-top pair production near the threshold in $e^+e^-$ collisions 
not only provides the place for a precise study of perturbative QCD, but also 
was known to be the best environment to determine the top quark mass $m_t$.  
This is because 
Monte Carlo detector simulations~\cite{F94,A98,FMS94} based on NLO 
calculations~\cite{SFHMN93,MS93,S93} 
showed that $m_t$ can be determined with statistical error%
\footnote{
  On the other hand, LHC may give $(\Delta\mt)^{\rm stat} \sim 2\GeV$.  
}
$(\Delta\mt)^{\rm stat} \simeq 0.2\GeV$ using 
the energy dependence of $R$-ratio and 
the top quark momentum distribution $\diffn{\sigma}{}/\diffn{p}{}$.  
It was also shown that QCD gauge coupling $\alpha_s(m_Z)$ can be measured 
with statistical error $0.005$.  
It was assumed that 11 energy points are sampled with $1\fb^{-1}$ for each 
point.  A similar MC study is shown in~\cite{A98}.  
These MC studies are based on Next-to-Leading Order 
(NLO = $\Order{1/c}$) theoretical calculations and assume no 
theoretical uncertainties.  
%
%Monte Carlo detector simulation studies 
%based on the NLO theoretical calculations are reported in~\cite{F94,A98}.  
%It was shown that statistical errors are estimated to be 
%$\Delta m_t \simeq 0.2\GeV$ and $\Delta \alpha_s(m_Z) \simeq 0.005$ 
%taken into account a realistic experimental environment.  
%Thus there was a consensus that the mass $\mt$ of top quark is 
%best measured in $e^+e^-$ (linear) colliders with \scCM\ energy 
%near the threshold of $t\tB$ pair production, $\sqrt{s}\simeq 2\mt$.  
%-------------------------------------------------------------------------
\subsection{NNLO calculations of $\sigma_{\rm tot}$}
However recently 
%a part of 
the NNLO corrections to the total cross section 
$\sigma_{\rm tot}$ has been calculated by several groups~\cite{HT98,MY98,Y99}%
\footnote{
  Their results are reproduced in \SecRef{sec:R_NNLO_fixed}.  
  Comprehensive summary (including our work) to the end of 1999 is 
  available in~\cite{Hetal00}.  
}, 
and it turned out to be unexpectedly large.  
In fact the magnitude of the NNLO correction is similar to that of 
the NLO correction.  This is problematic since 
it is desirable to determine the top-quark mass in high precision 
in order to pin down contributions from new physics (and/or Higgs).  
For example~\cite{A98} for wide range of Higgs mass $m_H$, 
the uncertainty ($\Delta m_H/m_H$) of Higgs mass 
extracted from radiative corrections depends on the uncertainty of 
$\mt$, and is $57\%$ for the present precision, 
while it reduces to $17\%$ for Linear Colliders (LCs) 
assuming $(\Delta\mt)^{\rm stat} \simeq 0.4\GeV$.  
%-------------------------------------------------------------------------
\subsection{Theoretical progress related to $\sigma_{\rm tot}$}
After the calculations of the NLO corrections to $\sigma_{\rm tot}$, 
several pieces of theoretical progress have been made with 
the precise calculation for heavy quark-antiquark system~\cite{B99a}.  
One is the use of low energy effective theories called 
NRQCD~\cite{CL86,BBL95} and pNRQCD~\cite{PS98,BPSV99}, 
which enabled the calculations of the NNLO corrections explained above, 
and the other is the cancellation of the ambiguities in the pole mass 
of a quark and in the coordinate space QCD potential; 
those ambiguities are due to the infrared structure of QCD, 
which is called ``renormalon ambiguity''~\cite{T77,M92,Z98,B99,B98,HSSW99}.  
The latter enables us to improve the perturbative convergence of 
$\sigma_{\rm tot}$, which is made in \cite{NOS99}, 
on which a part of the thesis is based.  
Here we explain these two topics briefly.  

At NNLO and beyond, non-relativistic calculations become complicated.  
There are several reasons.  Typical one is that the normalization of 
the $t\tB$ current $j_{t\tB}$ depends on the momentum transfer; 
in other words, anomalous dimension of $j_{t\tB}$ is non-zero.  
This makes the use of low-energy effective theories particularly powerful.  
Schematically, the reduction goes as follows.  
The starting point is QCD, which is fully relativistic.  
By expanding the fields for non-relativistic particles in $1/c$, 
one obtains NRQCD~\cite{CL86,BBL95}.  
At NNLO (and lower orders) further reduction is possible, since 
no real gluon is emitted%
\footnote{
  This may be understood by the fact that the vertex for transverse 
  gluon $g_T$ in Coulomb gauge is suppressed by $\Order{1/c}$ 
  compared to the vertex for Coulomb gluon $g_C$.  
}
at these orders.  Thus ``soft'' gluons can be integrated out 
to result in a $t\tB$ potential~\cite{PS98,BPSV99}.  
It is known that this formalism, called potential NRQCD or pNRQCD, 
reproduces correctly the energy shift of the hydrogen atom and 
positronium to $m\alpha^2 \times \alpha^3$~\cite{PS98a}. 
%; it includes the Lamb shift: $E_{2s\frac{1}{2}}-E_{2p\frac{1}{2}}$, 
%where the subscripts are $(n,\ell,j)$.  
%
For actual calculations, 
one does not need to know the precise ``normalization'', or the 
matching coefficients, of the each operator in pNRQCD Lagrangian, a priori.  
Only a combination of them is relevant for, say, a total production cross 
section $\sigma_{\rm tot}$.  
It can be determined by calculating $\sigma_{\rm tot}$ in both 
perturbative QCD and pNRQCD, and by demanding those two results coincide.  
This procedure called ``direct matching''~\cite{H97} is possible%
\footnote{
  Of course one needs the result for perturbative QCD.  
  For the case we deal, $\Order{\alpha_s^2}$ corrections to 
  $\sigma(e^+e^- \to \gamma^* \to t\tB)$ are calculated in~\cite{CM98}.  
}
since for $\alpha_s \ll \beta_t \ll 1$ 
both perturbative QCD and pNRQCD are applicable.  

On the other hand, ``renormalon ambiguity''~\cite{T77,M92,Z98,B99} 
is related to the convergence of a perturbative expansion in QFT, 
which is in fact an asymptotic expansion.  
One of the techniques to sum up an asymptotic series $S(a)$ 
with respect to $a$ is Borel summation%
\footnote{
  Here $a$ may be considered as a coupling.  
  Asymptotic behavior of $S(a)$ can be inferred by 
  using renormalization group, for example.  
  See \SecRef{sec:Borel_sum} for detail.  
}, 
where the sum is expressed in terms of integration of 
$\expo^{-t/a}\tilde{S}(t)$ from $t=0$ to $t=\infty$.  
Here $\tilde{S}(t)$ is the function related to $S(a)$.  
The fact that $S(a)$ diverges is expressed by the poles of $\tilde{S}(t)$ on 
the integration path $t=0$--$\infty$, which are called 
``renormalon poles''.  One can step aside those poles by modifying the 
integration path; there may be two ways for each pole.  
Ambiguities due to those poles may be estimated by the difference between 
the two choices of the modification.  
Due to the factor $\expo^{-t/a}$, the severest ambiguity comes from 
the ``renormalon pole'' nearest to the origin ($t=0$).  
Now, it is known that the pole mass $m_{\rm pole}$ suffers from 
the renormalon poles~\cite{BB94,BSUV94,B95,SW97}, which 
originate from the IR structure of QCD.  
However it was shown~\cite{B98,HSSW99} that in the combination 
$2m_{\rm pole}+V(r)$, where $V(r)$ is the coordinate space QCD potential, 
the severest renormalon pole is cancelled; this means $V(r)$ also 
suffer from the renormalon pole, and the location of the pole is the same 
as for $m_{\rm pole}$ and the residue is twice and the sign is opposite.  
This suggests that one should not attempt to determine pole mass.  
Instead another mass scheme that do not sensitive to IR physics of QCD, 
such as \MSbar\ mass, should be used.  
Several such ``short-distance'' mass schemes are proposed in literature; 
potential-subtracted mass~\cite{B98}, $1S$ mass~\cite{HLM99}, 
and kinetic mass~\cite{BSU97}.  
%-------------------------------------------------------------------------
\subsection{Our contributions}
As was explained above, 
the previous NNLO calculations of $\sigma_{\rm tot}$ showed 
large corrections.  
However, since the convergence of a perturbative series changes depending on 
the renormalization scheme%
\footnote{
  A different renormalization scheme (and renormalization scale) 
  corresponds to a different way of summing up a perturbation series.  
}, 
there is a possibility for finding a scheme where $\sigma_{\rm tot}$ 
converges better.  
This is the subject of the latter half of \ChRef{ch:total_cr}.  
There we implement two prescriptions; one is to use the mass scheme 
that do not suffer from IR renormalon ambiguity; and the other is 
to sum up (potentially large) leading logarithms by using renormalization 
group.  
Both of these reduce the theoretical uncertainty of $\sigma_{\rm tot}$.  
With this improvement, 
we estimate 
the theoretical uncertainty $(\Delta\mt)^{\rm th}$ of the top quark mass 
determination is reduced to $\sim 0.1\GeV$, which is smaller than 
the expected statistical error.  
%This is one of our results in the thesis.  

We also calculate the NNLO correction for ${\rm d}\sigma/{\rm d}p$ 
from the rescattering between $t$ and $\tB$.  
However it is not considered that the rescattering between $t$ and 
$\bB$ etc., which may modify the distribution to the similar extent.  
%------------------------------------------------------------------------
\section{Overview of top-quark anomalous EDMs}
Of course top quark is (potentially) sensitive not only to the SM dynamics 
but also to the physics beyond the SM.  One particularly clear signal is 
\itCP\ violation in top-quark sector.  
This is because the SM contribution is suppressed to many orders below the 
current or near-future experimental reach%
\footnote{
  This is because 
  the CKM matrix elements for the third generation are almost diagonal: 
  $|V_{tb}| \simeq 1$.  
  See \SecRef{sec:EDM-pred} for more quantitative statements.  
}.  
Thus a non-zero expectation value of a CP-odd observable means contribution 
from new physics immediately%
\footnote{
  Provided that the initial state is a $\ChargeCOp\ParityOp$ eigenstate.  
}.  
In fact there are plenty of new sources for \itCP\ violation 
once the particle content of the SM is extended.  
Many models, including Minimal SuperSymmetric SM (MSSM) 
or multi Higgs-doublets model in general, 
induce \itCP\ violating $t$-$\tB$-gauge 
boson interactions at one-loop level.  
In the \scCM\ frame of $t\tB$ pair, or a particle--anti-particle system 
in general, a CP-odd observable related to them is odd under 
$\sBI_t \leftrightarrow \sBBI_t$, and vice versa, 
since $\ChargeCOp\ParityOp$ transformation exchanges their spins $\sBI$ 
at the frame.  
Also in this respect top quark is excellent since the polarization $\PolBR_t$ 
of $t$ is not disturbed by hadronization, and the information of 
$\PolBR_t$ is inherited to its decay products; especially, 
the charged lepton $\ell^+$ is emitted with the angular distribution%
\footnote{
  See \SecRef{sec:pol_t=p_ell}.  
}
$\propto 1+|\PolBR_t|\cos\theta$, where $\theta$ is the angle between 
the polarization of $t$ and the momentum of $\ell^+$.  
Thus the \itCP\ violations in the top-quark sector can be measured by the 
difference $\PolBR_t-\PolBBR_t$ 
between the polarizations of $t$ and $\tB$, which can be 
measured statistically from the directions of the charged leptons.  
%
%One of the features of the threshold region is that 
%``anti-hermitian part'' (= absorptive part) of an amplitude is as large as 
%``hermitian part''.  This is due to Coulomb rescattering between $t\tB$, 
%which is enhanced near the threshold.  
%This enables one to measure both ``hermitian'' and ``anti-hermitian'' parts%
%\footnote{
%  Note that the anti-hermitian part of an effective coupling parametrizes 
%  absorptive part \SecRef{sec:absorp-CPTT}.  
%}
%of EDM couplings simultaneously.  

Since many extensions of the SM contain new sources of \itCP\ violation, 
we parameterize their effects by effective couplings 
in a model independent way.  
Among CP violating interactions of $t$, 
Electric Dipole Moment (EDM) interaction is the lowest dimensional 
operator%
\footnote{
  See \SecRef{sec:Gordon-id}.  
}.  
Thus in terms of an effective Lagrangian, 
\itCP\ violating effects in the top quark sector are 
parameterized by several anomalous EDMs of top quark 
at the first approximation.  
There are several of them: 
$t\tB g$-, $t\tB\gamma$-, $t\tB Z$-, and $t\bar{b}W$-EDM.  
The first three affect the production (and rescattering) process of $t\tB$, 
while the last one the decay process.  
%------------------------------------------------------------------------
\subsection{Previous studies}
Many studies%
\footnote{
  See Sections~\ref{sec:EDM-pred} and~\ref{sec:EDMstudy_open} for 
  the references.  
}
have been done for the anomalous EDMs of top quarks.  
One major topic is how well those EDMs are measured in future colliders.  
Some study for $e^+e^-$ colliders with $\sqrt{s} \gg 2\mt$, and 
the others for hadron colliders.  
The former is suited for the study of EW-EDMs, $t\tB\gamma$ and $t\tB Z$, 
while the latter for Chromo-EDM, $t\tB g$.  
Both of these results are summarized in \SecRef{sec:EDMstudy_open}.  
%------------------------------------------------------------------------
\subsection{Our contributions}
Although many studies have done for $e^+e^-$ colliders with 
$\sqrt{s} \gg 2\mt$ and for hadron colliders, there was no one 
for $e^+e^-$ colliders near the threshold to date.  
Thus we study this case.  
We concentrate on the anomalous EDMs in the production process 
($t\tB$-gauge bosons) here, and 
leave $t\bar{b}W$-EDM for a future study.  
Among the three flavor-diagonal EDMs, our prime concern is on 
Chromo Electric Dipole Moment (CEDM): $t\tB g$-EDM.  
The reason is as follows.  
As shall be shown in \SecRef{sec:corr_from_Coulomb=EDM}, 
two ElectroWeak EDMs, 
$t\tB\gamma$- and $t\tB Z$-EDM, directly modify the $t\tB$ production vertex, 
and their interactions are proportional to the relative momentum 
$|\pBI|\simeq\sqrt{\mt E}$ between $t\tB$.  This means open top region 
($\sqrt{s}\gg 2\mt$) is more appropriate to study them.  
On the other hand CEDM modifies the rescattering of $t\tB$, 
which can is suppressed when $\sqrt{s}\gg 2\mt$.  
Thus in open top region at lepton colliders, 
it can be studied only by real gluon emissions.  
We find that the threshold region is more sensitive to CEDM 
than the open top region.  
As for hadron colliders, they are sensitive to the existence of a non-zero 
CEDM, since $t\tB$ pairs are copiously produced in gluon fusions.  
In fact our result for CEDM is that the sensitivity of 
$e^+e^-$ colliders at the threshold are worse than that of hadron colliders.  
However precise measurement of CEDM may be difficult in hadron colliders.  
Moreover, while detailed detector simulation seems to be indispensable for 
the serious study of the sensitivity at hadron colliders, 
it has not been done so far%
\footnote{
  At least to our knowledge.  
}.  
On the other hand, 
a full detector simulation with realistic experimental setup 
is in progress~\cite{IF99} based on our study presented here.  
Thanks to the clean environment of $e^+e^-$ collisions, 
their first result for the sensitivity is consistent with 
our naive estimate.  
%------------------------------------------------------------------------
\section{Organization of the thesis}
In the thesis I study%
\footnote{
  This work is based on the collaborations with 
  Y.~Sumino and A.~Ota~\cite{NOS99}, 
  and with M.~Je\.{z}abek and Y.~Sumino~\cite{JNS99}.  
}
the physics of $t\tB$ threshold in $e^+e^-$ collisions for 
the study of both in the SM and beyond the SM dynamics.  

Chapters~\ref{ch:total_cr} and~\ref{ch:diff_cr} are devoted to the 
SM dynamics.  
Total production cross section $\sigma_{\rm tot}$ is studied 
in \ChRef{ch:total_cr}.  
By reorganizing perturbation series, we obtain better convergence 
of $\sigma_{\rm tot}$ to NNLO than ever obtained.  
With this result, theoretical uncertainty to measure $m_t$ becomes 
smaller than experimental uncertainties.  
The chapter also includes a review of previous studies, of theoretical 
set-up such as the Green function method to calculate $\sigma_{\rm tot}$.  
A derivation of non-relativistic Hamiltonian to NNLO is also given.  
\ChRef{ch:diff_cr} deals with the top-quark momentum distribution.  
Coulomb rescattering between $t\tB$ is treated to NNLO.  

On the other hand, \ChRef{ch:EDMs} is devoted to the contributions from 
new physics, especially to new sources of $\ChargeCOp\ParityOp$ violation.  
It is explained that, the $\ChargeCOp\ParityOp$ violation in 
top-quark sector within the SM is so small that if it is discovered 
it would immediately imply new physics.  
$\ChargeCOp\ParityOp$ violation in top-quark sector is parameterized by 
Electric Dipole Moment (EDM) interactions of top quark, since it is 
the lowest dimension operator that violates $\ChargeCOp\ParityOp$ 
symmetry.  Many studies have been done on EDMs of top quark, but 
they are all in the open-top region $\sqrt{s} \gg 2m_t$.  
We analyze them in the threshold region $\sqrt{s} \simeq 2m_t$ 
for the first time.  
Due to the multiple Coulomb rescattering near the threshold, 
differential production cross section is sensitive to $t$-$\tB$-gluon 
anomalous EDM even when $t\tB$ is produced in $e^+e^-$ collisions.  
We find that the sensitivity of future $e^+e^-$ Linear Colliders (LC) is 
worse than that at hadron colliders by factor $\Order{1/10}$ or less.  
%where top quark are produced about $10^3$ times more than in LC.  
Sensitivities for $t\tB\gamma$- and $t\tB Z$-EDMs are also studied, 
and are found to be comparable to those in the open-top region at LC.  

Summary and discussions are given in Chapter 5.  
\clearemptydoublepage
%
%
%-------------------------------------------------------------------------
\chapter{Total Production Cross Section}
\label{ch:total_cr}
%-------------------------------------------------------------------------
\section{Overview}
As was explained in \ChRef{ch:intro}, 
it was considered that the mass $m_t$ of top quark $t$ can be 
determined most accurately by the energy scan of $t\tB$ total production 
cross section near the threshold in $e^+e^-$ collisions.  
However there emerge the fear that this way of determination 
suffer from a large theoretical uncertainty, 
due to the large NNLO (= $\Order{1/c^2}$) correction to the $t\tB$ 
production cross section $\sigma_{\rm tot}(s)$, where $\sqrt{s}$ denotes the 
\scCM\ energy.  

In the thesis, we follow closely the procedure used in~\cite{MY98}.  
There the Green function $G$ for a $t\tB$ system is rewritten 
in a form convenient for numerical calculations.  
By implementing ``renormalon cancellation'' and ``$\log(q/\mu)$ resummation'', 
we improve the convergence of $\sigma_{\rm tot}$ obtained 
in~\cite{HT98,MY98,Y99}.  

This chapter is organized as follows.  
In \SecRef{sec:NRdescr_ttB}, it is explained that how an non-relativistic 
(color-singlet) $t\tB$ pair is described in terms of the Green function 
of the pair.  In particular, the total production cross section is 
given by the optical theorem, which is explained in \SecRef{sec:optical}.  
In \SecRef{sec:order_count}, order counting is explained.  
Also shown is that the Coulomb potential is a part of leading Hamiltonian.  
At NNLO, there arise several difficulties, which is given 
in \SecRef{sec:complications_at_NNLO}.  In \SecRef{sec:NR-H}, the Hamiltonian 
to NNLO is shown.  The Hamiltonian is looked at more closely in 
\SecRef{sec:NR-pot}.  
We obtain the Green function $G$ by solving the Schr\"odinger equation 
numerically.  For this purpose, we express $G$ in terms of reduced 
Green function $G'$, which satisfies simpler Schr\"odinger equation than 
$G$ itself.  This is done in \SecRef{sec:reduced_G}.  
At that stage, we are equipped to reproduce the previously obtained 
results for the total cross section, including NNLO ones.  
They are collected in \SecRef{sec:prev_res_R}, and exhibit large 
correction at NNLO.  Hereafter, we try to improve the convergence of 
the perturbative expansion.  It is known that (certain) perturbative 
expansions involving the pole mass of a quark are less converging than 
those involving \MSbar\ mass, or ``short-distance'' mass.  
It is suspected that this is because pole mass suffer from non-perturbative 
dynamics of QCD.  \SecRef{sec:renormalon} is devoted to this issue.  
There we adopt potential-subtracted mass scheme, which is proposed 
in~\cite{B98}.  This improves the convergence of binding energies.  
In \SecRef{sec:RGimp_R}, certain kind of logarithm in the 
Coulombic potential is summed up using renormalization group.  
Tentatively, the convergence of $R$ ratio is best with these two 
prescriptions.  The section contains the estimate of theoretical 
uncertainties for this case; in terms of top quark mass, 
it is $(\Delta \mt)^{\rm th} \sim 0.1\GeV$.  
%
%
%

%
%
%-------------------------------------------------------------------------
\section{Non-relativistic description of $t\tB$ system}
\label{sec:NRdescr_ttB}
%----------------------------------------------------------------------
\subsection{Optical Theorem}
\label{sec:optical}
In terms of an amplitude $\calM$, the unitarity [\EqRef{eq:unitarity_T}] 
of $S$-matrix is expressed by optical theorem, 
which relates the total cross section of a two-body scattering to the 
imaginary part of the forward scattering amplitude: 
\begin{align*}
  \sigma_{\rm tot} (k_1,k_2 \to \mbox{anything})
  = 
  \frac{1}{s\betaB_i} 
  \Im \calM (k_1,k_2 \to k_1, k_2)
  \sepcomma
\end{align*}
where $\betaB_i = 2|\kBI_{\rm CM}|/\sqrt{s}$, and 
$|\kBI_{\rm CM}| \equiv |\kBI_1| = |\kBI_2|$ at the \scCM-frame of 
two particles with momenta $k_1$ and $k_2$.  
If we restrict the intermediate states appropriately on the RHS, 
we have the total cross section for $k_1,k_2 \to \mbox{such-and-such}$.  
For the process 
$e^+(\pB_e) e^-(p_e) \to \gamma^* \to t\tB$, 
the total cross section of this process can be obtained from the imaginary 
part of the amplitude for 
$e^+(\pB_e) e^-(p_e) \to \gamma^* \to t\tB 
  \to \gamma^* \to e^+(\pB_e) e^-(p_e)$.  
Let $(-ieQ_t)^2 \cdot i\Pi^{\mu\nu}(q)$ be the $t\tB$ contribution to the 
vacuum polarization of a photon with momentum $q^\mu$: 
\begin{align*}
  i\Pi^{\mu\nu}(q) 
  &= 
  i(q^2 g^{\mu\nu} - q^\mu q^\nu) \Pi(q^2)  \\
  &= 
  \int\!\!\diffn{x}{4} \expo^{iq{\cdot}x} \bra{0}
  \rmT [ j^{\mu}(x) j^{\nu}(0) ] \ket{0}  \\
  &=
  \bra{0} \rmT [ \tilde{\jmath}^{\mu}(q) 
                 \tilde{\jmath}^{\nu}(-q) ] \ket{0}  \\
  &=
  (-1) N_C \int\!\!\frac{\diffn{p}{4}}{(2\pi)^4}
  \frac{i}{(p-q/2)^2-m^2+i\epsilon} \frac{i}{(p+q/2)^2-m^2+i\epsilon} 
  \\
  &\qquad{}\times
  \btr{(\pS-\qS/2+m)\gamma^\mu
       (\pS+\qS/2+m)\gamma^\mu}
  \sepcomma
\end{align*}
where $j^{\mu}(x) = \tB(x)\gamma^\mu t(x)$, 
and the last expression corresponds to one-loop.  
By contracting indices, we have 
\begin{align*}
  \Pi(q^2)
  = 
  \frac{-i}{3q^2}
  \bra{0} \rmT [ \tilde{\jmath}^{\mu}(q) 
                 \tilde{\jmath}_\mu(-q) ] \ket{0}  
  =
  \frac{1}{3q^2} g_{\mu\nu} \Pi^{\mu\nu}(q)
  \sepperiod
\end{align*}
Thus for unpolarized $e^+ e^-$, 
\begin{align*}
  i\calM (p_e,\pB_e \to \cdots \to p_e,\pB_e) 
  &=
  (-ieQ_t)^2 \cdot
  i(q^2 g^{\mu\nu} - q^\mu q^\nu) \Pi(q^2) \pfrac{-i}{q^2}^2 \times \\
  &\qquad{}\times
  \frac{1}{2^2}\sum_{\rm spins}
  \btr{u(p_e)\uB(p_e)\gamma_\mu v(\pB_e)\vB(\pB_e)\gamma_\nu}  \\
%  &=
%  i(4\pi\alpha)^2 Q_e^2 Q_t^2 N_C \cdot
%  q^2 \Pi(q^2) \pfrac{-i}{q^2}^2 (-q^2)  \\
  &=
  i (4\pi\alpha)^2 Q_e^2 Q_t^2 \, \Pi(q^2)
  \sepperiod
\end{align*}
Here we neglect electron mass.  
Note that the $q^\mu q^\nu$-part of $\Pi^{\mu\nu}$ do not contribute, 
since $e^+ e^-$-current is conserved: 
$\vB(\pB_e)\qS u(p_e) = 0$, where $q^\mu = (p_e+\pB_e)^\mu$.  
Thus, with $s = q^2$, 
\begin{align}
  R(s) &\equiv 
  \frac{\sigma_{\rm tot}(e^+ e^- \to \gamma^* \to t\tB)}{\sigma_{\rm pt}}  
  \label{eq:R_def}\\
  &=
  Q_t^2 \, 12\pi \Im\Pi(q^2) 
  =
  Q_t^2 \, \frac{4\pi}{s}
    \Im \Bigl[ -i \bra{0} \rmT [ \tilde{\jmath}^\mu(q) 
                 \tilde{\jmath}_\mu(-q) ] \ket{0} \Bigr]
  \sepcomma\nonumber
\end{align}
where 
\begin{align}
  \sigma_{\rm pt}(s) \equiv \frac{4\pi\alpha^2}{3s} 
  = 0.817 \pb \left( \frac{\sqrt{s}}{2\times 175\GeV} \right)^{-2}
  \label{eq:sigma_pt}
\end{align}
is ``point cross section'', the total cross section 
for $e^+ e^- \to \gamma^* \to \mu^+\mu^-$ with $\sqrt{s} \gg 2m_\mu$.  

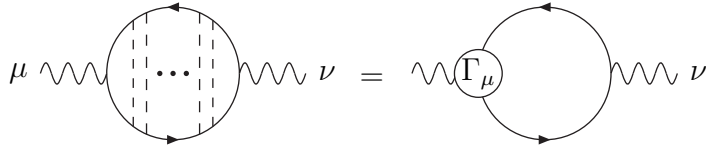
\begin{figure}[htbp]
  \hspace*{\fill}
  \lower 3cm \hbox{
  \begin{picture}(120,50)(-60,-50)
    \normalsize
    \ArrowArc(0,0)(25,0,180)
    \ArrowArc(0,0)(25,180,360)
    \Photon(-50,0)(-25,0){4}{3}
    \Photon(25,0)(50,0){4}{3}
    \DashLine(-15,-20)(-15,20){4}
    \DashLine(-10,-23)(-10,23){4}
    \Vertex(-5,0){1}
    \Vertex(0,0){1}
    \Vertex(5,0){1}
    \DashLine(10,-23)(10,23){4}
    \DashLine(15,-20)(15,20){4}
    \Text(-55,0)[r]{$\mu$}
    \Text(55,0)[l]{$\nu$}
  \end{picture}
%  \hspace*{\fill}
  \hspace{0.1cm}
  \raise 1.6cm\hbox{$=$}
  \hspace{0.1cm}
%  \hspace*{\fill}
  \begin{picture}(110,50)(-50,-50)
    \normalsize
    \ArrowArc(0,0)(25,0,180)
    \ArrowArc(0,0)(25,180,360)
    \Photon(-50,0)(-25,0){4}{3}
    \Photon(25,0)(50,0){4}{3}
    \BCirc(-25,0){9}
    \Text(-25,0)[]{$\Gamma_\mu$}
    \Text(55,0)[l]{$\nu$}
  \end{picture}
  }
  \hspace*{\fill}
  \\[-1cm]
  \hspace*{\fill}
  \begin{Caption}\caption{\small
      Lowest-order (=$\Order{1/c}$) contribution of gluon exchange between 
      $t\tB$ to the total cross section.  
      \label{fig:optical-LS}
  }\end{Caption}
  \hspace*{\fill}
\end{figure}
\begin{figure}[htbp]
  \hspace*{\fill}
  \begin{picture}(50,50)(-25,-25)
    \ArrowLine(0,0)(25,25)
    \ArrowLine(25,-25)(0,0)
    \Photon(-25,0)(0,0){4}{3}
    \BCirc(0,0){9}
    \Text(0,0)[]{$\Gamma_\mu$}
  \end{picture}
%  \hspace*{\fill}
  \raise 0.8cm\hbox{$=$}
%  \hspace*{\fill}
  \begin{picture}(60,50)(-35,-25)
    \ArrowLine(0,0)(25,25)
    \ArrowLine(25,-25)(0,0)
    \Photon(-25,0)(0,0){4}{3}
    \Text(-30,0)[r]{$\mu$}
  \end{picture}
%  \hspace*{\fill}
  \raise 0.8cm\hbox{$+$}
%  \hspace*{\fill}
  \begin{picture}(70,50)(-25,-25)
    \ArrowLine(0,0)(21,21)
    \ArrowLine(21,-21)(0,0)
    \Photon(-25,0)(0,0){4}{3}
    \BCirc(0,0){9}
    \Text(0,0)[]{$\Gamma_\mu$}
    \Oval(25,0)(30,7)(360)
    \Text(25,0)[]{$V$}
    \ArrowLine(29,21)(45,21)
    \ArrowLine(45,-21)(29,-21)
  \end{picture}
  \hspace*{\fill}
  \\
  \hspace*{\fill}
  \begin{Caption}\caption{\small
      Ladder approximation of vertex function $\Gamma$.  
      This is sufficient for Coulomb gluon, which is instantaneous.  
      \label{fig:ladder}
  }\end{Caption}
  \hspace*{\fill}
\end{figure}
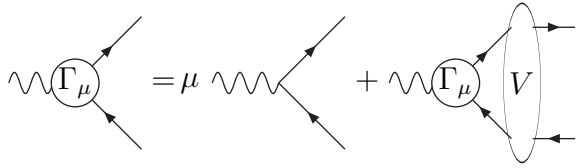
As will be shown in \SecRef{sec:rescat_corr_to_vertex}, Coulomb rescattering%
\footnote{
  We have in mind the rescattering due to the exchange of Coulomb gluons.  }
between $t$ and $\tB$ can be taken into account%
\footnote{
  Only the effect of ``soft'' gluons can be taken into account by 
  the Green function $G$.  See \SecRef{sec:soft_corr_and_hard_corr}.  
}
by modifying (one of the) $t\tB\gamma$ vertex as [\FigRef{fig:optical-LS}]
\begin{align*}
  \gamma^\mu \to 
  [\gamma^\mu]_{\rm rescat} = 
  \gamma^\mu \, \GT(|\pBI|,E) \left[ \frac{\pBI^2}{\mt} - (E+i\Gamma) \right]
  \sepcomma
\end{align*}
where 
\begin{align*}
  \GT(|\pBI|,E) 
  \equiv
  \int\!\!\frac{\diffn{\Omega_{\vec{p}}}{}}{4\pi}
  \bra{\pBI} G \ket{\xBI'=0}
  \sepcomma\quad
  G \equiv \frac{1}{H-(E+i\Gamma_t)}
  \sepcomma
\end{align*}
and $H$ is a Non-Relativistic Hamiltonian.  
Thus to the lowest order of $1/c$-expansion, 
\begin{align*}
  i\Pi^{\mu\nu}(q)
  &=
  (-1) N_C \int\!\!\frac{\diffn{p}{4}}{(2\pi)^4}\,
  \frac{i}{   \left(\dfrac{E}{2}+p^0+i\dfrac{\Gamma_t}{2} \right) 
            - \dfrac{\pBI^2}{2\mt} }\,
  \frac{i}{   \left(\dfrac{E}{2}-p^0+i\dfrac{\Gamma_t}{2} \right) 
            - \dfrac{\pBI^2}{2\mt} }
  \times\\
  &\qquad\qquad{}\times
  \btr{\dfrac{1+\gamma^0}{2} [\gamma^\mu]_{\rm rescat}
       \dfrac{1-\gamma^0}{2} \gamma^\nu }  \\
  &=
  i \, 2(g^{\mu\nu}-g^{\mu 0}g^{\nu 0})
  N_C \int\!\!\frac{\diffn{p}{3}}{(2\pi)^3}
  \GT(|\pBI|,E)  \\
  &=
  i \, 2(g^{\mu\nu}-g^{\mu 0}g^{\nu 0}) \, N_C \, G(\xBI=0,E)
  \sepcomma
\end{align*}
or
\begin{align*}
  \Pi(q^2) = \frac{2N_C}{q^2} G(\xBI=0,E)  \sepcomma\quad
  \Im \Bigl[ -i \bra{0} \rmT [ \tilde{\jmath}^\mu(q) 
    \tilde{\jmath}_\mu(-q) ] \ket{0} \Bigr]
  = 6 N_C G(\xBI=0,E)
  \sepcomma
\end{align*}
where $G(\xBI,E) \equiv \bra{\xBI} G \ket{\xBI'=0}$.  
Thus up to short distance corrections and relativistic corrections, 
$R$ ratio near the threshold is given as follows: 
\begin{align}
  R(s) 
  &= \frac{3}{2} N_C Q_t^2 \frac{4m^2}{s} \cdot 
     \frac{4\pi}{m^2c} \Im G(\xBI=0,E)  
  \label{eq:R_with_ImG_LO}
  \sepperiod
\end{align}
Note that the velocity $\beta$ of $t$ at $t\tB$ \scCM\ frame is 
\begin{align}
  \beta = \sqrt{1-\frac{4\mt^2}{s}}
  \label{eq:beta_def}
  \sepcomma\quad\mbox{ or, }\quad
  \frac{4\mt^2}{s} = 1-\beta^2 = \frac{1}{\gamma^2}
  \sepperiod
\end{align}
If one expands as amplitude $\calM$ into eigenstates of orbital angular 
momentum $L$, $\calM = \sum_L \calM^{(L)}$, 
then $|\calM^{(L)}|\propto \beta^L$.  
Thus $S$ wave ($L=0$) is the leading contribution near the threshold.  
We sometimes use $S$-wave projected notation of Green function $G(r,r')$ 
defined in \SecRef{sec:S-wave_proj}: 
\begin{align}
  G(r=0,r'=0) \,=\, G(\xBI=0,E)  \sepperiod
\end{align}
Here the energy dependence is implicit.  
%-------------------------------------------------------------------------
\subsection{Order counting}
\label{sec:order_count}
The non-relativistic expression of a quantity is obtained by expanding it 
with respect to $1/c$.  
Noting that 
\begin{align*}
  \alpha_s = \frac{g_s^2}{4\pi \hbar c} = \Order{\frac{1}{c}}  \sepcomma
\end{align*}
it is useful to introduce a dimensionful (but order unity) coupling 
$a_s \equiv \alpha_s^\MSbar(\mu_s) c$ for order counting; 
here $\mu_s$ is a ``soft scale'', which means the typical scale for 
the $t\tB$ potential.  
Thus the QCD correction $\alpha_s$ is the same order to 
the relativistic correction $\beta$, where $\beta = \sqrt{1-4\mt^2/s}$ 
is the velocity of $t$ and $\tB$ at their \scCM-frame: 
\begin{align}
  \beta = \frac{v}{c}
  \sepcomma\quad
  \alpha_s(\mu_s) = \frac{a_s}{c}
  \sepcomma\quad
  \frac{\alpha_s(\mu_s)}{\beta} = \frac{a_s}{v} = \Order{1}
  \sepperiod
\end{align}
In fact as we shall see near \EqRef{eq:R_free_LO_withGamma}, 
$\beta_t \sim \sqrt{\Gamma_t/m_t} \simeq 0.1 \sim \alpha_s$ 
near the threshold.  
One can clearly see that 
one need to sum up $\alpha_s(\mu_s)/\beta$ to all order.  
The Leading Order (LO) is $\Order{1}$, the Next-to-Leading Order (NLO) is 
$\Order{1/c}$, and the Next-to-Next-to-Leading Order (NNLO) is 
$\Order{1/c^2}$.  

The following relations may be useful to count the order of $1/c$: 
\begin{align*}
  \dim[mrc] = \dim[rp] = \dim[\hbar] = \dim[1]  \sepcomma\quad
  \dim[1/r] = \dim[p]  \sepperiod
\end{align*}
Since $\dim[a_s] = \dim[c]$ or $\dim[a_s/r] = \dim[E]$, 
the Coulomb potential is the same order to the kinetic term.  
Thus it is a part of the leading-order Hamiltonian $H_0$.  
%-----------------------------------------------------------------------
\subsection{Complications at $\Order{1/c^2}$}
\label{sec:complications_at_NNLO}
There are several complications at $\Order{1/c^2}$ and beyond.  
Some are already in freely-propagating $t\tB$ pairs, 
some are in the $t\tB$ potential $V$, and 
some are especially in the Non-Abelian part $V_{\rm NA}$ of $V$.  

For the free $t\tB$ propagation, 
(i--1) the Hamiltonian contains $p^4$ term besides $p^2$ term, 
which is the standard kinetic term.  
(i--2) The wave-function normalization of the production $t\tB$ current 
$j_{t\tB}$ depends on the momentum transfer $q^2$, 
or in other words, anomalous dimension of $j_{t\tB}$ is non-zero.  
(i--3) At $\Order{1/c}$ and below, the decay width $\Gamma_t$ of $t$ quark 
can be incorporated by replacing $H \to H - i\Gamma$~\cite{FK87,S93}.  
At $\Order{1/c^2}$ and higher, 
this is no longer justified.  
These three issues are discussed just below.  

For the $t\tB$ potential $V$, 
there are $\log(q^2/\mu_s^2)$ corrections~\cite{F77,P97,S99}.  
This is the only correction to $\Order{1/c}$ aside from a finite term.  
However to $\Order{1/c^2}$, besides the $\log$ corrections, 
(ii--1) there are also corrections called Breit-Fermi potential 
$V_{\rm BF}$~\cite{LL74}.  
These corrections are common to the potential due to 
both Abelian ($\gamma$) and non-Abelian gauge ($g$) boson exchange.  
(ii--2) There is also a correction~\cite{SN81} especially due to 
non-Abelian nature of a gluon, 
which is absent at $\Order{1/c}$ and below.  
These two issues are discussed in \SecRef{sec:NR-pot}.  
%-----------------------------------------------------------------------
\subsubsection{Current normalization $C_1^{\rm (cur)}$, $C_2^{\rm (cur)}$}
For $R$-ratio, the lowest in both $\beta$ and $\alpha_s$ expansion 
is $\Order{\beta^1,\alpha_s^0}$.  
There are several corrections beyond the lowest order.  
Only relativistic corrections are considered here%
\footnote{
  Others are considered in \SecRef{sec:matching}.  
}.  
For $\alpha_s=0$ and to all-order in $\beta$, 
the $t\tB$ production cross section is 
\begin{align}
  R = \frac{3}{2} N_C Q_t^2 \, 
    \beta\left(1-\frac{\beta^2}{3}\right)
  \label{eq:R_in_pQCD_LO}
  \sepcomma
\end{align}
where $\beta$ is velocity of $t$ at $t\tB$ \scCM\ frame [\EqRef{eq:beta_def}]. 
The Green function formalism [\EqRef{eq:R_with_ImG_LO}] should reproduce 
this result.  
The Non-Relativistic Hamiltonian $H_{\rm free}^{\rm (NR)}$ 
for free $t\tB$ is 
\begin{align*}
  H_{\rm free}^{\rm (NR)} 
  &= H_{\rm free}^{\rm (R)} - 2mc^2  \\
  &= \frac{\pBI^2}{m} - \frac{\pBI^4}{4m^3c^2} + \Order{\frac{1}{c^4}}
  \sepcomma 
\end{align*}
where $H_{\rm free}^{\rm (R)}$ is defined in \EqRef{eq:H_free_R}.  
A form factor depends on the momentum-transfer in general, and so does 
the wave-function normalization for $t\tB$ vector current $j^\mu(q)$.  
A Taylor expansion, or derivative expansion in coordinate space, of 
$R(s)$ reads 
\begin{align}
  R(s) 
  &= Q_t^2 \frac{4\pi}{s} 
  \Im \Bigl[ -i \bra{0} \rmT [ \tilde{\jmath}^\mu(q) 
    \tilde{\jmath}_\mu(-q) ] \ket{0} \Bigr]  
  \nonumber\\
  &= \frac{3}{2} N_C Q_t^2 \, \frac{4m^2}{s} \, \frac{4\pi}{m^2c} 
     \left.
     \left\{ C_1^{\rm (cur)} 
       + C_2^{\rm (cur)} \frac{\triangle_r+\triangle_{r'}}{2m^2c^2} \right\}
     \Im G(r,r')  \right|_{r,r'\to r_0}  
   + \Order{\frac{1}{c^3}}
  \label{eq:R_with_ImG_and_C}
  \sepcomma
\end{align}
where $C_1^{\rm (cur)} = 1 + \Order{\alpha_s}$ and 
$C_2^{\rm (cur)} = \mbox{const.} + \Order{\alpha_s}$; 
the constant of $C_2^{\rm (cur)}$ is determined below in this section.  
The reason for introducing the cutoff parameter $r_0 (\neq 0)$ shall be 
explained also below in this section.  
If the potential between $t\tB$ is neglected, 
a part of the expression above can be rewritten as follows: 
\begin{align}
  &
  \bIm{
     \left\{ C_1^{\rm (cur)} 
       + C_2^{\rm (cur)} \frac{\triangle_r+\triangle_{r'}}{2m^2c^2} 
       + \Order{\frac{1}{c^4}}
     \right\}
  \bra{r} \frac{1}{\ds 
    \frac{\pBI^2}{m} - \frac{\pBI^4}{4m^3c^2} + \Order{\frac{1}{c^4}} 
    - (E+i\Gamma_t)} \ket{r'}
  }  \nonumber\\ 
  &=
  \bIm{
    \left( 1 - C_2^{\rm (cur)} \frac{E+i\Gamma_t}{mc^2} \right)
  \bra{r} \frac{1}{\ds 
    \frac{\pBI^2}{m} - (E+i\Gamma_t)} \ket{r'}
  }  \nonumber\\
  &\qquad{}
  + \Im{
    \bra{r} \frac{1}{\ds \frac{\pBI^2}{m} - (E+i\Gamma_t)} \,
      \frac{\pBI^4}{4m^3c^2} \,
      \frac{1}{\ds \frac{\pBI^2}{m} - (E+i\Gamma_t)} 
    \ket{r'}
    } 
  + \Order{\frac{1}{c^4}}  \sepperiod
  \label{eq:Gfree-with-cur}
\end{align}
Here we used 
\begin{align}
  (\triangle_r + \triangle_{r'}) \braket{\rBI}{G}{\rBI'}
  = - \braket{\rBI}{(\pBI^2 G + G \pBI^2)}{\rBI'}
  \label{eq:d'Alembertian_to_p2}
  \sepperiod
\end{align}
First, consider the case $\Gamma_t = 0$.  
With the explicit analytic calculations%
\footnote{
  The relevants are $\Im G_0$ and $\Im G_2$.  
}
of the Green functions in \SecRef{sec:ImG_exp}, 
the $\beta$-dependence of $R$ in Green function method is 
\begin{align*}
  R \, / \left( \frac{3}{2} N_C Q_t^2 \right)
  &= 
  (1-\beta^2) \cdot 
  \frac{u}{c} 
  \left\{ \left( 1 - C_2^{\rm (cur)} \pfrac{u}{c}^2 \right) 
    + \frac{5}{8}\pfrac{u}{c}^2 + \Order{\frac{1}{c^4}}
  \right\}  \\
  &=
  \beta \Bigl( 1-C_2^{\rm (cur)}\beta^2 + \Order{\beta^4} \Bigr)
  \sepcomma
\end{align*}
while relativistic QFT says [\EqRef{eq:R_in_pQCD_LO}] 
it is $\beta(1-\beta^2/3)$.  
Note that $u = \beta + \Order{\beta^3}$ is defined in 
\SecRef{sec:matching-exp}.  
Thus we obtain 
\begin{align}
  C_2^{\rm (cur)} = \frac{1}{3} + \Order{\alpha_s}
  \label{eq:C_2}
  \sepperiod
\end{align}
As shall be seen in \SecRef{sec:order_count}, 
soft gluon corrections are suppressed by 
$\alpha_s(\mu_s) = \Order{1/c}$, the lowest matching coefficient is sufficient 
for the $\Order{1/c^2}$ calculations, since $C_2^{\rm (cur)}$ is a 
coefficient of $\Order{1/c^2}$ correction.  
%-----------------------------------------------------------------------
\subsubsection{Treatment of the finite width $\Gamma_t$}
Concentrating on freely-propagating $t\tB$, 
the Green function%
\footnote{
  This is derived in \SecRef{sec:ImG_exp}.  
  The relevant one is $\Im G_0$.  
}
at the origin is 
\begin{align}
  \Im G(0,0) = \frac{m_t^2}{4\pi} \Re\sqrt{\frac{E+i\Gamma_t}{m_t}} 
  + \Order{\frac{1}{c}}
  \label{eq:R_free_LO_withGamma}
  \sepperiod
\end{align}
Note that $\sqrt{E/\mt} = \beta+\Order{\beta^3}$.  
Thus effectively $\beta\sim\sqrt{\Gamma_t/\mt}\simeq 0.1$ near the threshold.  
By using 
\begin{align*}
  \sqrt{E+i\Gamma_t} = 
      \left[ \frac{\sqrt{E^2+\Gamma_t^2} + E}{2} \right]^{1/2}
  + i \left[ \frac{\sqrt{E^2+\Gamma_t^2} - E}{2} \right]^{1/2}
    \times\sgn[\Gamma_t]
 \sepcomma
\end{align*}
in general, we have 
\begin{align}
  R = \frac{3}{2} N_C Q_t^2 
      \left[ \frac{\sqrt{E^2+\Gamma_t^2} + E}{2m_t} \right]^{1/2}
      + \Order{\frac{1}{c}}
  \label{eq:R_for_free_with_Gamma}
  \sepperiod
\end{align}

It is known that the effect of finite decay width $\Gamma_t$ can be taken 
into account by the replacement $E \to E + i\Gamma_t$ to 
LO~\cite{FK87} and NLO~\cite{S93}.  
However to NNLO with non-zero $\Gamma_t$, 
imaginary part of the Green function is no longer 
finite when one takes the limit $r=r' \to 0$, which is required by 
the optical theorem: 
\begin{align*}
  {\rm \EqRef{eq:Gfree-with-cur}}
  =
  \left( - C_2^{\rm (cur)} + \frac{1}{2} \right)\frac{\Gamma_t}{4\pi c^2 r} 
%  + \mbox{(const. to $r,r'$)}
  + \Order{r^0=1,{r'}^0=1,\frac{1}{c^4}} 
  \sepcomma
\end{align*}
for $r>r'$.  
This means that the effect of the finite width $\Gamma_t$ is not properly 
treated by the replacement $E \to E + i\Gamma_t$ 
at $\Order{1/c^2}$.  
Here following~\cite{MY98}, we regulate this singularity by putting 
$r=r' \to r_0 \neq 0$, as shown in \EqRef{eq:R_with_ImG_and_C}.  
%***How this treatment is justified?***
%
%
%

%
%
%-------------------------------------------------------------------------
\subsection{Non-relativistic NNLO Hamiltonian $H$}
\label{sec:NR-H}
The NNLO Hamiltonian%
\footnote{
  When we say simply ``Hamiltonian'' $H$, 
  it means ``Non-Relativistic Hamiltonian'' $H^{\rm (NR)}$.  
}
$H$, which is relevant to us, is 
\begin{align}
  H = H_0 + V_1(r) + U(\pBI,\rBI)  
  \label{eq:NNLO_H}
  \sepcomma\quad
  H_0 = \frac{\pBI^2}{\mt} - \frac{C_F a_s}{r}  \sepcomma
\end{align}
where $H_0$ is the leading-order Hamiltonian, 
$V_1(r)$ is radiative corrections to the Coulomb potential%
\footnote{
  We may sometimes use the word ``Coulombic potential'' $V_\rmC$ 
  for the Coulomb potential with radiative corrections: 
  \begin{align*}
    V_\rmC(r) = \frac{-C_F a_s}{r} + V_1(r)  \sepperiod
  \end{align*}
  Coefficients $a_2$ etc.\ are defined in \SecRef{sec:log_corr=V1}.  
}, 
\begin{align}
  V_1(r) 
  &= - \frac{C_F a_s}{r} \Biggl[
       \frac{a_s}{4\pi c} 
       \left\{ 2 \beta_0 \ln(\mu'r) + a_1 \right\}  
  \label{eq:V1_def}
  \\
  &\qquad{}
     + \pfrac{a_s}{4\pi c}^2
       \left\{ \beta_0^2 \left(4\ln^2(\mu'r) + \frac{\pi^2}{3} \right)
         + 2(\beta_1+2\beta_0 a_1) \ln(\mu'r) + a_2
       \right\} \Biggr]  \sepcomma\nonumber
\end{align}
and $U(\pBI,\rBI)$ is a momentum-dependent potential, which includes only 
$\Order{1/c^2}$ terms: 
\begin{align*}
  U(\pBI,\rBI) 
  &= -\frac{\pBI^4}{4\mt^3 c^2}
     - \frac{C_A C_F a_s^2}{2\mt r^2 c^2}
  \\
  &\qquad{}
     - \frac{C_Fa_s}{2\mt^2 c^2} \left\{ \frac{1}{r}, \pBI^2 \right\}
     + \frac{C_Fa_s}{2\mt^2 c^2} \frac{\LBI^2}{r^3}
     + \frac{\pi C_Fa_s}{\mt^2 c^2} 
         \left( 1 + \frac{4}{3}\SBI^2 \right) \delta^{(3)}(\rBI)
  \\
  &\qquad{}
     + \frac{3 C_F a_s}{2\mt^2 c^2} \frac{\dprod{\LBI}{\SBI}}{r^3}
     - \frac{C_F a_s}{2\mt^2 c^2} \frac{1}{r^3}
      \left( \SBI^2 - 3\frac{\pdprod{\SBI}{\rBI}^2}{r^2} \right)
  \\
  &= -\frac{\pBI^4}{4\mt^3 c^2}
     - \frac{C_A C_F a_s^2}{2\mt r^2 c^2}
%  \\
%  &\qquad{}
     - \frac{C_Fa_s}{2\mt^2 c^2} \left\{ \frac{1}{r}, \pBI^2 \right\}
     + \frac{11\pi C_Fa_s}{3\mt^2 c^2} \delta^{(3)}(\rBI)
  \sepcomma
\end{align*}
where $\anticommutator{~}{~}$ means anticommutator, 
and the last expression is for the $S$-wave state, 
$L = 0, \SBI^2 = 2, \overline{r^i r^j} = \delta^{ij} r^2/3$.  
The first term in $U(\pBI,\rBI)$ is the relativistic correction to the 
free Hamiltonian, and the second term $V_{\rm NA}(r)$ is due to 
Non-Abelian nature of a gluon.  
The rest is Breit-Fermi potential $V_{\rm BF}(\pBI,\rBI)$.  
In the next section, we shall see each of them more closely.  

We need to calculate the Green function $G(r,r')$ for the Hamiltonian $H$.  
Because of the logarithmic corrections to Coulomb potential, 
$G(r,r')$ cannot be obtained analytically%
\footnote{
  Actually, the Green function without $V_1$ can be obtained analytically.  
  See \SecRef{sec:coulomb+1/r2}.  
}.  
For numerical calculations, there are two obvious complications 
for the $H$ above.  
One is $p^4$, which makes the Schr\"odinger equation 
a fourth-rank differential equation, 
and the other is the $\delta$-function.  
These two difficulties can be circumvented by rewriting $H$ appropriately.  
We do this in \SecRef{sec:reduced_G}.  
%-------------------------------------------------------------------------
\subsection{Soft corrections and hard corrections}
\label{sec:soft_corr_and_hard_corr}
It is known that the matrix element $|\calM|^2$ is 
singular near the threshold due to ``soft'' gluons, if one 
calculate to a fixed order of $\alpha_s$.  This is called 
``threshold singularity''.  
Here the ``soft'' gluon means that its momentum is $\sim \alpha_s\mt$.  
However it is also known that this singularity 
is absent once the terms of $(\alpha_s/\beta)^n$ are summed over, and 
this summation is achieved in Non-Relativistic Green function $G$ with 
Coulomb potential%
\footnote{
  Note that $V$ represents gluons of 
  its momentum is $\sim \alpha_s\mt$.  The relevant scale for these 
  soft gluons (or for the potential) is denoted by $mu_s$.  
}
$V$.  
Thus Green function method [\EqRef{eq:R_with_ImG_and_C}] is convenient 
for the threshold physics.  

On the other hand, 
if the gluon, exchanged between $t\tB$, is ``hard'', 
which means its momentum is $\sim \mt$, 
it is not treated by 
the non-relativistic potential, but by the matching coefficients of 
$t\tB$ production current.  
It is called ``hard vertex correction''~\cite{KK52}, 
which is the same as usual radiative correction to the vector vertex: 
\begin{align}
  \gamma^\mu \to \gamma^\mu 
  \left( 1 - 2 C_F \frac{\alpha_s(\mu_h)}{\pi} + \Order{\alpha_s^2} \right)
  \label{eq:hard_vert_corr_to_NLO}
\end{align}
for each vertex.  
Here $\mu_h\simeq \mt$ is a ``hard scale'' which is a typical scale for 
short distance physics.  
We limit ourselves to the contributions from the ``soft'' gluons for the 
moment.  The matching coefficients, which represent the contributions from 
the ``hard'' gluons, are determined in \SecRef{sec:matching}.  
%------------------------------------------------------------------------
\section{Non-relativistic $t\tB$ potential $V$ to NNLO}
\label{sec:NR-pot}
%------------------------------------------------------------------------
\subsection{Normalization of wave functions}
In QFT, fields of spin-$0$ are dim-$1$, while 
those of spin-$1/2$ are dim-$3/2$.  
Thus if one uses the convention%
\footnote{
  This is the convention that is usually used in 
  (modern) relativistic calculations.  
}
that the dimensions of the creation-annihilation 
operators are the same for spin-$0$ fields and spin-$1/2$ fields, 
the dimension of c-number wave functions for spin-$1/2$ fields is 
higher by $1/2$ than that for spin-$0$.  Since the dimensions of 
states with different spins are the same%
\footnote{
  Since $\braket{\psi}{}{\psi}$ is probability, $\ket{\psi}$ is dimensionless. 
  While the dimensions of bases are determined by orthnormality relations.  
  For momentum, $\braket{\pBI}{}{\pBI'} = (2\pi)^3\delta^{(3)}(\pBI-\pBI')$ 
  means $\dim[\ket{\pBI}] = \dim[p^{-3/2}]$.  
  For spins, $\ket{\sBI}$ is dimensionless in usual convention.  
}
in NR-Quantum Mechanics, 
one has to rescale the wave functions for spin-$1/2$ fields by hand.  
We may proceed as follows.  
Free Relativistic Hamiltonian $H_{\rm free}^{(\rm R)}$ of $t\tB$ pair is 
\begin{align}
  H_{\rm free}^{(\rm R)}
  &= \sqrt{(mc^2)^2+(\pBI c)^2} \times 2  
  \label{eq:H_free_R}  \\
  &= 2 mc^2 + \frac{\pBI^2}{m} - \frac{\pBI^4}{4m^3c^2} + \Order{\frac{1}{c^4}}
  \sepperiod\nonumber
\end{align}
Thus, the propagator for one-particle state in a non-Lorentz-covariant QM is 
\begin{align*}
  \frac{i}{p^0 - \sqrt{(mc^2)^2+(\pBI c)^2}}
  = 
  \frac{i}{\ds p^0 - mc^2 - \frac{\pBI^2}{2m} + \frac{\pBI^4}{8m^3c^2} 
    + \Order{\frac{1}{c^4}}}
  \sepperiod
\end{align*}
On the other hand, in Lorentz-covariant QFT, it is 
\begin{align*}
  \frac{1}{p^2-m^2+i\epsilon}
%  =
%  \frac{1}{(p^0-\sqrt{\pBI^2+m^2}+i\epsilon)
%           (p^0+\sqrt{\pBI^2+m^2}-i\epsilon)}
  =
  \frac{1}{2p^0} \left[
      \frac{1}{p^0-\sqrt{\pBI^2+m^2}+i\epsilon}
    + \frac{1}{p^0+\sqrt{\pBI^2+m^2}-i\epsilon}
  \right]
\end{align*}
or 
\begin{align*}
  \frac{i(\pS+m)}{p^2-m^2+i\epsilon}
  &\simeq \frac{i}{p^0-\sqrt{\pBI^2+m^2}+i\epsilon}
          \frac{\pS+m}{2p^0}
  \sepcomma\\
  \frac{i(\pS-m)}{p^2-m^2+i\epsilon}
  &\simeq \frac{i}{p^0-\sqrt{\pBI^2+m^2}+i\epsilon}
          \frac{\pS-m}{2p^0}
  \sepperiod
\end{align*}
Thus the non-covariant normalization of spinors, 
which is used for non-relativistic calculations hereafter, is 
\begin{align*}
  \sum_s u(p,s) \uB(p,s) = \frac{\pS+m}{2p^0}
  \sepcomma\quad
  \sum_s v(p,s) \vB(p,s) = \frac{\pS-m}{2p^0}
  \sepcomma
\end{align*}
or in Dirac representation, 
\begin{align*}
  u(p,s) =
  \sqrt{\frac{p^0+m}{2p^0}}
  \begin{pmatrix}
    \chi^s  \rule[-3ex]{0em}{0ex}  \\
    \dfrac{\dprod{\pBI}{\sigmaBI}}{p^0+m}\chi^s
  \end{pmatrix}
  \sepcomma\qquad
  v(p,s) =u^c(p,s) = C\uB(p,s)^\sfT
  \sepcomma
\end{align*}
where $C$ is charge conjugation matrix defined in \SecRef{sec:severel_conj}.  
%------------------------------------------------------------------------
\subsection{Breit-Fermi potential $V_{\rm BF}$}
The potential $\VT$ (in momentum space) 
between $t\tB$ can be obtained by calculating the 
matrix element $\calM$ for the elastic scattering of $t\tB$, 
$t(p)\,\tB(\pB) \to t(p')\,\tB(\pB')$, 
in QFT and comparing this with the formula in Born approximation, 
which is proportional to $\VT(\kBI)$, where $\kBI$ is the momentum transfer.  
Working in the \scCM\ frame of $t\tB$, the system can be dealt with 
one particle QM [\SecRef{sec:relative_coord}].  
Let us define%
\footnote{
  Note that $k$ is not relative momentum but momentum transfer; 
  likewise $K$ is not total momentum.  
}
$k \equiv p'-p$, $K \equiv (p'+p)/2$, or
$p' = K + \frac{1}{2}k$, $p = K - \frac{1}{2}k$.  
Using $p^0 = mc^2 + \pBI^2/(2m) + \Order{1/c^2}$ etc., 
\begin{align*}
  \uB(p')\gamma^0 u(p)
  &=
  {\chi_t'}^\dagger \left[
    1 + \frac{1}{c^2}
    \frac{-\kBI^2 + 2i\dprod{\sigmaBI_t}{\pcprod{\kBI}{\KBI}}}{8m^2}
    + \Order{\frac{1}{c^4}}
  \right]
  \chi_t
  \sepcomma\\
  \uB(p')\gamma^i u(p)
  &=
  {\chi_t'}^\dagger \left[
    \frac{1}{c}\,
    \frac{2K^i-i\epsilon^{ijk}k^j\sigma_t^k}{2m}
    + \Order{\frac{1}{c^3}}
  \right]
  \chi_t
  \sepperiod
\end{align*}
The corresponding expressions for 
\begin{align*}
  \vB(\pB) \gamma^\mu v(\pB') 
  = - \overline{v^c}(\pB') (\gamma^\mu)^c v^c(\pB)
  = + \uB(\pB') \gamma^\mu u(\pB)
\end{align*}
can be obtained by replacing 
$\kBI \to -\kBI$, $\KBI \to -\KBI$, $\sigmaBI_t \to \sigmaBI_\tB$, 
$\chi_t^{(\prime)} \to \chi_\tB^{(\prime)}$.  
Thus in Coulomb gauge%
\footnote{
  In fact $V_{\rm BF}$ depends on the choice of gauge, and is not 
  hermitian in general.  
  However Coulomb gauge gives hermitian result.  
}
[\SecRef{sec:gluon_prop}], 
\begin{align*}
  i\calM
  &=
  (-C_F)(-ig_s)^2 \Bigl[ 
      \uB(p') \gamma^0 u(p) \cdot \uB(\pB') \gamma^0 u(\pB) \cdot D^{00}  
    + \uB(p') \gamma^i u(p) \cdot \uB(\pB') \gamma^j u(\pB) \cdot D^{ij}
  \Bigr] \\
  &=
  (-i){\chi_t'}^\dagger{\chi_\tB'}^\dagger\,\VT_{\rm BF}(\pBI',\pBI)\,
  \chi_t'\chi_\tB'
  \sepcomma
\end{align*}
where $V_{\rm BF}$ is the Breit-Fermi potential~\cite{LL74}: 
\begin{align*}
  &\, \VT_{\rm BF}(\pBI',\pBI)
  =
  \braket{\pBI'}{V_{\rm BF}}{\pBI}  \\
%  &=
%  \frac{-4\pi C_F\alpha_s}{\kBI^2}
%  \biggl[ 1 
%    + \frac{1}{4m^2c^2} \biggl(
%        2{\pBI'}^2 + 2\pBI^2 - \kBI^2
%      - \frac{2\pdprod{\pBI'}{\kBI}^2 + 2\pdprod{\kBI}{\pBI}^2}{\kBI^2}
%  \\
%  &\qquad{}
%      + 3i \dprod{(\SBI_t+\SBI_\tB)}
%          {(-\cprod{\pBI'}{\kBI}+\cprod{\kBI}{\pBI})}
%      - 4\kBI^2\pdprod{\SBI_t}{\SBI_\tB}
%      + 4\pdprod{\SBI_t}{\kBI} \pdprod{\SBI_\tB}{\kBI}
%      \biggr)
%    + \Order{\frac{1}{c^4}}
%  \biggr]  \\
%  &=
%  \frac{-4\pi C_F\alpha_s}{\kBI^2}
%  \biggl[ 1 
%    + \frac{1}{4m^2c^2} \biggl(
%        2{\pBI'}^2 + 2\pBI^2 + \kBI^2
%      - \frac{2\pdprod{\pBI'}{\kBI}^2 + 2\pdprod{\kBI}{\pBI}^2}{\kBI^2}
%  \\
%  &\qquad\qquad\qquad{}
%      + 3i \dprod{\SBI}{(-\cprod{\pBI'}{\kBI}+\cprod{\kBI}{\pBI})}
%      - 2\kBI^2\SBI^2
%      + 2\pdprod{\SBI}{\kBI}^2
%      \biggr)
%    + \Order{\frac{1}{c^4}}
%  \biggr]  \\
  &=
  -4\pi C_F\alpha_s
  \biggl[ \frac{1}{\kBI^2}
      + \frac{1}{2m^2c^2}
        \left( 
            \frac{{\pBI'}^2}{\kBI^2} 
          + \frac{\pBI^2}{\kBI^2}
        \right)
      - \frac{1}{2m^2c^2}
        \left( 
            \frac{\pdprod{\pBI'}{\kBI}^2}{(\kBI^2)^2}
          + \frac{\pdprod{\kBI}{\pBI}^2}{(\kBI^2)^2} 
        \right)
      + \frac{1}{4m^2c^2}
  \\
  &\qquad\qquad\qquad{}
      + \frac{3i}{4m^2c^2} \left( \frac{\cprod{\kBI}{\pBI'}}{\kBI^2}\cdot\SBI
          + \frac{\cprod{\kBI}{\pBI}}{\kBI^2}\cdot\SBI \right)
      - \frac{1}{2m^2c^2}\left( 
          \SBI^2 - \frac{\pdprod{\SBI}{\kBI}^2}{\kBI^2} \right)
    + \Order{\frac{1}{c^4}}
  \biggr]
  \sepcomma
\end{align*}
where $\kBI = \pBI'-\pBI$, $\SBI = \SBI_t + \SBI_\tB$ 
and $\SBI_t = \sigmaBI_t/2$, $\SBI_\tB = \sigmaBI_\tB/2$.  
There are several remarks.  
Since the overall phase of $\calM$ is unphysical, relation between $i\calM$ 
and $\VT$ is determined so as to reproduce the Coulomb potential at the 
leading order.  
We rewrote $\KBI$ with $\pBI,\pBI',\kBI$ in a symmetric way: 
\begin{align*}
  &  4\KBI^2 = 2 {\pBI'}^2 + 2\pBI^2 - \kBI^2  \sepcomma\\
  &  4\pdprod{\kBI}{\KBI}^2 
    = 2\pdprod{\pBI'}{\kBI}^2 + \pdprod{\kBI}{\pBI}^2 - (\kBI^2)^2
  \sepcomma\\
  &  \cprod{\kBI}{\KBI} = \frac{1}{2} \left( 
       -\cprod{\pBI'}{\kBI} + \cprod{\kBI}{\pBI}
     \right)
  \sepperiod
\end{align*}
This makes the hermiticity of $V_{\rm BF}$ transparent.  
Due to $\ChargeCOp\ParityOp$ invariance of QCD, 
$V_{\rm BF}$ is symmetric under $\SBI_t \leftrightarrow \SBI_\tB$ 
[Equations~\ref{eq:CPTTfor_ffB} and \ref{eq:jj*_CP}].  
They are conveniently expressed in terms of the total spin $\SBI$: 
\begin{align*}
    \dprod{\SBI_t}{\SBI_\tB} = \frac{1}{2}\left(\SBI^2-\frac{3}{2}\right)  
  \sepcomma\qquad
     \pdprod{\SBI_t}{\kBI}\pdprod{\SBI_\tB}{\kBI}
    = \frac{1}{2}\left(
        \pdprod{\SBI}{\kBI}^2 - \frac{\kBI^2}{2} \right)  \sepperiod
\end{align*}
The $\VT_{\rm BF}$ depends not only on $\kBI$ but also on $\KBI$.  
This means the potential is momentum dependent 
[\SecRef{sec:mom_and_coord_space}].  
This can be seen clearly in the coordinate-space representation given below.  

The Breit-Fermi potential $V_{\rm BF}(r)$ in coordinate-space can 
be obtained by Fourier transform: 
\begin{align*}
  V_{\rm BF}(\rBI)\,\delta^{(3)}(\rBI'-\rBI)
  = \braket{\rBI'}{V_{\rm BF}}{\rBI}  
  = \int\!\!\frac{\diffn{p'}{3}}{(2\pi)^3}\frac{\diffn{p}{3}}{(2\pi)^3}
     \expo^{i(\vec{p}'\cdot\vec{r}'-\vec{p}\cdot\vec{r})}
     \VT_{\rm BF}(\pBI',\pBI)
  \sepcomma
\end{align*}
where $\rBI$ ($\rBI'$) is the relative coordinate for the initial (final) 
state.  With 
\begin{align*}
   \;& \diffn{p'}{3}\,\diffn{p}{3} &
   \;& \dprod{\pBI'}{\rBI'} - \dprod{\pBI}{\rBI}  \\
  =\;& \diffn{k}{3}\,\diffn{p}{3} &
  =\;& \dprod{\kBI}{\rBI'} + \dprod{\pBI}{(\rBI'-\rBI)}  \\
  =\;& \diffn{p'}{3}\,\diffn{k}{3} &
  =\;& \dprod{\pBI'}{(\rBI'-\rBI)} + \dprod{\kBI}{\rBI}  \\
  =\;& \diffn{K}{3}\,\diffn{k}{3} \sepcomma &
  =\;& \dprod{\KBI}{(\rBI'+\rBI)}/2 + \dprod{\kBI}{(\rBI'-\rBI)} \sepcomma
\end{align*}
and 
\begin{align*}
  &
  \int\!\!\frac{\diffn{k}{3}}{(2\pi)^3}
  \expo^{i\vec{k}\cdot\vec{r}} 4\pi
  \Biggl\{
    1 \, ,\, 
    \frac{k^i k^j}{\kBI^2}\, ;\, 
    \frac{1}{|\kBI|}\, ,\, 
    \frac{k^i}{\kBI^2}\, ;\, 
    \frac{1}{\kBI^2}\, ,\, 
    \frac{k^i k^j}{(\kBI^2)^2}
  \Biggr\}
  \\
  &=
  \Biggl\{
    4\pi\delta^{(3)}(\rBI)\, , \, 
    \frac{1}{r^3}\left( \delta^{ij}-3\frac{r^i r^j}{r^2} \right)
      + \frac{4\pi}{3}\delta^{ij}\delta^{(3)}(\rBI)\, ; \, 
    \frac{2}{\pi} \frac{1}{r^2}\, , \, 
    \frac{ir^i}{r^3}\, ; \, 
    \frac{1}{r}\, ,\, 
    \frac{1}{2r}\left( \delta^{ij}-\frac{r^i r^j}{r^2} \right)
  \Biggr\}
  \sepcomma
\end{align*}
we have 
\begin{align}
  V_{\rm BF}(\rBI)
  &=
  - \frac{C_F\alpha_s}{r}
  - \frac{C_F\alpha_s}{2m^2 c^2} 
    \anticommutator{\frac{1}{r}}{\pBI^2}
  + \frac{C_F\alpha_s}{2m^2 c^2} \frac{\LBI^2}{r^3}
  + \frac{\pi C_F\alpha_s}{m^2 c^2} 
      \left( 1 + \frac{4}{3}\SBI^2 \right) \delta^{(3)}(\rBI)
  \nonumber\\
  &\qquad\qquad{}
  + \frac{3 C_F \alpha_s}{2m^2 c^2} \frac{\dprod{\LBI}{\SBI}}{r^3}
  - \frac{C_F \alpha_s}{2m^2 c^2} \frac{1}{r^3}
    \left( \SBI^2 - 3\frac{\pdprod{\SBI}{\rBI}^2}{r^2} \right)
  + \Order{\frac{1}{c^4}}
  \label{eq:BF-coord}
  \sepcomma
\end{align}
where $\anticommutator{~}{~}$ means anti-commutator, and 
$\pBI$ and $\LBI=\cprod{\rBI}{\pBI}$ are differential operators that 
correspond to linear and orbital-angular momentum, respectively.  
Note that $\LBI$ and $r$ commutes [\SecRef{sec:polar_coord}].  
In this form, one can clearly see that $V_{\rm BF}(\rBI)$ is hermitian.  
The following relations may be useful to compare with 
the formulas in other literatures: 
\begin{gather*}
    \dprod{\SBI_t}{\SBI_\tB} 
    - 3\frac{\pdprod{\SBI_t}{\rBI}\pdprod{\SBI_\tB}{\rBI}}{\rBI^2}
    = \frac{1}{2}\left( \SBI^2-3\frac{\pdprod{\SBI}{\rBI}}{\rBI^2} \right)
 \sepcomma\\
  -\left\{ \frac{1}{r} \, , \, \pBI^2 \right\}
  + \frac{\LBI^2}{r^3} + 4\pi\delta^{(3)}(\rBI)
  =
  \frac{-1}{r} \left(
    \pBI^2 + \frac{1}{r}
    \dprod{\rBI}{\pdprod{\rBI}{\pBI}\pBI} \right)
  \sepcomma\\
  \left\{
    p^i p^j \, , \, 
    \frac{1}{2r} \left( \delta^{ij} - \frac{r^i r^j}{r^2} \right)
  \right\}
  =
  \frac{\LBI^2}{r^3} + 4\pi\delta^{(3)}(\rBI)
  \sepperiod
\end{gather*}
%
%------------------------------------------------------------------------
\subsection{Non-Abelian effect $V_{\rm NA}$}
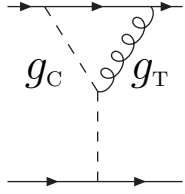
\begin{figure}[htbp]
  \hspace*{\fill}
  \begin{picture}(66,66)(0,0)
    \Large
    \ArrowLine(0,0)(34,0)
    \ArrowLine(34,0)(66,0)
    \ArrowLine(0,66)(14,66)
    \ArrowLine(14,66)(54,66)
    \ArrowLine(54,66)(66,66)
    \DashLine(34,0)(34,34){4}
    \DashLine(34,34)(14,66){4}
    \Gluon(34,34)(54,66){-3}{4}
    \Text(14,46)[t]{$g_{\mbox{\scriptsize C}}$}
    \Text(54,46)[t]{$g_{\mbox{\scriptsize T}}$}
  \end{picture}
  \hspace*{\fill}
  \\
  \hspace*{\fill}
  \begin{Caption}\caption{\small
      Feynman diagram that contributes to $V_{\rm NA}$.  
      \label{fig:V_NA_diagram}
  }\end{Caption}
  \hspace*{\fill}
\end{figure}
There is also a potential~\cite{SN81} proportional to $C_A$, which is zero 
for Abelian gauge group (like QED) [\FigRef{fig:V_NA_diagram}]: 
\begin{align}
  \VT_{\rm NA}(k) = - \frac{\pi^2 C_A C_F a_s^2}{\mt |\kBI| c^2} 
  \sepcomma\quad
  V_{\rm NA}(r) = - \frac{C_A C_F a_s^2}{2\mt r^2 c^2}
  \sepperiod
\end{align}
One can see that the potential in momentum space is non-analytic, 
and in coordinate space it is a strong attractive potential near the origin.  
%------------------------------------------------------------------------
\subsection{Radiative corrections $V_1$ for Coulomb potential}
\label{sec:log_corr=V1}
There are also radiative corrections to Coulomb potential.  
We call this potential Coulombic potential $V_\rmC$: 
\begin{align}
  V_\rmC(\kBI^2) = - C_F \frac{4\pi\alpha_s^V(\kBI^2;\mu^2)}{\kBI^2}
  \label{eq:VC_mom}
  \sepcomma
\end{align}
where the coupling in $V$-scheme is~\cite{F77,P97,S99} 
\begin{align}
  \alpha_s^V(\kBI^2;\mu^2) 
  &= \alpha_s^\MSbar(\mu^2) \Biggl[
      1 + \frac{\alpha_s^\MSbar(\mu^2)}{4\pi} \left(
        - \beta_0 \ln\frac{\kBI^2}{\mu^2} + a_1 \right)
  \label{eq:alphaV_mom}
  \\&\qquad\qquad{}
      + \pfrac{\alpha_s^\MSbar(\mu^2)}{4\pi}^2 \left( 
          \left( \beta_0 \ln\frac{\kBI^2}{\mu^2} \right)^2 
          - \left( 2 \beta_0 a_1 + \beta_1 \right) \ln\frac{\kBI^2}{\mu^2} 
          + a_2 \right)
      + \cdots \Biggr]
  \nonumber\\
  &= \alpha_s^\MSbar(\kBI^2) \Biggl[ 
      1 + \frac{\alpha_s^\MSbar(\kBI^2)}{4\pi} a_1 
        + \pfrac{\alpha_s^\MSbar(\kBI^2)}{4\pi}^2 a_2 
        + \cdots \Biggr]
  \sepcomma\quad\mbox{ for $\mu^2 = \kBI^2$.  }  \nonumber
\end{align}
Here the first expression is for fixed order series, and the second for 
Renormalization-Group (RG) improved series, $\mu^2 = \kBI^2$.  
In \SecRef{sec:NR-H} we wrote the log corrections as $V_1$: 
\begin{align*}
  V_\rmC(\kBI^2) 
  = - C_F \frac{4\pi\alpha_s^\MSbar}{\kBI^2} + V_1(\kBI^2) 
  \sepperiod
\end{align*}

The potential $V_\rmC(r)$ for coordinate space can be obtained analytically 
only for the fixed-order series: 
\begin{align}
  V_\rmC(r) 
  = - C_F \frac{\alphaB_s^V(1/r;\mu)}{r}
  = - \frac{C_F \alpha_s^\MSbar}{r} + V_1(r) 
  \sepcomma
\end{align}
where
\begin{align}
  \alphaB_s^V(1/r;\mu)
  &= \alpha_s^\MSbar(\mu^2) \Biggl[ 1 + 
       \frac{\alpha_s^\MSbar(\mu^2)}{4\pi}
       \left\{ 2 \beta_0 \ln(\mu r \expo^\Egamma) + a_1 \right\}  
  \label{eq:alphaV_coord}
  \\
  &\qquad{}
     + \pfrac{\alpha_s^\MSbar(\mu^2)}{4\pi}^2
       \left\{ \beta_0^2 \left(4\ln^2(\mu r \expo^\Egamma) 
             + \frac{\pi^2}{3} \right)
         + 2(\beta_1+2\beta_0 a_1) \ln(\mu r \expo^\Egamma) + a_2
       \right\} \Biggr]  \sepperiod  \nonumber
\end{align}
We can see 
\begin{align*}
  \alphaB_s^V(1/r;\mu) \simeq
  \alpha_s^V(\kBI^2 = 1/{r'}^2;\mu^2) 
  \sepcomma\quad
  r' \equiv r \expo^\Egamma
\end{align*}
aside from $\pi^2/3$.  Here $\Egamma = 0.5772\cdots$ is Euler-Mascheroni 
constant, and originates from the Fourier transform: 
\begin{align*}
  \int\!\!\frac{\diffn{k}{3}}{(2\pi)^3}
  \frac{\ln(\kBI^2)\expo^{i\vec{k}\cdot\vec{r}}}{\kBI^2}
  = \frac{-1}{4\pi r} \left[ \ln(r^2) + 2\Egamma \right]
  \sepperiod
\end{align*}

Coefficients $a_1$ and $a_2$ are coefficients for $\Order{1/c}$ and 
$\Order{1/c^2}$ corrections, respectively: 
\begin{align}
  a_1 &= \frac{31}{9}C_A - \frac{20}{9} T_F n_f  
    = \frac{31}{3} - \frac{10}{9} n_f = 4.77778\cdots
  \sepcomma\nonumber\\
  a_2 &= \left( \frac{4343}{162} + 4\pi^2 - \frac{\pi^4}{4} 
          + \frac{22}{3}\zeta_3 \right) C_A^2 
        - \left( \frac{1798}{81} + \frac{56}{3}\zeta_3 \right) C_A T_F n_f 
  \nonumber\\
  &\qquad{}
        - \left( \frac{55}{3} - 16\zeta_3 \right) C_F T_F n_f 
        + \left( \frac{20}{9} T_F n_f \right)^2  
  \\
  &={}
    456.749 - 66.3542\,n_f + 1.23457\,n_f^2  =  155.842\cdots
  \sepperiod\nonumber
\end{align}
Group theoretic factors $C_F$ etc.\ are given in \SecRef{sec:group_factor}.  
In the first calculation of $\Order{\alpha_s^2}$ correction~\cite{P97}, 
the $4\pi^2$ in $C_A^2$ term of $a_2$ was incorrectly calculated 
to be $6\pi^2$.  With this, 
\begin{align*}
  a_2^{\rm (old)}
  = 634.402 - 66.3542\,n_f + 1.23457\,n_f^2
  = 333.495\cdots  \sepperiod
\end{align*}
Thus the correct value~\cite{S99} of $a_2$ is less than half of 
$a_2^{\rm (old)}$.  

Since $a_2$ is a coefficient of $\Order{1/c^2}$ correction, 
one can expect better convergence with the correction to $a_2$.  
However for fixed order calculation, $\ln(k/\mu_s)$ term disturbs this 
naive expectation.  While for RG improved calculation, the size of 
$\Order{1/c^2}$ correction is diminished by 2.  

Coefficients $\beta_0$ and $\beta_1$ determine the RG flow of 
QCD coupling $\alpha_s(\mu)$.  These first two coefficients are 
independent to renormalization scheme: 
\begin{align*}
  &
  \pdiff{}{\ln\mu^2} \pfrac{4\pi}{\alpha_s(\mu^2)} 
  = \beta_0 + \beta_1 \frac{\alpha_s(\mu^2)}{4\pi} 
    + \beta_2 \pfrac{\alpha_s(\mu^2)}{4\pi}^2 + \cdots
  \sepcomma\quad\mbox{ or }\\
  &
  \mu \pdiff{\alpha_s}{\mu} = - \frac{\beta_0}{2\pi} \alpha_s^2 
    - \frac{\beta_1}{8\pi^2}\alpha_s^3 
    - \frac{\beta_2}{32\pi^3}\alpha_s^4 - \cdots 
  \sepcomma
\end{align*}
where 
\begin{align}
  &
  \beta_0 = \frac{11}{3} C_A - \frac{4}{3} T_F n_f 
    = 11 - \frac{2}{3} n_f = 7.66667\cdots
  \sepcomma\\
  &
  \beta_1 = \frac{34}{3} C_A^2 - \frac{20}{3} C_A T_F n_f - 4 C_F T_F n_f 
    = 2 \left( 51 - \frac{19}{3} n_f \right) 
    = 38.6667\cdots
  \sepperiod\nonumber
\end{align}
At the leading order of running, one have 
\begin{align*}
  \alpha_s(\mu^2) = 
  \frac{\alpha_s(M^2)}
       {\ds 1+\frac{\beta_0\alpha_s(M^2)}{4\pi}\log\frac{\mu^2}{M^2}}
  = \frac{4\pi}{\ds \beta_0\log\frac{\mu^2}{\Lambda^2}}
  \sepcomma
\end{align*}
where 
$\Lambda^2 = M^2\exp\left( \frac{-4\pi}{\beta_0\alpha_s(M^2)} \right)$.  
If we identify $\alpha_s(\mu)/\pi$ and 
$(\alpha_s(\mu)/\pi)\ln(\mu^2/k^2)$ to be the expansion parameters, 
then the coefficients are 
\begin{align*}
  &
  \frac{\beta_0}{4} = 1.91667\cdots  \sepcomma\quad
  \frac{\beta_1}{16} = 2.41667\cdots  \sepcomma\\
  &
  \frac{a_1}{4} = 1.19444\cdots  \sepcomma\quad
  \frac{a_2}{16} = 9.74014\cdots  \sepperiod
\end{align*}
One can see that the finite term of $\Order{\alpha_s^2}$ correction 
$a_2/16$ is large compared to the other.  
%------------------------------------------------------------------------
\subsubsection{RG improved Coulombic potential}
As can be seen in Eqs.~(\ref{eq:alphaV_mom}) and~(\ref{eq:alphaV_coord}), 
leading log's, $\sum (\alpha_s \ln)^n$, in $V_{\rm C}$ can be summed up 
either in momentum space ($\mu = k$) or in coordinate space ($\mu = 1/r'$).  
However as we saw above, there is an extra term $\pi^2/3$ in 
$\alpha_s^2$ term for the coordinate space potential.  
Thus it seems that RG improvement in momentum space is better.  
The potential in coordinate space is obtained by numerical Fourier transform.  
There is another advantage%
\footnote{
  It was argued in~\cite{JKPST98} that large theoretical uncertainty 
  remains even after the RG improvement of $V_{\rm C}$.  
  This claim was based on a large discrepancy 
  between results of renormalization-group improvements in momentum 
  space and in coordinate space.  
  However in view of ``renormalon cancellation'', momentum space is better.  
}
for this prescription, which is related to the 
``renormalon cancellation'', where the infrared part of $\VT_\rmC(k)$ plays 
an important role.  
However let us postpone this subject until \SecRef{sec:renormalon}.  

\begin{figure}[tbp]
  \hspace*{\fill}
  \begin{minipage}{6cm}
    \includegraphics[width=7.5cm]{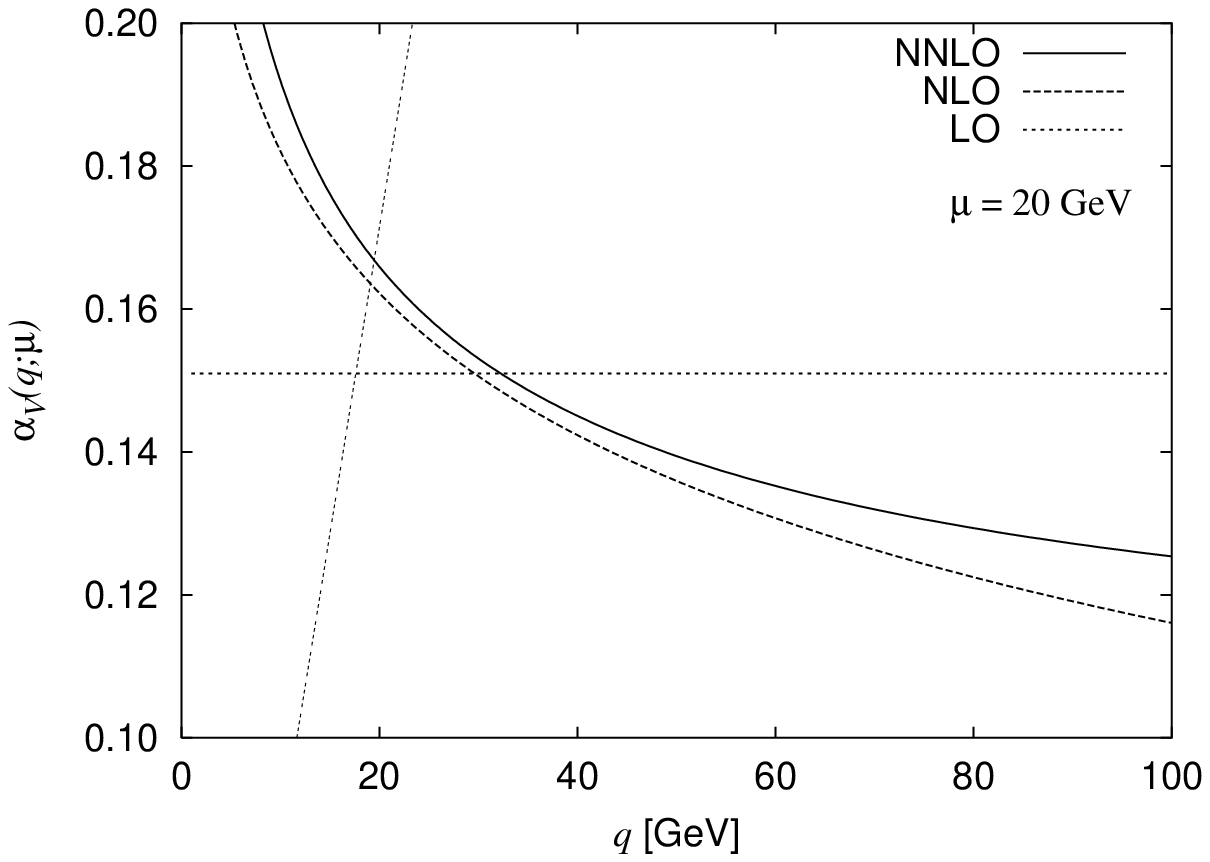}
  \end{minipage}
  \hspace*{\fill}
  \begin{minipage}{6cm}
    \includegraphics[width=7.5cm]{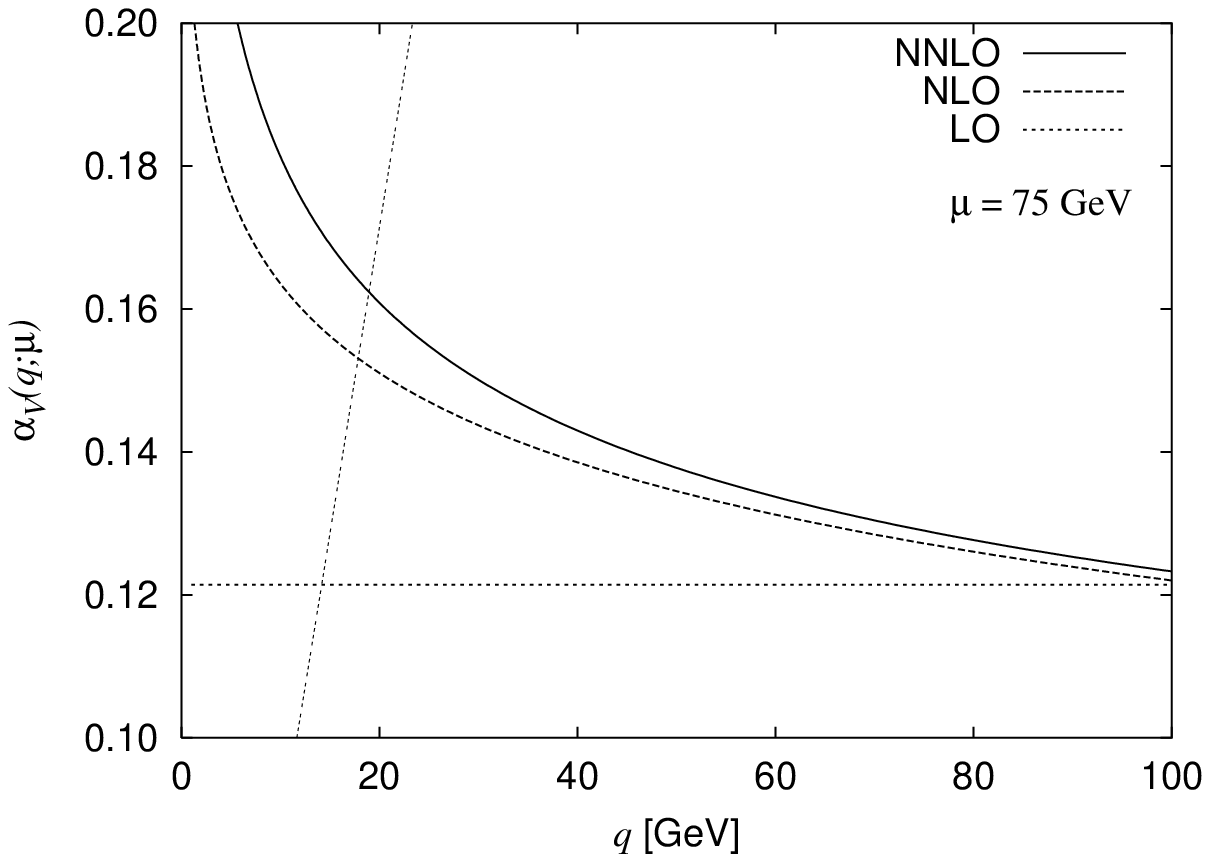}
  \end{minipage}
  \hspace*{\fill}
   \\
  \hspace*{\fill}
  \begin{minipage}{6cm}
    \includegraphics[width=7.5cm]{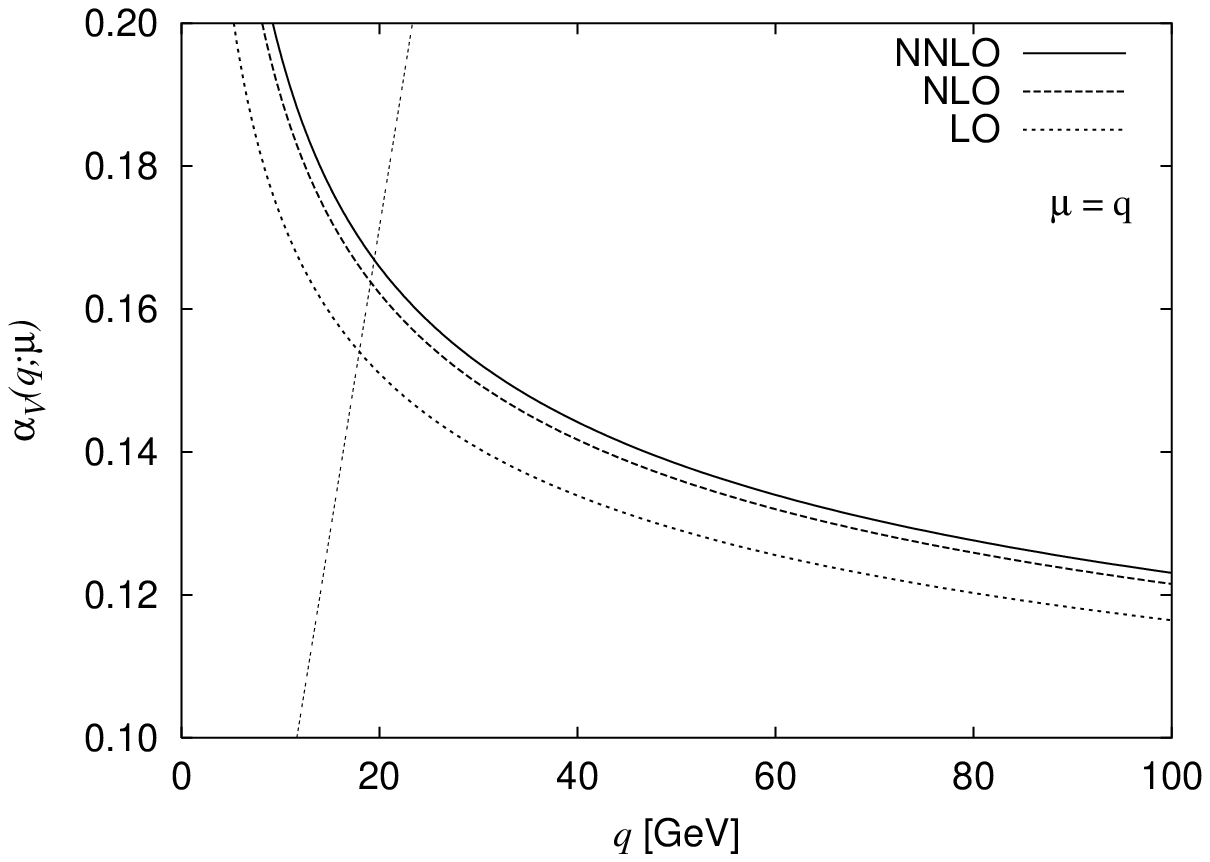}
  \end{minipage}
  \hspace*{\fill}
  \begin{minipage}{6cm}
    \includegraphics[width=7.5cm]{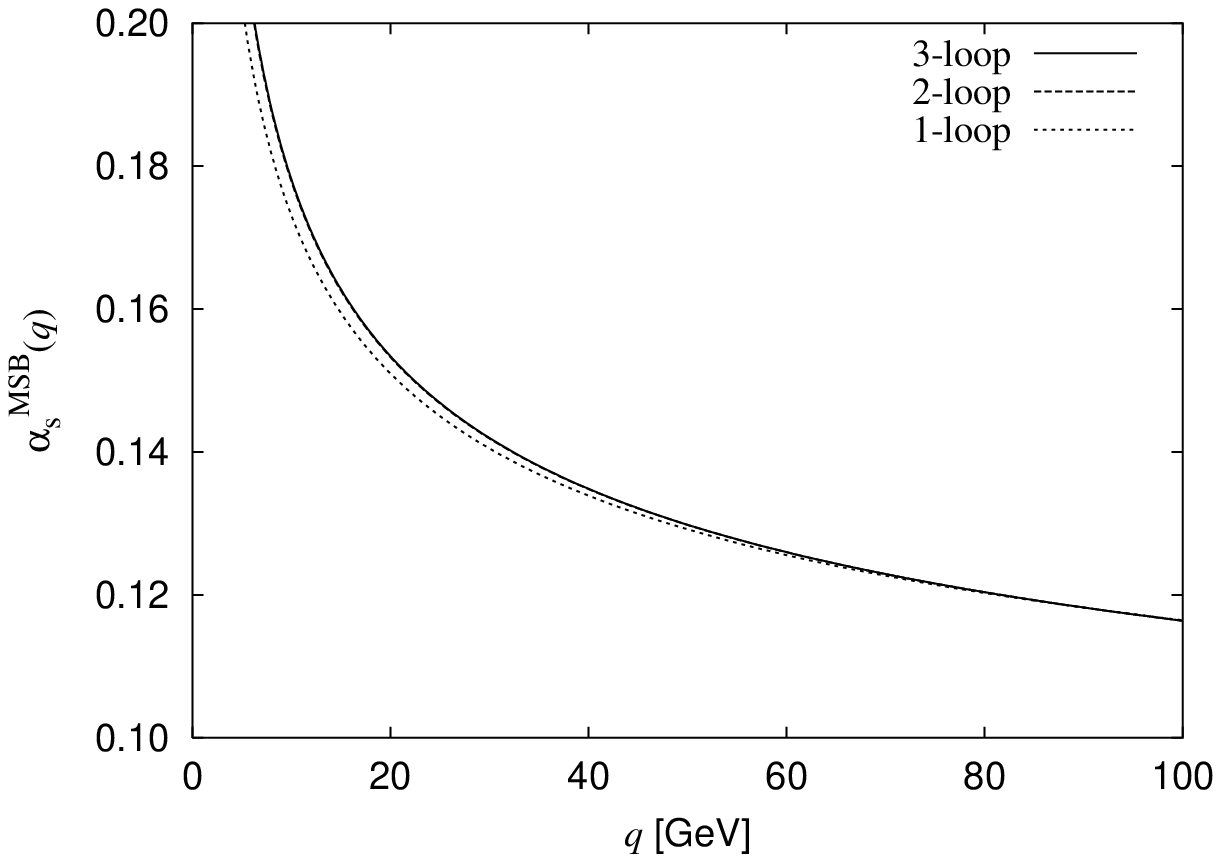}
  \end{minipage}
  \hspace*{\fill}
   \\
  \hspace*{\fill}
  \begin{Caption}\caption{\small
      The momentum-space couplings $\alpha_V$ vs.\ momentum transfer
      $q$ at LO (dot-dashed), NLO (dashed), and 
      NNLO (solid).  
      Two are fixed order ($\mu_s = 20\GeV, 75\GeV$), and 
      one is RG improved ($\mu_s = q$).  
      Intersection of $\alpha_V$ and the straight line $q/(C_F\mt/2)$, 
      which is also shown, gives Bohr momentum $p_\rmB$ and the coupling 
      $\alpha_V(p_\rmB)$ at the scale $p_\rmB$ 
      [\SecRef{sec:G_for_pure_Coulomb}].  
      Running of \MSbar\ coupling is also shown for reference.  
      \label{fig:alphaV_20-75-q}
  }\end{Caption}
  \hspace*{\fill}
\end{figure}

Scale dependence of $\alpha_V(q,\mu_s)$ and $\alpha_s^{\MSbar}(\mu_s)$ 
are shown in \FigRef{fig:alphaV_20-75-q} for 
$\mu_s = 20\GeV$, $75\GeV$, and $q$.  
One can see that perturbative convergence is improved by 
the RG prescription $\mu_s = q$ over the whole relevant momentum scale.  
Since the coupling becomes stronger as higher orders are taken into account, 
it is expected that the binding energy becomes larger with higher order 
corrections.  Scaling behavior of \MSbar\ coupling is also shown for 
comparison.  It is similar but higher order corrections are much smaller.  
In fact one can hardly distinguish 2-loop and 3-loop running.  
This is because $a_2$ (and $a_1$) is large.  
See the relation of V-scheme coupling to \MSbar-scheme coupling 
[\EqRef{eq:alphaV_mom}].  
Also shown is the straight line $q/(C_F \mt/2)$.  
The solution $q/(C_F \mt/2) = \alpha_V(q)$, or 
\begin{align*}
  \mu_s = C_F \alpha_V(\mu_s) \mt/2
\end{align*}
gives Bohr momentum $p_\rmB = \mu_s$.  
We can see that $p_\rmB \simeq 20\GeV$ and $\alpha_V(p_\rmB) \simeq 0.16$.  

\begin{figure}[tbp]
  \hspace*{\fill}
  \begin{minipage}{6cm}
    \includegraphics[width=7.5cm]{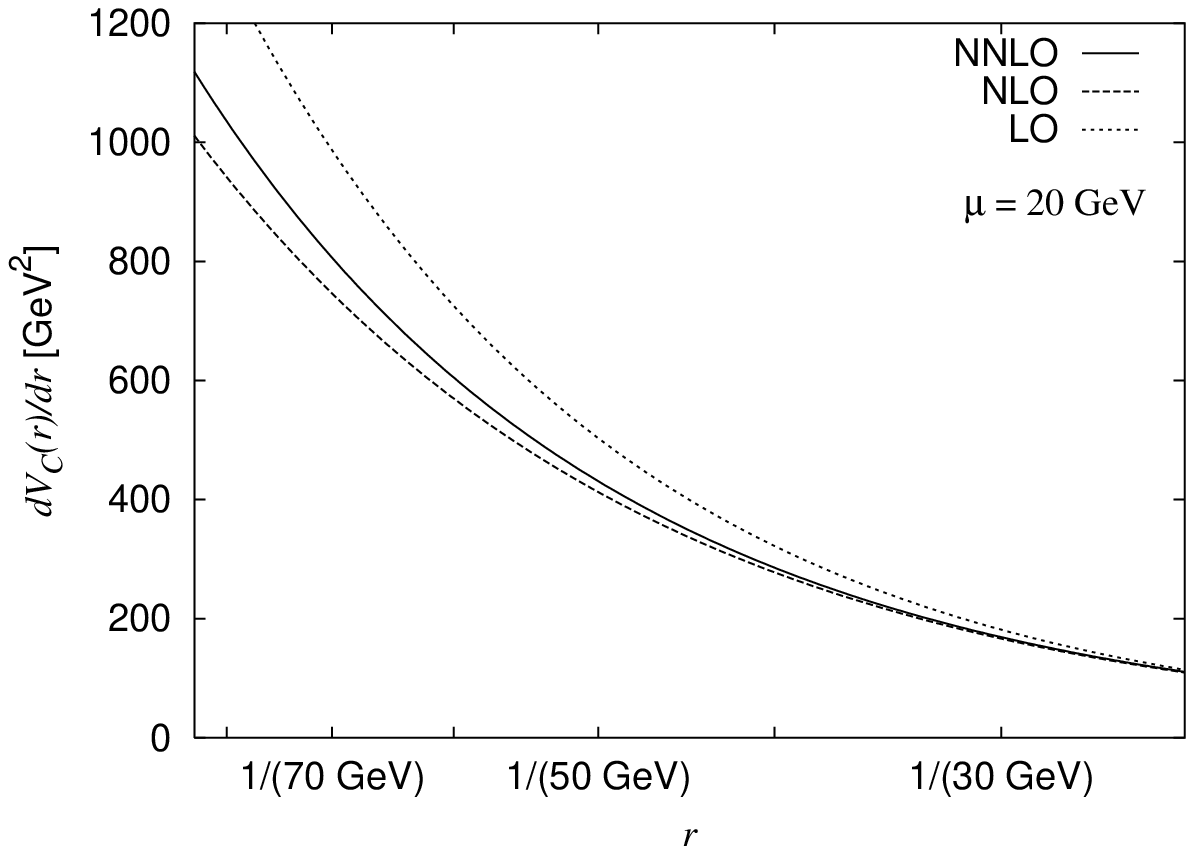}
  \end{minipage}
  \hspace*{\fill}
  \begin{minipage}{6cm}
    \includegraphics[width=7.5cm]{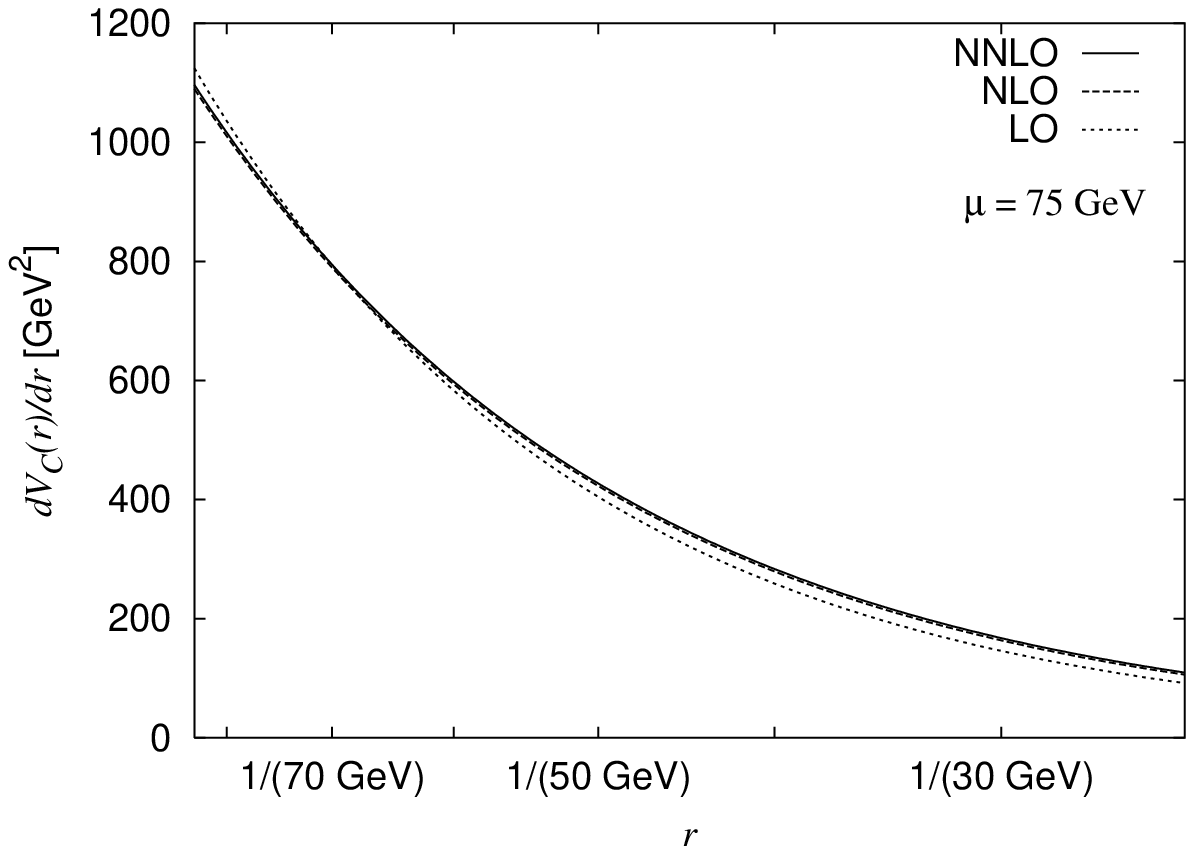}
  \end{minipage}
  \hspace*{\fill}
   \\
  \hspace*{\fill}
  \begin{minipage}{6cm}
    \includegraphics[width=7.5cm]{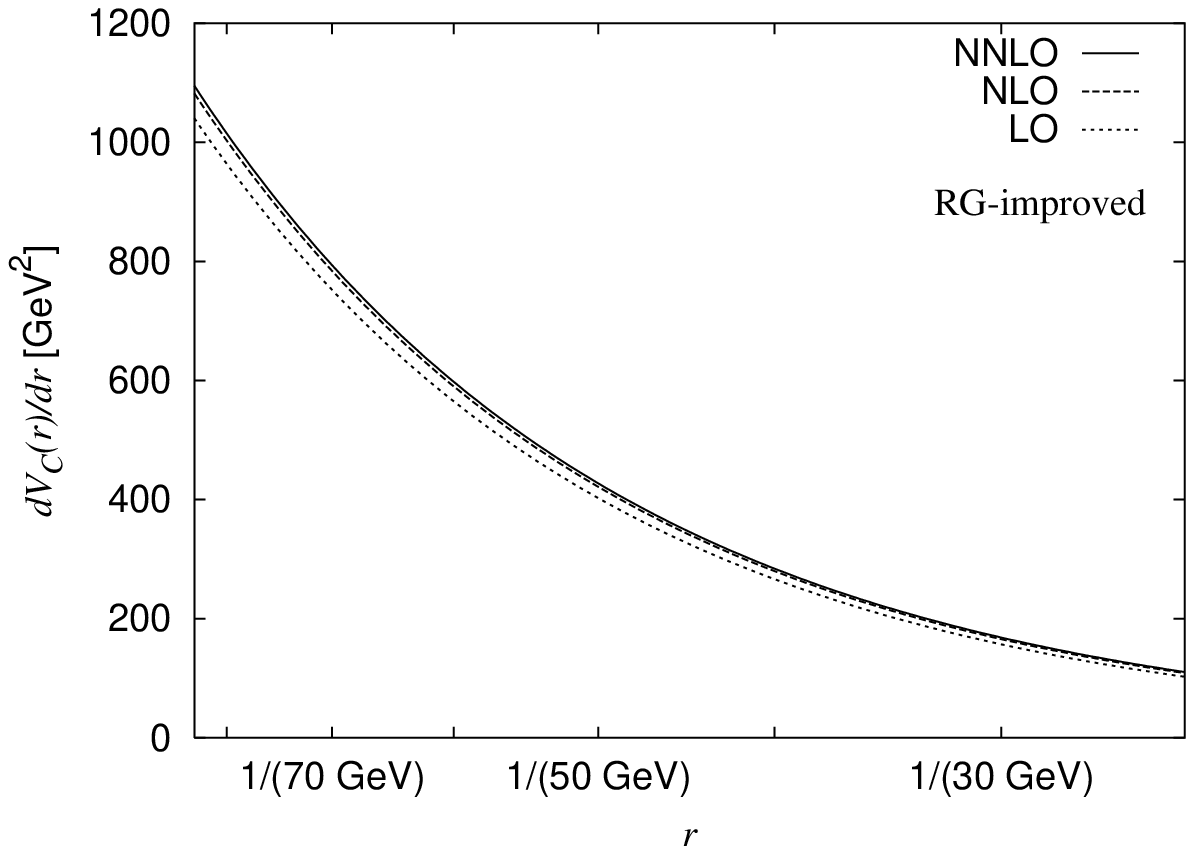}
  \end{minipage}
  \hspace*{\fill}
   \\
  \hspace*{\fill}
  \begin{Caption}\caption{\small
      Attractive force $\diffn{V}{}/\diffn{r}{}$ due to Coulombic 
      potential $V_{\rm C}$, with $\mu_s=20\GeV$, $75\GeV$, $q$.  
      \label{fig:dVdr_20-75-RG}
  }\end{Caption}
  \hspace*{\fill}
\end{figure}

As shall be explained in \SecRef{sec:psi(0)_V(r)}, 
the normalization of the total cross section is 
determined by the strength of the attractive force 
$\diffn{V}{}/\diffn{r}{}$ between $t\tB$.  
This is shown in \FigRef{fig:dVdr_20-75-RG}.  
One can see that the perturbative convergence is worse for 
$\mu_s = 20\GeV \simeq p_\rmB$ than for $\mu_s=75\GeV$ or $\mu_s=q$.  
%
%------------------------------------------------------------------------
\section{Reduced Green function $G'$}
\label{sec:reduced_G}
%------------------------------------------------------------------------
\subsection{$G$ in terms of $G'$}
As was explained above, the Hamiltonian $H$ to NNLO (= $\Order{1/c^2}$) 
[\EqRef{eq:NNLO_H}] is not convenient for numerical calculation.  
Thus following~\cite{MY98}, we rewrite the Green function $G$ for $H$ 
in this section.  

We are interested in a Green function $G(r,r')$ 
that is projected to the $S$-wave.  
In operator language, it is defined as follows [\SecRef{sec:S-wave_proj}]: 
\begin{gather}
  G \equiv \frac{1}{H-\omega}  \sepcomma\quad
  \omega \equiv E+i\Gamma_t
  \label{eq:G_and_H}
  \sepcomma\\
  G(\rBI,\rBI') \equiv \braket{\rBI}{G}{\rBI'}  \sepcomma\quad
  G(r,r') \equiv \int \frac{\diffn{\Omega_{r}}{}}{4\pi}
    \frac{\diffn{\Omega_{r'}}{}}{4\pi} G(\rBI,\rBI')
  \sepperiod\nonumber
\end{gather}
Here $H$ is defined in \EqRef{eq:NNLO_H}.  
Using operator identities derived in \SecRef{sec:polar_coord} 
\begin{align*}
  [ \pBI^2 , i p_r ] = 4\pi \delta^{(3)}(\rBI) + \frac{2\LBI^2}{r^3}
  \sepcomma\quad
  [ \frac{1}{r}, i p_r ] = \frac{1}{r^2}
  \sepcomma
\end{align*}
or 
\begin{align*}
  &
  \frac{4\pi}{\mt} \delta^{(3)}(\rBI) = 
    [ H_0 , i p_r ] 
  - \frac{2\LBI^2}{\mt r^3}
  + \frac{C_F a_s}{r^2}
  \sepcomma\\
  &
  \pfrac{\pBI^2}{\mt}^2 = H_0^2 
    + \left\{ H_0 \, , \, \frac{C_F a_s}{r} \right\}
    + \frac{C_F^2 a_s^2}{r^2}
  \sepcomma 
\end{align*}
the momentum dependent $\Order{1/c^2}$ potential $U(\pBI,\rBI)$ 
can be rewritten as follows: 
\begin{align*}
  U(\pBI,\rBI) 
  &={}
  - \frac{1}{4\mt c^2} H_0^2 
  - \frac{C_F^2 a_s^2}{2\mt c^2} \left(
      \frac{2}{3}(3-\SBI^2) + \frac{C_A}{C_F} \right)
    \frac{1}{r^2}
  \\
  &\qquad{}
  - \frac{3C_F a_s}{4\mt c^2} \left\{ H_0 \, , \, \frac{1}{r} \right\}
  + \frac{3+4\SBI^2}{12} \frac{C_F a_s}{\mt c^2} 
    \left[ H_0 \, , \, ip_r \right]
  \\
  &\qquad{}
  - \frac{2 C_F a_s}{3 \mt^2 c^2} \frac{\LBI^2 \SBI^2}{r^3}
  + \frac{3 C_F a_s}{2\mt^2 c^2} \frac{\dprod{\LBI}{\SBI}}{r^3}
  - \frac{C_F a_s}{2\mt^2 c^2} \frac{1}{r^3}
    \left( \SBI^2 - 3\frac{\pdprod{\SBI}{\rBI}^2}{r^2} \right)
  \\
  &={}
  - \frac{1}{4\mt c^2} H_0^2 
  - \frac{C_F^2 a_s^2}{2\mt c^2} \left(
      \frac{2}{3} + \frac{C_A}{C_F} \right)
    \frac{1}{r^2}
  \\
  &\qquad{}
  - \frac{3 C_F a_s}{4\mt c^2} \left\{ H_0 \, , \, \frac{1}{r} \right\}
  + \frac{11 C_F a_s}{12 \mt c^2} 
    \left[ H_0 \, , \, ip_r \right]
  \sepcomma
\end{align*}
where the last expression is for the $S$-wave.  
We concentrate on the $S$-wave hereafter.  
To summarize, 
\begin{align}
  &  H = H_0 + V_1 + U  \sepcomma\nonumber\\
  &  U = b H_0^2 + \left\{ H_0 \, , \, \calO_E \right\} 
           + \left[ H_0 \, , \, \calO_O \right] + W  
  \sepcomma\\
  &  b \equiv \frac{-1}{4\mt c^2}
  \sepcomma\quad
     \calO_E \equiv \frac{-3 C_F a_s}{4\mt c^2} \, \frac{1}{r}
  \sepcomma\quad
     \calO_O \equiv \frac{11 C_F a_s}{12\mt c^2} \, i p_r
  \sepcomma\nonumber\\
  &  W \equiv - \frac{C_F^2 a_s^2}{2\mt c^2}
           \left( \frac{2}{3} + \frac{C_A}{C_F} \right)
           \frac{1}{r^2}
       \equiv - \frac{\kappa}{m_tr^2c^2}
  \sepperiod\nonumber
\end{align}
Note that $\calO_O$ is anti-hermitian, and thus $U$ is hermitian.  
Using this expression for $U$, we can rewrite $G$ up to $\Order{1/c^3}$: 
\begin{align*}
  G 
  &= \frac{1}{H_0 + V_1 + U -\omega}  \\
  &\simeq 
    \frac{1}{H_0-\omega}
  - \frac{1}{H_0-\omega} (V_1+U) \frac{1}{H_0-\omega}
  +  \frac{1}{H_0-\omega} V_1 \frac{1}{H_0-\omega} V_1 \frac{1}{H_0-\omega}  \\
  &\simeq{}
  - b 
  + (1-b\omega -\calO_E - \calO_O) \frac{1}{H_0-\omega}
    (1-b\omega -\calO_E + \calO_O)  \\
  &\qquad{}
  - \frac{1}{H_0-\omega} (V_1+b\omega^2+2\omega\calO_E+W) \frac{1}{H_0-\omega}
  + \frac{1}{H_0-\omega} V_1 \frac{1}{H_0-\omega} V_1 \frac{1}{H_0-\omega}  \\
  &\simeq
  - b
  + (1-b\omega -\calO_E - \calO_O)
    \frac{1}{H_0 + V_1 + b\omega^2 + 2\omega\calO_E + W -\omega}
    (1-b\omega -\calO_E + \calO_O)
  \sepcomma
\end{align*}
or 
\begin{align}
  &
  G =
  G' - \left\{ b\omega+\calO_E \, , \, G' \right\}
  - \left[ \calO_O \, , \, G' \right] - b + \Order{\frac{1}{c^3}} 
  \label{eq:G_and_G'}
  \sepcomma\\
  &
  G' \equiv \frac{1}{H'-\calE}
  \sepcomma\quad
  H' \equiv H_0 + V_1 + 2\omega\calO_E + W
  \sepcomma\quad
  \calE \equiv \omega - b\omega^2
  \sepperiod\nonumber
\end{align}
Note that $H'$ does not contain neither $p^4$ nor $\delta(\rBI)$: 
\begin{align}
  H' = \frac{p^2}{m} - \frac{C_F \aB_s}{r} + V_1(r) - \frac{\kappa}{mr^2c^2}
  \label{eq:H'_exp_form}
  \sepcomma
\end{align}
where 
\begin{align*}
  \aB_s \equiv a_s \left( 1 + \frac{3}{2}\frac{E+i\Gamma}{mc^2} \right)
  \sepcomma\quad
  \calE \equiv (E+i\Gamma)\left(1+\frac{E+i\Gamma}{4mc^2}\right)
  \sepcomma
\end{align*}
and 
\begin{align}
  \kappa 
  &\equiv \frac{C_F^2}{2}\left(\frac{2}{3}+\frac{C_A}{C_F}\right)
    \pfrac{a_s}{c}^2  \\
  &= 2.5926\,a_s^2
  = 0.06637 \times \pfrac{a_s}{0.16}^2
  \sepperiod\nonumber
\end{align}
Thus numerical calculation of Green function is much easier 
for $G'$ than $G$ itself.  
In terms of c-number functions, they are related as 
\begin{align}
  G(r,r') 
  &=
  \biggl[ 1 + \frac{E+i\Gamma}{2mc^2} 
    + \frac{3C_F}{4}\pfrac{a_s}{c}^2 \left(
        \frac{1}{ma_sr} + \frac{1}{ma_sr'} \right) 
  \nonumber\\&\qquad\qquad{}
  - \frac{11C_F}{12}\pfrac{a_s}{c}^2 \left(
      \frac{1}{ma_sr}\odiff{}{r}r + \frac{1}{ma_sr'}\odiff{}{r'}r' 
    \right) \biggr] G'(r,r') 
  \nonumber\\&\qquad{}
  + \frac{1}{4mc^2} \frac{1}{4\pi rr'} \delta(r-r')
  + \Order{\frac{1}{c^3}}
  \nonumber\\&=
  \left[ 1 + \frac{E+i\Gamma}{2mc^2} 
    - \frac{C_F}{6}\pfrac{a_s}{c}^2 \left(
        \frac{1}{ma_sr} + \frac{1}{ma_sr'} \right) 
    + \frac{11 C_F^2}{12} \pfrac{a_s}{c}^2
  \right] G'(r,r') 
  \nonumber\\&\qquad{}
  + \frac{1}{4mc^2} \frac{1}{4\pi rr'} \delta(r-r')
  + \Order{r,r'\frac{1}{c^3}}
  \sepcomma
\end{align}
where we used the relation \EqRef{eq:diff_for_GC} 
for the Coulomb Green function.  Note that the difference is higher order.  
Thus, 
\begin{align}
  &  \Im G(r_0,r_0)  \nonumber\\
  &=
  \left[ 1+\pfrac{a_s}{c}^2\left(\frac{-C_F}{3ma_sr_0}+\frac{11C_F^2}{12}
    \right) \right] 
  \bIm{ \left( 1+\frac{E+i\Gamma}{2m^2c^2} \right) G'(r_0,r_0) }
  + \Order{r_0,\frac{1}{c^3}}
  \sepperiod
\end{align}
%
%
%
%

%
%
%----------------------------------------------------------------------
\subsection{Green function for Coulomb plus $1/r^2$ potentials}
\label{sec:coulomb+1/r2}
In this section, we consider the Hamiltonian%
%\footnote{
%  For notational simplicity, $H$ in this section means the Hamiltonian 
%  here.  Likewise, $G$ in this section means the Green function 
%  for $H$.  
%}%
, 
\begin{align}
  H = \frac{p^2}{m} - \frac{C_F a_s}{r} - \frac{\kappa}{mr^2c^2}
  \sepcomma
\end{align}
because (i) the Green function $G$ (projected to $S$-wave) 
can be obtained analytically; and (ii) the Green function $G'$ for $H'$ 
in \EqRef{eq:H'_exp_form} without $\log$ corrections $V_1(r)$ 
can be obtained by replacing $a_s \to \aB_s$, $E \to \calE$.  
The results for pure Coulomb potential, which we shall use several times, 
are also obtained by expanding in $\kappa$.  

The Schr\"odinger equation for $G(\vec{r},\vec{r}')$ is 
\begin{align*}
  \left[ E - \left( -\frac{\triangle_{\vec{r}}}{m} - \frac{C_F \alpha_s}{r}
        - \frac{\kappa}{mr^2} \right) \right] G(\vec{r},\vec{r}')
  = - \delta^{(3)}(\vec{r}-\vec{r}')
  \sepcomma
\end{align*}
where $r = |\vec{r}|$.  
We are interested in the Green function $G(r,r')$ projected to $S$-wave: 
\begin{align*}
  \left[ E - \left( \frac{-1}{mr} \pdiffn{}{r}{2} r 
      - \frac{C_F \alpha_s}{r} - \frac{\kappa}{mr^2} \right) \right] 
  G(r,r') = \frac{-1}{4\pi r r'} \delta(r-r')
  \sepperiod
\end{align*}
To simplify the equation, we introduce the following notations: 
\begin{align}
  G(r,r') = \frac{g(r,r')}{rr'} \sepcomma\quad 
  z = C_F \alpha_s m r \sepcomma\quad \nu^2 = \frac{(C_F\alpha_s)^2m}{4E}
  \sepperiod
\end{align}
Here $z$ is $r$ measured in unit of (twice) the Bohr radius $p_{\rmB}$, 
and $1/\nu^2$ is $E$ in the unit of the Bohr energy $E_\rmB$ 
[\EqRef{eq:pB_def}].  
Thus by dividing by $(C_F\alpha_s)^2m$, we have 
\begin{align*}
  \left( \frac{1}{4\nu^2} + \pdiffn{}{z}{2} + \frac{1}{z} + \frac{\kappa}{z^2}
  \right) g(z,z') = \frac{-1}{4\pi C_F\alpha_s} \delta(z-z')
  \sepcomma
\end{align*}
where $g(z,z')$ is just $g(r,r')$ without rescaling: 
\begin{align*}
  g(z,z') = g\bigl(r\!=\!z/(C_F\alpha_sm),r'\!=\!z'/(C_F\alpha_sm)\bigr)
  \sepcomma 
\end{align*}
while the normalization of $\delta$-function changes according to 
$\pi\delta(z) = \bIm{1/(z-i\epsilon)}$.  
Solutions with $\delta$-function source can be obtained by 
the solutions without $\delta$-functions: 
\begin{align}
  g(z,z') = \frac{1}{4\pi C_F\alpha_s \, W} \left[ 
    g_<(z) g_>(z') \theta(z'-z) + g_<(z') g_>(z) \theta(z-z') \right]
  \sepcomma
\end{align}
where $W$ is the Wronskian%
\footnote{
  One can easily show that the Wronskian $W$ for the two solutions of 
  a 2nd-rank homogeneous ordinary differential equation without 
  1st-derivative term, is independent of $z$: 
  \begin{align*}
    \odiff{W}{z} = 0  \sepperiod
  \end{align*}
  Thus one can calculate $W$ wherever he/she wants.  
}, 
\begin{align*}
  W \equiv W(g_>,g_<;z) \equiv 
  \begin{vmatrix}
    g_>(z)  &  g_<(z)  \\
    g_>'(z)  &  g_<'(z)  \\
  \end{vmatrix}
  = g_>(z)\,g_<'(z) - g_<(z)\,g_>'(z)
  \sepcomma
\end{align*}
and 
the function $g_>(z)$ ($g_<(z)$) is the 
solution to the homogeneous differential equation: 
\begin{align}
  \left( \frac{1}{4\nu^2} + \odiffn{}{z}{2} + \frac{1}{z} + \frac{\kappa}{z^2}
  \right) g(z) = 0
  \label{eq:diff_eq-for-g(z)}
  \sepcomma
\end{align}
which is ``well-behaved'' at $z\to\infty$ ($z\to 0$).  
Since we use the value of the Green function near the origin, 
it is important to know its behavior around there.  
For $\kappa > 0$, which is the case, the potential is dominated by 
$1/z^2$ near the origin.  
Assuming a solution of power behavior, 
\begin{align*}
  \left( \odiffn{}{z}{2} + \frac{\kappa}{z^2} \right) z^d \simeq 0
  \sepcomma
\end{align*}
one obtains 
\begin{align}
  d = \frac{1 \pm \sqrt{1-4\kappa}}{2} \equiv d_\pm
  \sepcomma\quad
  d_+ + d_- = 1  \sepcomma\quad d_+ d_- = \kappa
  \sepperiod
\end{align}
This may indicate that the wave function is collapsed into the origin 
when the attractive force is as strong as $\kappa>1/4$.  
However we don't have to worry about this because $\kappa \sim 0.06$ for 
our case.  
For small $\kappa$, $d_+ \simeq 1-\kappa$, $d_- \simeq \kappa$.  
Thus in order to obtain the results for Coulomb potential, 
it is convenient to write $d_+ (= 1-d_-)$ in terms of $d_-$.  
For both choice of $d_\pm$, 
%$G(r,r') = g(r,r')/rr'$ 
$g(z)/z$ 
diverges at the origin, but it's milder for $d_+$.  
Thus $g_<(z) \sim z^{d_+}$ near the origin.  
For Coulomb ($\kappa = 0$), two independent solutions are 
\begin{align*}
  z - \frac{1}{2}z^2 + \Order{z^3}  \sepcomma\quad
  1 - z \ln z + \Order{z^2\ln z}  \sepperiod
\end{align*}
In order for $g(z)/z$ to be finite at the origin, 
$g_<(z) \sim z - \frac{1}{2}z^2$.  

Further calculations are done in \SecRef{sec:coulomb+1/r2_app}.  
Especially, results for pure Coulomb potential, which we frequently refer to, 
is collected in \SecRef{sec:G_for_pure_Coulomb}.  
%
%
%

%
%
%-----------------------------------------------------------------------
\subsection{Explicit expression for $G'$}
\label{sec:exp_G'}
%Having determined the matching coefficient $C_1^{\rm (cur)}$, 
%or $C_1$ and $C_2(r_0)$, with Eqs.~(\ref{eq:R_with_ImG_and_C}) 
%and~(\ref{eq:C_ImG_to_C_ImG'}), all we have to do is to calculate 
%$G'(r,r')$ near the origin.  
The optical theorem shows that all we have to do is to calculate 
the Green function $G$, or $G'$, near the origin.  
The reduced Green function $G'$ is defined in 
Eqs.~(\ref{eq:G_and_G'}) and~(\ref{eq:H'_exp_form}).  
Thus we numerically solve the Schr\"odinger equation 
\begin{align}
  \left[ \calE 
    - \left( -\frac{\triangle}{m} - \frac{C_F \bar{a}_s}{r}
        + V_1(r) - \frac{\kappa}{mr^2} 
      \right) \right] \frac{g(r)}{r}
  = 0
  \label{eq:Schr_for_g<g>_NNLO}
  \sepcomma 
\end{align}
or, equivalently, 
\begin{align*}
  \left( \frac{1}{4\nu^2} + \odiffn{}{z}{2} + \frac{\kappa}{z^2}
    + \frac{1}{z} \left[
        \ell_2 \ln^2(\muT'z) + \ell_1 \ln(\muT'z) + \ell_0 \right]
  \right) g(z) = 0
  \sepcomma
\end{align*}
where 
\begin{align*}
  z = C_F \alpha_s m r
  \sepcomma\quad 
  \nu^2 = \frac{(C_F\alpha_s)^2m}{4\calE}
  \sepcomma\quad 
  \muT'z = \mu r \expo^\Egamma
  \sepcomma
\end{align*}
and 
\begin{align*}
  &  \ell_2 = \pfrac{a_s}{4\pi}^2 \, 4\beta_0^2  
  \sepcomma\\
  &  \ell_1 = \frac{a_s}{4\pi} \, 2\beta_0 
       + \pfrac{a_s}{4\pi}^2 \, 2\left( \beta_1 + 2\beta_0 a_1 \right)  
  \sepcomma\\
  &  \ell_0 = 1 + \frac{3(E+i\Gamma)}{2m} 
       + \frac{a_s}{4\pi} \, a_1 
       + \pfrac{a_s}{4\pi}^2 \, \left( \frac{\pi^2}{3}\beta_0^2 + a_2 \right)
  \sepperiod
\end{align*}
There are two independent solutions for this differential equation.  
Assuming their form near the origin as 
\begin{align}
  g_\pm(z) = z^{d_\pm} \left[ 1 + z \left( 
      L_2^\pm \ln^2(\muT'z) + L_1^\pm \ln(\muT'z) + L_0^\pm \right) 
    + \Order{z^2}
  \right]  
  \label{eq:g_pm_near_origin}
  \sepcomma
\end{align}
we have 
\begin{align*}
  &  L_2^\pm = -\frac{\ell_2}{2d_\pm}  
  \sepcomma\\
  &  L_1^\pm = \frac{2d_\pm+1}{2d_\pm^2} \ell_2 - \frac{1}{2d_\pm}\ell_1  
  \sepcomma\\
  &  L_0^\pm = -\frac{4d_\pm^2+2d_\pm+1}{4d_\pm^3}\ell_2
        + \frac{2d_\pm+1}{4d_\pm^2}\ell_1 - \frac{1}{2d_\pm}\ell_0
  \sepperiod
\end{align*}
With these two solutions, solutions (nearly) regular for 
$z \to \infty$ and $z\to 0$, respectively, are 
\begin{align*}
  g_>(r) = g_-(z) + B g_+(z) 
  \sepcomma\quad
  g_<(r) = g_+(z)
  \sepperiod
\end{align*}
With the normalization in \EqRef{eq:g_pm_near_origin}, 
the Wronskian $W$ with respect to $z$ is the same as 
\EqRef{eq:wronskian_for_g_pm}: $W = \sqrt{1-4\kappa}$.  
Thus we have 
\begin{align}
  \frac{4\pi}{m^2c} G'(r,r')
  =&\; \frac{4\pi}{m^2c} \frac{1}{4\pi C_F a_s W} \frac{g_>(z) g_<(z')}{rr'}
  \sepcomma\quad (z > z')  \nonumber\\
  =&\; \frac{C_F a_s}{\sqrt{1-4\kappa}} \left[ 
      \frac{g_+(z) g_-(z')}{zz'} + B \frac{g_+(z) g_+(z')}{zz'} \right]  
  \nonumber\\
  \to&\;
    \frac{C_F a_s}{\sqrt{1-4\kappa}} \Biggl[ 
      \frac{1}{z} \left\{ 1 + z \sum_{i=\pm} \left( 
            L_2^{(i)}\ln^2(\muT'z) + L_1^{(i)}\ln(\muT'z) + L_0^{(i)} \right)
          + \cdots \right\}  \nonumber\\
  &\quad{}
    + B z^{-2d_-} \left\{ 1 + 2z\left( 
            L_2^{(+)}\ln^2(\muT'z) + L_1^{(+)}\ln(\muT'z) + L_0^{(+)} \right)
          + \cdots \right\}  
    \Biggr]  \nonumber\\
  =&\;
    \frac{C_F a_s}{\sqrt{1-4\kappa}} \Biggl[ 
      \frac{1}{z} + \sum_{i=\pm} \left( 
            L_2^{(i)}\ln^2(\muT'z) + L_1^{(i)}\ln(\muT'z) + L_0^{(i)} \right)
    + B z^{-2d_-} \Biggr]  \nonumber\\
  &\quad{}
    + \mbox{ (vanish when $z\to 0$).  }
  \label{eq:ImG'(0,0)_detail}
\end{align}
where $\to$ means $r=r'=r_0 \to 0$.  
%
%
%

%
%
%-----------------------------------------------------------------------
\section{Matching of NRQCD with QCD}
\label{sec:matching}
Having taken into account the ``soft'' gluon contributions, 
next we consider the effect of ``hard'' gluons.  
It is determined by matching the NRQCD calculation to QCD one.  
Short distance coefficients $C_1^{\rm (cur)}$, $C_2^{\rm (cur)}$
for $t\tB$ production current, defined in \EqRef{eq:R_with_ImG_and_C}, 
are determined in the same section to the lowest order of $\alpha_s$.  
Higher orders can be 
determined by matching the result for Green function method 
with that for usual perturbative QCD.  
The Green function method, or NRQCD, can be applied when $\beta \ll 1$, 
while pQCD can be applied when $\alpha_s/\beta \ll 1$.  
Thus both formalism is valid when $\alpha_s \ll \beta \ll 1$.  
For this energy region, Green functions can be calculated perturbatively.  
While relativistic pQCD calculation was done in~\cite{CM98}.  
Both of these results are summarized in \SecRef{sec:matching-exp}.  
By demanding these two results to coincide, we obtain 
$C_1^{\rm (cur)}$ and $C_2^{\rm (cur)}$.  
Actually for $C_2^{\rm (cur)}$, we need only to the leading order, 
which was obtained already in free-propagating limit [\EqRef{eq:C_2}].  
With $C_2^{\rm (cur)} = 1/3$, 
\begin{align}
  &
  \left.
  \left\{ C_1^{\rm (cur)} 
    + C_2^{\rm (cur)} \frac{\triangle_r+\triangle_{r'}}{2m^2c^2} \right\}
  \Im G(r,r') \right|_{r,r'\to r_0}
  \label{eq:C_ImG_to_C_ImG'}
%  \\&=
%  C_1^{\rm (cur)} \Im G(r,r') 
%  + \frac{2}{6mc^2} \bIm{
%      -\left( \frac{C_F a_s}{r_0} + E + i\Gamma \right) G(r_0,r_0) }
  \\&=
  \left\{ C_1^{\rm (cur)} - \pfrac{a_s}{c}^2\frac{C_F}{3ma_sr_0} \right\}
  \bIm{ \left( 1 - \frac{E+i\Gamma}{3mc^2} \right) G(r_0,r_0) }
%  \\&=
%  \left\{ C_1^{\rm (cur)} - \pfrac{a_s}{c}^2\frac{C_F}{3ma_sr_0} \right\}
%  \left\{ 1 + \pfrac{a_s}{c}^2
%    \left( \frac{-C_F}{3ma_sr_0} + \frac{11}{12}C_F^2 \right) \right\} 
%  \times\\
%  &\qquad{}\times
%  \bIm{ \left( 1 - \frac{E+i\Gamma}{3mc^2} \right) 
%    \left( 1 + \frac{E+i\Gamma}{2mc^2} \right) G'(r_0,r_0) }
  \nonumber\\&=
  \left\{ C_1^{\rm (cur)} + \pfrac{a_s}{c}^2 \left(
      \frac{-2C_F}{3ma_sr_0} + \frac{11}{12}C_F^2 \right) \right\} 
  \bIm{ \left( 1 + \frac{E+i\Gamma}{6mc^2} \right) G'(r_0,r_0) }
  \nonumber
%  \\&=
%  \left\{ 1 + \pfrac{\alpha_s(\mu_h)}{\pi}C_F C_1
%    + \pfrac{\alpha_s(\mu_h)}{\pi}^2 C_F C_2(r_0) \right\}
%  \bIm{ \left( 1 + \frac{E+i\Gamma}{6mc^2} \right) G'(r_0,r_0) }
\end{align}
where the equalities are up to $\Order{1/c^3}$.  
Here we used Eqs.~(\ref{eq:d'Alembertian_to_p2}) and (\ref{eq:G_and_H}), or 
\begin{align*}
  \left[ -\frac{\triangle_r}{m} - \frac{C_Fa_s}{r} - (E+i\Gamma) 
    + \Order{\frac{1}{c}} \right] G(r,r') 
  = \frac{1}{4\pi rr'} \delta(r-r')
  \sepcomma\mbox{ or }\\
  \frac{\triangle_r}{m} \Im G(r,r') 
  = - \bIm{\left( \frac{C_Fa_s}{r} + E+i\Gamma \right) G(r,r')}
  + \Order{\frac{1}{c}}
  \sepperiod
\end{align*}
The same relation holds also for $G'(r,r')$ since their difference is 
$\Order{1/c^2}$.  
With the definition 
\begin{align}
  & \left\{ 1 + \pfrac{\alpha_s(\mu_h)}{\pi}C_F C_1
    + \pfrac{\alpha_s(\mu_h)}{\pi}^2 C_F C_2(r_0) \right\}  \nonumber\\
  &\equiv  \left\{ C_1^{\rm (cur)} + \pfrac{a_s}{c}^2 \left(
      \frac{-2C_F}{3m_ta_sr_0} + \frac{11}{12}C_F^2 \right) \right\}  
  \sepcomma
  \label{eq:C1C2-Ccur}
\end{align}
we have~\cite{MY98} 
\begin{align}
  C_1 &= -4 
  \label{eq:C1_C2}
  \sepcomma\\
  C_2(r_0) &= 
      C_F C^{A}_2
    + C_A C^{NA}_2
    + T_F n_f C^{L}_2
    + T_F n_H C^{H}_2  \nonumber\\
  &\quad{}
  + \pi^2 \left( \frac{2}{3}C_F + C_A \right) 
    \ln\pfrac{r_0}{r_0^{\rm (ref)}}
  + 2\beta_0 \ln\pfrac{m_t}{\mu_h}
  \nonumber\\
  &={}
  -19.17+0.611111\,n_f+0.251199\,n_H 
  \nonumber\\
  &\qquad{}
  + 38.3818\,\ln\pfrac{r_0}{r_0^{\rm (ref)}}
  + \left( 22-\frac{4}{3}n_f \right) \ln\pfrac{m_t}{\mu_h}
  \nonumber\\
  &={}
  -18.3077 + 38.3818\,\ln\pfrac{r_0}{r_0^{\rm (ref)}} 
  + 15.3333 \ln\pfrac{m_t}{\mu_h}
  \sepcomma\\
  &
  r_0^{\rm (ref)} \equiv \frac{\expo^{2-\Egamma}}{2m_tc} 
  = \frac{2.07433}{\mt c}
  \label{eq:r0ref}
\end{align}
where 
\begin{align*}
  &  C_2^A = \frac{39}{4} - \zeta_3 + \pi^2 \left( 
      \frac{4}{3}\ln 2 - \frac{35}{18} \right)
      = -1.5215\cdots
  \sepcomma\\
  &  C_2^{NA} = -\frac{151}{36} - \frac{13}{2}\zeta_3 + \pi^2 \left(
      \frac{179}{72} - \frac{8}{3}\ln 2 \right)
      = -5.71378\cdots
  \sepcomma\\
  &  C_2^L = \frac{11}{9}
      = 1.22222\cdots
  \sepcomma\\
  &  C_2^H = \frac{44}{9} - \frac{4}{9}\pi^2
      = 0.502398\cdots
  \sepperiod
\end{align*}
Note that the velocity dependence of the matching coefficient is determined 
by single number: $C_2^{\rm (cur)}=1/3$.  
Thus it is highly non-trivial that the matching is surely possible.  

Let us consider the $r_0$ dependence of $R$ ratio 
for the case $\Gamma_t = 0$.  
Since only $\ell_0$ and $L_0^\pm$, among $\ell_i$ and $L_i^\pm$, 
have imaginary part, we see from \EqRef{eq:ImG'(0,0)_detail}, 
\begin{align*}
  \frac{4\pi}{m^2c} \Im G'(r_0,r_0)
  &= \frac{C_F a_s}{\sqrt{1-4\kappa}} \left\{
      \bIm{B} z_0^{-2d_-} + \bIm{L_0^+ + L_0^-} \right\}  \\
  &= \frac{C_F a_s}{\sqrt{1-4\kappa}} \left\{
      \bIm{B} z_0^{-2d_-} - \frac{3\Gamma_t}{4\kappa\mt} \right\}  
  \sepperiod
\end{align*}
Here we omit the terms that vanish when $z_0\to 0$.  
Since $\kappa$ is a coefficient of $\Order{1/c^2}$ interaction, 
we need only the terms with the first (or less) order in $\kappa$.  
Thus 
\begin{align*}
  \Im G' \sim 
  z_0^{-2d_-} 
  &= \expo^{-2d_-\ln z_0} = 1 - 2d_- \ln(z_0) + \cdots  \\
  &= 1 - 2 \kappa \ln(z_0) - 2 \kappa^2 \ln(z_0)(1-\ln(z_0)) + \cdots 
\end{align*}
where $z_0 = C_F \alpha_s(\mu_s) \mt r_0$.  
While $r_0$ dependence of matching coefficients [\EqRef{eq:C1_C2}] is 
\begin{align*}
  C \sim 1 + 2 \kappa \ln(r_0/r_0^{\rm (ref)})  \sepperiod
\end{align*}
Thus we can see that $\ln r_0$ singularity is cancelled in 
$R$ ratio to the order we are concerning%
\footnote{
  The $\kappa$ in $G'$ is evaluated at the scale $\mu_s$, while 
  the $\kappa$ in $C$ is evaluated at $\mu_h$.  However, 
  the difference between $\alpha_s(\mu_s)^2$ and $\alpha_s(\mu_h)^2$ 
  is higher order, at least formally.  
}.  
However for finite width $\Gamma_t$, the expression for $R$ diverges 
with $1/r_0$ and $\ln(r_0)$.  
This is due to the incomplete treatment of the width $\Gamma_t$.  
Here we follow the treatment of~\cite{MY98}; that is, 
expand the expression with respect to $r_0$ as detail as possible, 
and drop the terms that vanish when $r_0 \to 0$.  
We choose $r_0 = r_0^{\rm (ref)}$.  
%
%
%

%
%
%------------------------------------------------------------------------
\section{Previously obtained results for $R$ ratio}
\label{sec:prev_res_R}
In this section, we show $R$ ratio for 
freely-propagating $t\tB$ with and without finite decay width $\Gamma_t$, 
Leading Order (LO) Coulomb rescattering with and without $\Gamma_t$, 
Next-to-Leading Order (NLO = $\Order{1/c}$) Coulomb rescattering 
with $\Gamma_t$, and 
Next-to-Next-to-Leading Order (NNLO = $\Order{1/c^2}$) Coulomb rescattering 
with $\Gamma_t$.  
All of these are for fixed order $\alpha_s(\mu_s)$ calculations%
\footnote{
  LO and NLO corrections with RG improvement were also calculated; 
  these are summarized in \FigRef{fig:R_75_3_Rf}.  
}, 
and are obtained before our work, except those with $a_2^{\rm new}$.  
%------------------------------------------------------------------------
\subsection{Analytic results}
Analytic formulas for $R$ ratios [\EqRef{eq:R_def}]
are available for the following cases: 
\begin{align}
  &  R = \frac{3}{2} N_C Q_t^2 \sqrt{\frac{E}{\mt}}
  \sepcomma&&\text{ for free; without $\Gamma_t$}  
  \sepcomma\nonumber\\
  &  R = \frac{3}{2} N_C Q_t^2 \sqrt{\frac{\sqrt{E^2+\Gamma_t^2}+E}{2\mt}}
  \sepcomma&&\text{ for free; with $\Gamma_t$}  
  \sepcomma\nonumber\\
  &  R = \frac{3}{2} N_C Q_t^2 \frac{\pi C_F \alpha_s}{1-\expo^{-z_E}}
  \sepcomma\quad z_E = \pi C_F \alpha_s \sqrt{\frac{\mt}{E}}
  \sepcomma&&\text{ for LO; without $\Gamma_t$}  
  \label{eq:R_LO_noGamma}
  \sepcomma
\end{align}
where $E = \sqrt{s}-2\mt$.  
The second one is given in \EqRef{eq:R_for_free_with_Gamma}, 
and the last one is in \EqRef{eq:ImGC_E>0}.  
\begin{figure}[tbp]
  \hspace*{\fill}
  \begin{minipage}{6cm}
    \includegraphics[width=8cm]{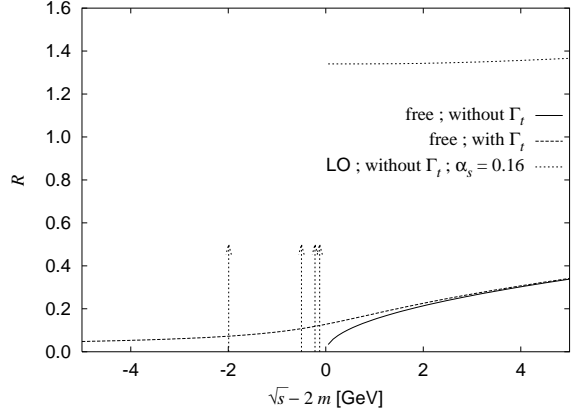}
  \end{minipage}
  \hspace*{\fill}
  \\
  \hspace*{\fill}
  \begin{Caption}\caption{\small
      $R$ ratios for the situations that analytic formulas are 
      available.  We used $\mt = 175\GeV$, $\Gamma_t = 1.43\GeV$, 
      $\alpha_s(\mu_s) = 0.16$, which is for $\mu_s \simeq 20\GeV$.  
      Shown in vertical lines are the first four resonances for 
      Coulomb potential: 
      $E_n = -E_\rmB/n^2$ where $E_\rmB \equiv p_\rmB^2/\mt$ and 
      $p_\rmB \equiv C_F \alpha_s(\mu_s) \mt/2$.  
      Note that the self-consistent solution for 
      $p_\rmB = C_F \alpha_s(\mu_s = p_\rmB) \mt/2$ is 
      $p_\rmB \simeq 20\GeV$.  
      \label{fig:Rfree}
  }\end{Caption}
  \hspace*{\fill}
\end{figure}
These results are shown in~\FigRef{fig:Rfree}.  
We can see several points.  Because of non-zero width $\Gamma_t$, 
the production cross section is non-zero also for $E<0$.  
More prominently, attractive Coulomb interaction between $t\tB$ 
enhance $t\tB$ production cross section by many times.  
This is because the $t\tB$ wave function $|\psi(0)|^2$ at the origin 
is enhanced by the attractive potential.  
%------------------------------------------------------------------------
\subsubsection{Wave function $\psi(0)$ at the origin and potential $V(r)$}
\label{sec:psi(0)_V(r)}
The magnitude of a wave function at the origin is related to 
(the expectation value of) the attractive force $\diffn{V(r)}{}/\diffn{r}{}$ 
~\cite{LSG91}.  
We are interested only in $S$ wave.  Higher orbital angular momentum states 
may not contribute to $\psi(0)$.  
Thus we define 
\begin{align*}
  \psi(\rBI) = \frac{1}{\sqrt{4\pi}} \frac{y(r)}{r}
\end{align*}
for $S$ wave, which means 
\begin{align*}
  \triangle \psi(\rBI) = \frac{1}{\sqrt{4\pi}} \frac{y''(r)}{r}
  \sepperiod
\end{align*}
With this and the Schr\"odinger equation 
\begin{align*}
  \left[ - \frac{\triangle}{2\mu} + V(\rBI) \right] \psi(\rBI) = E \psi(\rBI)
  \sepcomma
\end{align*}
where $\mu=\mt/2$ is the reduced mass, 
the following quantity can be evaluated in two ways: 
\begin{align*}
  - \int\!\!\diffn{r}{3} \frac{y'' y'}{4\pi r^2} 
  &= \frac{-1}{2} \int\!\! \diffn{r}{} (y^2)' 
  = \frac{-1}{2} \left. (y^2)' \right|_0^\infty
  = -2\pi \left. (\psi+r\psi')^2 \right|_0^\infty
  = 2\pi |\psi(0)|^2
  \sepcomma\quad\mbox{ while }\\
  &= 2\mu \int\!\!\diffn{r}{3} [E-V(r)] \frac{y y'}{4\pi r^2} 
  = \mu \int\!\! \diffn{r}{} [E-V(r)] (y^2)'
  = \mu \int\!\! \diffn{r}{} V'(r) \, y^2
  = \mu \mean{V'}
  \sepperiod
\end{align*}
Thus we have 
\begin{align}
  |\psi(0)|^2 = \frac{\mu}{2\pi} \mean{V'(r)}  
  \label{eq:psi0_V}
  \sepperiod
\end{align}

On the other hand, constant shift of the potential, 
$V(r) \to V(r) + \delta V$, can be compensated by the shift of the energy, 
$E \to E + \delta V$.  Thus it is related to the shift of the mass $\mt$, 
since $E = \sqrt{s} - 2\mt$.  This issue is discussed intensively in 
\SecRef{sec:renormalon} in connection with renormalon ambiguity of both 
pole mass $\mt^{\rm pole}$ of a quark and 
potential $V(r)$ in coordinate space.  
%------------------------------------------------------------------------
\subsection{Numerical results}
\label{sec:R_NNLO_fixed}
Analytic formulas for $R$ ratio cannot be obtained 
with non-zero width $\Gamma_t$ even for LO.  
For LO and NLO, the expression for $R$ ratio is 
\begin{align*}
  R(s) 
  &= \frac{3}{2} N_C Q_t^2 \, \frac{4\pi}{m^2c} 
     C_1^{\rm (cur)} \Im G(r=0,r'=0)
  \sepcomma
\end{align*}
where the ($S$ wave projected) Green function $G(r,r')$ is the solution for 
\begin{align*}
  \left[ (E +i\Gamma_t) 
    - \left( \frac{-1}{m_t}
      \left[ \odiffn{}{r}{2} + \frac{2}{r} \odiff{}{r} \right]
      - \frac{C_F a_s}{r} + V_1(r) \right) \right] 
  G(r,r') = \frac{-1}{4\pi r r'} \delta(r-r')
  \sepcomma
\end{align*}
with $E = \sqrt{s} - 2 m_t^{\rm (pole)}$.  
Note that $V_1(r)$ is zero for LO.  For NLO, only $\Order{1/c}$ terms are 
used in $V_1(r)$.  The same notice is applied for the 
short-distance current renormalization $C_1^{\rm (cur)}$.  
It can read from Eqs.~(\ref{eq:C1C2-Ccur}) and~(\ref{eq:C1_C2}); 
it is unity for LO, and 
$(1-2C_F\alpha_s(\mu_h)/\pi)^2 = (1-4C_F\alpha_s(\mu_h)/\pi)$ for NLO
[\EqRef{eq:hard_vert_corr_to_NLO}].  

The situation becomes more complicated for NNLO (and higher).  
The last expression is the one we actually used: 
\begin{align}
  R
  &= \frac{3}{2} N_C Q_t^2 \, \frac{4m^2}{s} \, \frac{4\pi}{m^2c} 
     \left.
     \left\{ C_1^{\rm (cur)} 
       + C_2^{\rm (cur)} \frac{\triangle_r+\triangle_{r'}}{2m^2c^2} \right\}
     \Im G(r,r')  \right|_{r,r'\to r_0}  \nonumber\\
  &= \frac{3}{2} N_C Q_t^2 \, \frac{4m^2}{s} \, \frac{4\pi}{m^2c}  
    \left\{ 1 + \pfrac{\alpha_s(\mu_h)}{\pi}C_F C_1
    + \pfrac{\alpha_s(\mu_h)}{\pi}^2 C_F C_2(r_0) \right\} 
  \times\nonumber\\
  &\quad{}\times
    \bIm{ \left( 1 + \frac{E+i\Gamma}{6mc^2} \right) G'(r_0,r_0) }
  \label{eq:R_NNLO_expl}
  \\
  &= \frac{3}{2} N_C Q_t^2 \, \frac{4m^2}{s} 
    \left\{ 1 + \pfrac{\alpha_s(\mu_h)}{\pi}C_F C_1
    + \pfrac{\alpha_s(\mu_h)}{\pi}^2 C_F C_2(r_0) \right\} 
    \frac{C_F a_s}{\sqrt{1-4\kappa}} 
  \times\nonumber\\
  &\quad{}\times
    \bIm{ \left( 1 + \frac{E+i\Gamma}{6mc^2} \right) \left\{
        B z_0^{-2d_-} 
        + \frac{1}{z_0} + \sum_{i=\pm} \left( 
          L_2^{(i)}\ln^2(\muT'z_0) + L_1^{(i)}\ln(\muT'z_0) + L_0^{(i)} \right)
  \right\} }
  \sepcomma\nonumber
\end{align}
where $z_0 = C_F \alpha_s m r_0$, $\muT'z_0 = \mu r_0 \expo^\Egamma$, 
and $r_0 = r_0^{\rm (ref)} \equiv \expo^{2-\Egamma}/(2m_t)$.  
Here we used Eqs.~(\ref{eq:R_with_ImG_and_C}), (\ref{eq:C_ImG_to_C_ImG'}), 
(\ref{eq:C1C2-Ccur}) and~(\ref{eq:ImG'(0,0)_detail}).  
Reduced Green function $G'(r,r')$ is the solution of 
\begin{align*}
  \left[ \calE 
    - \left( \frac{-1}{m_t}
      \left[ \odiffn{}{r}{2} + \frac{2}{r} \odiff{}{r} \right]
      - \frac{C_F \aB_s}{r} + V_1(r) - \frac{\kappa}{\mt r^2} \right) \right] 
  G'(r,r') = \frac{-1}{4\pi r r'} \delta(r-r')
  \sepcomma
\end{align*}
where 
\begin{align*}
  \aB_s \equiv a_s \left( 1 + \frac{3}{2}\frac{E+i\Gamma}{mc^2} \right)
  \sepcomma\quad
  \kappa = \frac{C_F^2}{2}\left(\frac{2}{3}+\frac{C_A}{C_F}\right)
           \pfrac{a_s}{c}^2
  \sepcomma\quad
  \calE = (E+i\Gamma)\left(1+\frac{E+i\Gamma}{4mc^2}\right)
  \sepperiod
\end{align*}
The coefficient $B$ is determined as 
\begin{align*}
  B = \left. - \frac{g_-(r)}{g_+(r)} \right|_{r \to \infty} 
  \sepcomma 
\end{align*}
where $g_\pm(r)$ are 
the solutions of homogeneous differential equation for $G'(r,r')$ 
with the boundary condition 
\begin{align*}
  g_\pm(z) = z^{d_\pm} \left[ 1 + z \left( 
      L_2^\pm \ln^2(\muT'z) + L_1^\pm \ln(\muT'z) + L_0^\pm \right) 
    + \Order{z^2}
  \right]  \sepperiod
\end{align*}

Our input is 
\begin{align*}
  \mt = 175\GeV  \sepcomma\quad
  \Gamma_t = 1.43\GeV  \sepcomma\quad
  \alpha_s^\MSbar(m_Z)=0.118  \sepperiod
\end{align*}
There are also another inputs.  
One is the ``hard scale'' $\mu_h = \mt$, which is for 
$\alpha_s$ in matching coefficients, 
and the ``soft scale'' $\mu_s = 20\GeV, 75\GeV$, which is for 
$\alpha_s$ in potential.  
The other is regulator $r_0 = r_0^{\rm (ref)}$ for the 
incomplete treatment of $\Gamma_t$.  

\begin{figure}[tbp]
  \hspace*{\fill}
  \begin{minipage}{6cm}
    \includegraphics[angle=-90,width=7.5cm]{./figure/R_75_0.eps}
  \end{minipage}
  \hspace*{\fill}
  \begin{minipage}{6cm}
    \includegraphics[angle=-90,width=7.5cm]{./figure/R_20_0.eps}
  \end{minipage}
  \hspace*{\fill}
   \\
  \hspace*{\fill}
  \begin{Caption}\caption{\small
      $R$-ratios for $e^+ e^- \to \gamma^* \to t\bar{t}$ 
      at LO (dot-dashed), NLO (dashed), and NNLO (solid) 
      as functions of the energy measured from 
      twice the pole mass, $E = \sqrt{s}-2 m_{\rm pole}$.
      Arrows indicate dislocations of the maximum point of $R$
      as the $\Order{1/c}$ and $\Order{1/c^2}$
      corrections are included, respectively.
      We put $m_{\rm pole} = m_t = 175\GeV$, 
      $\Gamma_t=1.43\GeV$, and $\alpha_s(m_Z)=0.118$.  
      Dotted lines show NNLO $R$-ratios calculated with an old 
      value of $a_2$ \protect\cite{P97}, which is one of the coefficients 
      in the two-loop perturbative QCD potential $V_1(r)$.  
      Figure (a) is for $\mu_s = 75\GeV$ and (b) is for $\mu_s = 20\GeV$.  
      \label{fig:R75-20_0}
  }\end{Caption}
  \hspace*{\fill}
\end{figure}%
\begin{figure}[tbp]
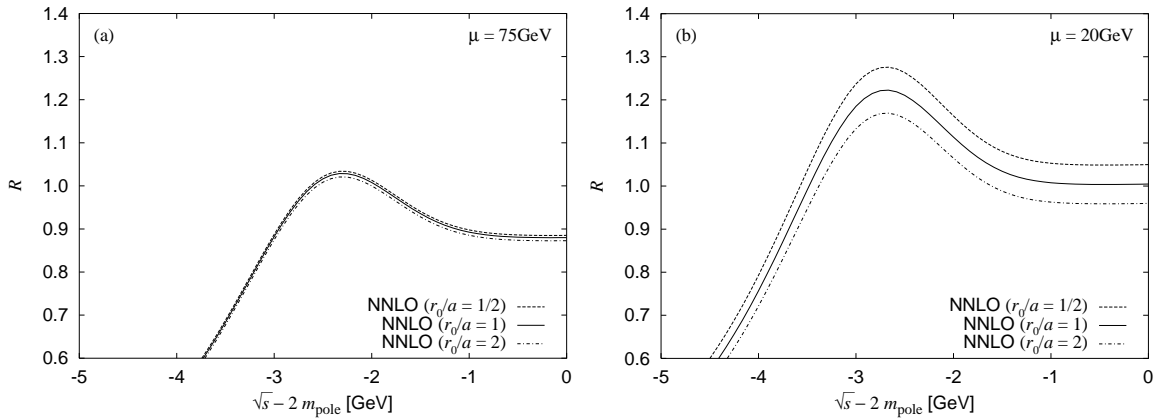

  \hspace*{\fill}
  \begin{minipage}{6cm}
    \includegraphics[angle=-90,width=7.5cm]{./figure/R_75_r0.eps}
  \end{minipage}
  \hspace*{\fill}
  \begin{minipage}{6cm}
    \includegraphics[angle=-90,width=7.5cm]{./figure/R_20_r0.eps}
  \end{minipage}
  \hspace*{\fill}
   \\
  \hspace*{\fill}
  \begin{Caption}\caption{\small
      $R$-ratios for $e^+ e^- \to \gamma^* \to t\bar{t}$ 
      at NNLO for several values of $r_0$: 
      $r_0$ = $a/2$ (dashed), $r_0=a$ (solid), and $r_0=2a$ (dot-dashed), 
      where $a \equiv r_0^{\rm (ref)} \equiv \expo^{2-\Egamma}/2m_t$.  
      Figure (a) is for $\mu_s = 75\GeV$, and (b) is for $\mu_s = 20\GeV$.  
      Other notations and parameters are same as in~\FigRef{fig:R75-20_0}.
      \label{fig:R75-20_r0}
  }\end{Caption}
  \hspace*{\fill}
\end{figure}%

Shown in Figure~\ref{fig:R75-20_0} is the $R$-ratios for 
LO, NLO (= $\Order{1/c}$) and NNLO (= $\Order{1/c^2}$), 
where the last one is calculated in~\cite{HT98,MY98,Y99} 
for $\mu_s=75\GeV$.  Our numerical results are consistent with them.  
As was already pointed in~\cite{HT98,MY98,Y99}, 
$\Order{1/c}$ and $\Order{1/c^2}$ corrections 
for $R$ ratio are not small at all.  
In fact this is the motivation for us to study this issue: 
how well can we improve the perturbative convergence of $R$.  
There are three possible sources of this bad convergence: 
(i) after the calculations in~\cite{HT98,MY98,Y99}, 
it was found~\cite{S99} that the previous calculation~\cite{P97} 
of a coefficient $a_2$ in $V_1$, which is one of the $\Order{1/c^2}$ 
corrections, turn out to wrong.  In fact the original value of $a_2$ is 
so huge that it was suspected to be wrong before the 
calculation in~\cite{S99}.  We see the effect of this modification 
just below in this section.  
(ii) One can see that the soft-scale ($\mu_s$) dependence of $R$ is also 
large.  This suggests the necessity of Renormalization-Group (RG) 
improvement.  We do this in \SecRef{sec:RGimp_R}.  
(iii) Yet another possible source is renormalon ambiguity 
of pole mass and potential.  We pursue this issue in \SecRef{sec:renormalon}.  

Before studying suspects (ii) and (iii) above, 
let us look at more closely these fixed order results for now.  
In Figures~\ref{fig:R75-20_0}, the difference of those two figures are 
the choice of the soft scale $\mu_s$, which is written simply as $\mu$.  
One choice is $\mu_s = 20\GeV \simeq p_\rmB$ and the other is $75\GeV$.  
The latter is chosen somewhat arbitrary.  
Naively, the natural scale for Coulomb potential is 
Bohr momentum $p_\rmB \simeq 20\GeV$ [\EqRef{eq:pB_def}].  
However one can see that 
the convergence of the normalization 
is better for $\mu_s = 75\GeV$ than that for $\mu_s = 20\GeV$, although 
the convergence of the peak position ($\simeq$ mass of the $1S$ resonance) 
is better for $\mu_s = 20\GeV$ than that for $\mu_s = 75\GeV$.  
This maybe because 
the normalization is determined by the wave function at the origin, 
which is much shorter than $1/p_\rmB$; $|\psi(0)|^2 (\propto \mean{V'(r)})$.  
More explicitly, from \FigRef{fig:dVdr_20-75-RG} one can see that 
the convergence of the attractive force due to Coulombic potential 
$V_{\rm C}$ is worse for $\mu = 20\GeV$.  
There is another attractive force due to $V_{\rm NA}$ for NNLO.  
This also increases the normalization%
\footnote{
  Short distance vertex corrections $C_1$ etc.\ 
  also affects the overall normalization.  
  For example, with $\alpha_s^\MSbar(m_t) = 0.107$, it is 
  $1-4C_F\alpha_s^\MSbar(\mu_h=m_t)/\pi = 0.818$ to NLO.  
}.
On the other hand, 
the position of $1S$ peak is determined by Bohr momentum $p_\rmB$.  
However corrections to the potential $V_1(r)$ are minimized 
near%
\footnote{
  For reference, 
  $\expo^{\gamma_E} = 1.781\cdots$.  
}
$1/r \simeq \mu_s'=\mu_s\expo^{\gamma_E}$.  
Interplay of these factors makes the estimate of the optimal scale 
for $\mu_s$ complicated.  
We consider this issue once more in \SecRef{sec:theo_err_NNLO_R}, 
where the theoretical ambiguities of our calculations are examined.  

Also shown in the same figures are the effects of 
correcting $a_2$~\cite{P97,S99}, which is a coefficient of NNLO correction 
to Coulomb potential.  
Numerically, the difference between 
$a_2^{\rm (old)}\simeq 333$ and $a_2^{\rm (new)}\simeq 156$ is large, 
but the effect of correcting it to $R$ turns out to be small.  
This is because $\log$ term in $V_1(r)$ [\EqRef{eq:V1_def}] are also large.  
In the following section, we give the results for RG improved potential, 
where large $\log$'s are summed.  For those cases, NNLO corrections 
diminishes by factor two when the correct value for $a_2$ is used.  

Shown in Figures~\ref{fig:R75-20_r0} are the $r_0$ dependence of $R$ ratio.  
Both vertical and horizontal directions are magnified by two, 
compared to Figures~\ref{fig:R75-20_0}.  
The value of $r_0$ is varied by factors 2 and 1/2.  
As was explained in \SecRef{sec:exp_G'}, 
the variation in $R$ is due to not only $\Order{1/c^3}$ corrections 
but also $\Order{\Gamma_t/m_t}$ terms in \EqRef{eq:R_NNLO_expl}. 
The size of this variation is a measure of 
one of the uncertainties of our theoretical prediction.  
We can see it is rather small compared to 
the $\Order{1/c^2}$ correction itself.  
Especially, side-ward shift is very small.  
%
%
%

%
%
%--------------------------------------------------------------------------
\section{Renormalon ambiguity and PS mass}
\label{sec:renormalon}
%------------------------------------------------------------------------
\subsection{Perturbative convergence}
It is known that the perturbative expansion in QFT is 
an asymptotic expansion.  The ambiguity due to this is called 
``renormalon ambiguity''~\cite{T77,M92,Z98,B99}.  
This is the subject of this section.  

For some quantities $Q$, explicit calculations show that 
perturbative convergence of $Q$ is much better when $Q$ are expressed 
in terms of \MSbar-mass $\mt^\MSbar$ than in pole-mass $\mt^\pole$.  
One example is the leading fermionic correction 
$(\Delta\rho)_f$ to the $\rho$-parameter~\cite{GS95}: 
\begin{align*}
  \Delta\rho^{\text{on-shell}} 
  &\simeq 
  N_C \frac{G_F M_t^2}{8\sqrt{2}\pi^2}
  \left[
    1 - \frac{\alpha_s(\mu)}{\pi} \times 2.9 
    - \pfrac{\alpha_s(\mu)}{\pi}^2
      \left( 5.0\ln\frac{\mu^2}{M_t^2} + 15. \right)
  \right]
  \sepcomma\\
  \Delta\rho^\MSbar
  &\simeq
  N_C \frac{G_F \bar{m}_t(\mu)^2}{8\sqrt{2}\pi^2}
  \left[
    1 + \frac{\alpha_s(\mu)}{\pi} (2 \ell - 0.19)
    + \pfrac{\alpha_s(\mu)}{\pi}^2
      \left( \frac{15}{4}\ell^2 + 2.0\ell - 4.0 \right)
  \right]
  \sepcomma
\end{align*}
where%
\footnote{See also~\cite{KS95}.  } 
$M_t \equiv m_t^{\rm pole}$, $\bar{m}_t(\mu) \equiv m_t^\MSbar(\mu)$, and 
$\ell \equiv \ln[\mu^2/\bar{m}_t(\mu)^2] = 0$ for 
$\mu = \mu_t \equiv \bar{m}_t(\mu_t)$.  
One can see that with $\mu = \mu_t$, 
the coefficient of $\alpha_s/\pi$ is smaller by factor 15 for \MSbar-mass, 
and factor 4 in the case $(\alpha_s/\pi)^2$.  
A similar behavior can also be seen in the QCD corrections to the 
interactions $\ell^+\ell^-H$, $W^+W^-H$, and $ZZH$~\cite{KS95}.  
It is suspected that these worse convergence may be a signal of 
renormalon contributions.  

Likewise, as we shall see below, 
the pole mass $\mt^{\rm pole}$ of a quark itself suffers from 
``renormalon ambiguity'' of $\Order{\Lambda_{\rm QCD}}$ even if 
$\Gamma_t > \Lambda_{\rm QCD}$~\cite{BB94,BSUV94,B95,SW97}.  
For example, the invariant mass $m_{jW}$ of the jet and $W$, which are 
the decay products of $t$, distributes around $m_t$.  
However since $b$ is not color-singlet, the $m_{jW}$-distribution 
do not directly represent $\mt^{\rm pole}$.  

On the other hand, there is also ``renormalon ambiguity'' for quark-antiquark 
potential $V(r)$ in coordinate space, and it can be shown~\cite{B98,HSSW99} 
that the severest 
ambiguity cancel in the combination $2\mt^{\rm pole} + V(r)$.  
We shall see this just below.  
%------------------------------------------------------------------------
\subsection{IR gluon contributions to self-energy $\Sigma$ and potential $V(r)$}
The mass $m$ of a particle $f$ is renormalized 
due to the self-energy $\Sigma$ of $f$ by $\delta m = m\Sigma_1$, 
since 
\begin{align*}
  \frac{i}{\pS-m-\Sigma}
  &= \frac{i}{\pS-m-m\Sigma_1-(\pS-m)\Sigma_2}  \\
  &\simeq \frac{1}{1-\Sigma_2} \frac{i}{\pS-m(1+\Sigma_1)}
  \sepcomma
\end{align*}
where 
\begin{align*}
  \Sigma(p,m) 
  &\equiv m \Sigma_1(p^2,m) + (\pS-m)\Sigma_2(p^2,m)  \\
  &= \pS \Sigma_2 + m(\Sigma_1-\Sigma_2)
  \sepperiod
\end{align*}
Let us consider the one-loop gluon contribution to $\Sigma$ of a quark.  
Since gluon is strongly coupled in IR region, we concentrate on those 
region.  For a static quark $p^\mu \simeq (m,\vec{0})$~\cite{SW97}, 
\begin{align*}
  -i\Sigma_{\rm IR}
  &= 
  (-ig_s)^2\,T^a T^a \int\limits_{\vec{k}\,=\,\text{soft}}\!\!
  \left.
  \frac{\diffn{k}{4}}{(2\pi)^4} \gamma^0 
  \frac{i(\pS+\kS+m)}{(p+k)^2-m^2+i\epsilon} \gamma_0
  \frac{-i}{-|\kBI|^2+i\epsilon}
  \right|_{p^2 \simeq m^2}
  \\
  &\simeq
  \frac{1+\gamma^0}{2}
  C_F g_s^2 \int\limits_{\vec{k}\,=\,\text{soft}}\!\!
  \frac{\diffn{k}{4}}{(2\pi)^4} 
  \frac{1}{(k^0+i\epsilon)(|\kBI|^2-i\epsilon)}
  \\
  &=
  \frac{1+\gamma^0}{2}
  \frac{-i}{2} \int\limits_{\vec{k}\,=\,\text{soft}}\!\!
  \frac{\diffn{k}{3}}{(2\pi)^3} 
  \frac{4\pi C_F \alpha_s}{|\kBI|^2}
  \\
  &\simeq
  (-i)
  \frac{1+\gamma^0}{2}
  \frac{1}{2} \int\limits_{\vec{k}\,=\,\text{soft}}\!\!
  \frac{\diffn{k}{3}}{(2\pi)^3} 
  \frac{4\pi C_F \alpha_s(|\kBI|^2)}{|\kBI|^2}
  \sepcomma
\end{align*}
where we used $k^i \sim 0$ in the denominator, and 
$\frac{1}{k^0 + i\epsilon} = \rmP\frac{1}{k^0}-i\pi\delta(k^0)$.  
Fermion bubbles in the gluon propagation alter the $|\kBI|$-dependence 
with the coefficient $n_f$, the number of (light) fermion flavor.  
By naively non-abelianizing $n_f$ to $\beta_0$, 
we have the last expression.  
Thus we have $\Sigma_2 = \Sigma_1-\Sigma_2$, or 
\begin{align*}
  \delta m_{\rm IR}
  &=
  m \Sigma_1 = m\cdot 2\Sigma_2 \\
  &=
  \frac{1}{2} \int\limits_{\vec{k}\,=\,\text{soft}}\!\!
  \frac{\diffn{k}{3}}{(2\pi)^3} 
  \frac{4\pi C_F \alpha_s(|\kBI|^2)}{|\kBI|^2}
  \sepcomma
\end{align*}
where $\delta m_{\rm IR}$ means the contribution from IR gluons.  
This expression is ill-defined due to the IR-pole of 
the QCD coupling $\alpha_s(\mu)$.  
On the other hand, the QCD-potential $V(r)$ in coordinate space is 
also ill-defined due to just the same reason: 
\begin{align*}
  \delta V(r)_{\rm IR}
  = 
  \int\limits_{\vec{k}\,=\,\text{soft}}\!\!
  \frac{\diffn{k}{3}}{(2\pi)^3} 
  \expo^{i\vec{k}\cdot\vec{r}}
  \frac{-4\pi C_F \alpha_s(|\kBI|^2)}{|\kBI|^2}
  \sepperiod
\end{align*}
However we can see that~\cite{B98,HSSW99} 
the ``severest IR renormalon pole'' is canceled in the combination 
$2m_{\rm pole}+V(r)$.  The precise meaning of this statement is 
explained in the next section.  
The origin of the difference of the sign can be understood as follows.  
Consider an Abelian gauge group for simplicity.  
For the self-energy $\Sigma$, both end of a ``gluon'' couple to 
the ``quarks'' of the same charge; thus $\Sigma \propto +C_F \alpha_s$, 
where $C_F \sim Q^2$ ($Q$ is Abelian charge).  
For the potential $V$, one end of a ``gluon'' couples to a ``quark'' while 
the other to a ``anti-quark''; thus $V \propto -C_F\alpha_s$.  
%------------------------------------------------------------------------
\subsection{Borel sum and IR renormalon pole}
\label{sec:Borel_sum}
IR renormalon in $m_{\rm pole}$ is discussed in~\cite{BPSV99,BSUV94,B95}.  
IR gluon contribution for a quark mass is 
\begin{align}
  \delta m_{\rm IR}(\mu)
  &\equiv
  \frac{1}{2} \int\limits_{|\vec{k}|<\mu}\!\!
  \frac{\diffn{k}{3}}{(2\pi)^3} 
  \frac{4\pi C_F \alpha_s(|\kBI|^2)}{|\kBI|^2}  
  = \frac{C_F}{\pi} 
  \int_0^\mu \diffn{k}{}\,\alpha_s(k^2)
  \nonumber\\
  &=
  \frac{C_F \alpha_s(\mu)}{\pi} \mu
  \sum_{n=0}^\infty n! 
  \left( 2 \frac{\beta_0 \alpha_s(\mu)}{4\pi} \right)^n  
  \label{eq:mIR_asym_ser}
  \\
  &=
  \frac{C_F}{\pi} \mu 
  \int_0^\infty\!\diffn{t}{} 
  \expo^{-t/\alpha_s(\mu)}
  \frac{1}{1-2\dfrac{\beta_0 t}{4\pi}}  
  \label{eq:mIR_Borel}
  \sepperiod
\end{align}
We can see several points.  
With the expression in the second line%
\footnote{
Here we used
$\displaystyle\int_0^1\diffn{x}{}\left(\ln\frac{1}{x^2}\right)^n = 2^n n!$.  
}, 
which is written in series, 
the coefficients grows with $n!$, thus the series is asymptotic one, 
which means it does not converge at all.  
The procedure that leads to the final expression is 
called ``Borel resummation''.  
One can easily check that this expression is the same as the above one 
at least formally, by expanding the denominator and 
integrating each term, which gives simply $\Gamma$-function or $n!$.  
The integration parameter $t$ is called the ``Borel parameter'' 
conjugate to $\alpha_s(\mu)$.  
It is sometimes convenient to use the rescaled Borel parameter 
$u \equiv \frac{\beta_0 t}{4\pi}$ 
conjugate to $\frac{\beta_0 \alpha_s(\mu)}{4\pi}$.  
The complex plane of the Borel parameter $t$ or $u$ is 
called ``Borel plane''.  
With the Borel-sum representation of $\delta m_{\rm IR}$, 
there is a pole on the path of integration, since $\beta_0 > 0$.  
The pole in the Borel plane that originates from the 
asymptotic behavior of a perturbative expansion is called 
a renormalon pole.  
We can see that the pole mass has a IR renormalon pole 
at $u = \frac{1}{2}$.  
The following expressions may be useful to 
follow some of the calculations here: 
\begin{align*}
  &  \alpha_s(k)
   = \frac{\alpha_s(\mu)}
        {1-\frac{\beta_0\alpha_s(\mu)}{4\pi}\ln\frac{\mu^2}{k^2}}  
   = \alpha_s(\mu)\cdot\sum_{n=0}^{\infty}
     \left( \frac{\beta_0\alpha_s(\mu)}{4\pi} \ln\frac{\mu^2}{k^2} 
     \right)^n  
  \sepcomma\\
  &  \alpha_s(\mu)
   = \frac{4\pi}{\beta_0\ln\pfrac{\mu^2}{\Lambda^2}}  
   = \int_0^\infty\!\!\diffn{t}{}
     \pfrac{\mu^2}{\Lambda^2}^{-\tfrac{\beta_0}{4\pi}t}  
  \sepcomma\\
  &  \exp\left( \frac{-4\pi}{\beta_0\alpha_s(\mu)} \right)
  = \frac{\Lambda^2}{\mu^2}
  \sepcomma\quad
    \expo^{-t/\alpha_s(\mu)}
  = \pfrac{\mu^2}{\Lambda^2}^{-\tfrac{\beta_0}{4\pi}t}  
  \sepperiod
\end{align*}
where $\Lambda = \Lambda_{\rm QCD}$.  
The most RHS of the second equation may be called 
Borel representation for the running coupling~\cite{Z98}.  
%------------------------------------------------------------------------
\subsubsection{Short distance mass and long distance mass}
IR renormalon pole disturbs pole mass $m_{\rm pole}$ but not \MSbar\ mass 
$m_{\MSbar}$.  This is because \MSbar\ scheme subtracts only 
the pole that originates from UV divergence.  Such a mass scheme may be 
called ``short distance mass''.  On the other hand, a mass scheme 
sensitive to IR physics may be called ``long distance mass''.  
For example, the peak of the invariant mass $m_{jW}$ of the jet and $W$ that 
are decay products of $t$, may be a kind of long distance mass of top quark.  
Those mass schemes suffer from IR renormalon ambiguity.  
%------------------------------------------------------------------------
\subsubsection{Ambiguity due to IR renormalon pole}
There are several ways to estimate the ambiguity of $m_{\rm pole}$ 
due to IR physics.  We show three ways here.  
It may be useful to see that both the Borel-summed expression and 
the asymptotic-series one give almost the same result.  
The symbol $\Delta$ denotes ambiguity.  

(i) In the Borel-summed expression \EqRef{eq:mIR_Borel}, 
there is a pole on the path of integration.  
There may be two choices to step aside this singularity; 
one is to around upper side of the pole, and the other is lower side.  
The ambiguity may be estimated by the difference of 
these two~\cite{BB94}: 
\begin{align*}
  \Delta\left( \delta m_{\rm IR}(\mu) \right)
  &\simeq
  \frac{1}{2} \bIm{
    \frac{C_F}{\pi}\mu \oint_{\rm C} \diffn{t}{} 
    \expo^{-t/\alpha_s(\mu)} \frac{1}{1-2\frac{\beta_0 t}{4\pi}}
  }  \\
  &=
  \frac{-1}{2} \frac{4\pi C_F}{\beta_0} 
  \mu \pfrac{\Lambda_{\rm QCD}^2}{\mu^2}^{1/2}  
%  \\&
  =
  \frac{-1}{2} \frac{4\pi C_F}{\beta_0} \Lambda_{\rm QCD}
  \sepperiod
\end{align*}
Note that the power $1/2$ of $\Lambda^2/\mu^2$ reflects 
the position of the renormalon pole $u = 1/2$.  
Because of the exponential dumping factor 
$\exp[-t/\alpha_s(\mu)]$, the severest ambiguity is due to 
the renormalon pole that is closest to the origin.  
For the case of $m_{\rm pole}$, it is $u=1/2$.  
However in the combination $2m_{\rm pole}+ V(r)$, this severest pole 
is cancelled each other.  The next severest pole, which may not be 
cancelled in the combination above, may be $u=1$.  
The ambiguity due to this renormalon pole is 
\begin{align*}
  \Delta\left( \delta m_{\rm IR} \right)
  \sim \mu\pfrac{\Lambda^2}{\mu^2}^1 
  = \Lambda\cdot\frac{\Lambda}{\mu}
  \quad\text{for}~u=1
  \sepcomma
\end{align*}
thus suppressed by factor $\Lambda/\mu \sim 1/175$ for $\mu \sim m_t$.  
Precise determination of the residue may be difficult.  

(ii) Let us consider a asymptotic series 
$\sum_{n=0}^{\infty} x^n n!$~\cite{BSUV94}.  
The best approximation may be obtained when one sums up until 
the ratio of a term and the next term becomes greater than unity: 
$[x^{n+1}(n+1)!]/[x^n n!] = xn < 1$, 
or $n \lesssim n_{\rm crit} \equiv 1/x$.  
The ambiguity may be estimated by the magnitude to the last term: 
\begin{align*}
  \Delta\left( \sum_{n=0}^{n_{\rm crit}} x^n n! \right)
  &\simeq \left. x^n n! \right|_{n = n_{\rm crit}}  
%  \\&
  \simeq \left. \frac{n!}{n^n} \right|_{n = n_{\rm crit}}  \\
  &\simeq \sqrt{2\pi n_{\rm crit}} \expo^{-n_{\rm crit}}
  \sepperiod
\end{align*}
For the case we deal, 
\begin{align*}
  n_{\rm crit} = \frac{1}{2}\frac{4\pi}{\beta_0\alpha_s(\mu)}
  \simeq 6 \text{~to~} 7
  \sepperiod
\end{align*}
Note that if the severest IR renormalon pole is not $u=1/2$ but $u=1$, 
then $n_{\rm crit} = 4\pi/(\beta_0\alpha_s)$.  
Thus from \EqRef{eq:mIR_asym_ser} we have 
\begin{align*}
  \Delta\left( \delta m_{\rm IR} \right)
  &\simeq \frac{C_F}{\pi}\mu\alpha_s(\mu) \cdot 
     \sqrt{2\pi n_{\rm crit}} \expo^{-n_{\rm crit}}  \\
  &= \frac{C_F \alpha_s(\mu)}{\pi}
     \frac{2\pi}{\sqrt{\beta_0\alpha_s(\mu)}} \Lambda_{\rm QCD}
  \sepperiod
\end{align*}

(iii) There is also an another way to estimate.  
It gives small imaginary part directly to the denominator of the 
integrand of momentum integration~\cite{BSUV94}: 
\begin{align*}
  \Delta\left( \delta m_{\rm IR}(\mu) \right)
  &\simeq \bIm{ \delta m_{\rm IR}(\mu) }  
%  \\&
  = \bIm{ \frac{C_F}{\pi} \frac{4\pi}{\beta_0} \int_0^\mu \diffn{k}{}
       \frac{1}{\ln\frac{k^2}{\Lambda^2}+i\epsilon} }  \\
  &= \frac{-1}{2} \frac{4\pi C_F}{\beta_0} \Lambda_{\rm QCD}
  \sepperiod
\end{align*}

For an asymptotic slavery theory $\beta_0 < 0$ such as QED, 
the IR renormalon pole is not on the path of integration, 
thus it may not be big problem.  
In terms of an asymptotic series, each term is sign altering, 
thus the singularity is less than the case for $\beta_0>0$.  
There are also UV renormalon poles in QCD (and in QED as well).  
However, they are situated at the negative region of the real axis 
(for $\beta_0>0$), thus cause no ambiguity at least 
with regard to the Borel resum.  
%--------------------------------------------------------------------------
\subsection{Potential-Subtracted mass $m_{\rm PS}(\mu_f)$}
Arguments by now indicate that one should attempt to extract 
the \MSbar\ mass $m_{\MSbar}$ directly, not the pole mass $m_{\rm pole}$, 
from experiments.  
However one crucial point is that \MSbar\ mass is not calculated 
to the order we need.  This is because the binding energy ($\sim -E_\rmB$)
is already $\Order{m\alpha_s^2}$ to the leading order 
[\EqRef{eq:pB_def}].  
Thus to NNLO, one needs $\Order{m\alpha_s^4}$ term, which is 
N${}^4$LO correction to the quark mass.  Recently N${}^3$LO correction to 
the quark mass was calculated~\cite{CS99}.  
But one more higher order may be very hard to calculate.  
One way to circumvent this dilemma is to introduce another 
``short-distance'' mass scheme.  
There may be a host of schemes, but one scheme which is convenient to our 
problem is Potential-Subtracted (PS) mass $m_{\rm PS}$ proposed in~\cite{B98}. 
The argument goes as follows.  
As we showed above, 
the combination $2m_{\rm pole}+V(r)$ is free from the (severest) 
IR renormalon pole, but each of $m_{\rm pole}$ and $V(r)$ are not.  
The problematic part is the IR part of $\VT(|\kBI|)$.  
The idea is to add that part to the mass.  Then the new potential is 
free from (severest) IR renormalon ambiguity.  
Thus so is the new mass scheme, since the sum is so: 
\begin{align}
  2 m_{\rm pole} + V_{\rm C}(r) =
  2 m_{\rm PS}(\mu_f) + V_{\rm C}(r;\mu_f) 
  \label{eq:2m+V_is_equal}
  \sepcomma
\end{align}
where%
\footnote{
  Note that our $\Delta m(\mu_f)$ is related to a corresponding quantity 
  in~\cite{B98} by $\Delta m(\mu_f) = - \delta m (\mu_f)$, 
  and is negative.  
}
\begin{align*}
  m_{\rm PS}(\mu_f) \equiv m_{\rm pole} + \Delta m (\mu_f)  
  \sepcomma\quad
  V_{\rm C}(r;\mu_f) \equiv V_{\rm C}(r) - 2 \Delta m(\mu_f)
\end{align*}
and 
\begin{align}
  \Delta m (\mu_f) 
  &\equiv \frac{1}{2} \int\limits_{|\vec{q}|<\mu_f}\!\!
  \frac{\diffn{q}{3}}{(2\pi)^3} \VT_\rmC(q)
  \label{eq:delta_mPS_def}
  \\
  &=
  \frac{-C_F a_s}{\pi} \mu_f \biggl[ 1 + 
    \frac{a_s}{4\pi} \left\{ a_1 
      - \beta_0 \left( \ln\frac{\mu_f^2}{\mu_s^2} - 2 \right) \right\} 
  \nonumber\\
  &\qquad{}
    + \pfrac{a_s}{4\pi}^2 \left\{ a_2 
      - (2a_1\beta_0+\beta_1) \left( \ln\frac{\mu_f^2}{\mu_s^2} - 2 \right) 
      + \beta_0^2 \left( \ln^2 \frac{\mu_f^2}{\mu_s^2} 
        - 4 \ln\frac{\mu_f^2}{\mu_s^2} + 8 \right) \right\}
  \biggr]
  \sepcomma\nonumber
\end{align}
where $\VT_{\rm C}(q)$ is the Fourier transform of the Coulombic
potential $V_{\rm C}(r)$ defined in \EqRef{eq:VC_mom}, 
and $a_s = \alpha_s^\MSbar(\mu_s)$.  
The scale $\mu_f$ for factorization should be larger than 
$\Lambda_{\rm QCD}$ in order to remove the IR ambiguity.  
On the other hand 
it seems that $\mu_f$ should not be as large as Bohr momentum $p_\rmB$ 
in order not to disturb the Coulombic potential near Bohr radius.  
Following the original paper~\cite{B98} for PS mass, 
we choose $\mu_f = 3\GeV$.  
Explicit values are given in \TableRef{table:delta_mPS}.  
For reference, $\Delta m(\mu_f=10\GeV)$ is also shown.  
\begin{table}[tbp]
\hspace*{\fill}
\begin{tabular}{r||cc|cc}
  \smash{\lower 1.5ex\hbox{$\Delta m(\mu_f)$}} 
  &  \multicolumn{2}{c|}{$\mu_f=3\GeV$} 
  &  \multicolumn{2}{c}{$\mu_f=10\GeV$}  \\
  &  $\mu_s = 20\GeV$  &  $\mu_s = 75\GeV$ 
  &  $\mu_s = 20\GeV$  &  $\mu_s = 75\GeV$  \\
  \hline
  LO  & $-0.1922\GeV$ & $-0.1546\GeV$ & $-0.6408\GeV$ & $-0.5153\GeV$ \\
  NLO  & $-0.3123\GeV$ & $-0.2589\GeV$ & $-0.8945\GeV$ & $-0.7707\GeV$ \\
  NNLO  & $-0.4003\GeV$ & $-0.3389\GeV$ & $-1.035~\GeV$ & $-0.9255\GeV$
\end{tabular}
\hspace*{\fill}
\\
\hspace*{\fill}
\begin{Caption}\caption{\small
    The difference $\Delta m(\mu_f) = m_{\rm PS}(\mu_f)-m_{\rm pole}$ 
    of PS mass and pole mass, defined in \EqRef{eq:delta_mPS_def}.  
    Although we use $\mu_f=3\GeV$ in the study, we also show 
    $\Delta m(\mu_f=10\GeV)$ for reference.  
    \label{table:delta_mPS}
}\end{Caption}
\hspace*{\fill}
\end{table}
Note that in the definition above, we dropped the $r$-dependence of 
$\Delta m (\mu_f)$.  Thus the equality in \EqRef{eq:2m+V_is_equal} is up to 
$\Order{\Lambda_{\rm QCD}^3 r^2}$ at least perturbatively%
\footnote{
  There is no $\Order{\Lambda_{\rm QCD}^2 r}$ term due to rotational 
  symmetry.  However I'm not sure if this is true also non-perturbatively.  
}.  
There may be renormalon ambiguity also in this $r$-dependent term.  
This ambiguity can not be cancelled in the combination 
$2m_{\rm pole}+V(r)$ since the pole mass is $r$-independent.  
However the ambiguity in this term may not be serious, because 
the typical scale $r$ for (would-be) $t\tB$ bound state is 
$1/p_\rmB \simeq 1/(20\GeV)$, which means the ambiguity is of order 
$\Lambda_{\rm QCD}^2 r \sim 0.05\GeV$.  
It is known that $V(r) \propto r$ outside the perturbative region.  
However as we shall see in \SecRef{sec:theo_err_NNLO_R}, 
this ambiguity is negligible in our case due to the large mass and 
decay width of the top quark~\cite{FK87,SP91,FY91,SFHMN93}.  
Thus the next-severest renormalon ambiguity above may also be negligible.  

%%--------------------------------------------------------------------------
%\subsubsection{Relation between $\MSbar$-mass and PS-mass}
Recently, the relation between $\MSbar$- and pole-mass 
was calculated to $\Order{\alpha_s^3}$~\cite{CSMv99}.  
With this, PS-mass can be related more accurately to \MSbar-mass than to 
pole-mass%
\footnote{
  See the reference above for details.  
}.  
%--------------------------------------------------------------------------
\subsection{Results for $R$ ratio with $m_{\rm PS}(\mu_f)$}
Following the arguments above, it is PS mass (or a short-distance mass, 
in general) that is well determined from experiments.  
Thus we fix PS mass but not pole mass.  
With the choice $\mu_f = 3\GeV$, the difference 
$\Delta m (\mu_f) \simeq -0.4\GeV$ 
of pole mass $m_{\rm pole}$ and PS mass $m_{\rm PS}$ is tiny 
compared to $m_{\rm pole}\simeq 175\GeV$.  
Thus the Schr\"odinger equation barely changes except the origin of 
non-relativistic energy $E$, 
since it is now $E = \sqrt{s}-2m_{\rm PS}$.  
Note that, as explained in \SecRef{sec:psi(0)_V(r)}, 
constant shift of potential $V$ simply means 
the shift of the origin of $E$.  
\begin{figure}[tbp]
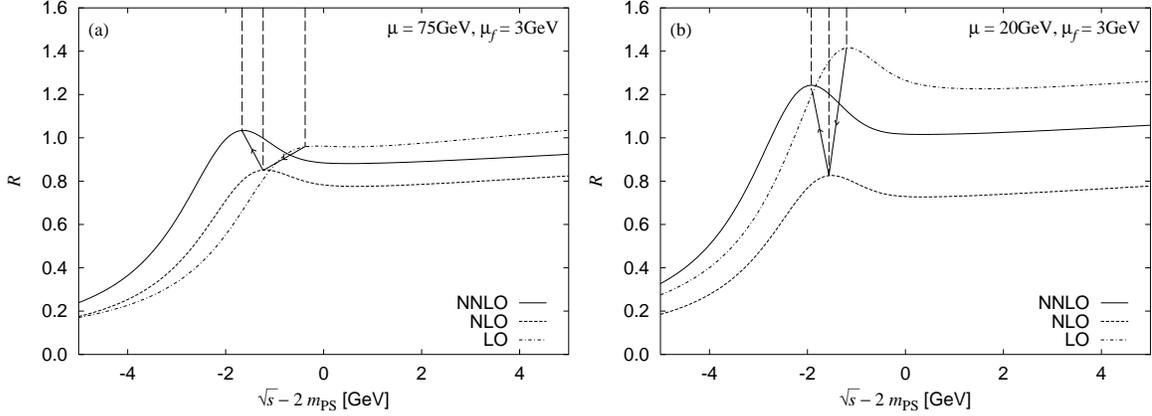

  \hspace*{\fill}
  \begin{minipage}{6cm}
    \includegraphics[angle=-90,width=7.5cm]{./figure/R_75_3.eps}
  \end{minipage}
  \hspace*{\fill}
  \begin{minipage}{6cm}
    \includegraphics[angle=-90,width=7.5cm]{./figure/R_20_3.eps}
  \end{minipage}
  \hspace*{\fill}
   \\
  \hspace*{\fill}
  \begin{Caption}\caption{\small
      $R$-ratios for $e^+ e^- \to \gamma^* \to t\bar{t}$ 
      at LO (dot-dashed), NLO (dashed), and NNLO (solid) 
      as functions of the energy measured from twice the
      potential-subtracted mass, $\sqrt{s}-2 m_{\rm PS}$.
      We set $\mu_f = 3\GeV$ and $m_{\rm PS}(\mu_f) = 175\GeV$.  
      Figure (a) is for $\mu_s = 75\GeV$, and (b) is for $\mu_s = 20\GeV$.  
      Other notations and parameters are same as in~\FigRef{fig:R75-20_0}.
      \label{fig:R75-20_3}
  }\end{Caption}
  \hspace*{\fill}
\end{figure}
Shown in \FigRef{fig:R75-20_3} are almost the same as 
those in \FigRef{fig:R75-20_0} except we fixed 
$m_{\rm PS}(\mu_f=3\GeV)=175\GeV$ rather than $m_{\rm pole}$.  
Since $|\Delta m (\mu_f)|$ is larger for higher order, 
the position of $1S$ peak moves more for higher order.  
We can see the sideward convergence becomes better by the difference 
$|(\Delta m)^{\rm NNLO} - (\Delta m)^{\rm NLO}| \simeq 0.1\GeV$ and 
$|(\Delta m)^{\rm NLO} - (\Delta m)^{\rm LO}| \simeq 0.1\GeV$, respectively.  
On the other hand the normalization barely changes, as is anticipated.  
In the next section we study RG improvement of the potential, 
which is expected to be effective to this problem.  
%
%
%

%
%
%------------------------------------------------------------------------
\section{Results for $R$ ratio with RG improved $V_{\rm C}$ and $m_{\rm PS}$}
\label{sec:RGimp_R}
As mentioned in \SecRef{sec:R_NNLO_fixed}, 
large renormalization-scale $\mu_s$ dependence of $R$ ratio 
indicates the necessity of $\log(\mu_s)$ resummation.  
As first step%
\footnote{
  A full resummation of logarithms up to NNLO requires a significant 
  modification of the formula \EqRef{eq:R_NNLO_expl}; 
  we will study its incorporation in our future work.  
}, 
we sum up $\log(\mu_s)$ in the 
Coulombic potential $V_{\rm C}(r)$.  
Thus our results here still depends on the soft scale $\mu_s$, 
which is in $V_{\rm BF}$ and $V_{\rm NA}$.  
As we saw in \FigRef{fig:alphaV_20-75-q}, 
couplings converge much better for RG-improved case 
than for fixed order, all over the relevant momentum scale
$C_F a_s \mt \lesssim 1/r < \mt$.  
Thus it is expected that the convergence is improved by the use of 
$V_{\rm C}^{\rm (RG)}(r)$ that is RG improved ($\mu_s=q$) in the 
momentum space and Fourier transformed to the coordinate space.  

\begin{figure}[tbp]
  \hspace*{\fill}
  \begin{minipage}{6cm}
    \includegraphics[angle=-90,width=7.5cm]{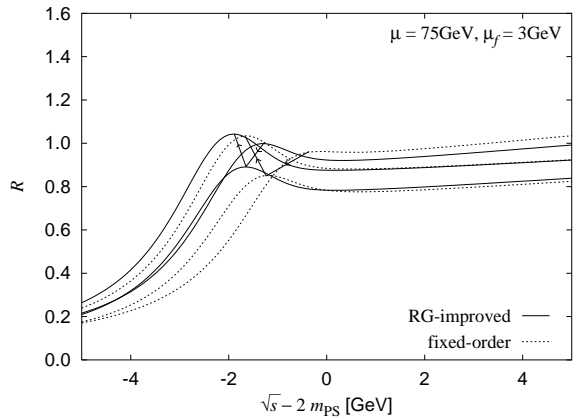}
  \end{minipage}
  \hspace*{\fill}
   \\
  \hspace*{\fill}
  \begin{Caption}\caption{\small
      Our final result for 
      $R$-ratios in this paper for $e^+ e^- \to \gamma^* \to t\bar{t}$ 
      at LO, NLO, and NNLO.  
      Solid lines show those with renormalization-group improved 
      Coulombic potentials, $V_{\rm C}^{\rm (RG)}(r;\mu_f)$.  
      Dashed lines are those with fixed-order Coulombic potentials 
      $V_{\rm C}(r)$.  
      Arrows indicate dislocations of the maximum point of $R$ 
      as the $\Order{1/c}$ and $\Order{1/c^2}$ 
      corrections are included, respectively.  
      We set $\mu_f = 3\GeV$, $m_{\rm PS}(\mu_f) = 175\GeV$, 
      $\mu_s=75\GeV$, $\Gamma_t=1.43\GeV$, and $\alpha_s(m_Z)=0.118$.  
      \label{fig:R_75_3_Rf}
  }\end{Caption}
  \hspace*{\fill}
\end{figure}

PS mass $m_{\rm PS}(\mu_f)$ with RG improved potential is ambiguous, 
since $\VT_\rmC^{\rm (RG)}$ diverges at IR region.  
Here we define as follows: 
\begin{gather*}
  m_{\rm PS}(\mu_f) \equiv m_{\rm pole} + \Delta m (\mu_f) 
  \sepcomma\quad
  V_\rmC^{\rm (RG)}(r;\mu_f) \equiv 
  \int\limits_{|\vec{q}|>\mu_f}\!\!\!\frac{\diffn{q}{3}}{(2\pi)^3} 
  \expo^{i\vec{q}{\cdot}\vec{r}} \VT_\rmC^{\rm (RG)}(q) 
  \sepcomma
  \\
  \Delta m (\mu_f) \equiv \frac{1}{2} \int\limits_{|\vec{q}|<\mu_f}\!\!\!
  \frac{\diffn{q}{3}}{(2\pi)^3}\!\VT_\rmC^{\rm (fixed)}(q)
  \sepperiod
\end{gather*}
Our final result for $R$-ratio in this paper is 
shown in~\FigRef{fig:R_75_3_Rf}.  
For comparison, $R$-ratios both with and without the 
RG improvement for $V_{\rm C}$ are given.  
We can see that convergence of both the normalization and the peak position 
is improved slightly.  
The latter is listed in~\TableRef{table:M1S} as 
the ``binding energies'' of the $1S$ resonance state
$2m_{\rm PS}(\mu_f)-M_{\rm 1S}$.  
They are determined by reducing the decay width $\Gamma_t$ tentatively.  
\begin{table}[tbp]
\hspace*{\fill}
\begin{minipage}{11cm}
\begin{tabular}{l|cccc} \hline\hline
&
\multicolumn{2}{c}{(fixed-order)} &
\multicolumn{2}{c}{(RG-improved)}
\\
      & $\mu_s=20\GeV$     & $\mu_s=75\GeV$
      & $\mu_s=20\GeV$     & $\mu_s=75\GeV$
\\ \hline
LO    & $1.390\GeV$                     & $0.838\GeV$
      & $1.573\GeV$                     & $1.573\GeV$  
\\
NLO   & $1.716\GeV$                     & $1.453\GeV$
      & $1.861\GeV$                     & $1.861\GeV$
\\
NNLO  & $2.062\GeV$                     & $1.817\GeV$
      & $2.136\GeV$                     & $2.058\GeV$
\\ \hline\hline
\end{tabular}
\end{minipage}
\hspace*{\fill}
\\
\hspace*{\fill}
\begin{Caption}\caption{\small
    ``Binding energies'' of the $1S$ resonance state 
    defined as $2m_{\rm PS}(\mu_f)-M_{1S}$ at LO, NLO, and 
    NNLO calculated with $V_{\rm C}(r)$ (fixed-order) and with 
    $V_{\rm C}^{(RG)}(r;\mu_f)$ (RG-improved).  
    These are determined by reducing the decay width $\Gamma_t$ tentatively.  
    We set $\mu_f = 3\GeV$, $m_{\rm PS}(\mu_f) = 175\GeV$, 
    $\Gamma_t=1.43\GeV$, and $\alpha_s(m_Z)=0.118$.  
    \label{table:M1S}
}\end{Caption}
\hspace*{\fill}
\end{table}

\begin{figure}[tbp]
  \hspace*{\fill}
  \begin{minipage}{6cm}
    \includegraphics[angle=-90,width=7.5cm]{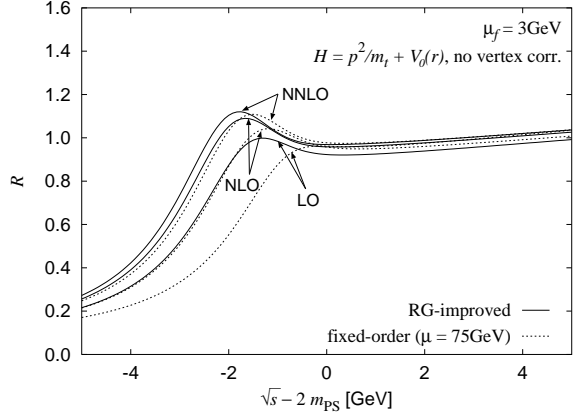}
  \end{minipage}
  \hspace*{\fill}
   \\
  \hspace*{\fill}
  \begin{Caption}\caption{\small
      $R$-ratios for $e^+ e^- \to \gamma^* \to t\bar{t}$ 
      calculated with a Hamiltonian 
      $H = p^2/m_t + V_0(r)$, where $V_0(r)$ includes 
      only the Coulombic part of the $t\bar{t}$ potential.  
      Other corrections (hard vertex corrections, 
      relativistic corrections, etc.) are not included.  
      Solid and dashed lines, respectively, show $R$-ratios with 
      ($V_0(r)=V_{\rm C}^{\rm (RG)}(r;\mu_f)$) and 
      without ($V_0(r)=V_{\rm C}^{\rm (fixed)}(r;\mu_f)$, $\mu_s=75\GeV$) a 
      renormalization-group improvement of the Coulombic potential.  
      We set $\mu_f = 3\GeV$, $m_{\rm PS}(\mu_f) = 175\GeV$, 
      $\Gamma_t=1.43\GeV$, and $\alpha_s(m_Z)=0.118$.  
      \label{fig:R_75_3_Rf_noU}
  }\end{Caption}
  \hspace*{\fill}
\end{figure}

Why is the convergence not improved much better?  
Let us drop the corrections other than those for $V_{\rm C}(r)$, 
for the moment.  
The $R$ ratio for this case is shown in~\FigRef{fig:R_75_3_Rf_noU}.  
The relevant formula is 
\begin{align*}
  R(s) = \frac{6\pi N_c Q_t^2}{m_t^2} \, \Im \, G(0,0) 
\end{align*}
with
\begin{align*}
  \left\{
    \frac{-1}{m_t} 
    \left[ \odiffn{}{r}{2} + \frac{2}{r}\odiff{}{r} \right]
    + V_0(r) - \omega
  \right\} G(r,r') = \frac{1}{4\pi r r'} \, \delta (r-r') ,
\end{align*}
both for fixed order $V_0(r)=V_{\rm C}^{\rm (fixed)}(r;\mu_f)$ and 
RG-improved $V_0(r)=V_{\rm C}^{\rm (RG)}(r;\mu_f)$.  
We can see clearly that the convergence is much improved 
by the $\log(\mu_s)$ resummations.  
Thus it is other corrections that disturb the perturbative convergence 
of $R$ ratio.  
Among them the $1/r^2$ potential $V_{\rm NA}(r)$, which is a strong 
attractive potential near the origin, is suspected to be the worst.  
It remains as our future task to gain better understandings of 
these residual large corrections.

\begin{figure}[tbp]
  \hspace*{\fill}
  \begin{minipage}{6cm}
    \includegraphics[width=7.5cm]{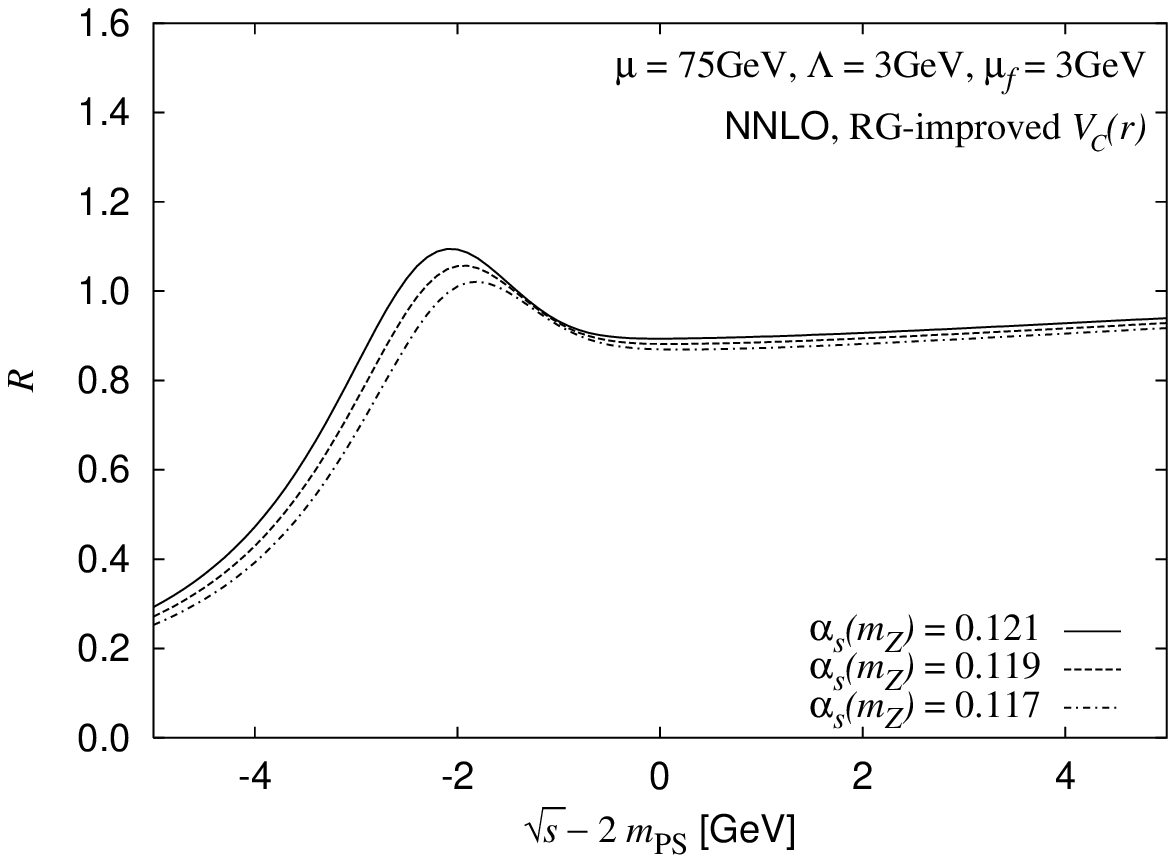}
  \end{minipage}
  \hspace*{\fill}
  \begin{minipage}{6cm}
    \includegraphics[width=7.5cm]{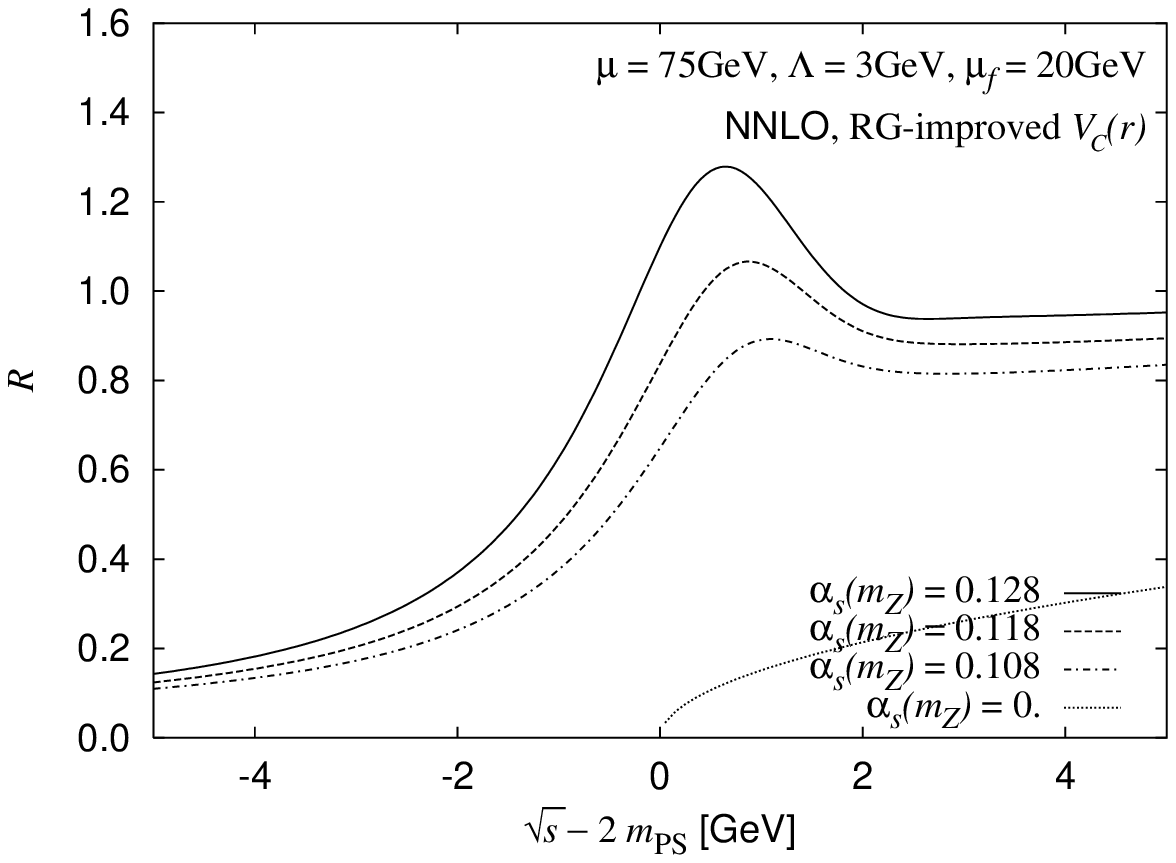}
  \end{minipage}
  \hspace*{\fill}
   \\
  \hspace*{\fill}
  \begin{Caption}\caption{\small
      Dependence of $R$ ratio with the variation of $\alpha_s^\MSbar(m_Z)$: 
      $0.117$, $0.119$, and $0.121$ for the left, and 
      $0.108$, $0.118$, and $0.128$ for the right.  
      For both figures, IR part $k <\Lambda = 3\GeV$ of Coulombic 
      potential $\VT_{\rm C}(k)$ is not taken into account for the 
      Fourier transform to $V_{\rm C}(r)$.  
      For the figure on the right, 
      twice the $\Delta m(\mu_f = 20\GeV) - \Delta m(\Lambda = 3\GeV)$ is 
      further subtracted from $V_{\rm C}(r)$ and added to $m_{\rm PS}$.  
      No width is included for $\alpha_s(m_Z)=0$.  
      We set $\mu_s = 75\GeV$, $m_{\rm PS}(\mu_f) = 175\GeV$, 
      $\Gamma_t=1.43\GeV$.  
      \label{fig:R_RG_as_PSRG_20-3}
  }\end{Caption}
  \hspace*{\fill}
\end{figure}

Shown in \FigRef{fig:R_RG_as_PSRG_20-3} is $\alpha_s^\MSbar(m_Z)$-dependence 
of $R$ ratio.  
%One can see that $\alpha_s^\MSbar(m_Z)$ may be determined with 
%uncertainty $\sim 0.005$.  
The peak position moves $0.12\GeV \times 2$ with the variation of 
$\alpha_s^\MSbar(m_Z) = 0.119 \pm 0.002$, 
which is the current uncertainty~\cite{C98}.  
%Thus we concentrate on the ability of $t\tB$ threshold to determine 
%top quark mass, whose uncertainty now is $\simeq 5\GeV$.  
%
%
%------------------------------------------------------------------------
\subsection{Theoretical uncertainties for NNLO calculation of $R$ ratio}
\label{sec:theo_err_NNLO_R}
There are several uncertainties in our NNLO calculation of $R$ ratio.  
In this section we study the size of them and estimate the 
theoretical uncertainty $(\Delta\mt^{\rm PS})^{\rm th}$ of 
top mass determination.  
Since the perturbative convergence of $R$ ratio is the best with 
the RG-improved potential and the Potential-Subtracted mass, 
we study the uncertainties for this case.  

\begin{figure}[tbp]
  \hspace*{\fill}
  \begin{minipage}{6cm}
    \includegraphics[width=7.5cm]{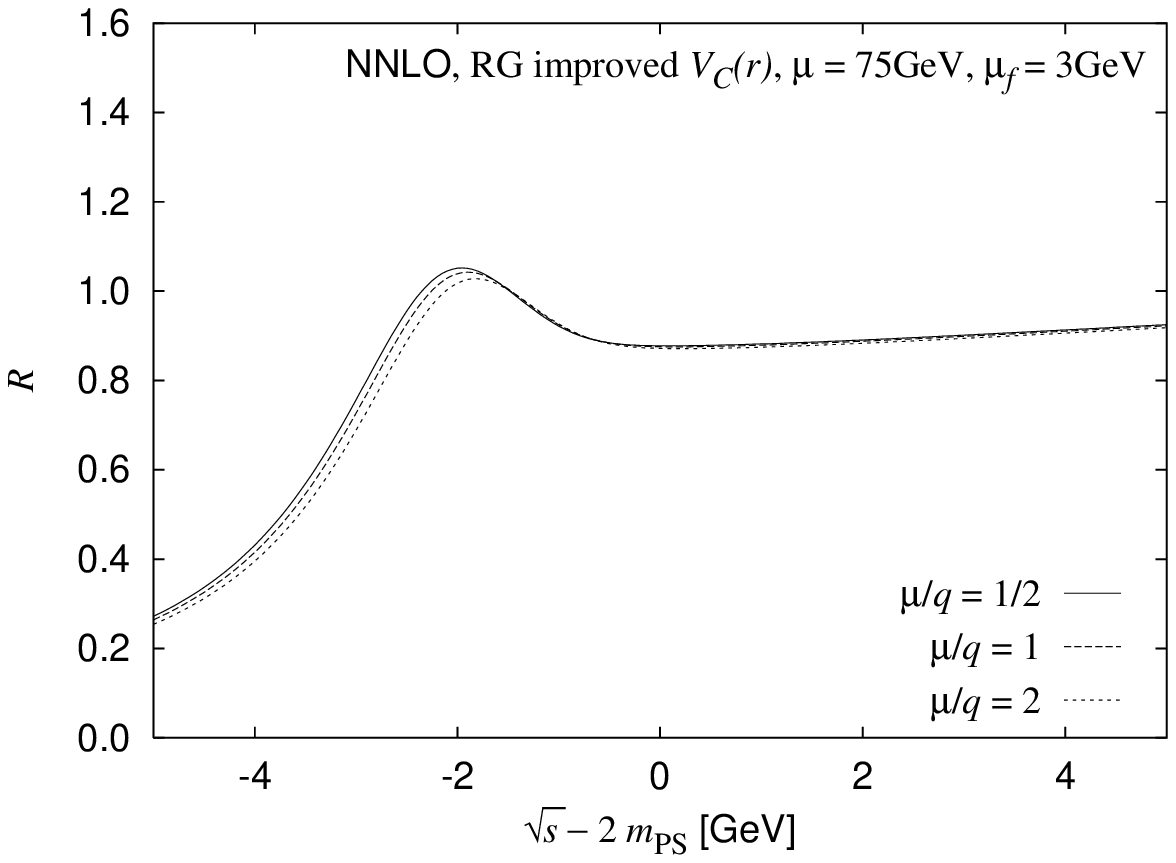}
  \end{minipage}
  \hspace*{\fill}
   \\
  \hspace*{\fill}
  \begin{minipage}{6cm}
    \includegraphics[width=7.5cm]{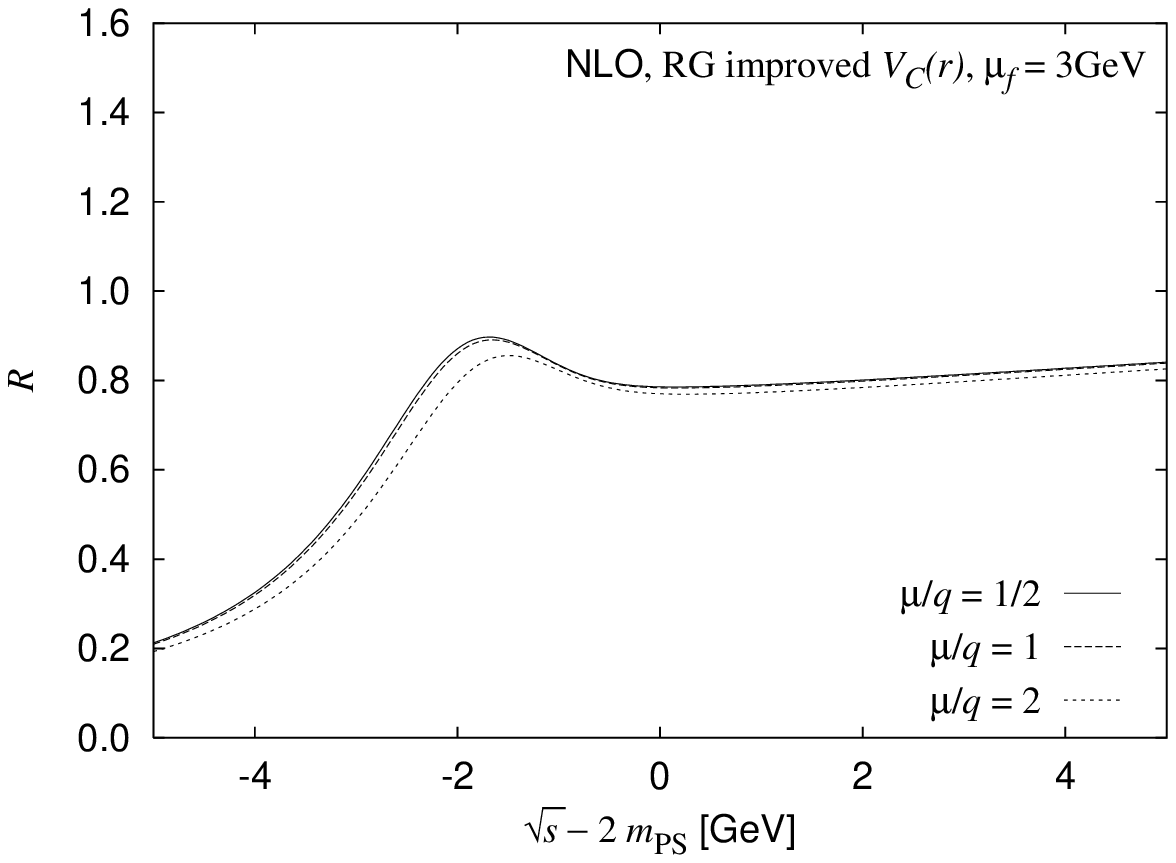}
  \end{minipage}
  \hspace*{\fill}
  \begin{minipage}{6cm}
    \includegraphics[width=7.5cm]{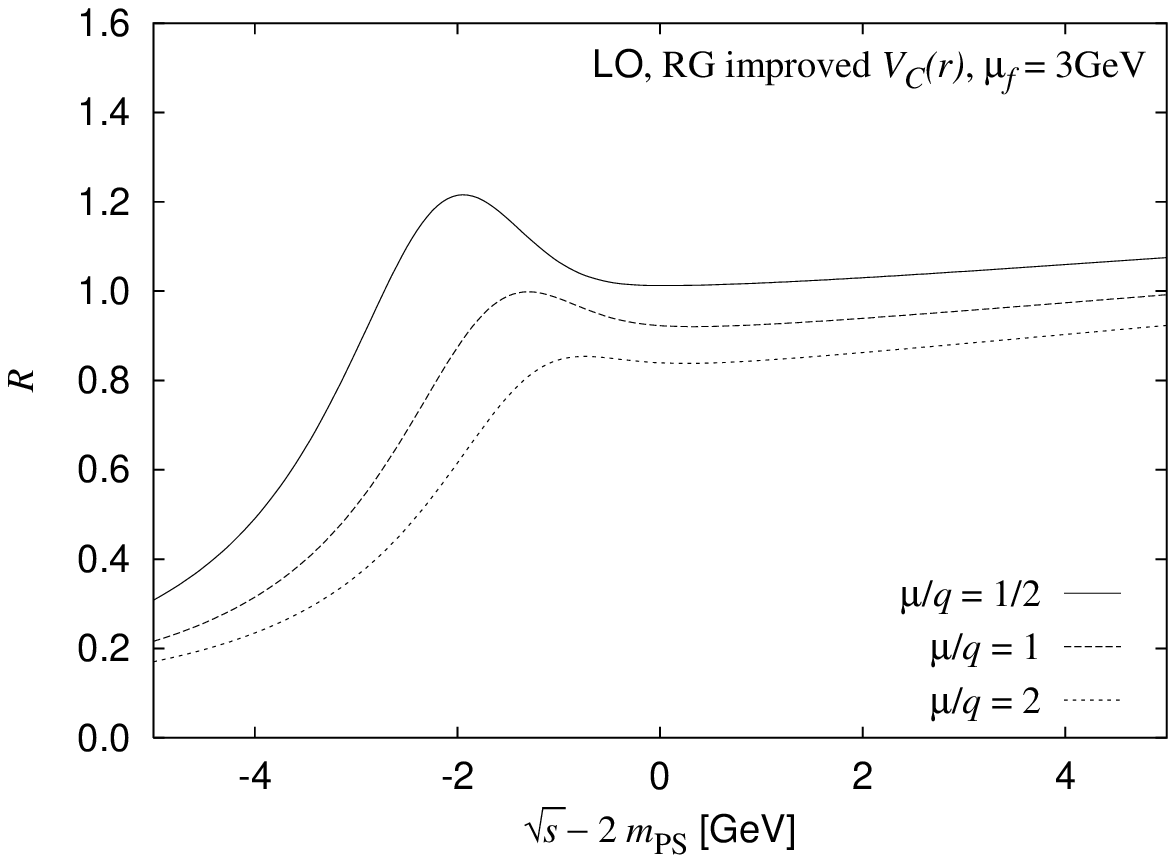}
  \end{minipage}
  \hspace*{\fill}
   \\
  \hspace*{\fill}
  \begin{Caption}\caption{\small
      Dependence of $R$ ratio on the prescription to sum-over 
      leading $\log(\mu_s/q)$'s of $\VT_{\rm C}(q)$.  
      Three choices are given for each with NNLO (top), 
      NLO (bottom--left), and LO (bottom-right) corrections: 
      $\mu_s=q/2$, $q$, $2q$.  
      We set $\mu_f = 3\GeV$, $m_{\rm PS}(\mu_f) = 175\GeV$, 
      $\Gamma_t=1.43\GeV$, and $\alpha_s(m_Z)=0.118$.  
      For NNLO $\mu_s = 75\GeV$.  
      \label{fig:RG_dep_NNLO-NLO-LO}
  }\end{Caption}
  \hspace*{\fill}
\end{figure}

Leading $\log(q/\mu_s)$'s of $\VT_{\rm C}(q)$ can be summed over not 
only with the choice $\mu_s=q$, but also $\mu_s=2q$ etc.  
Their difference is related to the summation 
of the next-to-leading log's (and higher).  
Shown in \FigRef{fig:RG_dep_NNLO-NLO-LO} are the uncertainties due to 
this freedom.  One can see that it is fairly small for both NLO and 
NNLO.  The dependence is large for LO, since there is no log that 
compensates the log dependence of the coupling.  

\begin{figure}[tbp]
  \hspace*{\fill}
  \begin{minipage}{6cm}
    \includegraphics[width=7.5cm]{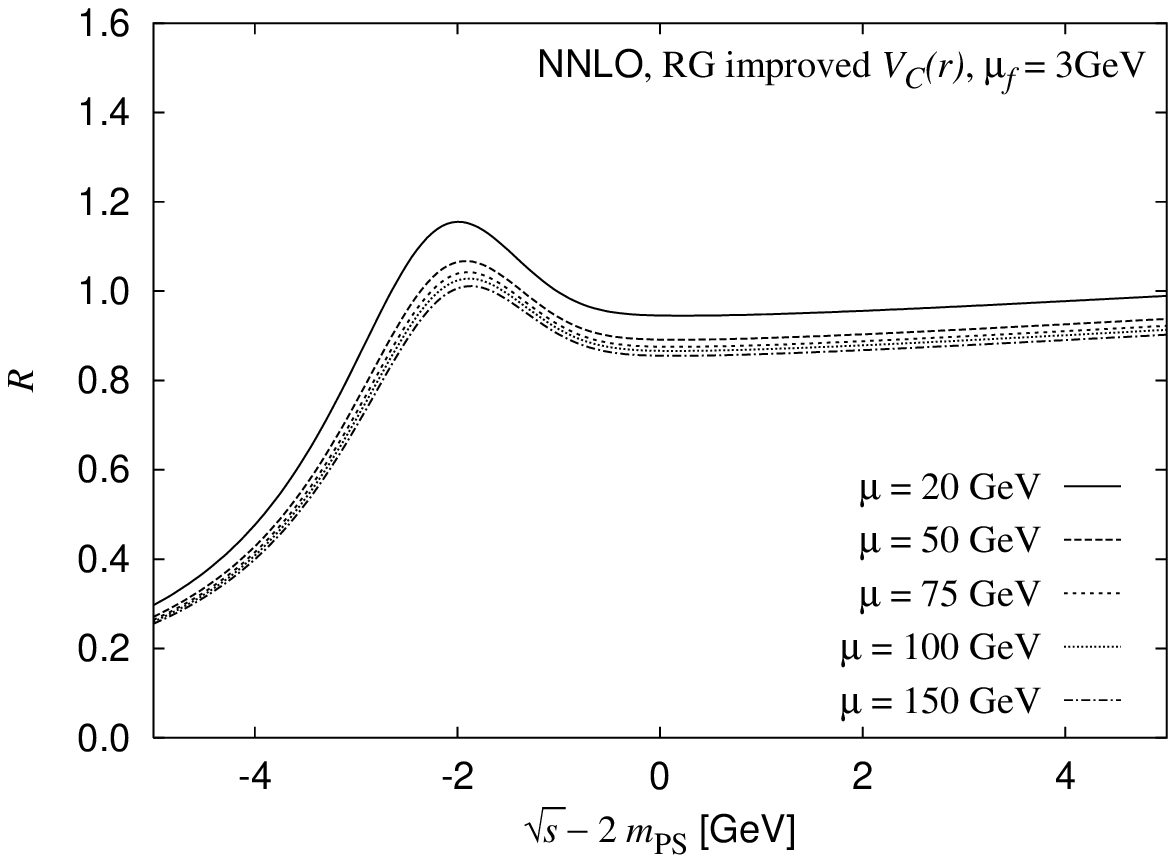}
  \end{minipage}
  \hspace*{\fill}
  \begin{minipage}{6cm}
    \includegraphics[width=7.5cm]{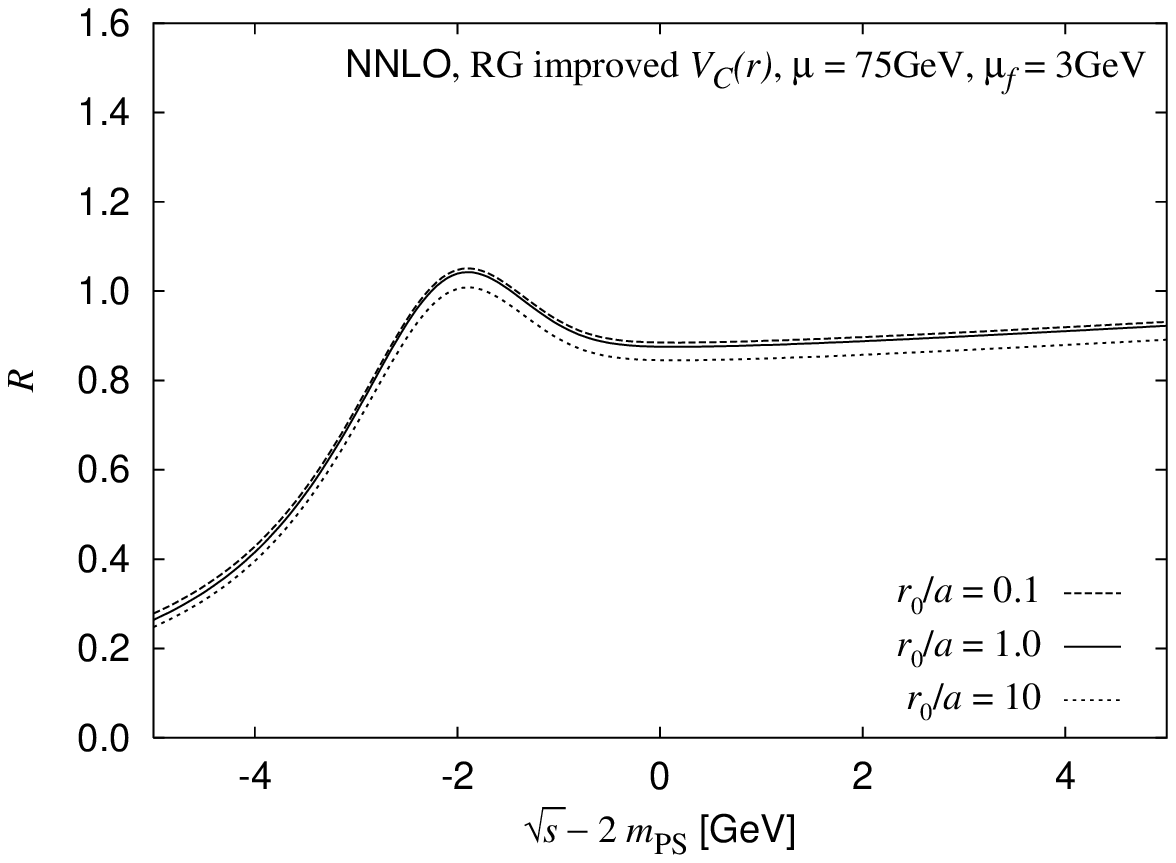}
  \end{minipage}
  \hspace*{\fill}
   \\
  \hspace*{\fill}
  \begin{Caption}\caption{\small
      (Left)
      The soft-scale ($\mu_s$) dependence of $R$ ratio with NNLO corrections.  
      $\mu_s$ is written as $\mu$ for simplicity.  
      The normalization changes monotonically in the choice of $\mu_s$ 
      shown in the figure: $\mu_s=20\GeV$ for highest and $\mu_s=150\GeV$ 
      for the lowest.  
      (Right)
      The cut-off ($r_0$) dependence of $R$ ratio with NNLO corrections.  
      $r_0^{\rm (ref)}$ is written as $a$ for simplicity.  
      In both figures, 
      we set $\mu_f = 3\GeV$, $m_{\rm PS}(\mu_f) = 175\GeV$, 
      $\Gamma_t=1.43\GeV$, and $\alpha_s(m_Z)=0.118$.  
      \label{fig:RG_mu_s=r0-dep}
  }\end{Caption}
  \hspace*{\fill}
\end{figure}

The soft-scale dependence still remains in the potential, since we RG improve 
only the Coulombic part.  It is shown in the left-half of 
\FigRef{fig:RG_mu_s=r0-dep}.  
A physical quantity $Q$ do not depend on the choice of 
renormalization scale $\mu$ provided $Q$ is calculated to all order.  
If perturbative expansion is terminated to certain order, $Q$ depends on 
$\mu$ since the discarded higher order terms depend on $\mu$.  
Thus it is expected that the truncated perturbative expansion is 
a ``good'' approximation when $\mu$ dependence is small, and 
the magnitude of $\mu$ dependence around there is a measure of 
(uncalculated) higher order corrections.  
With this view, one can see that the choice 
$\mu_s \simeq p_\rmB \simeq 20\GeV$ is not ``good'' for $R$ ratio.  
The right-half of \FigRef{fig:RG_mu_s=r0-dep} shows cut-off $r_0$ dependence, 
which is absent for NLO and LO, and also for NNLO if one treats 
the effect of top decay width $\Gamma_t$ properly.  
One can see that the dependence on $r_0$ is small for 
$r_0\simeq r_0^{\rm (ref)}$.  
Thus 
one can see both $\mu_s$ and $r_0$ dependence is small compared to 
the difference of NNLO $R$ ratio and NLO $R$ ratio.  
Besides the shifts are only horizontal, thus do not disturb the determination 
of top mass $\mt^{\rm PS}(\mu_f)$.  

More quantitatively, the variation of the peak position of 
$\sigma_{\rm tot}^{\rm NNLO}$ is $0.08\GeV \times 2$ both for 
$\mu_s=20\GeV$--$150\GeV$ and $\mu_s/q=1/2$--$2$.  
%
%------------------------------------------------------------------------
\subsubsection{Estimate of $(\Delta\mt)^{\rm th}$}
Now let us estimate the theoretical uncertainty of top quark mass 
$\mt^{\rm PS}(\mu_f)$ with $\mu_f=3\GeV$ based on our NNLO calculations.  
Since both $\mu_s$- and $r_0$-dependence are fairly small in the sideward 
direction, 
we can forget about them.  
RG-improve prescription dependence [\FigRef{fig:RG_dep_NNLO-NLO-LO}] 
may serve as a measure of theoretical uncertainty.  
One can see that the sideward NLO correction to LO $R$-ratio is 
similar to the uncertainty of LO $R$-ratio due to the choice of $\mu_s/q$.  
Likewise, 
NNLO correction of NLO $R$ is also similar to the uncertainty of NLO, 
but it is a little smaller.  
This may be because there are sources of NNLO correction 
other than those log's; that is, $V_{\rm BF}$ and $V_{\rm NA}$.  
Thus we estimate the NNLO theoretical ambiguity to be 
$(\Delta\mt)^{\rm th} \sim 0.1\GeV$.  
Here we are conservative, since there are corrections aside from 
log's, such as $V_{\rm VA}$.  

As for N${}^3$LO = $\Order{1/c^3}$ corrections, 
those terms for $V_{\rm BF}$ is $\sim C_F a_s/(m^3 c^3 r^4)$, 
since $m c r$ is dimensionless.  
Here only the dimension is considered.  
Thus translations like $1/r^2 \sim p^2$, $1/r^3 \sim \delta^{(3)}(\rBI)$, 
$1/r \sim \diffn{}{}/\diffn{r}{}$ are possible.  
Likewise, $\Order{1/c^3}$ correction to $V_{\rm NA}$ may be 
$\sim C_A C_F a_s^2/(m^2 r^3 c^3)$, since $a_s^4$ may be $\Order{1/c^4}$.  
Another significant modification is that 
real gluon emission is possible to $\Order{1/c^3}$.  

In our calculations, leading log's in $V_{\rm C}$ are summed with RG.  
Recently next-to-leading log's corrections of N${}^3$LO, 
$\alpha_s^3\ln^2(1/\alpha_s)$, to $R$ ratio is calculated in~\cite{KP99}.  
One can read that the size of N${}^3$LO corrections to the peak position 
is $\sim 2 \times 0.1\GeV$, which is consistent with the estimate above.  
%------------------------------------------------------------------------
\subsubsection{Non-perturbative effect}
Non-perturbative corrections modify QCD potential at $r \sim 1/(1\GeV)$ 
and far.  Thus their effects can be estimated by the dependence of 
Green function $G$ with respect to the variation of the potential around 
$r \sim 1/(1\GeV)$; and it turns out to be small.  
For $r > r'$, 
\begin{align*}
  G'(r,r') = \frac{1}{4\pi C_F a_s W} \frac{g_>(r) g_<(r')}{rr'}
  \sepcomma
\end{align*}
where $g_>(r)/r$ and $g_<(r)/r$ are the solutions of 
(reduced) Schr\"odinger equation [\EqRef{eq:Schr_for_g<g>_NNLO}] 
with the boundary conditions 
$\lim_{r\to\infty}g_>(r)=0$ and 
$\lim_{r\to 0}g_<(r)\simeq (C_F a_s \mt r)^{d_+}$, respectively.  
Since only $G'(r,r')|_{r=r'\simeq 0}$ is relevant for $R$ ratio, 
it is $g_>(r)$ near the origin that determines $R$ ratio.  
It may be affected by non-perturbative effect ($r \gtrsim 1/(1\GeV)$) 
through the boundary condition.  
However we can see in \FigRef{fig:gg1=5} 
that $g_>(r)$ dumps well within perturbative region 
$r/r_\rmB \lesssim 20$ with $r \sim 1/(1\GeV)$.  
Thus non-perturbative effect (to the QCD potential) do not alter 
the Green function for $t\tB$.  
We also show $g_<(r)$ in Figures~\ref{fig:gl-5=0} and~\ref{fig:gl1=5}, 
for reference.  
\begin{figure}[tbp]
  \hspace*{\fill}
  \begin{minipage}{6cm}
    \includegraphics[width=7.5cm]{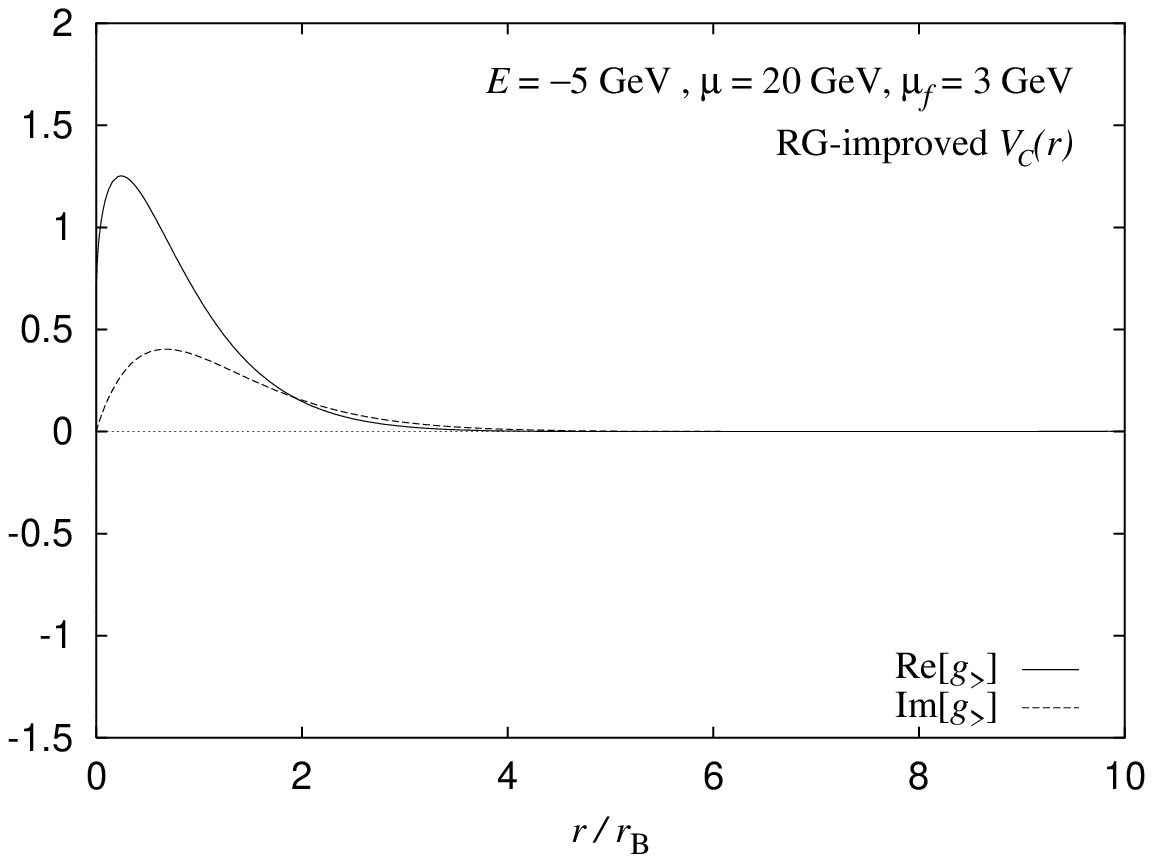}
  \end{minipage}
  \hspace*{\fill}
  \begin{minipage}{6cm}
    \includegraphics[width=7.5cm]{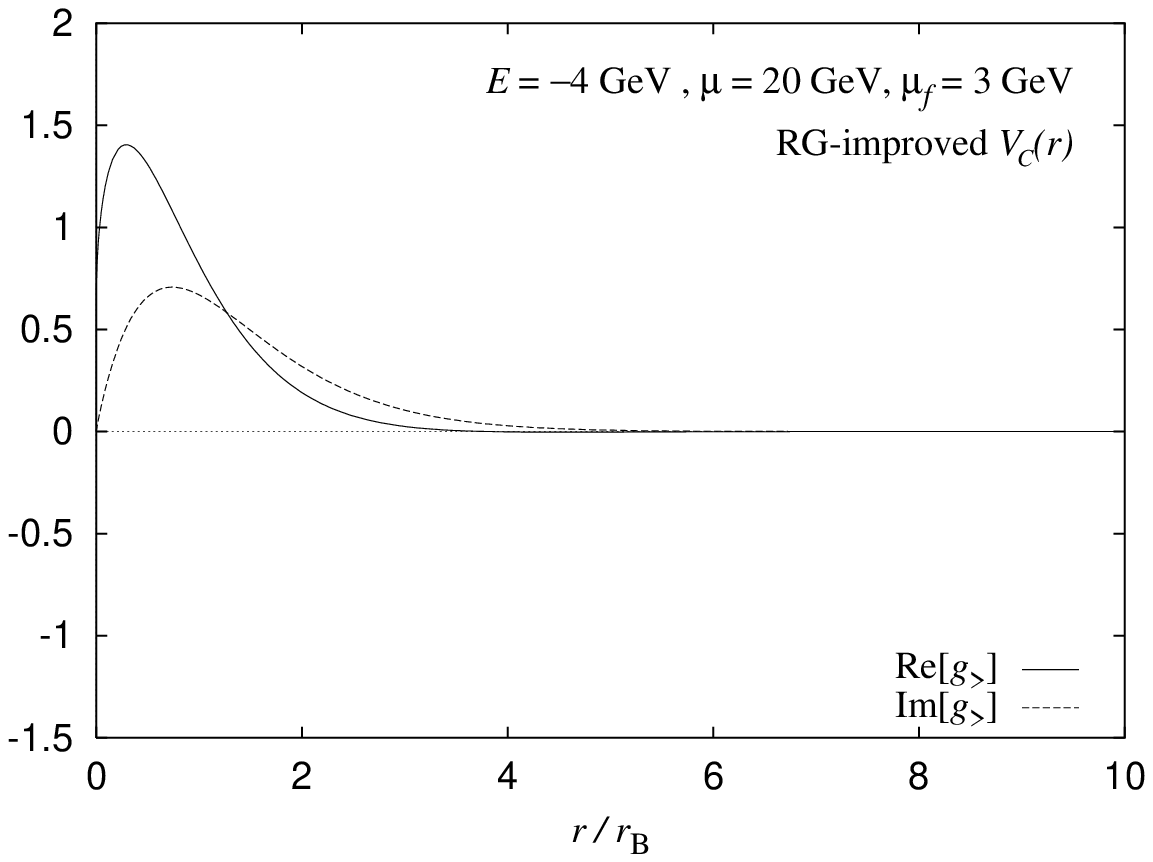}
  \end{minipage}
  \hspace*{\fill}
   \\
  \hspace*{\fill}
  \begin{minipage}{6cm}
    \includegraphics[width=7.5cm]{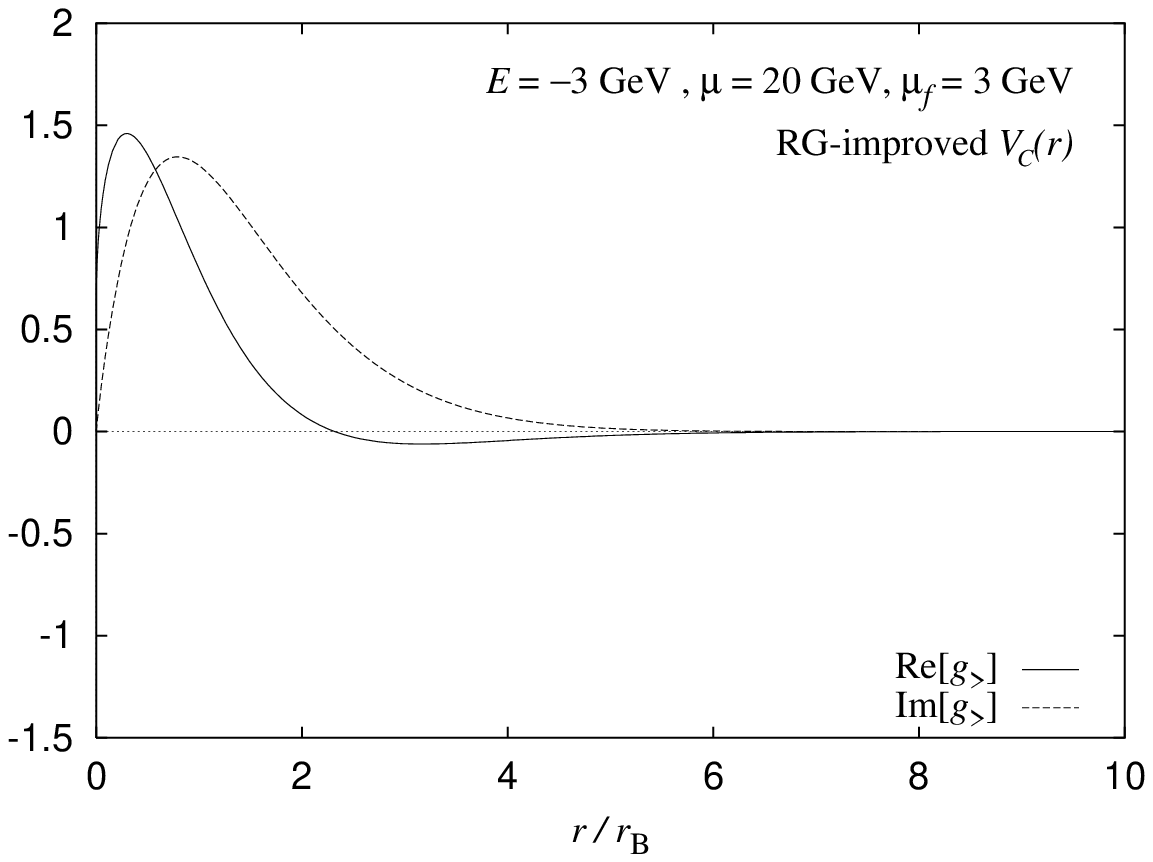}
  \end{minipage}
  \hspace*{\fill}
  \begin{minipage}{6cm}
    \includegraphics[width=7.5cm]{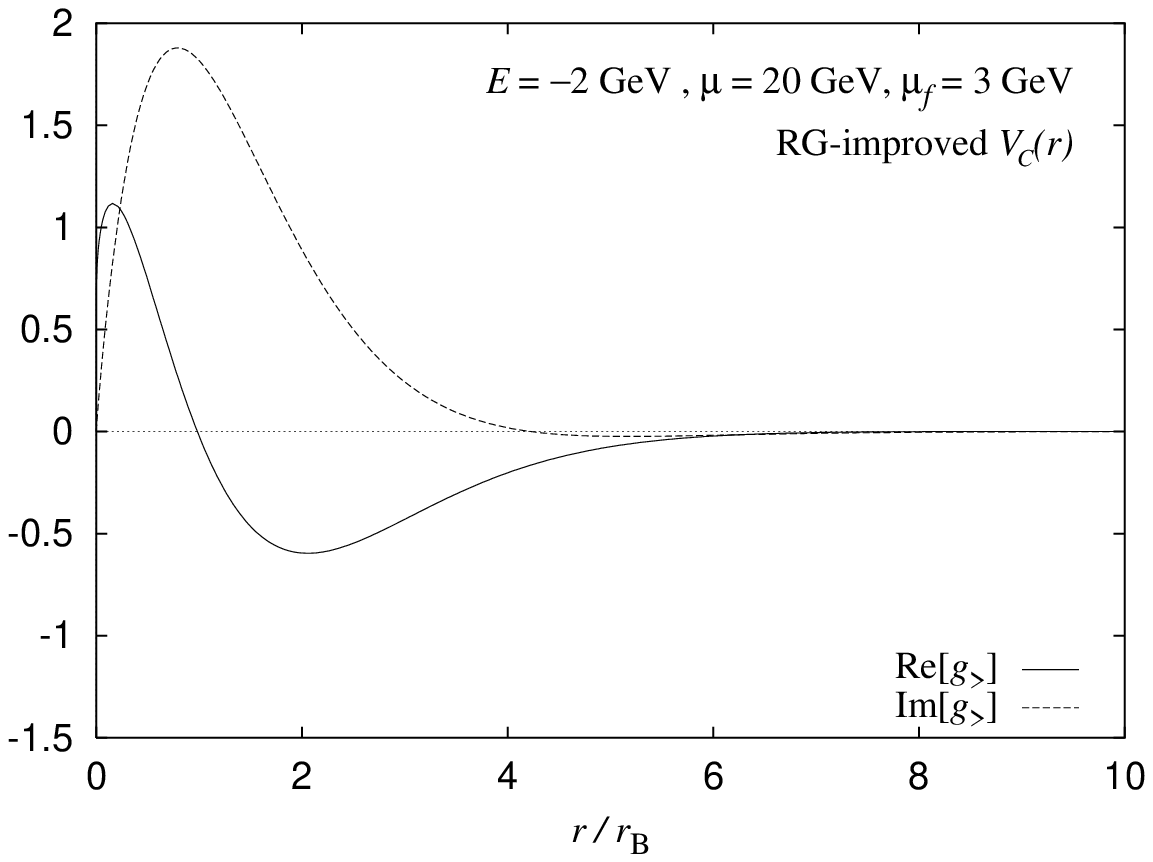}
  \end{minipage}
  \hspace*{\fill}
   \\
  \hspace*{\fill}
  \begin{minipage}{6cm}
    \includegraphics[width=7.5cm]{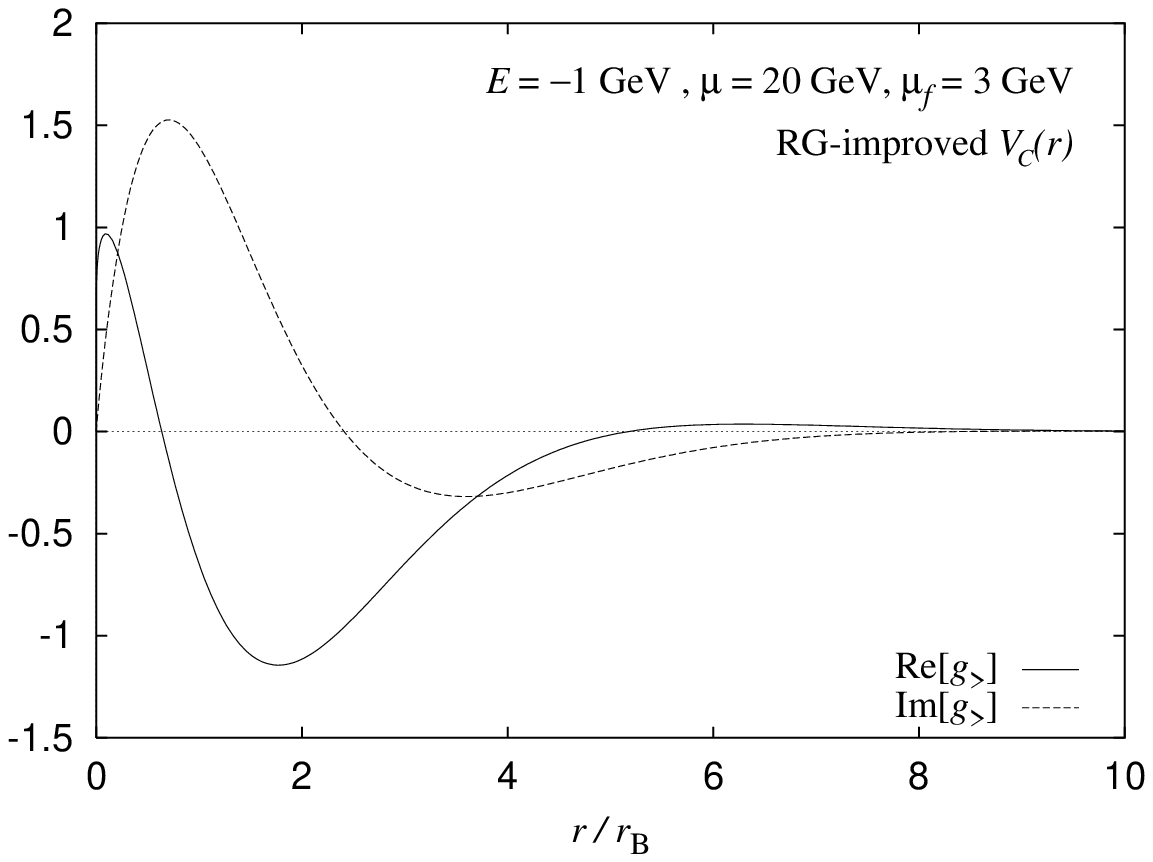}
  \end{minipage}
  \hspace*{\fill}
  \begin{minipage}{6cm}
    \includegraphics[width=7.5cm]{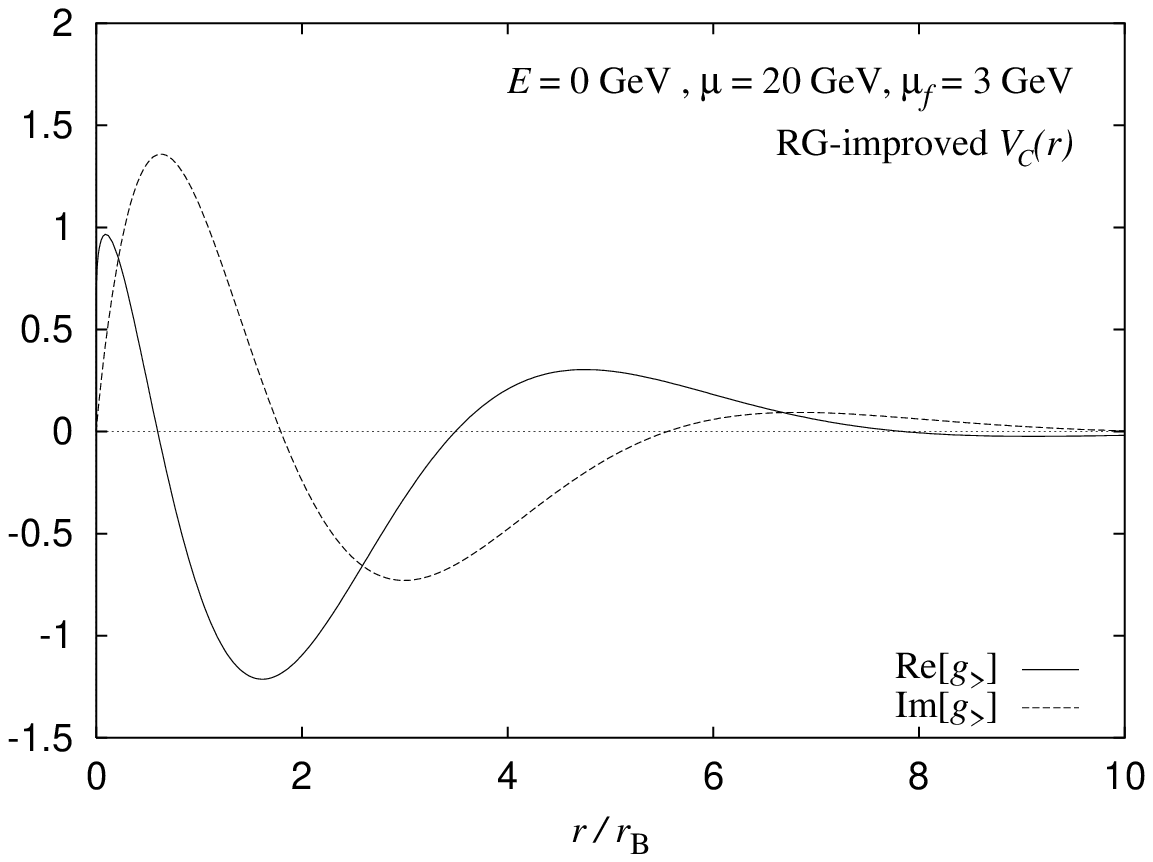}
  \end{minipage}
  \hspace*{\fill}
   \\
  \hspace*{\fill}
  \begin{Caption}\caption{\small
      The function $g_>(r)/r$ is a 
      solution of the (reduced) Schr\"odinger equation to NNLO 
      [\EqRef{eq:Schr_for_g<g>_NNLO}]
      that dumps at $r \to \infty$.  
      \scCM\ energy is changed from $E=-5\GeV$ to $0\GeV$.  
      They are plotted vs.\ $r/r_\rmB = p_\rmB\,r$, where 
      $1/r_\rmB = p_\rmB = C_F \alpha_s\mt/2 = 17.94\GeV$ for 
      $\alpha_s = \alpha_s^\MSbar(\mu_s=20\GeV) = 0.1534$.  
      Perturbative potential description is valid within the hard scale 
      $\sim p_\rmB / \mt = 0.1023$ and the $\Lambda_{\rm QCD}$ scale 
      $\sim p_\rmB / (1\GeV) = 17.94$.  
      We set $\mu_s = 20\GeV$, $\mu_f = 3\GeV$, $m_{\rm PS}(\mu_f) = 175\GeV$, 
      $\Gamma_t=1.43\GeV$, and $\alpha_s(m_Z)=0.118$.  
      \label{fig:gg-5=0}
  }\end{Caption}
  \hspace*{\fill}
\end{figure}
\begin{figure}[tbp]
  \hspace*{\fill}
  \begin{minipage}{6cm}
    \includegraphics[width=7.5cm]{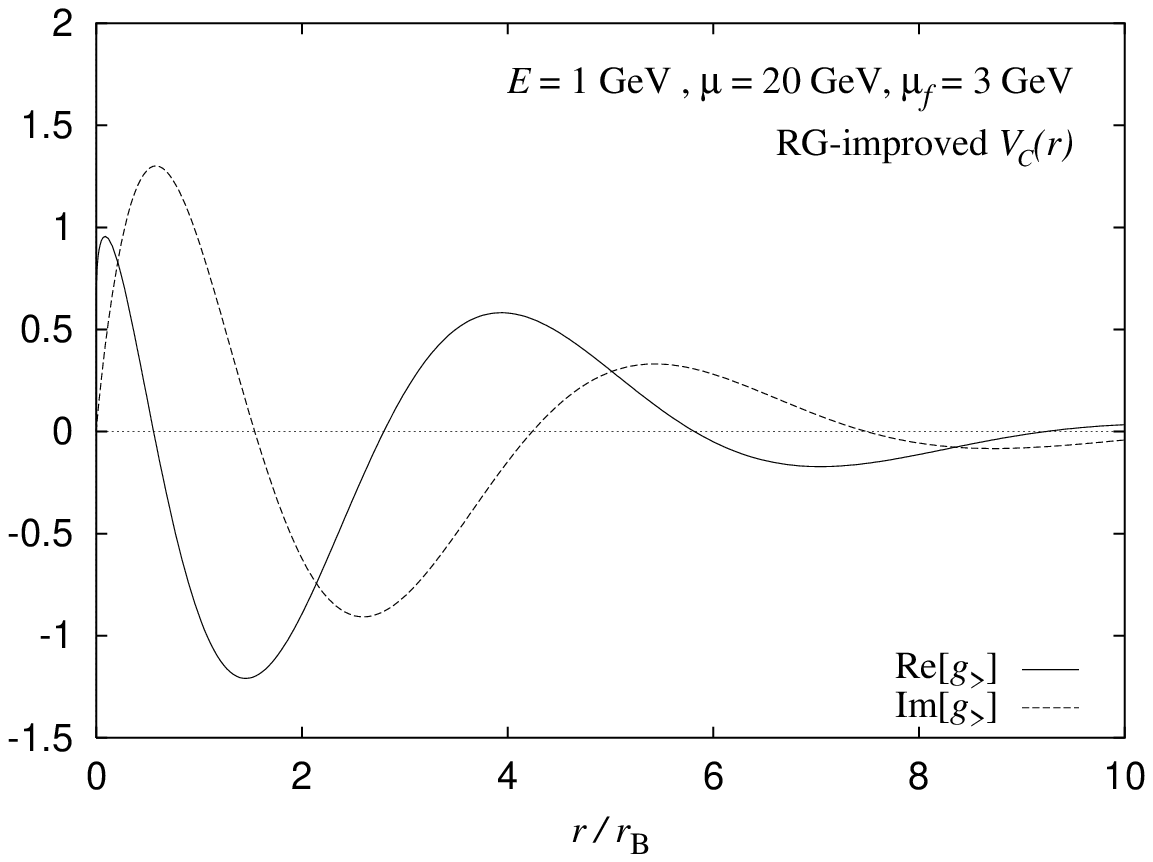}
  \end{minipage}
  \hspace*{\fill}
  \begin{minipage}{6cm}
    \includegraphics[width=7.5cm]{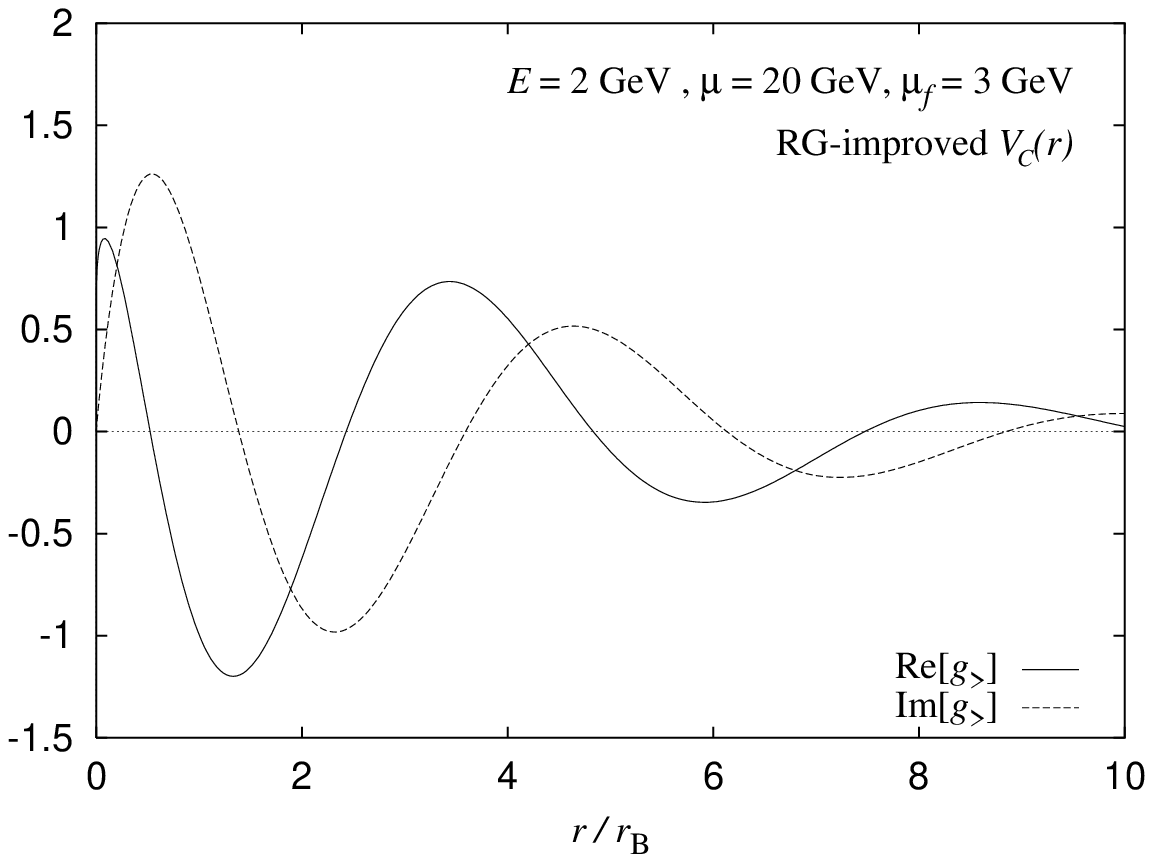}
  \end{minipage}
  \hspace*{\fill}
  \\
  \hspace*{\fill}
  \begin{minipage}{6cm}
    \includegraphics[width=7.5cm]{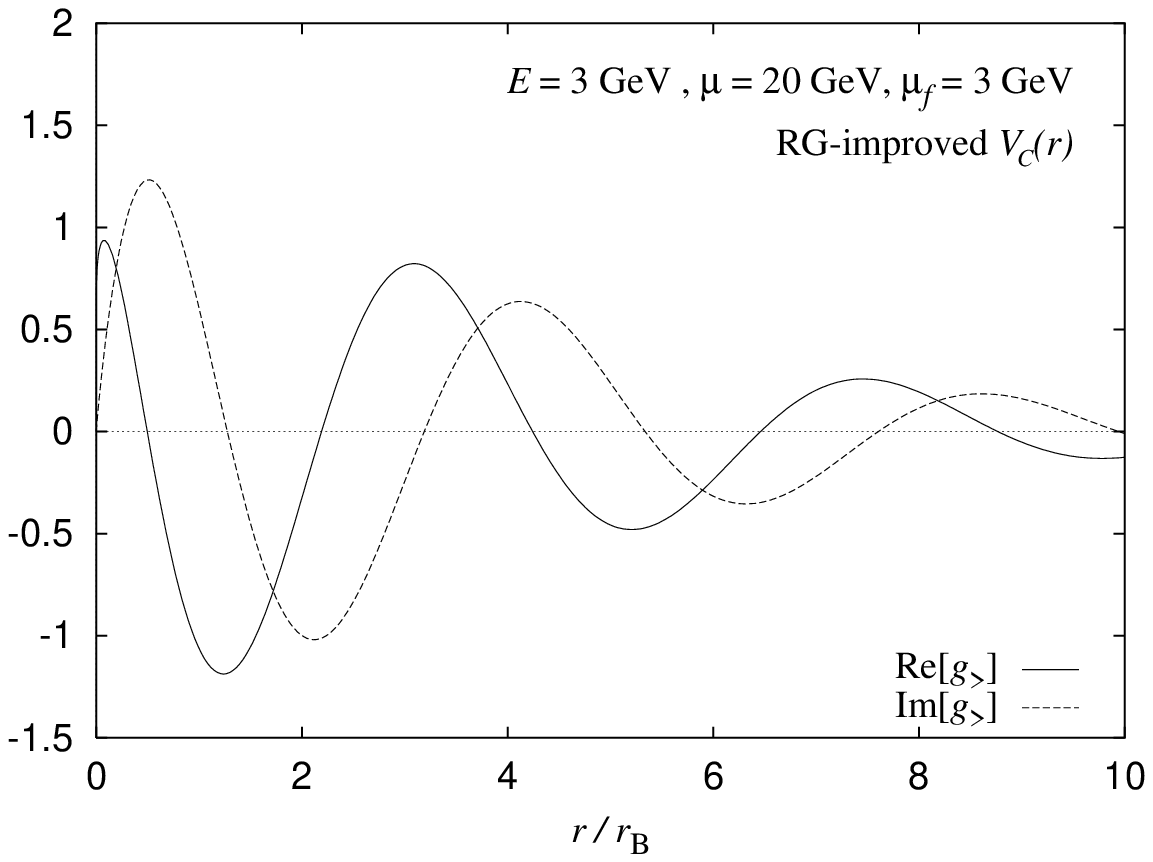}
  \end{minipage}
  \hspace*{\fill}
  \begin{minipage}{6cm}
    \includegraphics[width=7.5cm]{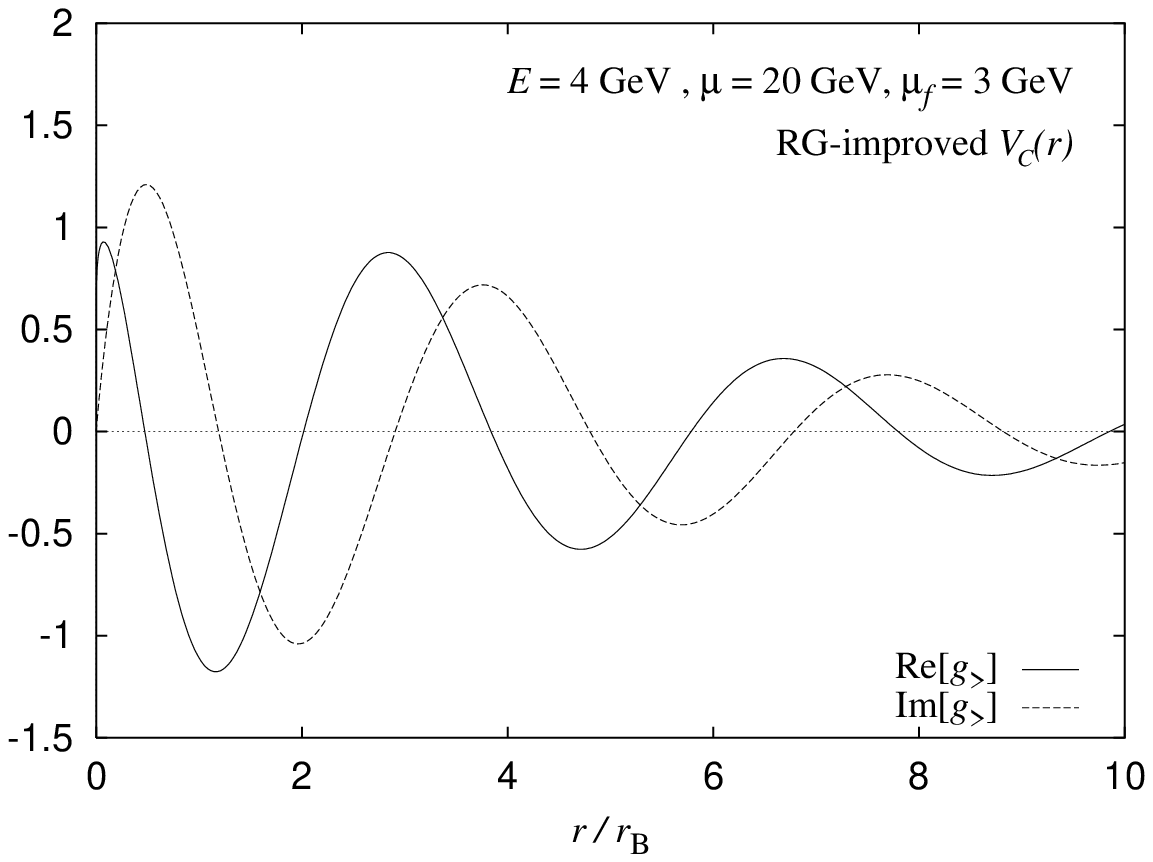}
  \end{minipage}
  \hspace*{\fill}
   \\
  \hspace*{\fill}
  \begin{minipage}{6cm}
    \includegraphics[width=7.5cm]{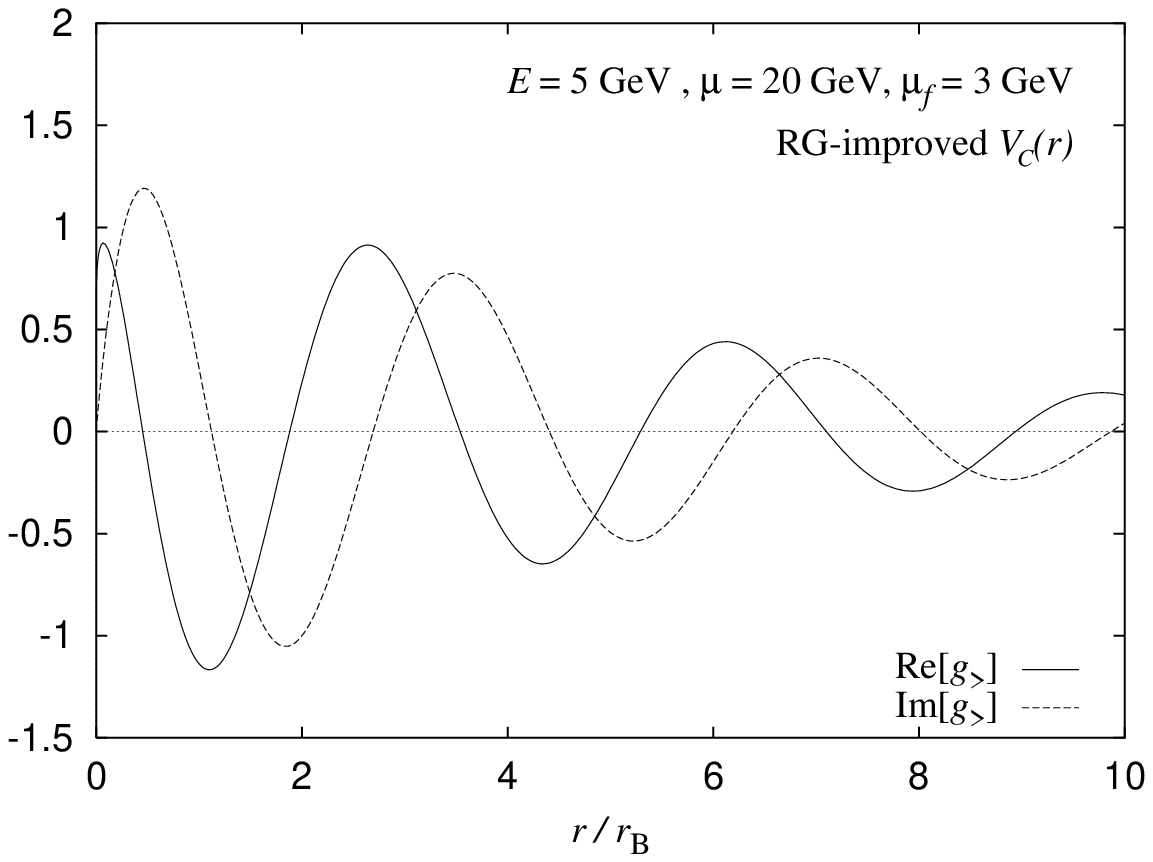}
  \end{minipage}
  \hspace*{\fill}
  \begin{minipage}{6cm}
    \includegraphics[width=7.5cm]{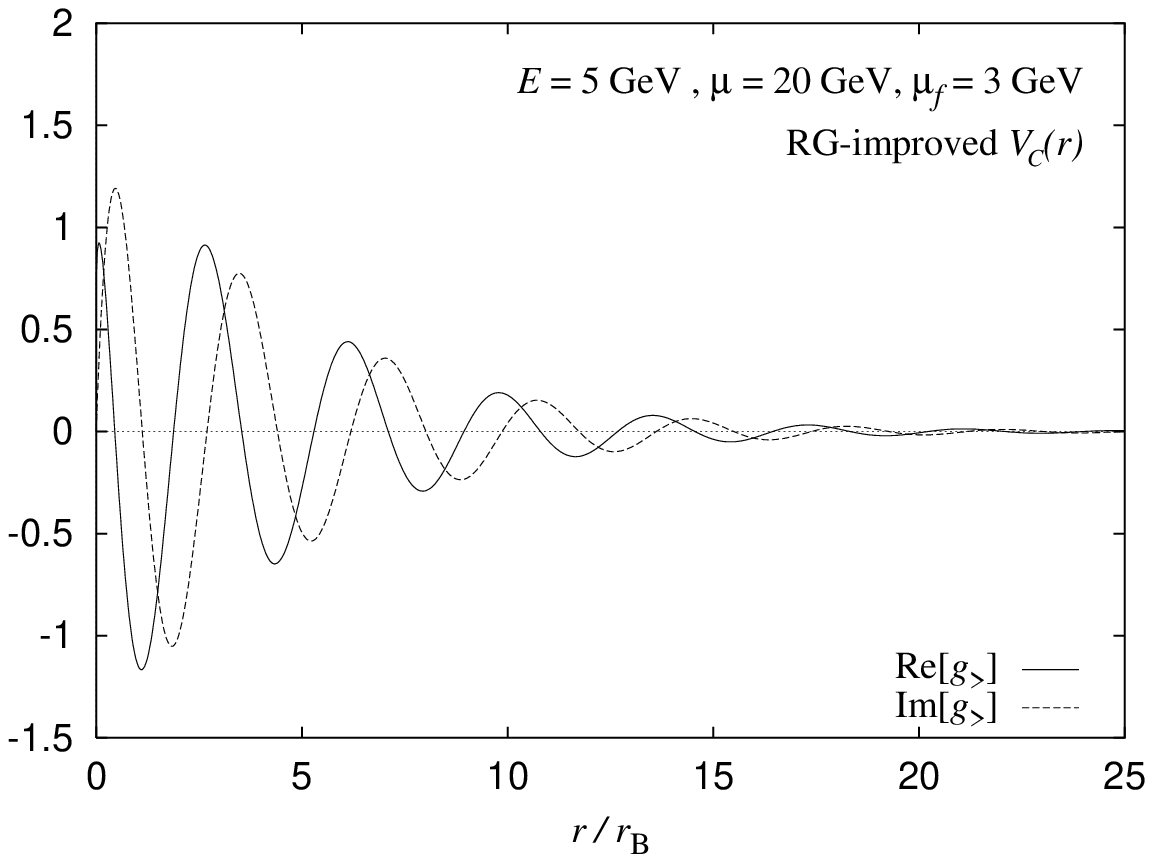}
  \end{minipage}
  \hspace*{\fill}
   \\
  \hspace*{\fill}
  \begin{Caption}\caption{\small
      The same as \FigRef{fig:gg-5=0}.  
      \scCM\ energy is chaged from $E=1\GeV$ to $5\GeV$.  
      The last two figures are both for $E=5\GeV$; 
      only the range of $r$ differs.  
      \label{fig:gg1=5}
  }\end{Caption}
  \hspace*{\fill}
\end{figure}
\begin{figure}[tbp]
  \hspace*{\fill}
  \begin{minipage}{6cm}
    \includegraphics[width=7.5cm]{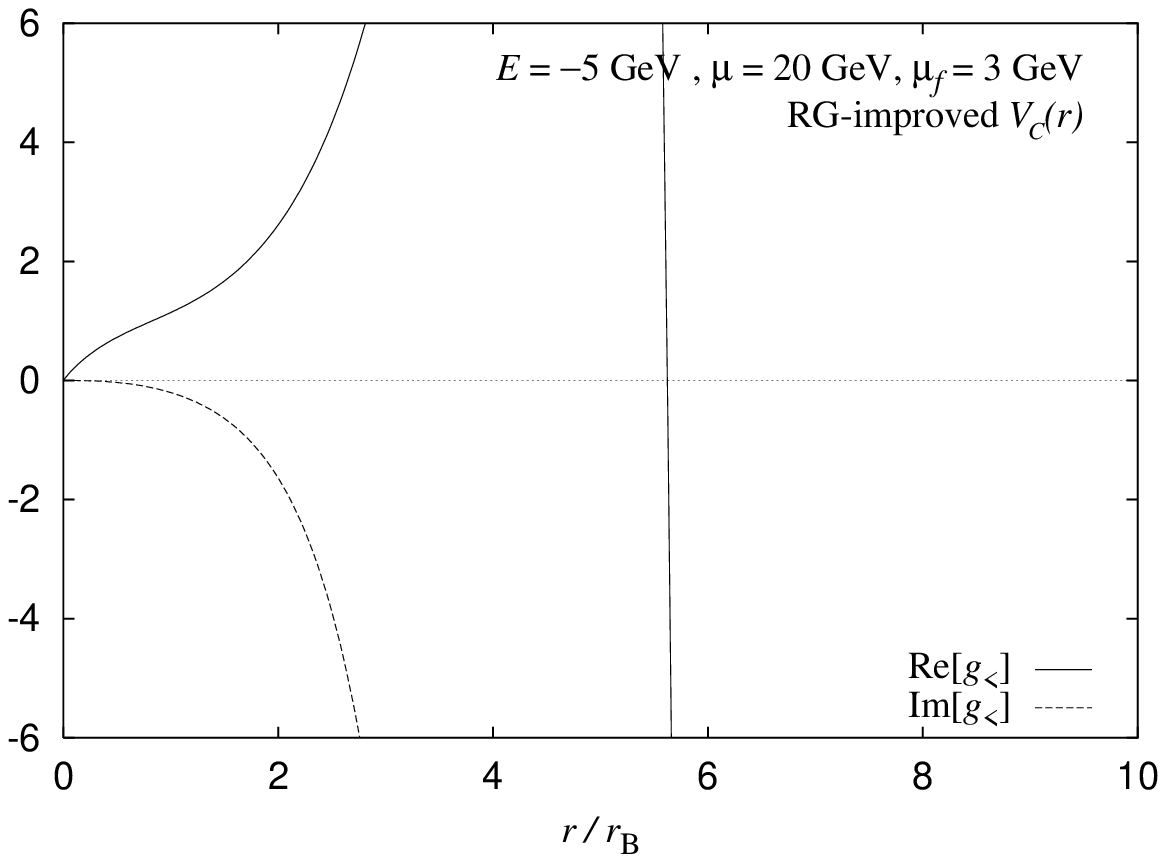}
  \end{minipage}
  \hspace*{\fill}
  \begin{minipage}{6cm}
    \includegraphics[width=7.5cm]{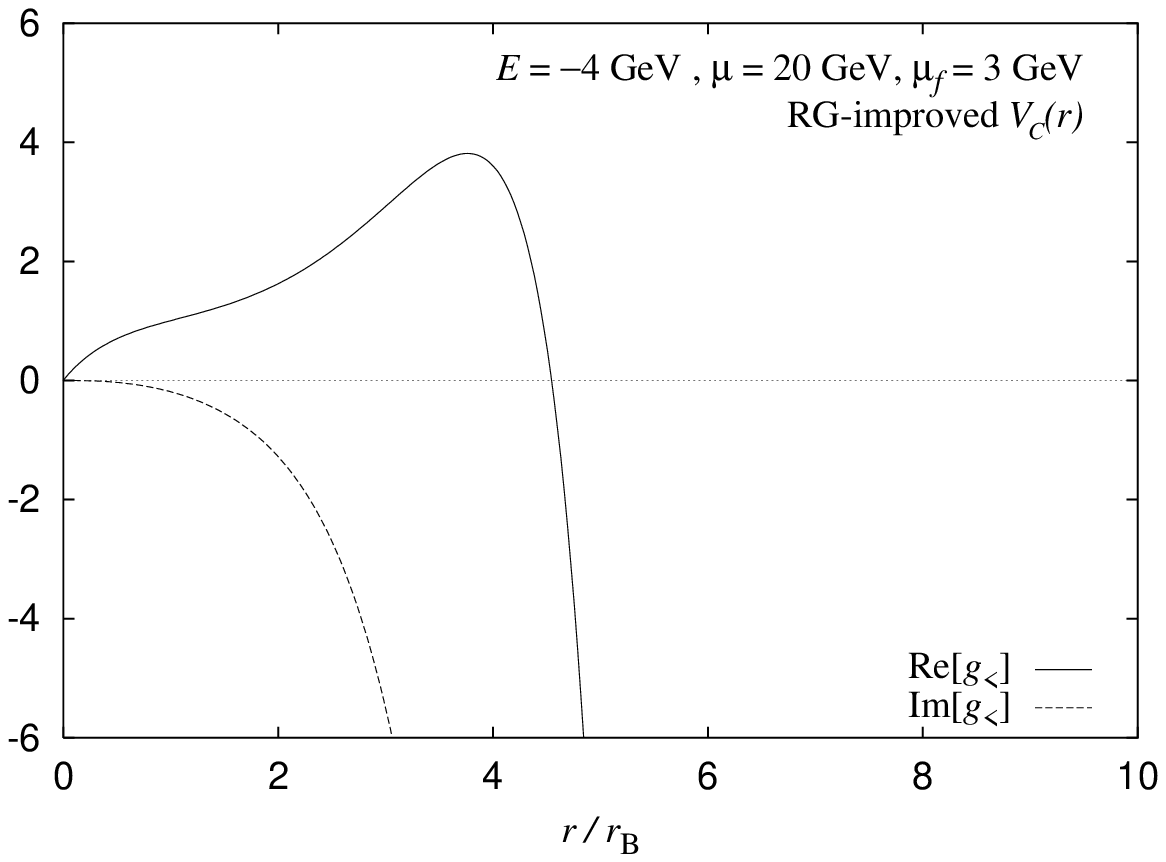}
  \end{minipage}
  \hspace*{\fill}
   \\
  \hspace*{\fill}
  \begin{minipage}{6cm}
    \includegraphics[width=7.5cm]{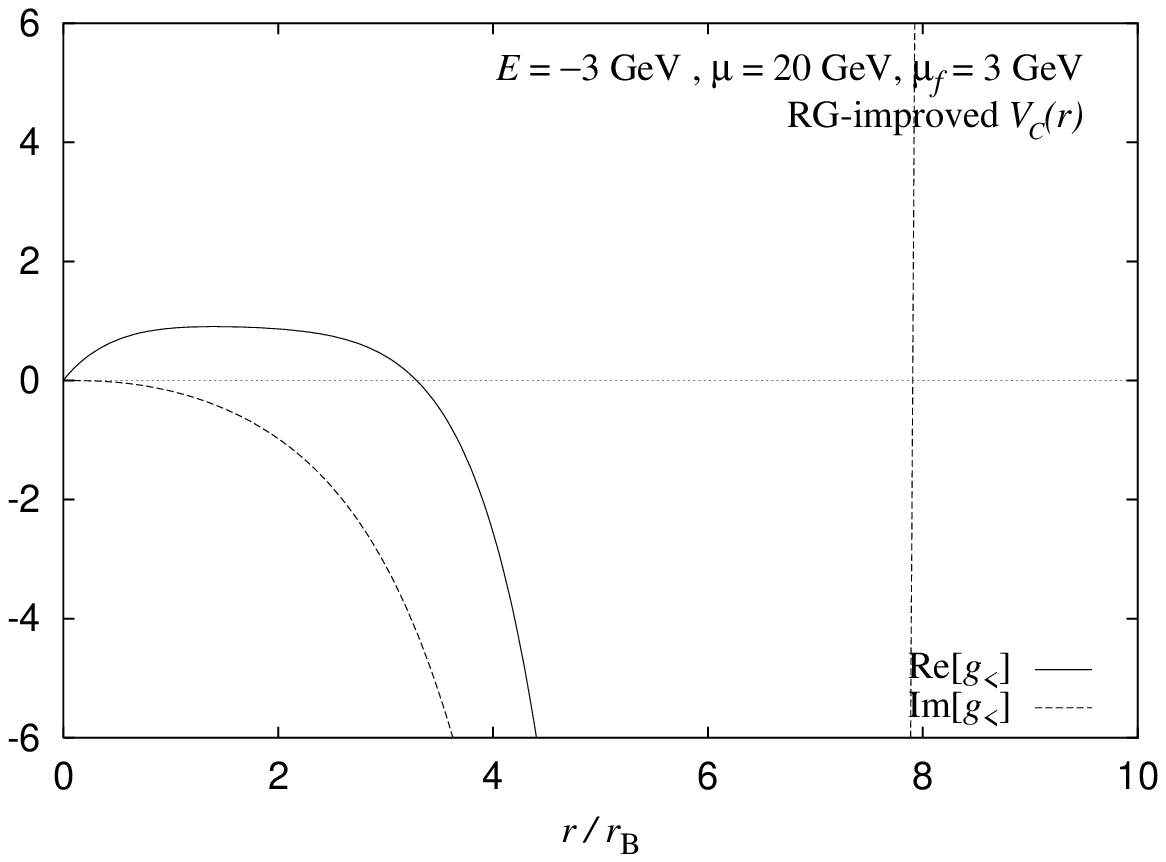}
  \end{minipage}
  \hspace*{\fill}
  \begin{minipage}{6cm}
    \includegraphics[width=7.5cm]{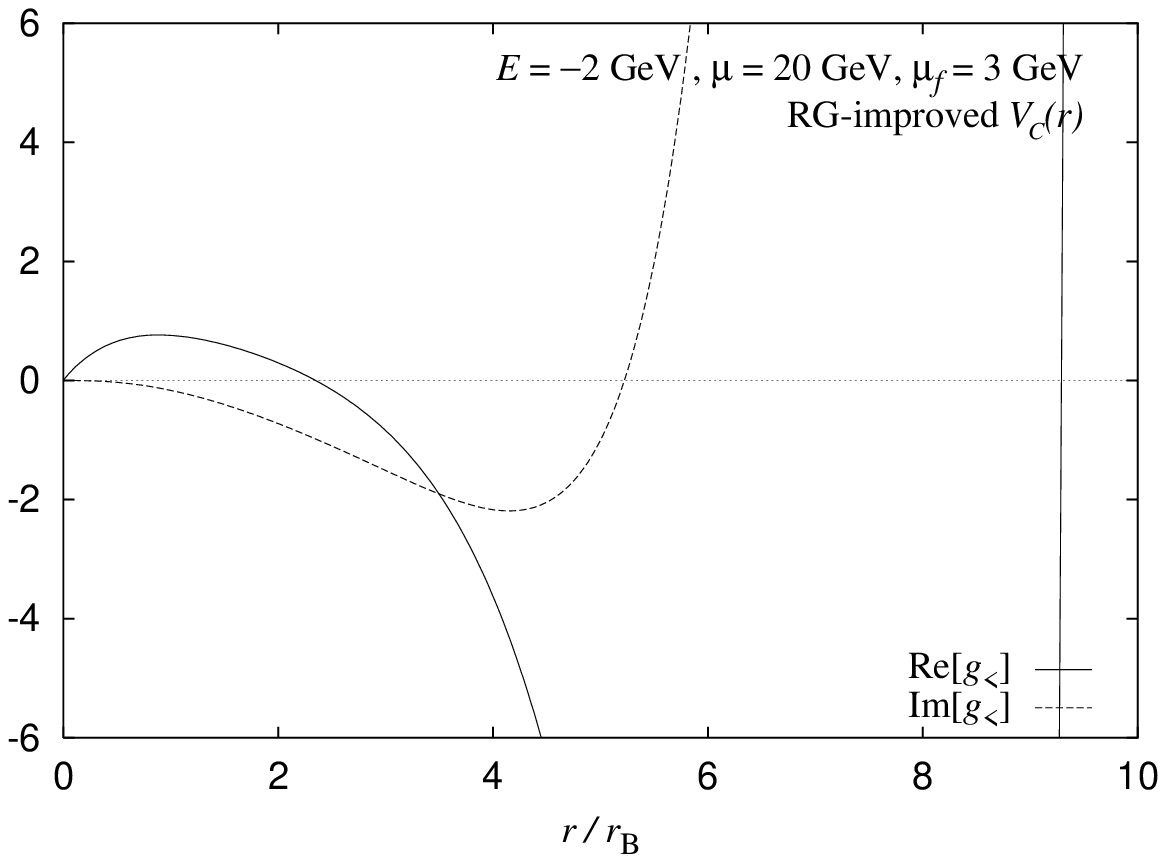}
  \end{minipage}
  \hspace*{\fill}
   \\
  \hspace*{\fill}
  \begin{minipage}{6cm}
    \includegraphics[width=7.5cm]{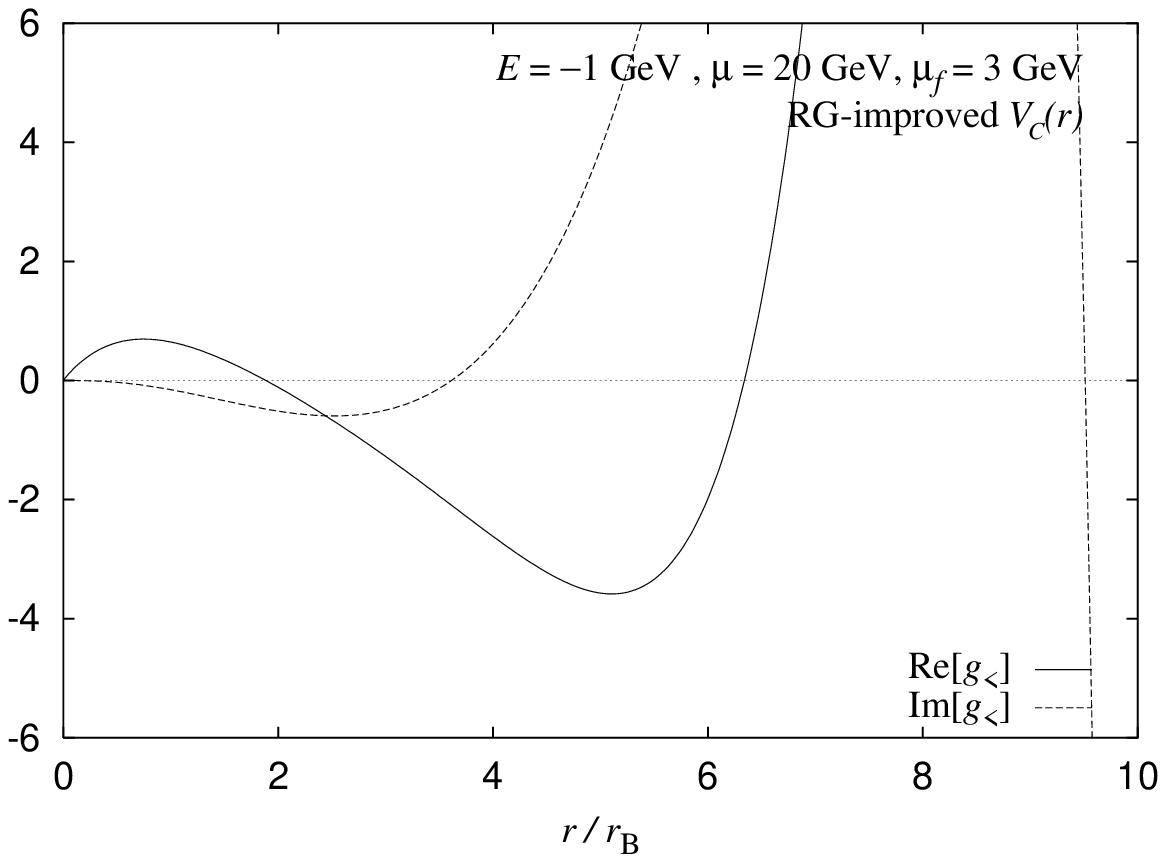}
  \end{minipage}
  \hspace*{\fill}
  \begin{minipage}{6cm}
    \includegraphics[width=7.5cm]{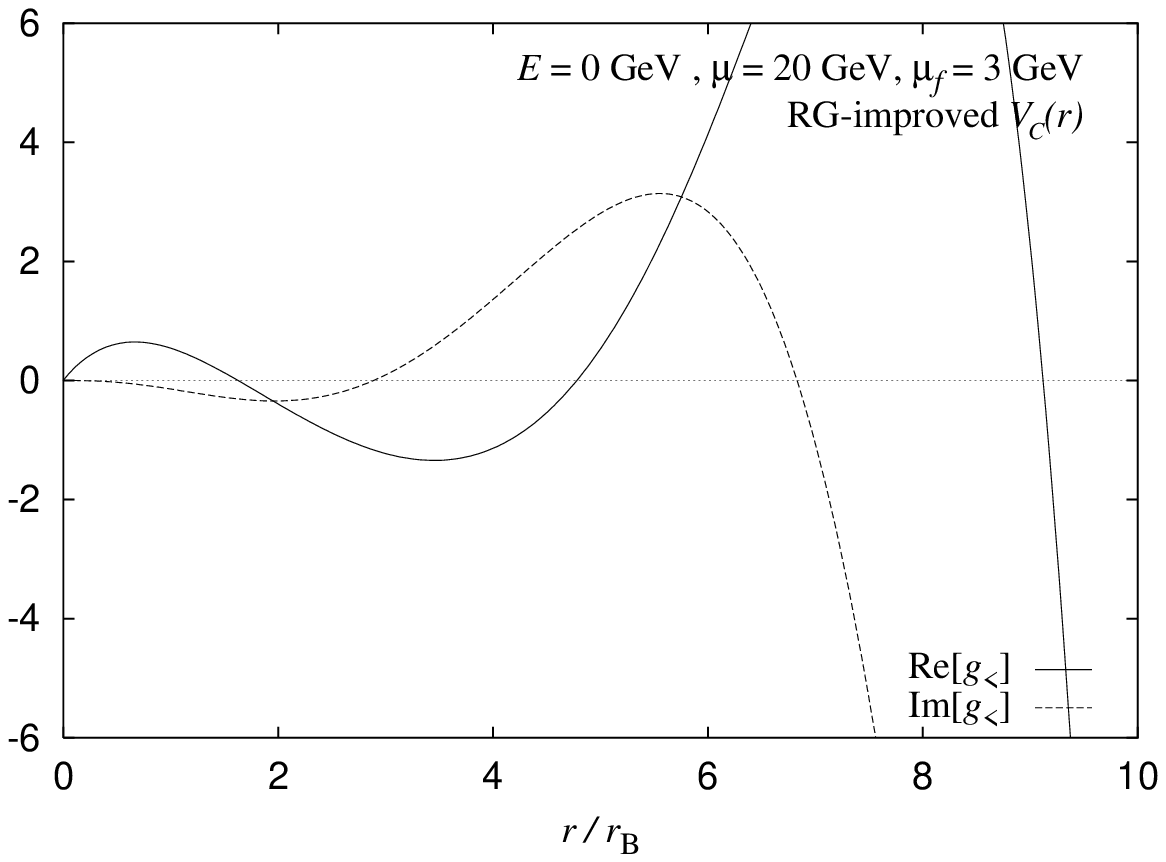}
  \end{minipage}
  \hspace*{\fill}
   \\
  \hspace*{\fill}
  \begin{Caption}\caption{\small
      The function $g_<(r)/r$ is a 
      solution of the (reduced) Schr\"odinger equation to NNLO 
      [\EqRef{eq:Schr_for_g<g>_NNLO}]
      that dumps at $r \to 0$.  
      Other situations are just the same as \FigRef{fig:gg-5=0}.  
      \label{fig:gl-5=0}
  }\end{Caption}
  \hspace*{\fill}
\end{figure}
\begin{figure}[tbp]
  \hspace*{\fill}
  \begin{minipage}{6cm}
    \includegraphics[width=7.5cm]{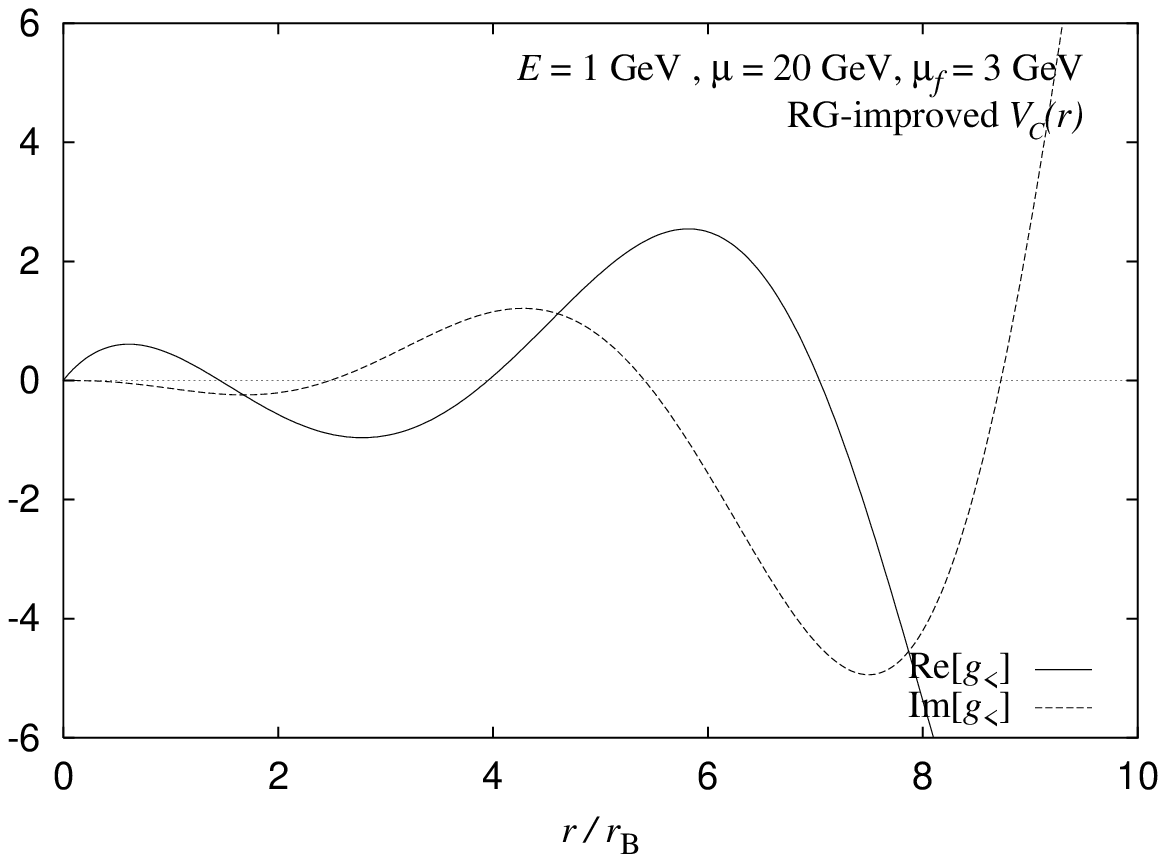}
  \end{minipage}
  \hspace*{\fill}
  \begin{minipage}{6cm}
    \includegraphics[width=7.5cm]{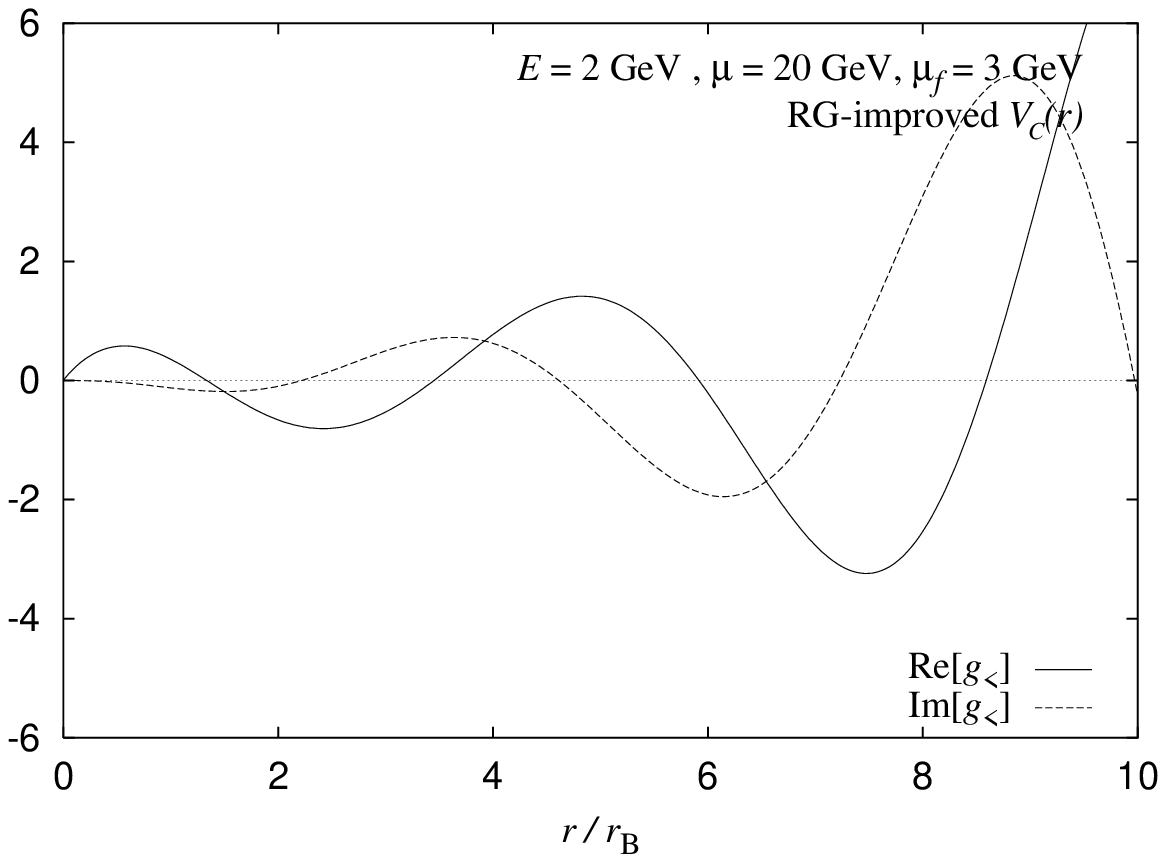}
  \end{minipage}
  \hspace*{\fill}
   \\
  \hspace*{\fill}
  \begin{minipage}{6cm}
    \includegraphics[width=7.5cm]{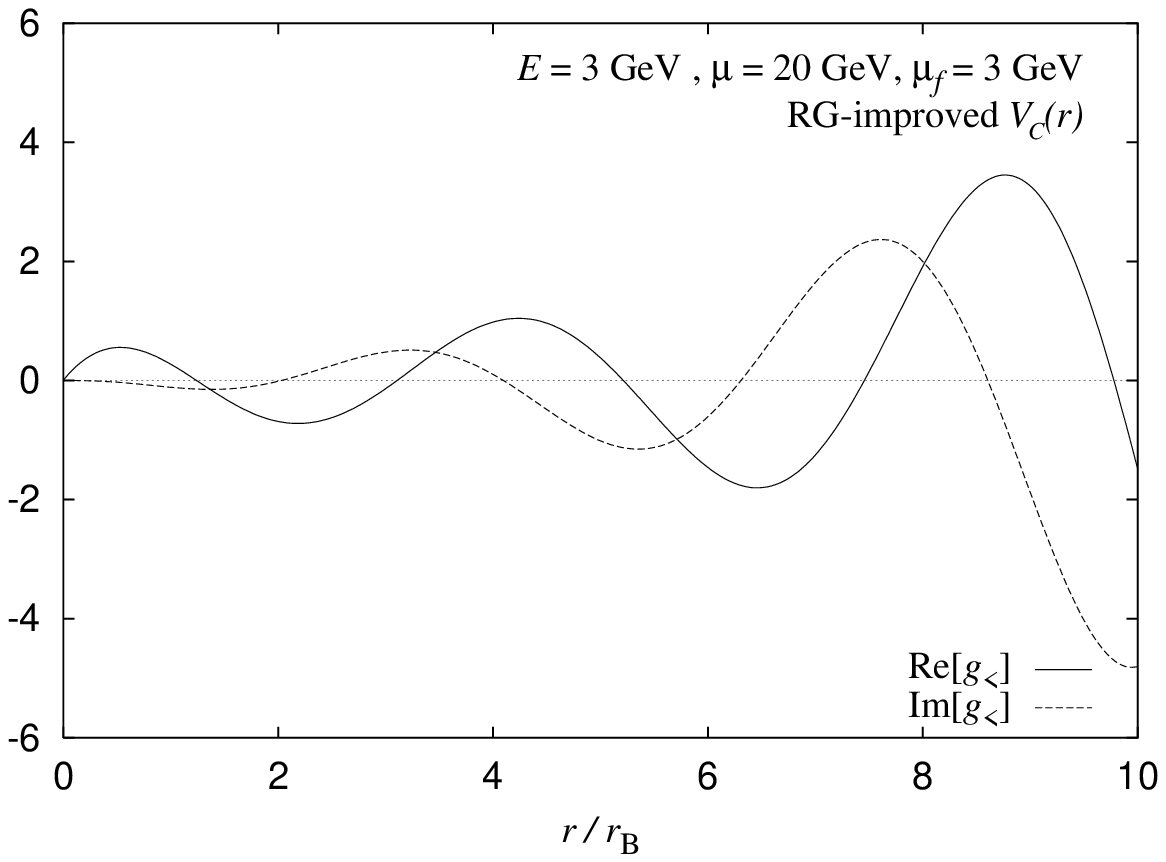}
  \end{minipage}
  \hspace*{\fill}
  \begin{minipage}{6cm}
    \includegraphics[width=7.5cm]{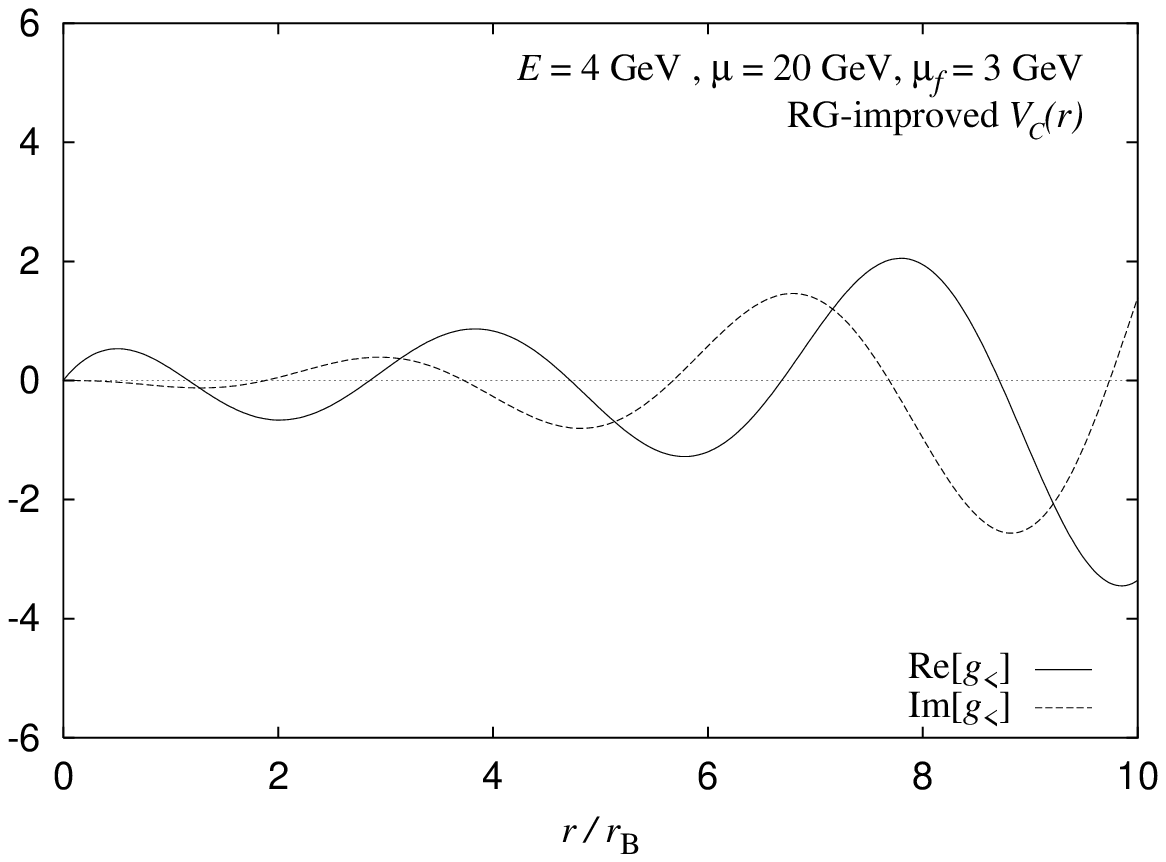}
  \end{minipage}
  \hspace*{\fill}
   \\
  \hspace*{\fill}
  \begin{minipage}{6cm}
    \includegraphics[width=7.5cm]{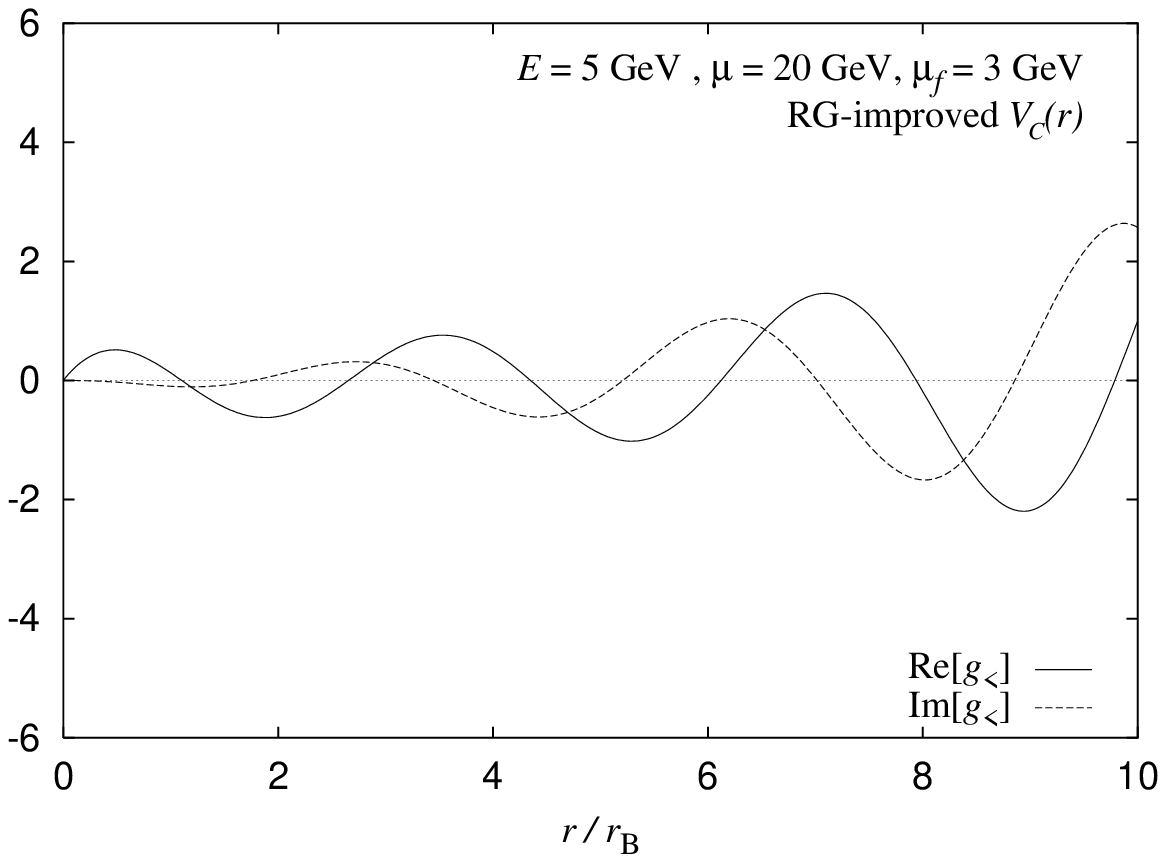}
  \end{minipage}
  \hspace*{\fill}
   \\
  \hspace*{\fill}
  \begin{Caption}\caption{\small
      Solutions of the (reduced) Schr\"odinger equation to NNLO 
      [\EqRef{eq:Schr_for_g<g>_NNLO}]
      that dumps at $r \to 0$.  
      Other situations are just the same as \FigRef{fig:gg1=5}.  
      \label{fig:gl1=5}
  }\end{Caption}
  \hspace*{\fill}
\end{figure}
%
%-------------------------------------------------------------------------
\subsubsection{Effect of $Z$ exchange}
To $\Order{1/c^2}$, axial-vector current $j_A^\mu$ also contributes 
since it is suppressed by $\Order{1/c}$.  
Note that the ``cross term'' $j_A^* j_V$ of axial-vector and vector 
($j_V^\mu$) current do not contribute to total cross section 
\SecRef{sec:exp-corr}.  
The effect of $Z$ exchange between $e^+e^-$ current and $t\tB$ current 
is incorporated by rescaling the vector--vector coupling and by adding 
the effect of axial-vector--axial-vector coupling.  
With the notation defined in \EqRef{eq:eff-coupling}, 
the former changes only overall normalization by the ratio 
$|[v^e v^t]|^2+|[a^e v^t]|^2$ to $Q_t^2$.  
The effective couplings $[v^e v^t]$ etc.\ are defined in 
\SecRef{sec:merge_Z_and_photon}.  
Numerically it is $0.480/0.444 = 1.080$ on the threshold.  
Especially, theoretical uncertainty $(\Delta m_t)^{\rm th}$ 
for determination of $m_t$ is not altered by this inclusion, since 
it is energy independent.  
While in~\cite{BSS99} it is shown that the effect of $j_A^* j_A$ is 
small and flat with respect to $E$ near the threshold%
\footnote{
  Note that the correspoding coupling is 
  $|[v^e a^t]|^2+|[a^e a^t]|^2 = 0.1430$ on the threshold.  
}, 
since it is suppressed by $\beta^2$ compared to $j_V^* j_V$.  
Note that only the leading order calculation of $j_A^* j_A$ is sufficient 
for the NNLO calculation of $R$ ratio.  
Thus $R$ ratio near the threshold is not altered when the effect of $Z$ 
exchange is taken into account, and so is $(\Delta m_t)^{\rm th}$.  
%-------------------------------------------------------------------------
\subsubsection{Effect of using $m_{\MSbar}$ instead of $m_{\rm pole}$}
\begin{figure}[tbp]
  \hspace*{\fill}
  \begin{minipage}{6cm}
    \includegraphics[width=7.5cm]{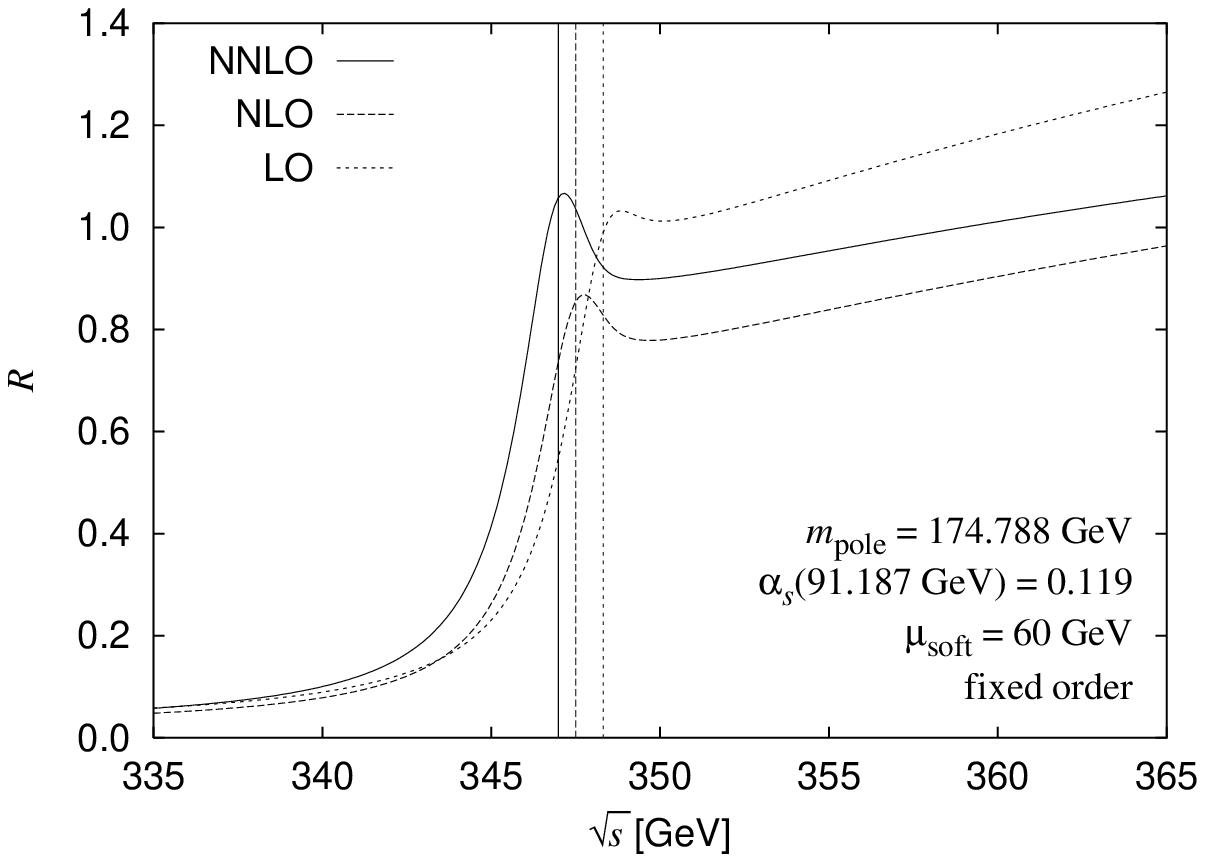}
  \end{minipage}
  \hspace*{\fill}
  \begin{minipage}{6cm}
    \includegraphics[width=7.5cm]{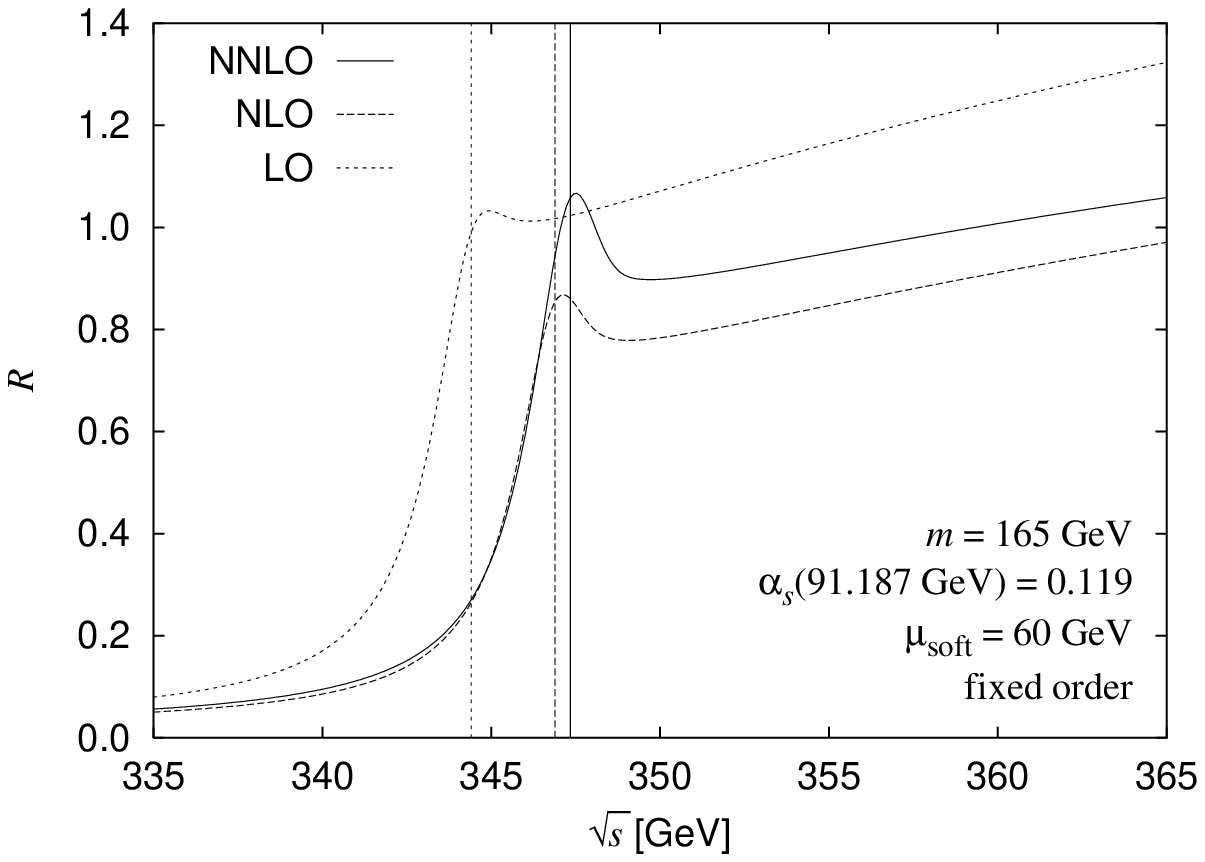}
  \end{minipage}
  \hspace*{\fill}
   \\
  \hspace*{\fill}
  \begin{Caption}\caption{\small
      $R$ ratio in $m_{\rm pole}$ scheme (left) and 
      $m_{\MSbar}$ scheme.  
      See text for detail.  
      \label{fig:mMSB-vs-mpole}
  }\end{Caption}
  \hspace*{\fill}
\end{figure}

Shown in \FigRef{fig:mMSB-vs-mpole} is the effect of using different 
mass schemes.  
In the left figure, pole-mass is fixed to $m_{\rm pole} = 174.788\GeV$; while 
in the right one, \MSbar-mass is fixed to $\bar{m}(\bar{m}) = 165\GeV$, 
which corresponds to $m_{\rm pole} = 174.788\GeV$ at NNNLO.  
In both figures, $\alpha_s^{\MSbar}(m_Z) = 0.119$ is used; 
this corresponds to $\alpha_s^{\MSbar}(165\GeV) = 0.109$.  
Note that the used potential is one in fixed-order; also PS-mass scheme is 
not used.  
By letting $\Gamma_t \to 0$, one can investigates the position of $1S$ peak.  
These positions are also shown in the figures by vertical lines.  
Numerically, they are $(348.321\GeV, 347.504\GeV, 346.986\GeV)$ for 
(LO, NLO, NNLO) when $m_{\rm pole}$ is fixed, 
while $(344.408\GeV, 346.892\GeV, 347.340\GeV)$ when $\bar{m}(\bar{m})$ 
is fixed.  
\clearemptydoublepage
%
%
%----------------------------------------------------------------------
\chapter{Top-Quark Momentum Distribution}
\label{ch:diff_cr}
Due to the large decay width $\Gamma_t$, the momentum $|\pBI_t|$ of 
top quark is not determined by the \scCM\ energy $\sqrt{s}=2\mt+E$, 
but distributes around $|\sqrt{m_t(E+1\GeV+i\Gamma_t)}|$, 
where $1\GeV\sim\mbox{binding energy}$.  
Momentum distribution $\diffn{\sigma}{}/\diffn{p}{}$ of top quark 
in $e^+e^- \to t\tB$ near the threshold is calculated 
in~\cite{JKT92} to LO, and in~\cite{MS93} to NLO.  
In this section we calculate a part of NNLO correction; 
that is, only gluon exchange between $t\tB$ is considered.  
Gluon exchange between $t$ and $\bB$ etc.\ are not considered.  
%
%
%

%
%
%
%-----------------------------------------------------------------------
\section{Derivation of the distribution}
Differential cross section of $t\tB$ pair production can be written 
as follows: 
%%
%\begin{align*}
%  \frac{\diffn{\sigma}{} (e^+ e^- \to t\tB \to b\ell^+\nu\bar{b}W^-) }
%    {\diffn{\ptBI}{3} \diffn{\pBI_{\!\ell}}{3}}
%  =
%  \frac{\diffn{\sigma}{} (e^+ e^- \to t\tB) }{\diffn{\ptBI}{3}}
%  \times
%  \frac{1}{\Gamma_t} \,
%  \frac{\diffn{\Gamma}{} (t_\Pol \to b\ell^+\nu)}{\diffn{\pBI_{\!\ell}}{3}}
%\end{align*}
%%
%
\begin{align*}
  &  \diffn{\sigma}{} (e^+ e^- \to t\tB \to bW^+\bar{b}W^-)
  \nonumber\\
  &=
  \frac{N_C}{2s} \sum_{X,Y=\gamma,Z}
  \frac{(2\mt)^4}{(s-m_X^2)(s-m_Y^2)}
  (E_{XY})^{\mu\nu} (H_{XY})^{\mu\nu}
  \nonumber\\
  &\qquad{}
  \times \left|
    \frac{1}{p_t^2-m_t^2+im_t\Gamma_t} \,
    \frac{1}{\pB_t^2-m_t^2+im_t\Gamma_t}
  \right|^2
  \diffn{\Phi_4}{} (bW^+\bar{b}W^-)
  \sepcomma
\end{align*}
where
\begin{align*}
  &  (E_{XY})^{\mu\nu}
  = \btr{\Lambda_X^\mu 
    \frac{(1-\Polep\Polem)-(\Polep-\Polem)}{4}
    \peS\LambdaB_Y^\nu\peBS}
  \sepcomma\nonumber\\
%  & L^{\rho\sigma}
%  = \pfrac{\gW}{\sqrt{2}}^{\!2} \btr{\gamma^\rho\frac{1-\gamma_5}{2}
%      \pS_{\!\nu}\gamma^\sigma\frac{1-\gamma_5}{2}\pBS_{\!\ell}}
%  \sepcomma\nonumber\\
  & (H_{XY})^{\mu\nu}
  = \pfrac{\gW}{\sqrt{2}}^{\!4} \tr\biggl[ 
      \frac{\ptS+\mt}{2\mt}\Gamma_X^\mu\frac{-\ptBS+\mt}{2\mt}
      \gamma^\alpha\frac{1-\gamma_5}{2}\pBS_{\!b}
      \gamma^\beta\frac{1-\gamma_5}{2}
    \times
  \nonumber\\
  &\qquad{}
    \times
      \frac{-\ptBS+\mt}{2\mt}\GammaB_Y^\nu\frac{\ptS+\mt}{2\mt}
    \gamma^\rho\frac{1-\gamma_5}{2}\pS_{\!b}\gamma^\sigma\frac{1-\gamma_5}{2}
    \biggr]
    \times
  \nonumber\\
  &\qquad{}
    \times
    \biggl[ -g_{\alpha\beta} + \frac{(\pB_W)_\alpha(\pB_W)_\beta}{m_W^2}
    \biggr]
    \biggl[ -g_{\rho\sigma} + \frac{(p_W)_\rho(p_W)_\sigma}{m_W^2}
    \biggr]
  \sepcomma
\end{align*}
and
\begin{align*}
  &  
  \Lambda_X^\mu = 
     g_{\!\mbox{\tiny $X$}} 
     \left(v^{eX}\gamma^\mu - a^{eX}\gamma^\mu\gamma_5 \right)
  \sepcomma\nonumber\\
  &  \Gamma_X^\mu = (|\ptBI|^2/\mt - (E+i\Gamma_t))
    \times
  \nonumber\\
  &\qquad{}
    \times g_{\!\mbox{\tiny $X$}} \left[
        v^{tX}(1+\delta_V)\GT(E,\ptBI)\gamma^\mu
      - a^{tX}(1+\delta_A)\FT(E,\ptBI)\gamma^\mu\gamma_5
    \right]
  \sepcomma\nonumber\\
  &  \delta_V = - \frac{8\alpha_s}{3\pi} \sepcomma\qquad
     \delta_A = - \frac{4\alpha_s}{3\pi}
  \sepperiod
\end{align*}
This can be obtain by usual Feynman rules with the modification of 
the $t\tB$ production vertex $\Gamma^\mu$ due to the Coulomb rescattering.  
The formula above can be reduced more by approximating the momenta of 
top and anti-top by their on-shell values%
\footnote{
  Except at the denominators of the top and anti-top propagators.  
}.  
By decomposing the phase space of $bW^+\bar{b}W^-$ into $t\tB$ 
\begin{align*}
  \diffn{\Phi_4}{} (bW^+\bar{b}W^-)
  =
  \frac{\diffn{p}{4}}{(2\pi)^4}
  \diffn{\Phi_2}{} (t^* \to bW^+) \,
  \diffn{\Phi_2}{} (\tB^* \to \bB W^-)
  \sepcomma
\end{align*}
where $p_t = q/2+p$ and $\pB_t = q/2-p$ ($q$ is the total momentum), 
both $bW^+$ and $\bar{b}W^-$ can integrated out: 
\begin{align*}
  &  \int\!\!\diffn{\Phi_2}{} (\tB \to \bB W^-) \,
    \pfrac{\gW}{\sqrt{2}}^{\!2} 
    \gamma^\alpha\frac{1-\gamma_5}{2}\pBS_{\!b}
    \gamma^\beta\frac{1-\gamma_5}{2}
    \biggl[ -g_{\alpha\beta} + \frac{(\pB_W)_\alpha(\pB_W)_\beta}{m_W^2}
    \biggr]
  \nonumber\\
  &=
    \frac{\Gamma_t}{\mt}\ptBS(1-\gamma_5)
  \sepperiod
\end{align*}
As for the remaining phase space $\diffn{p}{4}/(2\pi)^4$, 
the time-component can be integrated with the formulas 
\begin{align*}
  &  \int\!\!\frac{\diffn{z}{}}{2\pi} \,
    \frac{1}{z-(x+iy)} \, \frac{1}{z+(x+iy)}
    = \frac{i}{2(x+iy)}
  \sepcomma\nonumber\\
  &  \int\!\!\frac{\diffn{z}{}}{2\pi} \,
    \left| \frac{1}{z-(x+iy)} \, \frac{1}{z+(x+iy)} \right|^2
    = \left| \frac{i}{2(x+iy)} \right|^2 \frac{1}{y}
  \sepcomma
\end{align*}
since we put $p_t^0 = \sqrt{\pBI_t^2+\mt^2}$ everywhere aside from 
at the denominators of propagators.  
Thus we have 
\begin{align*}
  \frac{\diffn{\sigma}{} (e^+ e^- \to t\tB)}{\diffn{p}{3}}
  &=
  \frac{1}{(2\pi)^3}
  \frac{N_C}{2s} 
  \sum_{X,Y=\gamma,Z}
  \frac{2\Gamma_t}{(s-m_X^2)(s-m_Y^2)}
  \left| \frac{1}{E-\pBI^2/\mt+i\Gamma_t} \right|^2
  \times
  \\
  &\qquad\qquad{}\times
  \btr{ \Gamma_X^\mu\frac{\ptBS-\mt}{2\mt}
    \GammaB_Y^\nu\frac{\ptS+\mt}{2\mt}
  }  (E_{XY})_{\mu\nu}
  \sepperiod
\end{align*}

The momentum distributions of the decay products of $W$'s depend on 
the polarization of top-quark.  For such calculations, the following 
relations may be useful: 
\begin{align*}
  \frac{\pS+m}{2m} G \frac{\pS+m}{2m}
  = \frac{\pS+m}{2m} \, \frac{1-\PolS\gamma_5}{2} C
  \sepcomma\quad
  \frac{\pBS-m}{2m} \GB \frac{\pBS-m}{2m}
  = \frac{\pBS-m}{2m} \, \frac{1-\PolBS\gamma_5}{2} \CB
  \sepcomma
\end{align*}
where 
\begin{align*}
  \frac{1-\dprod{\Pol}{s}}{2} C
  = \btr{G \frac{\pS+m}{2m} \, \frac{1-\sS\gamma_5}{2}}
  \sepcomma\quad
  \frac{1-\dprod{\PolB}{\sB}}{2} \CB
  = \btr{\GB \frac{\pBS-m}{2m} \, \frac{1-\sBS\gamma_5}{2}}
  \sepperiod
\end{align*}
There relations hold when $p^2 = m^2$, $\dprod{\Pol}{p} = \dprod{s}{p} = 0$, 
and the likewise for $\pB$.  These relations can be easily verified at the 
rest frame of $p$.  Note that although $G$ is a general $4\times 4$ matrix, 
only $2\times 2$ sub-matrix contributes because of the projection operators 
$\pS+m/(2m)$ etc.  Those four degrees of freedom are expressed by one scalar 
$C$ and one 4-vector $\Pol$ with one constraint $\dprod{\Pol}{p}=0$.  

For the case of $\gamma$-mediated $t\tB$ production, 
the distribution $\diffn{\sigma}{}/\diffn{p}{}$ can be calculated 
in usual way with the vertices 
\begin{align*}
  \Lambda^\mu = e Q_e \gamma^\mu
  \sepcomma\quad
  \Gamma^\mu = (|\ptBI|^2/\mt - (E+i\Gamma_t))\,
    e Q_t \GT(E,\ptBI)\gamma^\mu
  \sepperiod
\end{align*}
At the leading order, the distribution can written as 
\begin{align*}
  \frac{1}{\sigma_{\rm pt}} \odiff{\sigma}{p} 
  = \frac{4\pi p^2}{(2\pi)^3} \cdot 
    \frac{3}{2} N_C Q_t^2 \frac{4m^2}{s} \, \frac{4\pi}{m^2c} \Gamma_t
    \left| \GT(r=0,p) \right|^2
  \sepperiod
\end{align*}
For $\Gamma_t \to 0$, this is zero unless $\GT(r=0,p)$ diverges.  
This is consistent with 
\begin{align*}
  \frac{1}{\sigma_{\rm tot}} \odiff{\sigma}{p} 
  = \delta(p-\sqrt{\mt E})
  \sepcomma\quad\mbox{ for $\Gamma_t=0$. }
\end{align*}
%

%
%
%

%
%
%-------------------------------------------------------------------------
\section{Unitarity relation}
Momentum integration of the differential cross section 
$\diffn{\sigma}{}/\diffn{p}{}$ should coincide to the total cross section 
$\sigma_{\rm tot}$.  This is relation is called unitarity.  
It is important to confirm oneself that unitarity holds both 
theoretically and numerically within the accuracy considered.  

For a Green function 
\begin{align*}
  G = \frac{1}{H -\omega}
  \sepcomma\quad \omega = E + i\Gamma
  \sepcomma
\end{align*}
its imaginary part can expressed as 
\begin{align*}
  \Im G 
  &\equiv \frac{1}{2i} (G - G^\dagger) 
  = G^\dagger \frac{(G^{-1})^\dagger - G^{-1}}{2i} G  
  = G \frac{(G^{-1})^\dagger - G^{-1}}{2i} G^\dagger  \\
  &= G^\dagger \, \bIm{(G^{-1})^\dagger} \, G 
  = G^\dagger \, \bIm{H-E+i\Gamma} \, G  \\
  &= G^\dagger G \Gamma = GG^\dagger \Gamma
  \sepcomma
\end{align*}
or
\begin{align*}
  \int\!\!\frac{\diffn{p}{3}}{(2\pi)^3} \,
  \Gamma_t \left| \GT(|\pBI|,E) \right|^2
  = \Im G(\xBI=0,E) 
  \sepperiod
\end{align*}
Thus the relation 
\begin{gather*}
  \frac{\diffn{\sigma}{}}{\diffn{p}{3}}
%  = \frac{N_C Q_t^2 \alpha^2}{4\pi \mt^4} \,
%    \Gamma_t \left| \GT(|\pBI|,E) \right|^2
  = \sigma_{\rm pt} \, N_C Q_t^2 \frac{3}{4\pi^2\mt^2} \, 
    \Gamma_t \left| \GT(|\pBI|,E) \right|^2
  \sepcomma\quad\mbox{ or }\quad
  \odiff{\sigma}{p}
%  = \frac{N_C Q_t^2 \alpha^2}{\mt^4} \,
%    p^2 \Gamma_t \left| \GT(|\pBI|,E) \right|^2
  = \sigma_{\rm pt} \, N_C Q_t^2 \frac{3}{\pi\mt^2} \, 
    p^2 \Gamma_t \left| \GT(|\pBI|,E) \right|^2
\end{gather*}
is consistent with the expression for the $R$ ratio in 
\EqRef{eq:R_with_ImG_LO}.  

Including the higher order corrections for $t\tB$ current, 
the unitarity relation above becomes 
\begin{align*}
  &
  \bIm{ \left\{ C_1^{\rm (cur)} 
      + C_2^{\rm (cur)}\frac{\triangle_r+\triangle_{r'}}{2m^2c^2} \right\}
    G(r,r') }  \\
  &=\; \int\!\!\frac{\diffn{p}{3}}{(2\pi)^2} \left\{ C_1^{\rm (cur)} 
    + C_2^{\rm (cur)}\frac{\triangle_r+\triangle_{r'}}{2m^2c^2} \right\}
      \GT^*(p,r) \GT(p,r') \Gamma  \\
  &=\; \int\!\!\frac{\diffn{p}{3}}{(2\pi)^2} \left\{ C_1^{\rm (cur)} 
    + C_2^{\rm (cur)}\frac{\triangle_r+\triangle_{r'}}{2m^2c^2} \right\}
      \GT(r,p) \GT^*(r',p) \Gamma
  \sepperiod
\end{align*}
Thus the momentum distribution is expected to be expressed as follows: 
\begin{align*}
  \frac{1}{\sigma_{\rm pt}} \odiff{\sigma}{p} 
  = \frac{4\pi p^2}{(2\pi)^3} \cdot 
    \frac{3}{2} N_C Q_t^2 \frac{4m^2}{s} \, \frac{4\pi}{m^2c} \Gamma
     \left.
     \left\{ C_1^{\rm (cur)} 
       + C_2^{\rm (cur)} \frac{\triangle_r+\triangle_{r'}}{2m^2c^2} \right\}
     \GT(r,p) \GT^*(r',p) \right|_{r,r'\to r_0}  
  \sepcomma
\end{align*}
With the relation 
\begin{align*}
  \frac{\triangle_r}{m} \GT(r,p) = - \left[ \frac{C_F a_s}{r} + (E+i\Gamma) 
    \right] \GT(r,p) - \frac{\sin(pr)}{pr} + \Order{\frac{1}{c}}
  \sepcomma
\end{align*}
which is just the Schr\"odinger equation with the Coulomb potential, 
the derivatives can be removed: 
\begin{align}
  &
  \left.
    \left\{ C_1^{\rm (cur)} 
      + C_2^{\rm (cur)} \frac{\triangle_r+\triangle_{r'}}{2m^2c^2} \right\}
    \GT(r,p) \GT(r',p)^* \right|_{r,r'\to r_0}  
  \label{eq:curGG}
  \\
  &= 
  \left\{ C_1^{\rm (cur)} - \frac{C_F}{3}\pfrac{a_s}{c}^2\frac{1}{ma_sr_0}
    - \frac{E}{3m^2c^2} \right\} 
  \left| \GT(r_0,p) \right|^2   \nonumber\\
  &\qquad{}
  - \frac{1}{3mc^2} \frac{\sin(pr_0)}{pr_0} \bRe{\GT(r_0,p)}  \nonumber
  \sepperiod
\end{align}
Here the equality is up to $\Order{1/c^3}$.  
%In terms of the reduced Green function $G'$, 
%the unitarity relation is 
%%
%\begin{align*}
%  \Im G &= 
%  \Im G' - \anticommutator{bE+\calO_E}{\Im G'} - \commutator{\calO_O}{\Im G'} 
%  - 2b\Gamma\,\Re G' + \Order{\frac{1}{c^3}}  \\
%  = G^\dagger G \,\Gamma &= \Biggl( 
%  {G'}^\dagger G' - 4bE \,{G'}^\dagger G' 
%  - 2 {G'}^\dagger \calO_E G' 
%  -  \anticommutator{\calO_E}{{G'}^\dagger G'} \\
%  &\qquad\qquad{}
%  - \commutator{\calO_O}{{G'}^\dagger G'} - 2b\,\Re G' 
%  + \Order{\frac{1}{c^3}} \Biggr)\, \Gamma
%  \sepperiod
%\end{align*}
%%
We want to express the unitary relation in terms of 
the reduced Green function $G'$: 
\begin{align*}
  G = G' + \delta G' + \Order{\frac{1}{c^3}}  
  \sepcomma\quad\mbox{ where }\quad
  \delta G' = - \left\{ b\omega+\calO_E \, , \, G' \right\}
  - \left[ \calO_O \, , \, G' \right] - b 
  \sepperiod
\end{align*}
Thus with 
\begin{align*}
  \GT(r,p) = \GT'(r,p) + \delta \GT'(r,p) + \Order{\frac{1}{c^3}}  
  \sepcomma
\end{align*}
where $\GT(r,p) = \bra{r}G\ket{p}$, we have 
\begin{align*}
  \left| \GT(r,p) \right|^2 
  = \left| \GT'(r,p) \right|^2 
  + 2\bRe{\delta\GT'(r,p) \, \GT'(r',p)^*} 
  + \Order{\frac{1}{c^3}}  
  \sepperiod
\end{align*}
Differential operators can be rewritten in terms of c-numbers as we did 
for the total cross section: 
%%
%\begin{align*}
%  \left[ \left( \frac{-1}{m}\frac{1}{r} \odiffn{}{r}{2} r 
%      - \frac{C_F a_s}{r} \right) - (E+i\Gamma) \right] \GT_C(r,p) 
%  = \frac{\sin(pr)}{pr} = 1 + \Order{r^2}
%  \sepcomma\\
%  \GT_C(r,p) = f(p) \left[ 1 -\frac{1}{2}C_Fa_sr \right] + \Order{r^2}
%\end{align*}
%%
%
\begin{align*}
  \frac{1}{ma_sr}\odiff{}{r}r \GT'(r,p) 
  = \left( \frac{1}{ma_sr}-\frac{C_F}{2} \right) \GT'(r,p) 
    + \Order{r,\frac{1}{c}}  
  \sepperiod
\end{align*}
Thus with the definitions 
\begin{align*}
  \GT_E'(r,p) &\equiv \bra{r} G'\frac{1}{ma_sr} \ket{p}  
%  \\&
  = \frac{4\pi}{ma_sp}\int_0^\infty\diffn{r'}{}\sin(pr')\,G'(r,r')
  \sepcomma\\
  \GT_O'(r,p) &\equiv \bra{r} G'\frac{1}{ma_s}ip_r\ket{p}  
%  \\&
  = \frac{-4\pi}{ma_s}\int_0^\infty\diffn{r'}{}\cos(pr')\,r'G'(r,r')
  \sepcomma
\end{align*}
we have, up to $\Order{1/c^3}$, 
\begin{align*}
  \delta\GT'(r,p) 
  &= \left( \frac{E+i\Gamma}{2mc^2} + \frac{3C_Fa_s}{4mc^2r} \right) 
  \GT'(r,p) + \frac{3C_F}{4} \pfrac{a_s}{c}^2 \GT_E'(r,p)  \\
  &\qquad{}
  - \frac{11C_Fa_s}{12mc^2}\frac{1}{r}\odiff{}{r}r\GT'(r,p) 
  - \frac{11C_F}{12}\pfrac{a_s}{c}^2\GT_O'(r,p)
  + \frac{1}{4mc^2}\frac{\sin(pr)}{pr}  \\
  &= \left\{ \frac{E+i\Gamma}{2mc^2} 
    + \frac{3C_F}{4}\pfrac{a_s}{c}^2\frac{1}{ma_sr}
    - \frac{11C_F}{12}\pfrac{a_s}{c}^2
      \left( \frac{1}{ma_sr}-\frac{C_F}{2} \right)
    \right\} \GT'(r,p)  \\
  &\qquad{}
    + \frac{3C_F}{4} \pfrac{a_s}{c}^2 \GT_E'(r,p) 
    - \frac{11C_F}{12}\pfrac{a_s}{c}^2\GT_O'(r,p)
    + \frac{1}{4mc^2}\frac{\sin(pr)}{pr} 
  \sepcomma
\end{align*}
\begin{align*}
  &  2 \bRe{\delta\GT'(r_0,p) \, \GT'(r_0,p)^*} \\
  &= \left\{ \frac{E}{mc^2}  + \pfrac{a_s}{c}^2 \left( 
        - \frac{C_F}{3ma_sr_0} + \frac{11C_F^2}{12} \right) \right\} 
     \left| \GT'(r_0,p) \right|^2   \\
  &\qquad{}
     + \frac{3C_F}{2}\pfrac{a_s}{c}^2 \bRe{\GT_E'(r_0,p) \, \GT'(r_0,p)^*}  
     - \frac{11C_F}{6}\pfrac{a_s}{c}^2 \bRe{\GT_O'(r_0,p) \, \GT'(r_0,p)^*} \\
  &\qquad{}
     + \frac{1}{2mc^2}\frac{\sin(pr_0)}{pr_0} \bRe{\GT'(r_0,p)}
  \sepperiod
\end{align*}
Thus
\begin{align*}
  &  {\rm \EqRef{eq:curGG}} \\
  &= \left\{ 1 + \pfrac{\alpha_s(\mu_h)}{\pi}C_F C_1
    + \pfrac{\alpha_s(\mu_h)}{\pi}^2 C_F C_2(r_0) \right\} \Biggl[ 
  \left( 1 + \frac{2E}{3mc^2} \right) \left| \GT'(r_0,p) \right|^2  \\
  &\qquad{}
   + \frac{3C_F}{2}\pfrac{a_s}{c}^2 \bRe{\GT_E'(r_0,p) \, \GT'(r_0,p)^*}  
   - \frac{11C_F}{6}\pfrac{a_s}{c}^2 \bRe{\GT_O'(r_0,p) \, \GT'(r_0,p)^*} \\
  &\qquad{}
   + \frac{1}{6mc^2}\frac{\sin(pr_0)}{pr_0} \bRe{\GT'(r_0,p)} \Biggr]
\end{align*}
Here we used~\EqRef{eq:C1C2-Ccur}.  
Up to the terms that vanish when $r_0 \to 0$, 
\begin{align*}
  &  \GT'(r_0,p) = 
    \frac{1}{4\pi C_Fa_s\sqrt{1-4\kappa}} \frac{4\pi}{p} \, 
    r_0^{-d_-} \int_0^\infty\diffn{r}{}\sin(pr)\,g_>(r)  \sepcomma\\
  &  \GT_E'(r_0,p) = 
    \frac{1}{4\pi C_Fa_s\sqrt{1-4\kappa}} \frac{4\pi}{ma_sp} \, 
    r_0^{-d_-} \int_0^\infty\diffn{r}{}\frac{\sin(pr)}{r}\,g_>(r)  \sepcomma\\
  &  \GT_O'(r_0,p) = 
    \frac{1}{4\pi C_Fa_s\sqrt{1-4\kappa}} \frac{-4\pi}{ma_s} \, 
    r_0^{-d_-} \int_0^\infty\diffn{r}{}\cos(pr)\,g_>(r)  
\end{align*}
%
%For consistency with our analyses of the total cross section, 
%we expand [EqRef] in terms of the cutoff $r_0$, 
%omit terms regular as $r_0 \to 0$, and set its value 
%as in [EqRef]%
%\footnote{
%  Note that strictly speaking the unitarity relation is violated
%  after this expansion,
%  because $\int dp$ integration and expansion in $r_0$ do not
%  commute for $\Gamma_t > 0$.
%  Practically the unitarity relation holds to a sufficient accuracy by
%  cutting off the momentum integration at some appropriately
%  large scale.
%}. 
%
%
%

%
%
%-----------------------------------------------------------------------
\section{Results for $\diffn{\sigma}{}/\diffn{p}{}$}
To summarize, the momentum distribution of top quark near the threshold 
can be written as follows: 
\begin{align*}
  \odiff{\sigma}{p} = \frac{16\alpha^2}{s^2} \,
  N_c Q_q^2 \, 
  \left\{
    1 
    + \left( \frac{\alpha_s(\mu_h)}{\pi} \right) C_F C_1 
    + \left( \frac{\alpha_s(\mu_h)}{\pi} \right)^2 C_F C_2(r_0) 
  \right\}
  \times p^2 \Gamma_t  \,  f(p,r_0) 
  \sepcomma
\end{align*}
where
\begin{align*}
  f(p,r_0) 
  &= \left\{
    \left( 1 + \frac{2E}{3m_t} \right) | \GT'(p,r_0) |^2 
    + \frac{3}{2} C_F \alpha_s(\mu_s)^2 \,
    \Re \left[
      \GT_{1/r}'(p,r_0) \, \GT'(p,r_0)^*
    \right]
  \right.\\
  &\quad{}\left.
    -\frac{11}{6}C_F \alpha_s(\mu_s)^2  \,
    \Re \left[
      \GT_{ip_r}'(p,r_0) \, \GT'(p,r_0)^*
    \right]
    + \frac{1}{6m_t} \, \frac{\sin (pr_0)}{pr_0} \, 
    \Re \left[
      \GT'(p,r_0) 
    \right] \right\} 
  \sepperiod
\end{align*}
Here $p=|\pBI_t|$.  
We changed a notation of momentum-space Green functions a little bit: 
\begin{align*}
  &
  \tilde{G}(p,r_0) =
  \int\!\!\diffn{r}{3} \, \expo^{i \vec{p}{\cdot}\vec{r}} \, G(r,r_0)
  \sepcomma\\ 
  &
  \tilde{G}_{1/r}(p,r_0) =
  \int\!\!\diffn{r}{3} \, \expo^{i \vec{p}{\cdot}\vec{r}} \, 
  \frac{1}{\alpha_s(\mu) m_t r}\,  G(r,r_0) 
  \sepcomma\\ 
  &
  \tilde{G}_{ip_r}(p,r_0) =
  \int\!\!\diffn{r}{3} \, \expo^{i \vec{p}{\cdot}\vec{r}} \, 
  \frac{i p_r}{\alpha_s(\mu) m_t}  \,  G(r,r_0) 
  \sepperiod
\end{align*}

The unitarity relation %[EqRef] 
holds numerically, 
to the degree we are concerning.  

Our choice for the soft scale is $\mu_s=20\GeV$ since a relevant scale 
around the distribution peak is the scale of Bohr momentum $p_\rmB$.  

\begin{figure}[tbp]
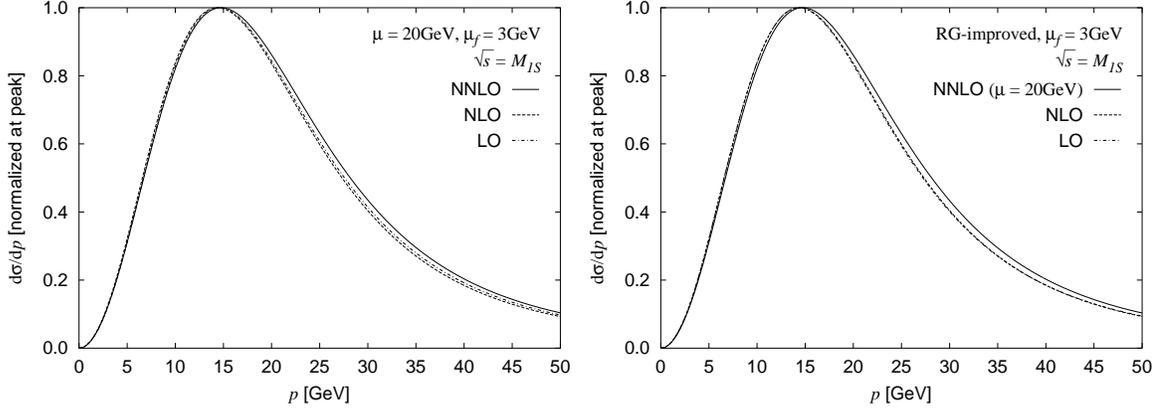

  \hspace*{\fill}
  \begin{minipage}{6cm}
    \includegraphics[angle=-90,width=7.5cm]{./figure/dsdp_fix_20+0.eps}
  \end{minipage}
  \hspace*{\fill}
  \begin{minipage}{6cm}
    \includegraphics[angle=-90,width=7.5cm]{./figure/dsdp_RGE_20+0.eps}
  \end{minipage}
  \hspace*{\fill}
   \\
  \hspace*{\fill}
  \begin{Caption}\caption{\small
      Top quark momentum distributions at LO (dot-dashed), NLO (dashed), 
      and NNLO (solid) for $\mu_s = 20\GeV$ (left) and $\mu_s = q$ (right).  
      For each curve, we set the c.m.\ energy on the $1S$ resonance
      state, $\sqrt{s}=M_{1S}$.  
      \label{fig:dsdp_fix-RG_20+0}
  }\end{Caption}
  \hspace*{\fill}
\end{figure}%
\begin{figure}[tbp]
  \hspace*{\fill}
  \begin{minipage}{6cm}
    \includegraphics[angle=-90,width=7.5cm]{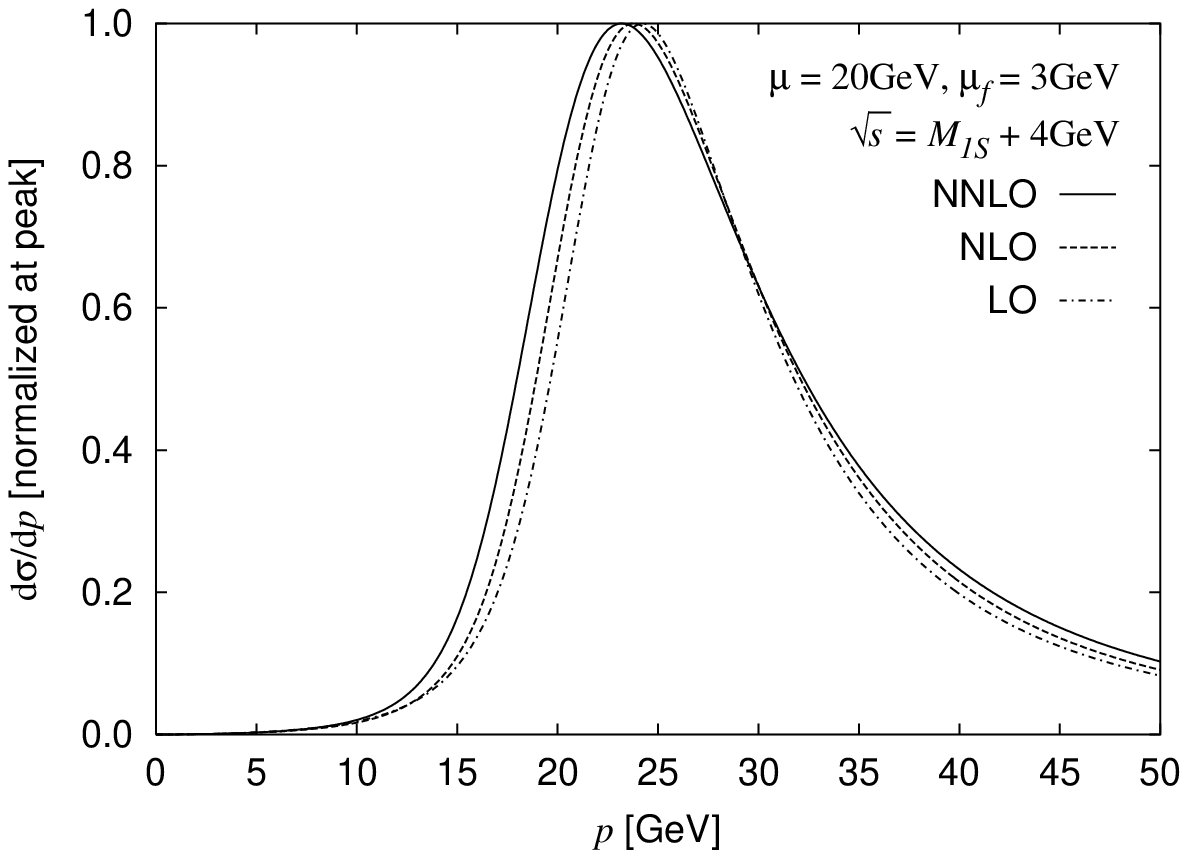}
  \end{minipage}
  \hspace*{\fill}
  \begin{minipage}{6cm}
    \includegraphics[angle=-90,width=7.5cm]{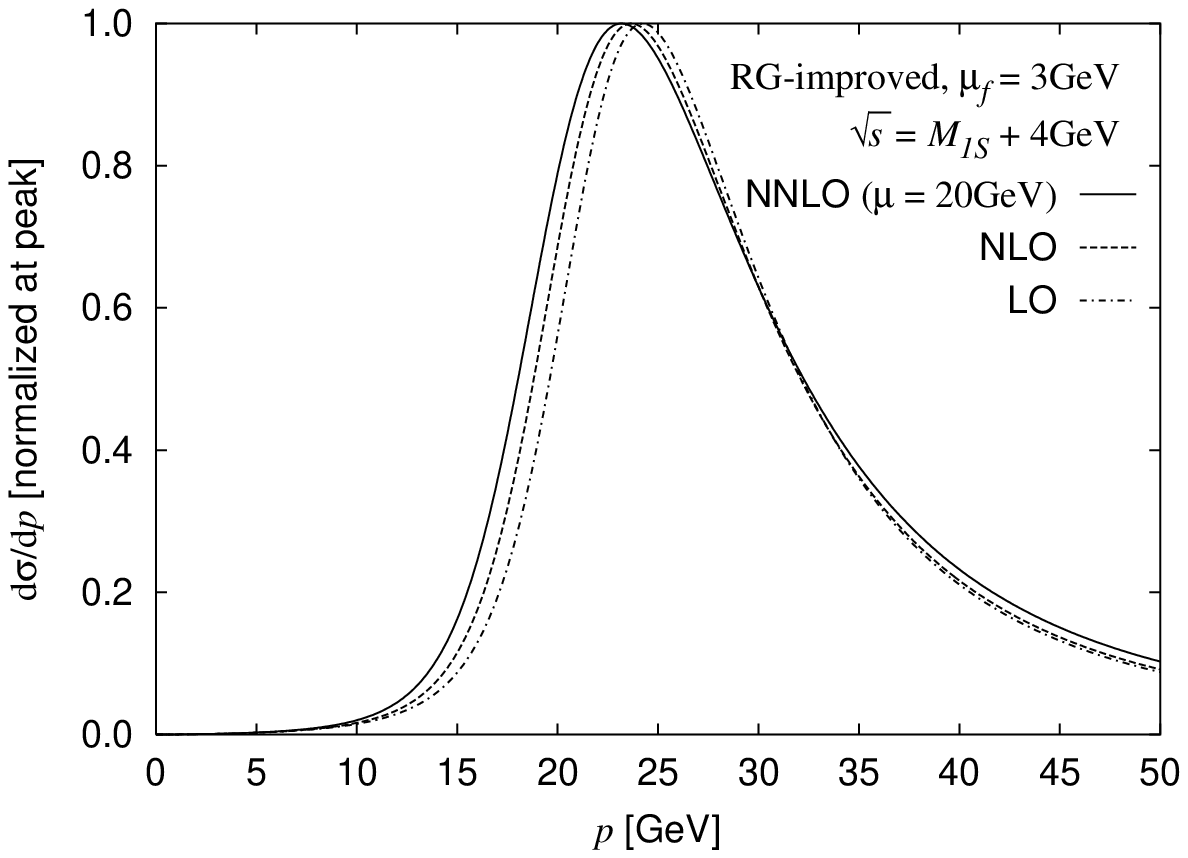}
  \end{minipage}
  \hspace*{\fill}
   \\
  \hspace*{\fill}
  \begin{Caption}\caption{\small
      Top quark momentum distributions at LO (dot-dashed), NLO (dashed), 
      and NNLO (solid) for $\mu_s = 20\GeV$ (left) and $mu_s = q$ (right).  
      For each curve, we set the c.m.\ energy at 4\GeV above the 
      $1S$ resonance mass.  
      \label{fig:dsdp_fix-RG_20+4}
  }\end{Caption}
  \hspace*{\fill}
\end{figure}%
Shown in Figures~\ref{fig:dsdp_fix-RG_20+0} and \ref{fig:dsdp_fix-RG_20+4} 
are the top quark momentum distributions.  
We normalize the distribution to unity at each distribution peak, 
because overall normalization do not have extra information other than 
the total cross section $R$ has.  
This distribution depends on \scCM\ energy.  
The~\FigRef{fig:dsdp_fix-RG_20+0} is on $1S$ resonance 
and~\FigRef{fig:dsdp_fix-RG_20+4} is $4\GeV$ above the $1S$.  
We use the notation $\Delta E = \sqrt{s} - M_{1S}$.  
One can see that corrections are fairly small compared to those 
for total cross section $R$.  
Using RG improved Coulombic potential also do not change qualitative feature.  
Both $\Order{1/c}$ and $\Order{1/c^2}$ corrections are 
similar qualitatively, while quantitatively, the latter is larger.  
The most prominent correction is the shifts of peak momenta $p_{\rm peak}$ 
for $\Delta E = 4\GeV$~\FigRef{fig:dsdp_fix-RG_20+4}.  
However this is mainly due to the shift of $1S$ resonance, 
which is already visible in $R$ ratio.  
That is, for energies $\Delta E > 1\mbox{--}2\GeV$, 
the peak momentum of the distribution 
tends to be determined only by kinematics, 
$p_{\rm peak} \simeq \frac{1}{2} \sqrt{s-4m_t^2}$.  
Note that binding energy becomes deeper with higher order 
correction~\TableRef{table:M1S}.  
Correspondingly, the peak momentum in~\FigRef{fig:dsdp_fix-RG_20+4} 
lower with higher corrections.  
On the other hand, this kind of correction is small 
on the $1S$ resonance ($\Delta E =0$, \FigRef{fig:dsdp_fix-RG_20+0}).  
Quantitatively, 
$\Order{1/c}$ and $\Order{1/c^2}$ corrections shift $p_{\rm peak}$ 
is $\delta p_{\rm peak}/p_{\rm peak} = -0.8\%$ and $+2.5\%$, respectively, 
for fixed order, while $+0.5\%$ and $+2.2\%$ for RG improved.  
Behavior in intermediate energy is as follows.  
At $\Delta E =0$ the corrections are positive $\sim + \, \mbox{few} \, \%$; 
between $\Delta E =0$ and $\Delta E = 1\mbox{--}2\GeV$, the corrections 
decrease and change sign from $+ \, \mbox{few} \, \%$ to 
$- \, \mbox{few} \, \%$; 
at higher energies, $\Delta E > 1\mbox{--}2\GeV$, 
the corrections stay negative, but their 
magnitude $|\delta p_{\rm peak}/p_{\rm peak}|$ 
decrease with energy.  

Aside from the correction mentioned above, 
the next prominent correction is for higher momentum region.  
This is because higher order corrections 
in $1/c$-expansion are less suppressed for higher energy and momentum.  
%-----------------------------------------------------------------------
\subsubsection{Other corrections}
\begin{figure}[tbp]
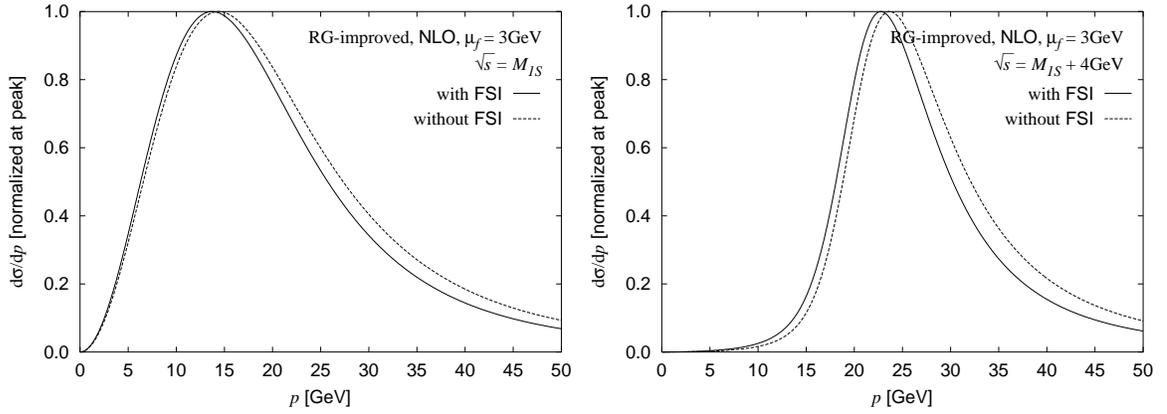

  \hspace*{\fill}
  \begin{minipage}{6cm}
    \includegraphics[angle=-90,width=7.5cm]{./figure/RG+0_1.eps}
  \end{minipage}
  \hspace*{\fill}
  \begin{minipage}{6cm}
    \includegraphics[angle=-90,width=7.5cm]{./figure/RG+4_1.eps}
  \end{minipage}
  \hspace*{\fill}
   \\
  \hspace*{\fill}
  \begin{Caption}\caption{\small
      Top quark momentum distributions at NLO with the renormalization
      group improvement for the Coulomb part of the potential.
      The c.m.\ energy is set on the $1S$ resonance state (left) 
      and 4\GeV\ above it (right).  
      The solid (dashed) line is calculated with (without) the
      $\Order{1/c}$ final-state interaction corrections.
      \label{fig:RG+04_1}
  }\end{Caption}
  \hspace*{\fill}
\end{figure}

Although the momentum distributions above are calculated with 
$\Order{1/c^2}$ corrections, they are not complete.  
This is mainly because we do not treat the decay process of $t$ and 
$\tB$ properly.  
In \SecRef{sec:complications_at_NNLO} 
we mentioned that treating the decay width $\Gamma_t$ with 
the replacement $E \to E+i\Gamma$ is valid for $\Order{1/c}$ and 
below.  In fact, this is only for total cross section $R$.  
Final-State Interactions (FSI) between the decay products of $t$ and $\tB$ 
modify the momentum distribution of $t$ already at $\Order{1/c}$.  
In the analysis above, we dropped this effect.  
The effect of FSI to $\Order{1/c^2}$ is not calculated yet.  
Shown in \FigRef{fig:RG+04_1} are $\Order{1/c}$ momentum distributions 
with and without FSI.  
We can see that the effect of NLO FSI is somewhat larger than 
that of NNLO potential modifications etc.  
Quantitatively, the final-state interactions reduce 
the peak momentum about 5\% almost independent to the energy~\cite{FMS94}.  
Note that the energy dependence of the corrections are different for 
NLO FSI and NNLO potential modifications etc.  
\clearemptydoublepage
%
%
%---------------------------------------------------------------------
\chapter{Anomalous Electric Dipole Moments of Top Quark}
\label{ch:EDMs}
In the early stage of studies on $e^+ e^- \to t\tB$ near the threshold, 
the effects of a Higgs exchange between $t\tB$ were also 
examined~\cite{SP91}.  
It showed that the mass $m_H$ and the coupling $g_{t\bar{t}H}$ of 
the Higgs can be probed, if the Higgs is light.  
Many of extensions of the Standard Model (SM), 
including Supersymmetric SM (SSM), 
contain extra Higgs multiplets.   
They in general introduce complex couplings that are unremovable 
by field redefinitions, or ``physical'', 
thus provide sources of \itCP\ violation other than 
those in CKM matrix.  
Not only in extra Higgs multiplets, but also in other extensions, 
there are plenty of sources for \itCP\ violation in general, 
once we expand the particle content of a model.  
For example in SSM, mass matrices for squarks and sleptons can have 
complex phases.  
These extra sources for \itCP\ violation induce EDM for 
various particles to one-loop level, while the leading SM contribution 
is three-loop.  The magnitude of EDM of a fermion is proportional to 
its mass, in general.  Thus we study the EDMs of top quark in this section.  
Because EDMs are \itCP\ violating, their effect can be seen in the 
difference of the polarization of $t$ and $\tB$, which is zero if 
all the relevant interaction is \itCP\ conserving.  

In~\SecRef{sec:EDMinQFT} 
it is explained that a non-zero value of (intrinsic) EDM violates 
both \itP\ and \itT.  Field theoretic description of EDM is also given.  
In~\SecRef{sec:EDM-pred} 
expected values of EDMs in various models are summarized.  
%---------------------------------------------------------------------
%
%
%

%
%
%
%---------------------------------------------------------------------
\section{Electric Dipole Moment (EDM)}
\label{sec:EDMinQFT}
Total angular momentum is a sum of orbital angular momentum and 
(intrinsic) spin.  
Just the same situation holds for Electric Dipole Moment (EDM) $\muBI_E$ and 
Magnetic Dipole Moment (MDM) $\muBI_B$: 
\begin{align}
  \muBI_E = \muBI_{\rm EDM} + \DBI  \sepcomma\quad
  \muBI_B = \muBI_{\rm MDM} + \MBI  \sepperiod
\end{align}
Here we mean intrinsic dipole moments by $\muBI_{\rm EDM}$ and 
$\muBI_{\rm MDM}$, and ``orbital'' ones by $\DBI$ and $\MBI$.  
As is well known, the transformation properties of electric field 
$\EBI$ etc.\ under, say, parity can be defined so that the classical 
electromagnetism is invariant under any combinations of 
Parity \itP, Charge conjugation \itC, and Time reversal \itT.  
However this is not the case for quantum field theories, 
as we shall see below.  

The EDM $\muBI_{\!E}$ and MDM $\muBI_{\!B}$ are defined by the following 
interactions: 
\begin{align}
  V = - \dprod{\muBI_{\!E}}{\EBI} - \dprod{\muBI_{\!B}}{\BBI}
  \sepperiod
\end{align}
Within the classical electromagnetism, only the ``orbital'' contributions 
($\DBI$,$\MBI$) to the dipole moments can be calculated: 
\begin{align*}
  \DBI(t) = \int\!\!\diffn{x}{3}\, \xBI \rho(\xBI,t)  \sepcomma\quad
  \MBI(t) = \frac{1}{2}\int\!\!\diffn{x}{3}\, \cprod{\xBI}{\jBI}(\xBI,t)  
    \sepcomma
\end{align*}
where
\begin{align*}
  \rho(\xBI,t) = \sum_i q_i \delta^{(3)}\!\bigl(\xBI-\xBI_i(t)\bigr)
    \sepcomma\quad
  \jBI(\xBI,t) = \sum_i q_i \odiff{\xBI_i(t)}{t} 
     \delta^{(3)}\!\bigl(\xBI-\xBI_i(t)\bigr)
    \sepcomma
\end{align*}
are charge density and current density, respectively.  
From these definitions, we can see the transformation properties under 
Parity and Time reversal.  
For example, 
\begin{align*}
  \rho(\xBI,t) \,\transform{\itP}\,
%  &\qquad
  \sum_i q_i \delta^{(3)}\!\bigl(\xBI-(-\xBI_i(t))\bigr)
%  \\
%  &=
%  + \sum_i q_i \delta^{(3)}\bigl((-\xBI)-\xBI_i(t)\bigr)
%  \qquad \text{($\because$ $\delta$-function is an even function)}
%  \\&
  =
  + \rho(-\xBI,t)
  \sepcomma
\end{align*}
because $\delta$-function is an even function, and then 
\begin{align*}
  \DBI(t) \,\transform{\itP}\,
%  &\qquad
  \int\!\!\diffn{x}{3} \cdot (-\xBI)\rho(-\xBI,t)
%  \\
%  &=
%  -\int\!\!\diffn{x'}{3} \xBI' \rho(\xBI',t)
%  \qquad \text{($\xBI \to \xBI' = -\xBI$)}
%  \\&
  =
  - \DBI(t)
  \sepperiod
\end{align*}
Here we changed the integration variable $\xBI \to \xBI' = -\xBI$.  
The transformation properties of other quantities are given 
in~\TableRef{table:PCTforEDM}.  
\begin{table}[tbp]
\centering
\begin{align*}
\begin{array}{c|ccccccccc}
     & \rho & \jBI & \EBI & \BBI & \DBI & \MBI & \sBI & \xBI & \pBI \\
    \hline
    \Parity  & + & - & - & + & - & + & + & - & - \\
    \ChargeC & - & - & - & - & - & - &   &   &   \\
    \TimeR   & + & - & + & - & + & - & - & + & - 
\end{array}
\end{align*}
\begin{Caption}\caption{\small
  Transformation properties under Parity $P$ etc.\ 
  of the quantities related to classical electromagnetism; 
  $\rho$ is charge density, $\jBI$ current density, 
  $\EBI$ electric field, $\BBI$ magnetic field, 
  $\DBI$ ``orbital'' part of electric dipole moment, 
  $\MBI$ ``orbital'' part of magnetic dipole moment.  
  These are fields.  $\sBI$ is spin (or angular momentum, in general), 
  $\xBI$ is coordinate, $\pBI$ is linear momentum.  
  \Parity\ etc.\ mean the eigenvalues under \itP\ etc.  
  For \itP\ and \itT, arguments of fields are changed 
  from $(\xBI,t)$ to $(-\xBI,t)$ and $(\xBI,-t)$, respectively.  
\label{table:PCTforEDM}
}\end{Caption}
\end{table}

Within the framework of QFT, spin $\sBI$ is the only vector 
available for a particle at rest, thus 
\begin{align*}
  \muBI_{\rm EDM} ~,~ \muBI_{\rm MDM} \Parallel \sBI \sepperiod
\end{align*}
Thus $\dprod{\muBI_{\rm MDM}}{\BBI}$ is ($\Parity$,$\TimeR$) = ($+$,$+$) 
while $\dprod{\muBI_{\rm EDM}}{\EBI}$ is ($\Parity$,$\TimeR$) = ($-$,$-$), 
which means non-zero value of (intrinsic) EDM $\muBI_{\rm EDM}$ is allowed 
only when both \itP\ and \itT\ (or \itCP) 
are violated.  

As is well known, when the interaction of fermion-fermion-gauge boson 
has the form of 
\begin{align*}
  \Gamma^\mu
  = 
  \gamma^\mu F_1(q^2) + \frac{i}{2m} \sigma^{\mu\nu} q_\nu F_2(q^2)
  \sepcomma
\end{align*}
MDM is expressed as 
\begin{align*}
  \muBI_{\rm MDM} = 
  \frac{e}{m} \left[ F_1(q^2) + F_2(q^2) \right] \frac{\sigmaBI}{2}
  \stackrel{q^2 \to 0}{=}
  \frac{e}{m} \, \frac{g}{2} \, \frac{\sigmaBI}{2}
  \sepcomma
\end{align*}
where 
$m$ is the mass of the fermion, 
$F_1(q^2)$ and $F_2(q^2)$ are form factors%
\footnote{
  $F_1(q^2=0)$ is the electric charge of the relevant particle.  
}, 
which depend on 
the momentum of the gauge boson $q_{\mu}$ 
(flows into the fermion current), and 
$g$ is Land\'e $g$-factor: $(g-2)/2 = \alpha/(2\pi) + \cdots$.  
Thus anomalous MDM is expressed by the second term of the following 
effective Lagrangian: 
\begin{align*}
  \calL
  =
  - e F_1(q^2) \psiB \gamma^\mu \psi A_\mu
  - \frac{e}{2m} F_2(q^2) \psiB \sigma^{\mu\nu} \psi \partial_\mu A_\nu
  \sepperiod
\end{align*}
From this we can obtain the corresponding term for EDM by replacing 
$\BBI$ by $\EBI$.  This can be done by using dual field strength: 
\begin{align*}
  F_{\mu\nu} = (F_{0i},\frac{-1}{2}\epsilon_{ijk}F_{jk})
  = (\EBI, \BBI)
  \sepcomma\quad
  \frac{1}{2!} \epsilon_{\mu\nu\rho\sigma} F^{\rho\sigma}
  = (-\BBI,\EBI)
  \sepperiod
\end{align*}
Our convention for the totally antisymmetric tensor is 
$\epsilon_{0123}=+1$.  
Thus with 
\begin{align*}
  \sigma^{\mu\nu} \partial_{\mu}A_{\nu}
  = \frac{1}{2} \sigma^{\mu\nu} F_{\mu\nu}
  \,\longrightarrow\,
  \frac{1}{2} \sigma^{\mu\nu} \cdot 
  \frac{1}{2}\epsilon_{\mu\nu\rho\sigma} F^{\rho\sigma}
%  \\
  =
  \frac{i}{2}  \sigma^{\mu\nu}\gamma_5 F_{\mu\nu}
%  \\
  =
  i\sigma^{\mu\nu}\gamma_5 \partial_{\mu}A_{\nu}
  \sepcomma
\end{align*}
we have 
\begin{align*}
  \Gamma^\mu
  = 
  \gamma^\mu F_1(q^2) 
  + \frac{i}{2m} \sigma^{\mu\nu} q_\nu 
    \left[ F_2(q^2) + i\gamma_5 \, d(q^2) \right] 
  \sepcomma
\end{align*}
or the following effective Lagrangian for EDM: 
\begin{align}
  \calL
  &=
  - e F_1(q^2) \psiB \gamma^\mu \psi A_\mu
  - \frac{e d(q^2)}{2m} \psiB i\sigma^{\mu\nu}\gamma_5 \psi \partial_\mu A_\nu
  \nonumber\\
  &\simeq
  - e F_1(q^2) \psiB \gamma^\mu \psi A_\mu
  + \frac{e d(q^2)}{2m} (\psiB i\gamma_5 i\lrpartial^\mu \psi) A_\mu
  \sepcomma
\end{align}
where $d(q^2)$ is a dimensionless form factor for anomalous EDM: 
\begin{align*}
  \muBI_{\rm EDM} = 
  \frac{ed}{m} \frac{\sigmaBI}{2}
  \sepperiod
\end{align*}
The ``$\simeq$'' in the last step means they coincide 
with the use of the eqs.~of motion [\SecRef{sec:Gordon-id}].  
The operator 
$\frac{1}{2} \psiB i\sigma^{\mu\nu}\gamma_5 \psi F_{\mu\nu} \simeq
-(\psiB i\gamma_5 i\lrpartial^\mu \psi) A_\mu$ has 
($\Parity$,$\TimeR$) = ($-$,$-$) 
[Tables \ref{table:PCT-bilinear}, \ref{table:PCT-EMfield}].  
The anomalous EDM interactions with $Z$ and gluon $g$ are defined 
in the same way.  
Since the interactions are flavor-diagonal, its existence immediately means 
\itCP\ violation.  

In fact, as shall be shown in~\SecRef{sec:Gordon-id}, 
the EDM-interaction above, which is dim-5, is the lowest dimension 
operator for $t$-$\tB$-gauge boson that violates \itCP\ invariance.  
Thus \itCP-violating effect is taken into account by the EDM 
to the leading order.  
The form factors $d(q^2)$ depend on the momentum transfer $q^2$ in general.  
However to the lowest order of derivative expansion, it is constant.  
To our knowledge, all studies to date 
concerning sensitivities of collider experiments 
to the EDMs assumed that the momentum dependence of $d(q^2)$ can be 
neglected.  
Also, as we shall see in the following section, 
the $q^2$-dependence is shown to be moderate, by 
explicit calculations of EDMs based on specific models like 
Minimum Supersymmetric Standard Model (MSSM).  
Thus we also follow this assumption.  
The magnitude of EDMs are 
\begin{align}
  t\tB\gamma\text{-EDM} =
  \frac{e\dtp}{\mt}
  &= \dtp \times 1.13 \times 10^{-16} e\cm
  \sepcomma\nonumber\\
  t\tB Z\text{-EDM} =
  \frac{\gZ\dtZ}{\mt}
  &= \dtZ \times 1.13 \times 10^{-16} \gZ\cm
  \label{eq:d_to_EDM}
  \sepcomma\\
  t\tB g\text{-EDM} =
  \frac{\gs\dtg}{\mt}
  &= \dtg \times 1.13 \times 10^{-16} \gs\cm
  \sepperiod\nonumber
\end{align}
Note that $\gZ/e = (g/\cos\thetaW)/(g\sin\thetaW) = 
1/(\cos\thetaW\,\sin\thetaW) = 2.372\cdots$.  

As can easily be checked, 
the term $d\,\psiB i\sigma^{\mu\nu}\gamma_5\psi F_{\mu\nu}$ is hermitian 
only if the EDM-coupling $d$ is real.  
However as explained in~\SecRef{sec:absorp-CPTT}, 
an effective Lagrangian need not be hermitian; 
its anti-hermitian part simply parameterizes an absorptive part $T-T^\dagger$, 
where $T=(S-1)/i$ is a $T$-matrix defined from the $S$-matrix.  
Thus we assume $d$'s are complex.  
%
%
%

%
%
%
%---------------------------------------------------------------------
\section{Predictions of EDMs in various models}
\label{sec:EDM-pred}
In all models with \itCP\ violation, 
EDMs of top quark may not be zero.  
For example in the Standard Model, $t\tB\gamma$-EDM is estimated to be 
$\sim 10^{-30} e\cm$~\cite{D78,CK97}.  
Those for $t\tB Z$- and $t\tB g$-EDM are obtained by 
changing the unit to $\gZ\cm$ and $\gs\cm$, respectively%
\footnote{
  These can be obtained by assuming that the result for 
  $d_{u\bar{u}\gamma}$ in~\cite{CK97} can applied also for 
  $d_{t\bar{t}\gamma}$ etc.  
}.  
Compared with the ``$\Order{1}$-coupling'' result [\EqRef{eq:d_to_EDM}] 
$e/\mt \simeq 10^{-16} e\cm$, 
it is very small.  
This is because in the SM, the lowest contribution arises from three-loop 
diagrams, which is proportional to $G_F^2\alpha_s$~\cite{D78,CK97}.  

Since the operator $\psiB i\sigma^{\mu\nu}\gamma^5\psi$ responsible for EDM 
is chirality flipping, one need to pick up a mass of a fermion at least once.  
Large mass may be advantageous to obtain large EDM, but it may also be a 
suppression factor when the relevant fermion is in a loop.  
Thus in various models, 
EDM of a fermion is typically proportional to its mass.  
This makes the EDM of top quark very interesting.  

There are several models that induce EDM at 1-loop.  
These include models with extra-Higgs multiples and supersymmetry, 
which are natural candidates for beyond the SM.  
See~\cite{EDM-review} for reviews on early studies.  
Here we summarize studies made after these reviews%
\footnote{
  See also~\cite{EDM-review-mini}.  
}.  
The conclusion of this section is as follows: 
the size of EDMs are expected to be 
\begin{align*}
  t\tB g\mbox{-EDM} \sim 10^{-19}\mbox{--}10^{-20} \gs\cm  \sepcomma
\end{align*}
and the same for other EDMs with appropriate changes of the unit, 
if \itCP-violating parameters in the models are $\Order{1}$.  

In a model with two Higgs doublets, it is estimated by 
W.~Bernreuther, T.~Schr\"oder and T.~N.~Pham (1992)~\cite{BSP92} that 
\begin{align*}
  t\tB \gamma\mbox{-EDM} 
  \sim \frac{e G_F m_t^3}{4\pi^2 m_\phi^2} 
  = 3.1\times 10^{-18} e \cm \, \pfrac{m_\phi}{100\GeV}^{-2}
  \sepcomma
\end{align*}
where the extra $m_t^2$ comes from the Higgs-fermion couplings.  
Numerical calculations show 
the real part and the imaginary part of the ElectroWeak-EDMs are 
the similar order each other.  
For $t\tB\gamma$-EDM it is $\sim 2\times 10^{-19} e\cm$ for 
$m_t = 150\GeV$, $m_H = 200\GeV$, $\sqrt{s} = 400\GeV$; and 
$t\tB Z\mbox{-EDM} \sim 0.2\times t\tB \gamma\mbox{-EDM}$.  
They assumed \itCP-violating parameters in the model to be of order unity.  
EDMs induced by neutral Higgs boson exchange are considered also 
in A.~Soni and R.~M.~Xu (1992)~\cite{SX92}.  They estimated 
\begin{align*}
  &  t\tB\gamma \mbox{-EDM} 
     \simeq 4\times 10^{-20} e\cm
  \sepcomma\\
  &  t\tB g \mbox{-EDM} 
     \simeq \frac{\gs}{Q_t e} \times t\tB\gamma \mbox{-EDM} 
     \simeq 6\times 10^{-20} \gs\cm
  \sepcomma
\end{align*}
for $m_H = 2m_t$ and $m_t = 125\GeV$.  
\itCP-violating parameters in the model are assumed to be of order unity.  
They argued that the effects due to charged-Higgs-boson exchange 
may be suppressed by a factor $m_b^2/M^2$, where $M^2$ is a linear 
combination of $m_t^2$ and $m_H^2$.  Thus the charged Higgs boson 
contribution is likely to be subleading to the neutral Higgs boson 
exchange.  
They also calculated the $q^2$ dependence of the EDM form factors, 
which turn out to be moderate.  
The $t\tB\gamma$- and $t\tB Z$-EDM induced by Higgs exchange 
are calculated also in ~\cite{BCK97,CKP93,SP92}.  

In the MSSM, one can introduce (unremovable) complex couplings 
with only one generation~\cite{DGH85}.  
Induced EDMs in MSSM is considered in~\cite{BCGM97}.  
%A.~Bartl, E.~Christova, T.~Gajdosik and W.~Majerotto (1997)
They find that both $t\tB\gamma$- and $t\tB Z$-EDM can reach 
$10^{-20} e\cm$.  
Exchanged particles are gluino, charginos, neutralinos and squarks.  
In most cases gluino $\tilde{g}$ gives the leading contribution.  
However when $m_{\gT} \gtrsim 500\GeV$, 
they find that chargino contribution can be larger than that from 
gluino due to the large Yukawa coupling of top quark.  
They assume the unification of gauge couplings and gaugino masses 
but others.  In particular, they do not assume unification of the 
scalar mass parameters and the trilinear scalar coupling parameter 
$A_q$ of the different generations.  
Their reference set of parameters are such that squark are 
lighter than 400\GeV, and complex phases in squark mixing matrices 
and in higgsino mixing parameter $\mu$ are of order unity.  
They vary one parameter while others are fixed to the reference values.  
The $t\tB\gamma$- and $t\tB Z$-EDMs in Supersymmetric Standard Model (SSM) 
was studied also in~\cite{BO94,prod-decay=SUSY}.  
The $t\bar{b}W$-EDM in SSM was studied in~\cite{decay=SUSY,BO94}.  

There are also other models.  
For example in the model~\cite{HL95} with 
$\SU(3)_C \times \SU(3)_L \times \U(1)_X$, 
contribution of charged and neutral Higgs exchange at one-loop is 
estimated to $10^{-19} e\cm$ and $10^{-19} \gs\cm$ for the values 
of relative phases of the vev's such that \itCP\ violation is maximal.  
%Others include left-right symmetric and composite model.  
%
See also~\cite{W76,W89,BGTW90,CKLY90} for works related to top-quark EDM, 
\cite{EDM-in-SUSY} for EDMs of $b$-quark etc.\ in supersymmetric models, and 
\cite{EDM-in-leptoquark} for EDMs of $b$-quark etc.\ in leptoquark model.  
%
%
%

%
%
%-------------------------------------------------------------------------
\section{CP violating observables}
At the Lagrangian level, 
$\ChargeCOp\ParityOp$-symmetry is violated if the field-dependent phases 
$\eta_{C}\eta_{P}$ in~\EqRef{eq:PCT-Dirac} cannot be tuned so that 
the all term in $\calL$ transform with the same phase%
\footnote{  
  One usually redefines the fields to include $(\eta_{C}\eta_{P})^{-1/2}$.  
  See \SecRef{sec:CPviol}.  
}.  
Since kinetic terms are invariant, or ``even'', under any combinations of 
$\ParityOp$, $\ChargeCOp$, and $\TimeROp$, 
a $\ChargeCOp\ParityOp$-odd interaction is synonymous with 
$\ChargeCOp\ParityOp$-violating one.  
We are concerned with CP-odd observables, which 
needs the interference of 
\itCP-even and \itCP-odd amplitudes: 
$\calM_{\text{\itCP-even}} \calM_{\text{\itCP-odd}}^*$.  
Transformation properties of $\calM$ or $\calL$ are 
discussed in \SecRef{sec:PCT=general}.  
Here we apply it to $|\calM|^2$, or $\diffn{\sigma}{}$.  
As we shall see in \SecRef{sec:absorp-CPTT}, 
observables are classified into \rmCPTT-even and -odd ones.  
The latter can be non-zero if and only if there is a finite absorptive part, 
which is defined as the anti-hermitian part $T-T^\dagger$ of $T$-matrix, 
$S = 1 + iT$.  
Sources for an absorptive part are, for example, decay width and 
final- and initial-state rescattering.  
For the case we are considering, $e^+ e^- \to t\tB$ near the threshold, 
there is plenty of Coulomb rescattering between $t\tB$.  
Thus we can probe both \rmCPTT-even and -odd observables simultaneously.  

A word for notation may be in order.  
$\ParityOp$ is a field-theoretic operator acting on 
creation-annihilation operators [\EqRef{eq:PCT-Dirac}]; 
\itP\ is an operation for $\calM$ defined by $\gamma$-matrices, 
which is c-number [\EqRef{eq:PCT-amp}]; 
\Parity\ is an eigenvalue with respect to $\ParityOp$ or \itP\ 
[Tables~\ref{table:PCTforEDM}, \ref{table:PCT-bilinear}, 
\ref{table:PCT-EMfield}]; 
P is an operation for quantities such as spin $\sBI$ or momentum $\pBI$ 
[\EqRef{eq:CPTTfor_ffB}, \TableRef{table:PCT=E,p,s}].  
However there may be no confusion even if these distinctions are not made, 
aside from Time Reversal.  
%----------------------------------------------------------------------
\subsection{P, C and \rmTT\ for $|\calM|^2$}
\label{sec:PCT=|amp|2}
Basic discrete symmetry operations in QFT are $\ParityOp$, $\ChargeCOp$ and 
$\TimeROp$.  These operate on creation-annihilation operators, 
and are summarized in \SecRef{sec:PCT=op}.  
We can define corresponding operations $P$, $C$ and $T$ 
for (a part of) an amplitude $\calM$.  These are represented not 
by operators of QFT, but by combinations of $\gamma$ matrices.  
These operations are summarized in \SecRef{sec:PCT=amp}.  
We can also define related operations \rmP, \rmC\ and \rmTT\ for 
the each term of $|\calM|^2$, or $\diffn{\sigma}{}$.  
For definiteness, we limit ourselves to the case we are considering: 
$e^-(\peBI,\sBI_{\!e})\,e^+(\peBBI,\sBBI_{\!e}) \to \gamma^*/Z^* \to
 t(\ptBI,\stBI)\,\tB(\ptBBI,\stBBI)$, 
followed by $t \to bW^+$ ($\tB \to \bar{b}W^-$) etc.  
For particle-antiparticle pair, these operations are defined by 
\begin{align}
\centering
\begin{array}{ccll}
  \rmP          &\cdots  &(\pBI,\sBI,\pBBI,\sBBI) \to (-\pBI,\sBI,-\pBBI,\sBBI)
                  &
  \\
  \rmC          &\cdots  &(\pBI,\sBI,\pBBI,\sBBI) \to 
                  (\pBBI,\sBBI,\pBI,\sBI) 
                  &= (-\pBI,\sBBI,-\pBBI,\sBI)
  \\
  \rmTT         &\cdots  &(\pBI,\sBI,\pBBI,\sBBI) \to 
                  (-\pBI,-\sBI,-\pBBI,-\sBBI)
                  &
  \\
  \rmC\rmP      &\cdots  &(\pBI,\sBI,\pBBI,\sBBI) \to 
                  (-\pBBI,\sBBI,-\pBI,\sBI) 
                  &= (\pBI,\sBBI,\pBBI,\sBI)
  \\
  \rmC\rmP\rmTT &\cdots  &(\pBI,\sBI,\pBBI,\sBBI) \to 
                  (\pBBI,-\sBBI,\pBI,-\sBI) 
                  &= (-\pBI,-\sBBI,-\pBBI,-\sBI)
  \sepperiod
\end{array}
\label{eq:CPTTfor_ffB}
\end{align}
The equalities are for the case $\pBI+\pBBI=0$.  
\begin{table}[tbp]
\centering
\begin{align*}
\begin{array}{c|cccccccccc}
  & E-\EB & E+\EB & \pBI-\pBBI & \pBI+\pBBI 
  & \cprod{\pBI}{\pBBI} & \dprod{\pBI}{\pBBI} 
  & \sBI-\sBBI & \sBI+\sBBI 
  & \cprod{\sBI}{\sBBI} & \dprod{\sBI}{\sBBI} 
  \\
  \hline
  \rmC\rmP      & - & + & + & - & - & + & - & + & - & + 
  \\
  \rmC\rmP\rmTT & - & + & - & + & - & + & + & - & - & + 
  \\
  \rmP          & + & + & - & - & + & + & + & + & + & + 
  \\
  \rmC          & - & + & - & + & - & + & - & + & - & + 
  \\
  \rmTT         & + & + & - & - & + & + & - & - & + & + 
\end{array}
\end{align*}
\begin{Caption}\caption{\small
    Transformation properties of the energies, the momenta and 
    the spins of particle-antiparticle system.  
    In their \scCM\ frame, $E-\EB=0$, $E+\EB=2E$, $\pBI-\pBBI = 2\pBI$, 
    $\pBI+\pBBI = 0$, $\cprod{\pBI}{\pBBI} = 0$ and 
    $\dprod{\pBI}{\pBBI} = -|\pBI|^2$.  
%    For other combinations, 
%    $|\pBI|^2 - |\pBBI|^2 = \dprod{(\pBI-\pBBI)}{(\pBI+\pBBI)}/2$ 
%    is $(\rmC\rmP,\rmC\rmP\rmTT) = (-,-)$.  
\label{table:PCT=E,p,s}
}\end{Caption}
\end{table}
Some combinations are given in~\TableRef{table:PCT=E,p,s}, 
whose contents can also be organized as follows: 
\begin{align}
\begin{array}{ccccc}
  (\rmC\rmP,\rmC\rmP\rmTT) = (-,+)
  & \cdots &
  \sBI-\sBBI \sepcomma &
  \pBI+\pBBI \sepcomma & 
  \\
  (\rmC\rmP,\rmC\rmP\rmTT) = (-,-)
  & \cdots & 
  \cprod{\sBI}{\sBBI} \sepcomma & 
  \cprod{\pBI}{\pBBI} \sepcomma & 
  E-\EB               \sepcomma
  \\
  (\rmC\rmP,\rmC\rmP\rmTT) = (+,+)
  & \cdots & 
  \dprod{\sBI}{\sBBI} \sepcomma & 
  \dprod{\pBI}{\pBBI} \sepcomma & 
  E+\EB               \sepcomma
  \\
  (\rmC\rmP,\rmC\rmP\rmTT) = (+,-)
  & \cdots & 
  \sBI+\sBBI \sepcomma & 
  \pBI-\pBBI \sepperiod & 
\end{array}
\label{eq:PCT=E,p,s}
\end{align}
Candidates of the particle-antiparticle pair are 
$(e^-,e^+)$, $(t,\tB)$, $(\ell^-,\ell^+)$ and $(b,\bB)$.  
The last two originate from the decays of $t$ and $\tB$.  

Note that the operation \rmTT\ do not interchange initial and final states.  
The tilde $\tilde{}$ on T emphasis this point.  This operation \rmTT\ 
is sometimes called ``naive T''%
\footnote{In my opinion, ``naive time reversal'' seems to mean 
  ``just flip initial and final states''; just the opposite.  
}.  
It may be difficult, if not impossible, 
to built the field theoretic operator $\tilde{\TimeROp}$ for \rmTT.  
For example, 
the operator with the same properties as $\TimeROp$ but without 
the anti-unitarity nature does not do the job.  
This is because the transformation for creation-annihilation 
operators $\tilde{\TimeROp}a(\pBI,\sBI)\tilde{\TimeROp}^\dagger 
= a(-\pBI,-\sBI)$ etc.\ with 
an unitary operator $\tilde{\TimeROp}$, 
do not lead to the same transformation for $\psi$ and $\psiB$ as $\TimeROp$, 
since $\psi$ (and $\psiB$) contains $u$ and $v$, which are c-numbers.  

With the knowledge of \SecRef{sec:PCT=amp}, 
it is sensible to operate P and/or C to a part of a process 
when all relevant Feynman diagrams are topologically the same.  
In our case, we can apply P etc.\ separately to the final or 
the initial current.  
On the other hand, one should operate \rmTT\ to the whole process.  

Symmetry property of $|\calM|^2$ is directly related to that of 
the expectation values of observables, which we now turn to.  
%----------------------------------------------------------------------
\subsection{Expectation values}
\label{sec:exp-corr}
Since an expectation value is obtained by integration over the phase space, 
not all terms in $|calM|^2$ contribute to it.  
The transformation properties under Parity etc.\ can be used to determine 
a relevant combination of interactions.  

Consider for example, an expectation value 
\begin{align*} 
  \mean{\dprod{(\peBI-\peBBI)}{(\ptBI-\ptBBI)}}
  = 4 \mean{\dprod{\peBI}{\ptBI}}
  = \frac{4}{\sigma_{\rm tot}} \int\!\diffn{\Phi}{}\, \dprod{\peBI}{\ptBI}\,
  \frac{\diffn{\sigma}{}(\peBI,\ptBI)}{\diffn{\Phi}{}}
  \sepcomma
\end{align*}
which is P-odd with respect to both the final and the initial currents.  
Thus only a part of $\diffn{\sigma}{}$ contributes to the expectation 
value; that is, which have the same transformation property to 
the relevant observable.  
There are four combinations for the SM vertices: 
\begin{align*}
  &
  [v^e v^t]^*\cdot[a^e a^t] \times
  \bigl((\vB \gamma_\mu u)_e (\uB \gamma^\mu v)_t\bigr)^* \cdot
  \bigl((\vB \gamma_\nu\gamma_5 u)_e (\uB \gamma^\nu\gamma_5 v)_t\bigr)
  +\text{c.c.}
  \\
  {}+{}&
  [v^e a^t]^*\cdot[a^e v^t] \times
  \bigl((\vB \gamma_\nu u)_e (\uB \gamma^\nu\gamma_5 v)_t\bigr)^* \cdot
  \bigl((\vB \gamma_\mu\gamma_5 u)_e (\uB \gamma^\mu v)_t\bigr)
  +\text{c.c.}
\end{align*}
Here $[v^e v^t]$ etc.\ are (energy dependent) couplings 
that combine both $\gamma$ and $Z$ contributions.  
These are defined in~\EqRef{eq:eff-coupling}.  
Note that $\Parity$ of both initial and final current should be opposite.  
Thus only the combination 
$a_3 \equiv [v^e v^t]^*\,[a^e a^t] + [a^e v^t]^*\,[v^e a^t]$ 
contributes to the forward-backward asymmetry (after one sum over spins)%
\footnote{
  Before summing the spin $\stBI$ of $t$, 
  $\pdprod{\stBI}{\peBI}\pdprod{\peBI}{\ptBI} \propto \cos\theta_{te}$
  contributes to ``spin-direction-specified FB asymmetry''.  
  This observable is P-odd w.r.t.\ the final $t\tB$ current 
  but P-even w.r.t.\ the initial $e^- e^+$ current.  
  Thus the combination 
  $a_4 \equiv [v^e v^t]^*\,[v^e a^t] + [a^e v^t]^*\,[a^e a^t]$ 
  also contributes to this asymmetry.  
  Note that non-zero polarization of initial $e^+ e^-$ effectively 
  mixes the vector vertex and the axial-vector vertex of $e^+ e^-$.  
}.  
The effective couplings $[v^e v^t]$ etc.\ are defined in 
\SecRef{sec:merge_Z_and_photon}, and the combinations $a_3$ etc.\ are 
defined in \EqRef{eq:def-a1}.  

Another example is the total cross section $\sigma_{\rm tot}$.  
After summing over spins, momenta are the only vectors.  
Thus each term in $|\calM|^2$ should be composed of even number of 
momenta.  
From~\EqRef{eq:oppositeP=no-total}, we know that the product of 
the Parity of the final-currents should be even, 
which means even number of momenta of final state.  
Thus the number of momenta of initial state should also be even, 
which means the product of the Parity eigenvalues of the initial-current 
should be even.  To summarize, 
an interference of amplitudes $\calM_1\calM_2^*$ contributes 
to the total cross section if and only if 
the Parities of the final-currents in $\calM_1$ and $\calM_2$ are the same, 
and so are initial-currents.  

CP transformation interchanges the spins of a particle and its anti-particle 
$(\stBI,\stBBI) \to (\stBBI,\stBI)$ at the \scCM\ frame of them.  
Thus the interference 
($\calM_{\ChargeC\Parity=\pm}\calM_{\ChargeC\Parity=\pm}^{~*}$)
of the amplitudes with the same \ChargeC\Parity\ 
contributes in the same sign to the polarizations%
\footnote{See~\SecRef{sec:pol=concept} for polarization.} 
of a particle and 
its anti-particle.  
On the other hand, the interference 
($\calM_{\ChargeC\Parity=\pm}\calM_{\ChargeC\Parity=\mp}^{~*}$)
of the amplitudes with the opposite \ChargeC\Parity\ 
contributes in the opposite sign to the polarizations of a particle and 
its anti-particle.  
We see the example latter.  
%-------------------------------------------------------------------------
\subsection{Several CP violating observables}
\label{sec:sev-CPv-obs}
From~\TableRef{table:PCT=E,p,s} or~\EqRef{eq:PCT=E,p,s}, 
we can see that 
$\dprod{(\stBI\mp\stBBI)}{\pcprod{\peBI}{\ptBI}}$
is $(\rmC\rmP,\rmC\rmP\rmTT) = (\mp,\pm)$ and 
$\dprod{(\stBI\mp\stBBI)}{(\ptBI-\peBI\pdprod{\peBI}{\ptBI}/|\peBI|^2)}$
is $(\rmC\rmP,\rmC\rmP\rmTT) = (\mp,\mp)$.  
Here the CP operation can be limited to the final current.  
Within the Standard Model, where \itCP-odd contributions to 
both $t\tB$ and $e^+ e^-$ vertices are highly suppressed, 
we see that 
\begin{gather*}
    \mean{\dprod{\stBI}{\pcprod{\peBI}{\ptBI}}}
  - \mean{\dprod{\stBBI}{\pcprod{\peBI}{\ptBI}}}
  = 0
  \sepcomma\\
    \mean{\dprod{\stBI}{(\ptBI-\peBI\pdprod{\peBI}{\ptBI}/|\peBI|^2)}}
  - \mean{\dprod{\stBBI}{(\ptBI-\peBI\pdprod{\peBI}{\ptBI}/|\peBI|^2)}}
  = 0
  \sepcomma
\end{gather*}
because they are CP-odd combinations.  
The former is suppressed in the open top region, since 
$\dprod{(\stBI+\stBBI)}{\pcprod{\peBI}{\ptBI}}$ is 
$\rmC\rmP\rmTT$-odd, which means the expectation value is 
proportional to the absorptive part of the amplitude, 
which come from EW and QCD radiative corrections.  
It is known~\cite{TopPol} that the polarization 
$\Polnorm$ ($\sim \mean{\dprod{\stBI}{\pcprod{\peBI}{\ptBI}}}$)
of top quark 
normal to the $e^+e^-$--$t\tB$ plane is $\Order{10^{-2}}$ for the SM.  
However for the SM, the difference $\Polnorm - \PolBnorm$ is 
far smaller than the current experimental reach, since it is CP-odd.  

The following are some of CP-odd combinations of observables%
\footnote{
  We follow the notations of~\cite{HJKP97,BN89,BNOS92}.  
}
(at the \scCM\ frame of $e^+ e^-$): 
\begin{align}
      \spara - \sBpara &\propto 
      \dprod{(\stBI-\stBBI)}{\peBI}
  &&  \rmC\rmP\rmTT = - 
  \nonumber\\
      \sperp - \sBperp &\propto 
      \dprod{(\stBI-\stBBI)}{(\ptBI-\peBI\pdprod{\peBI}{\ptBI}/|\peBI|^2)}
  &&  \rmC\rmP\rmTT = - 
  \nonumber\\
      \snorm - \sBnorm &\propto 
      \dprod{(\stBI-\stBBI)}{\pcprod{\peBI}{\ptBI}}
  &&  \rmC\rmP\rmTT = + 
  \nonumber\\
      A_E^f &= E_f - E_\fB \qquad (f = \ell^+, b)
  &&  \rmC\rmP\rmTT = - 
  \nonumber\\
      A_1^f &= \dprod{\pBIH_{\!e}}{\pcprod{\pBI_{\!f}}{\pBI_{\!\fB}}}
  &&  \rmC\rmP\rmTT = + 
  \label{eq:CPo-obs}
  \\
      A_2^f &= \dprod{\pBIH_{\!e}}{(\pBI_{\!f}+\pBI_{\!\fB})}
  &&  \rmC\rmP\rmTT = - 
  \nonumber\\
      T_{33}^f &= 2(\pBI_{\!f}-\pBI_{\!\fB})_3\,
        \pcprod{\pBI_{\!f}}{\pBI_{\!\fB}}_3
  &&  \rmC\rmP\rmTT = + 
  \nonumber\\
      Q_{33}^f &= 2(\pBI_{\!f}+\pBI_{\!\fB})_3\,
        (\pBI_{\!f}-\pBI_{\!\fB})_3
        -\tfrac{2}{3}(|\pBI_{\!f}|^2-|\pBI_{\!\fB}|^2)
  &&  \rmC\rmP\rmTT = - 
  \nonumber
\end{align}
Note that $\pBI_{\!e} = (\pBI_{\!e}-\pBBI_{\!e})/2$ and 
$\pBI_{\!t} = (\pBI_{\!t}-\pBBI_{\!t})/2$.  
Also, kinematical cuts should respect CP-invariance.  

Expectation values of CP-odd observables are not 
the only CP-violating observables.  
Event rate asymmetries also do the job.  However, these two kinds of 
observables may be related closely.  
For example, Forward-Backward asymmetry $A_{\rm FB}$ of top quark 
is similar to the correlation 
$\mean{\dprod{\peBI}{\ptBI}} \propto \mean{\cos\theta_{te}}$.  
More precisely, 
$A_{\rm FB} = \mean{\sgn(\dprod{\peBI}{\ptBI})}/\sigma_{\rm tot}$.  
Several kinds of event rate asymmetries are considered in literatures.  
One is the difference of production cross sections for $t_L \tB_L$ and 
$t_R \tB_R$~\cite{SP92,CKP93}%
\footnote{
  References in this section are not meant to be extensive but only examples.  
}
\begin{align*}
  \sigma(t_L \tB_L) - \sigma(t_R \tB_R)
  \qquad\qquad\qquad \rmC\rmP\rmTT = - 
  \sepcomma
\end{align*}
where $\tB_L = \overline{t_R}$.  
These are chirality-flipping, and thus are small for relativistic 
top quarks.  
Note that from~\EqRef{eq:CPTTfor_ffB}, 
$\sigma(t_L \tB_L) \leftrightarrow \sigma(t_R \tB_R)$ under \rmC\rmP\rmTT, 
while the other two chirality-non-flipping's, 
$\sigma(t_L \tB_R)$ and $\sigma(t_R \tB_L)$, 
are self-conjugate under \rmC\rmP.  
As we shall see in~\SecRef{sec:pol_t=p_ell}, the charged leptons $\ell^+$ 
produced in $t$ (or more directly, $W^+$) decays are 
emitted preferentially in the direction of 
top-quark spin.  Thus $\ell^+$ produced in $t_L$ decay is less 
energetic than that in $t_R$ decay.  
Thus the event asymmetry above is similar to $A_E^\ell$ defined above, 
which is also \rmC\rmP\rmTT-odd.  
Another event rate asymmetry is 
up-down asymmetry $A_{\text{up-down}}$~\cite{CKP93}, 
which is the rate difference between the events with 
$\ell^\pm$ above the reaction plane and the events 
with $\ell^\pm$ below the $e^+e^-$--$t\tB$ plane: 
\begin{align*}
  A_{\text{up-down}}(f)
  =
  \frac{N_{\rm up}(f^\pm)-N_{\rm down}(f^\pm)}
       {N_{\rm up}(f^\pm)+N_{\rm down}(f^\pm)}
  \qquad\qquad\qquad \rmC\rmP\rmTT = + 
  \sepcomma
\end{align*}
where $f = \ell^+, b$, and 
$N_{\rm up}(f^\pm) = N_{\rm up}(f) + N_{\rm up}(\fB)$ 
etc.  
This quantity is ``similar'' to the correlation 
$\mean{\dprod{(\pBI_{\ell^+}+\pBI_{\ell^-})}{\pcprod{\peBI}{\ptBI}}}$, 
which is \CPTT-even.  
Yet another one is the particle-antiparticle asymmetry~\cite{PR95,BCGM98}: 
\begin{align*}
  N(f) - N(\fB)
  \qquad\qquad\qquad \rmC\rmP\rmTT = - 
  \sepcomma
\end{align*}
where $f = \ell^+, b$.  
Since both momentum and spin are summed%
\footnote{
  In this case, both CP and \rmCPTT\ just interchange 
  particle and antiparticle.  
}%
, CP-odd means \rmC\rmP\rmTT-odd.  
Still another one is~\cite{PR95}
\begin{align*}
  A_{\rm FB}(f) + A_{\rm FB}(\fB) 
  \,\propto\,
  (N_{\rm F}(f) - N_{\rm B}(\fB)) + (N_{\rm F}(\fB) - N_{\rm B}(f))
  \qquad\qquad \rmC\rmP\rmTT = + 
  \sepcomma
\end{align*}
where $f = \ell^+, b$.  
This quantity is $(\rmC\rmP,\rmC\rmP\rmTT) = (-,+)$, because 
$\pBI \leftrightarrow -\pBBI$ for \rmC\rmP\ and 
$\pBI \leftrightarrow +\pBBI$ for \rmC\rmP\rmTT.  

For $pp$ collisions, 
there may be non-zero CP-odd asymmetry 
that is not related to CP-violating interactions, 
since the initial state is not a CP eigenstate.  
However such an effect is estimated to be small~\cite{SP92,AAS92}.  
%
%
%

%
%
%-------------------------------------------------------------------------
\section{Studies of EDMs at open top region}
\label{sec:EDMstudy_open}
There has been many studies on EDMs of top quark.  
These exploit ``open'' $t$'s; 
that is, relativistic $t$ and $\tB$: $\beta\simeq 1$.  
Although our concern is at the threshold region $\beta \lesssim 0.1$, 
we summarize results obtained from the analyses in open top region, 
for comparison.  
%---------------------------------------------------------------------
\subsection{Present bounds for EDMs}
There are several EDMs for top-quark; 
$t\tB\gamma$, $t\tB Z$, $t\tB g$, and $t\bar{b}W$.  
Experimental bounds on these EDMs at present are summarized here.  
Conclusion is that ``$\Order{1}$'' couplings are allowed: 
\begin{align*}
  t\tB g\mbox{-EDM} \lesssim 10^{-16} \gs\cm
  \qquad\text{etc.}
\end{align*}

P.~Haberl, O.~Nachtmann and A.~Wilch (1996)~\cite{HNW96} 
studied the present bound on $t\tB g$-EDM and -MDM 
from the $p\pB \to t\tB X$ total production rate 
at Tevatron $\sqrt{s} = 1.8\TeV$.  
They found the coupling of order unity is allowed, 
which means $t\tB g\mbox{-EDM} = 10^{-16}\gs\cm$ is allowed, 
and so is for MDM.  

K.~Cheung and D.~Silverman (1997)~\cite{CS97} 
also studied the present bounds on $q\qB g$-EDM and -MDM, 
and are both calculated to be $0.89\times 10^{-16}\gs\cm$.  
They used the transverse momentum distribution of prompt photon produced in 
$qg \to q\gamma$, $\qB g \to \qB\gamma$, $q\qB \to g\gamma$ 
at Tevatron $\sqrt{s} = 1.8\TeV$, $0.1\fb^{-1}$ (Run I).  

EW-EDMs are constrained indirectly from $B \to X_s\gamma$~\cite{HTT96}; 
it shows that $\Order(1)$ coupling is allowed.  

A.~De~R\'ujula, M.~B.~Gavela, O.~P\`ene and F.~J.~Vegas (1991)~\cite{DRGPV91} 
studied bounds on EDMs of quarks from neutron $\gamma$-EDM: 
$|t\tB\gamma\mbox{-EDM}| < 2.0\times 10^{-17}e\cm$, 
$|t\tB Z\mbox{-EDM}| < 8.7\times 10^{-18}\gZ\cm$, and 
$|t\tB g\mbox{-EDM}| < 5.8\times 10^{-20}\gs\cm$.  
They used $\mt = 100\GeV$; this may change the limit above.  
As they themselves pointed out, 
it is doubtful whether their calculation is applicable to top quark.  
Thus we don't take these limits seriously.  

In~\cite{DV96}, {\it CP} conserving anomalous EW couplings of top quark 
is constrained from LEP/SLC data.  

For $\tau$, present bound from LEP is obtained in~\cite{EDM-bound-no-top}.  
%---------------------------------------------------------------------
\subsection{Expected future bounds on EDMs (Open top region)}
There are already several studies on the top quark EDMs.  
Some are for $e^+ e^-$ colliders, others are for hadron colliders.  
We choose two representative studies for each of these, 
and report the results of their analyses.  
In $e^+ e^-$ collisions, $t\tB\gamma$- and $t\tB Z$-EDM can be 
studied in the production process, 
while in hadron collisions, $t\tB g$-EDM can be studied.  
The summary of this section is, 
\begin{itemize}
  \item 
    Bounds of $10^{-18}e\cm$ on $t\tB\gamma$- and $t\tB Z$-EDM 
    are possible at a $e^+e^-$ colliders at $\sqrt{s} = 500\GeV$.  
    With the use of ``optimal observable'', this can be improved by 
    factor 2.  
    However this observable may have larger systematic error than 
    simpler observables.  
  \item
    Bound of $10^{-18}\mbox{--}10^{-19}\gs\cm$ on $t\tB g$-EDM 
    is possible for hadron collisions at $\sqrt{s} = 14\TeV$.  
\end{itemize}
%
%---------------------------------------------------------------------
\subsubsection{(i) Anomalous ElectroWeak vertices in 
$e^+ e^-$ or $\gamma\gamma$ collisions}

(i--0) Detector simulation \cite{F96} was performed with 
$\int\!\calL = 50\fb^{-1}$, $\mt=180\GeV$, $\sqrt{s}=500\GeV$, 
polarization of $e^-$ beam $=\pm 80\%$.  
They used the decay chain $t\tB \to b\bar{b}WW \to b\bar{b}q\bar{q}'\ell\nu$, 
where $\ell = e,\mu$.  The relevant branching fraction is $8/27$.  
%Efficiency of the event reconstruction $\simeq 70\%$.  
They obtained overall efficiency of the analysis, 
including branching fractions, reconstruction efficiency, and 
acceptance, $\simeq 18\%$.  
Their results is that the sensitivity for EW-EDM is $\Order{0.1}$ 
with $90\%$ CL.  

%(i--2) \cite{LY94}.  
%Dilepton mode: $e^+e^- \to t\bar{t} \to b\bar{b}\ell^-\ell^+\nu\bar{\nu}$.  
%Branching fraction $=2(1/9)(1/9)=2/81$ for $\ell = e,\mu$.  $\mt=140\GeV$.  
%Use $(LL-RR)/(LL+RR)$, where $LL$ denotes the production rate 
%of $t_L\bar{t}_L$.  Efficiency of finding $LL$ or $RR$ mode 
%$=\epsilon \simeq 0.7 \cdot (1/4) \simeq 0.18$, where $0.7$ for 
%the reconstruction of kinematics, and $1/4$ for the acceptance of 
%the kinematic cuts in selecting $LL$ and $RR$ events.  
%Sensitivity is $\Order{0.1}$.  

(i--1) D.~Atwood and A.~Soni (1992)~\cite{AS92}.  
They studied the sensitivities for $t\tB\gamma$- and $t\tB Z$-EDM, 
(and -MDM) of $e^+ e^-\to t\tB$ with $\sqrt{s}=0.5\TeV$. 
No specific model is assumed.  
Assuming that the top decay vertex $t\bar{b}W$ is the SM one, 
they claimed $1\sigma$-bounds are $\sim 10^{-18}e\cm$ for $10^4$ events.  
They introduce the notion of an ``optimal observable'' 
that have the maximum sensitivity to the relevant coupling.  
This observable turn out to be powerful as we shall see.  
However in general, this observable is rather complicated combination of 
momenta (and spins, maybe).  
Thus as themselves noted~\cite{AAS92}, 
there may be larger systematic error than that for simple observables.  
The events they use are that with both $W$ decay leptonically: 
$b\ell^+ \nu_\ell \bB\ell^-\nuB_\ell$.  
In this case, one can measure the polarization $E_W^\mu$ of $W$, 
statistically by the distribution of $\ell^\pm$: 
\begin{align*}
  &  E_W^\mu = \frac{\btr{\pS_{\!\nu}\SlashIt{\omega}_{\!0}\pS_{\!\ell}
      \gamma^\mu(1-\gamma_5)}}
  {4\sqrt{\dprod{p_\nu}{\omega_0}\,\dprod{p_\ell}{\omega_0}}}
  \sepcomma
\end{align*}
where its ``gauge'' is fixed by an arbitrary lightlike vector $\omega_0$.  
As summarized in~\TableRef{table:AS92-sensitivity}, 
the use of $E_W^\mu$ improves the sensitivity about $\Order{10^2}$.  
They propose several observables that are expectation values of 
certain operators.  
Explicit formulas of the operators for optimal observables are quite lengthy.  
They also studied simple operators.  
Those with best sensitivities are 
\begin{align*}
  &  \epsilon_{\mu\nu\rho\sigma} P_b^\mu Q_z^\nu H^{+\rho} H^{-\sigma}
  &&\text{for Re[$t\tB\gamma$-EDM]}
  \sepcomma\\
  &  \epsilon_{\mu\nu\rho\sigma} P_e^\mu Q_z^\nu H^{+\rho} H^{-\sigma}
  &&\text{for Re[$t\tB Z$-EDM]}
  \sepcomma\\
  &  \dprod{H^{-}}{Q_z}
  &&\text{for both Im[$t\tB\gamma$-EDM] and Im[$t\tB Z$-EDM]}
  \sepcomma
\end{align*}
where
$P_b = p_{\bB} - p_{b}$, $P_e = p_{e^+} - p_{e^-}$, 
$Q_z = p_{e^+} + p_{e^-}$, 
$H^{\pm} = 2 \pdprod{E_{W^+}}{p_t} E_{W^+} 
  \pm 2 \pdprod{E_{W^-}}{p_t} E_{W^-}$.  
\begin{table}[tbp]
\centering
\begin{align*}
\begin{array}{c|cccccccc}
  & \bRe{t\tB\gamma\mbox{-EDM}} & \bIm{t\tB\gamma\mbox{-EDM}} 
  & \bRe{t\tB Z\mbox{-EDM}} & \bIm{t\tB Z\mbox{-EDM}}  \\
  \hline
  \mbox{without $E_W$, with optimal} 
  & 4\times 10^{-17} & 1\times 10^{-17} 
  & 1\times 10^{-17} & 4\times 10^{-17} \\
  \mbox{with $E_W$, with simple} 
  & 4\times 10^{-18} & 1\times 10^{-18} 
  & 4\times 10^{-18} & 4\times 10^{-18} \\
  \mbox{with $E_W$, with optimal} 
  & 4\times 10^{-19} & 4\times 10^{-19} 
  & 4\times 10^{-19} & 4\times 10^{-19} 
\end{array}
\end{align*}
\begin{Caption}\caption{\small
    D.~Atwood and A.~Soni (1992)~\cite{AS92}.  
    $1\sigma$ sensitivities in units of $e\cm$.  
    ``With $E_W$'' means ``to use polarization of $W$''.  
    Simple observables are explained in the text.  
    See the original paper for optimal observables.  
\label{table:AS92-sensitivity}
}\end{Caption}
\end{table}

(i--2) M.~S.~Baek, S.~Y.~Choi and C.~S.~Kim (1997)~\cite{BCK97}.  
They studied the sensitivities for $t\tB\gamma$-EDM and $t\tB Z$-EDM 
of $e^+ e^-$ collisions and $\gamma\gamma$ collisions at NLC with 
$\sqrt{s} = 0.5\TeV$, $\int\!\calL_{ee} = 10\fb^{-1}$ for polarized electrons 
and the twice for unpolarized electrons.  
No specific model is assumed.  
The real part of EDMs are bounded by \CPTT-even observables 
(when there is no absorptive part).  
With polarized electrons, the tightest bound is 
obtained through $A_1^b$, defined in~\EqRef{eq:CPo-obs}, 
in the inclusive top-quark decay mode: 
\begin{align*}
  &  |\Re(t\tB\gamma\mbox{-EDM})| \leq 1.4\times10^{-17}e \cm
  \sepcomma\quad
  |\Re(t\tB Z\mbox{-EDM})| \leq 2.3\times10^{-17}e \cm
  \sepperiod
\end{align*}
Likewise, the imaginary part is bounded by \CPTT-odd observables. 
With polarized electrons, the tightest bound is 
obtained through $A_E^b$, defined in~\EqRef{eq:CPo-obs}, 
in the inclusive top-quark decay mode: 
\begin{align*}
  &  |\Im(t\tB\gamma\mbox{-EDM})| \leq 1.8\times10^{-17}e \cm
  \sepcomma\quad
  |\Im(t\tB Z\mbox{-EDM})| \leq 3.1\times10^{-17}e \cm
  \sepperiod
\end{align*}
For polarized $\gamma\gamma$ collisions, 
$|\Re(t\tB\gamma\mbox{-EDM})| \leq 1.8\times10^{-17}e \cm$ 
for $\sqrt{s} = 0.5\TeV$, 
and $0.2\times10^{-17}e \cm$ for $1.0\TeV$.  
They argued that the sensitivity is much sensitive to $\sqrt{s}$ than 
$e^+ e^-$ collisions.  

(i--3) 
These couplings are also studied in~\cite{LY94,CR95,PR95}.  
In~\cite{CKP93,EDM-prod-epem=2HDM}, $t\tB\gamma$-EDM and $t\tB Z$-EDM 
are consider within two Higgs doublets model.  
QCD corrections to this process well above the threshold are 
considered in~\cite{S96}.  
Studies in $\gamma\gamma$ colliders are given in~\cite{BBO96,gamma-gamma}; 
sensitivity is $\sim 10^{-17}e \cm$ 
with $\int\!\calL=10\fb^{-1}$ and $\sqrt{s}=500\GeV$.  
%---------------------------------------------------------------------
\subsubsection{(ii) Anomalous Chromomagnetic vertex in hadron collisions}
(ii--1) D.~Atwood, A.~Aeppli and A.~Soni (1992)~\cite{AAS92}.  
It was studied that the sensitivities of $gg\to t\tB$ at SSC and LHC 
for the real part of $t\tB g$-EDM.  
$1\sigma$ sensitivity is $0.3\times 10^{-19}\gs\cm$ with 
optimal observable $f_{\rm opt}$, 
$0.5\times 10^{-19}\gs\cm$ with $f_{1}$, and 
$0.6\times 10^{-19}\gs\cm$ with $f_{2}$.  
We do not reproduce their explicit form.  
Assumed experimental setup is mainly, SSC, $\sqrt{s}=40\TeV$, 
$10\fb^{-1}$, $10^7$ leptonic $t\tB$ events.  
However, they say that the difference is only a few percent by 
changing SSC $\sqrt{s} = 40\TeV$ to LHC $\sqrt{s} = 16\TeV$.  
Correlations between the $t$ and $\tB$ polarizations are used: 
$f_{\rm opt}$, $f_1$, $f_2$, $f_3$. 
The $f_{\rm opt}$ and $f_1$ are tailored to maximize the sensitivity for 
Re($t\tB g$-EDM), 
and are fairly complicated combination of momenta and spins.  
As they pointed out, there may be larger systematic error than 
simpler observables such as $f_2$ and $f_3$: 
\begin{align*} 
  f_2 = \frac{\epsilon_{\mu\nu\rho\sigma}p_{e1}^\mu p_{e2}^\nu
    p_{b1}^\rho p_{b2}^\sigma}
  {(\dprod{p_{e1}}{p_{e2}}\,\dprod{p_{b1}}{p_{b2}})^{1/2}}
  \sepperiod
\end{align*}
$f_3$ turn out to be less sensitive than $f_2$ by $\Order{10}$.  

(ii--2) S.~Y.~Choi, C.~S.~Kim and Jake Lee (1997)~\cite{CKL97}.  
The sensitivity of $gg\to t\tB$ (LHC) for $t\tB g$-EDM was studied.  
Assumed setup is acceptance efficiency $\epsilon = 10\%$, 
$B_\ell = B_\ellB = 20\%$ for $\ell = e,\mu$, 
$\sqrt{s} = 14\TeV$, and $\int\!\calL_{pp} = 10\fb^{-1}$.  
\CP-odd energy and angular correlations are used 
for unpolarized proton beam.  $1\sigma$ sensitivity is 
\begin{align*}
  &   |\Re(\mbox{$t\tB g$-EDM})| = 0.899\times10^{-17}\gs\cm
  &&  \cdots
  &&  T_{33}^\ell
  \sepcomma\\
  &   |\Im(\mbox{$t\tB g$-EDM})| = 0.858\times10^{-18}\gs\cm
  &&  \cdots
  &&  A_E^\ell
  \sepcomma\\
  &   |\Im(\mbox{$t\tB g$-EDM})| = 0.205\times10^{-17}\gs\cm
  &&  \cdots
  &&  Q_{33}^\ell
  \sepcomma
\end{align*}
where written in the RHS are relevant observables defined 
in~\EqRef{eq:CPo-obs}.  
There may be (at least) two more operators that are CP-odd: 
$A_1^\ell$ and $A_2^\ell$, also defined in~\EqRef{eq:CPo-obs}.  
However these correlation are zero, since the cross section is 
symmetric under $(\pBI_{\!g},\pBBI_{\!g}) \to (\pBBI_{\!g},\pBI_{\!g})$ 
due to Bose symmetry: 
$\ket{\pBI_{\!g},\pBBI_{\!g}} 
= a^\dagger(\pBI_{\!g}) a^\dagger(\pBBI_{\!g}) \ket{0}
= +\ket{\pBBI_{\!g},\pBI_{\!g}}$.  
For polarized proton beam, one can probe \CP-violating effect without 
detailed reconstruction of the final states.  
Thus they used \CP-odd rate asymmetry, 
\begin{align*}
  A = \frac{\sigma_+ - \sigma_-}{\sigma_+ + \sigma_-}
  \sepcomma
\end{align*}
where $\sigma_\pm$ is the cross section for $t\tB$ production in 
collision of an unpolarized proton to a polarized proton of 
helicity $\pm$.  
They obtained $|\Im(\mbox{$t\tB g$-EDM})| \leq 10^{-20}\gs\cm$.  
They found sensitivity depends crucially on the 
degree of polarization of gluons.  
Berger-Qiu parameterization~\cite{BQ89} for polarized 
gluon distribution functions was used.  

(ii--3) 
In~\cite{SP92}, $N(t_L \tB_L) - N(t_R \tB_R)$ was studied.  
In~\cite{C96} Tevatron Run~II ($\sqrt{s} = 2\TeV$) was considered.  
In~\cite{BBO96}, $\gamma\gamma \to t\tB$ is also considered; 
a list of sensitivities is also given.  
In~\cite{prod-had=twoHDM}, two-doublets Higgs models was considered.  
In~\cite{R96} the process $gW^+ \to t\bB$ was studied, and the contribution 
turned to be small; the reason is also given.  

(ii--4) 
Here are comments for the uncertainties in the theoretical analyses of 
sensitivity studies at hadron colliders.  
Some of the existing sensitivity studies make use of some complicated 
combinations of momenta of leptons and b's; 
those include ``optimal observable''.  
Whether these analyses can be taken seriously was questioned; 
there are several reasons.  
Firstly the \scCM\ energy of a reaction cannot be measured in 
hadron collisions.  
In view of event reconstruction, more questionable is the efficiencies 
for the assignments of jets to the parent partons.  
A jet, which is assigned to a parent parton, must be well 
separated from other jets in order for it to be used as a signal jet; 
in other words, not all hadrons are assigned to one of the signal jets, 
which means the jets used for the analysis.  
This is in contrast to e+e- collisions, where every hadrons are assigned 
to one of the jets.  
Generally certain amount of energy is not correctly assigned but included 
into other (non-signal) jets.  Usually the missing energy will be 
supplemented artificially by multiplying some factor to the observed 
energy.  Under such circumstances it should be examined carefully, 
incorporating realistic detector simulations, how well the required 
parton four-momenta can be reconstructed.
%---------------------------------------------------------------------
\subsubsection{(iii) Other situations}
Anomalous chromo-EDM vertex in $e^+e^-$ collisions is studied 
in~\cite{R96a,RT98}.  
In~\cite{R96a}, gluon jet energy distribution for the process 
$e^+e^- \to t\tB g$ is studied.  They concluded that the sensitivity is 
$\Order(1)$ with $\sqrt{s}=500\GeV$, $\int\!\calL=50\fb^{-1}$, 
identification-efficiency for top-quark pair-production event $\simeq 100\%$, 
and $E_g^{\rm min}=25\GeV$.  
The last one is the cut for the minimum gluon-jet energy, which is 
necessary because (1) the cross section is IR singular, and (2) 
and to avoid the contamination due to the gluon radiation from $b$-quark; 
the cross section grows rapidly as $E_g^{\rm min}$ is lowered.  
On the other hand, in~\cite{RT98}, {\it CP}-odd spin correlation of 
$t$ and $\tB$ is studied.  They concluded that the sensitivity is 
$\Order{10}$ with $\sqrt{s}=500\GeV$, $\int\!\calL=50\fb^{-1}$, 
and top-detection efficiency $\epsilon=0.1$.  

Anomalous EW vertices in hadron collisions are studied in~\cite{SP92}.  

Anomalous decay vertex is studied 
in~\cite{decay=epem-had,decay=epem,decay=SUSY,decay=no-initial}.  
Both production and decay anomalous vertices are studied simultaneously 
in~\cite{F96,prod-decay=review,BNOS92,prod-decay-ee=general,BCGM98,BO94,prod-decay=SUSY} for $e^+ e^-$ collisions, and 
in~~\cite{prod-decay-had=general} for hadron collisions.  
CP violation in top sector is studied also 
in~\cite{CP-violation=misc}.  
CP violation in other sectors (EDM of $\tau$ etc.) are studied 
in~\cite{BN89,BLMN89,CP-violation-no-top}.  
%
%
%

%
%
%-------------------------------------------------------------------------
\section{Polarizations}
\label{sec:pol=concept}
In the previous section, we saw that the spin-direction of $t$ and $\tB$ 
may serve as a probe of EDMs.  
In this section, 
we see how the effects of polarization can be incorporated to 
actual calculations, 
and how the polarizations can be measured.  
%-------------------------------------------------------------------------
\subsection{Polarizations of $e^+ e^-$}
\label{sec:pol=eeB-concept}
\begin{figure}[tbp]
  \hspace*{\fill}
  \begin{minipage}{7cm}
    \includegraphics[width=9cm]{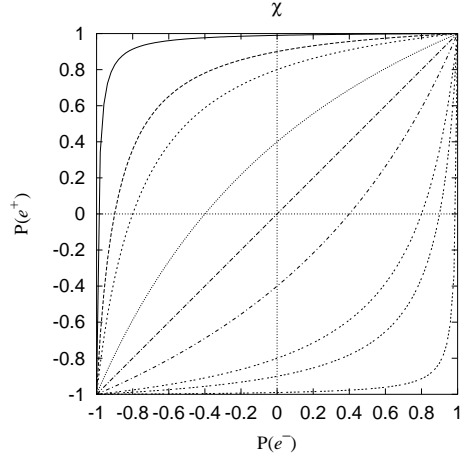}
  \end{minipage}
  \hspace*{\fill}
  \\
  \hspace*{\fill}
  \begin{Caption}\caption{\small
      Contours of the measure of initial polarization: 
      $\chi = (\PolB_e-\Pol_e)/(1-\Pol_e\PolB_e)$.  
      In the figure $\Pol(e^-)=\Pol_e$ and $\Pol(e^+)=\PolB_e$.  
      From left-top to right-bottom, the contours are for 
      $\chi=0.99$, $0.9$, $0.8$, $0.4$, $0.0$, $-0.4$, $-0.8$, $-0.9$, 
      and $-0.99$.  
      For unpolarized positron beam $\PolB_e=0$, $\chi = -\Pol_e$.  
      \label{fig:chi}
  }\end{Caption}
  \hspace*{\fill}
\end{figure}

Polarization is a incoherent sum of states with definite spin direction.  
Thus this can be treated by a density matrix.  
For spin-$1/2$ particle, polarization $\vec{\Pol}$ is defined to twice 
the expectation value of spin $\vec{S}$: 
$\vec{\Pol} = 2\langle\vec{S}\rangle$.  
Thus the density matrix $\rho$ is 
\begin{align}
  \rho 
  &= \frac{1}{2} \Bigl( \mathbf{1} + \sum_i \Pol_i \sigma_i \Bigr)
  = \frac{1}{2}
     \begin{pmatrix}
       1 + \Pol_z  &  \Pol_x - i \Pol_y  \\
       \Pol_x + i \Pol_y  &  1 - \Pol_z
     \end{pmatrix}
  \sepcomma\nonumber
\end{align}
because
$\ptr{\rho} = 1$, 
$\ptr{\rho S_i} = \langle S_i \rangle$.  
Then for longitudinal polarization, we just sum ``spin-up'' and ``spin-down'' 
with the weight $\frac{1+\Pol}{2}$ and $\frac{1-\Pol}{2}$, respectively.  
Since electron mass is negligible, chirality is the same to helicity 
for electron, and the opposite for positron: $(\psi_L)^c = (\psi^c)_R$.  
Thus the matrix element 
$E^{\mu\nu} \simeq \sum_{\rm spins}j_{\bar{e}e}^\mu j_{\bar{e}e}^{\nu*}$ 
for initial $e^+ e^-$ current 
with incomplete polarizations can be written as follows: 
\begin{align}
  E_{\mu\nu} 
  &=
  \tr\Bigl[ \Lambda_\mu 
    \left( \frac{1+\Pol_e}{2}\frac{1+\gamma_5}{2}
      + \frac{1-\Pol_e}{2}\frac{1-\gamma_5}{2}
    \right) \peS \times
  \nonumber\\
  &\qquad{}
  \times
  \LambdaB_\nu \peBS
    \left( \frac{1+\PolB_e}{2}\frac{1+\gamma_5}{2}
      + \frac{1-\PolB_e}{2}\frac{1-\gamma_5}{2}
    \right)
  \Bigr]
%  \nonumber\\
%  &=
%  \btr{\Lambda_\mu \frac{(1-\Pol_e\PolB_e)-(\PolB_e-\Pol_e)\gamma_5}{4} 
%    \peS \LambdaB_\nu \peBS}
  \nonumber\\
  &=
  \tfrac{1-\Pol_e\PolB_e}{4}
  \btr{\Lambda_\mu (1-\chi\gamma_5) \peS \LambdaB_\nu \peBS}
  \sepperiod\nonumber
\end{align}
Here $\Pol_e = 1$ ($\PolB_e = 1$) means the complete polarization 
with right-handed helicity for electron (positron), and
\begin{align}
  \chi = \frac{\PolB_e-\Pol_e}{1-\Pol_e\PolB_e}
  \label{eq:chi-def}
\end{align}
is a measure of initial polarization; 
$\chi = \pm 1$ means $(\Pol_e,\PolB_e) = (\mp 1,\pm 1)$.  
The contours of $\chi$ are shown in \FigRef{fig:chi}.  
We can see that non-zero polarization effectively mixes 
vector- and axial-vector-vertices, 
or $\chi \neq 0$ means that the initial state is parity violating.  
Thus when $\chi \neq 0$, an expectation value of parity-odd observable 
can be non-zero, even if all the relevant interactions are parity conserving.  

The effect of non-zero $\chi$ can easily be recovered from the results 
with $\chi=0$, in the following way.  
Since we don't assume anomalous vertices for initial $e^+ e^-$ current, 
\begin{align}
  E_{\mu\nu}
  \propto&\quad
  \tr[ (v^{eX}\gamma_{\mu}-a^{eX}\gamma_{\mu}\gamma_{5})
    (1-\chi\gamma_5) \peS
    ({v^{eY}}^*\gamma_{\nu}-{a^{eY}}^*\gamma_{\nu}\gamma_{5}) \peBS ]
  \nonumber\\
  &={}
  \tr[ \gamma_{\mu}\peS\gamma_{\nu}\peBS
    \left\{ 
             (v^{eX}{v^{eY}}^*+a^{eX}{a^{eY}}^*) 
      + \chi (v^{eX}{a^{eY}}^*+a^{eX}{v^{eY}}^*)
    \right\}
  \nonumber\\
  &\qquad{}
    + \gamma_{\mu}\peS\gamma_{\nu}\peBS\gamma_5
    \left\{
             (v^{eX}{a^{eY}}^*+a^{eX}{v^{eY}}^*)
      + \chi (v^{eX}{v^{eY}}^*+a^{eX}{a^{eY}}^*)
    \right\}
  ]
  \sepperiod\nonumber
\end{align}
Thus we can recover terms proportional to $\chi$ from those without $\chi$ 
as follows: 
every terms in $|\calM|^2$ contain two factors of $v^e$ or $a^e$; 
exchange each one of them with the rule $v^e \leftrightarrow a^e$ 
and divide by two.  
The product of couplings in Eqs.~(\ref{eq:def-a1}) and (\ref{eq:def-a5}) 
are defined such that each pair of 
($a_1$, $a_2$), ($a_3$, $a_4$) and ($a_5$, $a_6$) 
are ``conjugate'' with the procedure above.  
Thus we can recover $\chi$ dependence simply by replacing 
($a_1$, $a_2$) with ($a_1+\chi a_2$, $a_2+\chi a_1$) and similarly for
($a_3$, $a_4$) and ($a_5$, $a_6$).  
%-------------------------------------------------------------------------
\subsection{Polarizations of $t\tB$}
\label{sec:pol=ttB-concept}
Although the spin $\sBI$ of, say, $t$ is surely a 3-vector, 
but one should be careful to its quantum nature.  
That is, two spins that are orthogonal 
in a sense of 3-vector, are not ``orthogonal''.  
For a spin-1/2 state, 
\begin{align}
  \left| \braket{\theta,\phi}{}{\hat{z}} \right|^2
  = \cos^2\frac{\theta}{2} = \frac{1+\cos\theta}{2}
  \sepcomma\nonumber
\end{align}
where $\theta$ and $\phi$ denote polar coordinates, and 
$\ket{\hat{z}} = \ket{\theta=0}$.  
Thus, when an amplitude vanishes for some choice of spin 
$\vec{s} = - \vec{n}$, it means the spin-direction is $+\vec{n}$, 
not perpendicular to $-\vec{n}$.  

Let us look at them more closer.  
In quantum mechanics, the polarization $\vec{\Pol}$ along some 
direction $\vec{n}$ ($|\vec{n}|^2 = 1$) may be calculated as follows: 
\begin{align}
  \dprod{\vec{n}}{\vec{\Pol}}
  &= \bra{\Psi} \dprod{\vec{n}}{(2\hat{\vec{S}})} \ket{\Psi}
    \,/\,\braket{\Psi}{}{\Psi}
  \nonumber\\
  &= \sum_{\Pol=\pm 1} \Pol 
     \left| \braket{\vec{s}=\Pol\vec{n}}{}{\Psi} \right|^2 
     \,/\,\braket{\Psi}{}{\Psi}
  \sepcomma\nonumber
\end{align}
Here we used 
\begin{align}
  1 = \sum_{\Pol=\pm 1} \ket{\vec{s}=\Pol\vec{n}} \bra{\vec{s}=\Pol\vec{n}}
  \sepcomma\quad
  \dprod{\vec{n}}{\hat{\vec{S}}} \ket{\Pol\vec{n}}
     = \frac{\Pol}{2} \ket{\Pol\vec{n}}
  \sepperiod\nonumber
\end{align}
Note that $\hat{S}_i$ means spin operator, while
$s_i$ means spin-direction: $|\vec{s}|^2 = 1$.  

With this experience in mind, next we consider the process 
$e^+ e^- \to t(s) \tB(\sB)$, where $s$ ($\sB$) is a 3-vector for 
the spin-direction of $t$ ($\tB$) at its rest frame.  
Let us write the spin dependence of 
the matrix element $\calM$ as follows: 
\begin{align}
  |\calM(s^i,\sB^j)|^2
  &= \frac{C}{4} \left[
      1 + s^i \Pol^i + \sB^j \PolB^j 
      + s^i \sB^j(\Pol^i\PolB^j + \mathrm{M}^{ij})
     \right]
  \nonumber\\
  &= C \left[
      \pfrac{1+s^i\Pol^i}{2} \pfrac{1+\sB^j\PolB^j}{2}
      + \frac{1}{4} s^i \sB^j \mathrm{M}^{ij}
     \right]
  \sepperiod
  \label{eq:MtoPPB}
\end{align}
As the notation indicates, 
$\Pol_i$ can be interpreted as the polarization vector for $t$.  
This can be seen as follows.  
Total cross section $\sigma$ is 
\begin{align}
  \sigma
  \propto
  \int\!\diffn{\Omega}{} 
  \sum_{\stackrel{\scriptstyle s^i = \pm n^i}{\sB^j = \pm \nB^j}}
  |\calM(s^i,\sB^j)|^2
  =
  \int\!\diffn{\Omega}{}\,C
  \sepperiod\nonumber
\end{align}
Note that momentum 3-vectors are summed over all direction, 
while spin 3-vectors are summed only ``up'' and ``down'', 
because this is the way they form a complete set.  
The directions $\vec{n},\vec{\nB}$ of quantization-axes are arbitrary, here.  
While, 
the polarization vector $\Pol^i$ of $t$ for a fixed momentum configuration is 
\begin{align}
  \mean{n^i{\cdot}2S^i}
  \propto
  \sum_{\stackrel{\scriptstyle \Pol=\pm 1}{\PolB=\pm 1}}
  \Pol |\calM(s^i=\Pol n^i,\sB^j=\PolB \nB^j)|^2
  = 
  n^i \Pol^i \, C
  \sepperiod\nonumber
\end{align}
The spin quantization-axis $\vec{\nB}$ for $\tB$ is arbitrary.  
Dividing by $\sigma$, we see that $\Pol^i$ defined at~\EqRef{eq:MtoPPB} 
is surely a polarization vector for $t$, when one sum over 
the spin of $\tB$.  
Likewise, 
\begin{align}
  \mean{n^i{\cdot}2S^i\,\nB^j{\cdot}2\SB^j}
  \propto
  \sum_{\stackrel{\scriptstyle \Pol=\pm 1}{\PolB=\pm 1}}
  \Pol\PolB |\calM(s^i=\Pol n^i,\sB^j=\PolB \nB^j)|^2
  = 
  n^i \nB^j (\Pol^i\PolB^j+\mathrm{M}^{ij}) \, C
  \sepperiod\nonumber
\end{align}
Thus we see 
\begin{align}
  \mean{S^i \SB^j} 
  = \frac{\Pol^i}{2}\frac{\PolB^j}{2} + \frac{\mathrm{M}^{ij}}{4}
  \neq \frac{\Pol^i}{2}\frac{\PolB^j}{2}
  = \mean{S^i} \mean{\SB^j}
  \sepcomma\nonumber
\end{align}
which means the spins of $t$ and $\tB$ are correlated when 
$\mathrm{M}^{ij} \neq 0$.  
This can be seen already in the definition~\EqRef{eq:MtoPPB}; 
when $\mathrm{M}^{ij} = 0$, 
$|\calM(s^i=-\Pol^i,\sB^j)|^2 = 0$ irrespective of $\sB^j$ 
(when $|\vec{\Pol}|^2=1$).  
This means the spin-direction $\vec{s}$ is $\vec{\Pol}$, and is determined 
solely by the momentum configuration, because so is $\vec{\Pol}$.  
In general $\mathrm{M}^{ij} \neq 0$, because 
momenta and spins have independent degrees of freedom.  
However in some cases, the two spin-directions factorize.  
For example, for the process 
 $e^-(L)\,e^+(R) \to \gamma^* \to t(s)\,\tB(\sB)$ near the threshold, 
\begin{align}
  |\calM|^2
  \propto
  \left( 1 - \dprod{\sBI_t}{\frac{\pBI_{e^-}}{\mt}} \right)
  \left( 1 - \dprod{\sBBI_t}{\frac{\pBI_{e^-}}{\mt}} \right)
  + \Order{\beta_t^2}
  \sepcomma\nonumber
\end{align}
where $\beta_t = 2|\ptBI|/\sqrt{s}$.  
Here $e^+(R)$ means positron of right-handed helicity, not chirality.  
This shows that the direction of spins of both $t$ and $\tB$ are 
anti-parallel to the direction of (left-handed) initial electron.  
This is due to the conservation of total angular momentum; 
on the threshold, there is no orbital angular momentum.  

The polarization vectors $\vec{\Pol}$,$\vec{\PolB}$ should be 
composed of odd number of $\ptBI$ and $\peBI$, 
since $|\calM|^2$ is a scalar under 3-dimensional rotation.  
This means the polarization vectors are odd under ``over-all'' parity 
transformation [\SecRef{sec:PC-amp}], 
which means parity violating.  These vectors are 
non-zero even for zero initial polarization $\chi$, because the 
coupling of $Z$ to fermion-antifermion is parity violating.  

A $|\calM|^2$ for the definite momenta and spin-directions of 
$t$ and $\tB$ can be calculated with the usual Trace technique 
with spin-projection operators [\SecRef{sec:proj-E,s}].  
One can easily see that the general spin structure for $|\calM|^2$ is 
that given in~\EqRef{eq:MtoPPB}.  
In the next section, 
it is given that the simple example of a calculation with a spin-projection.  
%---------------------------------------------------------------------
\subsection{Spin direction of top quark}
\label{sec:pol_t=p_ell}
The direction of the spin of top quark can be measured statistically 
for the decay process $t \to b \bar{\ell} \nu_{\ell}$.  
This is because~\cite{TopPol} at the rest frame of $t$, 
the charged lepton $\bar{\ell}$ is emitted 
preferentially to the direction of the spin of top quark.  
This can be seen as follows.  
Using 
$\uB_1 \Gamma v_2 = -\overline{v_2^c}\Gamma^c u_1^c = -\uB_2 \Gamma^c v_1$ 
for commuting spinors, and Fierz identity, 
\begin{align}
  \calM \sim
  &\phantom{=}{}
  \uB(p_b) \gamma^{\mu}\tfrac{1-\gamma_5}{2} u(p_t)
  ~ \uB(p_\nu) \gamma_{\mu}\tfrac{1-\gamma_5}{2} v(p_{\bar{\ell}})
  \nonumber\\
  &=
  + \uB(p_b) \gamma^{\mu}\tfrac{1-\gamma_5}{2} u(p_t)
  ~ \uB(p_{\bar{\ell}}) \gamma_{\mu}\tfrac{1+\gamma_5}{2} v(p_\nu)
  \nonumber\\
  &=
  \tfrac{1}{2} \uB(p_b) (1+\gamma_5) v(p_\nu)
  ~ \uB(p_{\bar{\ell}}) (1-\gamma_5) u(p_t)
  \sepcomma\nonumber
\end{align}
which means 
\begin{align}
  |\calM|^2 \propto
  &\phantom{=}{}
  \sum_{s_{\bar{\ell}}} 
  |\uB(p_{\bar{\ell}},s_{\bar{\ell}}) (1-\gamma_5) u(p_t,s_t)|^2
  \nonumber\\
  &=
  \btr{\plBS(1-\gamma_5)(\ptS+m_t)\tfrac{1-\stS\gamma_5}{2}(1+\gamma_5)}
  \nonumber\\
  &=
  4 p_{\bar{\ell}} \cdot (p_t-m_t s_t)
  \nonumber\\
  &=
  4 m_t E_{\bar{\ell}} (1+\dprod{\stBI}{\betaBI_{\bar{\ell}}})
  \qquad\text{at the rest frame of top.}
  \nonumber
\end{align}
That is, roughly speaking, $\sBI_t \Parallel +\mean{\pBI_{\ell^+}}$ and 
$\sBI_{\tB} \Parallel -\mean{\pBI_{\ell^-}}$.  
Thus the component $\dprod{n}{s_t}$ of the spin $s_t$ of top quark 
parallel to certain direction $n$ can be obtained statistically by 
$\mean{\dprod{n}{p_{\bar{\ell}}}}$.  
Then we treat the spin (or polarization) of top quark as if it were 
a observable.  
%
%
%

%
%
%
%---------------------------------------------------------------------
\section{Corrections due to Coulomb rescattering}
\label{sec:corr_from_Coulomb=EDM}
%---------------------------------------------------------------------
\subsection{Relevant Lagrangian}
\label{sec:EDMs_Lagrangian}
From now on, we concentrate to the $t\tB$ production near the threshold.  
This also means to use $e^+ e^-$ collisions, 
since for hadron colliders $\sqrt{s}$ of each subprocess is not fixed.  
As emphasized in \ChRef{ch:total_cr}, 
the most prominent feature of threshold region is 
leading-order multiple Coulomb rescattering%
\footnote{
  Since the Coulomb rescattering is indispensable for the threshold region, 
  in some case its effect is implicit, and 
  the word ``rescattering'' is saved for the other corrections; 
  that is, gluon exchange between $t$ and $\bB$ for example.  
  However here we do not care about those corrections involving 
  decay products.  
}
, which means 
higher orders of $\alpha_s$ are not suppressed compared to 
the propagation without gluon exchange.  
This drastically changes the behavior of $t\tB$ production cross section, 
compared to the tree-level expectation.  
For example, the magnitude of the ``imaginary part'' (= absorptive part) 
of the amplitude is similar to the ``real part''.  

The relevant part of the Lagrangian is as follows: 
\begin{align*}
  \calL
  =&{}
  - \gs (\tB\gamma^{\mu}T^a t) G_\mu^a
  - \frac{\gs\dtg}{2\mt} (\tB\,i\sigma^{\mu\nu}\gamma_5 T^a t)
    \partial_{\mu}G_{\nu}^a
  \\
  &{}
  - \gZ (\tB\gamma^\mu (v^{tZ}-a^{tZ}\gamma_5)t) Z_\mu
  - \frac{\gZ\dtZ}{2\mt} (\tB\,i\sigma^{\mu\nu}\gamma_5 t)
    \partial_{\mu}Z_{\nu}
  \\
  &{}
  - e \Qt (\tB\gamma^\mu t) A_\mu
  - \frac{e\dtp}{2\mt} (\tB\,i\sigma^{\mu\nu}\gamma_5 t)
    \partial_{\mu}A_{\nu}
  \\
  &{}
  + \mbox{terms required by gauge invariance}
  + \cdots
  \sepcomma
\end{align*}
where the couplings are defined in~\EqRef{eq:def=pZ-coupl}.  
Feynman rule for, say, $t\tB\gamma$ vertex is proportional to 
$Q_t\gamma^\mu + \dtp/(2\mt)\sigma^{\nu\mu}\gamma_5 q_\nu$, where 
$q_\nu$ is a momentum flows into $t\tB$ current.  
Note that 
\begin{align*}
  \frac{d}{2m} (\tB\,i\sigma^{\mu\nu}\gamma_5 t) \partial_\mu A_\nu 
  \simeq \frac{-d}{2m} (\tB\,i\gamma_5 i\lrpartial^\mu t) A_\mu
  \simeq -d \frac{(\pt-\pBt)^\mu}{2m} 
         \uB(\pt)\,i\gamma_5\,v(\pBt)\;\epsilon_\mu
  \sepcomma
\end{align*}
where the first ``$\simeq$'' means 
they coincide with the use of eqs.\ of motion, 
and the second means to sandwich with $\bra{\pt,\pBt}$ and $\ket{q}$ 
and to integrate over $x$ [\SecRef{sec:Gordon-id}].  
We checked that 
both of these forms of the interaction give the same result, 
at least for the case we deal with.  

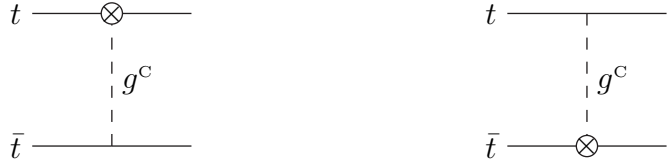
\begin{figure}[htbp]
  \hspace*{\fill}
  \begin{picture}(50,60)(0,0)
    \normalsize
    \Line(0,50)(30,50)
    \Line(30,50)(60,50)
    \Line(60,0)(30,0)
    \Line(30,0)(0,0)
    \DashLine(30,50)(30,0){5}
    \BCirc(30,50){4}
    \Text(30,50)[]{$\times$}
    \Text(-4,50)[r]{$t$}
    \Text(-4,0)[r]{$\bar{t}$}
    \Text(34,25)[l]{$g^{\mbox{\tiny C}}$}
  \end{picture}
  \hspace*{\fill}
  \begin{picture}(50,60)(0,0)
    \normalsize
    \Line(0,50)(30,50)
    \Line(30,50)(60,50)
    \Line(60,0)(30,0)
    \Line(30,0)(0,0)
    \DashLine(30,50)(30,0){5}
    \BCirc(30,0){4}
    \Text(30,0)[]{$\times$}
    \Text(-4,50)[r]{$t$}
    \Text(-4,0)[r]{$\bar{t}$}
    \Text(34,25)[l]{$g^{\mbox{\tiny C}}$}
  \end{picture}
  \hspace*{\fill}
  \\
  \hspace*{\fill}
  \begin{Caption}\caption{\small
      Contribution of anomalous $t\bar{t}g$-EDM to $t\tB$ potential.  
      \label{fig:CEDM-contr-to-V}
  }\end{Caption}
  \hspace*{\fill}
\end{figure}

As we saw in \SecRef{sec:NR-pot}, the potential $V_\rmC$ between $t\tB$ 
due to $\calL_{\rm int} = - \gs (\tB\gamma^{\mu}T^a t) G_\mu^a$ is 
$\VT_\rmC = -C_F \gs^2/|\qBI|^2$, where $\qBI$ is momentum transfer from 
$\tB$ to $t$.  Likewise, one can show that 
the potential $V_{\rm CEDM}$ due to the $t\tB g$-EDM interaction above is 
[\FigRef{fig:CEDM-contr-to-V}]
\begin{align}
  \VT_{\rm CEDM} 
  = \frac{-C_F \gs^2}{|\qBI|^2} \, \frac{-d_{tg}}{m_t} 
    i \qBI {\cdot} 
    \left( \frac{\sigmaBI_t}{2} - \frac{\sigmaBI_{\tB}}{2} \right)
  \sepperiod
  \label{eq:CEDM-pot}
\end{align}
This potential is $(\Parity,\ChargeC) = (-,+)$, as it should be.  

Note that the magnitude of the EDM-interactions 
are proportional to the momentum transfer between $t\tB$.  
This can clearly be seen in the modified vertices 
\EqRef{eq:vertex-after-rescat} in the next section.  
Thus the effects of two EW-EDMs becomes larger 
for higher \scCM\ energy $\sqrt{s}$, because they affect directly the $t\tB$ 
production vertex.  
While the effect of CEDM do not become larger 
for higher $\sqrt{s}$, because%
\footnote{
  More concretely speaking, because the combination of the Green function 
  $D = F-G$, which is relevant to CEDM, becomes zero for $\sqrt{s}\gg 2\mt$.  
}
it affects the Coulomb rescattering 
between $t\tB$, whose relative significance becomes smaller for $\sqrt{s}$.  
Thus the threshold region does not suit for the measurement of EW-EDMs.  
On the other hand, CEDM can be measured only near the threshold for 
lepton colliders.  
Note that even though the sensitivity to the existence of finite 
CEDM may be better for hadron colliders, the precise measurement of 
its magnitude may only be achieved at lepton colliders.  
%---------------------------------------------------------------------
\subsection{Corrections to vertices}
\label{sec:rescat_corr_to_vertex}
In the next section, we show that 
the effect of Coulomb rescattering between $t$ and $\tB$ 
is to change the vertices for $t\tB\gamma$ and $t\tB Z$.  
The result is as follows.  
To the tree-level (and without anomalous vertices), 
the electroweak $t\tB$ vertices are 
\begin{align}
  \left(\Gamma^{X}\right)^\mu 
  = v^{tX} \Gamma^\mu_V - a^{tX} \Gamma^\mu_A 
  \sepcomma\text{ where}\quad
  \Gamma^\mu_V = \gamma^\mu   \sepcomma\quad
  \Gamma^\mu_A = \gamma^\mu\gamma_5  
  \qquad (X = \gamma,Z)
  \sepcomma\nonumber
\end{align}
times $-ig_X$.  See \SecRef{sec:merge_Z_and_photon} for the definition of 
the couplings.  
Rescattering and anomalous vertices modify the vertices: 
\begin{align}
  &
  \left(\Gamma^{X}\right)^\mu 
    = v^{tX} \Gamma^\mu_V - a^{tX} \Gamma^\mu_A 
      + d_{tX}\Gamma^\mu_{\text{EW-EDM}}
  \sepcomma
\end{align}
where 
\begin{align}
  &
  \Gamma^\mu_V = 
    \left[
        \gamma^\mu G(E,|\pBI|) 
%      + i\gamma_5 \frac{-p^\mu}{\mt} \dtg D(E,|\pBI|)
      - i\gamma_5 \frac{-p^\mu}{\mt} \dtg D(E,|\pBI|)
    \right] \times
    \left( \frac{|\pBI|^2}{\mt}-(E+i\Gamma_t) \right)
  \sepcomma\nonumber\\
  &
  \Gamma^\mu_A = 
    \gamma^\mu\gamma_5 F(E,|\pBI|) 
    \times
    \left( \frac{|\pBI|^2}{\mt}-(E+i\Gamma_t) \right)
  \label{eq:vertex-after-rescat}
  \sepcomma\\
  &
  \Gamma^\mu_{\text{EW-EDM}} = 
    i\gamma_5 \frac{-p^\mu}{\mt} F(E,|\pBI|) 
    \times
    \left( \frac{|\pBI|^2}{\mt}-(E+i\Gamma_t) \right)
  \sepcomma\nonumber
\end{align}
where $\pBI = (\ptBI-\ptBBI)/2$ is the relative momentum of $t\tB$, and 
$E = \sqrt{s} - 2\mt$ is non-relativistic \scCM\ energy; 
$G(E,|\pBI|)$ and $F(E,|\pBI|)$ are $S$- and $P$-wave 
Green functions, respectively 
\begin{align}
  G(E,|\pBI|) = \langle \pBI |G|\xBI'=\vec{0} \rangle  
  \sepcomma\quad
  \pBI F(E,|\pBI|) = \langle \pBI |\pBI F| \xBI'=\vec{0} \rangle
  \sepcomma
\end{align}
and 
\begin{align}
  \left( \frac{|\pBI|^2}{\mt} - \omega + V_\rmC \right) G = 1
  \sepcomma\quad
  \left( \frac{|\pBI|^2}{\mt} - \omega + V_\rmC \right) \pBI F = \pBI
  \sepcomma
\end{align}
%
%and $D(E,|\pBI|) = F(E,|\pBI|) - G(E,|\pBI|)$.  
and $D(E,|\pBI|) = G(E,|\pBI|) - F(E,|\pBI|)$.  
We can see from the equations for Green functions that 
\begin{align*}
  G(E,|\pBI|) = F(E,|\pBI|) = 
  1 / \left( \frac{|\pBI|^2}{\mt} - (E+i\Gamma_t) \right)
  \qquad\text{for $V \to 0$}
  \sepcomma
\end{align*}
and $D(E,|\pBI|) = 0$.  
Thus in the weak potential limit, or for $\sqrt{s}\gg 2\mt$ where 
the effect of Coulomb rescattering is diminished, 
the vertices reduces the tree-level ones with anomalous 
EW-EDMs but QCD-EDM.  

We can see from~\EqRef{eq:vertex-after-rescat} that 
the effects of anomalous $\ChargeCOp\ParityOp$-violating interactions 
are suppressed%
\footnote{
  Note that the time-component of the vertex do not contribute to 
  $|\calM|^2$, since the initial $e^+e^-$ current $j_{e^+e^-}^\mu$ 
  is conserved $q_\mu j_{e^+e^-}^\mu = 0$ 
  (because the mass of $e^-$ can be neglected).  
  Thus at the \scCM-frame $q_\mu = (\sqrt{s},\vec{0})$, 
  the time-component of $e^+e^-$ current is zero $j_{e^+e^-}^0 = 0$.  
  Propagators of gauge bosons are proportional to $g_{\mu\nu}$, 
  aside from $q_\mu q_\nu$.  
  Thus only the space-component of $t\tB$ current $j_{t\tB}^\mu$ 
  contributes.  
}
only by $\beta_t =|\ptBI|/\mt$.  
Thus we take into account $\Order{\beta_t}$ corrections, and 
neglect the higher orders%
\footnote{
  We also neglect the $\Order{\beta_t}$ corrections from FSIs.  
}.  
%---------------------------------------------------------------------
\subsection{Sketch of the calculation}
Using Non-Relativistic form of the propagators, 
\begin{align}
  &  S_F(k+\frac{q}{2})
  = \frac{1+\gamma^0}{2} \frac{i}{E/2 + k^0 - |\qBI|^2/(2\mt)+i\Gamma_t/2}
    + \Order{\frac{1}{c}}
  \sepcomma\nonumber\\
  &  S_F(k-\frac{q}{2})
  = \frac{1-\gamma^0}{2} \frac{i}{E/2 - k^0 - |\qBI|^2/(2\mt)+i\Gamma_t/2}
    + \Order{\frac{1}{c}}
  \sepcomma
\end{align}
Lippmann-Schwinger equation for the vector vertex is written as follows: 
\begin{align}
  \Gamma_V^i(E,p)
  &=
  \gamma^i
  + C_F (-i\gs)^2 \int\!\!\!\frac{\diffn{k}{4}}{(2\pi)^4}
    \frac{i}{E/2 + k^0 - |\kBI|^2/(2\mt)+i\Gamma_t/2}
    \times
  \nonumber\\
  &\qquad{}
    \times
    \frac{i}{E/2 - k^0 - |\kBI|^2/(2\mt)+i\Gamma_t/2}
    \times
  \nonumber\\
  &\qquad{}
    \times
    \biggl[ \quad
      \gamma^0\frac{1+\gamma^0}{2} \Gamma_V^i(E,k) 
      \frac{1-\gamma^0}{2}\gamma^0
  \nonumber\\
  &\qquad\qquad{}
    + \pfrac{\dtg}{2\mt} \sigma^{j0}\gamma_5 \frac{-(p-k)^j}{c}
      \frac{1+\gamma^0}{2} \Gamma_V^i(E,k) \frac{1-\gamma^0}{2}\gamma^0
  \nonumber\\
  &\qquad\qquad{}
    + \pfrac{\dtg}{2\mt} \gamma^0\frac{1+\gamma^0}{2} \Gamma_V^i(E,k) 
      \frac{1-\gamma^0}{2} \sigma^{j0}\gamma_5 \frac{+(p-k)^j}{c}
  \nonumber\\
  &\qquad{}
    \biggr]
    \times \frac{i}{|\pBI-\kBI|^2}
  \sepperiod
\end{align}
A similar equation holds also for the axial vertex.  
Since there is no $p^0$-dependence on the right-hand side, 
consistency requires $\Gamma_V^i(E,p) = \Gamma_V^i(E,\pBI)$.  
Thus we can trivially integrate $p^0$, and obtain 
\begin{align}
  \Gamma_V^i(E,\pBI)
  &=
  \gamma^i
  - \int\!\!\!\frac{\diffn{k}{3}}{(2\pi)^3} 
    \frac{-\CF\gs^2}{|\pBI-\kBI|^2}
    \frac{1}{|\kBI|^2/\mt-(E+i\Gamma_t)}
    \times
  \\
  &\qquad{}
    \times
    \frac{1+\gamma^0}{2} \left[
      \Gamma_V^i(E,\kBI)
    - \frac{\dtg}{2\mt} \frac{(\pBI-\kBI)^j}{c} 
      \left\{ 
          \sigma^{j0}\gamma_5 \Gamma_V^i(E,\kBI)
        + \Gamma_V^i(E,\kBI) \sigma^{j0}\gamma_5
      \right\}
    \right] \frac{1-\gamma^0}{2} 
  \sepperiod\nonumber
\end{align}
In view of the structure of gamma matrix, we decompose the vertex function 
$\Gamma_V^i(E,\pBI)$ as follows: 
\begin{align}
%  &\Shift{-3em}
  \frac{1+\gamma^0}{2} \Gamma_V^i(E,\pBI) \frac{1-\gamma^0}{2}
%  \nonumber\\
  &=
  \frac{1+\gamma^0}{2} \biggl[ \quad
    \gamma^j \Bigl( 
        \delta^{ij} \Gamma_G(E,\pBI)
      + \Gamma_B^{ij}(E,\pBI)
    \Bigr)
  \nonumber\\
  &\qquad\qquad{}
    + i\sigma^{ij}\gamma_5 \Gamma_F^j(E,\pBI)
%    + i\gamma_5 \Gamma_D^i(E,\pBI)
    - i\gamma_5 \Gamma_D^i(E,\pBI)
    \biggr] \frac{1-\gamma^0}{2}
\end{align}
where $\sum_{ij} \Gamma_B^{ij}(E,\pBI) = 0$.  
By plugging this expression into the integral equation above, 
one obtains the integral equations for each (reduced) vertex functions 
$\Gamma_G(E,\pBI)$ etc.  
One can see that $\Gamma_D^i(E,\pBI) = \Order{\dtg}$, 
$\Gamma_B^{ij}(E,\pBI) = \Order{\dtg^{~2}}$.  
Thus we forget about $\Gamma_B^{ij}(E,\pBI)$ hereafter.  
One can do just the same procedure for the axial vertex $\Gamma_A^i(E,\pBI)$, 
and finds that anomalous $t\tB g$ vertex we inserted do not alter 
the axial vertex to the order we consider now.  
We extract Lorentz structure of $\Gamma_F^i(E,\pBI)$ etc.~as follows: 
\begin{align}
  &  \Gamma_G(E,\pBI) = 
    \left( \frac{|\pBI|^2}{\mt} - (E+i\Gamma_t) \right) 
    G(E,|\pBI|)
  \sepcomma\nonumber\\
  &  \Gamma_F^i(E,\pBI) = 
    \left( \frac{|\pBI|^2}{\mt} - (E+i\Gamma_t) \right) 
    \frac{-p^i}{\mt c} F(E,|\pBI|)
  \sepcomma\nonumber\\
  &  \Gamma_D^i(E,\pBI) = 
    \left( \frac{|\pBI|^2}{\mt} - (E+i\Gamma_t) \right) 
    \frac{-p^i}{\mt c} \dtg D(E,|\pBI|)
  \sepperiod
\end{align}
In terms of $G(E,|\pBI|)$, the Lippmann-Schwinger eqs.\ become 
NR Schr\"odinger equations: 
\begin{align}
  &  \left( \frac{|\pBI|^2}{\mt} - (E+i\Gamma_t) \right) G(E,|\pBI|)
  + \int\!\!\!\frac{\diffn{k}{3}}{(2\pi)^3}
    \left[ \VT_\rmC(|\pBI-\kBI|) G(E,|\kBI|) \right]
  = 1
  \sepcomma
  \nonumber\\
  &  \left( \frac{|\pBI|^2}{\mt} - (E+i\Gamma_t) \right) \pBI^i F(E,|\pBI|)
  + \int\!\!\!\frac{\diffn{k}{3}}{(2\pi)^3} 
    \left[ \VT_\rmC(|\pBI-\kBI|) \kBI^i F(E,|\kBI|) \right]
  = \pBI^i
  \sepperiod
  \nonumber\\
  &  \left( \frac{|\pBI|^2}{\mt} - (E+i\Gamma_t) \right) \pBI^i D(E,|\pBI|)
  + \int\!\!\!\frac{\diffn{k}{3}}{(2\pi)^3} 
    \left[ \VT_\rmC(|\pBI-\kBI|) \kBI^i D(E,|\kBI|) \right]
  \nonumber\\
  &\qquad\qquad
  = \int\!\!\!\frac{\diffn{k}{3}}{(2\pi)^3} 
%    \left[ \VT_\rmC(|\pBI-\kBI|) (\pBI-\kBI)^i G(E,|\kBI|) \right]
    \left[ \VT_\rmC(|\kBI-\pBI|) (\pBI-\kBI)^i G(E,|\kBI|) \right]
  \sepperiod
  \nonumber
\end{align}
%
%We find that $D = F - G$.  
We find that $D = G - F$.  
As for EW-EDMs, since 
\begin{align}
  \frac{1}{2\mt} \sigma^{\lambda\mu}\gamma_5 q_{\lambda}
  &\simeq \sigma^{0\mu} \gamma_5
   \simeq i\gamma_5 \frac{-p^\mu}{\mt c}
  \sepcomma
\end{align}
we can put as 
\begin{align}
  \Gamma_{\text{EW-EDM}}^i(E,\pBI)
  =
  i\gamma_5 \frac{-p^i}{\mt c} K(E,|\pBI|) 
  \left( \frac{|\pBI|^2}{\mt} - (E+i\Gamma_t) \right)
  \sepperiod
\end{align}
By comparing with 
\begin{align*}
  \gamma^i\gamma_5 \simeq i\sigma^{ij}\gamma_5\frac{-p^j}{\mt c}
  \sepcomma
\end{align*}
one can see that $p^i K(E,|\pBI|)$ satisfies the same 
Lippmann-Schwinger eq.\ as $p^i F(E,|\pBI|)$, 
thus $K(E,|\pBI|) = F(E,|\pBI|)$.  

Likewise, 
\begin{align*}
  \Gamma_A^i \simeq i\sigma^{ij}\gamma_5 \Gamma_F^j
  \simeq 
  \left( \frac{|\pBI|^2}{\mt} - (E+i\Gamma_t) \right) 
  \gamma^i\gamma_5 F(E,|\pBI|)
  \sepperiod
\end{align*}
%
%---------------------------------------------------------------------
%
%
%

%
%
%----------------------------------------------------------------------
\subsection{Polarization vector of top and antitop}
Including $\mathcal{O}(\beta_t )$ contribution, 
the production cross section for $t\tB$ pair can be written as follows: 
\begin{align*}
  \frac{\diffn{\sigma}{}}{\diffn{\pBI_t}{3}}
  =
  \frac{N_C \alpha^2 \Gamma_t}{2\pi m_t^4}
  \frac{1-\Pol_{e^+}\Pol_{e^-}}{2}
%  \times
  |G|^2 (a_1+\chi a_2) \left\{
    1 + 2\pRe{\CFB\frac{F}{G}}\beta_t \cos\theta_{te}
  \right\}
  \sepcomma
\end{align*}
where 
\begin{align*}
  \beta_t = \frac{|\ptBI|}{m_t}
  \sepcomma\quad
  \cos\theta_{te} = \frac{\dprod{\peBI}{\ptBI}}{|\peBI|\,|\ptBI|} 
  \sepcomma
\end{align*}
and $\chi$ is a measure of the polarization of initial $e^+ e^-$ defined 
in~\EqRef{eq:chi-def}.  
The coefficient $\CFB$ is defined in \EqRef{eq:Cnorm}.  
For unpolarized positron beam ($\PolB_e = 0$), 
$\chi = -\Pol_e$, where $\Pol_e = -1$ for the electron beam 
that is completely polarized to left-handed helicity.  
Hard vertex corrections are not included, since we are working with 
non-renormalizable interactions.  
In the formula above we summed over spins of $t$ and $\tB$.  
Before summing over the spins of $t\tB$, 
the production cross section for them can be written 
as follows: 
\begin{align*}
  \frac{\diffn{\sigma}{}(\stBI,\stBBI)}{\diffn{\pBI_t}{3}}
  =
  \frac{\diffn{\sigma}{}}{\diffn{\pBI_t}{3}}
  \frac{%
      1
    + \dprod{\PolBR}{\stBI}
    + \dprod{\PolBBR}{\stBBI}
    + (\stBI)_i(\stBBI)_j\PolQBR_{ij} }%
  {4}
  \sepcomma
\end{align*}
As was explained in~\SecRef{sec:pol=ttB-concept}, 
$\PolBR$ and $\PolBBR$ can be interpreted as polarization vectors 
for $t$ and $\tB$, respectively.  
Both SM and anomalous interactions contribute to the polarizations: 
\begin{align*}
  \PolBR  = \PolBR_\rmSM  + \delta\PolBR  \sepcomma\quad
  \PolBBR = \PolBBR_\rmSM + \delta\PolBBR  
  \sepperiod
\end{align*}
With the knowledge of~\SecRef{sec:exp-corr}, 
the interference of an anomalous vertex, which is CP-odd, and 
the standard one, which is CP-even, contributes in the opposite sign 
to the polarizations of $t$ and $\tB$.  
On the other hand, the interference among the standard interactions 
contribute in the same sign%
%\footnote{We checked these by explicit calculation.}%
: 
\begin{align}
  \PolBBR_\rmSM =   \PolBR_\rmSM  \sepcomma\quad
  \delta\PolBBR = - \delta\PolBR  \sepcomma
  \label{eq:rel-between-P-and-PB}
\end{align}
or 
\begin{align*}
  \PolBR_\rmSM = (\PolBR + \PolBBR)/2  \sepcomma\quad
  \delta\PolBR = (\PolBR - \PolBBR)/2  \sepperiod
\end{align*}
Note that we are working to the leading order of the anomalous couplings.  
Hereafter we express those vector by components: 
\begin{align*}
  \PolBR = \Polpara\nBIpara + \Polperp\nBIperp + \Polnorm\nBInorm
  \sepcomma\quad
  \stBI = \spara\nBIpara + \sperp\nBIperp + \snorm\nBInorm
  \sepcomma
\end{align*}
where the axes are determined by the beam direction and the production plane: 
\begin{align*}
  \nBIpara \equiv \frac{\peBI}{|\peBI|}  
  \sepcomma\quad
  \nBInorm \equiv \frac{\cprod{\peBI}{\ptBI}}{|\cprod{\peBI}{\ptBI}|}  
  \sepcomma\quad
  \nBIperp \equiv \cprod{\nBInorm}{\nBIpara}
  \sepperiod
\end{align*}
Now the polarization of $t$ is as follows%
\footnote{
  We use notation similar to~\cite{HJKP97}.  
  The are two differences: (i) $a_3$ and $a_4$ of them are twice of ours; 
  and (ii) $C_N$ of them is $-C_N$ of ours.  
}:
\begin{align*}
  & \PolSMpara
  = \Cpara^0 + \pRe{\Cpara^{1} \frac{F}{G}} \beta_t \cos\theta_{te}
  \sepcomma\\
  & \PolSMperp
  = \pRe{\Cperp \frac{F}{G}} \beta_t \sin\theta_{te}
  \sepcomma\\
  & \PolSMnorm
  = \pIm{\Cnorm \frac{F}{G}} \beta_t \sin\theta_{te}
\end{align*}
for the contributions from the standard interactions%
\footnote{
  These results coincide with those in~\cite{HJKP97}.  
}
, and 
\begin{align*}
  & \delta\Polpara = 0
  \sepcomma\\
  & 
  \delta\Polperp
  =
  \left[
      \pIm{\Bperp^{g}      d_{tg}      \frac{D}{G}}
    + \pIm{\Bperp^{\gamma} d_{t\gamma} \frac{F}{G}}
    + \pIm{\Bperp^{Z}      d_{tZ}      \frac{F}{G}}
  \right] \beta_t \sin\theta_{te}
  \sepcomma\\
  &
  \delta\Polnorm
  =
  \left[
      \pRe{\Bnorm^g        d_{tg}      \frac{D}{G}}
    + \pRe{\Bnorm^{\gamma} d_{t\gamma} \frac{F}{G}}
    + \pRe{\Bnorm^{Z}      d_{tZ}      \frac{F}{G}}
  \right] \beta_t \sin\theta_{te}
\end{align*}
for the contribution from the anomalous interactions.  
Coefficients $\Cpara^0$ etc.\ are defined below.  
There is also correlation between $\stBI$ and $\stBBI$: 
\begin{align*}
  (\stBI)_i(\stBBI)_j\PolQBR_{ij}
  =
    (\stBI)_i(\stBBI)_j {\PolQBR_{ij}}_{\rmSM}
  + (\stBI)_i(\stBBI)_j \,\delta\PolQBR_{ij}
  \sepcomma
\end{align*}
\begin{align*}
  (\stBI)_i(\stBBI)_j {\PolQBR_{ij}}_{\rmSM}
  &=
    \spara\sBpara
  + (\spara\sBperp+\sperp\sBpara) \pRe{\Cnorm \frac{F}{G}}
    \beta_t \sin\theta_{te}
  \\
  &\qquad{}
  + (\spara\sBnorm+\snorm\sBpara) \pIm{\Cperp \frac{F}{G}}
    \beta_t \sin\theta_{te}
  \sepcomma
\end{align*}
\begin{align*}
  &
  (\stBI)_i(\stBBI)_j \,\delta\PolQBR_{ij}
  =
  \\
  &{}
  + (\spara\sBperp-\sperp\sBpara)\left[
        \Im\left(\Bnorm^g        d_{tg}      \frac{D}{G}\right)
      + \Im\left(\Bnorm^{\gamma} d_{t\gamma} \frac{F}{G}\right)
      + \Im\left(\Bnorm^{Z}      d_{tZ}      \frac{F}{G}\right)
    \right]
    \beta_t \sin\theta_{te}
  \\
  &{}
  + (\spara\sBnorm-\snorm\sBpara)\left[
      \Re\left(\Bperp^{g}      d_{tg}      \frac{D}{G}\right)
    + \Re\left(\Bperp^{\gamma} d_{t\gamma} \frac{F}{G}\right)
    + \Re\left(\Bperp^{Z}      d_{tZ}      \frac{F}{G}\right)
    \right]
    \beta_t \sin\theta_{te}
  \sepperiod
\end{align*}
We do not use these correlations in this work.  

Coefficients of the polarization are defined as follows: 
\begin{align}
  &  \Cpara^0(\chi) = - \frac{a_2+\chi a_1}{a_1+\chi a_2} 
  \sepcomma\quad
  && \Cpara^1(\chi) = 2\Cpara^0\Cnorm+\Cperp
       = 2(1-\chi^2)\frac{a_2 a_3-a_1 a_4}{(a_1+\chi a_2)^2} 
  \sepcomma\nonumber\\
  &  \Cperp(\chi) = - \frac{a_4+\chi a_3}{a_1+\chi a_2} 
  \sepcomma\quad
  && \Cnorm(\chi) = \frac{a_3+\chi a_4}{a_1+\chi a_2}
       = \CFB 
  \label{eq:Cnorm}
\end{align}
for the standard vertices, and 
\begin{align}
%  & \Bperp^g(\chi) = +1
  & \Bperp^g(\chi) = -1
    \sepcomma\quad
%    \Bnorm^g(\chi) = - \Cpara^0(\chi) \sepcomma\nonumber\\
    \Bnorm^g(\chi) = + \Cpara^0(\chi) \sepcomma\nonumber\\
  & \Bperp^{\rm EW}(\chi) = \frac{a_5+\chi a_6}{a_1+\chi a_2}
       =   \Bperp^\gamma(\chi) \, {d_{t\gamma}} 
         + \Bperp^Z(\chi) \, {d_{tZ}} 
  \label{eq:Bnorm}
  \sepcomma\\
  & \Bnorm^{\rm EW}(\chi) = \frac{a_6+\chi a_5}{a_1+\chi a_2}
       = \Bnorm^\gamma(\chi) \, {d_{t\gamma}} + \Bnorm^Z(\chi) \, {d_{tZ}}
  \sepcomma\nonumber
\end{align}
where 
\begin{align}
  \Bperp^\gamma(\chi)
  &= \frac{1}{a_1+\chi a_2}
    \left\{
        \left( [v^e v^t]^*\,{v^{e\gamma}} + [a^e v^t]^*\,{a^{e\gamma}} \right)
      + \chi
        \left( [v^e v^t]^*\,{a^{e\gamma}} + [a^e v^t]^*\,{v^{e\gamma}} \right)
    \right\}
  \nonumber\\
  &= \frac{1}{a_1+\chi a_2}
    \left(
        [v^e v^t]^*\,{v^{e\gamma}} + \chi [a^e v^t]^*\,{v^{e\gamma}}
    \right)
  \sepcomma\nonumber\\[2ex]
  \Bperp^Z(\chi)
  &= \frac{1}{a_1+\chi a_2}
    \left\{
        \left( [v^e v^t]^*\,{v^{eZ}} + [a^e v^t]^*\,{a^{eZ}} \right)
      + \chi
        \left( [v^e v^t]^*\,{a^{eZ}} + [a^e v^t]^*\,{v^{eZ}} \right)
    \right\}
    d(s)
  \sepcomma\nonumber\\[2ex]
  \Bnorm^\gamma(\chi)
  &= \frac{1}{a_1+\chi a_2}
    \left\{
        \chi
        \left( [v^e v^t]^*\,{v^{e\gamma}} + [a^e v^t]^*\,{a^{e\gamma}} \right)
      + \left( [v^e v^t]^*\,{a^{e\gamma}} + [a^e v^t]^*\,{v^{e\gamma}} \right)
    \right\}
  \label{eq:BnormZ}
  \\
  &= \frac{1}{a_1+\chi a_2}
    \left(
        \chi [v^e v^t]^*\,{v^{e\gamma}} + [a^e v^t]^*\,{v^{e\gamma}}
    \right)
  \sepcomma\nonumber\\[2ex]
  \Bnorm^Z(\chi)
  &= \frac{1}{a_1+\chi a_2}
    \left\{
        \chi
        \left( [v^e v^t]^*\,{v^{eZ}} + [a^e v^t]^*\,{a^{eZ}} \right)
      + \left( [v^e v^t]^*\,{a^{eZ}} + [a^e v^t]^*\,{v^{eZ}} \right)
    \right\}
    d(s)
  \nonumber
\end{align}
for the anomalous vertices.  
Symbols $a_{1\sim 6}$ are the combinations of EW couplings: 
\begin{align}
     a_1 &= \left|[v^e v^t]\right|^2+\left|[a^e v^t]\right|^2 
  \sepcomma\quad
  &  a_2 &= 2 \pRe{[v^e v^t]^*\,[a^e v^t]} 
  \label{eq:def-a1}
  \sepcomma\\
     a_3 &= [v^e v^t]^*\,[a^e a^t] + [a^e v^t]^*\,[v^e a^t] 
  \sepcomma\quad
  &  a_4 &= [v^e v^t]^*\,[v^e a^t] + [a^e v^t]^*\,[a^e a^t] 
  \sepcomma\nonumber
\end{align}
for the standard vertices and the anomalous $t\tB g$-vertex, and 
\begin{align}
  a_5 
  &= [v^e v^t]^*\,[v^e d^t] + [a^e v^t]^*\,[a^e d^t] 
  \nonumber\\
  &=  [v^e v^t]^*\,{v^{e\gamma}} {d^{t\gamma}} 
    + [a^e v^t]^*\,{a^{e\gamma}} {d^{t\gamma}} 
    + [v^e v^t]^*\,{v^{eZ}} d(s)\,{d^{tZ}} 
    + [a^e v^t]^*\,{a^{eZ}} d(s)\,{d^{tZ}} 
  \nonumber\\
  &=  [v^e v^t]^*\,{v^{e\gamma}} {d^{t\gamma}} 
    + [v^e v^t]^*\,{v^{eZ}} d(s)\,{d^{tZ}} 
    + [a^e v^t]^*\,{a^{eZ}} d(s)\,{d^{tZ}} 
  \sepcomma\nonumber\\[2ex]
  a_6 
  &= [v^e v^t]^*\,[a^e d^t] + [a^e v^t]^*\,[v^e d^t] 
  \label{eq:def-a5}
  \\
  &=  [v^e v^t]^*\,{a^{e\gamma}} {d^{t\gamma}} 
    + [a^e v^t]^*\,{v^{e\gamma}} {d^{t\gamma}} 
    + [v^e v^t]^*\,{a^{eZ}} d(s)\,{d^{tZ}}
    + [a^e v^t]^*\,{v^{eZ}} d(s)\,{d^{tZ}}
  \nonumber\\
  &=  [a^e v^t]^*\,{v^{e\gamma}} {d^{t\gamma}} 
    + [v^e v^t]^*\,{a^{eZ}} d(s)\,{d^{tZ}}
    + [a^e v^t]^*\,{v^{eZ}} d(s)\,{d^{tZ}}
  \nonumber
\end{align}
for the anomalous $t\tB\gamma$- and $t\tB Z$-vertices.  
Symbols $[v^e a^t]$ etc.\ are defined in~\EqRef{eq:eff-coupling}, 
and $d(s)$ is a ratio of $Z$-propagator to 
$\gamma$-propagator [\EqRef{eq:ds-def=prop-ratio}].  
These combinations $a_{1\sim 6}$ are chosen so that the 
dependence on the initial polarization can be seen easily; 
each pair of $(a_1,a_2)$, $(a_3,a_4)$, $(a_5,a_6)$ is ``conjugate'' 
each other with respect to the exchange $e^+e^-$ couplings 
[\SecRef{sec:pol=eeB-concept}].  

\TableRef{table:P_for_a} is 
``eigenvalues'' of the products of couplings $a_1$ etc., defined in 
Eqs.~(\ref{eq:def-a1}) and (\ref{eq:def-a5}) 
under Parity transformations P for 
the final ($t\tB$) and initial ($e^-e^+$) currents.  
    It read as follows.  For example, the coupling 
    $a_5 = [v^e v^t]^*\,[v^e d^t] + [a^e v^t]^*\,[a^e d^t]$ 
    is a coefficient of the cross term of SM contribution and EW-EDM 
    contribution, $\calM_{\rm SM}^* \calM_{\text{EW-EDM}}$.  
    Thus in $|\calM|^2$, $a_5$ is accompanied by $G^*F$, 
    since 
    \begin{align*}
        \calM_{\text{SM}} \sim G + F &= \Order{1} + \Order{\beta_t}
      && \cdots
      && \text{CP-even}  \\
        \calM_{\text{C-EDM}} \sim D &= \Order{\beta_t}
      && \cdots
      && \text{CP-odd}  \\
        \calM_{\text{EW-EDM}} \sim F &= \Order{\beta_t}
      && \cdots
      && \text{CP-odd}  \sepperiod
    \end{align*}
%
%    since $\calM_{\rm SM} \propto G$ and $\calM_{\text{EW-EDM}} \propto F$.  
    Note that we are working to $\Order{\beta_t}$.  
    Under $\Parity_{t\tB}$, $v^t$-part of $\calM_{\rm SM}$ is transformed 
    with the eigenvalue $+1$, while $d^t$-part of $\calM_{\text{EW-EDM}}$ 
    is $-1$; thus in $|\calM|^2$, the term with $a_5$ should be $-1$ under 
    $\Parity_{t\tB}$.  
    A complication occurs for $a_1$ and $a_2$, which contain $v_t$, 
    which is a coefficient of both $\gamma^\mu G$, which is P-even, and 
    $i\gamma_5(p^\mu/m) D$, which is P-odd.  
    The former is $\Order{\beta^0}$, while 
    the others are $\Order{\beta}$.  
    Since now we are working to $\Order{\beta}$, 
    only the interference $G^*D$ contributes, as far as $D$ is concerned.  
    Thus $v^t v^t$ can be $\Parity_{t\tB} = \pm 1$, 
    where $+1$ for $|G|^2$, while $-1$ for $G^*D$.  
    Thus denoted in parentheses for $a_1$ and $a_2$ are 
    for $G^*D$.  
    On the other hand, $v_t$ in $a_{\text{3--6}}$ should be SM contribution, 
    not chromo-EDM contribution.  
    This is can be seem from the order counting for the velocity $\beta$; 
    all anomalous contributions to $\calM$ are $\Order{\beta}$, 
    while SM contributions are $\Order{1}$.  
    This table can be used as follows.  
    For example in $|\calM|^2$, 
    a term with $(\Parity_{t\tB},\Parity_{e^- e^+}) = (+,+)$ 
    should be accompanied by $a_1 |G|^2$.  
%    Only the term with this transformation 
%    property contributes to total cross section [\SecRef{sec:exp-corr}].  
    Likewise, 
    a term with $(\Parity_{t\tB},\Parity_{e^- e^+}) = (-,-)$ 
    should be $a_3 G^*F$, which is SM contribution, 
    or $a_2 G^*D$, which is the interference between 
    chromo-EDM contribution and SM one.  
\begin{table}[tbp]
\centering
\begin{align*}
\begin{tabular}{l||cc|cc|cc}
  & \multicolumn{2}{c}{$|G|^2$ ($G^*D$)} \vline
  & \multicolumn{4}{c}{$G^*F$} \\
  & \multicolumn{2}{c}{} \vline
  & \multicolumn{2}{c}{\scriptsize (SM)$^*$ (SM)} \vline
  & \multicolumn{2}{c}{\scriptsize (SM)$^*$ (EW-EDM)} \\
  \cline{2-7}
  & $a_1$ & $a_2$ & $a_3$ & $a_4$ & $a_5$ & $a_6$  \\
  \hline
  $\Parity_{t\tB}$    & $+$($-$) & $+$($-$) & $-$ & $-$ & $-$ & $-$  \\
  $\Parity_{e^- e^+}$ & $+$      & $-$      & $-$ & $+$ & $+$ & $-$  
\end{tabular}
\end{align*}
\begin{Caption}\caption{\small
    ``Eigenvalues'' of the products of couplings $a_1$ etc., defined in 
    Eqs.~(\ref{eq:def-a1}) and (\ref{eq:def-a5}) 
    under Parity transformations P for 
    the final ($t\tB$) and initial ($e^-e^+$) currents.  
    See text for the details.  
\label{table:P_for_a}
}\end{Caption}
\end{table}
%
%----------------------------------------------------------------------
\subsubsection{Physical considerations}
Let us see if our results for
$\diffn{\sigma}{}(\stBI,\stBBI)/\diffn{\pBI_t}{3}$ 
is plausible.  
For this purpose, it is convenient to express it in terms of vectors 
$\pBI$ and $\sBI$, not the components of them; see \SecRef{sec:top-pol-expl}.  
We already saw in~\EqRef{eq:rel-between-P-and-PB} that CP-property 
of the terms in $|\calM|^2$ relates the $\sBI$-dependence and 
$\sBBI$-dependence [See also \EqRef{eq:CPTTfor_ffB}].  
Here we shall see that Parity-property can be used to derive 
momentum-dependence.  
As was explained in~\SecRef{sec:pol=eeB-concept}, 
non-zero polarization $\chi$ for initial $e^+e^-$ current 
mixes the vector and axial vertices, thus it complicates 
symmetry considerations.  
However as explained in \SecRef{sec:pol=eeB-concept}, 
the results for non-zero $\chi$ can easily be recovered 
from those for $\chi=0$ by replacing couplings appropriately: 
$a_1 \to a_1 + \chi a_2$, for example.  
Thus in this section, we consider the case when the spins of 
$e^+ e^-$ are summed with even weights: $\chi = 0$.  
Before calculating the traces for $|\calM|^2$, it is easy to see that 
each momentum%
\footnote{We are working at the \scCM-frame of $e^+e^-$.  }
$(\ptBI,\peBI)$ 
can appear only twice or less in $|\calM|^2$, 
and each spin-vector $(\stBI,\stBBI)$ can appear only once or less.  
However since we are working to $\Order{\beta_t}$ or less, 
momenta of $t$ and $\tB$ cannot appear twice: 
$|\ptBI| = \beta_t\gamma_t\mt$, 

Let us start with 
the unpolarized part $\diffn{\sigma_{\rm unpol}}{}$.  
As is shown in~\EqRef{eq:oppositeCP=no-total}, 
\itCP-odd part of the (product of the) $t\tB$ current 
$(jj^\dagger)_{t\tB} \equiv j_{t\tB}j_{t\tB}^\dagger$ 
do not contribute to 
spin-summed cross section.  Since the anomalous interactions are 
$\ChargeCOp\ParityOp$-odd while the Standard ones are even, 
the interferences between them are \itCP-odd.  
This means those interferences ($\calM_{\rm SM}^* \calM_{\rm EDM}$) 
do not contribute to $\diffn{\sigma_{\rm unpol}}{}$; 
the interference among anomalous contributions 
($\calM_{\rm EDM}^* \calM_{\rm EDM}$) is $\Order{\beta_t^2}$, and thus is 
beyond our concern here; 
only $\calM_{\rm SM}^* \calM_{\rm SM}$ contributes to 
$\diffn{\sigma_{\rm unpol}}{}$.  
Since $\diffn{\sigma_{\rm unpol}}{}$ is a scalar, $(\peBI,\ptBI)$-dependence 
of it is either $|\peBI|^2$, $\dprod{\peBI}{\ptBI}$, or $|\ptBI|^2$; 
see~\TableRef{table:mom-to-aGD}.  
\begin{table}[tbp]
\centering
\begin{align*}
\begin{tabular}{l|ccccc}
  & & $(\Parity_{e^+e^-},\Parity_{t\tB})$ 
  & {\scriptsize (SM)$^*$ (SM)} 
  & {\scriptsize (SM)$^*$ (C-EDM)} & {\scriptsize (SM)$^*$ (EW-EDM)}
  \\ \hline
  $|\peBI|^2$            & $\Order{1}$       
  & $(+,+)$ & $a_1 |G|^2$ & & 
  \\
  $\dprod{\peBI}{\ptBI}$ & $\Order{\beta_t}$ 
  & $(-,-)$ & $\bRe{a_3 G^*F}$ & & 
%  \\
%  $|\ptBI|^2$            & $\Order{\beta_t^2}$ 
%  & & & & 
  \\[1.5ex]
  $\peBI$                                  & $\Order{1}$       
  & $(-,+)$ & $a_2 |G|^2$ & & 
  \\
  $\ptBI$\, ,~$\peBI\pdprod{\peBI}{\ptBI}$ & $\Order{\beta_t}$ 
  & $(+,-)$ & $a_4 G^*F$ & $a_1 G^*D$ & $a_5 G^*F$
  \\
  $\cprod{\peBI}{\ptBI}$                   & $\Order{\beta_t}$ 
  & $(-,-)$ & $a_3 G^*F$ & $a_2 G^*D$ & $a_6 G^*F$
  \\[1.5ex]
  $(\peBI)_i\,(\peBI)_j$               & $\Order{1}$       
  & $(+,+)$ & $a_1 |G|^2$ & & 
  \\
  $(\peBI)_i\,(\ptBI)_j$               & $\Order{\beta_t}$ 
  & $(-,-)$ & $a_3 G^*F$ & $a_2 G^*D$ & $a_6 G^*F$
  \\
  $(\peBI)_i\,(\cprod{\peBI}{\ptBI})_j$ & $\Order{\beta_t}$ 
  & $(+,-)$ & $a_4 G^*F$ & $a_1 G^*D$ & $a_5 G^*F$
\end{tabular}
\end{align*}
\begin{Caption}\caption{\small
    Note that spin ($\sBI$) is even under $\Parity$.  
    Thus $\Parity_{t\tB}$ etc.\ is determined solely by momenta 
    ($\peBI,\ptBI$).  On the other hand, \rmCPTT\ depends on $\sBI$.  
\label{table:mom-to-aGD}
}\end{Caption}
\end{table}
Among these, $|\ptBI|^2$ is $\Order{\beta_t^2}$, 
and thus is beyond our concern here.  
``Eigenvalues'' under $\Parity_{e^+e^-}$ etc.\ can be easily seen; 
for example, $\dprod{\peBI}{\ptBI}$ has odd number of $\peBI$ 
and odd number of $\ptBI$; thus it has 
$(\Parity_{e^+e^-},\Parity_{t\tB}) = (-,-)$.  
Accompanied couplings and Green functions can also be determined 
from~\TableRef{table:P_for_a}; for example, 
a term of $(\Parity_{e^+e^-},\Parity_{t\tB}) = (-,-)$ is accompanied by 
either $a_2 G^*D$ or $a_3 G^*F$.  But the former do not contributes to 
$\diffn{\sigma_{\rm unpol}}{}$, since it is 
$\calM_{\rm SM}^* \calM_{\rm EDM}$.  
Since $\dprod{\peBI}{\ptBI}$ is \rmCPTT-even, it is accompanied by 
$\pRe{a_3 G^*F}$; 
note that a \rmCPTT-even (-odd) term is proportional to ``real'' 
(``imaginary'') part [\SecRef{sec:absorp-CPTT}].  
Among $\diffn{\sigma_{\rm unpol}}{}$, only the part with $\Parity_{t\tB}=+1$ 
contributes to the total cross section 
$\sigma_{\rm tot}$ [\EqRef{eq:oppositeP=no-total}]; 
thus $\sigma_{\rm tot} \propto a_1|G|^2$.  
On the other hand, $\Parity_{t\tB}=-1$ part contributes to the 
Forward-Backward asymmetry 
$A_{\rm FB} = \mean{\sgn(\dprod{\peBI}{\ptBI})}/\sigma_{\rm tot}$ 
[\SecRef{sec:exp-corr}].  
Thus the ``coupling'' of the term 
$\dprod{\peBI}{\ptBI} \propto \beta_t\cos\theta_{te}$ is 
$\pRe{a_3 G^*F}$.  
This completes the analysis of $\diffn{\sigma_{\rm unpol}}{}$.  

Next we consider the terms proportional to the spin-vectors: 
$\dprod{\PolBR_t}{\stBI} + \dprod{\PolBBR_t}{\stBBI}$.  
Since the polarization vectors $\PolBR_t$ and $\PolBBR_t$ are 3-vectors, 
each term of them is proportional to either of $\peBI$, $\ptBI$ or 
$\cprod{\peBI}{\ptBI}$.  The number of $\ptBI$ should be zero or one, 
since we are working to $\Order{\beta_t}$; note that $|\peBI|^2 = E^2$, 
where $E$ is beam energy.  Thus the momentum dependence of 
$\PolBR_t$ and $\PolBBR_t$ is either $\peBI$, $\ptBI$, 
$\peBI\pdprod{\peBI}{\ptBI}$, or $\cprod{\peBI}{\ptBI}$.  
These%
\footnote{
  Here $\peBI\pdprod{\peBI}{\ptBI}$ is the vector $\peBI$ with the 
  coefficient $\dprod{\peBI}{\ptBI}$.  
}
are listed in~\TableRef{table:mom-to-aGD}.  
Also shown is the order of velocity; only the interference among 
SM contributions are $\Order{1}$, and the others are $\Order{\beta_t}$.  
Note that under \rmCPTT, ``eigenvalues'' of $(\pBI,\sBI+\sBBI,\sBI-\sBBI)$ are 
$(-,-,+)$ [\TableRef{table:PCT=E,p,s}].  
From this, for example, 
one can see that the chromo-EDM contribution to the CP-odd term%
\footnote{
  Of course there is no such contribution to the CP-even term 
  such as $(\stBI+\stBBI){\cdot}(\dprod{\peBI}{\ptBI})$.  
}
$(\stBI-\stBBI){\cdot}(\dprod{\peBI}{\ptBI})$ in $\diffn{\sigma}{}$ 
is proportional to $\pRe{a_2G^*D}$, since the term is \rmCPTT-even.  
This completes the analysis of the term 
$\dprod{\PolBR_t}{\stBI} + \dprod{\PolBBR_t}{\stBBI}$ 
in $\diffn{\sigma_{\rm unpol}}{}$, 
except the relative magnitude of each term, especially 
of $\ptBI$ and $\peBI\pdprod{\peBI}{\ptBI}$.  
Our result for $\delta\PolBR$ shows that they are proportional to 
either 
\begin{align*}
  \dprod{(\stBI-\stBBI)}{(\ptBI - \peBI\frac{\dprod{\peBI}{\ptBI}}{|\peBI|^2})}
  \quad\mbox{or}\quad
  \dprod{(\stBI-\stBBI)}{\pcprod{\peBI}{\ptBI}}
  \sepcomma
\end{align*}
which are both perpendicular to the beam direction $\peBI$.  
It seems that we miss some symmetry.  
However for the effect of chromo-EDM, it can be understood as follows.  
For that case, $t\tB$ is produced in $(L,S)=(0,1)$ or $J^{PC}=1^{--}$ 
at first.  Note that $\Parity = (-1)^{L+1}$ and $\ChargeC = (-1)^{L+S}$ 
for fermion--anti-fermion pair.  
Then they are modified by the anomalous $t\tB g$-EDM potential 
[\EqRef{eq:CEDM-pot}], which has $(\Parity,\ChargeC)=(-,+)$; 
thus the resulting state is $J^{PC}=1^{+-}$ or $(L,S)=(1,0)$.  
On the other hand, 
$\braket{\mbox{spin-1}}{(S_i-\bar{S}_i)}{\mbox{spin-0}}$ is zero 
for $i=z$, and non-zero for $i=x,y$.  
Thus there is no contribution to parallel component.  

One also see the terms proportional to $\stBI^i\stBBI^j$ 
can be understood in the same manner, 
except the anomalous contributions: 
\begin{align*}
  \dprod{\pcprod{\stBI}{\stBBI}}
        {(\ptBI - \peBI\frac{\dprod{\peBI}{\ptBI}}{|\peBI|^2})}
  \quad\mbox{or}\quad
  \dprod{\pcprod{\stBI}{\stBBI}}
        {\pcprod{\peBI}{\ptBI}}
  \sepperiod
\end{align*}

We can see that neglecting $\mathcal{O}(\beta_t )$ and higher, 
\begin{align*}
  \frac{\diffn{\sigma}{}(\stBI,\stBBI)}{\diffn{\pBI_t}{3}}
  \propto
  1 + \Cpara^0(\spara+\sBpara) + \spara\sBpara
\end{align*}
Thus if $\spara = 1$, which means the spin of $t$ is parallel to $\peBI$, 
then $\sBpara \neq -1$, which means the spin of $\tB$ cannot be anti-parallel 
to $\peBI$, or should be parallel to $\peBI$ [\SecRef{sec:pol=ttB-concept}].  
The situation is more clear for the case $\chi = \pm 1$, 
or $\Cpara^0=\mp 1$: 
\begin{align*}
  \frac{\diffn{\sigma}{}(\stBI,\stBBI)}{\diffn{\pBI_t}{3}}
  \propto
  (1\mp\spara)(1\mp\sBpara)
\end{align*}
This factorization shows that when $\chi=+1$ [$-1$], 
or when the helicity of initial $(e^-,e^+)$ is $(L,R)$ [$(R,L)$], 
the spins of both $t$ and $\tB$ are anti-parallel [parallel] to $\peBI$.  
For other choice of $\chi$, the spins of $t$ and $\tB$ have no definite 
direction.  
Including $\mathcal{O}(\beta_t )$ corrections, the factorization do not hold 
even for $\chi=\pm 1$.  
%
%---------------------------------------------------------------------
%
%
%

%
%
%--------------------------------------------------------------------
\section{Results for EDMs}
Here comes our numerical results.  
We shall see that it is possible to disentangle that 
the effects of each three anomalous interactions $\dtg$, $\dtp$, and $\dtZ$.  

\begin{figure}[tbp]
  \hspace*{\fill}
  \begin{minipage}{6cm}
    \includegraphics{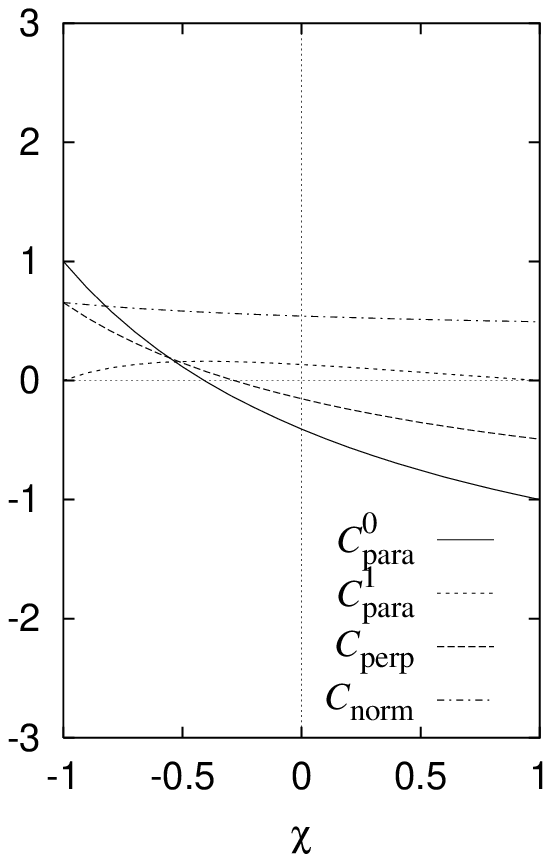}
  \end{minipage}
  \hspace*{\fill}
  \begin{minipage}{6cm}
    \includegraphics{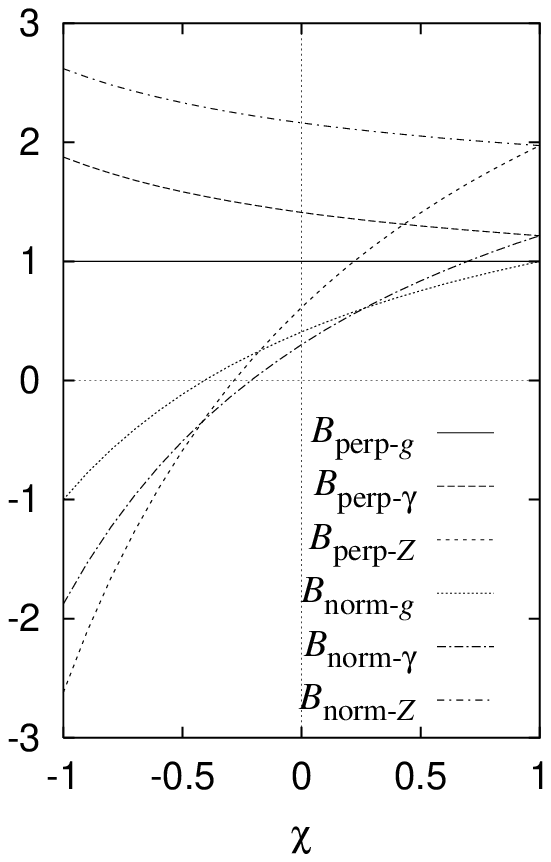}
  \end{minipage}
  \hspace*{\fill}
  \\
  \hspace*{\fill}
  \begin{minipage}{6cm}
    \includegraphics{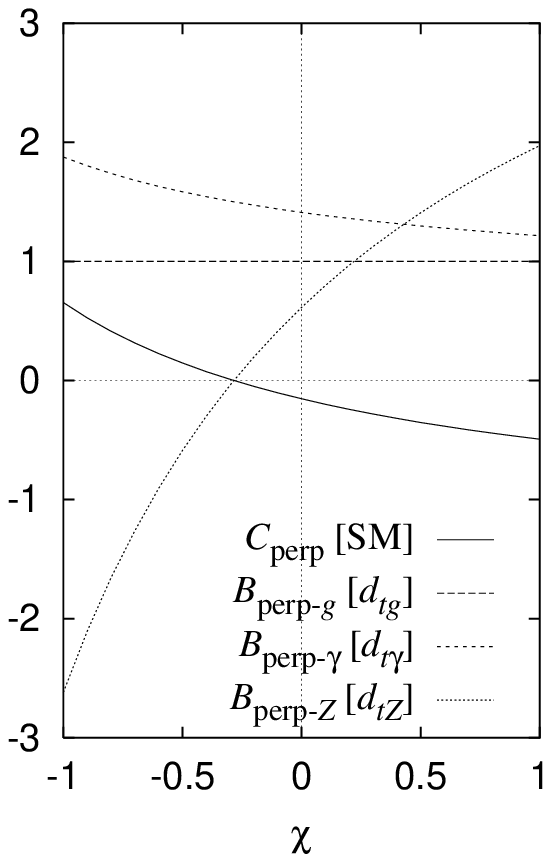}
  \end{minipage}
  \hspace*{\fill}
  \begin{minipage}{6cm}
    \includegraphics{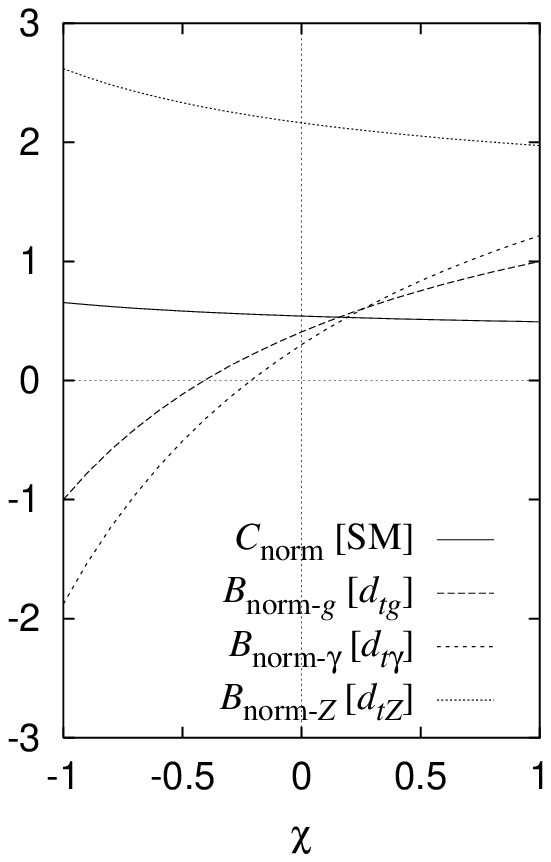}
  \end{minipage}
  \hspace*{\fill}
  \\
  \hspace*{\fill}
  \begin{Caption}\caption{\small
      Coefficients for polarizations as functions of initial polarization 
      $\chi$.  In the figures, 
      $C_{\rm para}=\Cpara$, $C_{\rm perp}=\Cperp$ and $C_{\rm norm}=\Cnorm$. 
      (Left--top) Coefficients for SM contributions.  
      (Right--top) Coefficients for anomalous EDM couplings.  
      Lines in the lower two figures the same to those in the upper two 
      figures: 
      (Left-bottom) Coefficients for the polarization parallel to $\nBIperp$. 
      (Right-bottom) Coefficients for the polarization parallel to $\nBInorm$. 
      \label{fig:Cpara}
  }\end{Caption}
  \hspace*{\fill}
\end{figure}

\begin{figure}[tbp]
  \hspace*{\fill}
  \begin{minipage}{8cm}
    \includegraphics[width=8cm]{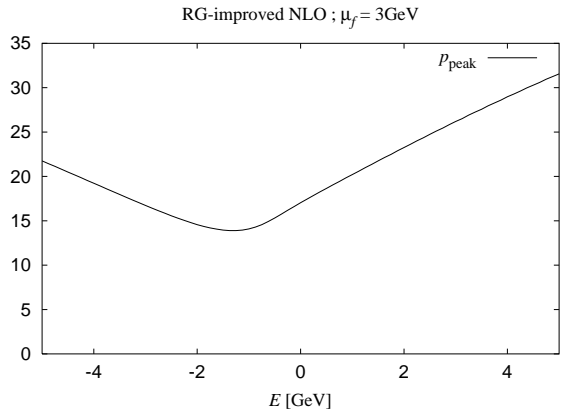}
  \end{minipage}
  \hspace*{\fill}
  \\
  \hspace*{\fill}
  \begin{Caption}\caption{\small
      Typical momentum of top quark, which is defined by the 
      peak momentum $p_{\rm peak}$ of the momentum distribution 
      $\diffn{\sigma}{}/\diffn{p}{} \propto |\pBI|^2 |G|^2$.  
      It is approximately 
      $p_{\rm peak} \simeq \left| \sqrt{m_t\,(E+1\GeV+i\Gamma_t)} \right|$.  
      Here $1\GeV \simeq 2m_t-M_{1S} = \mbox{binding energy}$.  
      Note that if $\Gamma_t=0$, the momentum of $t$ is uniquely determined.  
      \label{fig:FT_scanE_p_peak}
  }\end{Caption}
  \hspace*{\fill}
\end{figure}

If all the couplings $\dtg$ etc.\ are real and 
the decay width $\Gamma_Z$ of $Z$ is neglected, 
the effects of anomalous interactions for polarization are 
\begin{align*}
  & 
  \delta\Polperp
  =
  \left( 
%    d_{tg} + \Bperp^{\gamma} d_{t\gamma} + \Bperp^{Z} d_{tZ} 
    {} - \Bperp^g d_{tg} + \Bperp^{\gamma} d_{t\gamma} + \Bperp^{Z} d_{tZ} 
  \right)
  \pIm{\frac{F}{G}} \beta_t \sin\theta_{te}
  \sepcomma\\
  &
  \delta\Polnorm
  =
  \left[
      \left( 
%        \Bnorm^g d_{tg} + \Bnorm^{\gamma} d_{t\gamma} + \Bnorm^{Z} d_{tZ} 
        {} - \Bnorm^g d_{tg} + \Bnorm^{\gamma} d_{t\gamma} + \Bnorm^{Z} d_{tZ} 
      \right)
      \pRe{\frac{F}{G}}
%    - \Bnorm^g d_{tg}
    + \Bnorm^g d_{tg}
  \right] \beta_t \sin\theta_{te}
  \sepperiod
\end{align*}
This assumption is realistic, since the imaginary part is generated by 
physical threshold in loop diagram.  
From Figures~\ref{fig:FT_scanE_b} and \ref{fig:FT_scanE_p_peak}, 
we can see that 
Chromo-EDM $\dtg$ and EW-EDMs $\dtp,\dtZ$ 
can be separated through their energy dependence, 
since typical $\beta_t$ become larger for larger energy.  
The $\bRe{(F-G)/G}\beta_t$ becomes smaller for larger energy, 
because $F=G$ for the case $V=0$; 
note that the effect of rescattering becomes smaller as one leaves 
from the threshold.  
On the other hand, from~\FigRef{fig:Cpara}, we can see that 
the two EW-EDMs, $\dtp$ and $\dtZ$, can be separated through the 
response with respect to the initial polarization $\chi$.  
This is also pointed out some time ago~\cite{CR95} 
in the context of open top analysis.  
Note also that from the definitions, 
$\Bperp = \pm\Bnorm$ for $\chi=\pm 1$.  

The relation
\begin{align}
  p_{\rm peak} \simeq \left| \sqrt{m_t\,(E+1\GeV+i\Gamma_t)} \right|
  \label{eq:p_peak_appr}
\end{align}
agrees qualitatively with \FigRef{fig:FT_scanE_p_peak}.  
Here $1\GeV \simeq 2m_t-M_{1S} = \mbox{binding energy}$.  
Note that with $V \to 0$, the relation 
$p_{\rm peak} = \left| \sqrt{m_t\,(E+i\Gamma_t)} \right|$ 
holds%
\footnote{
  This can be obtained from 
  $G = 1/(p_t^2/m_t-(E+i\Gamma_t))$ for $V \to 0$, and 
  ${\rm d}\sigma/{\rm d}p_t \propto p_t^2|G|^2$.  
}.  

\begin{figure}[tbp]
  \hspace*{\fill}
  \begin{minipage}{8cm}
    \includegraphics[width=8cm]{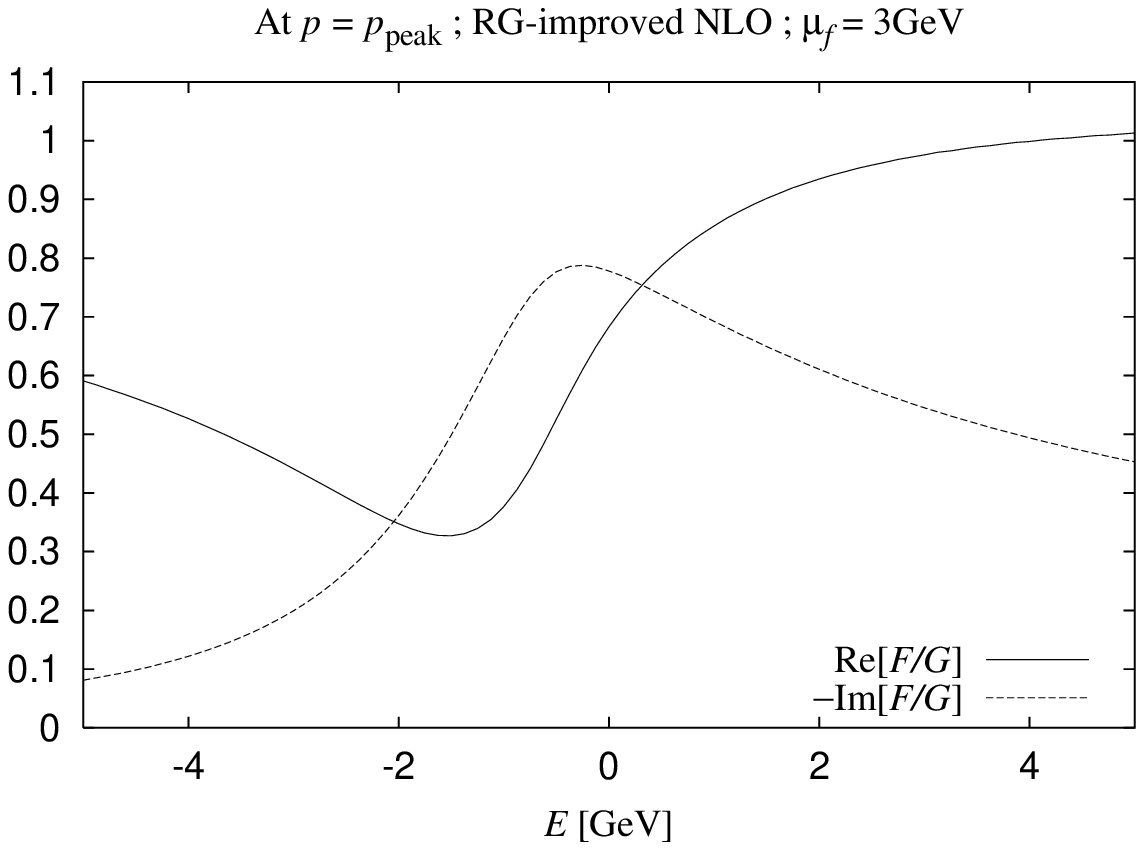}
  \end{minipage}
  \hspace*{\fill}
  \begin{minipage}{8cm}
    \includegraphics[width=8cm]{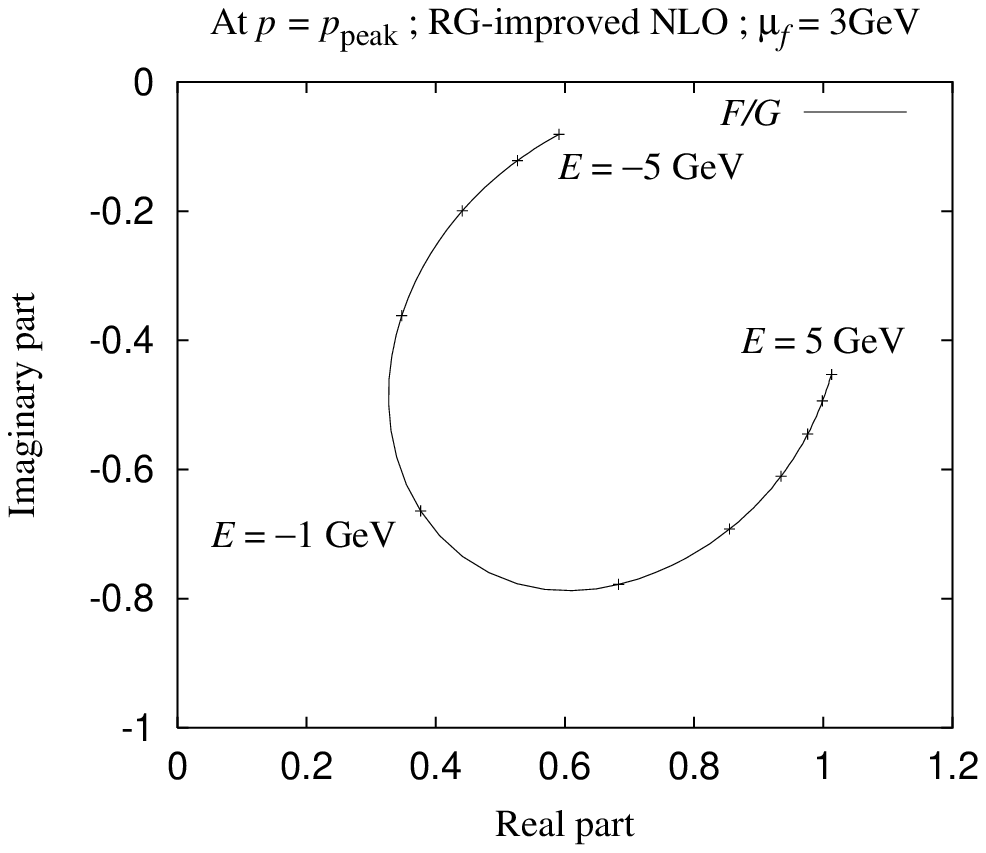}
  \end{minipage}
  \hspace*{\fill}
  \\
  \hspace*{\fill}
  \begin{Caption}\caption{\small
      The ratio of the Green functions $F(p)/G(p)$ at the 
      peak momentum $p_{\rm peak}$ of the momentum distribution 
      $\diffn{\sigma}{}/\diffn{p}{}$.  
      In operator language, $G(p) = \left<p\right|G\left|r'=0\right>$.  
      In order to avoid the complexity to NNLO (= $\Order{1/c^2}$), 
      we use NLO QCD potential with RG improved in momentum space.  
      We also use Potential-Subtracted (PS) mass 
      $m_{\rm PS}(\mu_f) = 175\GeV$ with $\mu_f=3\GeV$.  
      Other inputs are 
      $\Gamma_t=1.43\GeV$, and $\alpha_s(m_Z)=0.118$.  
      Two figures are exactly the same: 
      (Left) Real and imaginary parts of the ratio $F/G$ are 
      plotted with respect to $E=\sqrt{s}-2m_{\rm PS}$.  
      (Right) The ratio $F/G$ is plotted in the complex plane.  
      \label{fig:FT_scanE}
  }\end{Caption}
  \hspace*{\fill}
\end{figure}

\begin{figure}[tbp]
  \hspace*{\fill}
  \begin{minipage}{8cm}
    \includegraphics[width=8cm]{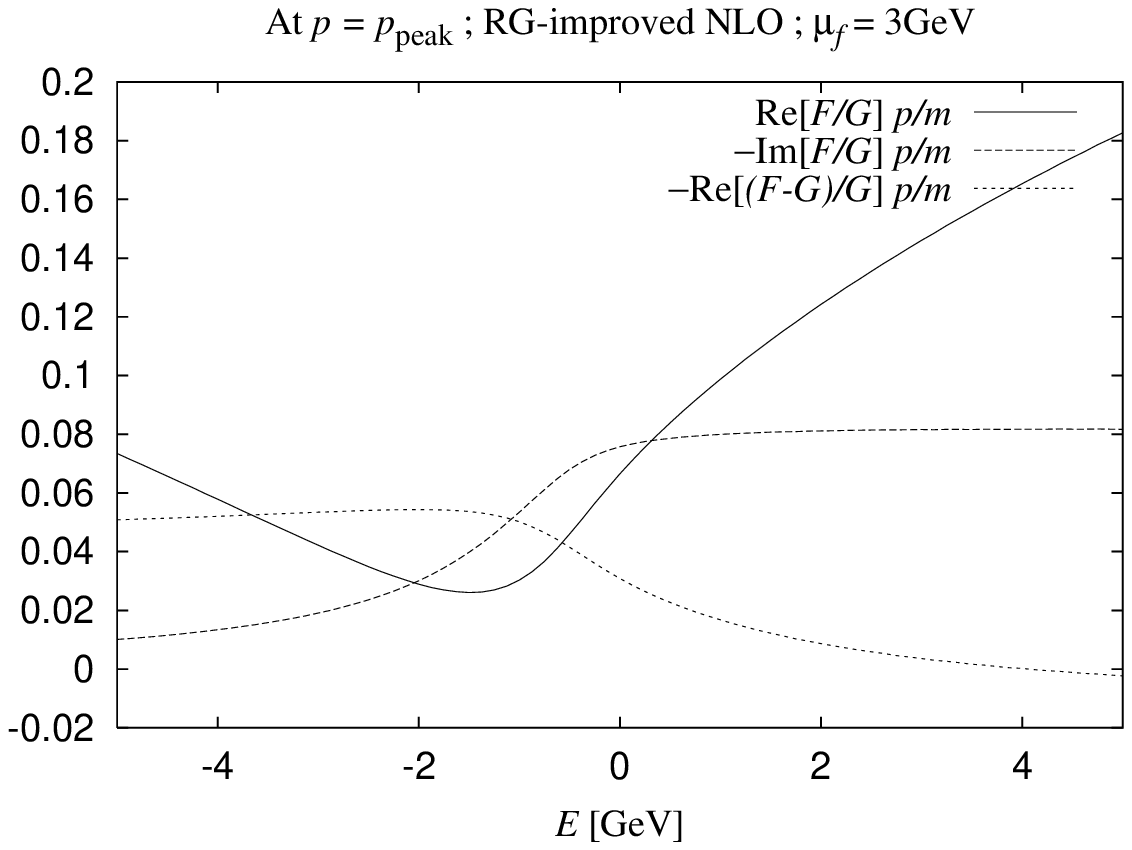}
  \end{minipage}
  \hspace*{\fill}
  \begin{minipage}{8cm}
    \includegraphics[width=8cm]{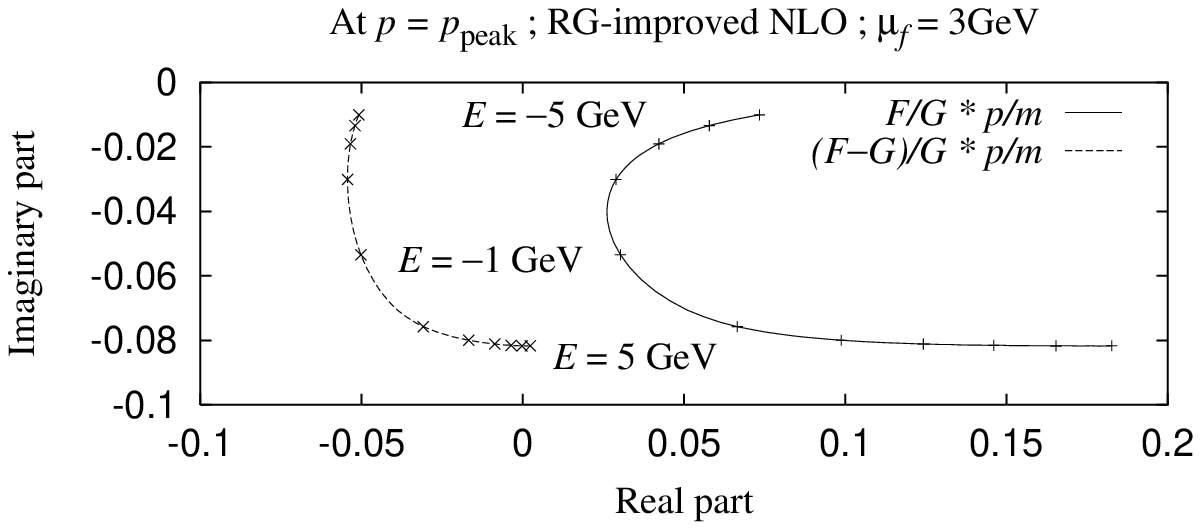}
  \end{minipage}
  \hspace*{\fill}
  \\
  \hspace*{\fill}
  \begin{Caption}\caption{\small
      The ratios of the Green functions times ``velocity'' of 
      top quark $F(p)/G(p) \cdot p/\mt$ and 
      $(F-G)/G \cdot p/\mt$ at the 
      peak momentum $p_{\rm peak}$ of the momentum distribution 
      $\diffn{\sigma}{}/\diffn{p}{}$.  
      Inputs are the same as in \FigRef{fig:FT_scanE}.  
      Two figures are exactly the same: 
      (Left) Real and imaginary parts of the ratios $F/G \cdot p/\mt$ 
      and $(F-G)/G \cdot p/\mt$ are plotted with respect to 
      $E=\sqrt{s}-2m_{\rm PS}$.  
      Note that imaginary part of them coincides each other.  
      (Right) The same ratios are plotted in the complex plane.  
      \label{fig:FT_scanE_b}
  }\end{Caption}
  \hspace*{\fill}
\end{figure}

For stable quark under LO Coulomb rescattering, $G$ and $F$ can be 
obtained analytically~\cite{HJKP97}: 
\begin{align*}
  &  \lim_{\stackrel{\scriptstyle \Gamma_t \to 0}
                    {E \to p^2/m_t}              }
  \left( E - \frac{\pBI^2}{\mt} + i\Gamma_t \right) G(\pBI,E) 
  = \exp\!\pfrac{\pi p_\rmB}{2p}\,
    \Gamma\!\left(1+i\frac{p_\rmB}{p}\right)
  \sepcomma\\
  &  \lim_{\stackrel{\scriptstyle \Gamma_t \to 0}
                    {E \to p^2/m_t}              }
  \left( E - \frac{\pBI^2}{\mt} + i\Gamma_t \right) F(\pBI,E) 
  = \left( 1-i\frac{p_\rmB}{p} \right) \, \exp\!\pfrac{\pi p_\rmB}{2p}\,
    \Gamma\!\left(1+i\frac{p_\rmB}{p}\right)
  \sepcomma
\end{align*}
where $p_\rmB = C_F \alpha_s m_t/2 \simeq 20\GeV$.  
Thus 
\begin{align*}
  \left. \frac{F}{G} \right|_{p=p_{\rm peak}}
  \simeq 1-i\frac{p_\rmB}{p_{\rm peak}}
  \simeq 1-i\frac{p_\rmB}{\sqrt{m_t\,(E+1\GeV+i\Gamma_t)}}
  \sepperiod
\end{align*}
This agrees qualitatively well with \FigRef{fig:FT_scanE}.  
Note that there is no energy dependence of $\bRe{F/G}$ if 
$\Gamma_t = 0$.  
Thus this may provide a way to measure the decay width of top quark.  

\begin{figure}[tbp]
  \hspace*{\fill}
  \begin{minipage}{8cm}
    \includegraphics[width=8cm]{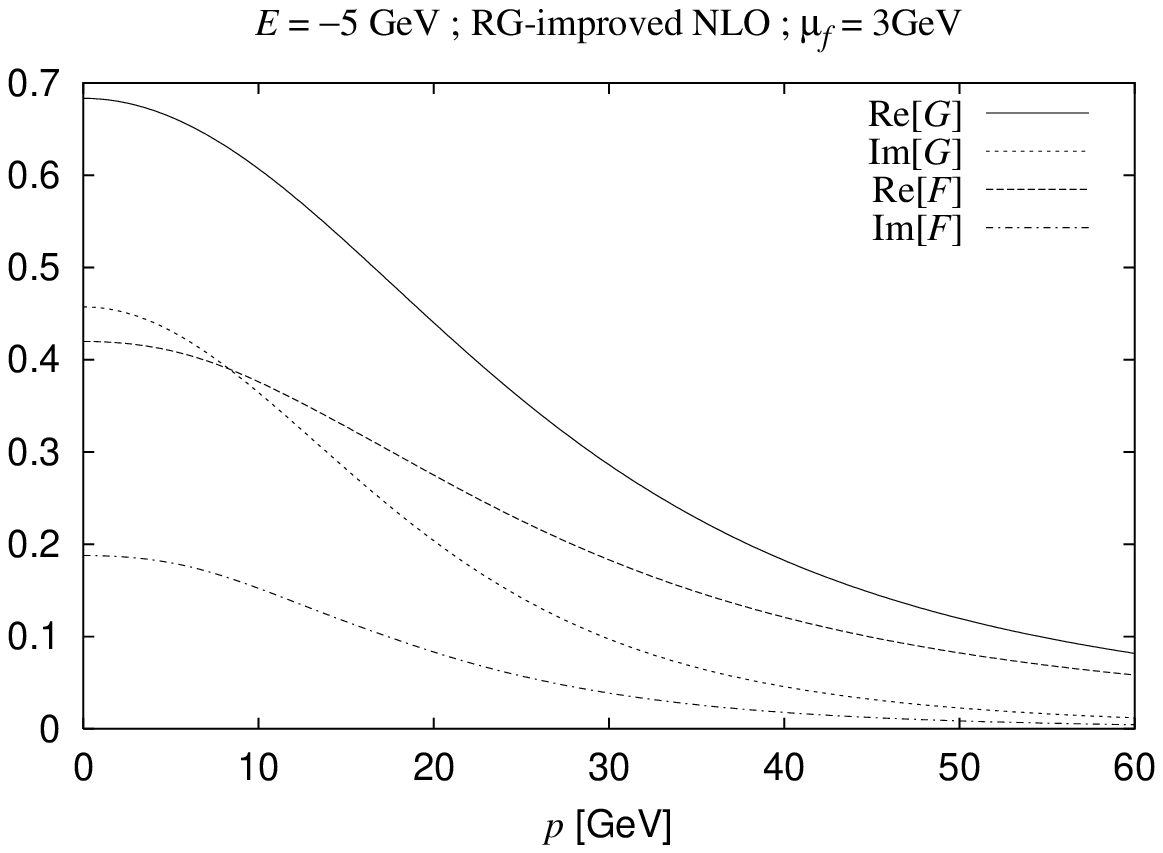}
  \end{minipage}
  \hspace*{\fill}
  \begin{minipage}{8cm}
    \includegraphics[width=8cm]{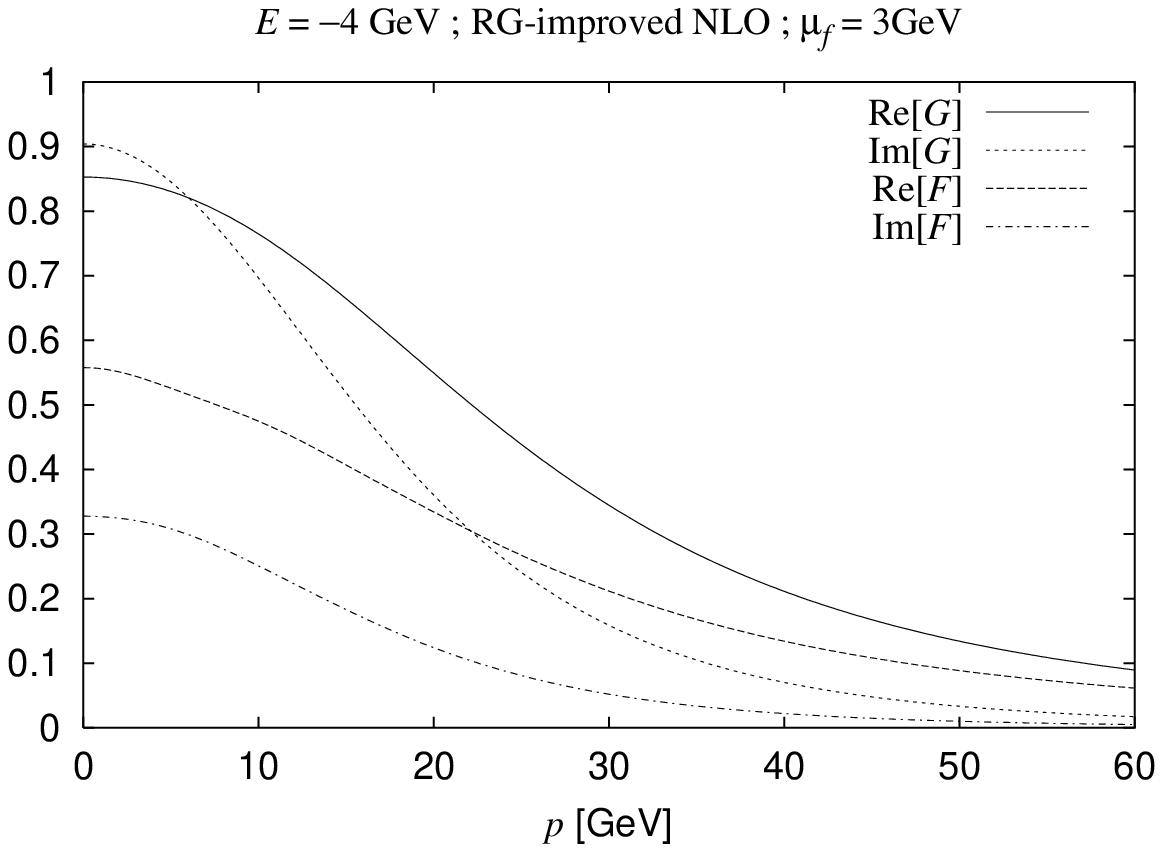}
  \end{minipage}
  \hspace*{\fill}
  \\
  \hspace*{\fill}
  \begin{minipage}{8cm}
    \includegraphics[width=8cm]{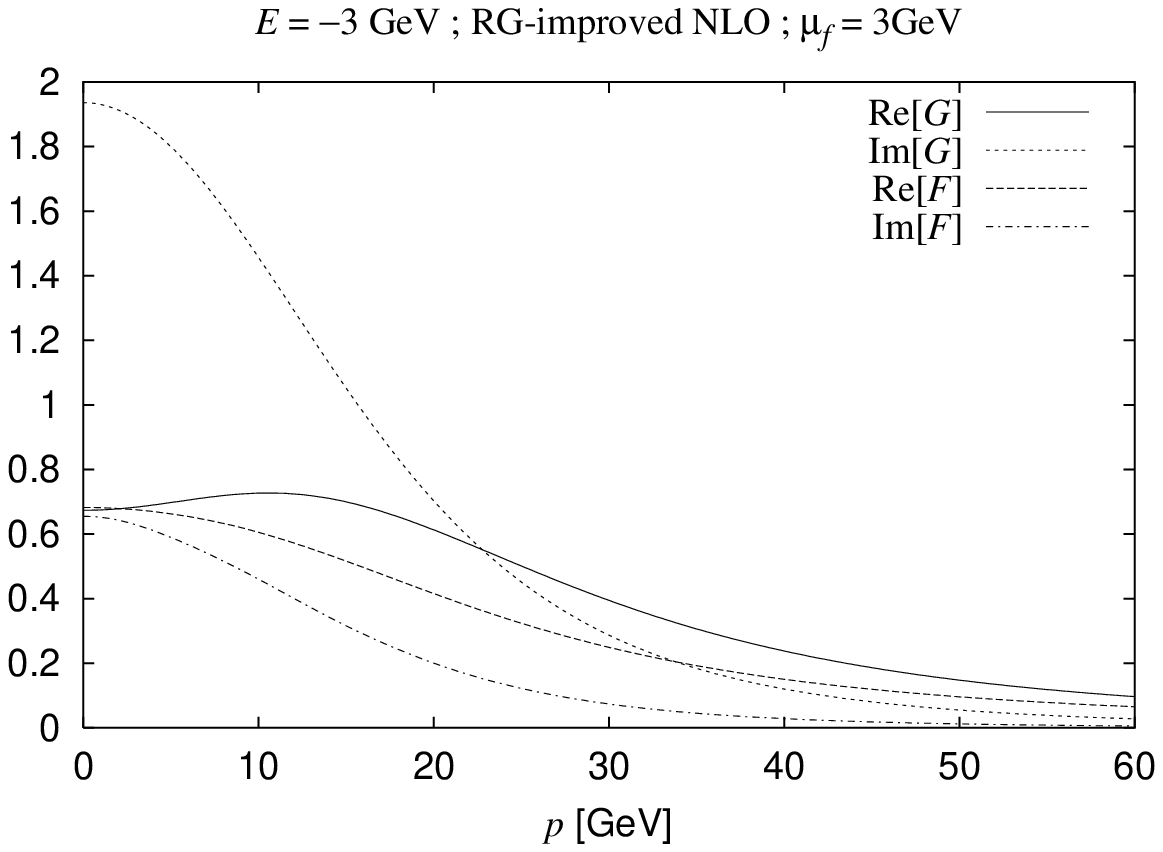}
  \end{minipage}
  \hspace*{\fill}
  \begin{minipage}{8cm}
    \includegraphics[width=8cm]{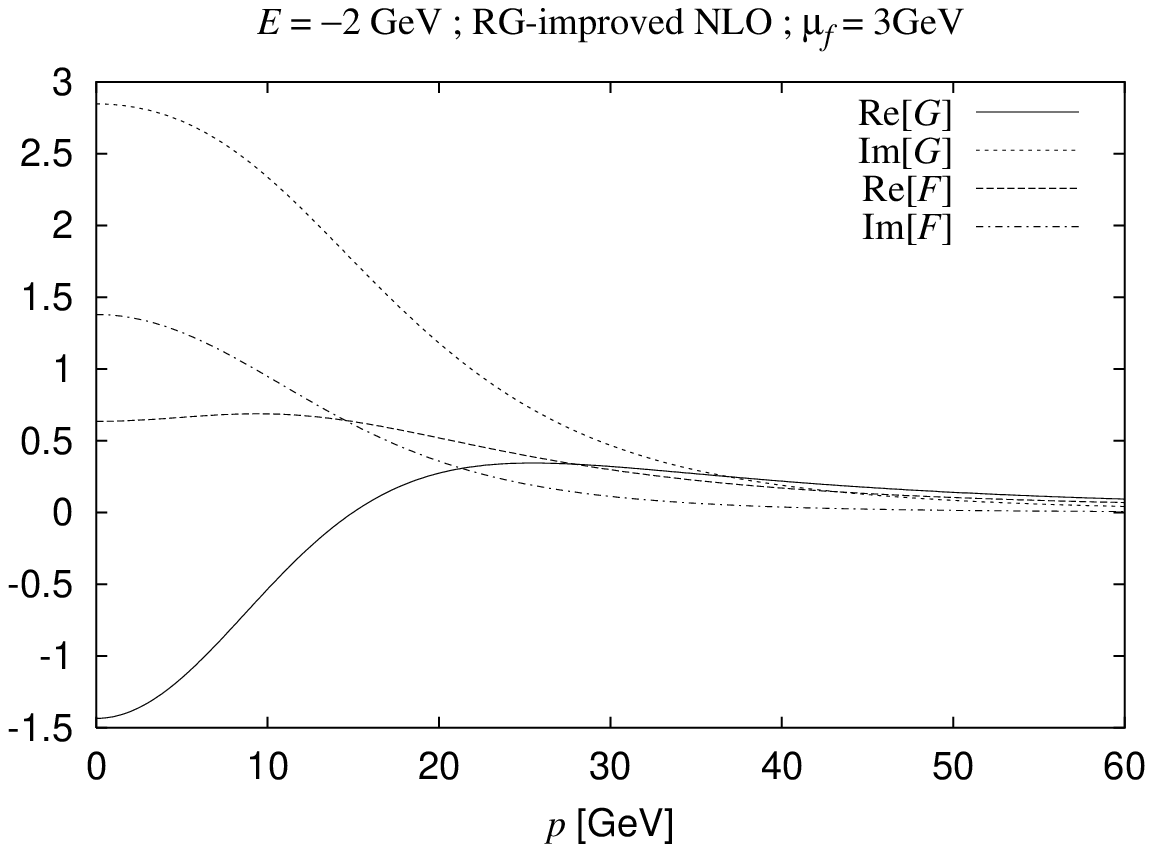}
  \end{minipage}
  \hspace*{\fill}
  \\
  \hspace*{\fill}
  \begin{minipage}{8cm}
    \includegraphics[width=8cm]{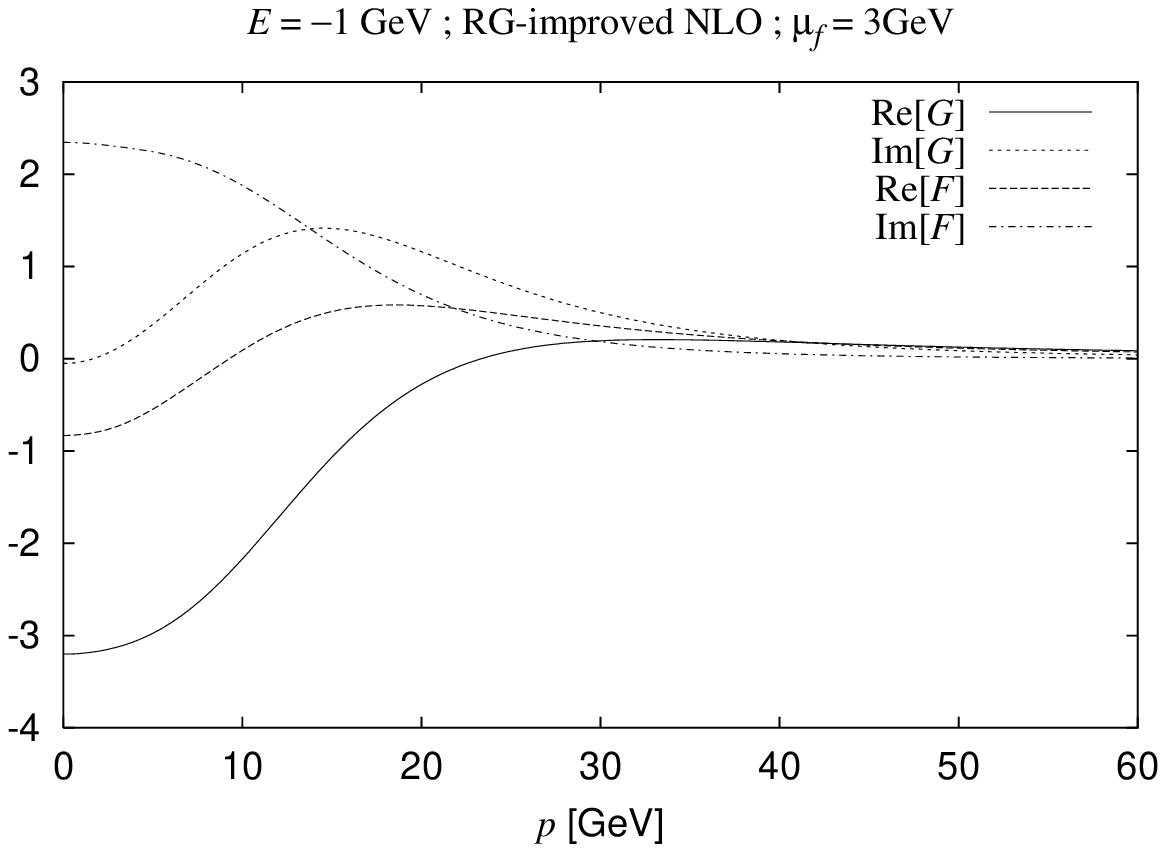}
  \end{minipage}
  \hspace*{\fill}
  \begin{minipage}{8cm}
    \includegraphics[width=8cm]{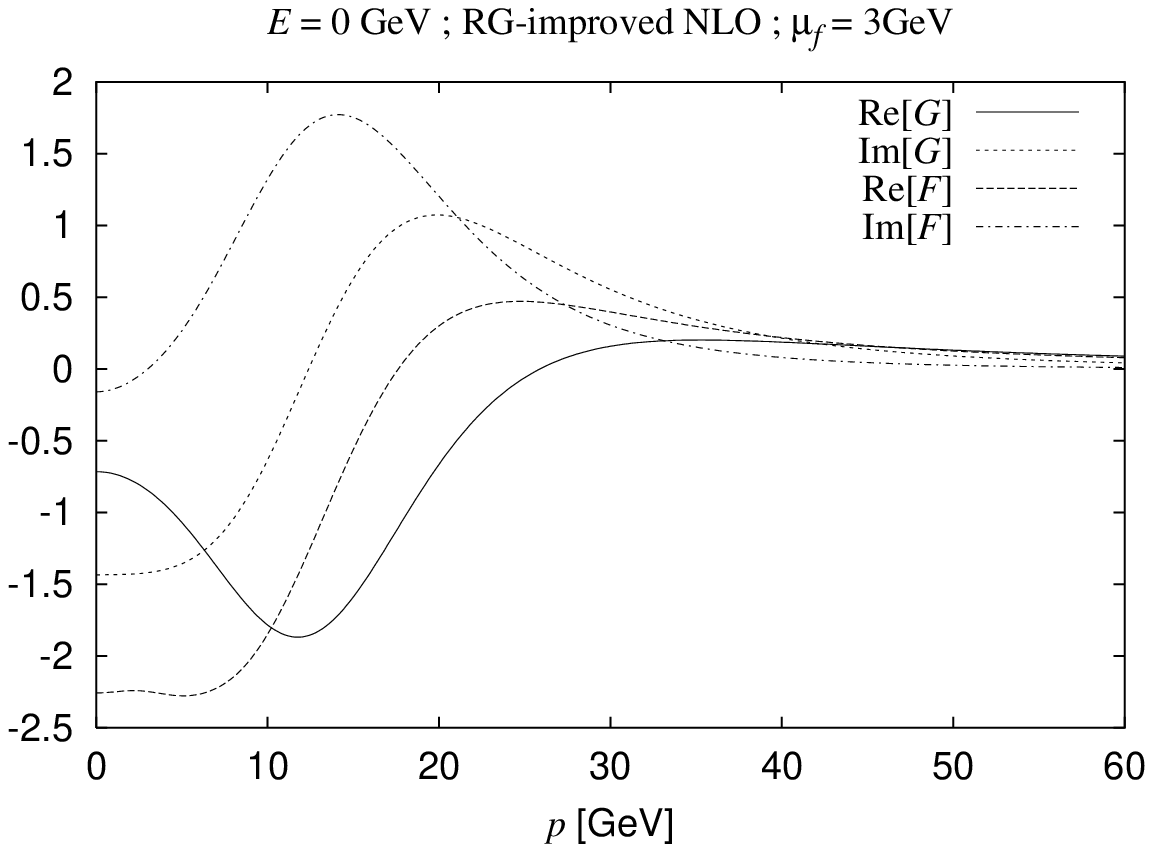}
  \end{minipage}
  \hspace*{\fill}
  \\
  \hspace*{\fill}
  \begin{Caption}\caption{\small
      Green functions
      \label{fig:FT_-5,0}
  }\end{Caption}
  \hspace*{\fill}
\end{figure}
\begin{figure}[tbp]
  \hspace*{\fill}
  \begin{minipage}{8cm}
    \includegraphics[width=8cm]{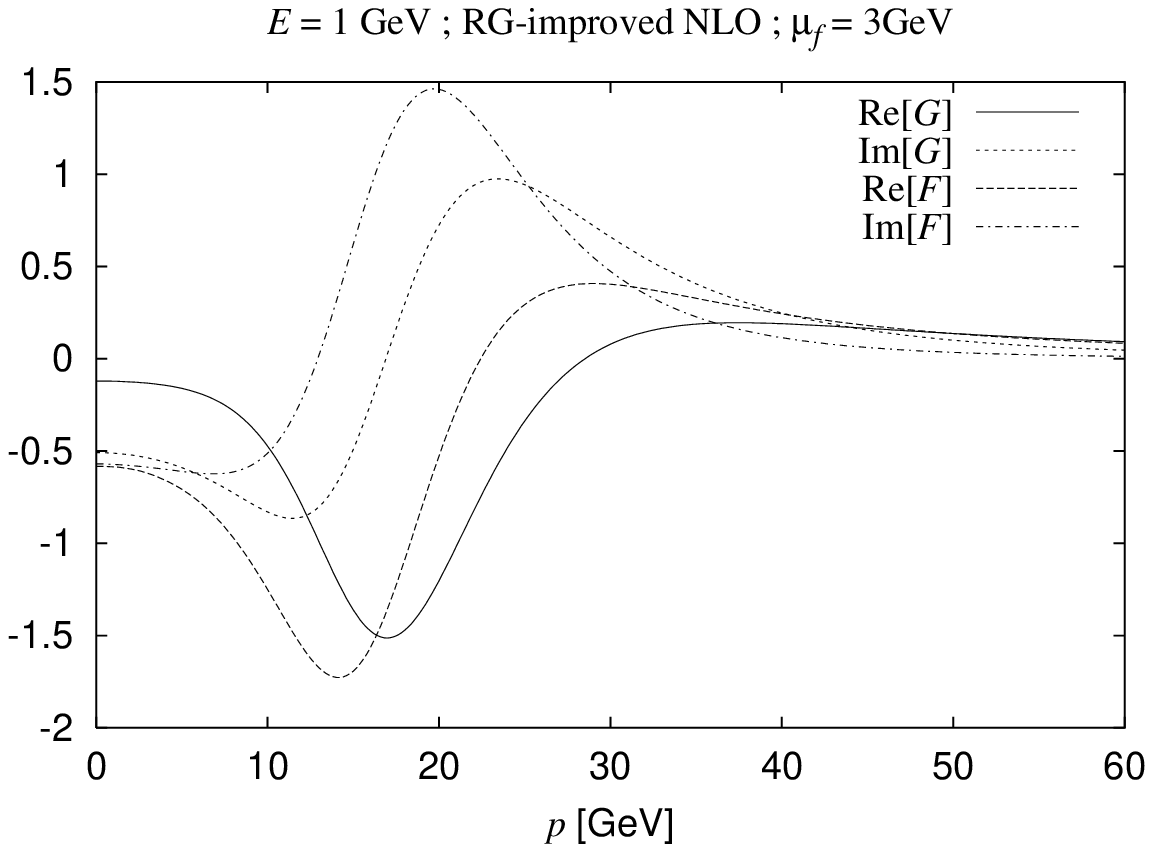}
  \end{minipage}
  \hspace*{\fill}
  \begin{minipage}{8cm}
    \includegraphics[width=8cm]{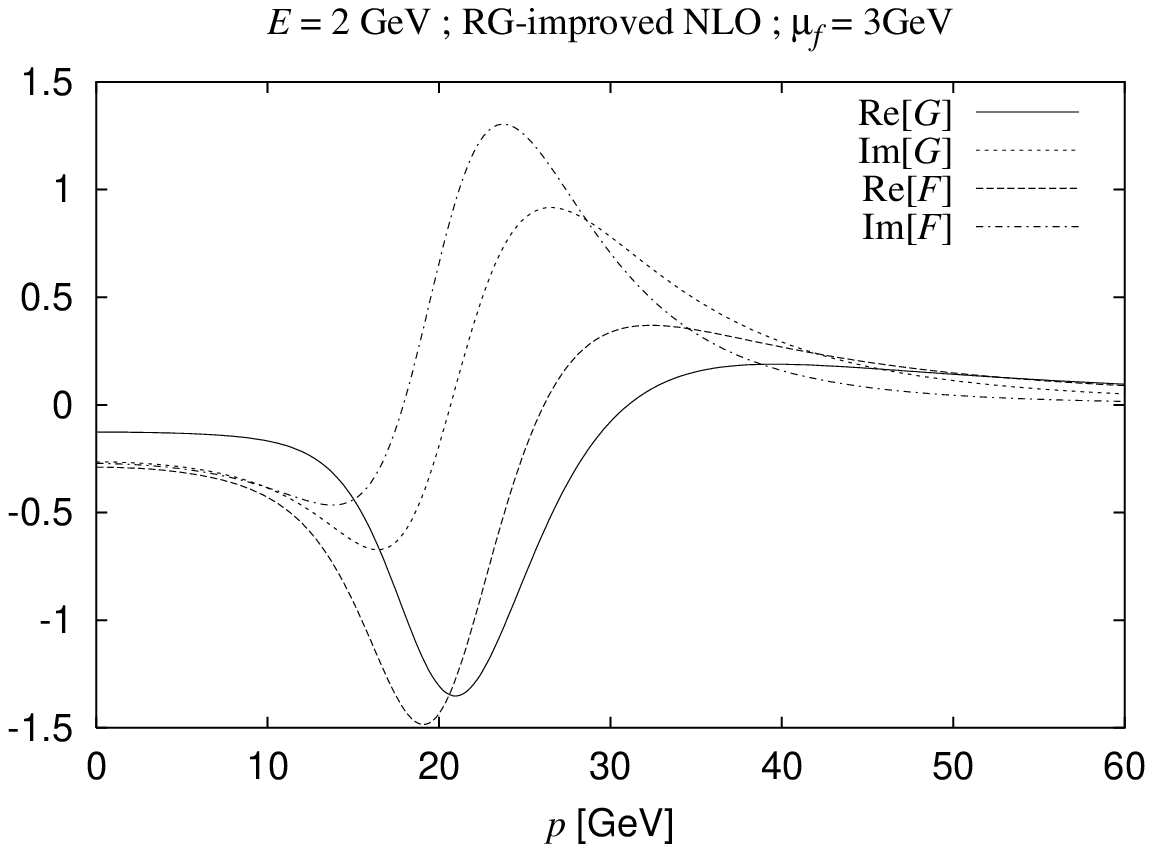}
  \end{minipage}
  \hspace*{\fill}
  \\
  \hspace*{\fill}
  \begin{minipage}{8cm}
    \includegraphics[width=8cm]{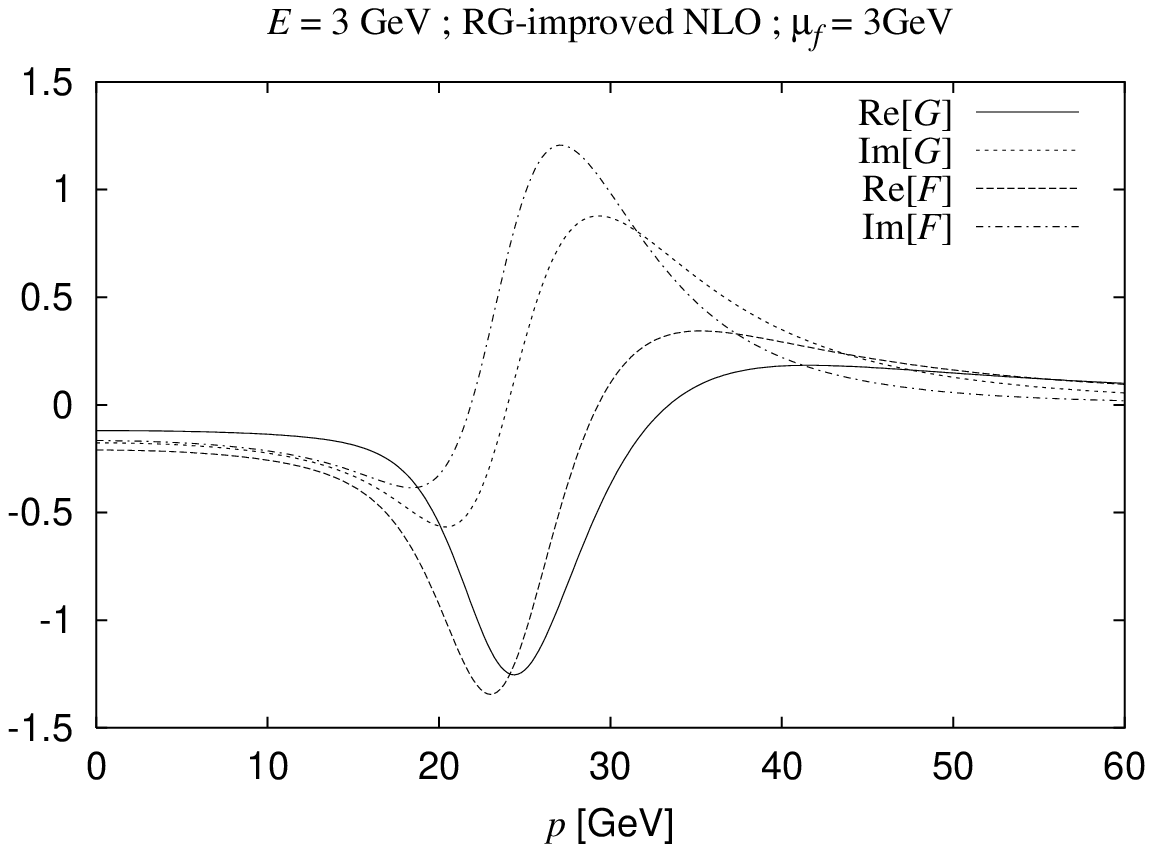}
  \end{minipage}
  \hspace*{\fill}
  \begin{minipage}{8cm}
    \includegraphics[width=8cm]{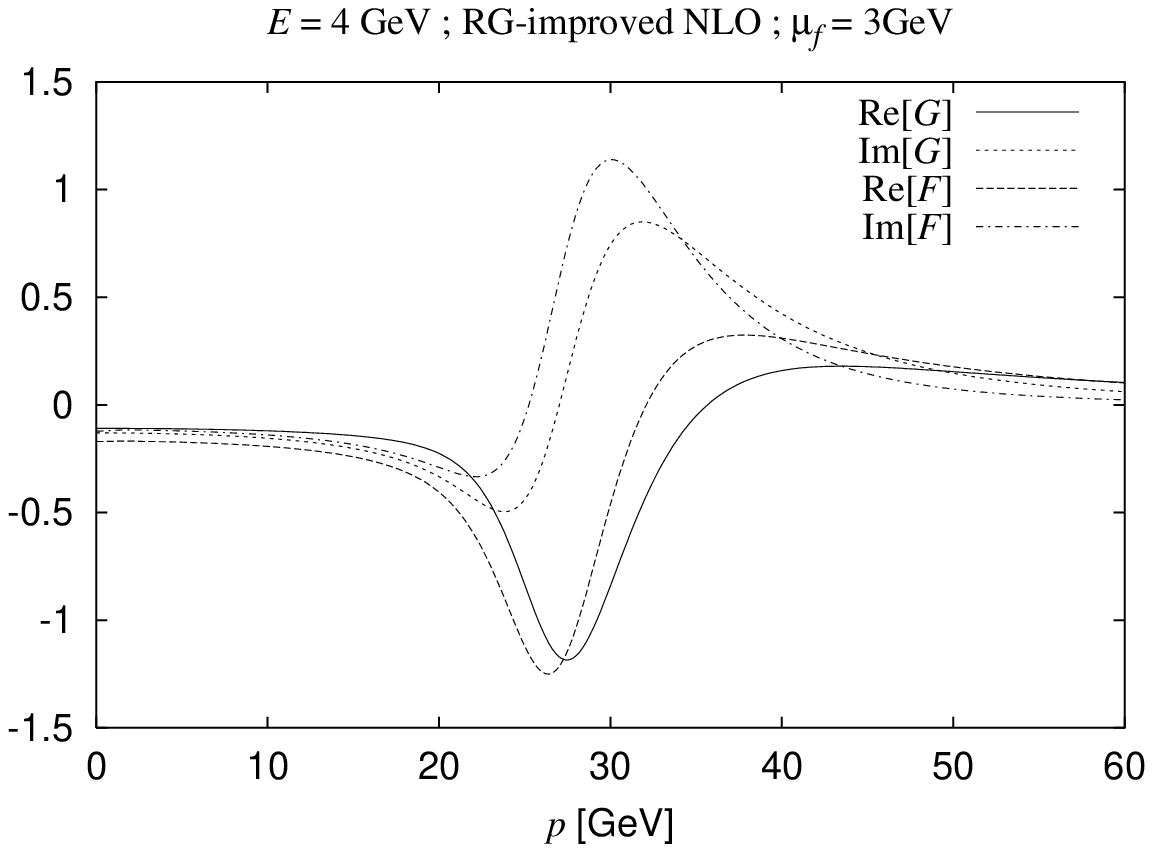}
  \end{minipage}
  \hspace*{\fill}
  \\
  \hspace*{\fill}
  \begin{minipage}{8cm}
    \includegraphics[width=8cm]{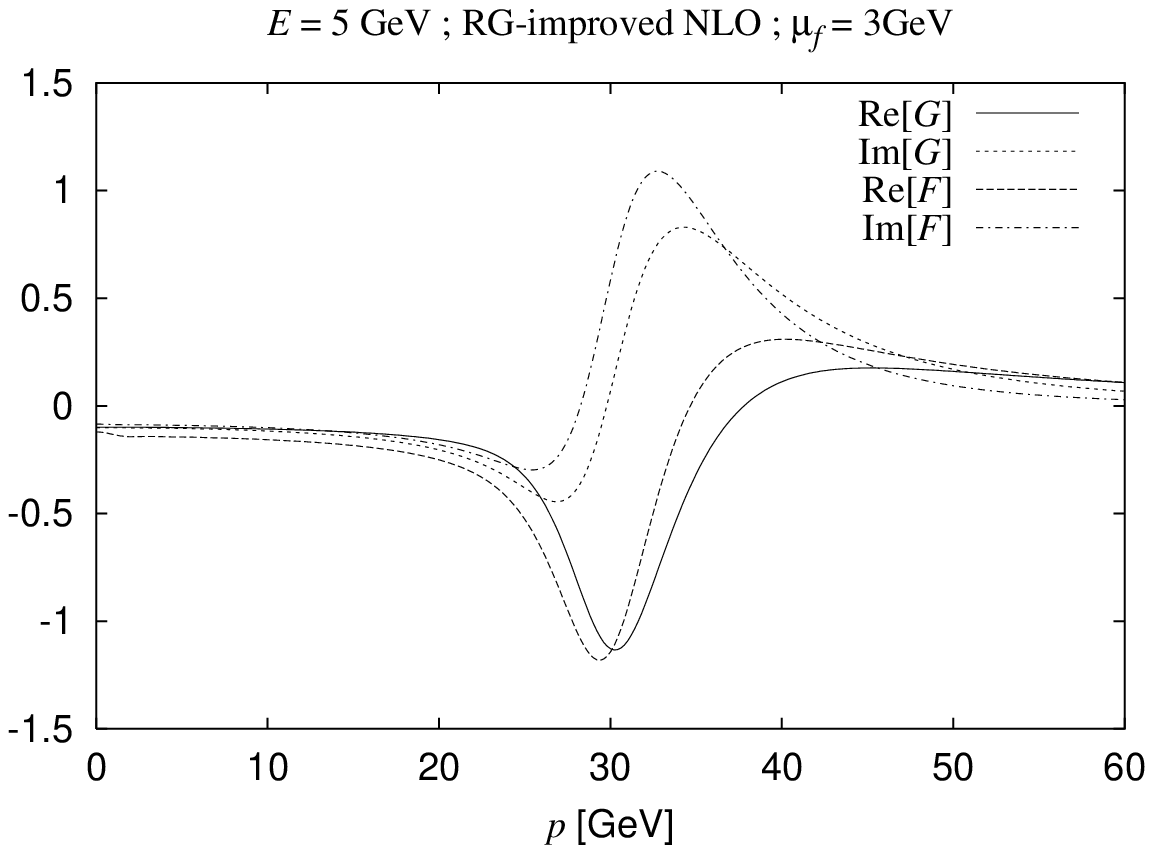}
  \end{minipage}
  \hspace*{\fill}
  \\
  \hspace*{\fill}
  \begin{Caption}\caption{\small
      Green functions
      \label{fig:FT_1,5}
  }\end{Caption}
  \hspace*{\fill}
\end{figure}
\begin{figure}[tbp]
  \hspace*{\fill}
  \begin{minipage}{8cm}
    \includegraphics[width=8cm]{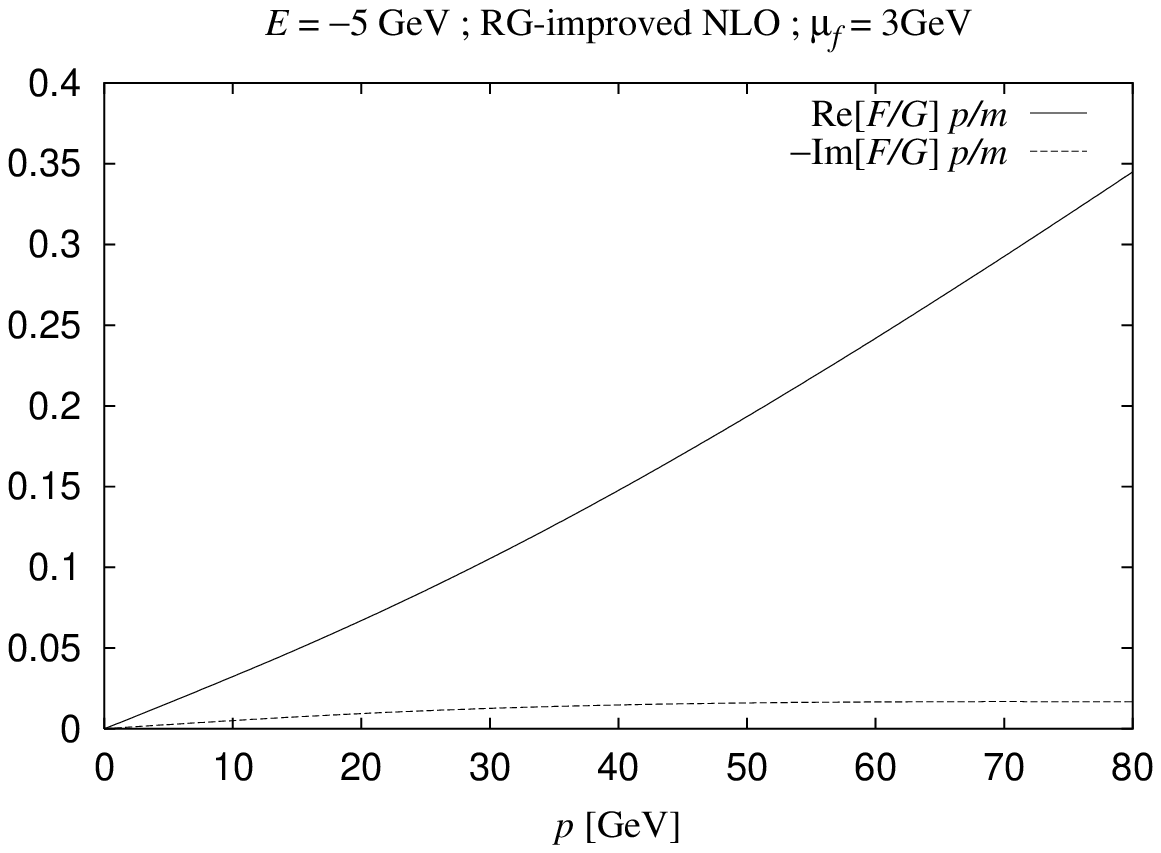}
  \end{minipage}
  \hspace*{\fill}
  \begin{minipage}{8cm}
    \includegraphics[width=8cm]{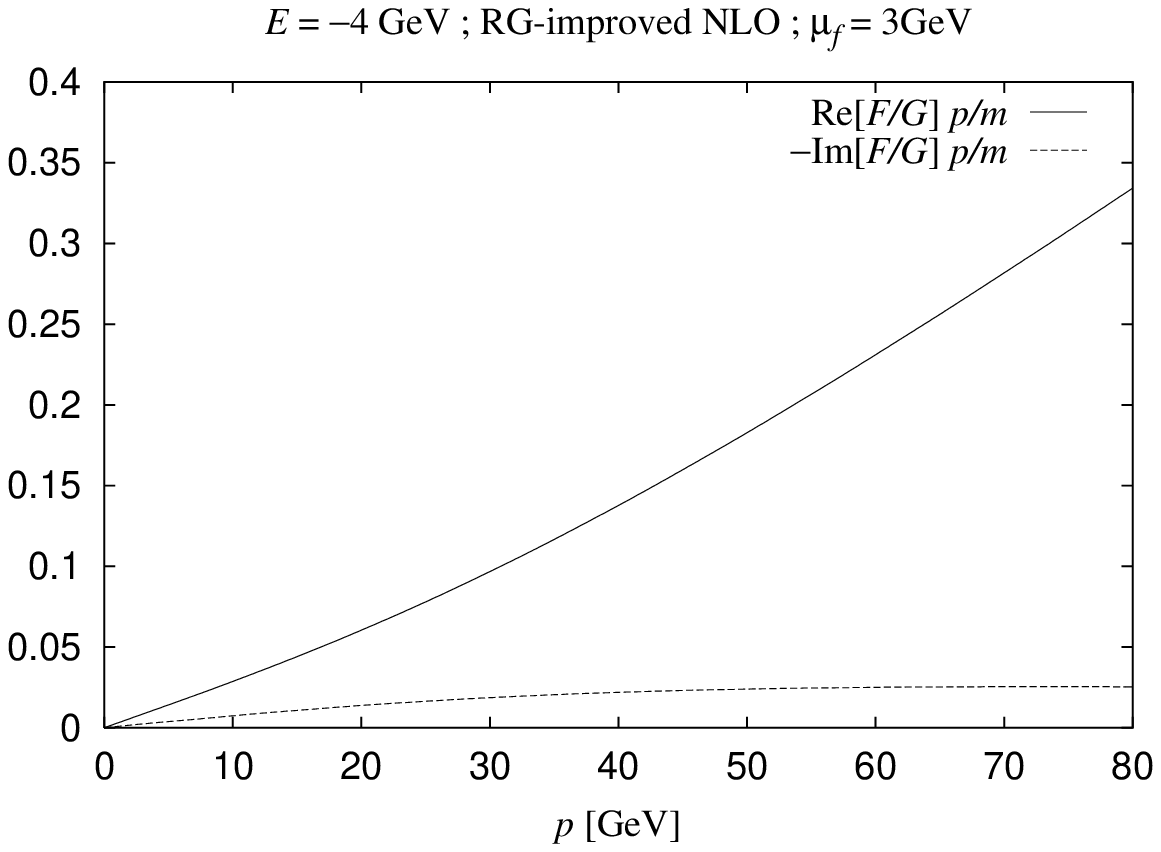}
  \end{minipage}
  \hspace*{\fill}
  \\
  \hspace*{\fill}
  \begin{minipage}{8cm}
    \includegraphics[width=8cm]{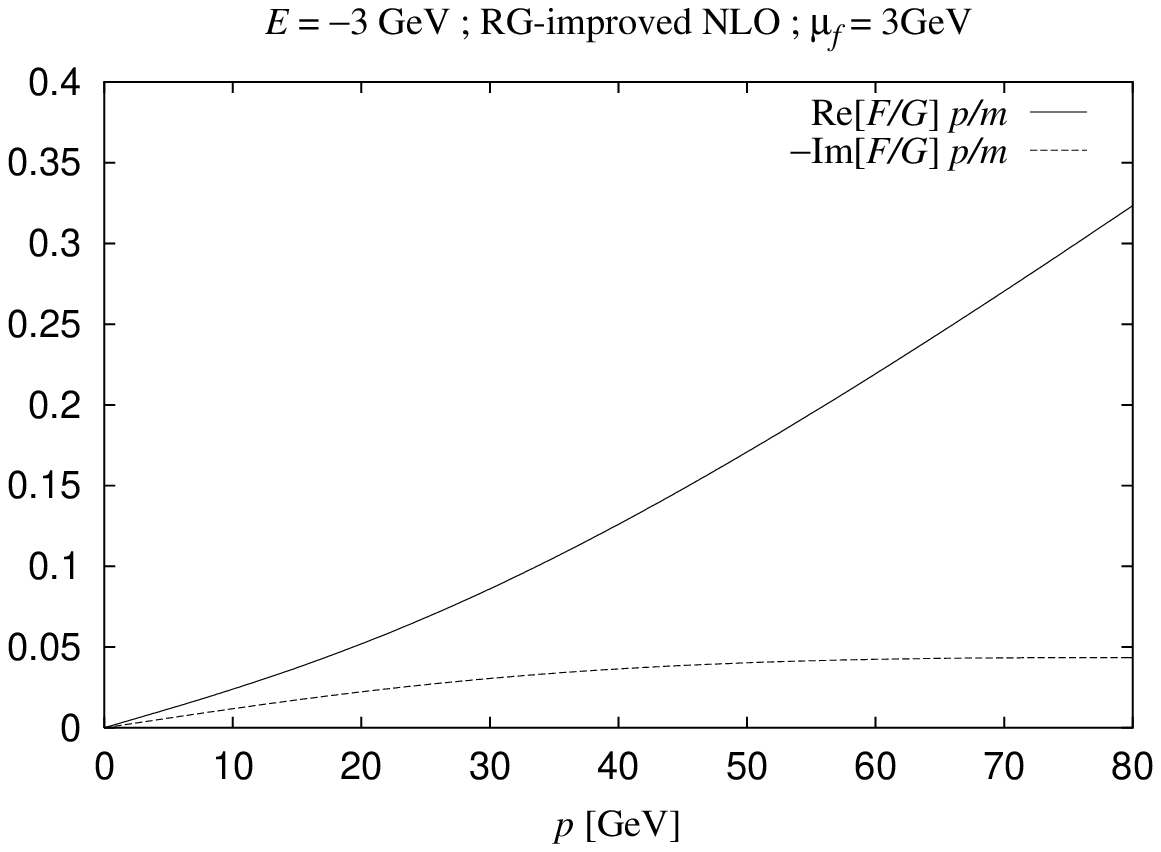}
  \end{minipage}
  \hspace*{\fill}
  \begin{minipage}{8cm}
    \includegraphics[width=8cm]{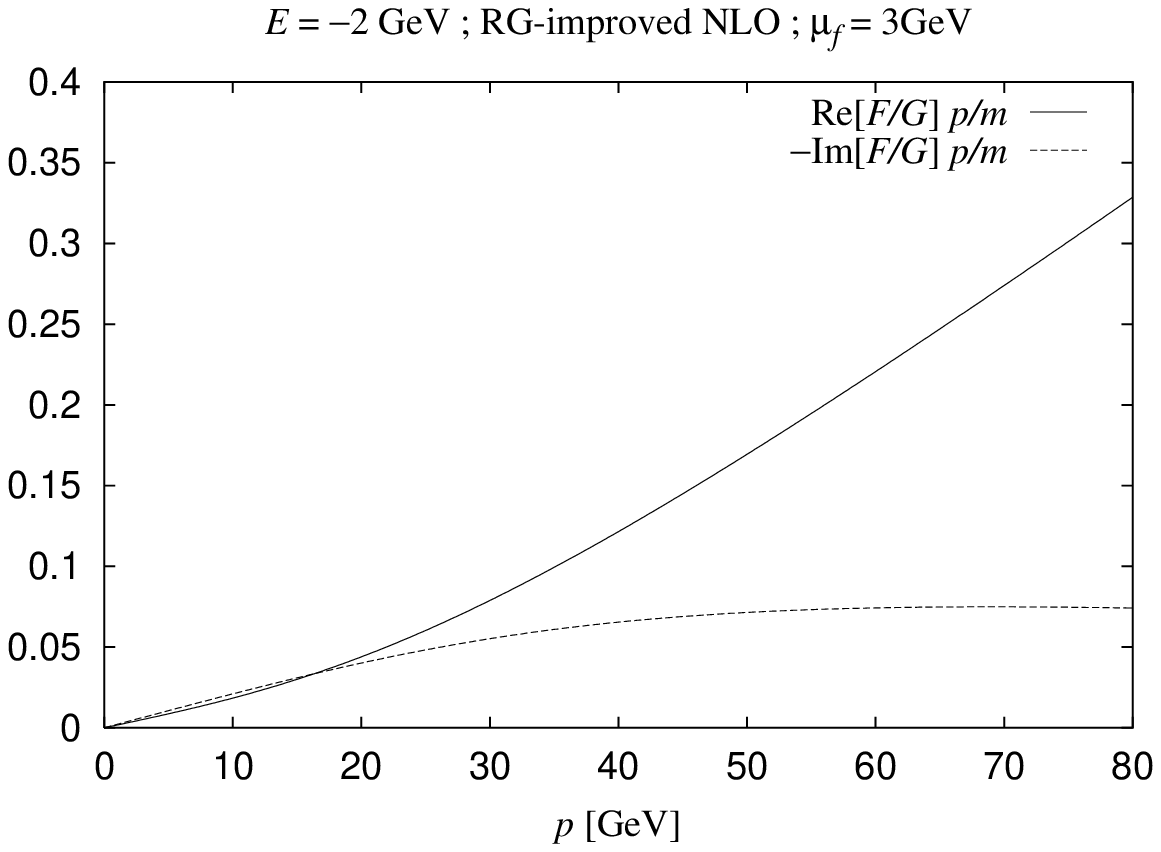}
  \end{minipage}
  \hspace*{\fill}
  \\
  \hspace*{\fill}
  \begin{minipage}{8cm}
    \includegraphics[width=8cm]{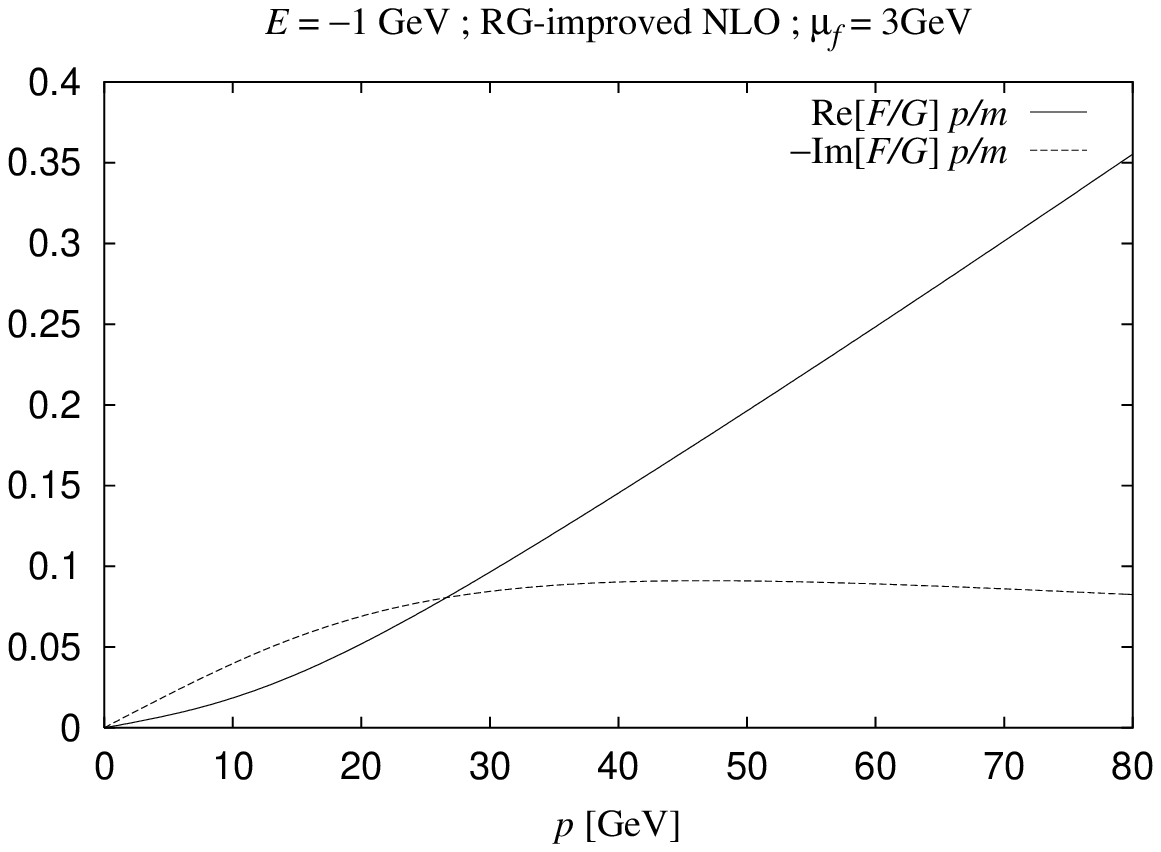}
  \end{minipage}
  \hspace*{\fill}
  \begin{minipage}{8cm}
    \includegraphics[width=8cm]{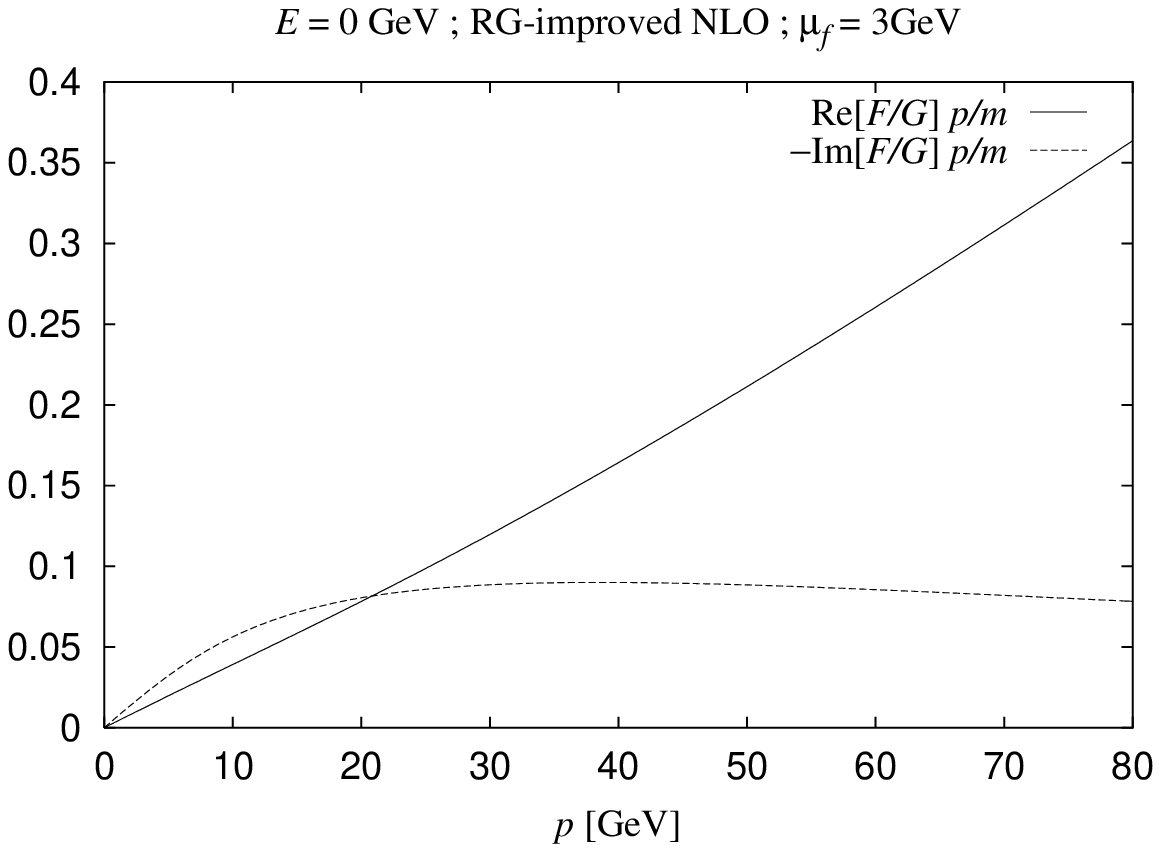}
  \end{minipage}
  \hspace*{\fill}
  \\
  \hspace*{\fill}
  \begin{Caption}\caption{\small
      Green functions
      \label{fig:FT3_-5,0}
  }\end{Caption}
  \hspace*{\fill}
\end{figure}
\begin{figure}[tbp]
  \hspace*{\fill}
  \begin{minipage}{8cm}
    \includegraphics[width=8cm]{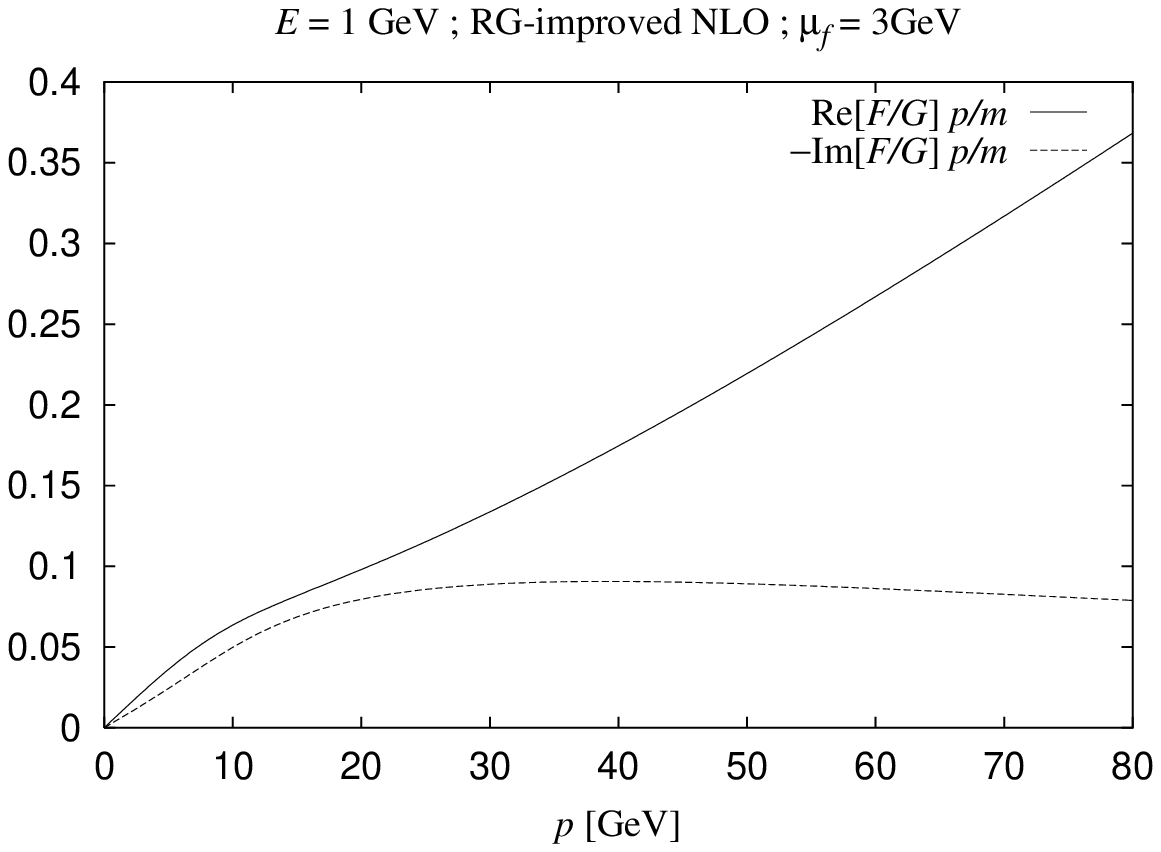}
  \end{minipage}
  \hspace*{\fill}
  \begin{minipage}{8cm}
    \includegraphics[width=8cm]{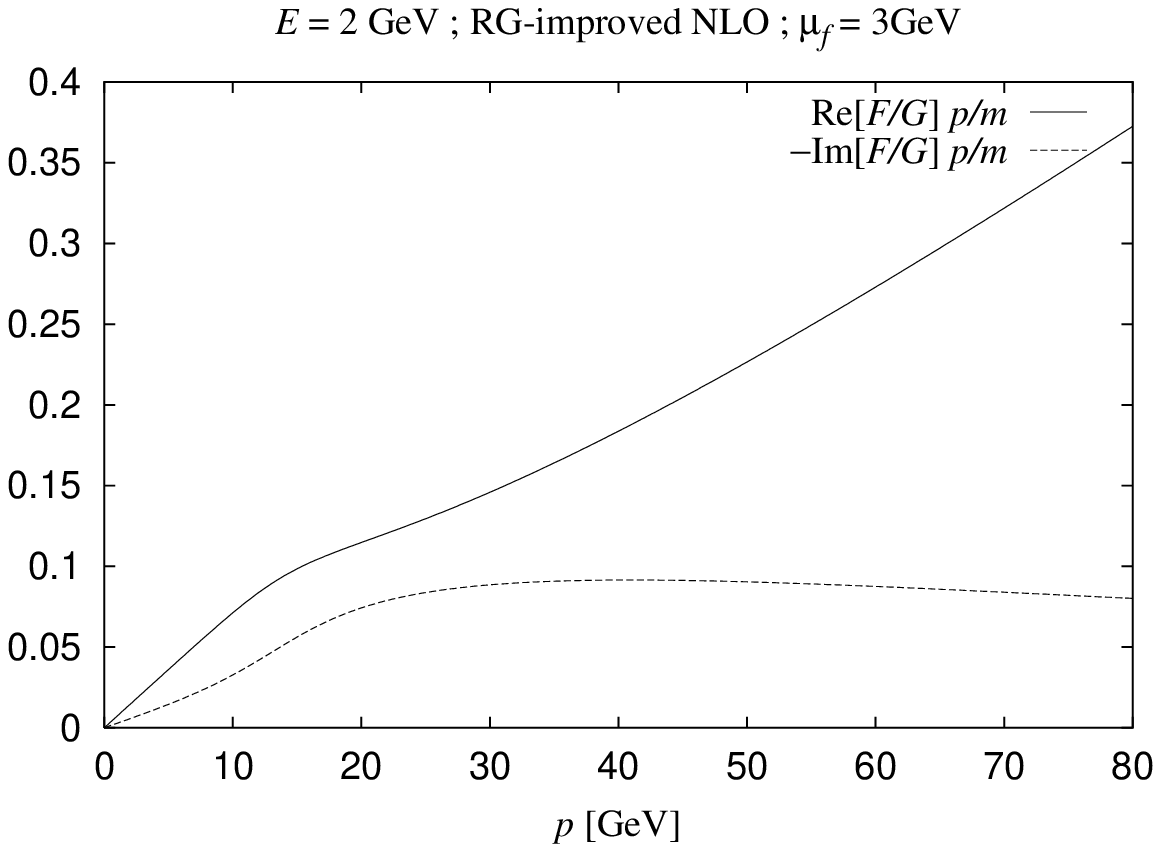}
  \end{minipage}
  \hspace*{\fill}
  \\
  \hspace*{\fill}
  \begin{minipage}{8cm}
    \includegraphics[width=8cm]{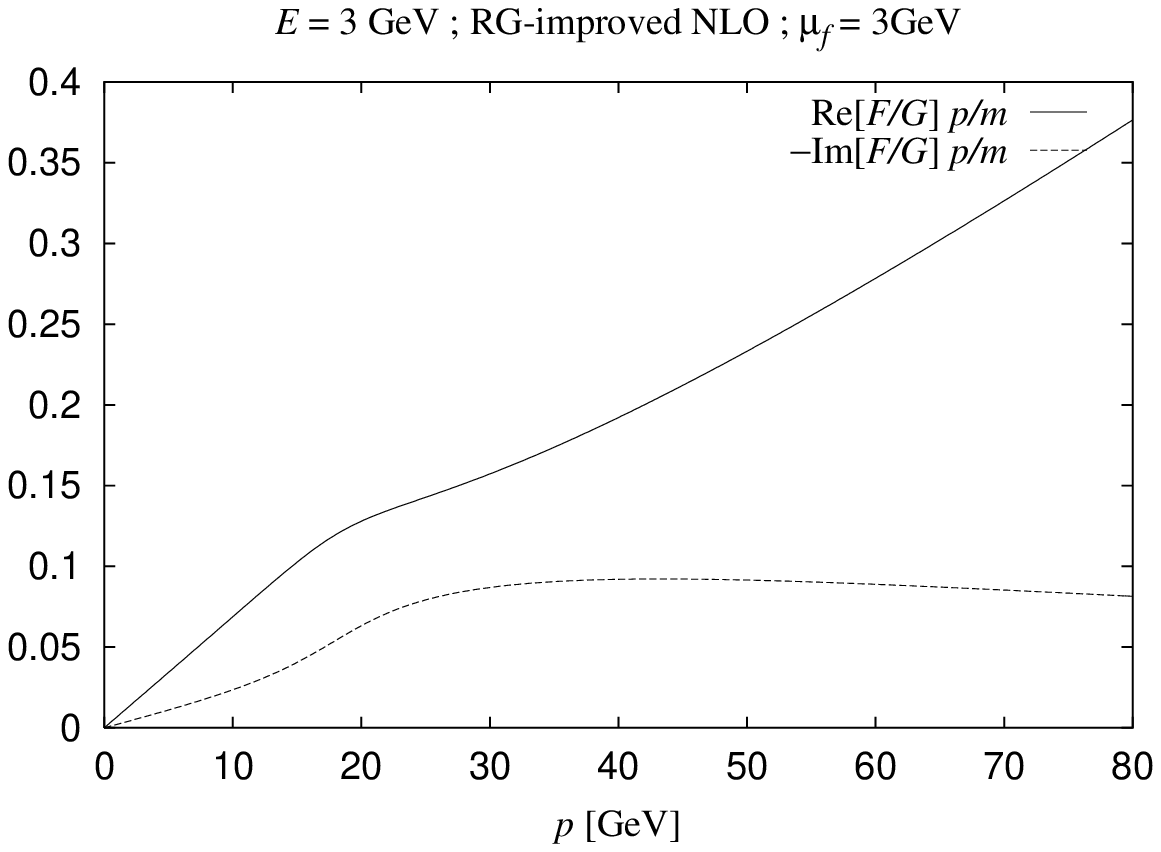}
  \end{minipage}
  \hspace*{\fill}
  \begin{minipage}{8cm}
    \includegraphics[width=8cm]{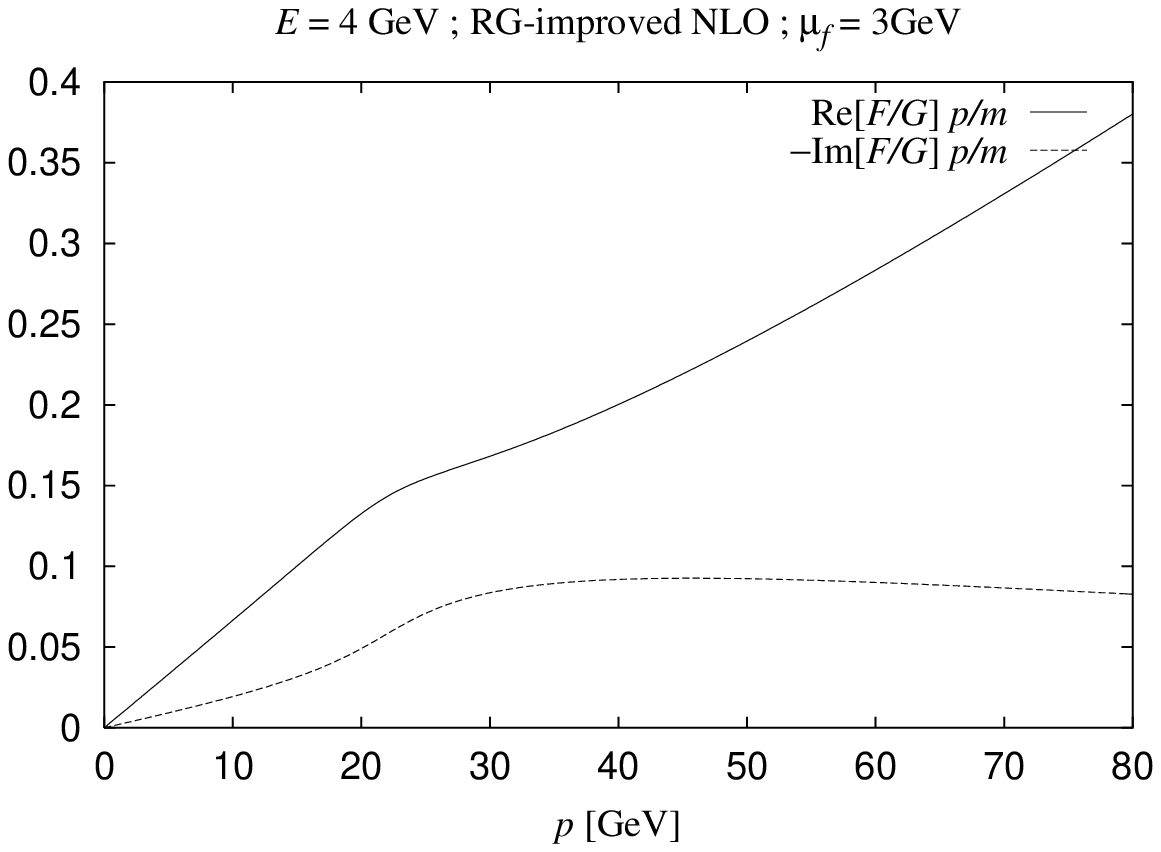}
  \end{minipage}
  \hspace*{\fill}
  \\
  \hspace*{\fill}
  \begin{minipage}{8cm}
    \includegraphics[width=8cm]{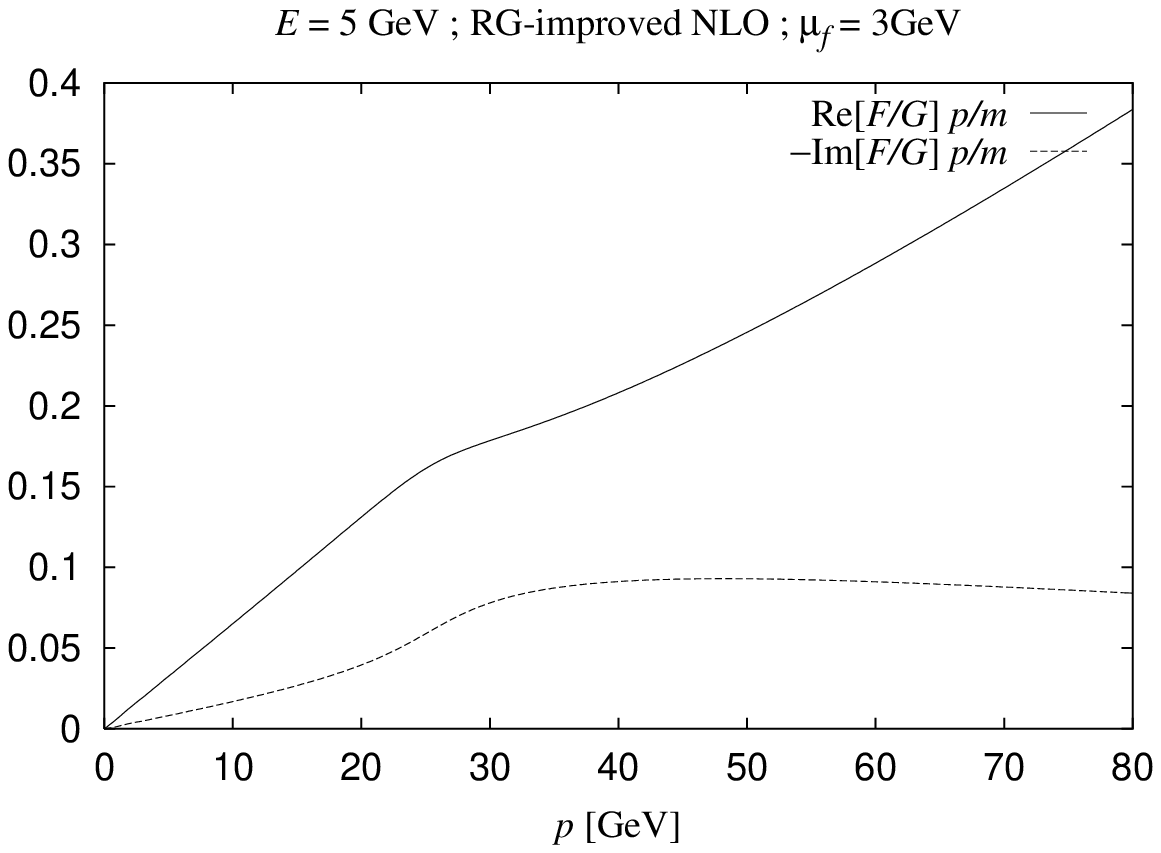}
  \end{minipage}
  \hspace*{\fill}
  \\
  \hspace*{\fill}
  \begin{Caption}\caption{\small
      Green functions
      \label{fig:FT3_1,5}
  }\end{Caption}
  \hspace*{\fill}
\end{figure}
\begin{figure}[tbp]
  \hspace*{\fill}
  \begin{minipage}{8cm}
    \includegraphics[width=8cm]{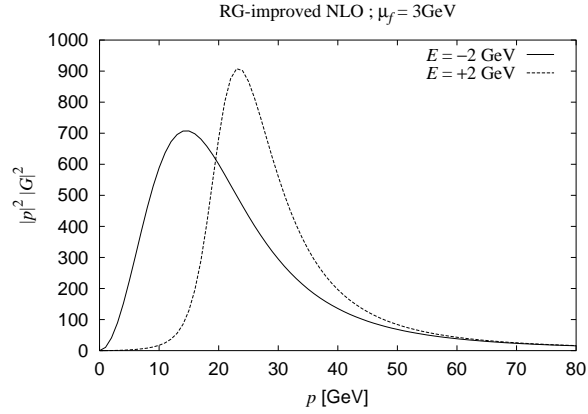}
  \end{minipage}
  \hspace*{\fill}
  \\
  \hspace*{\fill}
  \begin{Caption}\caption{\small
      $p^2 |G(E,p)|^2 \propto {\rm d}\sigma/{\rm d}p$ vs.\ top quark momentum 
      $p$ for fixed \scCM\ energy.  
      \label{fig:p2G2}
  }\end{Caption}
  \hspace*{\fill}
\end{figure}
%

%---------------------------------------------------------------------
\subsection{Sensitivity}
The polarization $\mean{\Pol}$ projected to certain direction can be 
measured as 
\begin{align*}
  \mean{\Pol} \simeq
  \frac{N_{\uparrow}-N_{\downarrow}}{N_{\uparrow}+N_{\downarrow}}
  \sepperiod
\end{align*}
Thus the error may be estimated roughly as 
$\delta\!\mean{\Pol} \simeq 1/\sqrt{N_{\rmeff}}$.  
Here $N_{\rmeff}$ means the number of events used for the analysis: 
\begin{align*}
  N_{\rmeff} 
  &= 
  \sigma_{t\tB} \times 
  \intlum \times (B_{\ell} B_{h}) \times \epsilon
  \nonumber\\
  &=
  0.5\pb \times 100\fb^{-1} \times \left(\frac{2}{9} \cdot \frac{2}{3}\right)
  \times 0.6
%  \nonumber\\
  = 4\times 10^3 \text{\,events,}
\end{align*}
which means $1/\sqrt{N_{\rm eff}} = 1.5 \times 10^{-2}$.  
We used 
%$\sigma_{\rmtot} \simeq \sigma_{\rmpt}$ for 
%$\sqrt{s} \simeq 2\times 175\GeV$, and 
the branching fraction $B_{\ell}$ ($B_{h}$) of $W^+$ 
to $e^+$ or $\mu^+$ (hadron) is $2/9$ ($2/3$).  
We assumed integrated luminosity $\intlum = 50\fb^{-1}$, 
and the detection efficiency $\epsilon = 0.6$.  

For example, with 
\begin{align*}
  \Polnorm
  = 
  \Bnorm^g \pRe{d_{tg}\frac{F-G}{G}} \beta_t \sin\theta_{te}
  \sepcomma
\end{align*}
statistical error for $\dtg$ is 
\begin{align*}
  \delta d_{tg}
  &\simeq
  \frac{1}{\Bnorm^g\times\pRe{d_{tg}\frac{F-G}{G}} \beta_t}
  \times \frac{\dint{-1}{1}\diffn{(\cos\theta_{te})}{}\,1}
    {\dint{-1}{1}\diffn{(\cos\theta_{te})}{}\,\sin\theta_{te}}
  \times \frac{1}{\sqrt{N_{\rmeff}}}
  \nonumber\\
  &\simeq
  \frac{1}{1\times 0.1} \times \frac{4}{\pi} \times 1.5 \times 10^{-2}
  \nonumber\\
  &\simeq
  0.2 \sim \Order{10\%} \sepperiod
\end{align*}
From this estimate, we have
\begin{align*}
  t\tB g\text{-EDM} = \frac{g_s}{m_t} d_{tg} \lesssim 10^{-17} g_s\cm
\end{align*}
for Chromo-EDM.  
Since all coefficients ($\Cperp$, $\Bnorm$, etc.) are similar in magnitude 
and so are the ratio of Green functions ($\pIm{F/G}$, etc.), 
bounds for electroweak EDMs are just the same order: 
\begin{align*}
  t\tB\gamma\text{-EDM}= 
  \frac{e}{m_t} d_{t\gamma} \lesssim 10^{-17} e\cm  \sepcomma\quad
  t\tB Z\text{-EDM} =
  \frac{\gZ}{m_t} \dtZ \lesssim 10^{-17} \gZ\cm  \sepperiod
\end{align*}

A realistic Monte Carlo detector simulation based on the calculation here 
is also in progress by experimentalists~\cite{IF99}.  
They show that high detection efficiency is possible with 
simple event selection criteria and $b$-tagging.  
Up to now, only lepton energy asymmetry $\mean{E_{\ell^+}-E_{\ell^-}}$ 
for dilepton+dijet event is obtained.  
Statistical error with $100\fb^{-1}$ is 
\begin{align*}
  \delta^{\rm(stat)} \left[ \mean{E_{\ell^+}-E_{\ell^-}} \right]
  = 0.649\GeV  \sepperiod
\end{align*}
Their results for anomalous couplings, based on our calculations, are 
\begin{align*}
  |\bRe{\dtp}| < 1.48  \sepcomma\quad
  |\bRe{\dtZ}| < 1.04  \sepcomma\quad
  |\bRe{\dtg}| < 3.93  
\end{align*}
to 95\% CL.  
%, which correspond to 
%%
%\begin{align*}
%  &  |t\tB\gamma\text{-EDM}| \simless 10^{-16} e\cm   \sepcomma\\
%  &  |t\tB Z\text{-EDM}|     \simless 10^{-16} \gZ\cm \sepcomma\\
%  &  |t\tB g\text{-EDM}|     \simless 10^{-16} \gs\cm \sepperiod
%\end{align*}
%%
In fact, lepton energy asymmetry is not a good observable for polarization.  
This is because the time component of spin 4-vector is suppressed by 
$\beta_t\sim 1/10$ compared to the space component.  
On the other hand event asymmetries are not suppressed in that way, 
since they can be obtained by integrating 3-momentum asymmetry.  
Moreover the branching fraction for single-lepton + 4-jets is 
larger than that for dilepton + dijet.  
Thus it is expected that 
the sensitivity for anomalous EDMs is improved by factor 10 or more.  
This is consistent with the naive estimate we made above.  
\clearemptydoublepage
%
%
%---------------------------------------------------------------------------
\chapter{Summary and Discussions}
We study how the $t\tB$ threshold region serves as a probe of physics at 
electroweak scale and beyond.  

In \ChRef{ch:total_cr}, we studied the convergence of the perturbative 
series of the total production cross section $\sigma_{\rm tot}$ for 
$e^+e^- \to t\bar{t}$ in the threshold region.  
Two efforts have been made to improve it.  
One is to use Potential-Subtracted mass $\mt^{\rm PS}(\mu_f)$, 
and the other is to resum leading $\log(q/\mu_s)$'s of 
the Coulombic potential $\VT_\rmC(q)$ in momentum space using 
Renormalization Group (RG).  
The former is to circumvent the (severest) renormalon ambiguity inherent to 
the pole mass and the coordinate-space potential of an IR-slave interaction.  
With this prescription, 
the convergence of the position of $1S$ resonance becomes better.  
The latter is to reduce the renormalization scale ($\mu_s$) dependence 
of $R$ ratio, where $\mu_s$ is the scale for the $t\tB$ potential $V$.  
Among various contributions to the potential 
$V=V_{\rm C}+V_{\rm BF}+V_{\rm NA}$, only the $\log(q/\mu_s)$'s 
in the Coulombic potential $V_{\rm C}$ are summed up in this work.  
RG improvement of the whole $V$ is a subject of a future study.  
With these two implementations, 
we estimate $(\Delta\mt)^{\rm th} \sim 0.1\GeV$, where 
$(\Delta\mt)^{\rm th}$ is the theoretical uncertainty 
for the determination of $\mt$ using the process $e^+e^- \to t\tB$ 
with $\sqrt{s}\simeq 2\mt$.  
This size is consistent with one of the N${}^3$LO corrections 
calculated recently~\cite{KP99}.  
Lower bound to the CKM element $|V_{tb}|$ may also be obtained.  
Note that $\Gamma_t \propto |V_{tb}|^2$, and the present bound~\cite{C98} 
for $|V_{tb}|$ is $0.06$--$0.9993$ when the unitary of CKM matrix 
within the first three generations is not assumed.  

In \ChRef{ch:diff_cr}, 
we calculated the momentum distributions $\diffn{\sigma}{}/\diffn{p}{}$ 
of top quark near the threshold.  
Corrections to $\Order{1/c^2}$ are included.  
However the Final-State Interactions%
\footnote{We call this FSI for brevity.  }
between $t$ and 
the decay products of $\tB$ etc.\ are not taken into account.  
The effect of the NNLO correction for 
$(1/\sigma_{\rm tot})\, {\rm d}\sigma/{\rm d}p$ 
from the rescattering between $t$ and $\tB$, 
which is calculated here, turns out to be rather small compared to 
the NNLO correction to $\sigma_{\rm tot}$.  
In fact the correction is smaller than the 
effect of the NLO correction from the FSI.  
Thus the NNLO correction from the FSI may contribute 
with the similar magnitude to the NNLO correction from $t\tB$ rescattering.  

In \ChRef{ch:EDMs}, 
anomalous EDMs of top quark are studied.  
They can be measured by CP-odd observables.  
Chromo-EDM modifies $t\tB$ rescattering, 
while ElectroWeak-EDMs modify $t\tB$ production vertex.  
Thus their effects can be distinguished from the energy dependences, 
since the latter is enhanced by raising \scCM\ energy, while 
the former is not.  
On the other hand two EW-EDMs, $t\tB\gamma$-EDM and $t\tB Z$-EDM, can be 
distinguished from the dependences on the polarization of the 
initial $e^-$ (and $e^+$).  
We estimate the sensitivity of $t\tB$ production near the threshold in 
future $e^+e^-$ colliders to be $\sim 10^{-17}\gs\cm$ ($e\cm$, $\gZ\cm$) 
for $t\tB g$-EDM ($t\tB\gamma$-, $t\tB Z$-EDM) with 
the integrated luminosity $\simeq 100\fb^{-1}$.  
It is worth mentioning that a realistic Monte Carlo detector simulation 
is in progress by experimentalists~\cite{IF99}, 
based on our theoretical calculations.  
Their first result is already available, and our naive estimate is 
consistent with it.  
We note that hadron colliders may be powerful to prove the existence 
of non-zero EDMs; however precise determination of their values would be 
difficult.  In this respect, $e^+e^-$ collider is much advantageous.

%After completion of this work, we received a paper 
%by Beneke, Signer and Smirnov [cite].  
%Their work has a significant overlap with [SecRef] of the present paper.
%Effects of introducing $m_{\rm PS}(\mu_f)$ on the cross section 
%are consistent between their results and ours.  
%We adopt a value of $\mu_f$ considerably smaller than that adopted 
%in their paper.  
%This is in view of our 
%application of the formalism to the renormalization-group improved potential; 
%see discussion below [EqRef].  
%
%
%

%----------------------------------------------------------------------------
\section*{Acknowledgment}
The author wishes to express his sincere gratitude to Y.~Sumino, 
who is a collaborator of the works, on which the thesis is based.  
He also would like to thank A.~Ota for collaboration, 
M.~Je\.{z}abek, K.~Fujii, Z.~Hioki for useful discussions 
at the Summer Institute '99 (August 1999, Yamanashi, Japan), 
M.~Yoshimura, K.~Hikasa, M.~Yamaguchi and M.~Tanabashi for several remarks.  

Finally the author wishes to express his thanks to the members of 
High Energy Theory Group at Tohoku University for their kind support 
while preparing the thesis.  
%-----------------------------------------------------------------------
\clearemptydoublepage
\rhead[]{{\bfseries\thepage}}
\appendix
%
%
%
%
%------------------------------------------------------------------------
\chapter{}
%------------------------------------------------------------------------
\section{Notes on group theory}
%------------------------------------------------------------------------
\subsection{Group theoretic factors}
\label{sec:group_factor}
Generators of a Lie group satisfy the following commutation relation: 
\begin{align*}
  \commutator{T^a}{T^b} = i f^{abc} T^c
  \sepperiod
\end{align*}
Trace normalization $T_R$ and quadratic Casimir $C_R$ for 
a representation $R$ is defined by 
\begin{align*}
  \btr{T_R^a T_R^b} = T_R \delta^{ab}
  \sepcomma\quad
  T_R^a T_R^a = C_R \mathbf{1}
  \sepperiod
\end{align*}
For $\SU(N=3)$ 
\begin{align}
  T_F = \frac{1}{2}
  \sepcomma\quad
  T_A = C_A = N = 3
  \sepcomma\quad
  C_F = \frac{N^2-1}{2N} = \frac{4}{3}
  \sepcomma
\end{align}
where $F$ and $A$ mean fundamental and adjoint representations, respectively.  

Coefficients $\beta_0$ etc.\ of renormalization group equations are 
given in \SecRef{sec:log_corr=V1}.  
When necessary, we use 
\begin{align}
  n_f = 5  \sepcomma\quad  n_H = 1  \sepcomma
\end{align}
where the former is the number of light quarks and the latter is the 
number of Higgs bosons.  
%------------------------------------------------------------------------
\subsection{Color charge summation for color-singlet state}
For a particle-antiparticle pair $\psi\psi^c$ in color-singlet state, 
the product of color charge is $-C_F$: 
\begin{align}
  \mean{T^a \bar{T}^a} = -C_F = - \frac{N_C^2-1}{2N_C}
  \sepperiod
\end{align}
This corresponds to $Q_\psi{\cdot}(-Q_\psi) = - Q_\psi^2$ 
for Abelian gauge group.  
Sum over the index $a$ is implicit.  
In this section, we show this relation.  
The relevant Lagrangian is 
\begin{align*}
  \calL = -g_s \psiB \gamma^\mu T^a \psi \, G_\mu^a  
        = -g_s \overline{\psi^c} \gamma^\mu \bar{T}^a\psi^c \, G_\mu^a
  \sepcomma
\end{align*}
where $\bar{T}^a = -{T^a}^\sfT$, and $\psi^c = C\psiB^\sfT$ with 
$C{\gamma^\mu}^\sfT C^\dagger = - \gamma^\mu$.  
Then 
\begin{align*}
  \mean{T^a \bar{T}^a}
  &=
  \frac{1}{N_C} \sum_{i,j}
  \left[ {\xi^j}^\dagger T^a \xi^i \right]
  \left[ \xi^j (-{T^a}^\sfT) {\xi^i}^\dagger \right]  \\
  &=
  \frac{-1}{N_C} \sum_{i,j}
  \btr{ \xi^j {\xi^j}^\dagger T^a \xi^i {\xi^i}^\dagger T^a }  \\
  &=
  \frac{-1}{N_C} C_F
  \btr{{\bf 1}}
   = - C_F
  \sepcomma
\end{align*}
where 
\begin{align*}
  \xi^{\rm R} = 
  \begin{pmatrix}
    1 \\ 0 \\ 0
  \end{pmatrix}
  \sepcomma\quad
  \xi^{\rm G} = 
  \begin{pmatrix}
    0 \\ 1 \\ 0
  \end{pmatrix}
  \sepcomma\quad
  \xi^{\rm B} = 
  \begin{pmatrix}
    0 \\ 0 \\ 1
  \end{pmatrix}
  \sepcomma
\end{align*}
for example.  The indices $i$ and $j$ mean the color of initial and 
final $\psi$, respectively.  

Although the calculation above is specific to fundamental representation $F$, 
generalization to arbitrary representation $R$ is obvious; 
$\mean{T^a \bar{T}^a} = -C_R$, where $C_R$ is quadratic Casimir for 
the representation $R$.  
%------------------------------------------------------------------------
\section{Notes on non-relativistic quantum mechanics}
%------------------------------------------------------------------------
\subsection{Relative coordinate}
\label{sec:relative_coord}
As is well known, a non-relativistic two-body system can be decomposed 
into relative and center-of-momentum motions: 
\begin{align*}
  H = \frac{\pBI_1^2}{2m_1} + \frac{\pBI_2^2}{2m_2}
    = \frac{\pBI^2}{2\mu} + \frac{\PBI^2}{2M}
  \sepcomma
\end{align*}
where
\begin{align*}
  \RBI &\equiv \frac{m_1}{m_1+m_2} \rBI_1 + \frac{m_2}{m_1+m_2} \rBI_2
  \sepcomma &
  \PBI &\equiv \pBI_1 + \pBI_2
  \sepcomma\\
  \rBI &\equiv \rBI_1 - \rBI_2 \sepcomma &
  \pBI &\equiv \frac{m_2}{m_1+m_2} \pBI_1 - \frac{m_1}{m_1+m_2} \pBI_2
  \sepcomma
\end{align*}
and 
\begin{align*}
  \frac{1}{\mu} \equiv \frac{1}{m_1} + \frac{1}{m_2}  \sepcomma\quad
  M \equiv m_1 + m_2  \quad (m_1 m_2 = \mu M)
  \sepperiod
\end{align*}
One can easily check that those coordinates and momenta satisfy the 
standard commutation relations $\commutator{R_i}{P_j} = i\delta_{ij}$ etc.  
Thus they are conjugate also in quantum mechanical sense.  
%------------------------------------------------------------------------
\subsection{Radial momentum $p_r$}
\label{sec:polar_coord}
The operator $p_r$ for radial momentum is defined by the Weyl-ordered 
product of $\rBI/r$ and $\pBI$: 
\begin{align*}
  p_r \equiv \frac{1}{2} \left( 
    \frac{1}{r} \dprod{\rBI}{\pBI} + \dprod{\pBI}{\rBI} \frac{1}{r} \right)
  \sepperiod
\end{align*}
Using 
\begin{align*}
  r \pdiff{}{r} = r^i \pdiff{}{r^i}  \sepcomma\quad
  \pdiff{}{r^i} \pfrac{r^i}{r} = \frac{n-1}{r}
  \quad\text{ with } n=3  \sepcomma
\end{align*}
one obtains 
\begin{align*}
  ip_r = \pdiff{}{r} + r = \frac{1}{r}\pdiff{}{r} r  \sepcomma\quad
  (ip_r)^2 = \pdiffn{}{r}{2} + \frac{2}{r}\pdiff{}{r}
     = \frac{1}{r}\pdiffn{}{r}{2} r
     = \frac{1}{r^2}\pdiff{}{r}\left(r^2\pdiff{}{r}\right)
  \sepperiod
\end{align*}
With these relations, 
one can show that this satisfy $\commutator{r}{p_r} = i$, and 
$\braket{\psi_1}{}{p_r \psi_2} = \braket{p_r \psi_1}{}{\psi_2}$ 
for the states of $\lim_{r\to 0}r\psi(r) = 0$.  
Note that $\braket{\psi_1}{}{\psi_2} 
= \int\diffn{\Omega}{}\int r^2\diffn{r}{}\,\psi_1^*(\rBI)\,\psi_2(\rBI)$.  
On the other hand, the angular part of $\pBI^2$ is given by 
orbital angular momentum $\LBI = \cprod{\rBI}{\pBI}$: 
\begin{align*}
  \pBI^2 = p_r^2 + \frac{\LBI^2}{r^2}  \sepcomma
\end{align*}
where 
\begin{align*}
  \LBI^2 
  = \rBI^2\pBI^2 - \dprod{\rBI}{\pdprod{\rBI}{\pBI}\pBI} 
    + 2i\dprod{\rBI}{\pBI}
  = \rBI^2 \pBI^2 + \pdiff{}{r}\left(r^2\pdiff{}{r}\right)
  \sepperiod
\end{align*}
Since $\commutator{r}{L^i}=0$, $\LBI^2$ and $r$ also commutes.  
The following relations may sometimes be useful: 
\begin{align*}
  \pdprod{\rBI}{\pBI}^2 \equiv r^i p^i r^j p^j  \sepcomma\quad
  \dprod{\rBI}{\pdprod{\rBI}{\pBI}\pBI} 
  \equiv r^i r^j p^j p^i 
  = \pdprod{\rBI}{\pBI}^2 + i\dprod{\rBI}{\pBI}
  = - r^2 \pdiffn{}{r}{2}
  \sepperiod
\end{align*}

One must be careful for the singularity at the origin $r=0$.  
For example, with $-\triangle = \pBI^2 = p_r^2$ for angle-independent 
function, naive calculation gives $\triangle(1/r) = 0$, while actually 
\begin{align*}
  \triangle \frac{1}{r} = -4\pi \delta^{(3)}(\rBI)
  \sepperiod
\end{align*}
The contributions of $\delta$-functions are easy to miss.  
They arise from differentiation of functions singular at the origin.  
It is safer to integrate by parts so that to avoid these subtle points.  
For example, with functions of well-behaving, 
\begin{align*}
  \bra{f} p_r^2 \frac{1}{r} \ket{g}
  &= \bra{p_r^2 f} \frac{1}{r} \ket{g}  \\
  &= -\int\!\!\diffn{\Omega}{} \int_0^\infty\! r^2 \diffn{r}{}
    \left\{ \frac{1}{r}\pdiffn{}{r}{2}(rf) \right\} \frac{1}{r} g  \\
%  &= -\int\!\!\diffn{\Omega}{} \int_0^\infty\! \diffn{r}{}
%    \pdiffn{}{r}{2}(rf) g  \\
  &= 4\pi f(0) g(0) 
    -\int\!\!\diffn{\Omega}{} \int_0^\infty\! r^2 \diffn{r}{}\,
      f \, \frac{1}{r} \pdiffn{}{r}{2}g
  \sepcomma
\end{align*}
or 
\begin{align*}
  p_r^2 \frac{1}{r}
  = 4\pi\delta^{(3)}(\rBI) - \frac{1}{r} \pdiffn{}{r}{2}
  \sepperiod
\end{align*}
We use following relations in the text: 
\begin{align*}
    [ p_r^2 , i p_r ] &= 4\pi \delta^{(3)}(\rBI)  \sepcomma\quad
  & [ \pBI^2 , i p_r ] &= 4\pi \delta^{(3)}(\rBI) + \frac{2\LBI^2}{r^3}
  \sepcomma\\
    [ p_r^2 , \frac{1}{r} ] 
    &= 4\pi \delta^{(3)}(\rBI) + \frac{2}{r^2}\pdiff{}{r}  \sepcomma\quad
  & [ \pBI^2 , \frac{1}{r} ]
    &= 4\pi \delta^{(3)}(\rBI) + \frac{2}{r^2}\pdiff{}{r}  \sepcomma\\
    \{ p_r^2 , \frac{1}{r} \} 
    &= 4\pi \delta^{(3)}(\rBI) + \frac{2}{r^2}\pdiff{}{r} + \frac{2}{r}p_r^2
  \sepcomma\quad
  & \{ \pBI^2 , \frac{1}{r} \}
    &= 4\pi \delta^{(3)}(\rBI) + \frac{2}{r^2}\pdiff{}{r} + \frac{2}{r}\pBI^2
  \sepcomma\\
    [ i p_r , \frac{1}{r} ] &= \frac{-1}{r^2}
  \sepperiod
\end{align*}
The first relation is highly anomalous; $p_r$ and $p_r^2$ do not commute.  
%------------------------------------------------------------------------
\subsection{Operators in momentum and coordinate space}
\label{sec:mom_and_coord_space}
A momentum independent potential $V$ is diagonal in coordinate space: 
\begin{align*}
  \braket{\xBI}{V(\hat{\xBI})}{\yBI} = V(\xBI) \, \delta^{(3)}(\xBI-\yBI)
  \sepcomma
\end{align*}
where hat $\hat{~}$ means operator.  
Such an operator in momentum space depends only on momentum transfer: 
\begin{align*}
  \braket{\pBI}{V(\hat{\xBI})}{\qBI} 
  &= \int\!\!\diffn{x}{3} \braket{\pBI}{}{\xBI} V(\xBI) 
      \braket{\xBI}{}{\qBI}  \\
  &= \int\!\!\diffn{x}{3} \expo^{-i(\vec{p}-\vec{q}){\cdot}\vec{x}} V(\xBI) \\
  &= \VT(\pBI-\qBI)  \sepperiod
\end{align*}
Using 
\begin{align*}
  \braket{\xBI}{\hat{\pBI}}{~} = - i \nablaBI_{\!x} \braket{\xBI}{}{~} 
  \sepcomma\quad
  \braket{~}{\hat{\pBI}}{\xBI} = + i \nablaBI_{\!x} \braket{~}{}{\xBI} 
  \sepcomma\quad
  \odiff{}{y} f(x-y) = - \odiff{}{x} f(x-y)
  \sepcomma
\end{align*}
we have 
\begin{align*}
  &
  \braket{\pBI}{ \commutator{\hat{\pBI}}{V(\hat{\xBI})} }{\qBI}
  = (\pBI-\qBI) \cdot \VT(\pBI-\qBI)  
  \sepcomma\\
  &
  \braket{\xBI}{ \commutator{\hat{\pBI}}{V(\hat{\xBI})} }{\yBI}
  = \bigr( -i\nablaBI_{\!x} V(\xBI) \bigr) \delta^{(3)}(\xBI-\yBI)
  \sepperiod
\end{align*}
More generally, 
\begin{align*}
  &  \braket{\pBI}{VG}{~} 
  = \int\!\!\frac{\diffn{q}{3}}{(2\pi)^3} \VT(\pBI-\qBI) \, \GT(\qBI)
  \sepcomma\quad&
  &  \braket{\xBI}{VG}{~} 
  = V(\xBI) \, G(\xBI)
  \sepcomma\\
  &  \braket{\pBI}{\hat{\pBI}VG}{~} 
  = \int\!\!\frac{\diffn{q}{3}}{(2\pi)^3} \pBI \VT(\pBI-\qBI) \, \GT(\qBI)
  \sepcomma\quad&
  &  \braket{\xBI}{\hat{\pBI}VG}{~} 
  = -i\nablaBI_{\!x} \bigl( V(\xBI) \, G(\xBI) \bigr)
  \sepcomma\\
  &  \braket{\pBI}{V\hat{\pBI}G}{~} 
  = \int\!\!\frac{\diffn{q}{3}}{(2\pi)^3} \VT(\pBI-\qBI) \, \qBI \,\GT(\qBI)
  \sepcomma\quad&
  &  \braket{\xBI}{V\hat{\pBI}G}{~} 
  = V(\xBI) \bigl( -i\nablaBI_{\!x} G(\xBI) \bigr)
  \sepcomma\\
  &  \braket{\pBI}{ [\hat{\pBI},V]G }{~} 
  = \int\!\!\frac{\diffn{q}{3}}{(2\pi)^3} 
      (\pBI-\qBI) \cdot \VT(\pBI-\qBI) \, \GT(\qBI)
  \sepcomma\quad&
  &  \braket{\xBI}{ [\hat{\pBI},V]G }{~} 
  = \bigl( -i\nablaBI_{\!x} V(\xBI) \bigr) G(\xBI) 
  \sepcomma
\end{align*}
where 
\begin{align*}
  \GT(\pBI) \equiv \braket{\pBI}{G}{~}  \sepcomma\quad
  G(\xBI) \equiv \braket{\xBI}{G}{~}  \sepcomma
\end{align*}
Our convention for Fourier transform is 
\begin{align*}
  &
  f(\xBI) = \braket{\xBI}{}{f} 
  = \int\!\!\frac{\diffn{p}{3}}{(2\pi)^3} 
      \braket{\xBI}{}{\pBI} \braket{\pBI}{}{f}
  = \int\!\!\frac{\diffn{p}{3}}{(2\pi)^3} 
      \expo^{i\vec{x}{\cdot}\vec{p}} \fT(\pBI)
  \sepcomma\\
  &
  \fT(\pBI) = \braket{\pBI}{}{f} 
  = \int\!\!\diffn{x}{3}
      \braket{\pBI}{}{\xBI} \braket{\xBI}{}{f}
  = \int\!\!\diffn{x}{3}
      \expo^{-i\vec{x}{\cdot}\vec{p}} f(\xBI)
  \sepperiod
\end{align*}
For $\delta$-functions, 
\begin{align*}
  \delta^{(3)}(\xBI-\yBI) = \braket{\xBI}{}{\yBI} 
  &= \int\!\!\frac{\diffn{p}{3}}{(2\pi)^3} 
      \expo^{i(\vec{x}-\vec{y}){\cdot}\vec{p}}
  \sepcomma\\
  (2\pi)^3 \delta^{(3)}(\pBI-\qBI) = \braket{\pBI}{}{\qBI} 
  &= \int\!\!\diffn{x}{3}
      \expo^{-i\vec{x}{\cdot}(\vec{p}-\vec{q})} 
  \sepperiod
\end{align*}
%
%------------------------------------------------------------------------
\subsection{$S$-wave projection}
\label{sec:S-wave_proj}
If a quantity $f$ depends on a 3-vector, say, $\rBI$, 
it can be decomposed into ``radial part'' and ``spherical part''.  
For example, a plain wave of momentum $\pBI$ can be written as

Sometimes we use the shorthand notation for $S$-wave projection: 
\begin{gather}
  G \equiv \frac{1}{H-\omega}  \sepcomma\quad
  \omega \equiv E+i\Gamma_t
%  \label{eq:G_and_H_app}
  \sepcomma\\
  G(\rBI,\rBI') \equiv \braket{\rBI}{G}{\rBI'}  \sepcomma\quad
  G(r,r') \equiv \int \frac{\diffn{\Omega_{r}}{}}{4\pi}
    \frac{\diffn{\Omega_{r'}}{}}{4\pi} G(\rBI,\rBI')
  \sepperiod\nonumber
\end{gather}
\begin{align*}
  \bra{r} \equiv \int \frac{\diffn{\Omega_{r}}{}}{4\pi} \bra{\rBI}
  \sepcomma\quad
  \bra{p} \equiv \int \frac{\diffn{\Omega_{p}}{}}{4\pi} \bra{\pBI}
  \sepcomma\quad\mbox{etc.  }
\end{align*}
$L = 0, \SBI^2 = 2, \overline{r^i r^j} = \delta^{ij} r^2/3$.  

In general, partial wave expansion can be defined as follows: 
\begin{align*}
  f(k,\theta) = \sum_{\ell=0}^{\infty} (2\ell+1) \, 
    f_\ell(k) \, P_\ell(\cos\theta) 
  \sepcomma\quad
  f_\ell(k) = \int\!\!\frac{\diffn{\Omega}{}}{4\pi} \, 
    P_\ell(\cos\theta) \, f(k,\theta)
  \sepperiod
\end{align*}
The second relation holds because 
\begin{align*}
  \int\!\!\frac{\diffn{\Omega}{}}{4\pi} 
    P_\ell(\cos\theta) P_{\ell'}(\cos\theta) 
  = \frac{\delta_{\ell\ell'}}{2\ell+1}
  \sepperiod
\end{align*}
Here $P_\ell(\cos\theta)$ is Legendre polynomials: 
\begin{align*}
  P_0(\cos\theta) = 1
  \sepcomma\quad
  P_1(\cos\theta) = \cos\theta
  \sepcomma\quad
  P_2(\cos\theta) = \frac{3}{2} \cos^2\theta - \frac{1}{2}
  \sepperiod
\end{align*}
Also the following relation holds: 
\begin{align*}
  \int\!\!\frac{\diffn{\Omega}{}}{4\pi} 
    \left| f(k,\theta) \right|^2 
  = \sum_{\ell=0}^{\infty} (2\ell+1) 
    \left| f_\ell(k) \right|^2
  \sepperiod
\end{align*}
An example may be in order.  A plane wave is expanded as follows: 
\begin{align*}
  \expo^{ikr\cos\theta} 
  = \sum_{\ell=0}^{\infty} (2\ell+1) \,
    i^\ell j_\ell(kr) \, P_\ell(\cos\theta)
  \sepcomma
\end{align*}
where $j_\ell(kr)$ is a spherical Bessel, which is a solution 
$R_\ell(r) = j_\ell(kr)$ of the radial Schr\"odinger equation: 
\begin{align*}
  \left[ \odiffn{}{r}{2} + \frac{2}{r} \odiff{}{r} - \frac{\ell(\ell+1)}{r^2}
  \right] R_\ell(r) + k^2 R_\ell(r) = 0
  \sepcomma
\end{align*}
where $k^2/(2m) = E$.  Explicitly, 
\begin{align*}
  j_0(\rho) = \frac{\sin\rho}{\rho}
  \sepcomma\quad
  j_1(\rho) = \frac{\sin\rho}{\rho^2} - \frac{\cos\rho}{\rho}
  \sepcomma\quad
  j_2(\rho) = \left( \frac{3}{\rho^3}-\frac{1}{\rho} \right) \sin\rho
    - \frac{3}{\rho^2}\cos\rho
  \sepperiod
\end{align*}
Fourier transform goes as follows: 
\begin{align*}
  f(\xBI',\xBI) &= \sum_{\ell=0}^{\infty} (2\ell+1) \,
    f_\ell(x',x) \, P_\ell(\cos\theta_{x'x})
  \sepcomma\\
  f(\pBI',\xBI) &\equiv 
  \int\!\!\diffn{x'}{3}\, \expo^{-i\vec{p}'{\cdot}\vec{x}'} f(\xBI',\xBI)  \\
  &=
  \sum_{\ell=0}^{\infty} (2\ell+1) \,
    f_\ell(p',x) \, P_\ell(\cos\theta_{p'x})
  \sepcomma
\end{align*}
where $\theta_{x'x}$ is the angle between $\xBI'$ and $\xBI$, and 
\begin{align*}
  &  f_\ell(p',x) \equiv 
    \int_0^\infty 4\pi {x'}^2 \diffn{x'}{}\, 
    (-i)^\ell j_\ell(p'x') \, f_\ell(x',x)
  \sepcomma\\
  &  f_\ell(x',x) = 
    \int_0^\infty \frac{4\pi{p'}^2\diffn{p'}{}}{(2\pi)^3} \, 
    i^\ell j_\ell(p'x') \, f_\ell(p',x)
  \sepperiod
\end{align*}
Here we used 
\begin{align*}
  &
  \int\!\!\frac{\diffn{\Omega_{x'}}{}}{4\pi}
  P_{\ell'}(\cos\theta_{p'x'}) \, P_{\ell}(\cos\theta_{x'x})
  = \frac{\delta_{\ell\ell'}}{2\ell+1} P_\ell(\cos\theta_{p'x})
  \sepcomma\\
  &
  \int_0^\infty \frac{4\pi{p}^2\diffn{p}{}}{(2\pi)^3} \, 
  j_\ell(px) \, j_\ell(px') 
  = \frac{1}{4\pi x^2} \delta^{(1)}(x-x')
  \sepcomma\quad\mbox{ for each $\ell$. }
\end{align*}
The following are completeness relations for the angular part: 
\begin{align*}
  &
  \frac{1}{2} \int_{-1}^{1} \diffn{(\cos\theta)}{}\,
  P_\ell(\cos\theta)\,P_{\ell'}(\cos\theta) 
  = \frac{\delta_{\ell\ell'}}{2\ell+1}
  \sepcomma\\
  &
  \frac{1}{2} \sum_{\ell=0}^\infty (2\ell+1) 
  P_\ell(\cos\theta)\,P_\ell(\cos\theta') 
  = \delta^{(1)}(\cos\theta-\cos\theta')
  \sepcomma\\
  &
  \frac{1}{2\pi} \int_0^{2\pi} \diffn{\phi}{} \, 
  \expo^{i(m-m')\phi}
  = \delta_{mm'}
  \sepcomma\\
  &
  \frac{1}{2\pi} \sum_{m=-\infty}^{\infty}
  \expo^{im(\phi-\phi')} 
  = \delta^{(1)}(\phi-\phi')
  \sepperiod
\end{align*}
Since 
\begin{align*}
  \int\!\!\diffn{x'}{3}\,\delta^{(3)}(\xBI-\xBI')\,
  f(x',\cos\theta',\phi')
  = f(x,\cos\theta,\phi)
  \sepcomma
\end{align*}
$\delta$-function in the polar coordinate can be written
\begin{align*}
  \delta^{(3)}(\xBI-\xBI') 
  &= 
  \frac{1}{x^2} \delta^{(1)}(x-x') \, 
  \delta^{(1)}(\cos\theta_{xz}-\cos\theta_{x'z}) \, 
  \delta^{(1)}(\phi_x-\phi_{x'})
  &,& \mbox{ for $f=f(x,\cos\theta,\phi)$, } \\
  &= 
  \frac{1}{2\pi x^2} \delta^{(1)}(x-x') \, 
  \delta^{(1)}(\cos\theta_{xz}-\cos\theta_{x'z}) 
  &,& \mbox{ for $f=f(x,\cos\theta)$, } \\
  &= 
  \frac{1}{4\pi x^2} \delta^{(1)}(x-x') 
  &,& \mbox{ for $f=f(x)$. }
\end{align*}
Here $\theta_{xz}$ is the angle between $\xBI$ and certain axis, 
and likewise for $\phi_x$.  
%
%
%

%
%
%
%------------------------------------------------------------------------
\section{Formulas for Dirac spinors}
References~\cite{SBDH64} are useful.  
%------------------------------------------------------------------------
\subsection{Mode expansions}
\label{sec:mode_exp}
The anticommutation rules for 
the creation-annihilation operators $a_\pBssI^s$ and $b_\pBssI^s$, are 
\begin{align}
  \anticommutator{a_\pBssI^r}{a_\qBssI^{s\dagger}} = 
  \anticommutator{b_\pBssI^r}{b_\qBssI^{s\dagger}} = 
  (2\pi)^3 \, 2E_\pBssI \, \delta^{(3)}(\pBI-\qBI) \, \delta^{rs}
  \sepcomma
\end{align}
and all the others are zero.  
Here $a_\pBssI^{s\dagger}$ ($b_\pBssI^{s\dagger}$) is the 
creation operator for fermion (anti-fermion) 
with the momentum $\pBI$ and the spin $s$.  
A quantized Dirac field $\psi(x)$ can be expanded in terms of them: 
\begin{align}
  \psi(x) &= \int\!\!\frac{\diffn{p}{3}}{(2\pi)^3\,2E_\pBssI}
    \sum_s \left(
        a_\pBssI^s u(p,s) \expo^{-i p{\cdot}x}
      + b_\pBssI^{s\dagger} v(p,s) \expo^{+i p{\cdot}x}
    \right)
  \label{eq:mode-exp}
  \sepcomma\\
  \psiB(x) &= \int\!\!\frac{\diffn{p}{3}}{(2\pi)^3\,2E_\pBssI}
    \sum_s \left(
        a_\pBssI^{s\dagger} \uB(p,s) \expo^{+i p{\cdot}x}
      + b_\pBssI^s \vB(p,s) \expo^{-i p{\cdot}x}
    \right)
  \sepperiod\nonumber
\end{align}
Since the anticommutator relations above are invariant under 
the phase rotation $a \to \expo^{i\delta} a$ (and the similar for $b$), 
the relative phase between $u$ and $v$ is not fixed, a priori.  
Our convention is explained in \SecRef{sec:antipariticle-field}.  

A one-particle state is $\ket{\pBI,s} = a_\pBssI^{s\dagger} \ket{0}$.  
Note that 
$\braket{\pBI,r}{}{\qBI,s} = 
  (2\pi)^3 \, 2E_\pBssI \, \delta^{(3)}(\pBI-\qBI) \, \delta^{rs}$
is Lorentz invariant.  

Note that all flavor-dependence is carried by the creation-annihilation 
operators $a_\pBssI^s$ and $b_\pBssI^s$; c-number spinors $u(p,s)$ and 
$v(p,s)$ is flavor-independent.  
%------------------------------------------------------------------------
\subsection{Several conjugations: 
  $\Gamma^\dagger$, $\GammaB$, $\Gamma^c$ and $\Gamma^b$}
\label{sec:severel_conj}
There are several conjugations%
\footnote{
  We limit ourselves to the spinors in 4-dimension.  
}
for $\gamma$-matrices $\Gamma$; 
that is, hermitian conjugation $\Gamma^\dagger$, 
Dirac conjugation $\GammaB$, 
Charge conjugation $\Gamma^c$, and 
the conjugation related to Time Reversal $\Gamma^b$: 
\begin{align}
  \GammaB \equiv \gamma^0 \Gamma^\dagger \gamma^0
  \sepcomma\quad
  \Gamma^c \equiv C \Gamma^\sfT C^\dagger
  \sepcomma\quad
  \Gamma^b \equiv B^\dagger \Gamma^\sfT B
  \sepcomma
\end{align}
where the Charge conjugation matrix $C$ and the Time Reversal matrix $B$ are 
defined by the following properties: 
\begin{gather}
  C^\dagger = C^{-1}
  \sepcomma\quad
  C^\sfT = - C
  \sepcomma\quad
  B^\dagger = B^{-1}
  \sepcomma\quad
  B^\sfT = - B
  \sepcomma\\
  C{\gamma^\mu}^\sfT C^\dagger = -\gamma^\mu
  \sepcomma\quad
  B^\dagger {\gamma^0}^* B = + \gamma^0
  \sepcomma\quad
  B^\dagger {\gamma^i}^* B = - \gamma^i
  \label{eq:ChargeC-matrix=def}
  \sepperiod
\end{gather}
The last two relations can be combined to the single relation 
$B^\dagger {\gamma^\mu}^\sfT B = + \gamma^\mu$.  
For a product of $\gamma$-matrices, one can show that 
$\overline{\Gamma_1\Gamma_2\cdots\Gamma_n} 
= \GammaB_n\cdots\GammaB_2\GammaB_1$ and 
$(\Gamma_1\Gamma_2\cdots\Gamma_n)^c = \Gamma_n^c\cdots\Gamma_2^c\Gamma_1^c$.  
Some of these are collected in \TableRef{table:conj's}.  
\begin{table}[tbp]
\centering
\begin{align*}
\begin{array}{c|ccccccc}
  & 1 & i\gamma_5 & \gamma^\mu & \gamma^\mu\gamma_5 & \sigma^{\mu\nu} 
  & i\sigma^{\mu\nu}\gamma_5 & i 
  \\
  \hline
  \Gamma^\dagger & + & - & (-1)^\mu & -(-1)^\mu & (-1)^\mu(-1)^\nu 
  & -(-1)^\mu(-1)^\nu & - 
  \\
  \GammaB  & + & + & + & + & + & + & - 
  \\
  \Gamma^c & + & + & - & + & - & - & + 
  \\
  \Gamma^b & + & + & + & - & - & - & + 
\end{array}
\end{align*}
\begin{Caption}\caption{\small
    ``Eigenvalues'' of $\gamma$-matrices under the several conjugations 
    defined in the text: 
    $\GammaB \equiv \gamma^0 \Gamma^\dagger \gamma^0$, 
    $\Gamma^c \equiv C \Gamma^\sfT C^\dagger$, and 
    $\Gamma^b \equiv B^\dagger \Gamma^\sfT B$.  
    The symbol $(-1)^\mu$ is defined to be $+1$ for $\mu=0$ and to be 
    $-1$ for $\mu=1,2,3$.  
    The index $\mu$ on $(-1)^\mu$ is always not summed; 
    besides there is no distinction whether it is covariant or contravariant.  
    The notations $\gammaT^\mu \equiv (-1)^\mu\gamma^\mu$ and 
    $\sigmaT^{\mu\nu} \equiv (-1)^\mu(-1)^\nu\sigma^{\mu\nu}$ may sometimes 
    be convenient.  
\label{table:conj's}
}\end{Caption}
\end{table}
By using these conjugations, 
the operations 
$\ParityOp$, $\ChargeCOp$ and $\TimeROp$ for a spinor bilinear, 
which are defined in Eqs.~(\ref{eq:PCT_for_bilinear}), (\ref{eq:PCT=g_matrix}) 
and~(\ref{eq:PCT=g_matrix-mom}), can be expressed as follows: 
\begin{align}
  \gamma^0 \Gamma \gamma^0 = (\GammaB)^\dagger
  \sepcomma\quad
  C \Gamma^\sfT C^\dagger = \Gamma^c
  \sepcomma\quad
  B^\dagger \Gamma^* B = (\Gamma^b)^\dagger
  \label{eq:Dirac-conj_Parity,etc}
  \sepperiod
\end{align}
One can see that 
``Parity'' and ``Time Reversal'' do not reverse the order of 
$\gamma$-matrices.  
For successive applications of the conjugations, one can show 
the following relations: 
\begin{align*}
  \overline{\Gamma^c} = (\GammaB)^c
  \sepcomma\quad
  \overline{\Gamma^b} = (\GammaB)^b
  \sepperiod
\end{align*}
Also one can show $(CB)\gamma^\mu(CB)^\dagger = - \gamma^\mu$, 
which implies $CB=\gamma_5$ up to phase.  
Note that $BC = (CB)^\sfT$.  

A Dirac conjugated $\gamma$-matrix $\GammaB$ appears when one takes the 
hermitian conjugation of a spinor bilinear: 
\begin{align}
  \left( \psiB_1 \Gamma \psi_2 \right)^\dagger 
  = \psiB_2 \GammaB \psi_1
  \sepcomma
\end{align}
for both anti-commuting and commuting spinors.  
%------------------------------------------------------------------------
\subsection{Antiparticle field $\psi^c$}
\label{sec:antipariticle-field}
The antiparticle field $\psi^c$ of a Dirac field $\psi$ is defined by 
\begin{align}
  \psi^c(x) \equiv C \psiB^\sfT(x)
  \sepcomma\qquad
  \overline{\psi^c}(x) = - \psi^\sfT(x) C^\dagger
  \label{eq:anti-fermion-field=def}
  \sepcomma
\end{align}
where $C$ is the charge conjugation matrix defined 
in \EqRef{eq:ChargeC-matrix=def}.  
It coincides with the charge-conjugated field 
$\ChargeCOp\psi\ChargeCOp^\dagger$ up to field-dependent phase.  
The following relations can be shown: 
\begin{align}
  (\psi^c)^c = \psi
  \sepcomma\quad
  (\psi_L)^c = (\psi^c)_R
  \sepcomma\quad
  (\psi_R)^c = (\psi^c)_L
  \sepcomma
\end{align}
where $\psi_{L/R} \equiv P_{L/R} \psi$ and 
$\overline{\psi_{L/R}} = \psiB P_{R/L}$, where 
$P_{L/R} \equiv (1\mp\gamma_5)/2$.  
Our convention%
\footnote{
  In general, $v = \exp^{i\delta}u^c$, which means $u = \exp^{i\delta}v^c$.  
}
for the relative phase between c-number wave-functions 
$u$ and $v$ are 
\begin{align*}
  v \equiv u^c = C\uB^\sfT
  \sepcomma\quad
  \vB = -u^\sfT C^\dagger
  \sepcomma\quad
  v^\sfT = - \uB C
  \sepperiod
\end{align*}
See the next section for details.  
With this convention, 
an anti-fermion field $\psi^c(x)$ is expanded 
in terms of the creation-annihilation operators as follows: 
\begin{align}
  \psi^c(x) &= \int\!\!\frac{\diffn{p}{3}}{(2\pi)^3\,2E_\pBssI}
    \sum_s \left(
        b_\pBssI^s u(p,s) \expo^{-i p{\cdot}x}
      + a_\pBssI^{s\dagger} v(p,s) \expo^{+i p{\cdot}x}
    \right)
  \label{eq:mode-exp_for_psi-c}
  \sepcomma\\
  \overline{\psi^c}(x) &= \int\!\!\frac{\diffn{p}{3}}{(2\pi)^3\,2E_\pBssI}
    \sum_s \left(
        b_\pBssI^{s\dagger} \uB(p,s) \expo^{+i p{\cdot}x}
      + a_\pBssI^s \vB(p,s) \expo^{-i p{\cdot}x}
    \right)
  \sepperiod\nonumber
\end{align}
One can see that $a$ and $b$ in $\psi$ are interchanged in $\psi^c$.  
Thus the role of particle and anti-particle is indeed interchanged.  

A spinor bilinear can be rewritten in terms of their anti-particle fields: 
\begin{align}
  \psiB_1 \Gamma \psi_2 = +\overline{\psi_2^c} \Gamma^c \psi_1^c
  \sepcomma\quad\mbox{ for anti-commuting spinors.  }
\end{align}
For commuting spinors, 
$\uB_1 \Gamma v_2 = -\overline{v_2^c}\Gamma^c u_1^c = -\uB_2\Gamma^c v_1$, 
for example.  The second equality holds irrespective of the phase convention 
between $u$ and $v$, since $v=\expo^{i\delta}u^c$ means 
$u=\expo^{i\delta}v^c$.  
%------------------------------------------------------------------------
\subsection{Dirac representation}
\label{sec:Dirac_rep}
For an non-relativistic particle, the time component of the 4-momentum 
is much larger than the space component.  
Thus it is convenient to choose the representation of $\gamma$-matrices 
such that the time component is ``special''.  
Dirac representation is one of them: 
\begin{gather}
  \gamma^0 = 
  \begin{pmatrix}
    1 &\\
      & -1
  \end{pmatrix}
  \sepcomma\quad
  \gamma^i = 
  \begin{pmatrix}
              &  \sigma^i \\
    -\sigma^i & 
  \end{pmatrix}
  \sepcomma\quad
  \sigma^{\mu\nu} \equiv \frac{i}{2}\commutator{\gamma^\mu}{\gamma^\nu}
  \label{eq:Dirac_gamma_def}
  \sepcomma\\
  \gamma^5 \equiv 
  \frac{i}{4!}\epsilon_{\mu\nu\rho\sigma}
  \gamma^\mu\gamma^\nu\gamma^\rho\gamma^\sigma = 
  i\gamma^0\gamma^1\gamma^2\gamma^3 = 
  \begin{pmatrix}
      & 1\\
    1 &
  \end{pmatrix}
  \sepcomma\nonumber
\end{gather}
where $\sigma^i$ are Pauli matrices: 
\begin{align*}
  \sigma^1 \equiv
  \begin{pmatrix}
      & 1\\
    1 &
  \end{pmatrix}
  \sepcomma\quad
  \sigma^2 \equiv
  \begin{pmatrix}
       & -i\\
    +i &
  \end{pmatrix}
  \sepcomma\quad
  \sigma^3 \equiv
  \begin{pmatrix}
    1 & \\
      & -1
  \end{pmatrix}
  \sepperiod
\end{align*}
These satisfy $\anticommutator{\gamma^\mu}{\gamma^\nu}=2g^{\mu\nu}$, where 
$g^{\mu\nu} = \mbox{diag}(+,-,-,-)$; and 
$\sigma^i \sigma^j = \delta^{ij} + i\epsilon^{ijk}\sigma^k$, where 
$\epsilon^{123} = +1$.  
Our convention for the 4-dimensional totally-antisymmetric 
tensor is that $\epsilon_{0123}=+1$.  
Note that the phase of $\gamma_5$ is chosen such that the chirality 
(= eigenvalue of $\gamma_5$) coincides with the helicity for a massless 
particle (or $u$); for a massless antiparticle (or $v$), 
the chirality is the opposite to the helicity [\SecRef{sec:spin_helicity}].  
Note also that $(\gamma_5)^2=+1$.  
Explicit forms of the charge conjugation matrix $C$ and the 
time-reversal matrix $B$ in the Dirac representation are 
\begin{align}
  C = i \gamma^2 \gamma^0 = 
  \begin{pmatrix}
               & -i\sigma^2 \\
    -i\sigma^2 & 
  \end{pmatrix}
  \sepcomma\qquad
  B = \gamma^1 \gamma^3 = 
  \begin{pmatrix}
    i\sigma^2 &           \\
              & i\sigma^2 
  \end{pmatrix}
  \sepperiod
\end{align}
Overall phases are not fixed by the defining properties 
in \EqRef{eq:ChargeC-matrix=def}.  
Our convention is such that $CB=\gamma^5$, which means 
$BC=(CB)^\sfT=\gamma_5^*=\gamma_5^\sfT$.  

Freely-propagating spinor states are specified by the 4-momentum 
$p^\mu = (E,\pBI)$ and 
the (3 dimensional) spin direction $\sBI$ ($|\sBI|^2=1$) at the rest frame.  
Note that $\sBI$ is not the space component of the spin 4-vector $s^\mu$ 
that is obtained by boosting $(0,\sBI)^\mu$ to the ``laboratory'' frame.  
However when the distinction between $\sBI$ and $s^\mu$ is not important, 
we write simply $s$.  
The c-number wave-functions are as follows: 
\begin{align}
  u(p,s) = \sqrt{E+m}
  \begin{pmatrix}
    \chi^s  \rule[-3ex]{0em}{0ex}  \\
    \dfrac{\dprod{\pBI}{\sigmaBI}}{E+m} \chi^s
  \end{pmatrix}
  \sepcomma\quad
  v(p,s) \equiv u^c(p,s) = \sqrt{E+m}
  \begin{pmatrix}
    \dfrac{\dprod{\pBI}{\sigmaBI}}{E+m} \chi^{-s}  \\
    \chi^{-s} \rule[+3ex]{0em}{0ex}  \\
  \end{pmatrix}
  \sepcomma
\end{align}
where $u^c(p,s) \equiv C\uB^\sfT (p,s)$.  The second relation defines the 
relative phase of $u$ and $v$.  From this it holds that $v^c(p,s) = u(p,s)$.  
One can see that the upper (lower) two components of $u$ ($v$) are 
$\Order{1}$, while the lower (upper) are $\Order{\beta}$.  
This is the reason why the Dirac representation is convenient for 
non-relativistic particles.  
%One can also see that 
%%
%\begin{gather*}
%  \uB \gamma^0 u = \Order{1}
%  \sepcomma\quad
%  \uB \gamma^i u = \Order{\beta}
%  \sepcomma\\
%  \uB \gamma^0\gamma_5 u = \Order{\beta}
%  \sepcomma\quad
%  \uB \gamma^i\gamma_5 u = \Order{1}
%  \sepcomma
%\end{gather*}
%%
%which mean the axial-vector interaction is suppressed by $\Order{\beta}$, 
%while the vector is not.  
%Thus it is convenient to separate interactions according to whether 
%it contains $\gamma_5$ or not.  
The expressions above can be obtained by boosting the expressions 
at the rest frame of a particle: 
\begin{align*}
  u(p,s) = \sqrt{2m}
           \begin{pmatrix}
             \chi^s \\ 0
           \end{pmatrix}
  \sepcomma\quad
  v(p,s) = \sqrt{2m}
           \begin{pmatrix}
             0  \\  \chi^{-s}
           \end{pmatrix}
  \sepperiod
\end{align*}
Note that the mass-dimension of a spinor field is $[\psi] = 3/2$, 
while that of a scalar field is $[\phi] = 1$.  
%------------------------------------------------------------------------
\subsubsection{Other representations}
Representations other than the Dirac representation can be obtained by 
unitary transformation%
\footnote{
  The fundamental relation 
  $\anticommutator{\gamma^\mu}{\gamma^\nu} = 2g^{\mu\nu}{\bf 1}$ 
  remains intact by the transformation $\gamma \to U \gamma U^{-1}$ 
  with invertible matrix $U$ that is not necessarily unitary.  
  However if $U$ is not unitary, $\psi^\dagger \psi$ changes when 
  one changes the representation.  
  It may be convenient if $\gamma^\mu$ is either hermitian or anti-hermitian.  
  Since $(\gamma^\mu)^2 = 2g^{\mu\mu}$ [no summation], 
  $\gamma^\mu$ should be hermitian (anti-hermitian) if $g^{\mu\mu}>0$ 
  ($<0$) [no summation].  
}
$U$.  
For example, 
so-called chiral (or Weyl) representation $\gamma^\mu_{\rm W}$ can be 
obtained from the Dirac representation $\gamma^\mu_{\rm D}$ as follows: 
\begin{align}
  \gamma^\mu_{\rm W} = U \gamma^\mu_{\rm D} U^\dagger
  \sepcomma\quad
  U = 
  \begin{pmatrix}
    1/\sqrt{2}  &  -1/\sqrt{2}  \\
    1/\sqrt{2}  &   1/\sqrt{2}
  \end{pmatrix}
  \sepperiod
\end{align}
Accordingly $\psi_{\rm W} = U \psi_{\rm D}$ and 
$\overline{\psi_W} = \overline{\psi_D} U^\dagger$.  
One can see that if $\gamma_{\rm D}$ is hermitian (anti-hermitian), 
then $\gamma_{\rm W}$ is also hermitian (anti-hermitian).  
As for the charge conjugation matrix $C$ and the time-reversal matrix $B$ 
\begin{align}
  C_{\rm W} = U C_{\rm D} U^\sfT
  \sepcomma\quad
  B_{\rm W} = U^* B_{\rm D} U^\dagger
  \sepcomma
\end{align}
in general.  These follow from the fundamental requirements 
$C\gamma^{\mu\sfT}C^\dagger=-\gamma^\mu$ and 
$B^\dagger\gamma^{\mu\sfT}B=\gamma^\mu$.  
Since $U$ for the Dirac-to-Weyl transformation is real, 
if one choose $C_{\rm D} = i\gamma_{\rm D}^2\gamma_{\rm D}^0$ for 
the Dirac representation, 
then $C_{\rm W} = i\gamma_{\rm W}^2\gamma_{\rm W}^0$ also for the 
Weyl representation.  
However in general, 
the representations of $C$ and $B$ in terms of $\gamma$-matrix 
depend on the representation of $\gamma$-matrix itself.  
On the other hand, since the fundamental requirement for the parity 
transformation is $A \gamma^{\mu\dagger} A^\dagger = \gamma^\mu$ 
($A=\gamma^0$), which means $A_{\rm W} = U A_{\rm D} U^\dagger$, 
the representation of ``parity matrix'' $A$ is independent to the 
representation of $\gamma$-matrix.  
One can also see that the relations $C^\sfT=-C$ and $B^\sfT=-B$ 
are representation independent.  On the other hand, 
a relation such as $C^\dagger = -C$ is representation dependent.  
%------------------------------------------------------------------------
\subsection{Pauli spin spinors $\chi^{\pm s}$ and helicity $h$}
\label{sec:spin_helicity}
%------------------------------------------------------------------------
\subsubsection{Spin $\protect\sBI$}
Pauli 2-component spinors $\chi^s$ and $\chi^{-s}$ for the spin direction%
\footnote{
  The subscript ${}_R$ emphasis that the $\sBI_R$ is not the space component 
  of the spin 4-vector $s^\mu$ in \EqRef{eq:spin4vector_def}.  
  Since we rarely refer only to the space component of $s^\mu$, 
  we sometimes write $\sBI_{\!R}$ simply $\sBI$.  
  In fact, the difference between $\sBI$ and the space component of 
  $s^\mu$ is $\Order{\beta^2}$, as can be seen in \EqRef{eq:spin4vector_def}.  
  When the distinction between $\sBI$ and $s^\mu$ is not important, 
  we may simply write them $s$.  
}
$\sBI_{\!R} = (\cos\phi\sin\theta,\sin\phi\sin\theta,\cos\theta)$ 
at the rest frame of a particle, is 
\begin{align}
  \chi^s = 
  \begin{pmatrix}
    \cos\dfrac{\theta}{2} \rule[-4ex]{0em}{0ex}  \\
    \sin\dfrac{\theta}{2} \expo^{i\phi}
  \end{pmatrix}
  \sepcomma\qquad
  \chi^{-s} \equiv
  (-i\sigma^2) (\chi^s)^* = 
  \begin{pmatrix}
    -\sin\dfrac{\theta}{2} \expo^{-i\phi}  \rule[-4ex]{0em}{0ex}  \\
    \cos\dfrac{\theta}{2}
  \end{pmatrix}
  \label{eq:pauli_spinor_def}
  \sepperiod
\end{align}
Note that 
$\chi^{-(-s)} \equiv (-i\sigma^2) (\chi^{-s})^* = - \chi^s$.  
The $\chi^s$ and $\chi^{-s}$ are eigenstates of spin operator 
for the direction $\sBI_{\!R}$: 
\begin{align*}
  \dprod{\sBI_{\!R}}{\sigmaBI}\chi^s = +\chi^s
  \sepcomma\quad
  \dprod{\sBI_{\!R}}{\sigmaBI}\chi^{-s} = -\chi^{-s}
  \sepperiod
\end{align*}
Thus 
$\chi^{\pm s}$ surely represent the states of $\mbox{spin}=\pm\sBI_R$, and 
$(1\pm\dprod{\sBI_{\!R}}{\sigmaBI})/2$ are 
the spin-projection operators for 2-component spinors.  

Spin 4-vector $s^\mu$ is obtained by boosting 
$(0,\sBI_{\!R})^\mu$ at the rest frame of the particle to 
the ``laboratory'' frame: 
\begin{align}
  s^\mu
  =
  \begin{pmatrix}
    \gamma \pdprod{\sBI_{\!R}}{\betaBI}  \rule[-3ex]{0em}{0ex}  \\
    \sBI_{\!R} + 
    \dfrac{\gamma^2}{\gamma+1} \pdprod{\sBI_{\!R}}{\betaBI} \betaBI
  \end{pmatrix}
  \label{eq:spin4vector_def}
  \sepcomma
\end{align}
where $\betaBI = \pBI/E$, $\beta = |\betaBI|$, 
$\gamma = 1/\sqrt{1-\beta^2}$.  
Note that $(p^\mu,s^\mu) \to (\pT^\mu,\mp\sT^\mu)$ is equivalent to 
$(\pBI,\sBI_{\!R}) \to (-\pBI,\pm\sBI_{\!R})$, 
that is, the Parity and Time Reversal transformations.  

The following relations are for the process 
$e^+(\pB_e)\,e^-(p_e) \to t(p_t,s_t)\,\tB(\pB_t,\sB_t)$ 
with $\sqrt{s} = 2\mt\gamma$, where $\gamma=1/\sqrt{1-\beta^2}$.  
We work in the \scCM\ frame of $t\tB$ (and $e^+e^-$).  
We neglect the electron mass, and dropped $\Order{\beta^2}$ and higher terms.  
Also note that $s_t$ and $\sB_t$ are spin 4-vectors at the 
``laboratory frame'', which is the \scCM\ frame of $t\tB$, while 
$\sBI_t$ and $\sBBI_t$ are spin 3-vectors at the rest frame of 
$t$ and $\tB$, respectively.  
Our convention for the totally antisymmetric tensor is 
$\epsilon_{0123} = +1$.  
Use $p_e + \pB_e = p_t + \pB_t$ and 
$\dprod{s_t}{p_t} = \dprod{\sB_t}{\pB_t} = 0$ 
to obtain other relations.  
\begin{gather*}
  \dprod{s_t}{p_e} 
  = 
%    + \dprod{\sBI_t}{\pBI_t} - \dprod{\sBI_t}{\pBI_e}
    + \dprod{\sBI_t}{(\pBI_t-\pBI_e)}
  \sepcomma\quad
  \dprod{\sB_t}{p_e} 
  = 
%    - \dprod{\sBBI_t}{\pBI_t} - \dprod{\sBBI_t}{\pBI_e}
    - \dprod{\sBBI_t}{(\pBI_t+\pBI_e)}
  \sepcomma\\
  \dprod{s_t}{\pB_e} 
  = 
%    + \dprod{\sBI_t}{\pBI_t} + \dprod{\sBI_t}{\pBI_e}
    + \dprod{\sBI_t}{(\pBI_t+\pBI_e)}
  \sepcomma\quad
  \dprod{\sB_t}{\pB_e} 
  = 
%    - \dprod{\sBBI_t}{\pBI_t} + \dprod{\sBBI_t}{\pBI_e}
    - \dprod{\sBBI_t}{(\pBI_t-\pBI_e)}
%
%  \sepcomma\\
%  \dprod{s_t}{p_t} = 0 \quad\mbox{ (exactly) }
%  \sepcomma\quad
%  \dprod{\sB_t}{p_t} = - 2 \dprod{\sBBI_t}{\pBI_t}
%  \sepcomma\\
%  \dprod{\sB_t}{\pB_t} = 0 \quad\mbox{ (exactly) }
%  \sepcomma\quad
%  \dprod{s_t}{\pB_t} = + 2 \dprod{\sBI_t}{\pBI_t}
%
  \sepcomma\\
  \epsilon_{\mu\nu\rho\sigma} (p_e)^\mu (\pB_e)^\nu (p_t)^\rho (s_t)^\sigma
  = -2\mt \dprod{\sBI_t}{\pcprod{\pBI_e}{\pBI_t}}
  \sepcomma\\
  \epsilon_{\mu\nu\rho\sigma} (p_e)^\mu (\pB_e)^\nu (p_t)^\rho (\sB_t)^\sigma
  = -2\mt \dprod{\sBBI_t}{\pcprod{\pBI_e}{\pBI_t}}
%
%  \sepcomma\\
%  \epsilon_{\mu\nu\rho\sigma} (p_e)^\mu (p_t)^\nu (\pB_t)^\rho (s_t)^\sigma
%  = +2\mt \dprod{\sBI_t}{\pcprod{\pBI_e}{\pBI_t}}
%  \sepcomma\quad
%  \epsilon_{\mu\nu\rho\sigma} (p_e)^\mu (p_t)^\nu (\pB_t)^\rho (\sB_t)^\sigma
%  = +2\mt \dprod{\sBBI_t}{\pcprod{\pBI_e}{\pBI_t}}
%
  \sepcomma\\
%  \epsilon_{\mu\nu\rho\sigma} (p_t)^\mu (\pB_t)^\nu (s_t)^\rho (\sB_t)^\sigma
%  = -2\mt \dprod{\pBI_t}{\pcprod{\sBI_t}{\sBBI_t}}
%  \sepcomma\quad
  \epsilon_{\mu\nu\rho\sigma} (p_e)^\mu (\pB_e)^\nu (s_t)^\rho (\sB_t)^\sigma
  = -2\mt \dprod{\pBI_e}{\pcprod{\sBI_t}{\sBBI_t}}
  \sepcomma\\
  \epsilon_{\mu\nu\rho\sigma} (p_e)^\mu (p_t)^\nu (s_t)^\rho (\sB_t)^\sigma
  = 
%    + \mt \dprod{\pcprod{\sBI_t}{\sBBI_t}}{\pBI_t}
%    - \mt \dprod{\pcprod{\sBI_t}{\sBBI_t}}{\pBI_e}
    + \mt \dprod{\pcprod{\sBI_t}{\sBBI_t}}{(\pBI_t-\pBI_e)}
%  \\
%  \hspace*{5cm}
%    + \frac{1}{\mt} \pdprod{\sBI_t}{\pBI_t} 
%      \dprod{\sBBI_t}{\pcprod{\pBI_e}{\pBI_t}}
%    + \frac{1}{\mt} \pdprod{\sBBI_t}{\pBI_t} 
%      \dprod{\sBI_t}{\pcprod{\pBI_e}{\pBI_t}}
    + \frac{1}{\mt} 
      \bigl( \pdprod{\sBI_t}{\pBI_t} \sBBI_t 
           + \pdprod{\sBBI_t}{\pBI_t} \sBI_t 
      \bigr) {\cdot} {\pcprod{\pBI_e}{\pBI_t}}
  \sepcomma\\
  \epsilon_{\mu\nu\rho\sigma} (\pB_e)^\mu (p_t)^\nu (s_t)^\rho (\sB_t)^\sigma
  = 
%    + \mt \dprod{\pcprod{\sBI_t}{\sBBI_t}}{\pBI_t}
%    + \mt \dprod{\pcprod{\sBI_t}{\sBBI_t}}{\pBI_e}
    + \mt \dprod{\pcprod{\sBI_t}{\sBBI_t}}{(\pBI_t+\pBI_e)}
%  \\
%  \hspace*{5cm}
%    - \frac{1}{\mt} \pdprod{\sBI_t}{\pBI_t} 
%      \dprod{\sBBI_t}{\pcprod{\pBI_e}{\pBI_t}}
%    - \frac{1}{\mt} \pdprod{\sBBI_t}{\pBI_t} 
%      \dprod{\sBI_t}{\pcprod{\pBI_e}{\pBI_t}}
    - \frac{1}{\mt} 
      \bigl( \pdprod{\sBI_t}{\pBI_t} \sBBI_t 
           + \pdprod{\sBBI_t}{\pBI_t} \sBI_t 
      \bigr) {\cdot} {\pcprod{\pBI_e}{\pBI_t}}
  \sepperiod
\end{gather*}
%
%------------------------------------------------------------------------
\subsubsection{Helicity $h$}
Spin degrees of freedom can also be specified by helicity%
\footnote{
  The helicity of massless particles may be defined by the 
  limiting procedure $m\to 0$.  
}
$h=\dprod{\sBI_{\!R}}{\pBI}/|\pBI|$.  
The state of $h=+1$ ($-1$) is called right-handed (left-handed) 
helicity state, for both particle and anti-particle.  
Thus the spin for helicity eigenstates are 
$\sBI = \pm\pBI/|\pBI| = \pm\betaBI/|\betaBI|$ for $h=\pm$.  
Since $\dprod{\sBI}{\sigmaBI}\chi^{\pm s} = \pm\chi^{\pm s}$, we have 
\begin{align*}
  \frac{\dprod{\pBI}{\sigmaBI}}{|\pBI|} \chi^{\pm} = \pm \chi^{\pm}
  \sepcomma
\end{align*}
where $\chi^{\pm} \equiv \chi^{\pm s}$ with $\sBI_{\!R} = +\pBI/|\pBI|$.  
By noting $\chi^{-(\pm)} = \pm \chi^{\mp}$, Dirac spinors of helicity 
eigenstates are 
\begin{align}
  u(p,h=\pm) = \sqrt{E+m}
  \begin{pmatrix}
    \chi^{\pm}  \rule[-3ex]{0em}{0ex}  \\
    \dfrac{\pm|\pBI|}{E+m} \chi^{\pm}
  \end{pmatrix}
  \sepcomma\quad
  v(p,h=\pm) = \sqrt{E+m}
  \begin{pmatrix}
    \dfrac{-|\pBI|}{E+m} \chi^{\mp}  \\
    \pm \chi^{\mp} \rule[+3ex]{0em}{0ex}  \\
  \end{pmatrix}
  \label{eq:uv_with_helicity}
  \sepperiod
\end{align}
Here $v(p,h) = u^c(p,h)$, 
since charge conjugation do not flip neither momentum nor spin.  
For a massless state, we have 
\begin{align*}
  u(p,h=\pm) = \sqrt{E}
  \begin{pmatrix}
    \chi^{\pm}  \\
    h \, \chi^{\pm}  
  \end{pmatrix}
  \sepcomma\quad
  v(p,h=\pm) = \sqrt{E}
  \begin{pmatrix}
    - \chi^{\mp}  \\
    h \, \chi^{\mp}  
  \end{pmatrix}
  \sepperiod
\end{align*}
With $\gamma_5$ in \EqRef{eq:Dirac_gamma_def}, 
we have 
\begin{align*}
  \gamma_5 \, u(p,h) = +h\,u(p,h)
  \sepcomma\quad
  \gamma_5 \, v(p,h) = -h\,v(p,h)
  \qquad\mbox{ for $m=0$,  }
\end{align*}
which means that our convention for the phase of $\gamma_5$ is such that 
it eigenvalue (= chirality) is the same as the helicity for $u$ 
($\sim$ fermion), 
while the opposite for $v$ ($\sim$ anti-fermion), for a massless case.  
Note that $u$ ($v$) can be used also for anti-fermion (fermion), since 
$\overline{u_1} \Gamma v_2 = - \overline{u_2} \Gamma^c v_1^c$, 
etc.\ [\SecRef{sec:antipariticle-field}].  

The spin 4-vector $s^\mu$ for helicity eigenstate is 
\begin{align*}
  s^\mu(h=\pm) 
  =
  \frac{\pm 1}{\beta}
  \begin{pmatrix}
    \gamma\beta^2  \rule[-3ex]{0em}{0ex}  \\
    \left( 1 + \frac{\gamma^2\beta^2}{\gamma+1} \right) \betaBI
  \end{pmatrix}
  =
  \frac{\pm \gamma}{\beta}
  \begin{pmatrix}
    \beta^2  \\
    \betaBI
  \end{pmatrix}
  \sepcomma\quad\mbox{ while }\quad
  \frac{p^\mu}{m\beta}
  =
  \frac{\gamma}{\beta}
  \begin{pmatrix}
    1  \\
    \betaBI
  \end{pmatrix}
  \sepperiod
\end{align*}
Thus $s^\mu(h=\pm 1) \to \pm p^\mu/m$ as $\beta \to 1$ both for 
a particle and an anti-particle.  
%------------------------------------------------------------------------
\subsection{Lorentz transformation for $\gamma$-matrices}
Parity operation etc.\ are defined to act on $\psi$'s, but 
in~\SecRef{sec:PCT=amp} we show that equivalent operations can be 
implemented as manipulations on $\gamma$-matrices.  
Likewise, although Lorentz transformation changes only 
fields and $(x^\mu,p^\mu,s^\mu)$, 
equivalent manipulations for $\gamma$-matrices, which are constant to 
Lorentz tr., are possible.  
Consider a Lorentz transformation 
\begin{align*}
  p^\mu \to {\Lambda^{\mu}}_{\nu} p^\nu
  \sepcomma\qquad
  \psi \to \Lambda_{\frac{1}{2}} \psi
  \sepperiod
\end{align*}
Since $\psiB\psi$ is a Lorentz scalar, 
$\psiB \to \psiB {\Lambda_{\frac{1}{2}}}^{-1}$.  
Since $\psiB p^{\mu} \gamma_\mu \psi$ is also a scalar, we have 
${\Lambda^{\mu}}_{\nu} {\Lambda_{\frac{1}{2}}}^{-1} 
  \gamma_\mu \Lambda_{\frac{1}{2}} = \gamma_\nu$, or 
\begin{align*}
  {\Lambda_{\frac{1}{2}}}^{-1} \gamma_\mu \Lambda_{\frac{1}{2}}
  = {(\Lambda^{-1})^\nu}_\mu \gamma_\nu
  \sepperiod
\end{align*}
This means $\gamma$-matrices are ``transformed backward'' by 
${\Lambda_{\frac{1}{2}}}^{-1}$ and $\Lambda_{\frac{1}{2}}$.  
For example, let us consider the Lorentz transformation 
$p^\mu = {\Lambda^{\mu}}_{\nu} (m,\vec{0})^\nu$, or 
${(\Lambda^{-1})^\nu}_\mu p^\mu = (m,\vec{0})^\nu$.  
Thus we have 
\begin{align*}
  {\Lambda_{\frac{1}{2}}}^{-1} \pS \Lambda_{\frac{1}{2}}
  = {(\Lambda^{-1})^\nu}_\mu p^\mu \gamma_\nu 
  = m \gamma^0
  \sepperiod
\end{align*}
%
%------------------------------------------------------------------------
\subsection{Projection operators for energy $\Lambda_\pm(p)$ and 
spin $\Sigma(s)$}
\label{sec:proj-E,s}
Thus energy projection operators $\Lambda_\pm(p)$ for positive 
and negative energy at the rest frame are 
\begin{align*}
  \Lambda_+ = 
  \begin{pmatrix}
    1 & \\
    & 0  
  \end{pmatrix}
  = \frac{1+\gamma^0}{2}
  = \frac{m+m\gamma^0}{2m}
  &
  \,\stackrel{\mbox{\small boost}}{\longrightarrow}\,
  \frac{m+\pS}{2m}
  \sepcomma\\
  \Lambda_- = 
  \begin{pmatrix}
    0 & \\
    & 1  
  \end{pmatrix}
  = \frac{1-\gamma^0}{2}
  = \frac{m-m\gamma^0}{2m}
  &
  \,\stackrel{\mbox{\small boost}}{\longrightarrow}\,
  \frac{m-\pS}{2m}
  \sepcomma
\end{align*}
or 
\begin{align}
  \Lambda_\pm(p) = \frac{\pm\pS+m}{2m}
  \sepperiod
\end{align}
It has a following properties: 
\begin{gather}
  \Lambda_+(p)\,u(p,s) = u(p,s)
  \sepcomma\quad
  \Lambda_-(p)\,v(p,s) = v(p,s)
  \sepcomma\nonumber\\
  \Lambda_-(p)\,u(p,s) = \Lambda_+(p)\,v(p,s) = 0 
  \sepcomma\nonumber\\
  \Lambda_\pm(p)\,\Lambda_\pm(p) = \Lambda_\pm(p)
  \sepcomma\quad
  \Lambda_+(p)\,\Lambda_-(p) = 0
  \quad\mbox{ for $p^2 = m^2$, }\\
  \Lambda_+(p) + \Lambda_-(p) = 1
  \sepcomma\nonumber\\
  \pS \, \Lambda_{\pm}(p) = \pm m \, \Lambda_{\pm}(p)
  \quad\mbox{ for $p^2 = m^2$.  }
  \nonumber
\end{gather}
Use the relation $(\uB_1 \Gamma v_2)^\dagger = \vB_2\GammaB u_1$ 
for $\uB(p,s)$ etc.  

Likewise, for the spin-projection operator $\Sigma(s)$ 
at the rest frame, and the spin parallel to $z$-axis ($s_z=\pm 1$), 
\begin{align}
  \Sigma(s) = 
  \begin{pmatrix}
    \frac{1+s_z\sigma_z}{2} & \\
    & \frac{1-s_z\sigma_z}{2}
  \end{pmatrix}
  \,\stackrel{\mbox{\small rotate}}{\longrightarrow}\,
  \begin{pmatrix}
    \frac{1+\sBI\cdot\sigmaBI}{2} & \\
    & \frac{1-\sBI\cdot\sigmaBI}{2}
  \end{pmatrix}
  = 
  \frac{1+\dprod{\sBI}{\gammaBI}\gamma_5}{2}
  \,\stackrel{\mbox{\small boost}}{\longrightarrow}\,
  \frac{1-\sS\gamma_5}{2}
  \sepperiod
\end{align}
This has the following properties: 
\begin{gather}
  \Sigma(s)\,u(p,s) = u(p,s)
  \sepcomma\quad
  \Sigma(s)\,v(p,s) = v(p,s)
  \sepcomma\nonumber\\
  \Sigma(-s)\,u(p,s) = \Sigma(-s)\,v(p,s) = 0
  \sepcomma\nonumber\\
  \Sigma(s)\,\Sigma(s) = \Sigma(s)
  \sepcomma\quad
  \Sigma(s)\,\Sigma(-s) = 0
  \sepcomma\\
  \Sigma(s) + \Sigma(-s) = 1
  \sepperiod\nonumber
\end{gather}
At the rest frame, $p^\mu = (m,\vec{0})$ and $s^\mu = (0,\vec{s})$.  
Thus $\dprod{p}{s} = 0$.  From this, projection operators for 
energy and spin commute: 
\begin{align}
  \Lambda_\pm(p)\,\Sigma(s) = \Sigma(s)\,\Lambda_\pm(p)
  \sepperiod
\end{align}

Just the same manipulations are possible also for $\chi^s$ and $\chi^{-s}$.  
We show only the results.  These can be verified also by direct calculations.  
For simplicity we mean $\sBI_R$ by $\sBI$, not the space component of 
$s^\mu$ in arbitrary frame.  
They are orthogonal 
${\chi^s}^\dagger \chi^{-s} = {\chi^{-s}}^\dagger \chi^{s} = 0$.  
\begin{align}
  \chi^s {\chi^s}^\dagger = \frac{1+\sBI{\cdot}\sigmaBI}{2}
  \sepcomma\quad
  \chi^{-s} {\chi^{-s}}^\dagger = \frac{1-\sBI{\cdot}\sigmaBI}{2}
  \sepperiod
\end{align}
Similarly for $u$ and $v$, 
\begin{align}
  &
  u(p,s)\,\uB(p,s) = \frac{1-\sS\gamma_5}{2}(\pS+m)
  = 2m \, \Sigma(s) \, \Lambda_+(p)  \sepcomma\\
  &
  v(p,s)\,\vB(p,s) = \frac{1-\sS\gamma_5}{2}(\pS-m)
  = - 2m \, \Sigma(s) \, \Lambda_-(p)  \sepperiod\nonumber
\end{align}

As we saw above, at high energy, the spin 4-vector of a particle becomes 
parallel to the 4-momentum of the particle: 
$s^\mu(h=\pm 1) \to \pm p^\mu/m$ as $\beta \to 1$.  
Here $h$ means helicity.  
Since $p^\mu\,(\pm\pS+m) = \pm m\,(\pm\pS+m)$ for $p^2=m^2$, 
we have for $\beta \simeq 1$, 
\begin{align}
  u(p,h)\,\uB(p,h) = \frac{1+h\gamma_5}{2}\pS  
  \sepcomma\quad
  v(p,h)\,\vB(p,h) = \frac{1-h\gamma_5}{2}\pS  
  \sepcomma
\end{align}
where $h=+1$ ($-1$) for right-handed (left-handed) helicity state.  
One can see that the relation between the helicity and the chirality 
(= eigenvalue of $\gamma_5$) is opposite between particle and antiparticle, 
or to be precise, $u$ and $v$.  
%-------------------------------------------------------------------------
\section{Gordon identities}
\label{sec:Gordon-id}
In this section, we show that 
the only \itCP-odd $t$-$\tB$-gauge boson interaction of dim-5 
is that of EDM interaction; others are not independent.  
We closely follow the Appendix.C of~\cite{BLMN89}.  
Let us see $t\tB Z$ interaction, for example.  
By integrating by parts, all derivatives on, say, $Z_\mu$ can be 
removed.  After this manipulation, dim-5 terms of effective Lagrangian 
can be written as follows: 
\begin{align*}
  \calL =& 
  \bigl( 
      g_1  \psiB\partial^\mu\psi 
    - g_2  \psiB\gamma_5\partial^\mu\psi
    + ig_3 \psiB\sigma^{\mu\nu}\partial_\nu\psi
    + g_4  \psiB i\sigma^{\mu\nu}\gamma_5\partial_\nu\psi
  \bigr) Z_\mu +\ \mbox{h.c.}
\end{align*}
Couplings $g_{1\sim 4}$ can be complex in general; but should be 
real in order for those terms are odd under \itCP.  
In interaction picture, 
where the field operators satisfy equations of motion for free fields 
\begin{align*}
  (i\Slash{\partial} - m_t)\psi = 0  \sepcomma\quad
  (\square + m_Z^2) Z_\mu = 0  \sepcomma\quad
  \partial^\mu Z_\mu = 0  \sepcomma
\end{align*}
the following relations hold: 
\begin{align*}
  \partial_\nu (\psiB \sigma^{\mu\nu} \psi)
  &= - (\psiB i\lrpartial^\mu \psi)
    + 2 m \psiB \gamma^\mu \psi
  \sepcomma\\
  (\psiB \sigma^{\mu\nu} i\lrpartial_\nu \psi)
  &= \partial^\mu (\psiB\psi)
  \sepcomma\\
  \partial_\nu (\psiB i\sigma^{\mu\nu}\gamma_5 \psi)
  &= - (\psiB i\gamma_5 i\lrpartial^\mu \psi)
  \sepcomma\\
  (\psiB i\sigma^{\mu\nu}\gamma_5 i\lrpartial_\nu \psi)
  &= - \partial^\mu (\psiB i\gamma_5 \psi)
    + 2 m \psiB \gamma^\mu\gamma_5 \psi
  \sepperiod
\end{align*}
In these relations, arguments of $\psi$ and $\psiB$ are all $x$, and 
$\partial_\nu = \partial/\partial x^\nu$.  
The first one is the celebrated Gordon decomposition: 
\begin{align*}
  \psiB \gamma^\mu \psi
  = 
  \frac{1}{2m} \left[
      (\psiB i\lrpartial^\mu \psi)
    + \partial_\nu (\psiB \sigma^{\mu\nu} \psi)
  \right]
  \sepperiod
\end{align*}
By using these ``Gordon identities'', 
the Lagrangian above can be organized as follows: 
\begin{align*}
  \calL 
  &={}
  \bigl(
      g_1' \partial^\mu(\psiB\psi)
    + g_2' \partial_\nu(\psiB i\sigma^{\mu\nu}\gamma_5 \psi)
  \bigr) Z_\mu
  \\
  &={}
  g_2' \psiB i\sigma^{\mu\nu}\gamma_5 \psi \, \partial_\mu Z_\nu
  \\
  &={}
  - g_2' (\psiB i\gamma_5 i\lrpartial^\mu \psi)Z_\mu
  \sepperiod
\end{align*}
Formulas for $u(p)$ etc.\ are obtained from 
\begin{gather*}
  \bra{p,s} \psiB(x) 
  = \bra{0} \uB(p,s)\,\expo^{ip\cdot x}
  \sepcomma\qquad
  \psi(x)\ket{p,s} 
  = u(p,s)\,\expo^{-ip\cdot x} \ket{0}
  \sepcomma\\
  \bra{\pB,\sB} \psi(x) 
  = \bra{0} v(\pB,\sB)\,\expo^{i\pB\cdot x}
  \sepcomma\qquad
  \psiB(x)\ket{\pB,\sB} 
  = \vB(\pB,\sB)\,\expo^{-i\pB\cdot x} \ket{0}
  \sepperiod
\end{gather*}
For example, with the convention 
$\bra{p,\pB} = (\ket{p,\pB})^\dagger 
= (a_p^\dagger a_\pB^\dagger \ket{0})^\dagger
= \bra{0} a_\pB a_p$, 
\begin{align*}
  \bra{p,\pB} \psiB \Gamma i\lrpartial^\mu \psi \ket{0}
  &= +(p-\pB)^\mu\,\uB(p)\Gamma v(\pB)\,\expo^{i(p+\pB)\cdot x}
  \sepcomma\\
  \bra{p'} \psiB \Gamma i\lrpartial^\mu \psi \ket{p}
  &= +(p'+p)^\mu\,\uB(p')\Gamma u(p)\,\expo^{i(p'-p)\cdot x}
  \sepcomma\\
  \bra{\pB'} \psiB \Gamma i\lrpartial^\mu \psi \ket{\pB}
  &= +(\pB'+\pB)^\mu\,\vB(\pB)\Gamma v(\pB')\,\expo^{i(\pB'-\pB)\cdot x}
  \sepperiod
\end{align*}
Here $p$ ($\pB$) means the momentum for a particle (anti-particle).  
Note that the last equation includes the minus sign from the 
anti-commutativity of fermion.  
The exponential factor becomes 4-momentum conservation for each vertex: 
\begin{align*}
  \int\!\!\diffn{x}{4} \expo^{i(p+\pB)\cdot x} \expo^{-iq\cdot x} 
  = (2\pi)^4 \delta^{(4)}(p+\pB-q)
  \sepperiod
\end{align*}
Thus for the process $\bra{p,\pB} \leftarrow \ket{q}$, 
\begin{align*}
  iq_\nu \,\uB(p) \sigma^{\mu\nu} v(\pB)
  &= -(p-\pB)^\mu \,\uB(p) v(\pB)
    + 2 m\,\uB(p) \gamma^\mu v(\pB)
  \sepcomma\\
  (p-\pB)_\nu\,
  \uB(p) \sigma^{\mu\nu} v(\pB)
  &= iq^\mu \,\uB(p)v(\pB)
  \sepcomma\\
  iq_\nu \,\uB(p) i\sigma^{\mu\nu}\gamma_5 v(\pB)
  &= -(p-\pB)^\mu \,\uB(p) i\gamma_5 v(\pB)
  \sepcomma\\
  (p-\pB)_\nu \,\uB(p) i\sigma^{\mu\nu}\gamma_5 v(\pB)
  &= - iq^\mu \,\uB(p) i\gamma_5 v(\pB)
    + 2 m\,\uB(p) \gamma^\mu\gamma_5 v(\pB)
  \sepcomma
\end{align*}
where $q = p+\pB$.  At the \scCM-frame of $q^\mu = (\sqrt{s},\vec{0})$, 
\begin{gather*}
  \uB(p) \gamma^i\gamma_5 v(\pB)
  = \uB(p)\,i\sigma^{ij}\gamma_5\frac{-p^j}{m} v(\pB)
  \sepcomma\\
  i\frac{q_\nu}{2m}\,\uB(p)\,i\sigma^{i\nu}\gamma_5 v(\pB)
  = i\frac{\sqrt{s}}{2m}\,\uB(p)\,i\sigma^{i0}\gamma_5 v(\pB)
  = \uB(p)\,i\gamma_5 \frac{-p^i}{m}v(\pB)
  \sepcomma
\end{gather*}
for example.  
%-------------------------------------------------------------------------
\section{Merging the effects of $\gamma$ and $Z$ exchange}
\label{sec:merge_Z_and_photon}
In $e^+ e^- \to t\tB$, 
both $\gamma$ and $Z$ are exchanged in $s$-channel.  
But their effect can be combined to single effective couplings.  
With the SM vertices 
\begin{align*}
  {\Lambda_X}_\mu = 
     g_{\!\mbox{\tiny $X$}} 
     \left(v^{eX}\gamma_\mu - a^{eX}\gamma_\mu\gamma_5 \right)
  \sepcomma\qquad
  \Gamma_X^\mu = 
     g_{\!\mbox{\tiny $X$}} 
     \left( v^{tX}\gamma^\mu - a^{tX}\gamma^\mu\gamma_5 \right)
  \qquad
  (X = \gamma,Z)
\end{align*}
where 
\begin{gather}
  g_\gamma = e = g \sin\thetaW \sepcomma\quad
  v^{f\gamma} = Q_f  \sepcomma\quad
  a^{f\gamma} = 0  \sepcomma\nonumber\\
  \gZ = \frac{g}{\cos\thetaW} \sepcomma\quad
  v^{fZ} = \frac{1}{2} T_{3L} - Q_f \sin^2\theta_W  \sepcomma\quad
  a^{fZ} = \frac{1}{2} T_{3L}  
  \label{eq:def=pZ-coupl}
  \sepcomma
\end{gather}
or
\begin{align*}
  &   v^{e\gamma} = -1             \sepcomma  
  &&  v^{eZ}      = \tfrac{-1}{4}+\sin^2\theta_W=-0.01876  \sepcomma
  &&  a^{e\gamma} = 0              \sepcomma
  &&  a^{eZ}      = \tfrac{-1}{4}  \sepcomma
  \\
  &   v^{t\gamma} = \tfrac{+2}{3}  \sepcomma  
  &&  v^{tZ}      = \tfrac{+1}{4}-\tfrac{2}{3}\sin^2\theta_W=0.09584  \sepcomma
  &&  a^{t\gamma} = 0              \sepcomma
  &&  a^{tZ}      = \tfrac{+1}{4}  \sepcomma
\end{align*}
we have 
\begin{align*}
%  &
%  \frac{1}{s}
%  \bigl( \vB(\pB_e) {\Lambda_\gamma}_\mu u(p_e) \bigr)
%  \bigl( \uB(p_t) \Gamma_\gamma^\mu v(\pB_t) \bigr)
%  +
%  \frac{1}{s-m_Z^2}
%  \bigl( \vB(\pB_e) {\Lambda_Z}_\mu u(p_e) \bigr)
%  \bigl( \uB(p_t) \Gamma_Z^\mu v(\pB_t) \bigr)
%  \\
%  =
  &
  \sum_{X = \gamma,Z}
  \frac{1}{s-m_X^2}
  \bigl( \vB(\pB_e) {\Lambda_X}_\mu u(p_e) \bigr)
  \bigl( \uB(p_t) \Gamma_X^\mu v(\pB_t) \bigr)
  \\
  =&
  \frac{e^2}{s}
  \Bigl[
  \quad{}
    [v^e v^t]\,
    \bigl( \vB(\pB_e) \gamma_\mu u(p_e) \bigr)
    \bigl( \uB(p_t) \gamma^\mu v(\pB_t) \bigr)
  \\
  &\qquad{}
    -
    [v^e a^t]\,
    \bigl( \vB(\pB_e) \gamma_\mu u(p_e) \bigr)
    \bigl( \uB(p_t) \gamma^\mu\gamma_5 v(\pB_t) \bigr)
  \\
  &\qquad{}
    -
    [a^e v^t]\,
    \bigl( \vB(\pB_e) \gamma_\mu\gamma_5 u(p_e) \bigr)
    \bigl( \uB(p_t) \gamma^\mu v(\pB_t) \bigr)
  \\
  &\qquad{}
    +
    [a^e a^t]\,
    \bigl( \vB(\pB_e) \gamma_\mu\gamma_5 u(p_e) \bigr)
    \bigl( \uB(p_t) \gamma^\mu\gamma_5 v(\pB_t) \bigr)
  \Bigr]
  \sepcomma
\end{align*}
where
\begin{align}
  &  [v^e v^t] =
     v^{e\gamma}v^{t\gamma} + d(s)\,v^{eZ}v^{tZ}
  \sepcomma\nonumber\\
  &  [v^e a^t] =
     v^{e\gamma}a^{t\gamma} + d(s)\,v^{eZ}a^{tZ}
     = d(s)\,v^{eZ}a^{tZ}
  \sepcomma\nonumber\\
  &  [a^e v^t] =
     a^{e\gamma}v^{t\gamma} + d(s)\,a^{eZ}v^{tZ}
     = d(s)\,a^{eZ}v^{tZ}
  \label{eq:eff-coupling}
  \sepcomma\\
  &  [a^e a^t] =
     a^{e\gamma}a^{t\gamma} + d(s)\,a^{eZ}a^{tZ}
     = d(s)\,a^{eZ}a^{tZ}
  \nonumber
\end{align}
are energy-dependent ``couplings'', 
and $d(s)$ is a ratio of $Z$-propagator to $\gamma$-propagator: 
\begin{align}
  d(s) \equiv \frac{\gZ^2}{e^2}\frac{s}{s-m_Z^2+im_Z\Gamma_Z}
  \label{eq:ds-def=prop-ratio}
  \sepcomma
\end{align}
where
\begin{align*}
  \frac{\gZ^2}{e^2} 
  = \frac{\ds\pfrac{g}{\cos\thetaW}^2}{(g\sin\thetaW)^2}
  = \frac{1}{\cos^2\thetaW\,\sin^2\thetaW} 
  = \frac{1}{(1-0.23124) \cdot 0.23124}
  = 5.625
  = (2.372)^2
  \sepperiod
\end{align*}
Extensions for anomalous vertices should be obvious.  
The width $\Gamma_Z$ of $Z$ introduces absorptive part: 
$\Gamma_Z/m_Z = 2.490\GeV/91.187\GeV = 0.0273$.  
For $\sqrt{s} = 2 \times 175\GeV$, its relative magnitude is typically 
$s/(s-m_Z^2+im_Z\Gamma_Z) = 1.073 - 0.002\,i$.  
Thus for the threshold region, 
the effect of Coulomb rescattering overwhelms the effect of $\Gamma_Z$.  
Also in the open top region, it is known~\cite{TopPol} that 
the QCD vertex correction is larger than $\Gamma_Z$, 
as far as the normal component $\Polnorm$ of the polarization of 
top quark is concerned.  
%--------------------------------------------------------------------------
\section{Propagators}
%--------------------------------------------------------------------------
\subsection{Fermion propagator}
Let us define the momentum configuration at \scCM-frame of $t\tB$ as follows: 
\begin{align}
  q^\mu = \pt^\mu + \pBt^\mu = (2\mt+E,\vec{0})  \sepcomma\quad
  p^\mu = (\pt^\mu - \pBt^\mu)/2  \sepcomma\quad
  \pt = \frac{q}{2} + p  \sepcomma\quad
  -\pBt = -\frac{q}{2} + p  \sepperiod
\end{align}
It is sometimes convenient to use $-\pBt$ instead of $+\pBt$, since 
it is the former that enters to the propagators.  
In other words, 
\begin{align}
    \sum_{\rm spin} u(\pt)\uB(\pt) = \pS_t +\mt \sepcomma\quad
  - \sum_{\rm spin} v(\pBt)\vB(\pBt) = -\pBS_t +\mt \sepperiod
  \label{eq:u-v=spin-sum}
\end{align}
The negative sign for $v$ reflects anticommuting property of fermion.  
%%
%\begin{align}
%  \frac{\ptS+\mt}{2\mt} = \frac{+\dfrac{\qS}{2}+\pS+\mt}{2\mt}
%  \sepcomma\quad
%  \frac{-\ptBS+\mt}{2\mt} = \frac{-\dfrac{\qS}{2}+\pS+\mt}{2\mt}
%\end{align}
%%
%
\begin{gather}
  \pm\frac{\qS}{2}+\pS+\mt
  =
  2\mt\left[
    \frac{1\pm\gamma^0}{2} - \frac{\dprod{\vec{p}}{\vec{\gamma}}}{2\mt c}
    + \frac{1}{2\mt c^2}\left(\pm\frac{E}{2}+p^0\right)\gamma^0
  \right]
  \\
  \left( \pm\frac{q}{2} +p \right)^2 - \mt^2 + i\mt\Gamma_t
  = 2\mt \left[
        \frac{E}{2} \pm p^0 - \frac{\pBI^2}{2\mt}
      + \frac{1}{2\mt c^2} \left(
          \frac{E^2}{4} \pm E p^0 + (p^0)^2 \right)
      + i \frac{\Gamma_t}{2}
    \right]
  \sepperiod\nonumber
\end{gather}
Thus the Feynman propagators 
$S_F(\pt) = i(\pS_t+\mt)/(\pt^2-\mt^2+i\mt\Gamma_t)$ for $t$ and $\tB$ are 
\begin{align*}
  S_F(\pm q/2 + p) 
%  &=
%  \frac{i\left( \pm \qS/2+\pS+\mt \right)}
%       {\left( \pm q/2 +p \right)^2 - \mt^2 + i\mt\Gamma_t}  \\
  &=
  \frac{\ds i\left( 
    \frac{1\pm\gamma^0}{2} - \frac{\dprod{\vec{p}}{\vec{\gamma}}}{2\mt c}
  \right) }{\ds
      \left( \frac{E}{2} \pm p^0 + i \frac{\Gamma_t}{2} \right) 
    - \frac{\pBI^2}{2\mt} 
  }
  + \Order{\frac{1}{c^2}}
  \sepperiod
\end{align*}
\begin{align}
  \frac{\ptS+\mt}{2\mt} \gamma^i \frac{-\ptBS+\mt}{2\mt}
  =
  \frac{1+\gamma^0}{2} \gamma^i
  \frac{1-\gamma^0}{2} + \Order{\frac{1}{c}}
\end{align}
\begin{align}
  \frac{\ptS+\mt}{2\mt} \gamma^i\gamma_5 \frac{-\ptBS+\mt}{2\mt}
  =
  \frac{1+\gamma^0}{2} \left[ i\sigma^{ij}\gamma_5\frac{-p^j}{\mt c} \right]
  \frac{1-\gamma^0}{2} + \Order{\frac{1}{c^2}}
\end{align}
\begin{align}
  \frac{\ptS+\mt}{2\mt} \sigma^{0i}\gamma_5 \frac{-\ptBS+\mt}{2\mt}
  =
  \frac{1+\gamma^0}{2} \left[ i\gamma_5\frac{-p^i}{\mt c} \right]
  \frac{1-\gamma^0}{2} + \Order{\frac{1}{c^2}}
\end{align}
Note that the last two (and~\EqRef{eq:u-v=spin-sum}) 
are consistent with ``Gordon identities'' in~\SecRef{sec:Gordon-id}.  
%--------------------------------------------------------------------------
\subsection{Gluon propagator}
\label{sec:gluon_prop}
Gluon propagator 
$\mean{G_\mu^a(q) \, G_\nu^b(-q)} = D_{\mu\nu} \delta^{ab}$
is given by 
\begin{align}
  D_{\mu\nu} = \frac{i}{q^2+i\epsilon}
  \left( - g_{\mu\nu}+(1-\xi)\frac{q_\mu q_\nu}{q^2} \right)
\end{align}
for covariant gauge, and
\begin{align}
  &  D^{00} = \frac{i}{|\qBI|^2-i\epsilon}
  \sepcomma \qquad D^{0i} = D^{i0} = 0 \sepcomma
  \nonumber\\
  &  D^{ij} = \frac{i}{(q^0/c)^2 - |\qBI|^2+i\epsilon}
     \left( \delta^{ij}-\frac{q^i q^j}{|\qBI|^2} \right)
\end{align}
for Coulomb gauge.  
The later is convenient for non-relativistic calculations.  
%
%
%

%
%
%
%----------------------------------------------------------------------
\section{Parity, Charge Conjugation and Time Reversal}
\label{sec:PCT=general}
Within QFT, Parity operation $\ParityOp$ etc.\ are defined on 
creation-annihilation operators.  
However sometimes it's convenient to consider those operations 
for c-numbers such as $\uB(p)\gamma^\mu v(\pB)$.  

For explicit calculations, 
it is convenient to introduce a shorthand notation $\tilde{~}$ 
for flipping the sign of space-components of 4-vectors and tensors:  
\begin{align}
  p^\mu &= (p^0,p^i) \sepcomma
  &
  \gamma^\mu &= (\gamma^0,\gamma^i) \sepcomma
  &
  \sigma^{\mu\nu} &= (\sigma^{0i},\sigma^{ij}) \sepcomma
  \nonumber\\[2ex]
  \pT^\mu &\equiv (p^0,-p^i)
  &
  \gammaT^\mu &\equiv (\gamma^0,-\gamma^i)
  &
  \sigmaT^{\mu\nu} &\equiv (-\sigma^{0i},+\sigma^{ij})
  \label{eq:tilde-parity}
  \\
  &= (-1)^\mu p^\mu \sepcomma
  &
  &= (\gamma^\mu)^\dagger \sepcomma
  &
  &= (\sigma^{\mu\nu})^\dagger
  \sepperiod\nonumber
\end{align}
Relations to another notations are 
\begin{align*}
  (-1)^\mu \gamma^\mu = \gammaT^\mu = \gamma_\mu
  \sepcomma\quad
  (-1)^\mu(-1)^\nu \sigma^{\mu\nu} = \sigmaT^{\mu\nu} = \sigma_{\mu\nu}
  \sepperiod
\end{align*}
Note that the index on $(-1)^\mu$ is never summed over.  
Note also that 
\begin{gather*}
  \pT^\mu \gamma_\mu = p^\mu \gammaT_\mu = \SlashIt{\pT}
  \sepcomma\quad
  \pT^\mu \gammaT_\mu = p^\mu \gamma_\mu = \pS
  \sepcomma\\
  (-1)^\mu(-1)^\nu(-1)^\rho(-1)^\sigma\epsilon_{\mu\nu\rho\sigma}
  = -\epsilon_{\mu\nu\rho\sigma}
  \sepperiod
\end{gather*}
%
%----------------------------------------------------------------------
\subsection{$\ParityOp$, $\ChargeCOp$ and $\TimeROp$ for field operators}
\label{sec:PCT=op}
Parity operation etc.\ for a Dirac field operator $\psi(x)$ 
[\EqRef{eq:mode-exp}] are 
\begin{align}
  \ParityOp \psiB(x) \ParityOp^{-1}
  &= \eta_P^* {\eta_P^0}^* \psiB(\xT)\gamma^0
  \sepcomma&
  \ParityOp \psi(x) \ParityOp^{-1}
  &= \eta_P \eta_P^0 \gamma_0 \psi(\xT)
  \sepcomma\nonumber\\
  \ChargeCOp \psiB(x) \ChargeCOp^{-1}
  &= \eta_C^* {\eta_C^0}^* \overline{\psi^c}(x)
  \sepcomma&
  \ChargeCOp \psi(x) \ChargeCOp^{-1}
  &= \eta_C \eta_C^0 \psi^c(x)
  \sepcomma
  \label{eq:PCT-Dirac}
  \\
  \TimeROp \psiB(x) \TimeROp^{-1}
  &= \eta_T^* {\eta_T^0}^* \psiB(-\xT)B^\dagger
  \sepcomma&
  \TimeROp \psi(x) \TimeROp^{-1}
  &= \eta_T \eta_T^0 B \psi(-\xT)
  \sepcomma\nonumber
\end{align}
where $\psi^c$ is an antiparticle field defined 
in \EqRef{eq:anti-fermion-field=def}.  
The symbol $\eta_P$ etc.\ are flavor-dependent phases
$|\eta_P|^2 = 1$, 
and $\eta_P^0$ etc.\ are flavor-independent ``over-all'' phases 
$|\eta_P^0|^2 = 1$.  
In terms of creation-annihilation operators, $a_\pBssI^s$ and $b_\pBssI^s$, 
these transformations are 
\begin{align}
  \ParityOp a_\pBssI^s \ParityOp^{-1} &= +\eta_P \eta_P^0 a_{-\pBssI}^s
  \sepcomma&
  \ParityOp b_\pBssI^s \ParityOp^{-1} &= -\eta_P^* {\eta_P^0}^* b_{-\pBssI}^s
  \sepcomma\nonumber\\
  \ChargeCOp a_\pBssI^s \ChargeCOp^{-1} &= \eta_C \eta_C^0 b_{\pBssI}^s
  \sepcomma&
  \ChargeCOp b_\pBssI^s \ChargeCOp^{-1} &= \eta_C^* {\eta_C^0}^* a_{\pBssI}^s
  \sepcomma\\
  \TimeROp a_\pBssI^s \TimeROp^{-1} &= \eta_T \eta_T^0 a_{-\pBssI}^{-s}
  \sepcomma&
  \TimeROp b_\pBssI^s \TimeROp^{-1} &= \eta_T^* {\eta_T^0}^* b_{-\pBssI}^{-s}
  \sepperiod\nonumber
\end{align}
Note that c-number spinors $u(p,s)$ and $v(p,s)$ are flavor-independent, and 
\begin{align}
  \uB(p,s) \gamma^0 &= +\uB(\pT,s)
  &
  \gamma^0 u(p,s) &= +u(\pT,s)
  \nonumber\\
  \vB(p,s) \gamma^0 &= -\vB(\pT,s)
  &
  \gamma^0 v(p,s) &= - v(\pT,s)
  \sepcomma\nonumber\\[1ex]
%  \intertext{for $\ParityOp$,}
%
  u(p,s)^\sfT C^\dagger &= -\vB(p,s)
  &
  C \uB(p,s)^\sfT &= v(p,s)
  \nonumber\\
  v(p,s)^\sfT C^\dagger &= -\uB(p,s)
  &
  C \vB(p,s)^\sfT &= u(p,s)
  \label{eq:PCT-amp}
  \sepcomma\\[1ex]
%  \intertext{for $\ChargeCOp$, and}
%
  \bigl(\uB(p,s)\bigr)^* B &= \uB(\pT,\sT)
  &
  B^\dagger \bigl(u(p,s)\bigr)^* &= u(\pT,\sT)
  \nonumber\\
  \bigl(\vB(p,s)\bigr)^* B &= \vB(\pT,\sT)
  &
  B^\dagger \bigl(v(p,s)\bigr)^* &= v(\pT,\sT)
  \sepperiod\nonumber
\end{align}
%
%for $\TimeROp$.  
In order for $\ChargeCOp\ParityOp\TimeROp$ (and the permutations) to be 
a symmetry of an action, these phases cannot to be arbitrary: 
\begin{align*}
  &  \eta_{Pi} = \pm 1
  \sepcomma\quad
     \eta_{Ti} = \eta_{Pi}^* \eta_{Ci}^*
  \sepperiod
\end{align*}
%
%Note that if a Lagrangian has $\U(1)_{*}$-symmetry, one can redefine 
%``$\ChargeCOp\ParityOp\TimeROp$ followed by $\U(1)_{*}$'' to be 
%$\ChargeCOp\ParityOp\TimeROp$.  
If one (or more) of fields is Majorana, the reality condition 
$\psi = \psi^c$ restricts the ``over-all'' phases: 
\begin{align*}
  \eta_P^0 = \pm i
  \sepcomma\quad
  \eta_C^0 = \pm 1
  \sepcomma\quad
  \eta_T^0 = \pm 1
  \sepperiod
\end{align*}
Note that $\ChargeCOp$ and $\ParityOp$ commutes if and only if 
$\eta_{Pi}^*{\eta_P^0}^* = - \eta_{Pi}\eta_P^0$, or%
\footnote{
  This follows since one can choose $\eta_{Pi}=+1$ for a certain 
  species $i$.  
}
$\eta_P^0 = \pm i$.  
Note also that $\TimeROp$ is anti-unitary: 
\begin{align}
  \bra{f} \TimeROp^{-1} \calO \TimeROp \ket{i}
  &= \bra{\TimeROp(f)} \calO \ket{\TimeROp(i)}^*
  &
  \TimeROp (c\calO) \TimeROp^{-1}
  = c^* \TimeROp \calO \TimeROp^{-1}
  \sepperiod\nonumber\\
  &= \bra{\TimeROp(i)} \calO^\dagger \ket{\TimeROp(f)}
  \sepcomma
\end{align}
Only spinor-bilinears are relevant for scattering matrices $\calM$, 
because they are Lorentz scalars: 
\begin{align}
  \ParityOp [ \psiB_1(x) \Gamma(x) \psi_2(x) ] \ParityOp^{-1} 
  &= \eta_{P1}^* \eta_{P2} \,\Parity(\Gamma) \;
          \psiB_1(x) \Gamma(x) \psi_2(x) \Bigr|_{x = \xT}
  \sepcomma\nonumber\\
  \ChargeCOp [ \psiB_1(x) \Gamma(x) \psi_2(x) ] \ChargeCOp^{-1} 
  &= \eta_{C1}^* \eta_{C2} \,\ChargeC(\Gamma) \;
          \psiB_2(x) \Gamma(x) \psi_1(x)
  \sepcomma
  \label{eq:PCT_for_bilinear}
  \\
  \TimeROp [ \psiB_1(x) \Gamma(x) \psi_2(x) ] \TimeROp^{-1} 
  &= \eta_{T1}^* \eta_{T2} \,\TimeR(\Gamma) \;
          \psiB_1(x) \Gamma(x) \psi_2(x) \Bigr|_{x = -\xT}
  \sepperiod\nonumber
\end{align}
Here the parity eigenvalue $\Parity(\Gamma)$ for $\Gamma$ etc.\ are 
defined through the following relations: 
\begin{align}
  \gamma^0\Gamma(x)\gamma^0 &\equiv \Parity(\Gamma) \; \Gamma(\xT)
  \nonumber\\
  C \Gamma^\sfT(x) C^\dagger &\equiv \ChargeC(\Gamma) \; \Gamma(x)
  \label{eq:PCT=g_matrix}
  \\
  B^\dagger \Gamma^*(x) B &\equiv \TimeR(\Gamma) \; \Gamma(-\xT)
  \nonumber
\end{align}
These eigenvalues are summarized in the~\TableRef{table:PCT-bilinear}.  
For example, $\gamma^0 \gamma^i \gamma^0 = -\gamma^i$ ($i=1,2,3$) means 
$\Parity(\gamma^i) = -1$, and 
$\gamma^0 \frac{\partial}{\partial x^i} \gamma^0 = 
- \frac{\partial}{\partial (-x^i)} $ means 
$\Parity(\partial_i) = -1$.  
Note that, since charge conjugation interchanges fields, 
even though $\ChargeC(\partial_\mu) = +1$, 
be sure that $\ChargeCOp$ changes 
$\rightvec{\partial}_\mu$ to $\leftvec{\partial}_\mu$, 
since $(\rpartial_\mu)^\sfT = \lpartial_\mu$.  
Thus $\ChargeC(\lrpartial_\mu) = -1$, where 
$A\lrpartial B 
\equiv A(\rightvec{\partial} - \leftvec{\partial})B
= A(\partial B) - (\partial A)B$.  
Note also that $\TimeR(i\Gamma) = - \TimeR(\Gamma)$, or 
$\TimeR(\expo^{i\delta}\Gamma) = \expo^{-2i\delta} \TimeR(\Gamma)$, 
in general.  
\begin{table}[tbp]
\centering
\begin{align*}
\begin{array}{c|cccccccc}
  & 1 & i\gamma_5 & \gamma^\mu & \gamma^\mu\gamma_5 & \sigma^{\mu\nu} 
  & i\sigma^{\mu\nu}\gamma_5 
  & \rightvec{\partial}_{\mu}+\leftvec{\partial}_{\mu} 
  & i\leftrightvec{\partial}_\mu
  \\
  \hline
  \Parity  & + & - & (-1)^\mu & -(-1)^\mu & (-1)^\mu(-1)^\nu 
  & -(-1)^\mu(-1)^\nu & (-1)^\mu & (-1)^\mu
  \\
  \ChargeC & + & + & - & + & - & - & + & -
  \\
  \TimeR   & + & - & (-1)^\mu & (-1)^\mu & -(-1)^\mu(-1)^\nu 
  & (-1)^\mu(-1)^\nu & -(-1)^\mu & (-1)^\mu
  \\
  \ChargeC\Parity  & + & - & -(-1)^\mu & -(-1)^\mu & -(-1)^\mu(-1)^\nu 
  & (-1)^\mu(-1)^\nu & (-1)^\mu & -(-1)^\mu
  \\
  \ChargeC\Parity\TimeR  & + & + & - & - & + 
  & + & - & - 
\end{array}
\end{align*}
\begin{Caption}\caption{\small
    Eigenvalues of $\gamma$-matrices and derivative.  
    They are not transformed under parity $\ParityOp$ etc., which are 
    defined to operate on creation-annihilation operators.  
    It should be understood that they are sandwiched by $\psiB$ and $\psi$.  
    However, one can define similar transformations for c-numbers.  
    See the text for the definitions.  
    Note that $i\sigma^{\mu\nu}\gamma_5 = 
    \frac{1}{2} \epsilon^{\mu\nu\rho\sigma} \sigma_{\rho\sigma}$ with 
    $\epsilon_{0123} = +1$.  
    Although $\psiB(\rightvec{\partial}_{\mu}+\leftvec{\partial}_{\mu})\psi
    = \partial_\mu(\psiB\psi)$, this is listed because each of 
    $\rpartial$ and $\lpartial$ is not the eigenstate of $\ChargeCOp$.  
    The symbol $(-1)^\mu$ is defined to be $+1$ for $\mu=0$ and to be 
    $-1$ for $\mu=1,2,3$.  
    The index on $(-1)^\mu$ is not summed always; 
    besides there is no distinction whether covariant or contravariant.  
\label{table:PCT-bilinear}
}\end{Caption}
\end{table}

In the momentum space, eigenvalues for Parity etc.\ are defined as follows: 
\begin{align}
  \gamma^0\Gamma(p,s,\pB,\sB)\gamma^0 
  &\equiv \Parity(\Gamma) \; \Gamma(\pT,s,\pBT,\sB)
  \nonumber\\
  C \Gamma^\sfT(p,s,\pB,\sB) C^\dagger 
  &\equiv \ChargeC(\Gamma) \; \Gamma(\pB,\sB,p,s)
  \label{eq:PCT=g_matrix-mom}
  \\
  B^\dagger \Gamma^*(p,s,\pB,\sB) B 
  &\equiv \TimeR(\Gamma) \; \Gamma(\pT,\sT,\pBT,\sBT)
  \nonumber
\end{align}
For example, 
$\gamma^0 \bigl(i \gamma_5 (p-\pB)^i\bigr) \gamma_0 
= i\gamma_5\bigl(-(p-\pB)^i\bigr)$ means 
$\Parity\bigl(i\gamma_5 (p-\pB)^i\bigr) = +1$.  
This is in accord with 
$\Parity\bigl(i\gamma_5\,i\lrpartial^i\bigr) = +1$.  
From the definitions above, we have the following for 
Dirac conjugates $\GammaB \equiv \gamma^0\Gamma^\dagger\gamma^0$: 
\begin{align*}
  \Parity(\GammaB) = \Parity(\Gamma)^*
  \sepcomma\quad
  \ChargeC(\GammaB) = \ChargeC(\Gamma)^*
  \sepcomma\quad
  \TimeR(\GammaB) = \TimeR(\Gamma)^*
  \sepperiod
\end{align*}
$\Parity(\Gamma)$ and $\ChargeC(\Gamma)$ are (usually) real, 
but $\TimeR(\Gamma)$ may be complex depending on 
the choice of the phase of $\Gamma$: 
$\TimeR(\expo^{i\delta}\Gamma) = \expo^{-2i\delta} \TimeR(\Gamma)$.  

All entries in~\TableRef{table:PCT-bilinear} are chosen so that 
$\GammaB = \Gamma$.  Note $(\rpartial_\mu)^\dagger = \lpartial_\mu$.  
For a product of $\gamma$-matrices, 
$\overline{\Gamma_1\Gamma_2\cdots\Gamma_n} 
= \GammaB_n\cdots\GammaB_2\GammaB_1$.  
Likewise for $\Gamma^c \equiv C\Gamma^\sfT C^\dagger$, 
$(\Gamma_1\Gamma_2\cdots\Gamma_n)^c
= \Gamma_n^c\cdots\Gamma_2^c\Gamma_1^c$.  
Thus, be careful to the eigenvalue $\ChargeC$ of a product; 
it is not just the product of each eigenvalue, in general; 
however $\gamma$-matrices and derivatives commutes, of course.  
On the other hand, $\ParityOp$ and $\TimeROp$ do not reverse the order of 
matrices.  

We also summarize the transformation properties of gauge fields 
in~\TableRef{table:PCT-EMfield}.  
For non-Abelian gauge, 
$A_\mu = A_\mu^a T^a$ and $F_{\mu\nu} = F_{\mu\nu}^a T^a$.  
For $\ParityOp$ and $\TimeROp$, the arguments $x$ of fields change to 
$\xT$ and $-\xT$, respectively.  
For Charge Conjugation, 
$\ChargeCOp A_\mu \ChargeCOp^{-1} = - A_{\mu}^* = - A_{\mu}^\sfT$.  
In general, field dependent phases should also be considered, 
as in the case of $\psi$.  But we do not go into detail here.  
\begin{table}[tbp]
\centering
\begin{align*}
\begin{array}{c|ccccc}
  & A_\mu & F_{\mu\nu} & \EBI & \BBI & j^\mu
  \\
  \hline
  \Parity  & (-1)^\mu & (-1)^\mu(-1)^\nu & - & + & (-1)^\mu
  \\
  \ChargeC & - & - & - & - & - 
  \\
  \TimeR   & (-1)^\mu & -(-1)^\mu(-1)^\nu & + & - & (-1)^\mu
  \\
  \ChargeC\Parity  & -(-1)^\mu & -(-1)^\mu(-1)^\nu & + & - & -(-1)^\mu
  \\
  \ChargeC\Parity\TimeR  & - & + & + & + & - 
\end{array}
\end{align*}
\begin{Caption}\caption{\small
    A gauge connection $A_\mu$ should transform the same way to 
    $i\partial_\mu$.  
    For charge conjugation, extra minus sign comes when changing 
    $\lpartial_\mu$ to $\rpartial_\mu$ by integration by 
    parts.  As for time reversal, (seemingly) extra minus sign comes from 
    the anti-unitary nature of $\TimeROp$.  
\label{table:PCT-EMfield}
}\end{Caption}
\end{table}
%
%----------------------------------------------------------------------
\subsection{$\ChargeCOp\ParityOp\TimeROp$ symmetry}
\label{sec:CPT-sym}
From the Tables~~\ref{table:PCT-bilinear} and~\ref{table:PCT-EMfield}, 
one can see that a Lorentz invariant interaction such as 
$\psiB\gamma^\mu\psi\,A_\mu$ is $\ChargeCOp\ParityOp\TimeROp$-even: 
$\ChargeC\Parity\TimeR=+1$.  
It's easy to remember, since 
one can assign $-1$ for each 4-vector index $\mu$; for example, 
$\ChargeCOp\ParityOp\TimeROp$-eigenvalue of a tensor such as 
$\sigma^{\mu\nu}$ is $(-1)^2=+1$.  
%----------------------------------------------------------------------
\subsection{$\ParityOp$, $\ChargeCOp$ and $\TimeROp$ for currents}
\label{sec:PCT=currents}
When the topology is the same for all the diagrams contributing to $\calM$, 
we can apply $\ParityOp$, $\ChargeCOp$, or $\ChargeCOp\ParityOp$ to, 
say, an initial current and a final current, independently.  
A (tree-level) amplitude can be written as 
\begin{align*}
  \calM
  &=
  \sum_{A,B} \bra{t\tB} (\tB\Gamma_A t) (ZZ) (\eB\Lambda_B e) \ket{e\eB}
  \\
  &=
  \sum_{A,B} \bra{t\tB} (\tB\Gamma_A t) \ket{0}
  \bra{0} (ZZ) \ket{0}
  \bra{0} (\eB\Lambda_B e) \ket{e\eB}
  \\
  &\sim
  j_{t\tB} \, j^\dagger_{e^+e^-}
  \sepperiod
\end{align*}
We can rewrite, say, the final $t\tB$ current $j_{t\tB}$ 
using $\ParityOp$ etc.: 
\begin{align*}
  \bra{t\tB} (\tB\Gamma_{\Parity=\pm} t) \ket{0}
  &=
  \bra{t\tB} (\ParityOp^\dagger\ParityOp) (\tB\Gamma_{\Parity=\pm} t) 
  (\ParityOp^\dagger\ParityOp) \ket{0}
  \\
  &=
  \pm \bra{\ParityOp(t\tB)} (\tB\Gamma_{\Parity=\pm} t) \ket{0}
  \sepcomma\\[2ex]
  \bra{t\tB} (\tB\Gamma_{\ChargeC\Parity=\pm} t) \ket{0}
  &=
  \bra{t\tB} ((\ChargeCOp\ParityOp)^\dagger\ChargeCOp\ParityOp) 
  (\tB\Gamma_{\ChargeC\Parity=\pm} t) 
  ((\ChargeCOp\ParityOp)^\dagger\ChargeCOp\ParityOp) \ket{0}
  \\
  &=
  \pm \bra{\ChargeCOp\ParityOp(t\tB)} (\tB\Gamma_{\ChargeC\Parity=\pm} t) 
  \ket{0}
  \sepcomma\\[2ex]
  \bra{t\tB} (\tB\Gamma_{\TimeR=\pm} t) \ket{0}
  &=
  \bra{t\tB} (\TimeROp^\dagger\TimeROp) (\tB\Gamma_{\TimeR=\pm} t) 
  (\TimeROp^\dagger\TimeROp) \ket{0}
  \\
  &=
  \pm \bra{\TimeROp(t\tB)} (\tB\Gamma_{\TimeR=\pm} t) \ket{0}^*
  \sepcomma
\end{align*}
where 
\begin{align*}
  \ket{\ParityOp (p,s,\pB,\sB)} &= -\ket{(\pT,s,\pBT,\sB)}
  \sepcomma\\
  \ket{\ChargeCOp (p,s,\pB,\sB)} &= \ket{(\pB,\sB,p,s)}
  \sepcomma\\
  \ket{\TimeROp (p,s,\pB,\sB)} &= \ket{(\pT,\sT,\pBT,\sBT)}
  \sepcomma
\end{align*}
because
\begin{align*}
  \ket{\ParityOp(p,s,\pB,\sB)}
  &=
  \bigl( \ParityOp a^\dagger(p,s) \ParityOp^\dagger \bigr)
  \bigl( \ParityOp b^\dagger(\pB,\sB) \ParityOp^\dagger \bigr)
  \bigl( \ParityOp \ket{0} \bigr)
  \\
  &=
  \bigl( +\eta_P \eta_P^0 a^\dagger(\pT,s) \bigr) 
  \bigl( -\eta_P^* {\eta_P^0}^* b^\dagger(\pBT,\sB) \bigr) \ket{0}
  \\
  &=
  -\ket{(\pT,s,\pBT,\sB)}
  \sepcomma\qquad\text{etc.}
\end{align*}
Due to the complex conjugation for $\TimeROp$, its not a good idea to 
consider only the final or initial current but the whole amplitude.  
%----------------------------------------------------------------------
\subsection{$P$, $C$ and $T$ for amplitudes or currents}
\label{sec:PCT=amp}
Although the parity transformation $\ParityOp$ etc.\ in the QFT are 
defined for creation-annihilation operators, the similar manipulations 
can be applied directly to c-numbers: 
Combined with the definitions~\EqRef{eq:PCT=g_matrix} 
[or~\EqRef{eq:PCT=g_matrix-mom}]
for the transformations of $\gamma$-matrices, 
we can analyze $\ParityOp$ properties etc.\ of matrix elements $\calM$ 
solely in terms of c-numbers.  
%----------------------------------------------------------------------
\subsubsection{Parity and Charge Conjugation}
\label{sec:PC-amp}
We consider a subprocess $Z^*(q) \to t(p,s)\,\tB(\pB,\sB)$ 
for definiteness.  Denoted in the parentheses are their momenta and 
their spins.  
The relevant vertex $\Gamma^\mu$ can be decomposed 
into pieces, so that each of which is an ``eigenstate'' of parity operation: 
\begin{align*}
  \Gamma^\mu = \Gamma_A^\mu + \Gamma_B^\mu + \cdots
\end{align*}
where $\Gamma_A^\mu = A(q^2) \gamma^\mu$, for example.  
Here $A(q^2)$ is a form factor.  
Hereafter, we consider a part of the $t\tB$ current
$j_{t\tB}^\mu = j_A^\mu + j_B^\mu + \cdots$:  
\begin{align*}
  j_A^\mu(\pBI,\sBI,\pBBI,\sBBI)
  &= \uB(p,s) \Gamma_A^\mu v(\pB,\sB)
  \\
  &= \bigl( \uB(p,s) \gamma^0 \bigr)
     \bigl( \gamma^0 \Gamma_A^\mu \gamma^0 \bigr)
     \bigl( \gamma^0 v(\pB,\sB) \bigr)
  \\
  &= - \Parity(\Gamma_A) \cdot j_A^\mu(-\pBI,\sBI,-\pBBI,\sBBI)
  \sepcomma
\end{align*}
where the minus sign can be understood by the fact that the product of 
the intrinsic parity of a particle and that of its anti-particle is $-1$.  

From the result above, or 
\begin{align*}
  j_A^\mu(\pBI,\pBBI) {j_B^\nu}^*(\pBI,\pBBI)
  = \Parity(\Gamma_A) \Parity(\Gamma_B)^* \cdot 
    j_A^\mu(-\pBI,-\pBBI) {j_B^\nu}^*(-\pBI,-\pBBI)
  \sepcomma
\end{align*}
we know that the terms in $|\calM|^2$ that are odd (even) under 
$(\ptBI,\ptBBI) \to (-\ptBI,-\ptBBI)$ should come from 
the interference of the $t\tB$ vertices with the opposite (same) parity, 
$\Parity(\Gamma_A) \Parity(\Gamma_B)^* = -1$ ($+1$), 
and vice versa%
\footnote{
  We may sometimes write $\Parity(\Gamma_A)$ as $\Parity(j)_{t\tB}$, and 
  $\Parity(\Gamma_A) \,\Parity(\Gamma_B)^*$ as $\Parity(jj^\dagger)_{t\tB}$.  
}.  

We can also see that 
\begin{align}
  \int\!\!\diffn{\Omega}{} j_A^\mu {j_B^\nu}^* = 0
  \qquad\text{for }
  \Parity(\Gamma_A) \,\Parity(\Gamma_B)^* = -1
  \sepperiod
  \label{eq:oppositeP=no-total}
\end{align}
This can be understood as follows.  
For a fermion-antifermion system, its parity is $\Parity = (-1)^{L+1}$.  
Then different parity eigenvalue means different orbital angular 
momentum $L$.  
And the states with different $L$ are orthogonal each other: 
\begin{align*}
  \int\!\!\frac{\diffn{\Omega}{}}{4\pi} 
  P_L(\cos\theta)\,P_{L'}(\cos\theta) = \frac{\delta_{LL'}}{2L+1}
  \sepcomma
\end{align*}
which means partial waves of different $L$ are orthogonal each other.  
Note that an eigenstate of partial wave is a spherical wave, 
not a plane wave that is a momentum eigenstate $\ket{\pBI}$.  
In other words, $\ket{\pBI}$ is a superposition of the states of 
definite $L$.  Thus before phase-space integration, they do not 
form eigenstates of definite $L$, and then need not be diagonal 
with respect to $L$.  

We can further operate Charge Conjugation: 
\begin{align*}
  j_A^\mu(\pBI,\sBI,\pBBI,\sBBI)
  &= - \Parity(\Gamma_A) \cdot 
       \bigl( v(\pBT,\sB)^\sfT C^\dagger \bigr)
       \bigl( C (\Gamma_A^\mu)^\sfT C^\dagger \bigr)
       \bigl( C \uB(\pT,s)^\sfT \bigr)
  \\
  &= + \Parity(\Gamma_A)\ChargeC(\Gamma_A) \cdot
     j_A^\mu(-\pBBI,\sBBI,-\pBI,\sBI)
  \\
  &= \ChargeC\Parity(\Gamma_A) \cdot
     j_A^\mu(\pBI,\sBBI,\pBBI,\sBI)
  \sepcomma
\end{align*}
or
\begin{align}
  j_A^\mu(\sBI,\sBBI) {j_B^\nu}^*(\sBI,\sBBI)
  =
  \ChargeC\Parity(\Gamma_A) \ChargeC\Parity(\Gamma_B)^* \cdot
  j_A^\mu(\sBBI,\sBI) {j_B^\nu}^*(\sBBI,\sBI)
  \label{eq:jj*_CP}
  \sepperiod
\end{align}
Thus the terms in $|\calM|^2$ that are odd (even) under 
$(\stBI,\stBBI) \to (\stBBI,\stBI)$ should come from 
the interference of the $t\tB$ vertices with the opposite (same) \CP, 
$\ChargeC\Parity(\Gamma_A) \ChargeC\Parity(\Gamma_B)^* = -1$ ($+1$), 
and vice versa%
\footnote{
  We may sometimes write $\ChargeC\Parity(\Gamma_A)$ as 
  $\ChargeC\Parity(j)_{t\tB}$, and 
  $\ChargeC\Parity(\Gamma_A) \,\ChargeC\Parity(\Gamma_B)^*$ as 
  $\ChargeC\Parity(jj^\dagger)_{t\tB}$.  
}.  
We can also see that 
\begin{align}
  \sum_{s,\sB} j_A^\mu {j_B^\nu}^* = 0
  \qquad\text{for }
  \ChargeC\Parity(\Gamma_A) \,\ChargeC\Parity(\Gamma_B)^* = -1
  \sepperiod
  \label{eq:oppositeCP=no-total}
\end{align}
This can be understood as follows.  
For a fermion-antifermion pair, $\ChargeC\Parity = (-1)^{S+1}$, 
which is a compact expression for the fact that 
spin-0 (spin-1) state is antisymmetric (symmetric) under the interchange 
of $s$ and $\sB$.  Explicit calculation may be useful: 
\begin{align*}
  \sum_{s,\sB}
  (\chi^{s\dagger} \chi^{-\sB})^\dagger \,
  (\chi^{s\dagger} \sigma^i \chi^{-\sB})
  =&
  \sum_{s,\sB}
  \btr{
    (\chi^{-\sB}\chi^{-\sB^\dagger})
    (\chi^{s}\chi^{s^\dagger}) \sigma^i
  }
  \\
  =&
  \btr{\sigma^i}
  = 0
  \sepcomma
\end{align*}
where we used the completeness relation for spin states: 
$\sum_{s}\chi^{s}\chi^{s^\dagger} = 1$.  
They need not be zero before spin-sum over both $s$ and $\sB$.  
With the definition $\chi^{-\sB} = -i\sigma^2(\chi^{\sB})^*$ and 
the explicit calculations, 
one can easily see that 
$\chi^{s\dagger} \chi^{-\sB}$ is a spin-0 state, 
and $\chi^{s\dagger} \sigma^i \chi^{-\sB}$ is a spin-1 state.  

It is easy to see that these analysis can equally be applied to 
initial current, say, $e^+ e^- \to Z^*$.  
We may use a term ``$t\tB$-current'' Parity transformation 
$\rmP_{t\tB}$ 
for $(\ptBI,\ptBBI) \to (-\ptBI,-\ptBBI)$, 
``$e^+e^-$-current'' for $(\peBI,\peBBI) \to (-\peBI,-\peBBI)$, 
and ``over-all'' for to flip all the momenta; 
and likewise for CP and C.  

The results of this subsection can be summarized as follows.  
First we express $\calM_i^*\,\calM_j$, which is a part of $|\calM|^2$, 
in terms of momenta $\pBI$ and spins $\sBI$.  
Then let us define Parity operation in terms of $\pBI$ and $\sBI$ as 
summarized in \TableRef{table:PCT-mom_spin}.  
Now the results in this subsection reads 
\begin{align*}
  &  \calM_{\Parity=\pm}^{~*}\,\calM_{\Parity=\pm}
  && \cdots  && \text{P-even}   \sepcomma\\
  &  \calM_{\Parity=\pm}^{~*}\,\calM_{\Parity=\mp}
  && \cdots  && \text{P-odd}    \sepcomma\\
  &  \calM_{\ChargeC\Parity=\pm}^{~*}\,\calM_{\ChargeC\Parity=\pm}
  && \cdots  && \text{CP-even}  \sepcomma\\
  &  \calM_{\ChargeC\Parity=\pm}^{~*}\,\calM_{\ChargeC\Parity=\mp}
  && \cdots  && \text{CP-odd}   \sepperiod
\end{align*}
These results are very plausible.  One might think that the same 
situation holds also for Time-Reversal.  However in the next subsection, 
we shall see that this is not the case.  
\begin{table}[tbp]
\centering
\begin{align*}
\begin{array}{c|cccc}
           &  \pBI  &  \pBBI &  \sBI  &  \sBBI 
  \\
  \hline
  \rmP     & -\pBI  & -\pBBI &  \sBI  &  \sBBI 
  \\
  \rmC     &  \pBBI \,(-\pBI) &  \pBI \,(-\pBBI)  &  \sBBI &  \sBI 
  \\
  \rmTT    & -\pBI  & -\pBBI & -\sBI  & -\sBBI 
  \\
  \rmC\rmP & -\pBBI \,(\pBI) & -\pBI \,(\pBBI)  &  \sBBI &  \sBI 
  \\
  \rmCPTT  &  \pBBI \,(-\pBI) &  \pBI \,(-\pBBI)  & -\sBBI & -\sBI 
\end{array}
\end{align*}
\begin{Caption}\caption{\small
    Transformation properties of momentum and spin.  
    Denoted in the parentheses are for when $\pBI+\pBBI=0$.  
\label{table:PCT-mom_spin}
}\end{Caption}
\end{table}
%
%----------------------------------------------------------------------
\subsubsection{Time reversal and \protect\rmTT}
As we saw, an initial current and a final current can separately be 
rewritten using either Parity and/or Charge Conjugation operations.  
This is because Charge Conjugation operation invokes transpose in the 
spinor space, while each current is $1\times 1$ in that space, 
which is invariant under transpose.  
As for Time Reversal, it is convenient to consider whole $\calM$, 
because each current is not real.  
Of course $\calM$ itself is not real, in general.  
We mention to this point later.  

For definiteness, we consider the process 
$e^-(p_e,r)\,e^+(\pB_e,\rB) \to Z^*(q) \to t(p_t,s)\,\tB(\pB_t,\sB)$ 
at the tree level, and decompose the final vertex $\Gamma^\mu$ and 
the initial vertex $\Lambda_\mu$ 
into pieces that are ``eigenstates'' of Time Reversal.  
Let $A(q^2)$ and $B(q^2)$ be the form factors for the final vertex, 
and $C(q^2)$ and $D(q^2)$ for the initial vertex: 
\begin{align*}
  \Gamma_A^\mu \equiv A \cdot \GammaH_A^\mu
  \sepcomma\quad
  {\Lambda_C}_\mu \equiv C \cdot \LambdaH_{C\mu}
  \sepcomma
\end{align*}
and 
\begin{align}
  \calM_{AC}
  &\equiv
  AC \cdot \calMH_{AC}
  \label{eq:forget_i_epsilon}
  \\
  &=
  AC \cdot 
  \uB(p_t,s) \GammaH_A^\mu v(\pB,\sB) \;
  \vB(\pB_e,\rB) \LambdaH_{C\mu} u(p_e,r)
  \sepcomma
  \nonumber
\end{align}
up to factors $e^2$ etc.  
%For convenience, we choose $\GammaH_A^\mu$ such that 
%$\TimeR(\GammaH_A^\mu)$ is real.  Note 
%$\TimeR(\expo^{i\delta}\GammaH) = \expo^{-2i\delta} \TimeR(\GammaH)$.  
Then 
\begin{align*}
  \calMH_{AC}^{~*}(\pBI,\sBI)
  &=
  \bigl( \uB(p_t,s)^* B \bigr)
  \bigl( B^\dagger \GammaH_A^{\mu*} B \bigr)
  \bigl( B^\dagger v(\pB,\sB)^* \bigr)
  \\
  &\qquad\times
  \bigl( \vB(\pB_e,\rB)^* B \bigr)
  \bigl( B^\dagger \LambdaH_{C\mu}^{~*} B \bigr)
  \bigl( B^\dagger u(p_e,r) \bigr)
  \\
  &=
  \TimeR(\GammaH_A)\TimeR(\LambdaH_C) \cdot \calMH_{AC}(-\pBI,-\sBI)
  \sepcomma
\end{align*}
or 
\begin{align*}
  & \calM_{AC}^{~*}(\pBI,\sBI) \calM_{BD}(\pBI,\sBI)
  \nonumber\\
  =& \TimeR(\GammaH_A)\TimeR(\GammaH_B)^*\TimeR(\LambdaH_C)\TimeR(\LambdaH_D)^*
    \cdot \calM_{AC}(-\pBI,-\sBI) \calM_{BD}^{~*}(-\pBI,-\sBI)
  \sepperiod
\end{align*}
Here $(-\pBI,-\sBI)$ means to flip all the momenta and the spins.  
Now 
\begin{align*}
  |\calM|^2
  &=
  |\calM_{AC} + \calM_{BD} + \cdots |^2
  \\
  &=
  \cdots + \calM_{AC}\calM_{BD}^{~*} 
         + \calM_{AC}^{~*}\calM_{BD}
  + \cdots
  \\
  &=
  \cdots + 
  \pRe{AB^*CD^*}
  \bigl[   \calMH_{AC}\calMH_{BD}^{~*} 
         + \calMH_{AC}^{~*}\calMH_{BD}
  \bigr]
  \\
  &\quad{}
  +i\pIm{AB^*CD^*}
  \bigl[   \calMH_{AC}\calMH_{BD}^{~*} 
         - \calMH_{AC}^{~*}\calMH_{BD}
  \bigr]
  + \cdots
\end{align*}
means that the terms in $|\calM|^2$ that are even (odd) under 
$(\pBI,\sBI) \to (-\pBI,-\sBI)$ should be accompanied by $\pRe{AB^*CD^*}$ 
when $\TimeR(\GammaH_A)\TimeR(\GammaH_B)^*
\TimeR(\LambdaH_C)\TimeR(\LambdaH_D)^* = +1$ ($-1$), or by $\pIm{AB^*CD^*}$ 
when $\TimeR(\GammaH_A)\TimeR(\GammaH_B)^*
\TimeR(\LambdaH_C)\TimeR(\LambdaH_D)^* = -1$ ($+1$).  
%means that the term proportional to $\pRe{AB^*CD^*}$ should be 
%even under $(\pBI,\sBI) \to (-\pBI,-\sBI)$ for 
% $\TimeR(\GammaH_A)\TimeR(\GammaH_B)^*
%\TimeR(\LambdaH_C)\TimeR(\LambdaH_D)^* = +1$, 
%and be odd for $-1$.  Just the opposite holds for $\pIm{AB^*CD^*}$.  
Note that 
one can always adjust the phases of $\GammaH$'s and $\LambdaH$'s so that 
$\TimeR(\GammaH_A)\TimeR(\GammaH_B)^*
\TimeR(\LambdaH_C)\TimeR(\LambdaH_D)^* = \pm1$.  
Correspondingly, the phase of $AB^*CD^*$ changes counterwise.  

The statement above can be simplified as follows.  
First let us write $\TimeR(\GammaH_A)\TimeR(\GammaH_B)^*
\TimeR(\LambdaH_C)\TimeR(\LambdaH_D)^*$ as $\TimeR(\calM\calM^*)$.  
Since $\calM\calM^*$ is Lorentz scalar, 
$\ChargeC\Parity\TimeR(\calM\calM^*) = +1$ [\SecRef{sec:CPT-sym}], 
which means $\TimeR(\calM\calM^*) = \ChargeC\Parity(\calM\calM^*)$.  
As we saw in the previous subsection, $\ChargeC\Parity(\calM\calM^*)$ 
is equivalent to the eigenvalue under CP-transformation 
$(\sBI,\sBBI)\to(\sBBI,\sBI)$ [\TableRef{table:PCT-mom_spin}].  
Thus the result in this subsection is that 
\begin{align*}
  (\calM\calM^*)_{\scriptsize \rmCPTT=+} &\propto 
  \Re(\text{product of relevant form factors})  \sepcomma\\
  (\calM\calM^*)_{\scriptsize \rmCPTT=-} &\propto 
  \Im(\text{product of relevant form factors})  \sepcomma
\end{align*}
where \rmTT-transformation is also defined in \TableRef{table:PCT-mom_spin}.  
Note that \rmTT\ do NOT interchange initial- and final-states, while 
the genuine $\TimeROp$-transformation do.  

In fact, the argument above holds only for tree-level diagrams.  
This is because we neglected $i\epsilon$ in propagators 
at~\EqRef{eq:forget_i_epsilon}.  The sign of this imaginary part becomes 
relevant once we across a physical threshold in a loop: 
$\log(-x\pm i\epsilon) = \log(x)\pm i\pi$, for example.  
Thus, to be precise, 
the argument above holds when there is no absorptive part.  
%the argument above holds when the couplings are 
%the only source of imaginary part; 
%when there is an absorptive part, the argument above do not hold.  
As we shall see below, the sources for absorptive part are rescattering and 
physical threshold, 
which are absent to tree level%
\footnote{Decay width is a part of physical threshold effect, 
  but can be treated to the tree level.  
  Likewise, the imaginary part of form factors originate from 
  physical thresholds, 
  but those are treated by effective vertices here.  
}.  
If there is an absorptive part, $\pRe{AB^*CD^*}$ and $\Im$ above are 
replaced by $\pRe{AB^*CD^*\expo^{i\delta}}$ etc., where $\expo^{i\delta}$ 
is an effect of absorptive parts.  

Symmetry considerations for $|\calM|^2$ continue to~\SecRef{sec:PCT=|amp|2}.  
%----------------------------------------------------------------------
\subsection{Absorptive parts and CP\rmTT}
\label{sec:absorp-CPTT}
Transformation properties of $S$-matrix and $T$-matrix, $S = 1 + iT$, 
is as follows: 
\begin{align*}
    \ParityOp S \ParityOp^{-1} &= S
  & \ParityOp T \ParityOp^{-1} &= T
  & &\text{if}
  & \ParityOp H \ParityOp^{-1} &= H
  \\
    (\ChargeCOp\ParityOp) S (\ChargeCOp\ParityOp)^{-1} &= S
  & (\ChargeCOp\ParityOp) T (\ChargeCOp\ParityOp)^{-1} &= T
  & &\text{if}
  & (\ChargeCOp\ParityOp) H (\ChargeCOp\ParityOp^{-1}) &= H
  \\
    \TimeROp S \TimeROp^{-1} &= S^\dagger
  & \TimeROp T \TimeROp^{-1} &= T^\dagger
  & &\text{if}
  & \TimeROp H \TimeROp^{-1} &= H
  \\
    (\ChargeCOp\ParityOp\TimeROp) S (\ChargeCOp\ParityOp\TimeROp)^{-1} &= S^\dagger
  & (\ChargeCOp\ParityOp\TimeROp) T (\ChargeCOp\ParityOp\TimeROp)^{-1} &= T^\dagger
  & &\text{if}
  & (\ChargeCOp\ParityOp\TimeROp) H (\ChargeCOp\ParityOp\TimeROp)^{-1} & = H
\end{align*}
The last two rows can be understood by $S \simeq \expo^{-iHt}$ and 
$\TimeROp$ is anti-unitary.  

In terms of matrix element, $\ChargeCOp\ParityOp\TimeROp$ invariance 
can be written as 
\begin{align*}
  T_{fi}
  &=
  \bra{f} T \ket{i}
  \\
  &=
  \bra{f} (\ChargeCOp\ParityOp\TimeROp)^{-1}
  (\ChargeCOp\ParityOp\TimeROp) T (\ChargeCOp\ParityOp\TimeROp)^{-1}
  (\ChargeCOp\ParityOp\TimeROp) \ket{i}
  \\
  &=
  \bra{\ChargeCOp\ParityOp\TimeROp(f)} T^\dagger
  \ket{\ChargeCOp\ParityOp\TimeROp(i)}^*
  \\
  &=
  \bra{\ChargeCOp\ParityOp\TimeROp(i)} T
  \ket{\ChargeCOp\ParityOp\TimeROp(f)}
  \\
  &=
  T_{\hat{i}\hat{f}}
  \sepcomma
\end{align*}
where $\hat{i}$ ($\hat{f}$) denotes CP\rmTT -conjugate 
state of a state $i$ ($f$).  
CP\rmTT\ is defined in~\EqRef{eq:CPTTfor_ffB}.  
Thus 
\begin{align*}
  \rule{10em}{0ex}
  T_{\hat{f}\hat{i}}^*
  &=
  T_{\hat{i}\hat{f}}^\dagger
  =
  T_{fi}^\dagger
  \sepcomma\quad
  &&\text{from $\ChargeCOp\ParityOp\TimeROp$ invariance,\rule{10em}{0ex}}
  &&\\
  &=
  T_{fi}
  \sepcomma\quad
  &&\text{if there is no absorptive part,}
\end{align*}
where the absorptive part of a $T$-matrix is defined by 
its anti-hermitian part%
\footnote{
  Hermitian may be called ``dispersive part''.  
}
$T-T^\dagger$ (or times $-i$ or something).  
The relation above $T_{\hat{f}\hat{i}}^* = T_{fi}$, or 
$|T_{\hat{f}\hat{i}}|^2 = |T_{fi}|^2$, for the case of no absorptive part 
has the following consequences: 
an expectation value of a CP\rmTT-odd observable is zero, unless 
there is an absorptive part, while that of CP\rmTT-even observable can be 
non-zero without an absorptive part.  
This is the basis of the classification of observables based on 
CP\rmTT\ transformation~\cite{HPZH87,CKP93}.  
In fact, as we saw in the previous subsection, \rmCPTT-even (-odd) term 
$\calM\calM^*$ is proportional to the real- (imaginary-) part of 
the relevant product of couplings.  

The dagger $\rule{0em}{1ex}^\dagger$ on $T$ (and $S$) for $\TimeROp$ 
complicates symmetry arguments.  
However $T$ is hermitian to tree level; 
that is, no absorptive part.  
This can be shown as follows.  
Unitarity $S^\dagger S = 1$ in terms of $T$-matrix is 
$-i(T-T^\dagger) = T^\dagger T$, or 
\begin{align}
  -i (T_{fi} - T_{fi}^\dagger)
  = \sum_n T_{fn}^\dagger T_{ni}
  \label{eq:unitarity_T}
  \sepperiod
\end{align}
This is a celebrated formula, which relates the absorptive part 
of an amplitude and the ``cut diagrams'' of the amplitude.  
If an intermediate state $n$ is a two- (or more) particle state, 
the RHS is loop effect.  If $n$ is a one-particle state, it hits 
the pole of the propagator for $n$: 
\begin{align*}
  \left. \frac{i}{p^2 - m_n^2 + i m_n\Gamma_n} \right|_{p^2 = m_n^2}
  = \frac{1}{m_n \Gamma_n}
  \sepcomma
\end{align*}
while $\Gamma_n \simeq \bIm{\text{self-energy of $n$}}$ is again loop effect.  
Thus anyway, absorptive part $(T - T^\dagger)$ comes from loop effect.  

A source of absorptive part can be traced back to the boundary condition for 
Green functions: a positive energy particle propagates forward in time, 
while a negative, backward.  This boundary condition is embodied by 
the Feynman prescription for the position of the pole of a propagator: 
\begin{align*}
  \frac{i}{p^2 - m^2 + i \epsilon}
  \sepperiod
\end{align*}
If one choose principal-value prescription 
\begin{align*}
  \mathrm{P}\frac{1}{x} 
  \equiv \frac{1}{2} 
    \left( \frac{1}{x+i\epsilon} + \frac{1}{x-i\epsilon} \right)
  \nonumber
\end{align*}
instead, there would be no absorptive part.  
We can see this by analyticity of an amplitude $\calM(s)$ 
as a function of \scCM\ energy $\sqrt{s}$.  
Below the threshold, there is no absorptive part, and thus 
$\calM(s)$ is real: 
\begin{align*}
  \calM(s) = [\calM(s^*)]^*
  \sepperiod
\end{align*}
Analyticity requires this to hold also above thresholds: 
\begin{align*}
  \bIm{\calM(s+i\epsilon)} = - \bIm{\calM(s-i\epsilon)}
  \sepperiod
\end{align*}
Thus, no absorptive part for principal-value prescription.  
We know by experience that the sign of $i\epsilon$ in propagators 
are irrelevant to tree-level.  This is in accord with the 
observation above that no absorptive part to tree-level.  

A word may be in order for ``tree-level''.  
Higher order effects can be treated by effective interactions, 
and so are absorptive parts.  They can effectively introduced as 
anti-hermitian parts of Lagrangian.  For example, a decay width $\Gamma$ 
can be introduced to ``tree'' level: 
\begin{align}
  \calL = \cdots - \left( m - i\frac{\Gamma}{2} \right) \psiB\psi + \cdots
  \sepcomma
\end{align}
which is consistent with 
$s - m^2 + im\Gamma \simeq s-(m-i\Gamma/2)^2$.  
Thus in the statement ``no absorptive part to tree-level'', 
anti-hermitian parts of Lagrangian are counted as higher order.  
It's interesting that 
it seems QFT with the Feynman prescription specifies 
the direction of time-flow: 
\begin{align*}
  |\psi|^2  \sim  |\expo^{-iEt}|^2  \sim  |\expo^{-i(m-i\Gamma/2)}|^2 
  \sim \expo^{-\Gamma t}
  \sepperiod
\end{align*}
That is, simple time reversal $t \to -t$ may not be a symmetry, 
if absorptive part is finite.  
This may be a ``conjugate statement'' to that \rmCPTT\ and 
$\ChargeCOp\ParityOp\TimeROp$ differs when absorptive part is finite.  
%----------------------------------------------------------------------
\subsection{$\ChargeCOp\ParityOp$ violation}
\label{sec:CPviol}
Roughly speaking, $\ChargeCOp\ParityOp$ transformation changes an 
operator to its hermitian conjugate, 
while the coupling intact [\EqRef{eq:CP=herm-ex}].  
$\TimeROp$ changes the coupling to its complex conjugate.  
Thus $\ChargeCOp\ParityOp\TimeROp$ changes a term in Lagrangian to its 
hermitian conjugate.

There two kinds of complex phase in $\calM$: one is weak phase $\epsilon$ 
and the other is strong phase $\delta$.  Non-zero $\sin\delta$ means 
non-zero absorptive part.  In this sense, $\delta$ can also be called 
``absorptive phase''.  On the other hand, $\epsilon$ is related to the 
hermitian contribution; thus it might also be called ``dispersive phase''; 
it is also sometimes called ``\itCP\ phase''.  
Both of these phases are important for CP-violating phenomena.  
For example, let us consider the amplitude $A_f$ for a certain process with 
a final state $f$, and its \itCP-conjugated amplitude $\bar{A}_{\bar{f}}$: 
\begin{align*}
  A_f &= A_1 \expo^{i\delta_1} + A_2 \expo^{i\delta_2}  \\
      &=  |A_1| \expo^{i\epsilon_1} \expo^{i\delta_1} 
        + |A_2| \expo^{i\epsilon_2} \expo^{i\delta_2}  \sepcomma\\[1.5ex]
  \bar{A}_{\bar{f}} &= A_1^* \expo^{i\delta_1} + A_2^* \expo^{i\delta_2}  \\
      &=  |A_1| \expo^{-i\epsilon_1} \expo^{i\delta_1} 
        + |A_2| \expo^{-i\epsilon_2} \expo^{i\delta_2}  \sepperiod
\end{align*}
Here $\bar{A}_{\bar{f}}$ is determined up to overall phase, which is 
irrelevant to $|\calM|^2$.  Now one can calculate the following 
\itCP-odd quantity: 
\begin{align*}
  \frac{|A_f|^2 - |\bar{A}_{\bar{f}}|^2}{|A_f|^2 + |\bar{A}_{\bar{f}}|^2}
  &=
  \frac{2\Im(A_1^*A_2) \sin(\delta_1-\delta_2)}
    {|A_1|^2+|A_2|^2+2\Re(A_1^*A_2)\cos(\delta_1-\delta_2)}  \\
  &=
  \frac{-2|A_1|\,|A_2| \sin(\epsilon_1-\epsilon_2) \sin(\delta_1-\delta_2)}
    {|A_1|^2+|A_2|^2+2|A_1|\,|A_2|
      \cos(\epsilon_1-\epsilon_2)\cos(\delta_1-\delta_2)}  \sepperiod
\end{align*}
One can clearly see that one needs both of two independent \itCP\ phases and 
two independent absorptive phases.  

The fact above that one needs an absorptive part, 
is well known from $K$ physics.  
However before sum-over final state momenta and/or spins, one 
do not need an absorptive part for $\ChargeCOp\ParityOp$ violating 
observables.  

In general, one can not say that a certain interaction $\calL_{\rm int}$ 
violates $\ChargeCOp\ParityOp$.  This is because of the degrees of freedom 
to redefine $\ChargeCOp\ParityOp$ by flavor-dependent complex phases.  
For flavor-off-diagonal interaction, vector-like phase transformation 
changes \itCP-property; while for flavor-diagonal interaction, chiral phase 
transformation does.  
%For flavor-off-diagonal, or, non-hermitian interactions, 
%$\ChargeCOp\ParityOp$ operation is a little bit complicated, because 
%it's ambiguous that such a term is $\ChargeCOp\ParityOp$-even or odd.  
A simple example is a Dirac mass term written 
in terms of two chiral fermions: 
\begin{align*}
  -\calL_m = m\,\psiB\psi 
  = m\,\overline{\psi_R}\psi_L + m\,\overline{\psi_L}\psi_R
  \sepperiod
\end{align*}
Here non-zero imaginary part of $m$ means (effective) absorptive part, not 
$\ChargeCOp\ParityOp$-violation.  
In terms of Dirac fermion, this is flavor-diagonal; while in terms of 
chiral fermion, this is flavor-off-diagonal.  
One can easily see that this term is $\ChargeCOp\ParityOp$-even.  
Now we make chiral rotation, which is a symmetry of the Lagrangian 
if it's anomaly-free: 
\begin{align*}
  \psi \to \expo^{i(\alpha/2)\gamma_5}\psi
  \sepcomma\text{ or}\quad
  \left\{
    \begin{array}{l}
      \psi_L \to \expo^{-i\alpha/2}\psi_L  \\
      \psi_R \to \expo^{+i\alpha/2}\psi_R
    \end{array}
  \right.
  \sepperiod
\end{align*}
Accordingly the mass term above changes its form: 
\begin{align*}
  -\calL_m 
  \to m\,\psiB \expo^{i\alpha\gamma_5}\psi
  &= m\cos\alpha\,\psiB\psi + m\sin\alpha\,\psiB\,i\gamma_5\psi
  \\
  &=   m\expo^{-i\alpha}\overline{\psi_R}\psi_L
     + m\expo^{+i\alpha}\overline{\psi_L}\psi_R
  \sepperiod
\end{align*}
Now the term $\psiB\,i\gamma_5\psi$ is $\ChargeCOp\ParityOp$-odd, 
while $\psiB\psi$ is even.  
Thus the mass term above violates $\ChargeCOp\ParityOp$ symmetry, 
although the original one do not.  
In terms of chiral fermions, $\ChargeCOp\ParityOp$ interchanges 
$\overline{\psi_R}\psi_L$ and $\overline{\psi_L}\psi_R$ with 
the transformation [\EqRef{eq:PCT-Dirac}] 
\begin{align*}
  (\ChargeCOp\ParityOp)\psi_{L/R}(\ChargeCOp\ParityOp)^{-1}
  = \eta_{CP}\,\gamma^0(\psi_{L/R})^c
  \sepcomma
\end{align*}
with%
\footnote{Here we dropped field-independent phases $\eta_P^0\eta_C^0$
which do not contribute to any fermion bilinears.}
$\eta_{CP}\equiv\eta_P\eta_C = 1$.  
With $\psiB_1\Gamma\psi_2 = \overline{\psi_2^c}\,\Gamma^c \psi_1^c$ 
and $(\psiB_1\Gamma\psi_2)^\dagger = \psiB_2\GammaB\psi_1$, 
one can see that the non-zero complex phase $\alpha$ seems to mean 
$\ChargeCOp\ParityOp$ violation.  
In fact, as can be expected from the context, 
this can be cured by taken into account of the 
degrees of freedom to choose the field-dependent phases $\eta_{CP}$ 
in the definition of $\ChargeCOp\ParityOp$ transformation: 
\begin{align*}
  (\ChargeCOp\ParityOp) \psi_L (\ChargeCOp\ParityOp)^{-1}
  = \expo^{+i\alpha} \gamma^0(\psi_L)^c
  \sepcomma\quad
  (\ChargeCOp\ParityOp) \psi_R (\ChargeCOp\ParityOp)^{-1}
  = \expo^{-i\alpha} \gamma^0(\psi_R)^c
  \sepperiod
\end{align*}
One can see that $\calL_m$ can indeed be 
a $\ChargeCOp\ParityOp$ eigenstate.  
Usually one redefines the field to include the phase $(\eta_{CP})^{-1/2}$.  
For the case above, 
\begin{align*}
  (\ChargeCOp\ParityOp) \expo^{-i\alpha/2}\psi_L (\ChargeCOp\ParityOp)^{-1}
  = \gamma^0(\expo^{-i\alpha/2}\psi_L)^c
  \sepcomma\quad
  (\ChargeCOp\ParityOp) \expo^{+i\alpha/2}\psi_R (\ChargeCOp\ParityOp)^{-1}
  = \gamma^0(\expo^{+i\alpha/2}\psi_R)^c
  \sepperiod
\end{align*}
Note that $\psi^c$ contains $\psi^\dagger$.  
The redefinition 
$(\expo^{-i\alpha/2}\psi_L,\expo^{+i\alpha/2}\psi_R) \to (\psi_L,\psi_R)$
takes the Lagrangian back to the original form, 
where the $\ChargeCOp\ParityOp$ property is transparent.  
To summarize, non-zero imaginary part in the coupling for 
a flavor-off-diagonal interaction may or may not be a source of 
$\ChargeCOp\ParityOp$ violation; it depends on how many 
explicitly broken U(1) symmetries are there%
\footnote{
  Unbroken U(1) symmetries, such as fermion-number or electric charge, 
  can not be used to redefine the phase of couplings, 
  since those symmetry do not alter the form of Lagrangians.  
  On the other hand, 
  axial U(1) symmetry is explicitly broken by a fermion mass, 
  thus can be used to absorb the phase of the fermion mass.  
}.  

While for flavor-diagonal such as 
\begin{align*}
  \psiB\,i\Slash{\partial}\psi
  =   \overline{\psi_L}\,i\Slash{\partial}\psi_L
    + \overline{\psi_R}\,i\Slash{\partial}\psi_R
  \sepcomma
\end{align*}
their $\ChargeCOp\ParityOp$ property is definite, since 
the flavor-dependent phases $\eta_{CP}$ cancel in each term.  
One can unambiguously say that a certain flavor-diagonal interaction is 
whether $\ChargeCOp\ParityOp$-even or odd.  

Some more examples may be useful.  
Let us consider the following $S-P$ interaction: 
\begin{align*}
  (\psiB\psi - \psiB\,i\gamma_5\psi)\phi  \sepperiod
\end{align*}
From \TableRef{table:PCT-bilinear}, one can see that $\psiB\psi$ is 
$\ChargeCOp\ParityOp$-even, while $\psiB\,i\gamma_5\psi$ is odd.  
Thus one can not define the $\ChargeCOp\ParityOp$ property of $\phi$ 
so as to preserve $\ChargeCOp\ParityOp$.  The situation is similar 
for the $V-A$ interaction, which violates both $\ChargeCOp$ and $\ParityOp$ 
maximally.  However there is a difference: $S-P$ interaction is off-diagonal 
in terms of chiral fermion, while $V-A$ is diagonal (and 
conserves $\ChargeCOp\ParityOp$).  Thus in principle, 
there is a possibility to ``rotate'' $\psi$ so that the interaction above 
preserves $\ChargeCOp\ParityOp$.  However such a degree of freedom may 
already be used to rewrite the mass term, as we did just above.  
Thus if $\phi$ is the Higgs that gives mass to $\psi$, the above $S-P$ 
interaction may be ``rotated'' to the $\ChargeCOp\ParityOp$-even interaction 
$\psiB\psi\phi$ at the same time.  On the other hand, if $\phi$ is not 
the Higgs that gives mass to $\psi$, the interaction of $\phi$ may 
violates $\ChargeCOp\ParityOp$, in general.  

Likewise, MDM and EDM interaction are also flavor-off-diagonal in terms of 
chiral fermion: 
\begin{align*}
  \tB \left( \sigma^{\mu\nu} F_2(q^2) 
    + \,i\sigma^{\mu\nu}\gamma_5 \, d(q^2) \right) t\,
  \partial_\mu A_\nu
  \sepperiod
\end{align*}
Since chiral rotation of $t$ is already fixed by the mass term of $t$, 
non-zero $d$ means $\ChargeCOp\ParityOp$ violation.  
Non-zero imaginary part of $d$ means (effective) absorptive part, 
in this case.  

Hermitian conjugation of a flavor-off-diagonal interaction is 
closely related to the $\ChargeCOp\ParityOp$ operation.  For example, 
\begin{align}
  \calL 
  &=
  \frac{-g}{\sqrt{2}} V_{tb} \cdot 
  \tB\gamma^\mu\frac{1-\gamma_5}{2}b \cdot W^+_\mu 
  + \text{h.c.} + \cdots
  \sepcomma\nonumber\\
  (\ChargeCOp\ParityOp) \calL (\ChargeCOp\ParityOp)^\dagger
  &=
  \frac{-g}{\sqrt{2}} V_{tb} \cdot 
  (-1)\bB\gammaT^\mu\frac{1-\gamma_5}{2}t \cdot
  (-1)\WT^{+*}_\mu + \text{h.c.} + \cdots 
  \label{eq:CP=herm-ex}
  \\
  &={}
  + \frac{-g}{\sqrt{2}} V_{tb} \cdot \bB\gamma^\mu\frac{1-\gamma_5}{2}t 
  \cdot W^-_\mu + \text{h.c.} + \cdots \nonumber\\
  &={}
  \left[ 
    \left( \frac{-g}{\sqrt{2}} V_{tb} \right)^*
    \tB\gamma^\mu\frac{1-\gamma_5}{2}b \cdot W^+_\mu
  \right]^\dagger + \text{h.c.} + \cdots
  \sepcomma\nonumber
\end{align}
where tilde $\tilde{}$ means to flip space components 
[\EqRef{eq:tilde-parity}].  
Since the gauge coupling $g$ is real, the term above is invariant 
under $\ChargeCOp\ParityOp$ with the phase convention that $V_{tb}$ is real.  
$\ChargeCOp\ParityOp$ transformation of gauge bosons are 
summarized in~\TableRef{table:PCT-EMfield}.  
%----------------------------------------------------------------------
\subsubsection{Effect of rephasing}
We saw that $\ChargeCOp\ParityOp$ transformation of $\calL_{\rm int}$ 
looks different for the different choice of the phase of fields.  
Let us consider how $\calM$ changes under the phase transformation of 
the fields.  The key observation is that $\calL_{\rm int}$ remains intact 
under field-redefinitions, since couplings are also redefined to absorb 
the difference.  Since 
$\calM \simeq \braket{f}{\bigl(\exp(i\int\calL_{\rm int})-1\bigr)}{i}$, 
rephasing affects only $\bra{f}$ and 
$\ket{i} = a^\dagger \cdots a^\dagger\ket{0}$.  This means each term in 
$\calM$ changes with the same phase.  Thus each term in $\calM\calM^*$ is 
rephasing invariant; thus it is sensible to consider the real or imaginary 
part of the product of the couplings in \rmCPTT-argument.  
%----------------------------------------------------------------------
%
%
%

%
\clearemptydoublepage
%
%
%
%------------------------------------------------------------------------
\chapter{}
\section{Top quark decay width $\Gamma_t$}
\label{sec:Gamma_t}
Decay width of top quark is calculated in~\cite{BDKKZ86,JK89}: 
\begin{align*}
  \Gamma(t\to bW^+) = 
  \Gamma_{\rm Born} \cdot \left( 1-\frac{2}{3}\frac{\alpha_s}{\pi}f \right)
  \sepcomma
\end{align*}
where 
\begin{align*}
  \Gamma_{\rm Born} 
  &= |V_{tb}|^2 \frac{G_F\mt^3}{8\pi\sqrt{2}} \frac{2p_W}{\mt}
      \left\{ \left[ 1-\pfrac{m_b}{m_t}^2 \right]^2 
        + \left[ 1+\pfrac{m_b}{m_t}^2 \right]^2 \pfrac{m_W}{m_t}^2
        - 2\pfrac{m_W}{m_t}^4 \right\}  \\
  &\simeq
  \frac{G_F\mt^3}{8\pi\sqrt{2}} \left( 1-\frac{m_W^2}{m_t^2} \right)^2
  \left( 1+2\frac{m_W^2}{m_t^2} \right)  
  \sepcomma\quad\mbox{ where }\\[1ex]
  p_W &= \text{$W$ momentum in the $t$ rest frame}  \\
  &= \sqrt{m_t^2-(m_W+m_b)^2} \sqrt{m_t^2-(m_W-m_b)^2} \;/\;(2m_t)
\end{align*}
and 
\begin{align*}
  f &= \left[ \pi^2 + 2\Li_2(y)-2\Li_2(1-y) \right]
  + \bigl[ 4y(1-y-2y^2)\ln y + 2(1-y)^2(5+4y)\ln(1-y) 
  \\&\qquad{}
  - (1-y)(5+9y-6y^2) \bigr] \, / \, 2(1-y)^2(1+2y)  \sepcomma
\end{align*}
where $y=m_W^2/m_t^2$.  
%
%
%

%
%
%
%----------------------------------------------------------------------
\section{Coulomb plus $1/r^2$ potential: explicit calculation}
\label{sec:coulomb+1/r2_app}
%----------------------------------------------------------------------
\subsubsection{Confluent hypergeometric function $F(\alpha,\gamma;z)$}
Confluent hypergeometric function $F(\alpha,\gamma;z)$, 
or Kummer's function, is defined by 
\begin{align*}
  F(\alpha,\gamma;z) 
  = \sum_{k=0}^{\infty} \frac{(\alpha)_k}{k!(\gamma)_k} z^k
  = 1 + \frac{\alpha}{\gamma} z 
      + \frac{\alpha(\alpha+1)}{2!\gamma(\gamma+1)} z^2 + \cdots 
  \sepcomma
\end{align*}
where $(\alpha)_k \equiv \alpha(\alpha+1)\cdots(\alpha+k-1)$, 
$(\alpha)_0 \equiv 1$.  
This is one of the solutions of 
\begin{align*}
  \odiffn{w}{z}{2} + \left(-1+\frac{\gamma}{z}\right)
  \odiff{w}{z} - \frac{\alpha}{z}w = 0
  \sepperiod
\end{align*}
Its asymptotic behavior is 
\begin{align}
  & F(\alpha,\gamma;\zeta) \to 1
  \sepcomma&&\mbox{ for }\zeta\to 0
  \sepcomma\nonumber\\
  & F(\alpha,\gamma;\zeta) \sim 
    \frac{\Gamma(\gamma)}{\Gamma(\gamma-\alpha)}(-\zeta)^{-\alpha}
  + \frac{\Gamma(\gamma)}{\Gamma(\alpha)}\expo^{\zeta} \zeta^{\alpha-\gamma}
  \sepcomma&&\mbox{ for }|\zeta| \to \infty
  \label{eq:F(a,g,z)-asym}
\end{align}
Note also that 
$F(\alpha,\gamma;\zeta) = \expo^{-\zeta} 
F(\gamma-\alpha,\gamma;-\zeta)$.  

Whittaker's differential equation is 
\begin{align*}
  \odiffn{W}{\zeta}{2} + \left( -\frac{1}{4} + \frac{k}{\zeta}
    + \frac{\mu^2-1/4}{\zeta^2} \right) W = 0
  \sepperiod
\end{align*}
One of the solution can be written by $F$: 
\begin{align*}
  W = \zeta^{\mu+1/2} \expo^{-\zeta/2} 
      F(\mu-k+\tfrac{1}{2},2\mu+1;\zeta)
  \sepperiod
\end{align*}
%
%----------------------------------------------------------------------
\subsubsection{Coulomb plus $1/r^2$ potential: explicit calculation}
This section is continued from \SecRef{sec:coulomb+1/r2}.  

Let us consider the case $E<0$ for the moment.  
The case when $E>0$ can be obtained from analytic continuation.  
The parameter $\nu$ is pure imaginary for $E<0$, thus we introduce 
another variable $\zeta$: 
\begin{align*}
  \zeta = \frac{iz}{\nu}
\end{align*}
By multiplying \EqRef{eq:diff_eq-for-g(z)} by $(-\nu^2)$, one obtains 
\begin{align}
  \left( \odiffn{}{\zeta}{2} - \frac{1}{4} - \frac{i\nu}{\zeta}
    + \frac{\kappa}{\zeta^2} \right) g(\zeta) = 0
  \label{eq:schr_for_g(zeta)}
\end{align}
By comparing this and Whittaker's differential equation, 
one can see the correspondence $k=-i\nu$, $\mu^2-1/4 = -\kappa$, 
or $\mu = \pm\sqrt{1/4-\kappa} = d_\pm-\frac{1}{2}$.  
Thus the solution%
\footnote{
  For Coulomb ($\kappa=0$), these two degenerate, 
  and independent solutions are 
  \begin{align*}
    \zeta \expo^{-\zeta/2} F(1+i\nu, 2 ; \zeta)  \sepcomma\quad
    \zeta \expo^{-\zeta/2} \left[ F(1+i\nu, 2 ; \zeta) \ln\zeta 
      + F^*(1+i\nu, 2 ; \zeta) \right]  
  \end{align*}
  where $F^*$ is some other function.  
  Thus it may not a good idea to let $\kappa=0$ at this stage.  
  However, wave function etc.\ can be 
  obtained with this method.  
}
is 
\begin{align}
  g(\zeta) = \zeta^{d_\pm} \expo^{-\zeta/2} 
         F(d_\pm+i\nu, 2 d_\pm ; \zeta)
  \sepcomma
\end{align}
or 
\begin{align*}
  g_\pm(z) \equiv z^{d_\pm} \expo^{z/(2\rho)} 
         F(d_\pm+\rho, 2 d_\pm ; -z/\rho)  \sepcomma\quad
  \rho \equiv i\nu   \sepcomma\quad
  z \equiv -\rho\zeta
  \sepperiod
\end{align*}
Any solution of \EqRef{eq:schr_for_g(zeta)} is expressed as 
a linear combination of $g_\pm$, 
and so are $g_>(z)$ and $g_<(z)$.  
Since $F(\alpha,\gamma;\zeta)\to 1$ for $\zeta\to 0$, we have 
\begin{align}
  g_<(z) = g_+(z)  \sepperiod
\end{align}
While the limit $\zeta\to \infty$ is more complicated.  
For $E<0$, $\Re\zeta < 0$, since 
\begin{align*}
  \nu = -\frac{C_F \alpha_s}{2} \sqrt{\frac{m}{E+i\epsilon}}
    \simeq -i|\nu|+\epsilon
  \sepcomma\quad\mbox{ or }\quad
  \rho \simeq |\nu| + i\epsilon
  \sepperiod
\end{align*}
Thus one can see the the first term in \EqRef{eq:F(a,g,z)-asym} dominates: 
\begin{align*}
  g_\pm(z) \stackrel{\scriptstyle z\to\infty}{\sim}&\;
  z^{d_\pm} \expo^{z/(2\rho)} \frac{\Gamma(2d_\pm)}{\Gamma(d_\pm-\rho)}
  \left( \frac{z}{\rho} \right)^{-d_\pm-\rho}  
  =\; \rho^{d_\pm+\rho} \, z^{-\rho} \, \expo^{z/(2\rho)} 
  \frac{\Gamma(2d_\pm)}{\Gamma(d_\pm-\rho)}
  \sepperiod
\end{align*}
With this expression, we have 
\begin{align}
  g_>(z) = g_-(z)
  - \rho^{d_- -d_+} \frac{\Gamma(2d_-)\Gamma(d_+ - \rho)}
    {\Gamma(2d_+)\Gamma(d_- - \rho)} g_+(z)
  \sepperiod
\end{align}
The Wronskian $W$ can be calculated 
from $g_\pm(z) \simeq z^{d_\pm}$ for $z\to 0$: 
\begin{align}
  W = 1 - 2d_- = \sqrt{1-4\kappa}  
  \label{eq:wronskian_for_g_pm}
  \sepcomma
\end{align}
Now all ingredients for the Green function in \SecRef{sec:coulomb+1/r2} 
are obtained.  

Next we want to obtain the location of pole and its residue 
$F(\alpha,\gamma;\zeta)$ diverges when $\gamma$ is zero or negative integer.  
However since $d_\pm > 0$, $g_\pm$ do not diverge.  
Thus the divergence of the Green function comes solely from 
$\Gamma(d_+-\rho)$.  
The poles sit at $d_+ -  \rho = - (n-1)$ $(n = 1,2,3, \cdots)$, or 
\begin{align*}
  \rho^2 &= (n - d_-)^2  \sepcomma\quad\text{ while } \\
  &= - \frac{(C_F \alpha_s)^2}{4} \frac{m}{E+i\epsilon}  \sepperiod
\end{align*}
Thus
\begin{align*}
  E_n = - \frac{(C_F \alpha_s)^2 m}{4(n-d_-)^2} - i\epsilon
  \quad (n = 1,2,3, \cdots)  \sepperiod
\end{align*}
Especially with $\kappa \to 0$ or $d_- \to 0$, 
we have energy levels for Coulomb potential 
\begin{align*}
  E_n = - \frac{(C_F \alpha_s)^2 m}{4n^2} 
      = - \frac{(C_F \alpha_s m/2)^2}{m} \frac{1}{n^2}
  \quad (n = 1,2,3, \cdots)  \sepperiod
\end{align*}

One can also read off energy eigenfunctions from the residue of the pole, 
since 
\begin{align*}
  G(\rBI,\rBI') = \bra{\rBI} \frac{1}{H-(E+i\Gamma)} \ket{\rBI'} 
  = \sumint \frac{\psi_n(\rBI)\,\psi_n^*(\rBI')}{E_n-(E+i\Gamma)}
  \sepcomma
\end{align*}
where the wave function for bound states are properly normalized: 
$\braket{\psi_n}{}{\psi_m} = \delta_{nm}$.  
With $n'=n-1$ ($d_+ + n' = n-d_-$)
\begin{align*}
  \Gamma(d_+ - \rho) \sim
  \frac{1}{d_+-\rho+n'}\frac{(-1)^{n'}}{n'!}
  \simeq 
  - \frac{(-1)^{n'}}{n'!} \frac{(C_F \alpha_s)^2m}{2(n-d_-)^3}
  \frac{1}{E-E_n+i\epsilon}
  \sepcomma\quad\mbox{ near the pole, }
\end{align*}
since with $\rho(E) = \frac{C_F \alpha_s}{2}\sqrt{\frac{m}{|E|}}$, 
\begin{align*}
  \pdiff{\rho}{|E|} \simeq -\frac{2(n-d_-)^3}{(C_F \alpha_s)^2m}
  \sepcomma\quad\mbox{ near the pole.  }
\end{align*}
Thus
\begin{align*}
  g_>(z) \sim 
  \frac{(-1)^{n'}}{n'!} \frac{(C_F \alpha_s)^2m}{2}
  \frac{1}{(n-d_-)^{4-2d_-}} 
  \frac{\Gamma(2d_-)}{\Gamma(2-2d_-)\Gamma(2d_--n)}
  \frac{1}{E-E_n+i\epsilon}g_+(z)
\end{align*}
or
\begin{align*}
  g(z,z') &\sim
  \frac{1}{4\pi(1-2d_-)} \frac{(-1)^{n'}}{n'!} \frac{C_F \alpha_s m}{2}
  \frac{1}{(n-d_-)^{4-2d_-}} 
  \frac{\Gamma(2d_-)}{\Gamma(2-2d_-)\Gamma(2d_--n)}
  \times\\
  &\qquad{}\times
  \frac{1}{E-E_n+i\epsilon} g_+(z) g_+(z')
  \sepcomma\quad\mbox{ while, }  \\
  = rr'G(r,r') 
  &\sim
  -\frac{r\psi_n(r) \, r'\psi_n^*(r')}{E-E_n+i\epsilon} 
  \sepcomma
\end{align*}
which means 
\begin{align*}
  &
  r\psi_n(r) = \left[ 
    \frac{C_F \alpha_s m}{8\pi} \frac{(-1)^{n}}{(n-1)!} 
    \frac{1}{(1-2d_-)(n-d_-)^{4-2d_-}} 
    \frac{\Gamma(2d_-)}{\Gamma(2-2d_-)\Gamma(2d_--n)} \right]^{1/2} 
  g_+(z)  \sepcomma\\
  &
  g_+(z) = z^{d_+} \expo^{-z/(2\rho)} F(1-n,2d_+;z/\rho)
  \sepcomma\quad \rho = n - d_- \sepcomma\quad d_+ = 1-d_-  \sepperiod
\end{align*}
Here we used $F(\alpha,\gamma;\zeta) = \expo^{-\zeta} 
F(\gamma-\alpha,\gamma;-\zeta)$.  
For Coulomb, using 
$\Gamma(\epsilon-n) = \frac{1}{\epsilon} \frac{(-1)^n}{n!}$, we have 
\begin{align*}
  r\psi_n(r) &= r\psi_n(0) \expo^{-z/(2n)} F(1-n,2;z/n)  \\
  &= r\psi_n(0) \left[ 1 - \frac{z}{2} + \frac{z^2}{12} \left(
      1+\frac{1}{2n^2} \right) + \Order{z^3} \right]
  \sepcomma
\end{align*}
where 
\begin{align*}
  \left| \psi_n(0) \right|^2 = \frac{(C_F \alpha_s m/2)^3}{\pi} \frac{1}{n^3}
  \sepperiod
\end{align*}
These are for $E<0$.  
For $E>0$, one obtains 
\begin{align*}
  \Im G(r,r') 
  &=
  \frac{m^2u}{4\pi} |\psi_u(0)|^2 \left[ 
    \Order{\kappa^0} + \Order{\kappa^1} + \Order{\kappa^2} \right]  
  \sepcomma\quad\mbox{ where }\\
  \Order{\kappa^0}
  &=
  1 - \frac{1}{2}C_F a_s m (r+r') + \Order{r^2}  \sepcomma\\
  \Order{\kappa^1}
  &=
  2\kappa \left[
    - \frac{1}{2} \left\{ 
      \ln\pfrac{ur}{r_0^{\rm (ref)}} + \ln\pfrac{ur'}{r_0^{\rm (ref)}}
    \right\}
    + f\!\pfrac{z_u}{2\pi} + \Order{r} 
  \right]
  \sepcomma
\end{align*}
where 
\begin{align*}
  f(x) = \bRe{\psi(ix)+\Egamma} 
  = \sum_{n=1}^\infty \zeta(2n+1) (-x^2)^n
  \sepcomma\quad x \in \mathbb{R}
  \sepcomma
\end{align*}
and $r_0^{\rm (ref)}$ is given in \EqRef{eq:r0ref}.  
Note $\Im G_0(0,0) = m^2 u/(4\pi)$.  
The factor $|\psi_u(0)|^2$ 
\begin{align*}
  |\psi_u(0)|^2 \equiv 
  \frac{z_u}{1-\expo^{-z_u}} 
  = 1 + \frac{z_u}{2} + \frac{z_u^2}{12} + \Order{z_u^4}
  \sepcomma\quad
  z_u \equiv \frac{\pi C_F \alpha_s}{u}
\end{align*}
is called Sommerfeld-Sakharov factor.  
One can see that for both $E<0$ and $E>0$ case, 
\begin{align*}
  \Im G_\rmC(r,r') \propto 1 - \frac{1}{2}C_F a_s m (r+r') 
  + \Order{r^2 \mbox{ etc.}}  \sepcomma
\end{align*}
irrespective of $E$.  

The Sommerfeld-Sakharov factor is derived as follows. 
Consider $\Im G(0,0)$ for Coulomb.  
By expanding to Taylor series, one have 
\begin{align*}
  &  g_<(z') = z'\left(1-\frac{1}{2}z'\right) + \Order{{z'}^3,\kappa}  
  \sepcomma\\
  &  g_>(z) = 1-z\ln z+\left[ 1-2\Egamma+\frac{1}{2\rho}+\ln\rho-\psi(-\rho)
      \right]z + \Order{z^2,\kappa}
  \sepcomma
\end{align*}
where $\psi(z) \equiv \Gamma'(z)/\Gamma(z)$ is digamma function.  
Thus
\begin{align*}
  G(r,r') = \frac{C_F \alpha_s m^2}{4\pi} \left\{ \frac{1}{z} - \ln z
    + \left[ -\frac{1}{2}\frac{z'}{z} + 1 - 2\Egamma + \ln\rho 
      + \frac{1}{2\rho} - \psi(-\rho) \right] +\Order{z} \right\}
  \sepcomma
\end{align*}
or 
\begin{align*}
  \Im G(0,0) = \frac{C_F \alpha_s m^2}{4\pi} \bIm{
    \ln\rho + \frac{1}{2\rho} - \psi(-\rho)}
  \sepperiod
\end{align*}
This can be reduced more.  Let us concentrate for $E>0$.  
Using 
\begin{align*}
  \rho^2 = - \frac{(C_F \alpha_s)^2}{4} \frac{m}{E+i\epsilon}
  \sepcomma\quad\mbox{ or } \quad
  \rho = \frac{i}{2\pi} \frac{\pi C_F \alpha_s}{u} + \epsilon
  \sepcomma\quad\mbox{ where } \quad
  u = \sqrt{\frac{E}{m}}
\end{align*}
and 
\begin{align*}
  -\frac{1}{2z} - \psi(z) 
  = \Egamma + \frac{\pi}{2} \cot(\pi z) + \sum_{n=1}^\infty
    \zeta(2n+1) \, z^{2n}
  \sepcomma
\end{align*}
one obtains 
\begin{align*}
  \Im\ln\rho = \frac{\pi}{2}
  \sepcomma\quad
  \frac{1}{2\rho} - \psi(-\rho) = i \frac{\pi}{2} 
  \frac{1+\expo^{-z_u}}{1-\expo^{-z_u}} + \mbox{real}
  \sepperiod
\end{align*}
Thus we obtain 
\begin{align*}
  \Im G(0,0) \
  = \frac{C_F \alpha_s m^2}{4} \frac{1}{1-\expo^{-z_u}}
  = \frac{m^2 u}{4\pi} \frac{z_u}{1-\expo^{-z_u}}
  \quad\mbox{ for }\quad E > 0
  \sepperiod
\end{align*}
The second factor in the last expression is the Sommerfeld-Sakharov factor.  
%----------------------------------------------------------------------
\section{Results for Coulomb potential}
\label{sec:G_for_pure_Coulomb}
Here we summarize some results for Coulomb potential $-C_F\alpha_s/r$.  
Noting that the reduced mass $\mu$ is $m/2$, we define 
\begin{align}
  \begin{array}{lll}
    \text{Bohr momentum}  &  \cdots  &  p_\rmB \equiv C_F \alpha_s m/2  
    \sepcomma\\[1ex]
    \text{Bohr radius}    &  \cdots  &  r_\rmB \equiv 1/p_\rmB  
    \sepcomma\\[1ex]
    \text{Bohr energy}    &  \cdots  &  E_\rmB \equiv p_\rmB^2/m  
    \sepperiod
  \end{array}
  \label{eq:pB_def}
\end{align}
Bohr energy may also be called Rydberg energy.  
Typically $p_\rmB \simeq 20\GeV$.  See \FigRef{fig:alphaV_20-75-q}.  
For $E<0$ $(n = 1,2,3, \cdots)$, 
\begin{align*}
  E_n = - E_\rmB \frac{1}{n^2}  \sepcomma\qquad
  |\psi_n(0)|^2 = \frac{p_\rmB^3}{\pi} \frac{1}{n^3}  
  \sepcomma
\end{align*}
and 
\begin{align*}
  G_C(r,r') \simeq - \frac{\psi_n(r)\,\psi_n^*(r')}{E-E_n+i\epsilon} 
  \sepcomma\quad\mbox{ near $E=E_n$,}
\end{align*}
where 
\begin{align*}
  \psi_n(r) &= \psi_n(0) \expo^{-r/(r_\rmB n)} F(1-n,2;2r/(r_\rmB n))  \\
  &= \psi_n(0) \left[ 1 - \frac{r}{r_\rmB} 
      + \frac{1}{3}\pfrac{r}{r_\rmB}^2 \left( 1 + \frac{1}{2n^2} \right) 
      + \Order{r^3} \right]
  \sepperiod
\end{align*}
While for $E>0$, 
\begin{align}
  &  \Im G_C(r,r') = \frac{m^2}{4\pi} \sqrt{\frac{E}{m}} 
  |\psi_E(0)|^2 \left[ 
    1 - \left( \frac{r}{r_\rmB} + \frac{r'}{r_\rmB} \right) 
    + \Order{r^2\mbox{ etc.}} \right]  
  \nonumber\sepcomma\\
  &  |\psi_E(0)|^2 = \frac{z_E}{1-\expo^{-z_E}}  \sepcomma\quad
  z_E = \pi C_F \alpha_s \sqrt{\frac{m}{E}} 
  \label{eq:ImGC_E>0}
  \sepperiod
\end{align}
Thus irrespective of energy $E$, 
\begin{align*}
  &  \Im G_C(r,r') \propto 
      1 - \left( \frac{r}{r_\rmB} + \frac{r'}{r_\rmB} \right) 
        + \Order{r^2\mbox{ etc.}}
\end{align*}
With this relation, we have 
\begin{align}
  \frac{1}{ma_sr}\odiff{}{r}r G_C(r) 
  = \left( \frac{1}{ma_sr}-\frac{C_F}{2} \right) G_C(r) 
    + \Order{r,\frac{1}{c}}  
  \label{eq:diff_for_GC}
\end{align}
where $G_C(r)$ may either be $G_C(r,r')$ or $G_C(r,p)$.  
%
%
%

%
%
%
%-----------------------------------------------------------------------
\section{Expressions needed for matching}
\label{sec:matching-exp}
As was explained in \SecRef{sec:matching}, perturbative expansions 
with respect to $\alpha_s$ and with respect to $\alpha_s/\beta$ are 
both valid when $\alpha_s \ll \beta \ll 1$.  Here we collect both of them.  
These are used to determine the matching coefficients for NRQCD calculations.  
In the following, we use the quantity $u$ that is equal to $\beta$ to 
the leading order.  Relation between them are as follows: 
\begin{align*}
  \frac{u}{c} &\equiv \sqrt{\frac{E}{\mt c^2}} 
   = \sqrt{\frac{\sqrt{s}-2\mt}{\mt}}  \\
  &= \sqrt{2(\gamma-1)} 
  = \beta \left( 1 + \frac{3}{8}\beta^2 + \Order{\beta^4} \right)
  \sepcomma
\end{align*}
\begin{align*}
  \frac{4m^2}{s} = \frac{1}{\gamma^2} = 1 - \beta^2
  \sepcomma\quad
  \gamma = \frac{1}{\sqrt{1-\beta^2}} = 1 + \frac{u^2}{2}
  \sepcomma
\end{align*}
\begin{align*}
  \frac{p^2}{m^2} &= \gamma^2\beta^2 = \frac{\beta^2}{1-\beta^2}
  = \beta^2 ( 1 + \beta^2 + \beta^4 + \cdots )  \\
  &= \gamma^2 - 1 = u^2 \left( 1 + \frac{u^2}{4} \right)
  \sepperiod
\end{align*}
%
%-----------------------------------------------------------------------
\subsection{NRQCD calculation of $R$ ratio}
\label{sec:ImG_exp}
When $\alpha_s \ll \beta \ll 1$, 
the Green function $G$ defined in \EqRef{eq:G_and_H} can be expanded 
perturbatively: 
\begin{align*}
  \Im G(r,r') = \sum_{i=0}^{10} \Im G_i(r,r')  \sepcomma
\end{align*}
where with 
\begin{align*}
  G_0 \equiv \frac{1}{\dfrac{p^2}{m}-E-i\epsilon}
  \sepcomma
\end{align*}
\begin{align*}
  &  \Im G_0(r_0,r_0)
   = \Im \bra{r_0} G_0 \ket{r_0}
  \\
  &\quad
   = \frac{m^2 u}{4\pi} + \Order{r_0^{~2}}  \sepcomma\\
  &  \Im G_1(r_0,r_0)
   = \Im \bra{r_0} G_0 \frac{C_F a_s}{r} 
     \left[ 1 + \frac{a_s}{4\pi c} \left\{
         2\beta_0\ln(\mu'r) + a_1 \right\} \right]
     G_0 \ket{r_0}  \\
  &\quad
   = \frac{m^2 u}{4\pi} \cdot
     \frac{C_F a_s}{u} \frac{\pi}{2} \left[
       1 - \frac{2}{\pi}m r_0 u
       + \frac{a_s}{c}\frac{1}{4\pi} \left\{
         a_1 -2\beta_0\ln\pfrac{2mu}{\mu} \right\} \right]
     + \Order{r_0}  \sepcomma\\
  &  \Im G_2(r_0,r_0)
   = \Im \bra{r_0} G_0 \frac{p^4}{4m^3c^2} G_0 \ket{r_0}  \\
  &\quad
   = \frac{m^2 u}{4\pi} \cdot
     \pfrac{u}{c}^2\frac{5}{8}
     + \Order{r_0}  \sepcomma\\
  &  \Im G_3(r_0,r_0)
   = \Im \bra{r_0} G_0
     \left[ \frac{-11\pi C_F a_s}{3m^2c^2} \delta^{(3)}(\rBI) \right]
     G_0 \ket{r_0}  \\
  &\quad
   = \frac{m^2 u}{4\pi} \cdot
     \frac{C_F a_s}{c} \frac{-11}{6m r_0 c}
     + \Order{r_0}  \sepcomma\\
  &  \Im G_4(r_0,r_0)
   = \Im \bra{r_0} G_0 \frac{C_F a_s}{2m^2c^2} 
       \left\{ p^2 \, , \, \frac{1}{r} \right\} G_0 \ket{r_0}  \\
  &\quad
   = \frac{m^2 u}{4\pi} \cdot
     \frac{C_F a_s}{c} 
     \left( \frac{1}{m r_0 c} + \frac{u}{c}\frac{\pi}{2} \right)
     + \Order{r_0}  \sepcomma\\
  &  \Im G_5(r_0,r_0)
   = \Im \bra{r_0} G_0 \frac{C_A C_F a_s^2}{2mr^2c^2} G_0 \ket{r_0}  \\
  &\quad
   = \frac{m^2 u}{4\pi} \cdot
     \frac{C_A C_F a_s^2}{c^2} 
     \left\{ -\ln\pfrac{2mur_0}{\expo^{2-\Egamma}} \right\}
     + \Order{r_0}  \sepcomma\\
  &  \Im G_6(r_0,r_0)
   = \Im \bra{r_0} G_0 \frac{C_F a_s}{r} G_0 \frac{C_F a_s}{r} G_0 \ket{r_0} \\
  &\quad
   = \frac{m^2 u}{4\pi} \cdot
     \frac{(C_F a_s)^2}{u^2} \frac{\pi^2}{12}
     + \Order{r_0}  \sepcomma\\
  &  \Im G_7(r_0,r_0)
   = \Im \bra{r_0} G_0 \frac{C_F a_s}{r} G_0 \frac{C_F a_s}{r} G_0 
     \frac{p^4}{4m^3c^2} G_0 \ket{r_0} + \text{ 2 permutations} \\
  &\quad
   = \frac{m^2 u}{4\pi} \cdot
     \frac{(C_F a_s)^2}{cu} \left[
         \frac{\pi}{4mr_0 c} 
       + \frac{u}{c} \left\{
             \frac{11}{96}\pi^2 - \frac{1}{2}
           - \frac{1}{2}\ln\pfrac{2mur_0}{\expo^{2-\Egamma}} \right\} \right]
     + \Order{r_0}  \sepcomma\\
  &  \Im G_8(r_0,r_0)
   = \Im \bra{r_0} G_0 \frac{C_F a_s}{r} G_0 
     \left[ \frac{-11\pi C_F a_s}{3m^2c^2} \delta^{(3)}(\rBI) \right] 
     G_0 \ket{r_0} 
     + \left[ \frac{1}{r} \leftrightarrow \delta^{(3)}(\rBI) \right] \\
  &\quad
   = \frac{m^2 u}{4\pi} \cdot
     \frac{(C_F a_s)^2}{cu} \left[
         \frac{-11\pi}{12mr_0 c} 
       + \frac{u}{c} \left\{
             \frac{11}{4}
           + \frac{11}{6}\ln\pfrac{2mur_0}{\expo^{2-\Egamma}} \right\} \right]
     + \Order{r_0}  \sepcomma\\
  &  \Im G_9(r_0,r_0)
   = \Im \bra{r_0} G_0 \frac{C_F a_s}{r} G_0 \frac{C_F a_s}{2m^2c^2} 
     \left\{ p^2 \, , \, \frac{1}{r} \right\} G_0 \ket{r_0} 
     + \left[ \frac{1}{r} \leftrightarrow 
         \left\{ p^2 \, , \, \frac{1}{r} \right\} \right]  \\
  &\quad
   = \frac{m^2 u}{4\pi} \cdot
     \frac{(C_F a_s)^2}{cu} \left[
         \frac{\pi}{2mr_0 c} 
       + \frac{u}{c} \left\{
             \frac{\pi^2}{6} - 1
           - 2\ln\pfrac{2mur_0}{\expo^{2-\Egamma}} \right\} \right]
     + \Order{r_0}  \sepcomma\\
  &  \Im G_{10}(r_0,r_0)
   = \Im \bra{r_0} G_0 \frac{C_F a_s}{r} G_0 
     \frac{p^4}{4m^3c^2} G_0 \ket{r_0} 
     + \left[ \frac{1}{r} \leftrightarrow p^4 \right]  \\
  &\quad
   = \frac{m^2 u}{4\pi} \cdot
     \frac{C_F a_s}{c} \left[
         \frac{1}{2mr_0 c} 
       + \frac{u}{c}\frac{\pi}{2} \right]
     + \Order{r_0}  \sepperiod
\end{align*}
Thus we have 
\begin{align}
  &  \frac{4\pi}{m^2c} \Im G(r_0, r_0) \,/ \pfrac{u}{c} \nonumber\\
  &=
  \left\{ 1+ \frac{a_s}{u} C_F \frac{\pi}{2} 
    - C_F a_s m r_0 + \pfrac{a_s}{u}^2 C_F^2 \frac{\pi^2}{12} \right\}
  \nonumber\\&\quad{}
  + \frac{a_s}{c} \left\{ \frac{a_s}{u} C_F 
      \frac{1}{8} \left[ a_1 - 2\beta_0\ln\pfrac{2mu}{\mu} \right] \right\}
  \nonumber\\&\quad{}
  + \pfrac{a_s}{c}^2 \biggl\{ \pfrac{u}{a_s}^2 \frac{5}{8}
      + \frac{u}{a_s} C_F \pi - C_F \left(C_A + \frac{2}{3}C_F \right)
    \ln\pfrac{2mur_0}{\expo^{2-\Egamma}}
  \nonumber\\&\qquad\qquad\qquad{}
      + C_F^2\left(\frac{9}{32}\pi^2 + \frac{5}{4} \right)
      + C_F\frac{1}{a_s mr_0}\pfrac{-1}{3}
      + \frac{a_s}{u} C_F^2 \frac{\pi}{a_smr^0}\pfrac{-1}{6} \biggr\}
  \nonumber\\&\quad{}
  + \Order{\frac{1}{c^3}}
  \nonumber\\
  &=
  \left\{ 1 + \pfrac{u}{c}^2\frac{5}{8} \right\}
  + C_F\frac{a_s}{c} \left\{
        \frac{c}{u}\frac{\pi}{2} \left[ 1-\frac{2}{\pi}mr_0u\right]
      - \frac{1}{3mr_0c} + \frac{u}{c}\pi \right\}
  \nonumber\\&\quad{}
  + C_F\pfrac{a_s}{c}^2 \biggl\{ \pfrac{c}{u}^2 C_F\frac{\pi^2}{12}
      + \frac{c}{u}\frac{1}{8} \left[ a_1 -2\beta_0\ln\pfrac{2mu}{\mu}\right]
      + \frac{c}{u}C_F\frac{\pi}{mr_0c}\pfrac{-1}{6}
  \nonumber\\&\qquad\qquad\qquad{}
      - \left( C_A+\frac{2}{3}C_F \right)\ln\pfrac{2mur_0}{\expo^{2-\Egamma}}
      + C_F\left(\frac{9}{32}\pi^2+\frac{5}{4}\right) \biggr\}
  \nonumber\\&\quad{}
  + \Order{a_s^3}
  \label{eq:ImG_pert_NRQCD}
  \sepperiod
\end{align}
The two expressions above are exactly the same each other.  
The $R$ ratio and $\Im G$ is related in \EqRef{eq:R_with_ImG_and_C}.  
By comparing the result in this section and that in the next section, 
one can determine the matching coefficients $C_1^{\rm (cur)}$ and 
$C_2^{\rm (cur)}$.  
For this purpose, it is convenient to rewrite $u$ with $\beta$.  
To each order in $\alpha_s$, their relations are 
\begin{gather*}
  u \left( 1 + \frac{5}{8}u^2 \right)
    = \beta ( 1 + \beta^2 + \Order{\beta^4} )  \sepcomma\\
  u \left( \frac{\pi}{2u}+\pi u \right)
    = \beta \left( \frac{\pi}{2\beta} + \pi\beta + \Order{\beta^3} \right)  
  \sepcomma\\
  u \left( \frac{\pi^2}{12u^2} + \frac{9}{32}\pi^2 \right)
    = \beta \left( \frac{\pi}{12\beta^2} + \frac{\pi^2}{4} + \Order{\beta^2} 
      \right)  \sepperiod
\end{gather*}
One can just substitute $\beta$ for $u$ in other cases, since there 
difference is $\Order{\beta^3}$ and higher.  
%-----------------------------------------------------------------------
\subsection{Perturbative QCD calculation of $R$ ratio}
Perturbative calculation of $R$ ratio 
for $e^+ e^- \to \gamma^* \to \qB q$ was done 
in~\cite{CM98}.  Here we summarize the result for reference: 
\begin{align*}
  R = \frac{3}{2} N_C Q_q^2 \, \beta \left( 1-\frac{\beta^2}{3} \right) \left[
    1 + C_F \pfrac{\alpha_s(\mu_h)}{\pi} \Delta^{(1)}
      + C_F \pfrac{\alpha_s(\mu_h)}{\pi}^2 \Delta^{(2)} 
      + \Order{\alpha_s^3} \right]
\end{align*}
where $m$ is mass of $q$, and 
\begin{align*}
  &  \Delta^{(1)} = \frac{\pi^2}{2\beta} - 4 + \frac{\pi^2}{2}\beta 
  + \Order{\beta^2}
  \sepcomma\\
  &  \Delta^{(2)} = C_F \Delta^{(2)}_A + C_A \Delta^{(2)}_{NA} 
  + T_F n_f \Delta^{(2)}_L + T_F n_H \Delta^{(2)}_H 
  + \Delta^{(2)}_{\ln 2\beta} + \Delta^{(2)}_{\mu_h} + \Order{\beta}
  \sepcomma
\end{align*}
where 
\begin{align*}
  &  \Delta^{(2)}_A 
  = \frac{\pi^4}{12\beta^2} - 2 \frac{\pi^2}{\beta} + \frac{\pi^4}{6}
    + \pi^2 \left( -\frac{35}{18} - \frac{2}{3}\ln\beta 
        + \frac{4}{3} \ln 2 \right) + \frac{39}{4} - \zeta_3
  \sepcomma\\
  &  \Delta^{(2)}_{NA} 
  = - \frac{31\pi^2}{72\beta} + \pi^2 \left( \frac{179}{72}
        - \ln\beta - \frac{8}{3} \ln 2 \right) - \frac{151}{36} 
    - \frac{13}{2}\zeta_3
  \sepcomma\\
  &  \Delta^{(2)}_L
  = - \frac{5\pi^2}{18\beta} + \frac{11}{9}
  \sepcomma\\
  &  \Delta^{(2)}_H 
  = \frac{44}{9} - \frac{4\pi^2}{9}
  \sepcomma\\
  &  \Delta^{(2)}_{\ln 2\beta}
  = - \frac{\pi^2}{4\beta}\beta_0\ln 2\beta
  \sepcomma\\
  &  \Delta^{(2)}_{\mu_h}
  = \left( - \frac{\pi^2}{4\beta} + 2 \right) \beta_0 \ln\frac{m}{\mu_h}
  = - \Delta^{(1)} \frac{\beta_0}{4} \ln\frac{m^2}{\mu_h^2}
  \sepperiod
\end{align*}
%
%
%
%

%
%
%-------------------------------------------------------------------------
\section{Top quark polarization}
\label{sec:top-pol-expl}
Neglecting $\mathcal{O}(\beta_t)$ and higher, 
the production cross section for $t\tB$ pair 
can be written as follows: 
\begin{align}
  \frac{d\sigma}{d^3\pBI_t}
  =
  \frac{N_C \alpha^2 \Gamma_t}{2\pi m_t^4}
  \frac{1-\Pol_{e^+}\Pol_{e^-}}{2}
  \times
  |G|^2 (a_1+\chi a_2)
\end{align}
Here we sum over the spins of $t$ and $\tB$.  
\begin{align}
%  D = F - G
  D = G - F
\end{align}
\begin{align}
  |G|^2 (a_1+\chi a_2)
  &\to{}
  \frac{1}{4} \Biggl( 
    [\mbox{no $\stBI$, $\stBBI$ dependence}]
  + [\mbox{$\stBI$ dependent}]
  \nonumber\\
  &\qquad{}
  + [\mbox{$\stBBI$ dependent}]
  + [\mbox{$\stBI$, $\stBBI$ dependent}]
  \Biggr)
\end{align}
\begin{align}
  & [\mbox{no $\stBI$, $\stBBI$ dependence}] \nonumber\\
  &={}
    |G|^2 (a_1+\chi a_2)
  + 2\bRe{G^*F(a_3+\chi a_4)} \frac{\dprod{\peBI}{\ptBI}}{m_t^2}
  \nonumber\\
  &={}
    |G|^2 (a_1+\chi a_2) \left\{
      1 + 2\pRe{\CFB\frac{F}{G}}\beta_t \cos\theta_{te}
    \right\}
\end{align}
\begin{align}
  & [\mbox{$\stBI$ dependent}]
  \nonumber\\
  =&{}
  - |G|^2 (a_2+\chi a_1) \frac{\dprod{\stBI}{\peBI}}{m_t}
  \nonumber\\
  &{}
  - \bRe{G^*F(a_4+\chi a_3)} \left[
      \frac{\dprod{\stBI}{\ptBI}}{m_t}
    + \frac{\dprod{\stBI}{\peBI}}{m_t} \frac{\dprod{\peBI}{\ptBI}}{m_t^2}
  \right]
  \nonumber\\
  &{}
  + \bIm{G^*F(a_3+\chi a_4)}
  \frac{\dprod{\stBI}{\pcprod{\peBI}{\ptBI}}}{m_t^2}
  \nonumber\\
  &{}
  + \left[
%      \Im(d_{tg} G^*D) (a_1+\chi a_2)
    {} - \Im(d_{tg} G^*D) (a_1+\chi a_2)
    + \bIm{G^*F(a_5+\chi a_6)}
  \right]
  \left[
      \frac{\dprod{\stBI}{\ptBI}}{m_t}
    - \frac{\dprod{\stBI}{\peBI}}{m_t} \frac{\dprod{\peBI}{\ptBI}}{m_t^2}
  \right]
  \nonumber\\
  &{}
  + \left[ 
%      \Re(d_{tg} G^*D) (a_2+\chi a_1)
    {} - \Re(d_{tg} G^*D) (a_2+\chi a_1)
    + \bRe{G^*F(a_6+\chi a_5)}
  \right] \frac{\dprod{\stBI}{\pcprod{\peBI}{\ptBI}}}{m_t^2}
  \nonumber\\
  =&\,
  |G|^2 (a_1+\chi a_2) \Biggl\{
    \Cpara^0 \spara
  + 2\pRe{\Cperp \frac{F}{G}}\spara\beta_t \cos\theta_{te}
  \nonumber\\
  &{}
  + \pRe{\Cperp \frac{F}{G}}
    \sperp\beta_t \sin\theta_{te}
%  \nonumber\\
%  &{}
  + \pIm{\Cnorm \frac{F}{G}}
    \snorm\beta_t \sin\theta_{te}
  \nonumber\\
  &{}
  + \left[
%      \pIm{d_{tg} \frac{D}{G}}
      \Bperp^g \pIm{d_{tg} \frac{D}{G}}
    + \pIm{\Bperp \frac{F}{G}}
  \right]
  \sperp\beta_t \sin\theta_{te}
  \nonumber\\
  &{}
  + \left[ 
      \Bnorm^g \pRe{d_{tg} \frac{D}{G}}
    + \pRe{\Bnorm \frac{F}{G}}
  \right] \snorm\beta_t \sin\theta_{te}
  \Biggr\}
\end{align}
\begin{align}
%  &
  [\mbox{$\stBBI$ dependent}] 
%  \nonumber\\
  &=
  \Bigl( \stBI \to \stBBI, \dtg \to -\dtg, 
    \dtp \to -\dtp, \dtZ \to -\dtZ
    ~\mbox{at}~[\mbox{$\stBI$ dependent}]
  \Bigr)
\end{align}
Note that $\dtp \to -\dtp$ and $\dtZ \to -\dtZ$ mean 
$\Bperp^{\gamma/Z} \to -\Bperp^{\gamma/Z}$ and 
$\Bnorm^{\gamma/Z} \to -\Bnorm^{\gamma/Z}$.  
\begin{align}
  & [\mbox{$\stBI$, $\stBBI$ dependent}]
  \nonumber\\
  =&\,
    |G|^2 (a_1+\chi a_2)
      \frac{\dprod{\stBI}{\peBI}}{m_t} \frac{\dprod{\stBBI}{\peBI}}{m_t}
  \nonumber\\
  &{}
  + \bRe{G^*F(a_3+\chi a_4)} \left[
      \frac{\dprod{\stBI}{\peBI}}{m_t} \frac{\dprod{\stBBI}{\ptBI}}{m_t}
    + \frac{\dprod{\stBBI}{\peBI}}{m_t} \frac{\dprod{\stBI}{\ptBI}}{m_t}
  \right]
  \nonumber\\
  &{}
  - \bIm{G^*F(a_4+\chi a_3)} \left[
      \frac{\dprod{\stBI}{\peBI}}{m_t} 
      \frac{\dprod{\stBBI}{\pcprod{\peBI}{\ptBI}}}{m_t^2}
    + \frac{\dprod{\stBBI}{\peBI}}{m_t} 
      \frac{\dprod{\stBI}{\pcprod{\peBI}{\ptBI}}}{m_t^2}
  \right]
  \nonumber\\
  &{}
  + \left[
%        \Im(d_{tg} G^*D) (a_2+\chi a_1)
      {}-  \Im(d_{tg} G^*D) (a_2+\chi a_1)
      + \bIm{G^*F(a_6+\chi a_5)}
    \right]
  \nonumber\\
  &{}\qquad\times
    \left[
        \frac{\dprod{\stBI}{\peBI}}{m_t} \frac{\dprod{\stBBI}{\ptBI}}{m_t}
      - \frac{\dprod{\stBBI}{\peBI}}{m_t} \frac{\dprod{\stBI}{\ptBI}}{m_t}
    \right]
  \nonumber\\
  &{}
  + \left[
%      \Re(d_{tg} G^*D) (a_1+\chi a_2)
    {}-  \Re(d_{tg} G^*D) (a_1+\chi a_2)
    + \bRe{G^*F(a_5+\chi a_6)}
    \right]
  \nonumber\\
  &{}\qquad\times
    \left[
        \frac{\dprod{\stBI}{\peBI}}{m_t} 
        \frac{\dprod{\stBBI}{\pcprod{\peBI}{\ptBI}}}{m_t^2}
      - \frac{\dprod{\stBBI}{\peBI}}{m_t} 
        \frac{\dprod{\stBI}{\pcprod{\peBI}{\ptBI}}}{m_t^2}
    \right]
  \nonumber\\
  =&\,
  |G|^2 (a_1+\chi a_2) \Biggl\{
    \spara\sBpara
  + 2\pRe{\Cnorm \frac{F}{G}} \spara\sBpara\beta_t \cos\theta_{te}
  \nonumber\\
  &{}
  + \pRe{\Cnorm \frac{F}{G}}
    (\spara\sBperp+\sBpara\sperp)\beta_t \sin\theta_{te}
  \nonumber\\
  &{}
  + \pIm{\Cperp \frac{F}{G}}
    (\spara\sBnorm+\sBpara\snorm)\beta_t \sin\theta_{te}
  \nonumber\\
  &{}
  + \left[
        \Bnorm^g \pIm{d_{tg} \frac{D}{G}}
      + \pIm{\Bnorm \frac{F}{G}}
    \right]
    (\spara\sBperp-\sperp\sBpara)\beta_t \sin\theta_{te}
  \nonumber\\
  &{}
  + \left[
%      \pRe{d_{tg} \frac{D}{G}}
      \Bperp^g \pRe{d_{tg} \frac{D}{G}}
    + \pRe{\Bperp \frac{F}{G}}
    \right]
    (\spara\sBnorm-\snorm\sBpara)\beta_t \sin\theta_{te}
\end{align}
In general, 
\begin{align}
    \pdprod{\stBI}{\peBI}\pdprod{\stBBI}{\ptBI}
  - \pdprod{\stBBI}{\peBI}\pdprod{\stBI}{\ptBI}
  =
  \dprod{\pcprod{\stBI}{\stBBI}}{\pcprod{\peBI}{\ptBI}}
\end{align}
%
%for arbitrary vectors.  
By replacing $\ptBI$ with $\pcprod{\peBI}{\ptBI}$ we have
\begin{align}
  &
    \pdprod{\stBI}{\peBI}\left[\dprod{\stBBI}{\pcprod{\peBI}{\ptBI}}\right]
  - \pdprod{\stBBI}{\peBI}\left[\dprod{\stBI}{\pcprod{\peBI}{\ptBI}}\right]
  \nonumber\\
  &\qquad{}
  + \dprod{\pcprod{\stBI}{\stBBI}}%
    {\left[\ptBI|\peBI|^2-\peBI\pdprod{\peBI}{\ptBI}\right]}
  \nonumber\\
  &=
  (p_e)_i (p_e)_j (p_t)_k (s_t)_l (\bar{s}_{t})_m
  \cdot \frac{1}{4!}\delta_{i[j}\epsilon_{klm]}
  = 0
\end{align}
Thus one can see that the anomalous contributions to 
the [\mbox{$\stBI$, $\stBBI$ dependent}]-part is proportional to 
$\cprod{\sBI_t}{\sBBI_t}$.  
\clearemptydoublepage
%
%
%
%-----------------------------------------------------------------------

%-----------------------------------------------------------------------
%
%
%

%
%\input{misc_memo}
\vspace*{\fill}
\hspace*{\fill}
%{\tiny update: \weekday{\year}{\month}{\day}, \today~\Daytime~+0900}
{\tiny update: 2 Apr 2000}
\end{document}